\documentclass[12pt, a4paper, twoside, openright]{report}
\usepackage{amsmath}
\usepackage{amssymb}
\usepackage{epsfig}
\usepackage{calc}
\usepackage[ansinew]{inputenc}
\usepackage{color}
\usepackage{fancyhdr}
\usepackage{graphicx}

\def\tilde{\widetilde}
\def\zf#1#2{z^{#1}_{{\rm fixed},#2}}
\newcommand{\shortexactseq}[5]{0 \longrightarrow #1 \stackrel{#2}{\longrightarrow} #3 \stackrel{#4}{\longrightarrow} #5 \longrightarrow 0}
\DeclareMathOperator{\ch}{c}
\def\Dt{\widetilde{D}}
\def\Et{\widetilde{E}}

\def\h {{1\over 2}}
\def\Dt{\widetilde{D}}
\def\Et{\widetilde{E}}

\def\ov {\overline}
\def\IC{\mathbb{C}}
\def\IF{\mathbb{F}}
\def\IP{\mathbb{P}}
\def\IR{\mathbb{R}}
\def\IZ{\mathbb{Z}}
\def\hat{\widehat}

\def\br{\hfill\break}

\def\det {{\rm det}}
\def\mod {{\rm mod}}
\def\lf {\left}
\def\ri {\right}
\def\ra {\rightarrow}

\def\re {{\rm Re}}
\def\im {{\rm Im}}
\def\p {\partial}

\def\Fc {{\cal F}} 
 \def\Oc {{\cal O}}
 \def\Sc {{\cal S}}
\def\Mc {{\cal M}}

\def\Ic {{\cal I}}
\def\Kc {{\cal K}}
\def\Tc{{\cal T}}
\def\Uc{{\cal U}}

\def\tb{type $IIB$\ }\def\ta{type $IIA$\ }
\def\ap{\alpha'}

\def\Bl#1{{\rm Bl}_{#1}}  
\def\zf#1#2{z^{#1}_{{\rm fixed},#2}}

\def\ie{{\it i.e.\ }}
\def\eg{{\it e.g.\ }}

\def\th{\theta}

\def\Om{\Omega}

\def\om{\omega}

\def\Om{\Omega}

\def\ov {\overline}
\def\o {\over}
\def\fc#1#2{{#1 \o #2}}
\def\half{\frac{1}{2}}
\newcommand{\ibar}{\bar{\imath}}
\newcommand{\jbar}{\bar{\jmath}}
\newcommand{\kbar}{\bar{k}}

\begin{document}

\setcounter{footnote}{0}
\setcounter{page}{0} 
\pagenumbering{roman}

\begin{titlepage}
\vskip-2cm
\rightline{MPP-2006-112}
\rightline{LMU-ASC 58/06}
\rightline{hep-th/0609040}


\begin{center}{\bf \large TOROIDAL ORBIFOLDS:\\ \vspace{2mm}
RESOLUTIONS, ORIENTIFOLDS\\ \vspace{2mm} 
AND APPLICATIONS IN STRING \\ \vspace{4mm} 
PHENOMENOLOGY}

\vskip1.15cm
\begin{figure}[h!]
\begin{center}
\includegraphics[width=30mm]{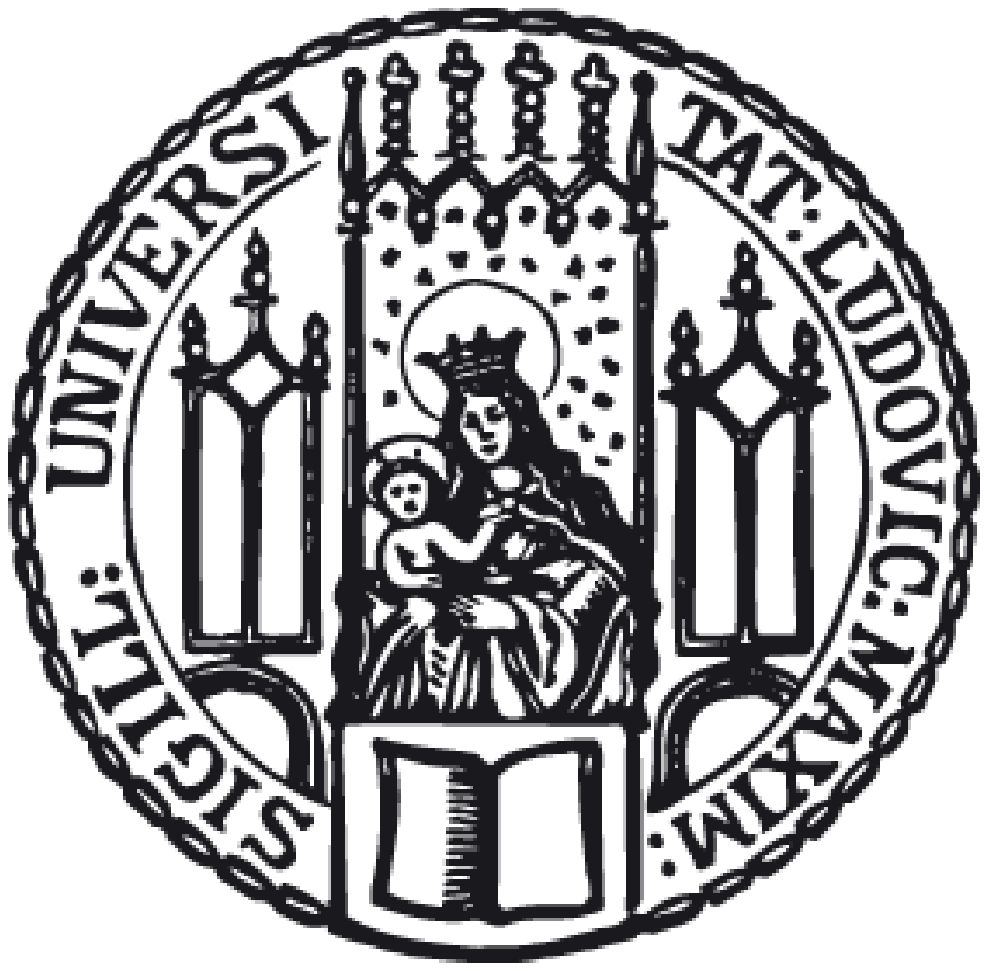}
\end{center}
\end{figure}

\vskip 0.1cm 
{ \large Dissertation\\[6pt]
 an der Fakult\"at f\"ur Physik der\\[6pt]
Ludwig-Maximilians-Universit\"at\\[6pt] 
M\"unchen}

\vskip 2.3cm 
{\large vorgelegt von\\[5pt]
 SUSANNE REFFERT\footnote{Present address: Institute for Theoretical Physics, University of Amsterdam, Valckenierstraat 65, 1018 XE Amsterdam, The Netherlands}\\[5pt]
 aus Z\"urich}

\vskip 2.5cm
{\large M\"unchen, Mai 2006}
\end{center}

\end{titlepage}
\newpage

\ \\

\vspace{19.6cm}
{\large \noindent{\bf 1. Gutachter:} Prof. Dr. Dieter L\"ust, LMU M\"unchen}\\[3pt]
{\large\noindent{\bf 2. Gutachter:} Prof. Dr. Wolfgang Lerche, CERN}\\[3pt]
{\large\noindent {Tag der m\"undlichen Pr\"ufung:} 21.7.2006}

\newpage

\ \
\vspace{7cm}
\begin{center}
{\bf \Large To my parents}\\
\end{center}

\cleardoublepage

\begin{center}
\
\vskip 0.7cm
{\bf \LARGE Summary}\\
\end{center}
\vspace{5mm}
\noindent
As of now, string theory is the best candidate for a theory of quantum gravity. Since it is anomaly--free only in ten space-time dimensions, the six surplus spatial dimensions must be compactified.

This thesis is concerned with the geometry of toroidal orbifolds and their applications in string theory. An orbifold is the quotient of a smooth manifold by a discrete group. In the present thesis, we restrict ourselves to orbifolds of the form $T^6/\IZ_N$ or $T^6/\IZ_N\times \IZ_M$. These so--called toroidal orbifolds are particularly popular as compactification manifolds in string theory. They present a good compromise between a trivial compactification manifold, such as the $T^6$ and one which is so complicated that explicit calculations are nearly impossible, which unfortunately is the case for many if not most Calabi--Yau manifolds. At the fixed points of the discrete group which is divided out, the orbifold develops quotient singularities. By resolving these singularities via blow--ups, one arrives at a smooth Calabi--Yau manifold. The systematic method to do so is explained in detail. Also the transition to the Orientifold quotient is explained. 

In string theory, toroidal orbifolds are popular because they combine the advantages of calculability and of incorporating many features of the standard model, such as non-Abelian gauge groups, chiral fermions and family repetition.

In the second part of this thesis, applications in string phenomenology are discussed. The applications belong to the framework of compactifications with fluxes in type $IIB$ string theory. Flux compactifications on the one hand provide a mechanism for supersymmetry breaking. One the other hand, they generically stabilize at least part of the geometric moduli. The geometric moduli, i.e. the deformation parameters of the compactification manifold correspond to massless scalar fields in the low energy effective theory. Since such massless fields are in conflict with experiment, mechanisms which generate a potential for them and like this fix the moduli to specific values must be investigated. After some preliminaries, two main examples are discussed. The first belongs to the category of model building, where concrete models with realistic properties are investigated. A brane model compactified on $T^6/\IZ_2\times\IZ_2$ is discussed. The flux-induced soft supersymmetry breaking parameters are worked out explicitly. The second example belongs to the subject of moduli stabilization along the lines of the proposal of Kachru, Kallosh, Linde and Trivedi (KKLT).  Here, in addition to the background fluxes, non-perturbative effects serve to stabilize all moduli. In a second step, a meta-stable vacuum with a small positive cosmological constant is achieved. Orientifold models which result from resolutions of toroidal orbifolds are discussed as possible candidate models for an explicit realization of the KKLT proposal.

The appendix collects the technical details for all commonly used toroidal orbifolds and constitutes 
a reference book for these models.

\cleardoublepage

\newpage\noindent 
\
\vskip 0.7cm
\begin{center}{\bf\large Acknowledgments}
\end{center}
\vskip 0.7cm

It is a pleasure to express my profound gratitude and appreciation to my thesis advisor Dieter L\"ust: For his continued support and encouragement, for letting me be a part of his wonderful work group, and especially for sharing his ideas and views of physics. 

I am deeply indebted to Stephan Stieberger, for countless hours of explanations, as well as his friendship and support. 

I am happy to extend my special thanks to Emanuel Scheidegger, who has taught me my first steps in algebraic geometry and much, much more. 

I would like to thank all those I had the pleasure to collaborate with: Dieter L\"ust, Peter Mayr, Emanuel Scheidegger, Waldemar Schulgin, Stephan Stieberger and Prasanta Tripathy. Furthermore, I would like to thank everyone who was or is part of the String Theory group at Humboldt University in Berlin and later in Munich, for contributing to the hospitable and lively atmosphere of the group I have enjoyed and have greatly benefitted from. In particular, I would like to thank two of my former office mates: George Kraniotis for sharing his great enthusiasm for physics, and Domenico Orlando, for pleasant company and lively discussions.

Finally, I would like to thank everyone who has taken the trouble to teach me or to share their insights, if only once and just for ten minutes. Sometimes small things make a big difference.

Outside of physics, I would like to thank my family and Bernhard for their support, and my landlords for taking care of the animals when I was traveling.

\cleardoublepage

\tableofcontents

\cleardoublepage
\newpage
\pagenumbering{arabic}

\chapter{Introduction and Overview}

String theory is as of now our best candidate for a theory of quantum gravity. The simple idea of taking a one-dimensional object instead of a point particle to be the fundamental building block of nature spawns a variety of consequences. Many of them were quite unexpected, such as the realization that the basic requirement that the theory be anomaly free leads to spacetime having ten dimensions. One of the main tasks of the string theorist consists therefore in reconciling these ten dimensions of string theory with the four dimensional world we live in. 

One possible solution to this puzzle is compactification: The six extra dimensions are curled up so small that none of the experiments conducted by humankind so far has been able to detect them or an effect related to their existence. 

The idea of compactification is most easily demonstrated by the example of a single extended dimension which is compactified to a circle. If only one dimension is compactified, turning it into a circle is the only available possibility. The circle has one parameter which can be deformed continuously without changing the defining properties of the circe: Its radius, or more generally speaking, its size. This is the first example of a modulus we encounter.

In string theory, not one, but six dimensions must be compactified, and it is evident that there are very many possibilities to do so. How to choose one over the other? To answer this question, one must explore how the particulars of the compactification manifold affect the physics of our theory. To put it naively, the "right" compactification manifold is the one which exactly reproduces the Standard Model.

Since we have to start looking for the right manifold somewhere, we restrict ourselves to a regime where we are likely to find what we are looking for. One of the main requirements which is usually imposed on the compactification manifold is that it allows for a supersymmetric theory in four dimensions. There are many reasons why one would want require supersymmetry, some of which date back to the days before the birth of string theory. One of the strongest arguments in favor of supersymmetry is that it solves the hierarchy problem by protecting a small Higgs mass from quantum corrections. The non--renormalization theorems which hold for supersymmetric theories simplify the concrete calculations immensely. Supersymmetry assumes a strong compatibility with the complex numbers. In supersymmetric theories, one can therefore make use of the powerful tools provided by complex analysis.

Requiring supersymmetry translates into certain requirements on the compactification manifold: It must allow for at least one covariantly constant spinor.  This leads to the manifold being complex: It must have an almost complex structure, i.e. a $(1,1)$--tensor $j$ such that $j^2=-1$ which fulfills certain integrability conditions.  Furthermore, it must be K\"ahler, i.e. be a Hermitian manifold whose K\"ahler form $J$ is closed. Requiring exactly one pair of covariantly constant spinors leads to the manifold having $SU(3)$ holonomy group. This is the same as the manifold allowing a Ricci-flat metric or having vanishing first Chern class.  Such a manifold is known under the name of {\it Calabi--Yau} manifold. 

Unfortunately, there are very many Calabi--Yau threefolds (three complex dimensions correspond to six real dimensions). As of today, it is unknown whether there are infinitely many of them or not. Finding the one which yields a theory compatible to our world is part of the big task of the string theorist. 

Unlike the circle, whose only free parameter is its radius, most Calabi--Yau manifolds come with a large number of possible deformations. For a Calabi--Yau manifold, the deformations fall into two types: Those parametrizing the shape of the manifold, or mathematically speaking, its complex structure, and those parametrizing its sizes, i.e. changes in the K\"ahler structure. 

This thesis treats a special class of compactification manifolds, the so-called toroidal orbifolds. 
Toroidal orbifolds are a happy compromise between a compactification manifold which is completely trivial, such as the six-torus $T^6$, and one so complicated that we know nearly nothing about it, as is unfortunately the case with most Calabi--Yau manifolds, where not even the metric is known explicitly. 

A toroidal orbifold is obtained by taking the quotient of a six-dimensional torus $T^6$ by a discrete abelian group $\Gamma$
. At the fixed points of $\Gamma$, the orbifold develops quotient singularities.
One way to look at orbifolds is to see them as a singular limit of a smooth Calabi--Yau manifold, one in which a number of four-cycles has been blown down to zero size. We can also go the other way, i.e. start with an orbifold and resolve its singularities via blow--ups. Like this we end up with a full-fledged, smooth Calabi--Yau, and more importantly, one we actually have an idea of what it looks like and on which we can do a number of explicit calculations.
This is the path which we will follow here. Two circumstances come to our aid in this construction. The first is that we can describe the singularities and their resolutions locally. For this, we need the machinery of toric geometry, of which all necessary basics will be introduced. The second is that we know what the orbifold looks like globally, knowing the $T^6$ and the configuration of fixed sets. This allows us to put the resolved local patches together correctly, determine divisor topologies and calculate the intersection ring.


The subject of toroidal orbifolds in the context of string theory was started off more than twenty years ago by Dixon, Harvey, Vafa and Witten \cite{DHVW}. Since then, a huge amount of literature has developed. A large part of it is devoted to world--sheet, i.e. conformal field theory calculations. String theory knows a lot about its target space, it even knows on the singular orbifold what the manifold looks like once its singularities are resolved. The present thesis does not enter the subject of CFT at all. All results are derived on a purely geometrical basis. In this sense, the approach taken here is complementary to much of the existing literature.


The first part of this thesis is devoted to the geometry of the toroidal orbifolds, the resolution of their singularities and the transition to the orientifold quotient.
After so much geometry, we will try to give the reader a flavor of why it was worth going through all this trouble by hinting at a number of applications in string phenomenology.

The use of toroidal orbifold models in the literature is so widespread that one cannot even begin to hope to do justice to all authors. Because of their phenomenologically interesting features, orbifold models are used extensively in model building efforts, where the goal is to reproduce the features of the (minimal supersymmetric) standard model (MSSM). In the past, a lot of attention has gone into heterotic orbifold compactifications (see for example \cite{Ibanez:1987pj}). There is a whole branch of literature concerned with $D$--branes on singularities, for which orbifolds again prove to be candidates of choice (see e.g. \cite{Douglas:1996sw}). Also intersecting brane models in type $IIA$ string theory are often studied in a setup with toroidal orbifolds as compactification manifolds.
Since we are unable to cover this wide variety of subjects in this thesis, we will confine ourselves to the more recent line of research in the context of compactifications with background fluxes in type $IIB$ string theory (see \cite{grana} for a recent review).
After introducing the preliminaries of flux compactifications, two main examples will be discussed. The first one belongs to the subject of model building, where several explicit string theory models are studied which aim to reproduce the Standard Model as well as possible. In the case at hand, the soft supersymmetry breaking terms induced by background fluxes are calculated for a type $IIB$ orientifold compactified on $T^6/\IZ_2\times\IZ_2$.

The other example belongs to the subject of moduli stabilization. The deformation parameters of the compactification manifold correspond to massless scalar fields in the low energy effective theory. Since such fields would give rise to a fifth force of about gravitational strength, they are clearly in conflict with experiment. For a string theory compactification to be realistic, its moduli must be stabilized to fixed values. Therefore, mechanisms which generate a potential for these massless scalars are investigated. Kachru, Kallosh, Linde and Trivedi \cite{KachruAW} (KKLT) have proposed a mechanism which not only freezes all moduli, but also gives rise to a meta--stable de Sitter vacuum with the small cosmological constant that is called for by experiment. Here, toroidal orbifolds and their resolutions are discussed as candidate models for a concrete realization of the KKLT proposal. 

These two examples independently address two of the main questions in string phenomenology. On the one hand, one tries to reproduce the spectrum and gauge group of the Standard Model, i.e. $SU(3)\times SU(2)\times U(1)$  to a high accuracy. The focus is mainly on the matter and gauge sector, and the aim is to derive estimates for certain measurable quantities, such as e.g. the masses of the scalar superpartners of the standard model matter fields, which might even allow to falsify specific string theory models in future experiments.

The question of moduli stabilization on the other hand focuses on the properties of the so--called hidden sector, which only couples to the Standard Model sector through gravitational interactions.

To address both sectors at the same time and to obtain models which yield realistic results on both these fronts is at the present stage of research still a long way off.

It should be stressed that string compactifications on toroidal orbifolds are essentially toy models. They reproduce a number of properties of the Standard model, such as non-abelian gauge groups, chiral fermions, and family repetition, but are otherwise far from realistic. Yet their phenomenologically interesting features make them worthwhile to study. By understanding these relatively simple models which allow for very concrete calculations, we hope to gain insights into the more general workings of string theory, insights into what string theory can do for us and what it cannot, insights into the big problems that still wait to be resolved. Apart from their applications in string phenomenology, the geometric constructions described in part I of this thesis captivate through their simplicity and elegance.

Although much of the present thesis is pure geometry, it is meant for the practical use of the working physicist. Therefore we have tried to keep explanations simple and geometrically intuitive, the idea being to provide a "How to"-sort of recipe.

While being very explicit with the geometry, a working knowledge of string theory will be assumed. Whenever direct reference to string theory calculations is made and not stated otherwise, we will be in type $IIB$, in a regime of large volume and weak string coupling which allows us to make use of the supergravity approximation. Furthermore, the basic notions of complex, K\"ahler and Calabi--Yau geometry will be assumed.

The geometric part of this thesis is based on material presented in \cite{Lust:2005dy} and \cite{geometry} but contains more geometric background material, in particular in the appendices. The chapter on applications in string phenomenology contains extracts from \cite{Lust:2005dy}, \cite{LustFI},  and \cite{pheno}.
Chapter two reviews the geometry of toroidal orbifolds in their singular limit. Possible point groups are discussed, the geometrical moduli are introduced, as well as their parametrization in terms of the radii and angles of the underlying torus lattice. Sub--tori with volumes larger than one are briefly discussed and the configuration of fixed sets and their equivalence classes which will be of paramount importance later on is explained.
Chapter three makes the transition to the resolved, smooth Calabi--Yau manifold. First, the resolution of the singularities via blow--ups in local, non--compact patches is described, along with the necessary background in toric geometry. Then, the procedure to put the resolved patches together to form a smooth manifold is discussed. We explain how the intersection ring is calculated and the divisor topologies in the compact geometry are determined.
Chapter four describes the systematic transition from the Calabi--Yau manifold to its orientifold quotient, the consequences of fixed sets which are not invariant under the orientifold involution, and the modified intersection ring.

In Chapter 5, the preliminaries of flux compactifications are reviewed: What it means to turn on background flux, which flux components are invariant under the orbifold twist, the low energy effective potential from fluxes.
Chapter six presents the first of the two examples: The soft supersymmetry breaking terms for a type $IIB$ model with D--branes on the orientifold of the singular orbifold $T^6/\IZ_2\times\IZ_2$ are calculated.
Chapter seven contains the second example. The KKLT proposal and the origin of the non--perturbative superpotential are reviewed, the conditions for the (non--)existence of stable vacua are discussed and the suitability of toroidal orbifolds and their resolutions as candidate models for the KKLT--construction is studied. Chapter eight contains the conclusions.

The appendices, which form the bulk of the present thesis collect the details to all orbifold models discussed here. They constitute a kind of reference book.
Appendix A gives the resolutions of the singular non--compact models. Appendix B collects the details for all compact models which were considered. Appendix C gives the Cartan matrices for the Lie--Algebra lattices which were used.

\part{The Geometry of Toroidal Orbifolds}

\chapter{At the orbifold point}

\section{What is an orbifold?}

An orbifold is obtained by dividing a smooth manifold by the non-free action of a discrete group: $X=Y/\Gamma$. The original mathematical definition is broader: Any algebraic variety whose only singularities are locally of the form of quotient singularities is taken to be an orbifold. Since our setup here is motivated by string theory, we will only be concerned with orbifolds of the form $T^6/\Gamma$, which, descending from a torus go by the name of {\it toroidal orbifolds}. While the torus is completely flat, the orbifold is not flat anymore. It is flat almost everywhere: Its curvature is concentrated in the fixed points of $\Gamma$. At these points, conical singularities appear. Only the simplest variety of toroidal orbifolds will be discussed here: $\Gamma$ is taken to be abelian, there will be no discrete torsion or vector structure.

Looking at string theory on $X=Y/\Gamma$, we must project onto the $\Gamma$-invariant states. But this is not the whole story yet. Since the points $x$ and $gx$ for $g\in\Gamma$ are identified on the quotient, we must not only consider strings whose coordinates fulfill $X^i(\sigma+2\pi)=X^i(\sigma)$, but also those with $X^i(\sigma+2\pi)=gX^i(\sigma)$. These new sectors are called twisted sectors, where we again have to project to invariant states. Physically, the twisted sector strings are only closed modulo a $\Gamma$ transformation. There are as many twisted sectors as group elements in $\Gamma$.

\section{Why should I care?}

As already mentioned in the introduction, toroidal orbifolds are simple, yet non-trivial. Their main asset is calculability, which holds for purely geometric as well as for string theoretic aspects. On the singular orbifold, string propagation is exactly solvable using its CFT description \cite{Dixon:1986qv}. What makes toroidal orbifolds especially interesting is that they allow for several phenomenologically interesting properties, such as non-abelian gauge groups, ${\cal N}=1$ supersymmetry and chiral fermions in heterotic string theory, and family repetition.

\section{Point groups and Coxeter elements}

A torus is specified by its underlying lattice $\Lambda$: Points which differ by a lattice vector are identified:
$$x \sim x+l,\quad l\in \Lambda.$$
The six-torus is therefore defined as quotient of $\IR^6$ with respect to the lattice $\Lambda$: $T^6=\IR^6/\Lambda$.
To define an orbifold of the torus, we divide as explained above by a discrete group $\Gamma$, which is called the {\it point group}, or simply the orbifold group. We cannot choose any random group as the point group $\Gamma$, it must be an automorphism of the torus lattice $\Lambda$, i.e. it must preserve the scalar product and fulfill $g\,l\in \Lambda$ if $l\in \Lambda,\ g\in\Gamma$.
To fully specify a toroidal orbifold, one must therefore specify both the torus lattice as well as the point group. In the context of string theory, a set-up with $SU(3)$--holonomy\footnote{This results in ${\cal N}=1$ supersymmetry for heterotic string theory and in ${\cal N}=2$ in type $II$ string theories in four dimensions.} is what is usually called for, which restricts the point group $\Gamma$ to be a subgroup of $SU(3)$. Since we restrict ourselves to abelian point groups, $\Gamma$ must belong to the Cartan subalgebra of $SO(6)$. On the complex coordinates of the torus, the orbifold twist will act as
\begin{equation}\label{cplxtwist}
\theta:\ (z^1,z^2,z^3) \to (e^{2\pi i \zeta_1}\,z^1, e^{2\pi i  \zeta_2}\,z^2, e^{2\pi i  \zeta_3}\,z^3), \quad 0\leq| \zeta^i|<1,\ i=1,2,3.
\end{equation}
The requirement of $SU(3)$--holonomy can also be phrased as requiring invariance of the $(3,0)$-form of the torus, $\Omega=dz^1\wedge dz^2\wedge dz^3$.
This leads to
\begin{equation}\label{req}
\pm  \zeta_1\pm  \zeta_2\pm  \zeta_3=0.
\end{equation}
We must furthermore require that $\Gamma$ acts crystallographically on the torus lattice. Together with the condition (\ref{req}), this amounts to $\Gamma$ being
either $\IZ_N$ with $N=3,4,6,7,8,12\,$ or $\IZ_N\times \IZ_M$ with $M$ a multiple of
$N$ and $N=2,3,4,6$. With the above, one is lead to the usual standard embeddings of the orbifold twists, which are given in Tables \ref{table:std} and \ref{table:zzstd}. The most convenient notation is 
$$( \zeta_1, \zeta_2, \zeta_3)=\frac{1}{n}(n_1,n_2,n_3)\ {\rm with}\ n_1+n_2+n_3=0\,\mod\, n.$$ 
Notice that $\IZ_6,\ \IZ_8$ and $\IZ_{12}$ have two inequivalent embeddings in $SO(6)$.

\begin{table}[h!]\begin{center}
\begin{tabular}{|l|c|}
\hline
Point group & $\frac{1}{n}(n_1,n_2,n_3)$ \\ [2pt]
\hline
$\ \IZ_{3}  $      & $\frac{1}{3}\,(1,1,-2)$\\ [2pt]
$\ \IZ_{4}  $      & $\frac{1}{4}\,(1,1,-2)$\\ [2pt]
$\ \IZ_{6-I}  $      & $\frac{1}{6}\,(1,1,-2)$\\ [2pt]
$\ \IZ_{6-II}  $      & $\frac{1}{6}\,(1,2,-3)$\\ [2pt]
$\ \IZ_{7}  $      & $\frac{1}{7}\,(1,2,-3)$\\ [2pt]
$\ \IZ_{8-I}  $      & $\frac{1}{8}\,(1,2,-3)$\\ [2pt]
$\ \IZ_{8-II}  $      & $\frac{1}{8}\,(1,3,-4)$\\ [2pt]
$\ \IZ_{12-I}  $      & $\frac{1}{12}\,(1,4,-5)$\\ [2pt]
$\ \IZ_{12-II}  $      & $\frac{1}{12}\,(1,5,-6)$\\ [2pt]
\hline
\end{tabular}
\caption{Group generators for $\IZ_N$-orbifolds.}\label{table:std}
\end{center}\end{table}

\begin{table}[h!]\begin{center}
\begin{tabular}{|l|c|c|}
\hline
Point group & $\frac{1}{n}(n_1,n_2,n_3)$ & $\frac{1}{m}(m_1,m_2,m_3)$ \\ [2pt]
\hline
$\ \IZ_{2} \times \IZ_{2} $      & $\frac{1}{2}\,(1,0,-1)$ & $\frac{1}{2}\,(0,1,-1)$\\ [2pt]
$\ \IZ_{2} \times \IZ_{4} $      & $\frac{1}{2}\,(1,0,-1)$ & $\frac{1}{4}\,(0,1,-1)$\\ [2pt]
$\ \IZ_{2} \times \IZ_{6} $      & $\frac{1}{2}\,(1,0,-1)$ & $\frac{1}{6}\,(0,1,-1)$\\ [2pt]
$\ \IZ_{2} \times \IZ_{6'} $      & $\frac{1}{2}\,(1,0,-1)$ & $\frac{1}{6}\,(1,1,-2)$\\ [2pt]
$\ \IZ_{3} \times \IZ_{3} $      & $\frac{1}{3}\,(1,0,-1)$ & $\frac{1}{3}\,(0,1,-1)$\\ [2pt]
$\ \IZ_{3} \times \IZ_{6} $      & $\frac{1}{3}\,(1,0,-1)$ & $\frac{1}{6}\,(0,1,-1)$\\ [2pt]
$\ \IZ_{4} \times \IZ_{4} $      & $\frac{1}{4}\,(1,0,-1)$ & $\frac{1}{4}\,(0,1,-1)$\\ [2pt]
$\ \IZ_{6} \times \IZ_{6} $      & $\frac{1}{6}\,(1,0,-1)$ & $\frac{1}{6}\,(0,1,-1)$\\ [2pt]
\hline
\end{tabular}
\caption{Group generators for $\IZ_N\times \IZ_M$-orbifolds.}\label{table:zzstd}
\end{center}\end{table}

For all point groups given in Tables \ref{table:std} and \ref{table:zzstd} it is possible to find a compatible torus lattice, in several cases even more than one. Here, we will consider the root lattices of semi-simple Lie-Algebras of rank 6. 
All one needs to know about such a lattice is contained in the Cartan matrix of the 
respective Lie algebra. The matrix elements of the Cartan matrix are defined as follows:
$$A_{ij}=2\ {\langle e_i, e_j\rangle\over \langle e_j, e_j\rangle}\,,$$
where the $e_i$ are the simple roots. The Cartan matrices of all lattices which are needed here can be found in Appendix \ref{AppC}. All necessary background material on Lie groups can be found in \cite{Slansky}.

The inner automorphisms of these root lattices are given by the Weyl-group of the Lie-algebra.  A Weyl reflection is a reflection on the hyperplane perpendicular to a given root:
\begin{equation}\label{Weyl}
{S_i({\bf x})= {\bf x}-2\frac{\langle{\bf x}, e_i\rangle}{ \langle e_i,e_i\rangle}e_i}.
\end{equation}
These reflections are not in $SU(3)$ and therefore are not suitable candidates for a point group, but the Weyl group does have a subgroup contained in $SU(3)$: The cyclic subgroup generated by the Coxeter element, which is given by successive Weyl reflections with respect to all 
simple roots: 
\begin{equation}\label{cox}
Q=S_1S_2...S_{rank}.
\end{equation}
The so-called outer automorphisms are those which are generated by transpositions of roots which are symmetries of the Dynkin diagram. By combining Weyl reflections with such outer automorphisms, we arrive at so-called generalized Coxeter elements.  $P_{ij}$ denotes the transposition of the $i$'th and $j$'th roots. 
The orbifold twist $\Gamma$ may be represented by a matrix $Q_{ij}$, which rotates
the six lattice basis vectors: $e_i\ra Q_{ji}\ e_j$\footnote{Different symbols for the orbifold twist are used according to whether we look at the quantity which acts on the real six-dimensional lattice $(Q)$ or on the complex coordinates $(\theta)$.}. The following discussion is restricted to cases in which the orbifold twist acts as the (generalized) Coxeter element of the group lattices, these are the so-called {\it Coxeter--orbifolds}\footnote{It is also possible to construct non--Coxeter orbifolds, such as e.g. $\IZ_4$ on $SO(4)^3$ as discussed in \cite{structure}.}.

\subsection{Example A: $\IZ_{6-I}$ on $G_2^2\times SU(3)$}\label{sec:exa}

We take the torus lattice to be the root lattice of $G_2^2\times SU(3)$, a direct product of three rank two root lattices, and explicitly construct its Coxeter element.
First, we look at the $SU(3)$--factor. With the Cartan matrix of $SU(3)$, see (\ref{sun}) and (\ref{Weyl}), the matrices of the two Weyl reflections can be constructed:
\begin{equation}
S_1=\left(\begin{array}{cc}\!\!-1&1\\ 0&1\end{array}\right),\quad S_2=\left(\begin{array}{cc}1&0\\ 1&\!\!-1\end{array}\right).
\end{equation}
The Coxeter element is obtained by multiplying the two:
\begin{equation}
Q_{SU(3)}=S_1S_2=\left(\begin{array}{cc}0&-1\\ 1&-1\end{array}\right).
\end{equation}
In the same way, we arrive at the Coxeter-element of $G_2$. The six-dimensional Coxeter element is built out of the three $2\times 2$--blocks:
\begin{equation}\label{Qsixi}
Q=\left(\begin{array}{cccccc}
2&-1&0&0&0&0\\ 
3&-1&0&0&0&0\\
0&0&2&-1&0&0\\
0&0&3&-1&0&0\\
0&0&0&0&0&-1\\
0&0&0&0&1&-1\end{array}\right).
\end{equation}
The eigenvalues of $Q$ are $e^{2\pi i/6},\, e^{-2\pi i/6},\,e^{2\pi i/6},\, e^{-2\pi i/6},\,e^{2\pi i/3},\, e^{-2\pi i/3}$, i.e. those of the $\IZ_{6-I}$--twist, see Table \ref{table:std}, and $Q$ fulfills $Q^6={\rm Id}$.


\section{The usual suspects}

In the following two tables, all orbifolds which will be discussed here are given. The list is the one given in \cite{Klemm}, other references such as \cite{Bailin:1999nk} give other lattices as well. 

\begin{table}[h!]
\begin{center}
\begin{center}
{\small
\begin{tabular}{|l|c|c|c|c|c|c|}
\hline
$\ \IZ_N$&Lattice &$h_{(1,1)}^{\rm untw.}$&$h_{(2,1)}^{\rm untw.}$&
$h_{(1,1)}^{\rm twist.}$&$h_{(2,1)}^{\rm twist.}$ \\ [2pt]
\hline
$\ \IZ_3 $    &\  $SU(3)^3 $          &9 & 0 & 27 & 0\cr
$\  \IZ_4  $   &\ $SU(4)^2  $      &5 & 1 & 20 & 0\cr
$\  \IZ_4 $    &\  $SU(2)\times SU(4)\times SO(5)$ &5 & 1 & 22 & 2\cr
$\  \IZ_4$      &\  $SU(2)^2\times SO(5)^2 $ &5 & 1 & 26 & 6\cr
$\  \IZ_{6-I} $   & $(G_2\times SU(3)^{2})^{\flat}$  &5 & 0 & 20 & 1\cr
$\  \IZ_{6-I}  $  &$ SU(3)\times G_2^2$ &5 & 0 & 24 & 5\cr
$\  \IZ_{6-II} $  &$ SU(2)\times SU(6)$    &3 & 1 & 22 & 0\cr
$\  \IZ_{6-II} $  &$SU(3)\times SO(8) $&3 & 1 & 26 & 4\cr
$\  \IZ_{6-II} $  &$(SU(2)^2\times SU(3)\times SU(3))^{\sharp} $ &3 & 1 & 28 & 6\cr
$\  \IZ_{6-II}  $ &$ SU(2)^2\times SU(3)\times G_2$  &3 & 1 & 32 & 10\cr
$\  \IZ_7  $      &$ SU(7) $                             &3 & 0 & 21 & 0\cr
$\  \IZ_{8-I}  $  &$ (SU(4)\times SU(4))^*$                              &3 & 0 & 21 & 0\cr
$\  \IZ_{8-I}  $  &$SO(5)\times SO(9)    $     &3 & 0 & 24 & 3\cr
$\  \IZ_{8-II} $  &$ SU(2)\times SO(10)  $     &3 & 1 & 24 & 2\cr
$\  \IZ_{8-II}  $   &$ SO(4)\times SO(9)$   &3 & 1 & 28 & 6\cr
$\  \IZ_{12-I}$   &$ E_6 $  &3 & 0 & 22 & 1\cr
$\  \IZ_{12-I} $  &$ SU(3)\times F_4$  &3 & 0 & 26 & 5\cr
$\  \IZ_{12-II} $  &$SO(4)\times F_4$  &3 & 1 & 28 & 6\cr
 \hline 
\end{tabular}}
\end{center}
\caption{Twists, lattices and Hodge numbers for $\IZ_N$ orbifolds.}\label{table:one}
\end{center}\end{table}

The tables give the torus lattices and the twisted and untwisted Hodge numbers.
The lattices marked with $\flat$, $\sharp$, and $*$ are realized as generalized Coxeter twists, the 
automorphism being in the first and second case $S_1S_2S_3S_4P_{36}P_{45}$ and in the third  $S_1S_2S_3P_{16}P_{25}P_{34}$.

\begin{table}[h!]\begin{center}
\begin{center}
\begin{tabular}{|l|c|c|c|c|c|c|}
\hline
$\ \IZ_N $&Lattice&
 $h_{(1,1)}^{\rm untw.} $& $h_{(2,1)}^{\rm untw.} $& 
 $h_{(1,1)}^{\rm twist.} $& $h_{(2,1)}^{\rm twist.} $\\ [2pt]
 \hline
$\  \IZ_2 \times\IZ_2$     &$SU(2)^6$  &3 & 3 & 48 & 0\cr
$\  \IZ_2 \times\IZ_4$    &$SU(2)^2\times SO(5)^2$ &3 & 1& 58 & 0\cr
$\  \IZ_2 \times\IZ_6 $    &$ SU(2)^2\times SU(3)\times G_2$&3 & 1 & 48 & 2\cr
$\  \IZ_2 \times\IZ_{6'}$&$ SU(3)\times G_2^2$&3 & 0 & 33 & 0\cr
$\  \IZ_3 \times\IZ_3  $   &$SU(3)^3$   &3 & 0 & 81 & 0\cr
$\  \IZ_3 \times\IZ_6 $    &$SU(3)\times G_2^2 $&3 & 0 & 70 & 1\cr
$\  \IZ_4 \times\IZ_4 $    &$SO(5)^3 $ &3 & 0 & 87 & 0\cr
$\  \IZ_6 \times\IZ_6  $   &$G_2^3$ &3 & 0 & 81 & 0
 \\ \hline 
\end{tabular}
\end{center}
\caption{Twists, lattices and Hodge numbers for $\IZ_N\times \IZ_M$ orbifolds.}\label{table:two}
\end{center}\end{table}


\section[Shape and size]{Shape and size: Introducing the geometrical moduli}\label{sec:shape}

A Calabi--Yau manifold, i.e. a K\"ahler manifold with vanishing first Chern class can be deformed in two ways: Either by varying its complex structure (its "shape"), or by varying its K\"ahler structure (its "size"). As explained in \cite{Candelas}, variations of the metric of mixed type $\delta g_{m \ov n}$ correspond to variations of the K\"ahler structure and give rise to $h_{1,1}$ parameters, whereas variations of pure type $\delta g_{mn},\,\delta g_{\ov m\ov n}$ correspond to variations of the complex structure and give rise to $h_{2,1}$ complex parameters. 
To metric variations of mixed type, a real $(1,1)$--form can be associated:
$$i\,\delta g_{m\ov n}\,dz^m\wedge d\ov z^{\ov n}.$$
To pure type metric variations, a complex $(2,1)$--form can be associated:
$$\Omega_{ijk}\,g^{k\ov n}\,\delta g_{\ov m\ov n}\,dz^i\wedge dz^j\wedge d\ov z^{\ov m},$$
where $\Omega$ is the Calabi--Yau $(3,0)$--form. Because of this correspondence, the number of untwisted moduli, i.e. $h^{(1,1)}_{untw.}$ and $h^{(2,1)}_{untw.}$ can be determined by counting the $(1,1)$-- and $(2,1)$--forms that are invariant under the orbifold twist. 
On $T^6$, there are nine independent $(1,1)$-forms. The three forms $dz^i\wedge d\ov z^i,\ i=1,2,3$ are invariant under all twists. For each pair $n_i=n_j$ in the twist (\ref{cplxtwist}), the forms $dz^i\wedge d\ov z^j$ and $dz^j\wedge d\ov z^i$ are invariant as well. For $\IZ_3$ where $n_1=n_2=n_3$, all nine $(1,1)$-forms of $T^6$ are invariant, while for $\IZ_4$ and $\IZ_{6-I}$ which have $n_1=n_2$, there are five invariant $(1,1)$--forms.

On the six-torus, there are also nine independent $(2,1)$-forms. The invariant $(2,1)$--forms correspond to sub-tori which are left completely unconstrained by the orbifold twist. Maximally three of the nine possible forms survive the twist, namely $dz^i\wedge dz^j\wedge d\ov z^k,\,\ i\neq j \neq k$.

To clarify the above, we will study the simple case of a single, unconstrained $T^2$, which has the metric
\begin{equation}\label{mm}{
g=\left(\begin{array}{cc}
g_{11}&g_{12}\cr g_{12}&g_{22}\end{array}\right)=\left(\begin{array}{cc}
R_1^2&R_1R_2\,\cos\theta_{12}\cr R_1R_2\,\cos\theta_{12}&R_2^2\end{array}\right).}
\end{equation}
A $T^2$ comes with one K\"ahler modulus ${\rm Im}({\cal T})$, which parametrizes its volume, and one complex structure modulus, which corresponds to its modular parameter ${\cal U}=\tau$.
\begin{figure}[h!]
\begin{center}
\includegraphics[width=70mm]{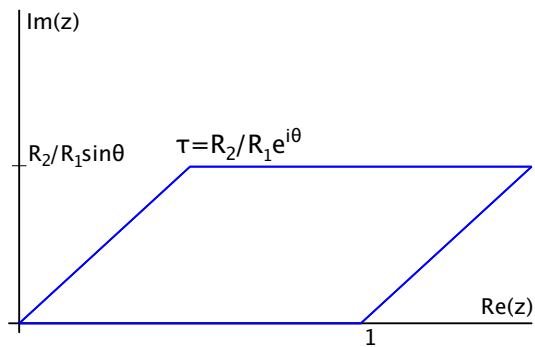}
\caption{Fundamental region of a $T^2$}
\label{fundamental}
\end{center}
\end{figure}
Figure \ref{fundamental} depicts the fundamental region of a $T^2$. The area of the torus is given by $R_1R_2\,\sin\theta$, expressed through the metric, we find 
\begin{equation}
{\rm Im}({\cal T})=\sqrt{\det\, g}=R_1R_2\,\sin\theta.
\end{equation}
In heterotic string theory, the K\"ahler moduli are complexified by pairing them up with the components of the anti-symmetric tensor $B$.
In type $IIB$ string theory, the K\"ahler moduli are paired with the components of the Ramond-Ramond four-form $C_4$.
The usual normalization of the fundamental region in string theory is such that the $a$--cycle is normalized to 1, while the modular parameter becomes $\tau=R_2/R_1\,e^{i\theta}$.
The complex structure modulus expressed through the metric is
\begin{equation}\label{CS}
\Uc=\frac{1}{{g_{11}}}\ (\,g_{12}+i\,\sqrt{\det\, g}\,).
\end{equation}
The twisted moduli will be discussed later on.
The same procedure naturally applies for all cases which have a $T^2$ factor, also when the $T^2$ is constrained by the orbifold twist and the complex structure is fixed.
When there is an untwisted complex structure modulus but the torus is not a direct product with a $T^2$ factor, the sub-torus which gives rise to the modulus must be identified and then, the same procedure can be applied, see the example in Section \ref{sec:exb}.


\section[The metric and its deformations]{The metric from the twist, deformations of the metric}

Once the Coxeter element $Q$ has been determined via (\ref{Weyl}) and the Cartan matrix of the lattice 
in question, the metric $g$ of the respective orbifold can be obtained through the requirement that the orbifold twist, being an isometry of the lattice, must leave the scalar product invariant \cite{Spalinski}:
\begin{equation}\label{findG}
{Q^tg\,Q=g\ .}
\end{equation}
The resulting metric can be conveniently parameterized in terms of the lengths of the vectors of the real lattice basis and the angles between them: 
$$g_{ij}=\langle e_i, e_j\rangle= R_i R_j \cos\theta_{ij}.$$ 
The metric of a Lie group lattice only leaves the overall scaling of the lattice vectors unfixed.
The orbifold twist allows in general more degrees of freedom, such as certain angles between the lattice vectors. This happens in particular when the torus lattice is a direct product of several root lattices. In such a case, the relative orientation of the different blocks is in general not completely fixed by the orbifold twist. The metric of the original Lie group root lattice can be recovered by setting the free parameters to certain fixed values.

To distinguish the deformation parameters of the metric which give rise to K\"ahler moduli from those which result in complex structure moduli, it is useful to study the anti-symmetric tensor, since it is paired up with the K\"ahler moduli. The form of the antisymmetric tensor $b$ is obtained in the same fashion, by solving
\begin{equation}\label{findB}
{Q^tb\,Q=b.}\end{equation}
The number of untwisted K\"ahler and complex structure moduli is obtained by counting the number of independent deformations $d$ allowed by the solutions of $Q^tg\,Q=g$ 
and $Q^tb\,Q=b$. The orbifold has $d_b$ untwisted K\"ahler moduli and $\half(d_g-d_b)$ untwisted complex 
structure moduli.

\subsection{Example A: $T^6/\IZ_{6-I}$ on $G_2^2\times SU(3)$}\label{sec:exai}

To find the metric for this example, (\ref{findG}) must be solved for the Coxeter element (\ref{Qsixi}). The result is
\begin{equation}\label{metricsixi}{
g=\left(\begin{array}{cccccc}
R_1^2&-{1\over 2}R_1^2&R_1R_3\cos\theta_{13}&y&0&0\cr
-{1\over 2}R_1^2&{1\over 3}R_1^2&{1\over\sqrt3}R_1R_3\cos\theta_{23} &{1\over 3}R_1R_3\cos\theta_{13}&0&0\cr
R_1R_3\cos\theta_{13}&{1\over\sqrt3}R_1R_3\cos\theta_{23}&R_3^2&-{1\over 2}R_3^2&0&0\cr
y&{1\over3}R_1R_3\cos\theta_{13}&-{1\over 2}R_3^2&{1\over 3} R_3^2&0&0\cr
0&0&0&0&R_5^2&-\half R_5^2\cr
0&0&0&0&-\half R_5^2&R_5^2\end{array}\right),
}\end{equation}
with $y=-{1\over3}(3R_1R_3\cos\theta_{13}-\sqrt3 R_1R_3\cos\theta_{23})$. There are five real continuous deformation parameters for the metric, $R_1^2,\ R_3^2,\ R_5^2,\ \theta_{13}$ and  $\theta_{23}$. One can immediately identify the three $2\times2$ blocks corresponding to the two $G_2$ factors and the $SU(3)$ factor. The angles between the two $G_2$ factors, $\theta_{13}$ and  $\theta_{23}$ are not constrained by the orbifold twist. For the choice $\theta_{13}=\theta_{23}=\pi/2$, the metric of the rigid root lattice $G_2^2\times SU(3)$ is recovered. 
Solving (\ref{findB}) we find
\begin{equation}\label{bsixi}{
b=\left(\begin{array}{cccccc}
0&b_1&3\,b_5&-3\,b_5-b_4&0&0\cr
-b_1&0&b_4&b_5&0&0\cr
-3\,b_5&-b_4&0&b_2&0&0\cr
3\,b_5+b_4&-b_5&-b_2&0&0&0\cr
0&0&0&0&0&b_3\cr
0&0&0&0&-b_3&0\end{array}\right)}\end{equation}
with the five real parameters $b_1,\ b_2,\ b_3,\ b_4,\ b_5$. We see that we get 5 untwisted K\"ahler moduli in this orbifold, while the complex structure is completely fixed.


\section{Parametrizing the geometrical moduli}\label{sec:moduli}

To find the dependence of the K\"ahler and complex structure moduli on the degrees of freedom
parametrized by the radii and angles in the lattice basis, one has to go 
to the complex basis $\{z^i\}_{i=1,2,3}$, 
where the twist $Q$ acts diagonally on the complex coordinates, i.e.
$$\theta\ :\ z^i\ra e^{2\pi i \zeta_i} z^i,$$ with the eigenvalues $2\pi i \zeta_i$ introduced above.
To find these complex coordinates we make the ansatz 
\begin{equation}\label{ansatz}{z^i=a^i_1\,x^1+a^i_2\,x^2+a^i_3\,x^3+a^i_4\,x^4+a^i_5\,x^5+a^i_6\,x^6}.\end{equation}
Knowing how the Coxeter twist acts on the root lattice and therefore on the real coordinates $x^i$, and knowing how the orbifold twist acts on the complex coordinates, see Tables \ref{table:one} and \ref{table:two}, we can determine the coefficients $a_j^i$ by solving
\begin{equation}\label{solveeq}
Q^t\, z^i=e^{2\pi i \zeta_i}\,z^i.
\end{equation}
The above equation constrains the coefficients up to an overall complex normalization factor. The transformation which takes us from the real to the complex basis must be unimodular. For convenience we choose a normalization such that the first term is real. The overall normalization of the complex coordinates does not influence the definition of the moduli. 
In addition, the ansatz (\ref{ansatz}) should yield a Hermitian metric, i.e. we require the
identity: 
$$ds^2=g_{ij}\ dx^i\otimes dx^j=g_{i\ov j}\ dz^i\otimes d\ov z^j.$$
After having introduced the complex coordinates $z^i$, which define the complex structure, we write down the K\"ahler form 
$J=i\,g_{i\ov j}\ dz^i\wedge d\ov z^j$. In the heterotic string, it is naturally paired with the anti-symmetric tensor 
$$B_2=b_{i\ov j}\ dz^i\wedge d\ov z^j\equiv \sum_{i=1}^{h_{(1,1)}} 
b^i\ \omega_i,$$
where the $\om_i$ form a basis of twist--invariant $2$--forms of the real cohomology $H^2(X,\IZ)$\footnote{In type $IIB$, the K\"ahler form is paired up with the anti--symmetric $4$--form 
$C_4=\sum\limits_{i=1}^{h_{(2,2)}} c^i\ d_i$, 
where the $d_i$ are a twist--invariant basis for $H^4(X,\IZ)$.}.

The K\"ahler moduli $\Tc^i$ are, as mentioned before, defined via the pairing 
\begin{equation}\label{pairing}
B+i\,J=\sum_{i=1}^{h_{(1,1)}} {\cal T}^i\,\om_i.
\end{equation}

How to find the complex coordinates and how to parametrize the K\"ahler moduli in a case with more than the usual three K\"ahler moduli is shown in detail in the following example.

In Section \ref{sec:exb}, the parametrization of a complex structure modulus is worked out.

\subsection{Example A: $T^6/\IZ_{6-I}$ on $G_2^2\times SU(3)$}\label{sec:exaii}

Solving (\ref{solveeq}) yields the following solution for the complex coordinates:
\begin{eqnarray}
z^1&=&a\,(-(1+e^{2\pi i/6})\,x^1+x^2)+b\,(-(1+e^{2\pi i/6})\,x^3+x^4),\cr
z^2&=&c\,(-(1+e^{2\pi i/6})\,x^1+x^2)+d\,(-(1+e^{2\pi i/6})\,x^3+x^4),\cr
z^3&=&e\,(e^{2\pi i/3}\,x^5+x^6),
\end{eqnarray}
where $a,\,b,\,c,\,d$ and $e$ are complex constants left unfixed by the twist alone. In the following, we will choose $a,\,d,\,e$ such that $x^1,\,x^3,\,x^5$ have a real coefficient and the transformation matrix is unimodular and set $b=c=0$, so the complex structure takes the following form:
\begin{eqnarray}\label{cplxzsixisim}
z^1&=&x^1+{1\over \sqrt3}\,e^{5\pi i/6}\,x^2,\cr
z^2&=&x^3+{1\over \sqrt3}\,e^{5\pi i/6}\,x^4,\cr
z^3&=&3^{1/4}\,(x^5+e^{2\pi i/3}\,x^6).
\end{eqnarray}
Now we proceed as outlined above. From the metric (\ref{metricsixi}) we can easily read off the K\"ahler form, which expressed in the complex coordinates (\ref{cplxzsixisim}) reads
\begin{eqnarray}
-i\,J&=&\,R_1^2\ dz^1\wedge d\ov z^1+R_3^2\ dz^2\wedge d\ov z^2+\frac{1}{\sqrt{3}}R_5^2\ dz^3\wedge d\ov z^3\cr
&&+2\,R_1R_3\,[(e^{2\pi i/6}\cos\theta_{13}+i\cos\theta_{23})\ dz^2\wedge d\ov z^1\nonumber\\[2pt]
&&+(e^{-2\pi i/6}\cos\theta_{13}-i\cos\theta_{23})\ dz^1\wedge d\ov z^2].
\end{eqnarray}
To be able to read off the K\"ahler moduli, we must look at the real cohomology. The five untwisted $(1,1)$--forms that are invariant under this orbifold twist are
\begin{eqnarray}
\om_1&=&dx^1\wedge dx^2,\quad \om_2=dx^3\wedge dx^4,\quad \om_3=dx^5\wedge dx^6,\cr
\om_4&=&dx^2\wedge dx^3-dx^1\wedge dx^4,\cr 
\om_5&=&3\ dx^1\wedge dx^3-3\ dx^1\wedge dx^4+dx^2\wedge dx^4.
\end{eqnarray}
The $B$--field (\ref{bsixi}) has the simple form
\begin{equation}{
B=b_1\ \om_1+b_2\ \om_2+b_3\ \om_3+b_4\ \om_4+b_5\ \om_5.
}\end{equation}
The K\"ahler form expanded in the real cohomology is
\begin{eqnarray}
J&=&{1\over2 \sqrt3}\ \{\, R_1^2\ \om_1+R_3^2\ \om_2+3\,R_5^2\ \om_3-2\,R_1R_3\,[\,\cos\theta_{13}\ \om_4\cr
&&-{2\over\sqrt3}(13\sqrt3\,\cos\theta_{13}+{29}\cos\theta_{23})\ \om_5]\}.
\end{eqnarray}
Via $B+i\,J={\cal T}^i\,\om_i$ the K\"ahler moduli can now be easily read off:
\begin{eqnarray}
{\cal T}^1&=&b_1+i\,{1\over2\sqrt3}\,R_1^2,\quad {\cal T}^2=b_2+i\,{1\over2\sqrt3}\,R_3^2,\quad {\cal T}^3=b_3+i\,{\sqrt3\over2}\,R_5^2,\nonumber\\[2pt]
{\cal T}^4&=&b_4-i\,{1\over\sqrt3}\,R_1R_3\cos\theta_{13},\nonumber\\[2pt] 
{\cal T}^5&=&b_5+i\,{1\over3}\,R_1R_3\,(13\sqrt3\,\cos\theta_{13}+{29}\,\cos\theta_{23}).
\end{eqnarray}





\section{Fixed tori with non-standard volumes}
\label{sec:nonstandardvols}

In the case of a torus lattice that does not factorize into $(T^2)^3$ it can happen that the sub-tori which are fixed under higher twists have a volume greater than one. Such a volume factor enters the definition of the complex coordinates and also shows up in physical quantities such as the partition function \cite{Klemm} or threshold corrections to gauge couplings \cite{Mayr:1993mq}.

Table \ref{table:vol} gives the volume factors for the $\IZ_n$--orbifolds for all sectors leading to fixed tori. Since the $\IZ_n \times \IZ_m$--orbifolds all factorize into $(T^2)^3$, all their fixed sub-tori have volume 1.

\begin{table}[h!]
\begin{center}
\begin{center}
{\small
\begin{tabular}{|l|c|c|}
\hline
$\ \IZ_N$&Lattice $T^6$&Vol. \\ [2pt]
\hline
$\ \IZ_3 $    &\  $SU(3)^3 $          &-\cr
$\  \IZ_4  $   &\ $SU(4)^2  $      &4\cr
$\  \IZ_4 $    &\  $SU(2)\times SU(4)\times SO(5)$ &2\cr
$\  \IZ_4$      &\  $SU(2)^2\times SO(5)^2 $ &1\cr
$\  \IZ_{6-I} $   & $(G_2\times SU(3)^{2})^{\flat}$  &4\cr
$\  \IZ_{6-I}  $  &$ SU(3)\times G_2^2$ &1\cr
$\  \IZ_{6-II} $  &$ SU(2)\times SU(6)$    &3,4\cr
$\  \IZ_{6-II} $  &$SU(3)\times SO(8) $&4,1\cr
$\  \IZ_{6-II} $  &$(SU(2)^2\times SU(3)\times SU(3))^{\sharp} $ &1,4\cr
$\  \IZ_{6-II}  $ &$ SU(2)^2\times SU(3)\times G_2$  &1,1\cr
$\  \IZ_7  $      &$ SU(7) $ &-\cr
$\  \IZ_{8-I}  $  &$ (SU(4)\times SU(4))^*$&4\cr
$\  \IZ_{8-I}  $  &$SO(5)\times SO(9)    $     &1\cr
$\  \IZ_{8-II} $  &$ SU(2)\times SO(10)  $     &2,2\cr
$\  \IZ_{8-II}  $   &$ SO(4)\times SO(9)$   &1,1\cr
$\  \IZ_{12-I}$   &$ E_6 $  &4,4\cr
$\  \IZ_{12-I} $  &$ SU(3)\times F_4$  &1,1\cr
$\  \IZ_{12-II} $  &$SO(4)\times F_4$  &1,1,1\cr
 \hline 
\end{tabular}}
\end{center}
\caption{Volume factors for $\IZ_N$ orbifolds.}\label{table:vol}
\end{center}\end{table}

The procedure of identifying the volume factors and the subsequent definition of the complex coordinates is best elucidated in an example. 

\subsection{Example B: $T^6/\IZ_{6-II}$ on $SU(2)\times SU(6)$}\label{sec:exb}

This example is illustrative not only because of its fixed tori with non-standard volumes, but also because it has a complex structure modulus, and moreover one which is associated to a sub-torus which is not simply a direct product factor in the lattice. The metric and anti-symmetric tensor are given in Appendix \ref{app:z6iisu2su6m}.

The Coxeter twist  on $SU(2)\times SU(6)$, given by (\ref{q6iisu2su6}), gives rise to two fixed tori: One is fixed under $Q^2$, the other under $Q^3$.
We identify them in the real basis by solving 
\begin{eqnarray}
Q^2\cdot n-n&=&0\,,\label{eq:one}\\
Q^3\cdot n-n&=&0\,,\label{eq:two}
\end{eqnarray}
where $n=\{n^i\}$ is a real 6-vector. The solution to (\ref{eq:one}) is
\begin{equation}\label{torus1}
t_1=(n^5,0,n^5,0,n^5,n^6),
\end{equation}
with $n^5,\,n^6$ free real parameters. 
The solution to (\ref{eq:two}) is
\begin{equation}\label{torus2}
t_2=(n^4,n^5,0,n^4,n^5,0),
\end{equation}
with $n^4,\,n^5$ free real parameters. Their volume is given by the scalar product with the dual tori, which are found via
\begin{eqnarray}
((Q^\dagger)^{-1})^2\cdot m-m&=&0\ \ \to\ \ t_1^*=(m^5,-m^5,m^5,-m^5,m^5,m^6),\label{eq:done}\\
((Q^\dagger)^{-1})^3\cdot m-m&=&0\ \ \to\ \ t_2^*=(m^4,m^5,-m^4-m^5,m^4,m^5,0).\label{eq:dtwo}
\end{eqnarray}
For the volumes, we find
\begin{eqnarray}
t_1^\dagger\cdot t_1^*&=& 3\,n^5m^5+n^6m^6,\label{vt1}\\
t_2^\dagger\cdot t_2^*&=& 2\,n^4m^4+2\,n^5m^5.\label{vt2}
\end{eqnarray}
The coefficients $k^i$ in front of the $n^im^i$ indicate that the respective real coordinates have a larger periodicity than the usual 1. This translates directly to the definition of the complex coordinate. Solving (\ref{solveeq}) with the ansatz (\ref{ansatz}) leads to
\begin{eqnarray}\label{z6iigen}
z^1&=&a\,(x^1+e^{2\pi i/6} x^2+e^{2\pi i/3}\, x^3-x^4+e^{2\pi i/3}\,x^5)),\cr
z^2&=&b\,(x^1+e^{2\pi i/3} x^2+e^{-2\pi i/3}\,x^3+x^4+e^{2\pi i/3}\,x^5) ,\cr
z^3&=&c\,(x^1-x^2+x^3-x^4+x^5)+d\,x^6.
\end{eqnarray}
The fact that in $z^3$ two instead of one free parameters appear is a sign that the complex structure is not fixed for this coordinate.
When we plug the fixed sub-tori into the expressions for the complex coordinates  (\ref{z6iigen}), we find
\begin{align}
z^1|_{t_1}&=0,     &z^1|_{t_2}&=0,\cr
z^2|_{t_1}&=0,     &z^2 |_{t_2}&=2\,b\,(n^4+e^{2\pi i/3}\,n^5),\cr
z^3|_{t_1}&=3\,c\,n^5+d\,n^6,   &z^3|_{t_2}&=0.
\end{align}
So $t_1$ obviously lies along the $z^3$ direction, while $t_2$ lies along the $z^2$ direction. As mentioned above, for these tori, the periodicity of the lattice is changed. In $t_2$, both complex shifts are scaled by a factor two. To compensate for this, we should choose $b\sim\tfrac{1}{2}$. In the case of $t_1$, it is only one of the shifts which is scaled, so we must choose $c\sim\tfrac{1}{3}$.
We now choose the normalization of (\ref{z6iigen}) such that $|\,\det\, Y|=1$, where $Y$ is the transformation from real to complex coordinates, which results in
\begin{eqnarray}\label{workiii}
z^1&=&\tfrac{1}{\sqrt3}\,(x^1+e^{2\pi i/6} x^2+e^{2\pi i/3}\, x^3-x^4+e^{2\pi i/3}\,x^5),\cr
z^2&=&\tfrac{1}{2\sqrt2}\,(x^1+e^{2\pi i/3} x^2+e^{-2\pi i/3}\,x^3+x^4+e^{2\pi i/3}\,x^5) ,\cr
z^3&=&\tfrac{1}{\sqrt{\im\,\Uc^3}}\ \lf[
\tfrac{1}{3}\,(x^1-x^2+x^3-x^4+x^5)+\Uc^3\, x^6\ \ri].\end{eqnarray}
What is left to do is parametrizing the complex structure modulus in terms of the metric. 
We first calculate
\begin{equation}\label{gtorus}
t_1^{\dagger} \cdot G\cdot t_1=3\,(n^5)^2\,R_1^2\,(1+2\,\cos\theta_{35})+6\,n^5n^6R_1R_6\,\cos\theta_{56}+(n^6)^2\,R_6^2.
\end{equation}
Now we express (\ref{gtorus}) in terms of a $2\times2$ metric, which is the metric of the sub-torus:
\begin{equation}
\left(\!\begin{array}{cc}n^5&\! n^6\end{array}\!\right)\!\left(\!\begin{array}{cc}\tilde g_{11}&\tilde g_{12}\\ \tilde g_{21}& \tilde g_{22}\end{array}\!\right)\!\left(\!\begin{array}{c}n^5\\ n^6\end{array}\!\right)=\left(\!\begin{array}{cc}n^5& \!n^6\end{array}\!\right)\!\left(\!\begin{array}{cc}3R_1^2\,(1+2\,\cos\theta_{35})&3R_1R_6\,\cos\theta_{56}\\ 3R_1R_6\,\cos\theta_{56}& R_6^2\end{array}\!\right)\!\left(\!\begin{array}{c}n^5\\ n^6\end{array}\!\right)
\end{equation}
With (\ref{CS}), we are now lead directly to the following complex structure modulus $\Uc^3$:
\begin{equation}\label{complexstr}{
\Uc^3=\fc{R_6}{R_1}\,\fc{\cos\theta_{56}+i\,\fc{1}{\sqrt3}
\sqrt{1+2\,\cos\theta_{35}-3\, \cos\theta_{56}^2}}{1+2\,\cos\theta_{35}}.}\end{equation}
The K\"ahler moduli present no further complications and are given in Appendix \ref{app:z6iisu2su6m}, see (\ref{z6iisu2su6k}).


\section{Fixed set configurations and conjugacy classes}\label{sec:schematic}

Many of the defining properties of an orbifold are encoded in its singularities. 
Not only the type (which group element they come from, whether they are isolated or not) and number of singularities is important, but also their spatial configuration. Here, it makes a big difference on which torus lattice a specific twist lives. The difference does not arise for the fixed points in the first twisted sector, i.e. those of the $\theta$-element which generates the group itself. But in the higher twisted sectors, in particular in those which give rise to fixed tori, the number of fixed sets differs for different lattices, which leads to differing Hodge numbers.

A point $f^{(n)}$ is fixed under $\theta^n\in \IZ_m,\ \ n=0,...,m-1,$ if it fulfills 
\begin{equation}\label{fix}{\theta^n\,f^{(n)}=f^{(n)}+l,\quad l\in \Lambda,
}\end{equation}
where $l$ is a vector of the torus lattice. In the real lattice basis, we have the identification $x^i \sim x^i + 1$. 
Like this, we obtain the sets that are fixed under the respective element of the orbifold group.
A twist $\frac{1}{n}(n_1,n_2,n_3)$ and its anti-twist $\frac{1}{n}(1-n_1,1-n_2,1-n_3)$ give rise to the same fixed sets, so do permutations of $(n_1,n_2,n_3)$. Therefore not all group elements of the point group need to be considered separately. The prime orbifolds, i.e. $\IZ_3$ and $\IZ_7$ have an especially simple fixed point configuration since all twisted sectors correspond to the same twist and so give rise to the same set of fixed points. 
Point groups containing subgroups generated by elements of the form 
$$\frac{1}{n}\,(n_1, 0, n_2),\ n_1+n_2=0\,\mod\,n$$
 give rise to fixed tori.

It is important to bear in mind that the fixed points were determined on the covering space. On the quotient, points which form an orbit under the orbifold group are identified. For this reason, not the individual fixed sets, but their conjugacy classes must be counted.

To form a notion of what the orbifold looks like, it is useful to have a schematic picture of the configuration, i.e. the intersection pattern of the singularities.

In some cases, not all information that will be needed later on is captured by looking at the fixed sets under a single group element. The points at the intersections of three $\IZ_2$ fixed lines will be relevant as well, which can be easily identified from the schematic figures provided here.
Interestingly, the case of three intersecting $\IZ_2$ fixed lines is the only instance of intersecting fixed lines where the intersection point itself is not fixed under a single group element. This case arises only for $\IZ_n\times\IZ_m$ orbifolds with both $n$ and $m$ even.

\subsection{Example A: $T^6/\IZ_{6-I}$ on $G_2^2\times SU(3)$}

In the following, we will identify the fixed sets under the $\theta$-, $\theta^2$- and $\theta^3$-elements. $\theta^4$ and $\theta^5$ yield no new information, since they are simply the anti-twists of $\theta^2$ and $\theta$. 
The $\IZ_{6-I}$--twist has only one fixed point in each torus, namely $z^i = 0$. The $\IZ_3$--twist has three fixed points in each direction, namely $z^1=z^2 = 0,1/3, 2/3$ and $z^3=0, 1/\sqrt3 \,e^{\pi i/6}, 1+i/\sqrt3$. The $\IZ_2$--twist, which arises in the $\theta^3$-twisted sector, has four fixed points, corresponding to $z^1=z^2 = 0,\half,\half \tau,\half(1+\tau)$ for the respective modular parameter $\tau$. As a general rule, we shall use red to denote the fixed set under $\theta$, blue to denote the fixed set under $\theta^2$ and pink to denote the fixed set under $\theta^3$. Note that the figure shows the covering space, not the quotient.

Table~\ref{tab:fssixi} summarizes the important data of the fixed sets. The invariant subtorus under $\theta^3$ is $(0,0,0,0,x^5,x^6)$ which corresponds simply to $z^3$ being invariant.

\begin{table}[h!]
\begin{center}
\begin{tabular}
{|c|c|c|c|}\hline
\ Group el.& Order &\ Fixed Set&Conj. Classes \cr
\noalign{\hrule}\noalign{\hrule}
$\ \theta$&6&\ 3 fixed points &\ 3\cr
$\ \theta^2$&3&\ 27 fixed points &\ 15\cr
$\ \theta^3$&2&\ 16 fixed lines &\ 6\cr
\hline
\end{tabular}
\end{center}
\caption{Fixed point set for $\IZ_{6-I}$ on $G_2^2\times SU(3)$}
\label{tab:fssixi}
\end{table}

\begin{figure}[h!]
\begin{center}
\includegraphics[width=80mm]{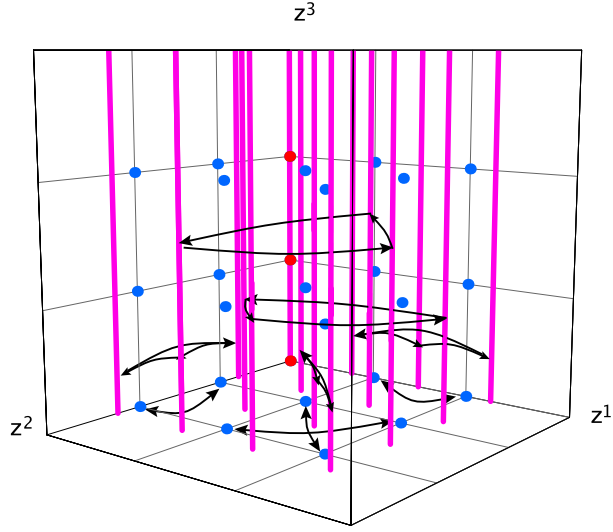}
\caption{Schematic picture of the fixed set configuration of $\IZ_{6-I}$ on $G_2^2\times SU(3)$}
\label{fig:ffixedi}
\end{center}
\end{figure}
Figure \ref{fig:ffixedi} shows the configuration of the fixed sets in a schematic way, where each complex coordinate is shown as a coordinate axis and the opposite faces of the resulting cube of length 1 are identified. Note that this figure again shows the whole six-torus and not the quotient.  The arrows indicate the orbits of the fixed sets under the action of the orbifold group, which we will now explain in detail.

We first look at the $z^1$ and $z^2$--directions. The two $\IZ_3$--fixed points at $1/3$ and $2/3$ are mapped to each other by $\theta$ and form orbits of length two. We choose to represent this orbit by $\zf{i}{2}$, $i=1,2$. The three $\IZ_3$ fixed points in the $z^3$--direction each form a separate conjugacy class. Therefore, we obtain the 15 conjugacy classes of $\IZ_3$--fixed points, 5 in each plane $z^3=\zf{3}{\gamma}$, $\gamma=1,2,3$:
\begin{align}
  \label{eq:zthreeconja}
  \mu=1:\; &(0,0, \zf{3}{\gamma}) & & &\notag\\  
  \mu=2:\; &(0,\tfrac{1}{3},\zf{3}{\gamma}),\ (0,\tfrac{2}{3},\zf{3}{\gamma}) & \mu=3:\; & (\tfrac{1}{3},0,\zf{3}{\gamma}),\ (\tfrac{2}{3},0,\zf{3}{\gamma})\notag\\ 
  \mu=4:\; &(\tfrac{1}{3},\tfrac{1}{3},\zf{3}{\gamma}),\ (\tfrac{2}{3},\tfrac{2}{3}, \zf{3}{\gamma}) & \mu=5:\; & (\tfrac{1}{3}, \tfrac{2}{3},\zf{3}{\gamma}),\ (\tfrac{2}{3},\tfrac{1}{3},\zf{3}{\gamma}). 
\end{align}
The 16 $\IZ_2$ fixed lines $(\zf{1}{\alpha},\zf{2}{\beta},z^3)$ fall into six conjugacy classes under the action of $\theta^2$:
\begin{align}
  \label{eq:ztwoconj}
  \nu=1:\; &(0,0, z^3) \cr
  \nu=2:\; &(\tfrac{1}{2},0,z^3),\ (\tfrac{1}{2}(1+\tau),0,z^3),\ (\tfrac{1}{2}\tau,0,z^3)\cr
  \nu=3:\; &(0,\tfrac{1}{2},z^3),\ (0,\tfrac{1}{2}(1+\tau),z^3),\ (0,\tfrac{1}{2}\tau,z^3) \cr
  \nu=4:\; &(\tfrac{1}{2},\tfrac{1}{2},z^3),(\tfrac{1}{2}(1+\tau),\tfrac{1}{2}(1+\tau),z^3),\ (\tfrac{1}{2}\tau,\tfrac{1}{2}\tau,z^3)\cr
  \nu=5:\; &(\tfrac{1}{2}, \tfrac{1}{2}(1+\tau),z^3),\ (\tfrac{1}{2}(1+\tau),\tfrac{1}{2}\tau,z^3),\ (\tfrac{1}{2}\tau,\tfrac{1}{2},z^3) \cr
  \nu=6:\; &(\tfrac{1}{2}, \tfrac{1}{2}\tau,z^3),\ (\tfrac{1}{2}(1+\tau), \tfrac{1}{2},z^3),\ (\tfrac{1}{2}\tau, \tfrac{1}{2}(1+\tau),z^3).
\end{align}
The configuration of fixed sets and their equivalence classes will become very important for the construction of the smooth Calabi--Yau later on.


\section[There's more than meets the eye]{There's more than meets the eye: Twisted moduli}

String theory compactified on an orbifold has twisted sectors - string states which are closed loops on the quotient space because their endpoints are identified under the orbifold twist. From the massless string spectrum we know that there are twisted K\"ahler and complex structure moduli. Obviously, these string states must have a geometric interpretation. It turns out that these are the moduli which arise from the resolution of the singularities.

Orbifold singularities are resolved via blow--ups, which correspond to a deformation of the K\"ahler structure. The singular point is replaced by an exceptional divisor\footnote{A divisor is a formal sum of codimension one submanifolds.}. The twisted K\"ahler moduli correspond to the volumes of these new four-cycles. 

Whereas isolated singularities do not give rise to complex structure deformations, fixed lines on which no fixed points sit do allow them, as we will see later. These are the twisted complex structure moduli.

In the case of orbifolds, the smooth Calabi--Yau manifold that results after the resolution of the singularities is in a way the more natural, since more general object to consider. The singular orbifold can be seen as a special, singular point in the moduli space of the smooth Calabi--Yau.
The following sections are devoted to the construction of smooth Calabi--Yau manifolds from singular toroidal orbifolds.



\chapter{The smooth Calabi-Yau}


\section[From singular to smooth]{From singular to smooth - A recipe}

To make the transition from the singular orbifold to the smooth Calabi--Yau manifold, clearly the singularities have to be resolved. But how is this done in practice? When we zoom in on a fixed point under a group $\Gamma$, space  in a small neighborhood around the singularity will look like $\IC^3/\Gamma$. Resolving this singularity of non-compact space is a well-known and straight-forward exercise. It can be done most conveniently by making use of the technology of toric geometry.

In the case of non-isolated singularities, i.e. fixed tori, space locally looks like $\IC^2/\Gamma^{(2)} \times \IC$, where $\Gamma^{(2)}$ is generated by $\frac{1}{n}\,(n_1, n_2)$. 

Once all singularities are resolved, we invoke our schematic picture of the fixed set configuration as introduced in Section \ref{sec:schematic}, which tells us how the local patches must be put together. We arrive at a full basis of $H_2$ and can calculate the intersection ring, from which we can in turn derive the volume of the manifold parameterized by the K\"ahler moduli. The topologies of the individual divisors can also be determined by a combination of local toric methods and global information descending from the torus. It is the combination of local knowledge about the neighborhoods of the resolved singularities, derived with the powerful methods of toric geometry, and global knowledge which descends from the six-torus itself which allows us to derive our knowledge about the smooth Calabi--Yaus.

A very readable introduction to toric geometry can be found in Chapter 7 of \cite{Horiab}, also \cite{DelaOssaXK} might be helpful.


\section[A lattice and a fan]{A lattice and a fan: Toric Geometry, the basics}

An $n$--dimensional toric variety has the form
\begin{equation}\label{variety}{X_\Sigma=({\IC}^N\setminus F_\Sigma)/({\IC}^*)^m,}\end{equation}
where $m<N,\ \ n=N-m$. $({\IC}^*)^m$ is the algebraic torus which lends the variety its name and acts via coordinatewise multiplication. $F_\Sigma$ is the subset that remains fixed under a continuous subgroup of $({\IC}^*)^m$ and must be subtracted for the variety to be well-defined. 

This toric variety $X_\Sigma$ can be encoded by a lattice $N$ which is isomorphic to $\IZ^n$ and its fan $\Sigma$. The fan is a collection of strongly convex rational cones in $N\otimes_{\IZ}{\bf R}$ with the property that each face of a cone in $\Sigma$ is also a cone in $\Sigma$ and the intersection of two cones in $\Sigma$ is a face of each. The $d$--dimensional cones in $\Sigma$ are in one-to-one correspondence with the codimension $d$--submanifolds of $X_\Sigma$. The one--dimensional cones in particular correspond to the divisors in $X_\Sigma$. 
The fan $\Sigma$ can be encoded by the generators of its edges or one--dimensional cones, i.e. by vectors $v_i\in N$. To each $v_i$ we associate a homogeneous coordinate $z^i$ of $X_\Sigma$.  To each of the $v_i$ corresponds the divisor $D_i$ which is determined by the equation $z^i=0$. The $(\IC^*)^r$ action is encoded on the $v_i$ in $r$ linear relations
\begin{equation}
  \label{eq:linrels}
  \sum_{i=1}^d l^{(a)}_i v_i = 0, \qquad a=1,\dots,r, \quad l^{(a)}_i \in \IZ.
\end{equation}
To each linear relation we assign a monomial $U^a = \prod_{i=1}^d z_i^{l^{(a)}_i}$. These monomials are the local coordinates of $X_\sigma$. 

We are uniquely interested in Calabi--Yau orbifolds, therefore we require $X_\Sigma$ to have trivial canonical class. The canonical divisor of  $X_\Sigma$ is given by $-D_1-...-D_n$, so for $X_\Sigma$ to be Calabi--Yau, $D_1+...+D_n$ must be trivial. This translates to requiring that the $v_i$ must all lie in the same affine hyperplane one unit away from the origin $v_0$. In our 3--dimensional case, this means that the third component of all the vectors $v_i$ (except $v_0$) equals one. The $v_i$ form a cone $C(\Delta^{(2)})$ over the triangle $\Delta^{(2)} = \langle v_1, v_2, v_3 \rangle$ with apex $v_0$. The Calabi--Yau condition therefore allows us to draw toric diagrams $\Delta^{(2)}$ in two dimensions only. 

How do we go about finding the fan of a specific ${\IC}^3/\IZ_n$--orbifold? We have just one three--dimensional cone in $\Sigma$, generated by $v_1,\,v_2,\,v_3$. The orbifold acts as follows on the coordinates of  ${\IC}^3$:
\begin{equation}\label{twistc}{\theta:\ (z^1,\, z^2,\, z^2) \to (\varepsilon\, z^1, \varepsilon^{n_1}\, z^2, \varepsilon^{n_2}\, z^3),\quad \varepsilon=e^{2 \pi i/n}.
}\end{equation}
For such an action we will use the shorthand notation $\frac{1}{n}(1,n_1,n_2)$. The coordinates of $X_\Sigma$ are given by
\begin{equation}\label{coord}{U^i=(z^1)^{(v_1)_i}(z^2)^{(v_2)_i}(z^3)^{(v_3)_i}.}\end{equation}
To find the coordinates of the generators $v_i$ of the fan, we require the $U^i$ to be invariant under the action of $\theta$. We end up looking for two linearly independent solutions of the equation
\begin{equation}\label{eqfan}{(v_1)_i+n_1\,(v_2)_i+n_2\,(v_3)_i= 0\ \mod\, n.}\end{equation}
The Calabi--Yau condition is trivially fulfilled since the orbifold actions are chosen such that $1+n_1+n_2=n$ and $\varepsilon^n=1$.

$X_\Sigma$ is smooth if all the top-dimensional cones in $\Sigma$ have volume one. By computing the determinant $\det(v_1,v_2,v_3)$, it can be easily checked that this is not the case in any of our orbifolds. We will therefore resolve the singularities by blowing them up.

\subsection{Example A.1: $\IC^3/\IZ_{6-I}$}\label{sec:exzsixi}

\begin{figure}[h!]
\begin{center}
\includegraphics[width=100mm]{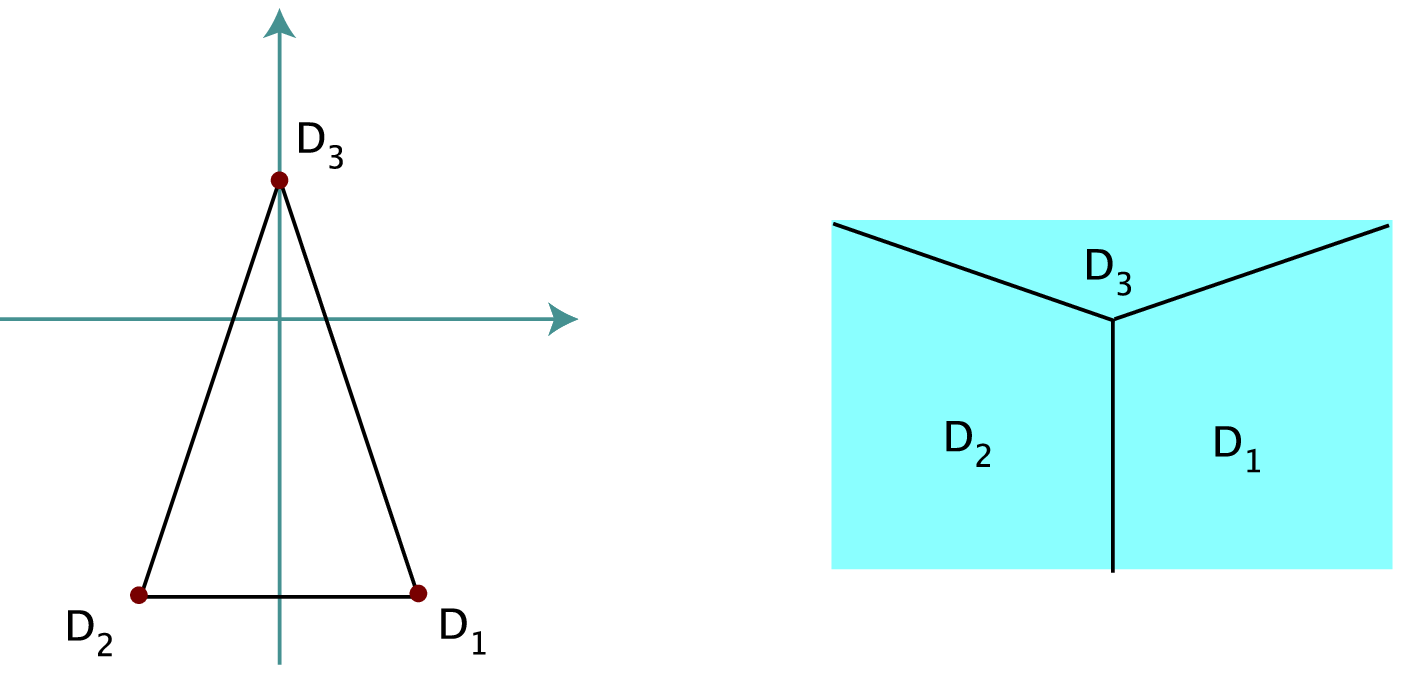}
\caption{Toric diagram of $\IC^3/\IZ_{6-I}$ and dual graph}
\label{fig:sixib}
\end{center}
\end{figure}

The group $\IZ_{6-I}$ acts as follows on $\IC^3$:
\begin{equation}
  \label{eq:twistsixi}
  \theta:\ (z^1,\, z^2,\, z^2) \to (\varepsilon\, z^1, \varepsilon\, z^2, \varepsilon^4\, z^3),\quad \varepsilon=e^{2 \pi i/6}.
\end{equation}
To find the components of the $v_i$, we have to solve $(v_1)_i+(v_2)_i+4\,(v_3)_i=0\ \mod\,6$. This leads to the following three generators of the fan (or some other linear combination thereof):
\begin{equation}
  v_1=(1,-2,1),\ v_2=(-1,-2,1),\ v_3=(0,1,1).
\end{equation}
The toric diagram of ${\IC}^3/\IZ_{6-I}$ and its dual diagram are depicted in Figure \ref{fig:sixib}.

\section{Resolving the singularities}\label{sec:resolve}

The process of blowing up a consists of two steps in toric geometry:
First, we must refine the fan, then subdivide it. Refining the fan means adding 1--dimensional cones. The subdivision corresponds to choosing a triangulation for the toric diagram. Together, this corresponds to replacing the point that is blown up by an exceptional divisor. We denote the refined fan by $\tilde\Sigma$.

We are interested in resolving the orbifold--singularities such that the canonical class of the manifold is not affected, i.e. the resulting manifold is still Calabi--Yau (in mathematics literature, this is called a crepant resolution). 
When adding points that lie in the intersection of the simplex with corners $v_i$ and the lattice $N$, the Calabi--Yau criterion is met. Aspinwall studies the resolution of singularities of type ${\IC}^d/G$ and gives a very simple prescription \cite{AspinwallEV}. We first write it down for the case of  ${\IC}^3/\IZ_n$.
For what follows, it is more convenient to write the orbifold twists in the form
\begin{equation}{\theta:\ (z^1,\, z^2,\, z^3) \to (e^{2 \pi i g_1}\, z^1,\,e^{2 \pi i g_2}\, z^2,\,e^{2 \pi i g_3}\, z^3).
}\end{equation}
The new generators $w_i$ are obtained via
\begin{equation}\label{prescription}{
w_i=g^{(i)}_1\,v_1+g^{(i)}_2\,v_2+g^{(i)}_3\,v_3,
}\end{equation}
where the $g^{(i)}=(g^{(i)}_1,\,g^{(i)}_2,\,g^{(i)}_3)\in \IZ_n=\{1, \theta, \theta^2,...\, , \theta^{n-1}\}$ such that 
\begin{equation}\label{eq:criterion}{\sum_{i=1}^3 g_i=1,\quad  0\leq g_i<1.}\end{equation}
$\theta$ always fulfills this criterion.  We denote the the exceptional divisors corresponding to the $w_i$ by $E_i$. To each of the new generators we associate a new coordinate which we denote by $y^i$, as opposed to the $z^i$ we associated to the original $v_i$.

Let us pause for a moment to think about what this method of resolution means. The obvious reason for enforcing the criterion (\ref{eq:criterion}) is that group elements which do not respect it fail to fulfill the Calabi--Yau condition: Their third component is no longer equal to one. But what is the interpretation of these group elements that do not contribute? Another way to phrase the question is: Why do not all twisted sectors contribute exceptional divisors? A closer look at the group elements shows that all those elements of the form $\frac{1}{n}\,(1,n_1,n_2)$ which fulfill (\ref{eq:criterion}) give rise to inner points of the toric diagram. Those of the form $\frac{1}{n}\,(1,0,n-1)$ lead to points on the edge of the diagram. They always fulfill (\ref{eq:criterion}) and each element which belongs to such a sub-group contributes a divisor to the respective edge, therefore there will be $n-1$ points on it. The elements which do not fulfill (\ref{eq:criterion}) are in fact anti-twists, i.e. they have the form $\frac{1}{n}\,(n-1, n-n_1, n-n_2)$. Since the anti-twist does not carry any information which was not contained already in the twist, there is no need to take it into account separately, so also from this point of view it makes sense that it does not contribute an exceptional divisor to the resolution.

The case $\IC^2/\IZ_n$ is even simpler. The singularity $\IC^2/\IZ_n$ is called a rational double point of type $A_{n-1}$ and its resolution is called a Hirzebruch--Jung sphere tree consisting of $n-1$ exceptional divisors intersecting themselves according to the Dynkin diagram of $A_{n-1}$. The corresponding polyhedron $\Delta^{(1)}$ consists of a single edge joining two vertices $v_1$ and $v_2$ with $n-1$ equally spaced lattice points $w_1,\dots,w_{n-1}$ in the interior of the edge, see discussion in Appendix \ref{app:rzctwo}. 

Now we subdivide the cone. For most groups $G$, several triangulations, and therefore several resolutions are possible (for large group orders even several thousands). They are all related via birational transformations, namely flop transitions. The diagram of the resolution of ${\IC}^3/G$ contains $n$ triangles, where $n$ is the order of $G$, yielding $n$ three-dimensional cones.

This treatment is easily extended to ${\IC}^3/\IZ_N\times\IZ_M$--orbifolds. When constructing the fan, the coordinates of the generators $v_i$ not only have to fulfill one equation (\ref{eqfan}) but three, coming from the twist $\theta^1$ associated to $\IZ_N$, the twist $\theta^2$ associated to $\IZ_M$ and from the combined twist $\theta^1\theta^2$. When blowing up the orbifold, the possible group elements $g^{(i)}$ are $\{(\theta^1)^i(\theta^2)^j,\ i=0,...,N-1,\ j=0,...,M-1\}$. The toric diagram of the blown--up geometry contains $N\cdot M$ triangles corresponding to the tree-dimensional cones. The remainder of the preceding discussion remains the same.

In the dual diagram, the geometry and intersection properties of a toric manifold are often easier to grasp than in the original toric diagram. The divisors, which are represented by vertices in the original toric diagram become faces in the dual diagram, the curves marking the intersections of two divisors remain curves and the intersections of three divisors which are represented by the faces of the original diagram become vertices. In the dual graph, it is immediately clear, which of the divisors and curves are compact. The curves at the intersection of two exceptional divisors are the exceptional curves.

We also want to settle the question to which toric variety the blown-up geometry corresponds. 
Applied to our case $X_\Sigma={\IC}^3/G$, the new blown up variety corresponds to
\begin{equation}\label{eq:blowup}{X_{\tilde\Sigma}={\IC}^{3+d}\setminus F_{\tilde\Sigma}/({\IC}^*)^d,}\end{equation}
where $d$ is the number of new generators $w_i$ of one--dimensional cones. The action of $(\IC^*)^d$ corresponds to the set of rescalings that leave the
\begin{equation}\label{newu}{\tilde U_i=(z^1)^{(v_1)_i}(z^2)^{(v_2)_i}(z^3)^{(v_3)_i}(z_4)^{(w_1)_i}\!...\,(z_{3+d})^{(w_d)_i}}\end{equation}
invariant.
The excluded set $F_{\tilde\Sigma}$ is determined as follows: Take the set of all combinations of generators $v_i$ of one--dimensional cones in $\Sigma$ that {\it do not span a cone} in $\Sigma$ and define for each such combination a linear space by setting the coordinates associated to the $v_i$ to zero. $F_\Sigma$ is the union of these linear spaces, i.e. the set of simultaneous zeros of coordinates not belonging to the same cone.  
In the case of several possible triangulations, it is the excluded set that distinguishes the different resulting geometries.


\subsection{Example A.1: $\IC^3/\IZ_{6-I}$}\label{sec:rzsixi}

\begin{figure}[h!]
\begin{center}
\includegraphics[width=120mm]{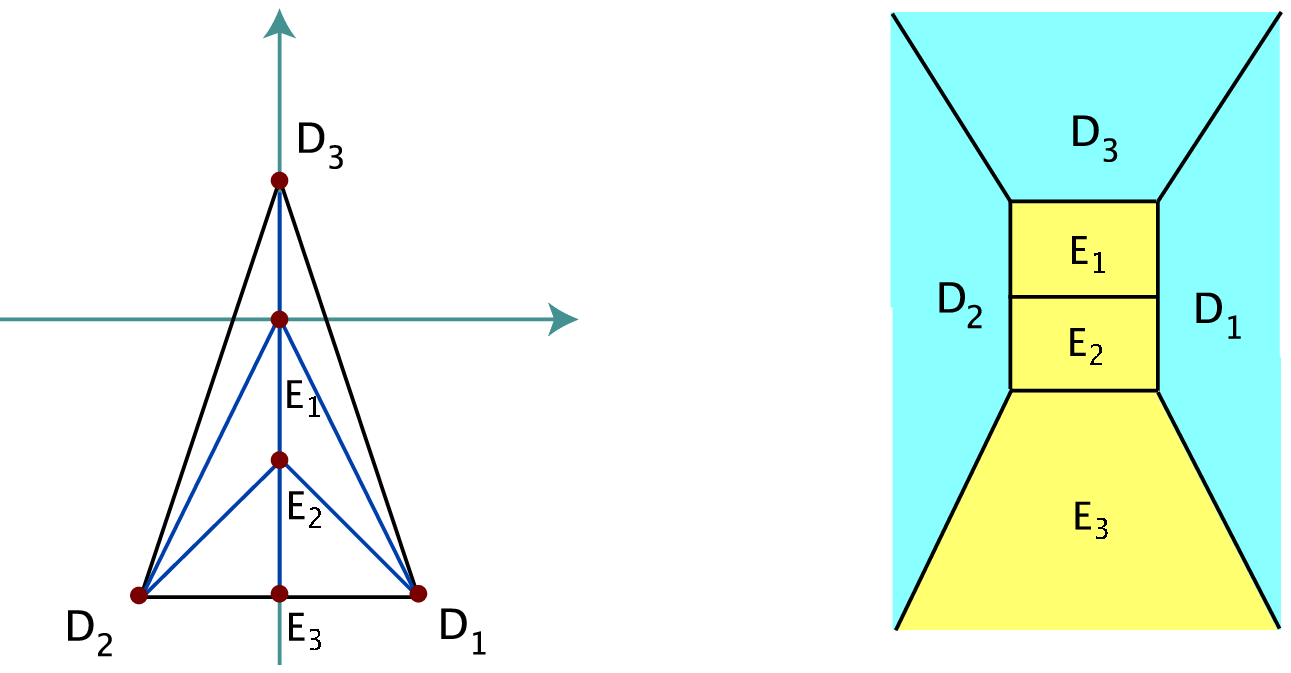}
\caption{Toric diagram of the resolution of $\IC^3/\IZ_{6-I}$ and dual graph}
\label{fig:fsixi}
\end{center}
\end{figure}

We will now resolve the singularity of $\IC^3/\IZ_{6-I}$.  $\theta,\,\theta^2$ and $\theta^3$ fulfill (\ref{eq:criterion}). This leads to the following new generators:
\begin{eqnarray}
w_1&=&\tfrac{1}{6}\,v_1+\tfrac{1}{6}\,v_2+\tfrac{4}{6}\,v_3=(0,0,1),\nonumber \\[2 pt]
w_2&=&\tfrac{3}{6}\,v_1+\tfrac{2}{6}\,v_2+\tfrac{2}{6}\,v_3=(0,-1,1),\nonumber \\[2 pt]
w_3&=&\tfrac{3}{6}\,v_1+\tfrac{3}{6}\,v_2=(0,-2,1).
\end{eqnarray}
In this case, the triangulation is unique. Figure \ref{fig:fsixi} shows the corresponding toric diagram and its dual graph.
Let us identify the new geometry. The $\tilde U_i$ are
\begin{eqnarray}
  \label{eq:tildeUsixi}
  \tilde U_1&=&{z^1\over z^2},\quad \tilde U_2={z^3\over(z^1)^2(z^2)^2y^2(y^3)^2},\quad \tilde U_3=z^1z^2z^2y^1y^2y^3.
\end{eqnarray}
The rescalings that leave the $\tilde U_i$ invariant are
\begin{equation}
  \label{eq:rescalessixi}
  (z^1,\,z^2,\,z^3,\,y^1,\,y^2,\,y^3) \to (\lambda_1\,z^1,\,\lambda_1\,z^2,\,\lambda_1^4\lambda_2\lambda_3\,z^3,\, {1\over\lambda_1^6\lambda_2^2\lambda_3^3}\,y^1,\,\lambda_2\,y^2,\,\lambda_3\,y^3).
\end{equation}
According to (\ref{eq:blowup}), the new blown-up geometry is
\begin{equation}
  \label{eq:blowupsixi}
  X_{\tilde\Sigma}=\,({\IC}^{6}\setminus F_{\tilde\Sigma})/({\IC}^*)^3,
\end{equation}
where the action of $(\IC^*)^3$ is given by (\ref{eq:rescalessixi}).  The excluded set is generated by
\begin{equation*}
  \label{eq:excludesixi}
  F_{\tilde\Sigma}=\{(z_3,y_2)=0,\,(z_3,y_3)=0,\,(y_1,y_3)=0,\,(z_1,z_2)=0\,\}.
\end{equation*}

As can readily be seen in the dual graph, we have seven compact curves in $X_{\tilde\Sigma}$. Two of them, $\{y^1=y^2=0\}$ and $\{y^2=y^3=0\}$ are exceptional. They both have the topology of ${\IP}^1$. Take for example $C_1$:  To avoid being on the excluded set, we must have $y^3\neq0,\ z^3\neq0$ and $(z^1,z^2)\neq0$. Therefore $C_1=\{(z^1,z^2, 1,0,0,1), (z^1,z^1)\neq0\}/(z^1,z^2)$, which corresponds to a ${\IP}^1$.

We have now six three-dimensional cones: $S_1=(D_1,\,E_2,\,E_3),\ S_2=( D_1,\,E_2,\,E_1),\ S_3=( D_1,\,E_1,\,D_3 ),$  $S_4=( D_2,\,E_2,\,E_3),\ S_5=( D_2,\,E_2,\,E_1),$ and $S_6=( D_2,\,E_1,\,D_3)$.


\section{Mori cone and intersection numbers}\label{sec:mori}

Since this will become important later, we want to investigate the intersection properties of the divisors of the resolved geometry. Note that the intersection number of two cycles $A,\ B$ only depends on the homology classes of $A$ and $B$.
Note also that $\sum b_i D_i$ and $\sum b_i' D_i$ (where the $D_i$ are the divisors corresponding to the one-dimensional cones) are linearly equivalent iff they are homologically equivalent.

First, we identify the linear relations between the divisors of the form 
$$a^i_1\,v_1+a^i_2\,v_2+a^i_3\,v_3+a^i_4\,w_1+...+a^i_{3+d}\,w_d=0.$$ 
These linear relations can be obtained either by direct examination of the generators or can be read off directly from the algebraic torus action $({\IC}^*)^m$. The exponents of the different scaling parameters yield the coefficients $a_i$. The divisors corresponding to such a linear combination are sliding divisors in the compact geometry.  It is very convenient to introduce a matrix $(\,P\, |\,Q\,)$: The rows of $P$ contain the coordinates of the vectors $v_i$ and $w_i$. The columns of $Q$ contain the linear relations between the divisors, i.e. the vectors $\{a^i\}$. From the rows of $Q$, which we denote by $C_i,\ i=1,...,d$, we can read off the linear equivalences in homology between the divisors which enable us to compute all triple intersection numbers. For most applications, it is most convenient to choose the $C_i$ to be the generators of the Mori cone. The Mori  cone is the space of effective curves, i.e. the space of all curves $C\in X_\Sigma$ with $C\cdot D\geq 0$ for all divisors $D\in X_\Sigma$. It is dual to the K\"ahler cone. In our cases, the Mori cone is spanned by curves corresponding to two-dimensional cones. The curves correspond to the linear relations for the vertices. The generators for the Mori cone correspond to those linear relations in terms of which all others can be expressed as positive, integer linear combinations.

We will briefly survey the method of finding the generators of the Mori cone. It can be found for example in \cite{BerglundGD}. We will present it here adapted to our context.

\begin{itemize}
\item[I.] In a given triangulation, take the three-dimensional simplices $S_k$ (corresponding to the three-dimensional cones). Take those pairs of simplices $(S_l,S_k)$ that share a two--dimensional simplex $S_k\cap S_l$.
\item[II.] For each such pair find the unique linear relation among the vertices in $S_k\cup S_l$ such that 
\begin{itemize}
\item[(i)] the coefficients are minimal integers and 
\item[(ii)] the coefficients for the points in $(S_k \cup S_l) \setminus (S_k \cap S_l)$ are positive.
\end{itemize}
\item[III.] Find the minimal integer relations among those obtained in step 2 such that each of them can be expressed as a positive integer linear combination of them. 
\end{itemize}
While the first two steps are very simple, step III. becomes increasingly tricky for larger groups.

The general rule for triple intersections is that the intersection number of three distinct divisors is 1 if they belong to the same cone and 0 otherwise. The set of collections of divisors which do not intersect because they do not lie in the same come forms a further characteristic quantity of a toric variety, the Stanley--Reisner ideal. It contains the same information as the exceptional set $F_{\Sigma}$. Intersection numbers for triple intersections of the form $D_i^2 D_j$ or $E_k^3$ can be obtained by making use of the linear equivalences between the divisors. Since we are working here with non--compact varieties at least one compact divisor has to be involved. For intersections in compact varieties there is no such condition. The intersection ring of a toric variety is -- up to a global normalization -- completely determined by the linear relations and the Stanley--Reisner ideal. The normalization is fixed by one intersection number of three distinct divisors.

The matrix elements of $Q$ are the intersection numbers between the curves $C_i$ and the divisors $D_i,\,E_i$. We can use this to determine how the compact curves of our blown--up geometry are related to the $C_i$. 


\subsection{Example A.1: $\IC^3/\IZ_{6-I}$}\label{sec:exsixiaa}

For this example, the method of working out the Mori generators is shown step by step.
We give the pairs, the sets $S_l\cup S_k$ (the points underlined are those who have to have positive coefficients) and the linear relations:
\begin{eqnarray}
  \label{eq:Moripairs}
1.\ \  S_6\cup S_3&=&\{\underline{D_1},\,\underline{D_2},\,D_3,\,E_1\},\quad D_1+D_2+4\, D_3-6\,E_1=0,\nonumber\\
2.\ \   S_5\cup S_2&=&\{\underline{D_1},\,\underline{D_2},\,E_1,\,E_2\},\quad D_1+D_2+2\, E_1-4\,E_2=0,\nonumber\\
3.\ \   S_4\cup S_1&=&\{\underline{D_1},\,\underline{D_2},\,E_2,\,E_3\},\quad D_1+D_2-2\,E_3=0,\nonumber\\
4.\ \   S_3\cup S_2&=&\{D_1,\,\underline{D_3},\,E_1,\,\underline{E_2}\},\quad D_3-2\,E_1+E_2=0,\nonumber\\
5.\ \   S_2\cup S_1&=&\{D_1,\,\underline{E_1},\,E_2,\,\underline{E_3}\},\quad E_1-2\,E_2+E_3=0,\nonumber\\
6.\ \  S_6\cup S_5&=&\{D_2,\,\underline{D_3},\,E_1,\,\underline{E_2}\},\quad D_3-2\,E_1+E_2=0,\nonumber\\
7.\ \   S_5\cup S_4&=&\{D_2,\,\underline{E_1},\,E_2,\,\underline{E_3}\},\quad E_1-2\,E_2+E_3=0.
\end{eqnarray}
With the relations 3, 4 and 5 all other relations can be expressed as a positive integer linear combination.
This leads to the following three Mori generators:
\begin{equation}
  \label{eq:Morigensixi}
  C_1=\{0,0,0,1,-2,1\},\quad C_2=\{1,1,0,0,0,-2\},\quad C_3=\{0,0,1,-2,1,0\}.
\end{equation}
With this, we are ready to write down $(P\,|\,Q)$:
\begin{equation}
  \label{eq:PQ}
  (P\,|\,Q)=\left(
  \begin{array}{cccccccc}
  D_1&1&-2&1&|&0&1&0\nonumber\\
  D_2&-1&-2&1&|&0&1&0\nonumber\\
  D_3&0&1&1&|&0&0&1\nonumber\\
  E_1&0&0&1&|&1&0&\!\!-2\nonumber\\
  E_2&0&-1&1&|&\!\!\!-2&0&1\nonumber\\
  E_3&0&-2&1&|&1&\!\!\!-2&0.
  \end{array}\right)
\end{equation}
From the rows of $Q$, we can read off directly the linear equivalences:
$$D_1 \sim D_2,\quad E_2 \sim -2\,E_1 - 3\,D_3,\quad E_3 \sim E_1-2\,D_1 + 2\,D_3.$$
For our later convenience, we recast them into a form in which each relation contains one $D$--divisor:
\begin{eqnarray}
  \label{eq:lineqsixI}
  0 &\sim& 6\,D_{{1}}+2\,E_{{2}}+E_{{1}}+3\,E_{{3}},\nonumber\\
  0 &\sim& 6\,D_{{2}}+2\,E_{{2}}+E_{{1}}+3\,E_{{3}},\nonumber\\
  0 &\sim& 3\,D_{{3}}+E_{{2}}+2\,E_{{1}}.
\end{eqnarray}
The matrix elements of $Q$ contain the intersection numbers of the $C_i$ with the $D_1,\,E_1$, \eg $E_1\cdot C_3=-2,\ D_3\cdot C_1=0$, etc.
We know that $E_1\cdot E_3=0$. From the linear equivalences between the divisors, we find the following relations between the curves $C_i$ and the seven compact curves of our geometry: 
$C_1=D_1\cdot E_2=D_2\cdot E_2,\ C_2=E_2\cdot E_3,\ C_3=D_1\cdot E_1=D_2\cdot E_1,\ E_1\cdot E_2=2\,C_1+C_2,\ D_3\cdot E_1=2\,C_1+C_2+4\,C_3$. From these relations and $(P\,|\,Q)$, we can get all triple intersection numbers, \eg $D_3 E_1^2=2\,C_1\cdot E_1+C_2\cdot E_1+4\,C_3\cdot E_1=-6$.
Table \ref{table:inter} gives the intersections of all compact curves with the divisors.
\begin{table}
\begin{center}
\begin{tabular}
{|c|cccccc|}\hline
{\rm Curve}&$D_1$&$D_2$&$D_3$&$E_1$&$E_2$&$E_3$\cr
\noalign{\hrule}\noalign{\hrule}
$E_1\cdot E_2$&1&1&0 &2&\!\!\!-4&0\cr
$E_2\cdot E_3$&1&1&0&0&0&\!\!\!-2\cr
$D_1\cdot E_1$&0&0&1&\!\!\!-2&1&0\cr
$D_1\cdot E_2$&0&0&0&1&\!\!\!-2&1\cr
$D_2\cdot E_1$&0&0&1&\!\!\!-2&1&0\cr
$D_2\cdot E_2$&0&0&0&1&\!\!\!-2&1\cr
$D_3\cdot E_1$&1&1&4&\!\!\!-6&0&0\cr
\noalign{\hrule}
\end{tabular}
\end{center}
\caption{Triple intersection numbers of the blow--up of $\IZ_{6-II}$}\label{table:inter}
\label{tab:inter}
\end{table}

Using the linear equivalences, we can also find the triple self--intersections of the compact exceptional divisors: $E_1^3=E_2^3=8$.

From the intersection numbers in $Q$, we find that $\{E_1+2\,D_3, D_2, D_3\}$ form a basis of the K\"ahler cone which is dual to the basis $\{C_1,C_2,C_3\}$ of the Mori cone.


\section{Divisor topologies, Part I}\label{sec:ExcepTop}

There are two types of exceptional divisors: The compact divisors, whose corresponding points lie in the interior of the toric diagram, and the semi-compact ones whose points sit on the boundary of the toric diagram. The latter case corresponds to the two--dimensional situation with an extra non--compact direction, hence it has the topology of $\IC \times \IP^1$ with possibly some blow--ups.

We first discuss the compact divisors. For this purpose we use the notion of the star of a cone $\sigma$, in terms of which the topology of the corresponding divisor is determined. The star, denoted ${\rm Star}(\sigma)$ is the set of all cones $\tau$ in the fan $\Sigma$ containing $\sigma$. This means that we simply remove from the fan $\Sigma$ all cones, i.e. points and lines in the toric diagram, which do not contain $w_i$. The diagram of the star is not necessarily convex anymore. Then we compute the linear relations and the Mori cone for the star. This means in particular that we drop all the simplices $S_k$ in the induced triangulation of the star which do not lie in its toric diagram. As a consequence, certain linear relations of the full diagram will be removed in the process of determining the Mori cone.  The generators of the Mori cone of the star will in general be different from those of $\Sigma$. 

Once we have obtained the Mori cone of the star, we can rely on the classification of compact toric surfaces: Any toric surface is either a $\IP^2$, a Hirzebruch surface $\IF_n$, or a toric blow--up thereof. The generator of the Mori cone of $\IP^2$ has the form 
$$Q^T = \left(\begin{array}{cccc}-3& 1& 1& 1\end{array}\right).$$ 
For $\IF_n$, the generators take the form $$Q^T = \left(\begin{array}{ccccc}-2 & 1 & 1 & 0 & 0\cr-n-2 & 0 & n & 1 & 1\end{array}\right) \qquad {\rm or} \qquad  Q^T = \left(\begin{array}{ccccc} -2 & 1 & 1 & 0 & 0\cr n-2 & 0 & -n & 1 & 1\end{array}\right)$$ since $\IF_{-n}$ is isomorphic to $\IF_n$. Finally, every toric blow--up of a point adds an additional independent relation whose form is $$Q^T = \left(\begin{array}{cccccc}0 & ... & 0 &1 &1 &-2\cr\end{array}\right).$$ We will denote the blow--up of a surface $S$ in $n$ points by $\Bl{n}S$. 

However, this is not yet the full story, since our toric variety $X_{\widetilde{\Sigma}}$ is actually three-dimensional. In particular, the stars are in fact cones over a polygon. Therefore, we have an additional possibility for a toric blow--up. We can add a point to the polygon such that the corresponding relation is of the form $$Q^T = \left(\begin{array}{ccccccc}0 & ... & 0 &1 &1 &-1 &-1\end{array}\right).$$ This corresponds to adding a cone over a lozenge and is well-known from the resolution of the conifold singularity. The lozenge has to be subdivided into two simplices, and there are two ways of doing this. The process of going from one way to the other is known as a flop and reverses the signs of the corresponding relation. It also affects some of the other relations. 
The curve $C_-$ that is flopped is the intersection of two divisors, say $E_1$ and $E_2$. If any other curve $C$ intersects one of these two divisors, i.e. $C\cdot E_i \not=0$, the new relation corresponding to $C$ is the sum of the relation of $C_-$ and $C$. Topologically, this means that an exceptional curve $C_-$ is blown down and another one, $C_+$, is blown up. As a consequence, we have to include these blow--ups in the list of surfaces given above. In addition, we have to include topologies that can be obtained by flopping a curve in a surface of this enlarged list of surfaces. 

Also the semi-compact exceptional divisors can be dealt with using the star. Since the geometry is effectively reduced by one dimension, the only compact toric manifold in one dimension is $\IP^1$ and the corresponding generator is $$Q^T = \left(\begin{array}{cccc}-2& 1& 1& 0\cr\end{array}\right),$$ where the 0 corresponds to the non-compact factor $\IC$.


\subsection{Example A.1: $\IC^3/\IZ_{6-I}$}\label{sec:exsixiaaa}

We now determine the topology of the exceptional divisors for our example $\IC^3/\IZ_{6-I}$.
\begin{figure}[h!]
\begin{center}
\includegraphics[width=140mm]{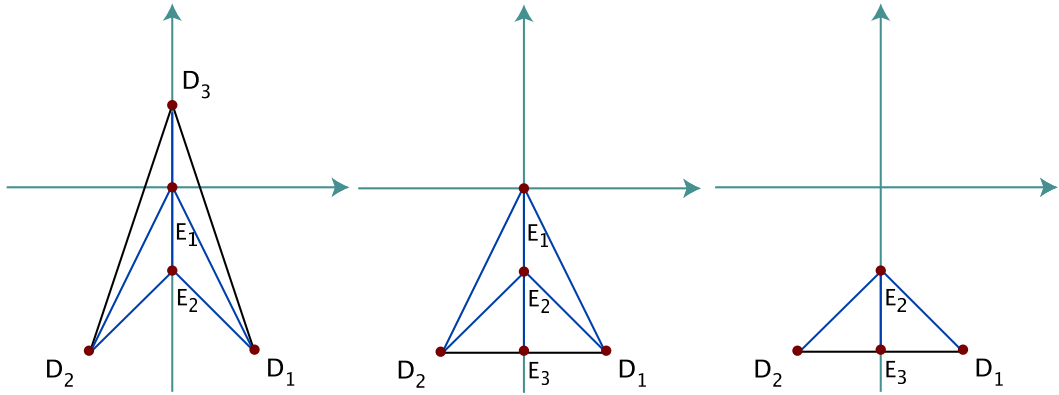}
\caption{The stars of the exceptional divisors $E_1$, $E_2$, and $E_3$, respectively.}
\label{fig:fstarsixi}
\end{center}
\end{figure}
As explained above, we need to look at the respective stars which are displayed in Figure \ref{fig:fstarsixi}.
In order to determine the Mori generators for the star of $E_1$, we have to drop the cones involving $E_3$ which are $S_1$ and $S_4$. From the seven relations in (\ref{eq:Moripairs}) only four remain, those corresponding to $C_3$, $2\,C_1 + C_2$ and $2\,C_1+C_2+4\,C_3$. These are generated by $2\,C_1+C_2=(1,1,0,2,-4,0)$ and $C_3=(0,0,1,-2,1,0)$ which are the Mori generators of $\IF_4$. Similarly, for the star of $E_2$ only the relations not involving $S_3$ and $S_6$ remain. These are generated by $C_1$ and $C_2$, and using (\ref{eq:Morigensixi}) we recognize them to be the Mori generators of $\IF_2$. Finally, the star of $E_3$ has only the relation corresponding to $C_3$. Hence, the topology of $E_3$ is $\IP^1\times \IC$, as it should be, since the point sits on the boundary of the toric diagram of $X_\Sigma$ and no extra exceptional curves end on it.

\subsection{Example B.1: $\IC^3/\IZ_{6-II}$}\label{sec:divsixii}

\begin{figure}[h!]
\begin{center}
\includegraphics[width=140mm]{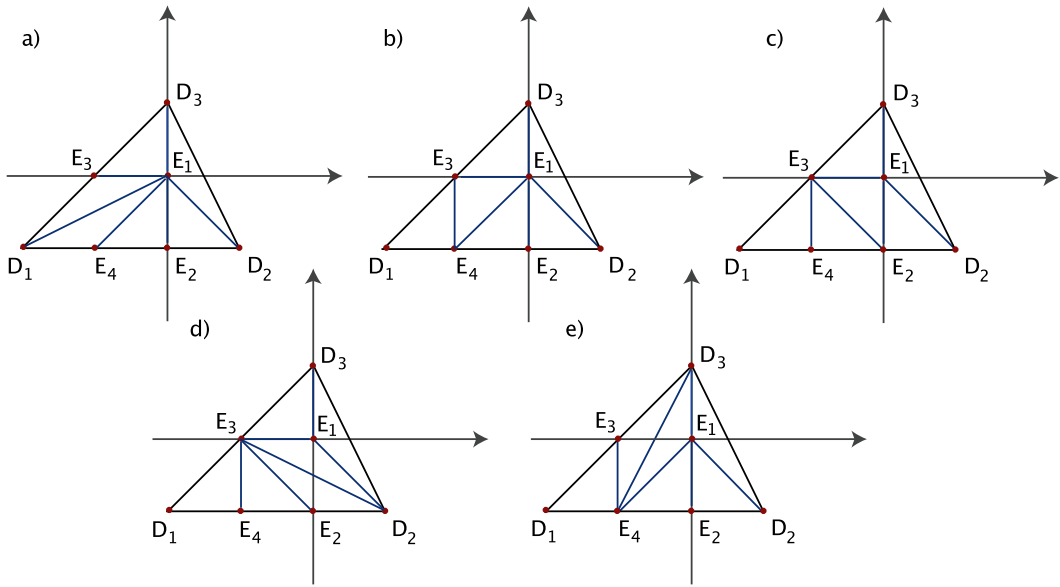}
\caption{The five different triangulations of the toric diagram of the resolution of $\IC^3/\IZ_{6-II}$}
\label{fig:sixiifive}
\end{center}
\end{figure}

We briefly give another example to illustrate the relation between different triangulations of a toric diagram. The resolution of $\IC^3/\IZ_{6-II}$ is given in Appendix \ref{app:rzsixii}. The toric diagram allows five different triangulations, i.e. five different resolutions. Figure \ref{fig:sixiifive} gives the five toric diagrams.

We start out with triangulation a). When the curve $D_1\cdot E_1$ is blown down and the curve $E_3\cdot E_4$ is blown up instead, we have gone through a flop transition and arrive at the triangulation b). From b) to c) we arrive by performing the flop $E_1\cdot E_4 \to E_2\cdot E_3$. From c) to d) takes us the flop $E_1\cdot E_2 \to D_2\cdot E_3$.
The last triangulation e) is produced from b) by flopping $E_1\cdot E_3 \to D_3\cdot E_4$. Thus, all triangulations are related to each other by a series of birational transformations.

We now identify the topologies of the exceptional divisors. The only compact exceptional divisor is $E_1$. In the triangulation c), we recognize its star to be the fan of an ${\IF}_1$. In the other triangulations, $E_1$ is birationally equivalent to ${\IF}_1$. In b), we have ${\IF}_1$ with one blow--up, in a) ${\IF}_1$ with two blow--ups. In d), the star of $E_1$ is the fan of $\IP^2$, while in e) $E_1$ again has the topology of ${\IF}_1$.

In triangulation a), all non-compact exceptional divisors have the topology of $\IP^1\times \IC$.


\section[The big picture]{The big picture: Gluing the patches}

In the easy cases, say in the prime orbifolds $\IZ_3$ and $\IZ_7$, it is obvious how the smooth manifold is obtained: Just put one resolved patch in the location of every fixed point and you are finished. Since these patches only have internal points, the corresponding exceptional divisors are compact, hence cannot see each other, and no complications arise from gluing.

Fixed lines which do not intersect any other fixed lines and on top of which no fixed points sit also pose no problem.

But what happens, when we have fixed lines on top of which fixed points are sitting? As discussed already, such a fixed point already knows it sits on a fixed line, since on the edge of the toric diagram of its resolution is the number of exceptional divisors appropriate to the fixed line the point sits on top of. Internal exceptional divisors are unproblematic in this case as well, since they do not feel the global surrounding. The exceptional divisors on the edges are identified or glued together with those of the corresponding resolved fixed lines.

The larger the order of the group, the more often it happens that a point or line is fixed under several group elements. How are we to know which of the patches we should use?
In the case of fixed lines answer is: Use the patch that belongs to the generator of the largest subgroup under which the patch is fixed, because the line is fixed under the whole sub-goup and its exceptional divisors already count the contributions from the other group elements. For fixed points, the question is a little more tricky. One possibility is to count the number of group elements this point is fixed under, not counting anti-twists and elements that generate fixed lines. Then choose the patch with the matching number of interior points. The other possibility is to rely on the schematic picture of the fixed set configuration and choose the patch according to the fixed lines the fixed point sits on. Isolated fixed points correspond to toric diagrams with only internal, compact exceptional divisors. When the fixed point sits on a fixed line of order $k$, its toric diagram has $k-1$ exceptional divisors on one of its boundaries. If the fixed point sits at the intersection of two (three) fixed lines, it has the appropriate number of exceptional divisors on two (three) of its boundaries. The right number of interior points together with the right number of exceptional divisors sitting on the edges uniquely determines the correct patch. Even though the intersection points of three $\IZ_2$ fixed lines are not fixed under a single group element, they must be resolved. The resolution of such a point is the resolution of $\IC^3/\IZ_2\times \IZ_2$ and its toric diagram is indeed the only one without interior points. 

\subsection{Example A: $\IZ_{6-I}$ on $G_2^2\times SU(3)$}\label{sec:exsixiglue}

This example is rather straightforward. We must again use the data of Table \ref{tab:fssixi} and the schematic picture of the fixed set configuration \ref{fig:ffixedi}. Furthermore, we need the resolved patches of $\IC^3/\IZ_{6-I}$ (see Section \ref{sec:rzsixi}, in particular Figure \ref{fig:fsixi}),  $\IC^3/\IZ_{3}$ (see Appendix \ref{app:rzthree}, in particular Figure \ref{fig:frthree}), and the resolution of the $\IZ_2$ fixed line, see Appendix \ref{app:rzctwo}.
The three $\IZ_6$--patches contribute two exceptional divisors each: $E_{1,\gamma}$, and $E_{2,1,\gamma}$,  where $\gamma=1,2,3$ labels the patches in the $z^3$--direction.
The exceptional divisor $E_3$ on the edge is identified with the one of the resolved fixed line the patch sits upon, as we will see.

There are furthermore 15 conjugacy classes of $\IZ_3$ fixed points. Blowing them up leads to a contribution of one exceptional divisor as can be seen from Figure~\ref{fig:frthree}. Since three of these fixed points sit at the location of the $\IZ_{6-I}$ fixed points which we have already taken into account ($E_{2,1,\gamma}$), we only count 12 of them, and denote the resulting divisors by $E_{2,\mu,\gamma},\ \mu = 2,\dots,5, \, \gamma=1,2,3$. The invariant divisors are built according to the conjugacy classes in~(\ref{eq:zthreeconja}): 
\begin{align}
  \label{eq:E2conj}
  E_{2,2,\gamma} &= \Et_{2,1,2,\gamma} +  \Et_{2,1,3,\gamma}, & E_{2,3,\gamma} &=  \Et_{2,3,1,\gamma} +  \Et_{2,5,1,\gamma},\notag\\
  E_{2,4,\gamma} &= \Et_{2,3,2,\gamma} +  \Et_{2,5,3,\gamma}, & E_{2,5,\gamma} &=  \Et_{2,3,3,\gamma} +  \Et_{2,5,2,\gamma}.
\end{align}
where $\Et_{2,\alpha,\beta,\gamma}$ are the representatives on the cover.
Finally, there are 6 conjugacy classes of fixed lines of the form $\IC^2/\IZ_2$. We see that after the resolution, each class contributes one exceptional divisor $E_{3,\alpha}, \alpha=1,2$. On the fixed line at $\zf{1}{1}=\zf{2}{1}=0$ sit the three $\IZ_{6-I}$ fixed points. The divisor coming from the blow--up of this fixed line, $E_{3,1}$, is identified with the three exceptional divisors corresponding to the points on the boundary of the toric diagram of the resolution of $\IC^3/\IZ_{6-I}$ that we mentioned above. 
The other exceptional divisors are built as invariant combinations according to the conjugacy classes in~(\ref{eq:ztwoconj}):
\begin{align}
  \label{eq:E3conj}
  E_{3,1} &= \Et_{3,1,1},                             & E_{3,2} &= \Et_{3,1,2} + \Et_{3,1,4} + \Et_{3,1,6}, \notag\\
  E_{3,3} &= \Et_{3,2,1} + \Et_{3,4,1} + \Et_{3,6,1}, & E_{3,4} &= \Et_{3,2,2} + \Et_{3,4,4} + \Et_{3,6,6}, \notag\\
  E_{3,5} &= \Et_{3,2,4} + \Et_{3,4,6} + \Et_{3,6,2}, & E_{3,6} &= \Et_{3,2,6} + \Et_{3,4,2} + \Et_{3,6,4}.
\end{align}

In total, this adds up to $3\cdot2+12\cdot1+6\cdot1=24$ exceptional divisors, which is the number which is given for $h^{(1,1)}_{twisted}$ in Table \ref{table:one}.

\subsection{Example C: $T^6/\IZ_{6}\times\IZ_6$}\label{sec:exsixsix}

This, being the point group of largest order, is the most tedious of all examples. It is presented here to show that the procedure is not so tedious after all.

First, the fixed sets must be identified. Table \ref{fssixsix} summarizes the results, some more details can be found in Appendix \ref{app:sixsix}.

\begin{table}[h!]\begin{center}
\begin{tabular}{|c|c|c|c|}
\hline
Group el.& Order &Fixed Set& Conj. Classes \cr
\hline
\noalign{\hrule}\noalign{\hrule}
$ \theta^1$&6    &1\ {\rm fixed\ line} &\ 1\cr
$ (\theta^1)^2$&3        &9\ {\rm fixed\ lines} &\ 4\cr
$ (\theta^1)^3$&2       &16\ {\rm fixed\ lines} &\ 4\cr
$ \theta^2$&6     &1\ {\rm fixed\ line} &\ 1\cr
$ (\theta^2)^2$&3     &9 \ {\rm fixed\ lines} &\ 4\cr
$ (\theta^2)^3$&2    &16 \ {\rm fixed\ lines} &\ 4\cr
$ \theta^1\theta^2$&$6\times6 $   &3\ {\rm fixed\ points} &\ 2\cr
$ \theta^1(\theta^2)^2$&$6\times 3$    &12\ {\rm fixed\ points} &\ 4 \cr
$ \theta^1(\theta^2)^3$&$6\times2  $  &12\ {\rm fixed\ points} &\ 4\cr
$ \theta^1(\theta^2)^4$&$6\times6  $ &3\ {\rm fixed\ points} &\ 2\cr
$ \theta^1(\theta^2)^5$&6  &1\ {\rm fixed\ line} &\ 1\cr
$ (\theta^1)^2\theta^2$&$3\times6$   &12\ {\rm fixed\ points} &\ 4\cr
$(\theta^1)^3\theta^2$&$2\times6 $   &12\ {\rm fixed\ points} &\ 4\cr
$ (\theta^1)^4\theta^2$&$6\times6 $     &3\ {\rm fixed\ points} &\ 2\cr
$ (\theta^1)^2(\theta^2)^2$&$3\times3$  &27\ {\rm fixed\ points} &\ 9\cr
$ (\theta^1)^2(\theta^2)^3$&$3\times2$  &12\ {\rm fixed\ points} &\ 4\cr
$ (\theta^1)^2(\theta^2)^4$&{3} &9\ {\rm fixed\ lines} &\ 4\cr
$ (\theta^1)^3(\theta^2)^2$&$2\times3$  &12\ {\rm fixed\ points} &\ 4\cr
$ (\theta^1)^3(\theta^2)^3$&{2}    &16\ {\rm fixed\ lines} &\ 4\cr
\hline
\end{tabular}
\caption{Fixed point set for $\IZ_6\times \IZ_6$.}\label{fssixsix}
\end{center}\end{table}

\begin{figure}[p]
\begin{center}
\includegraphics[width=140mm]{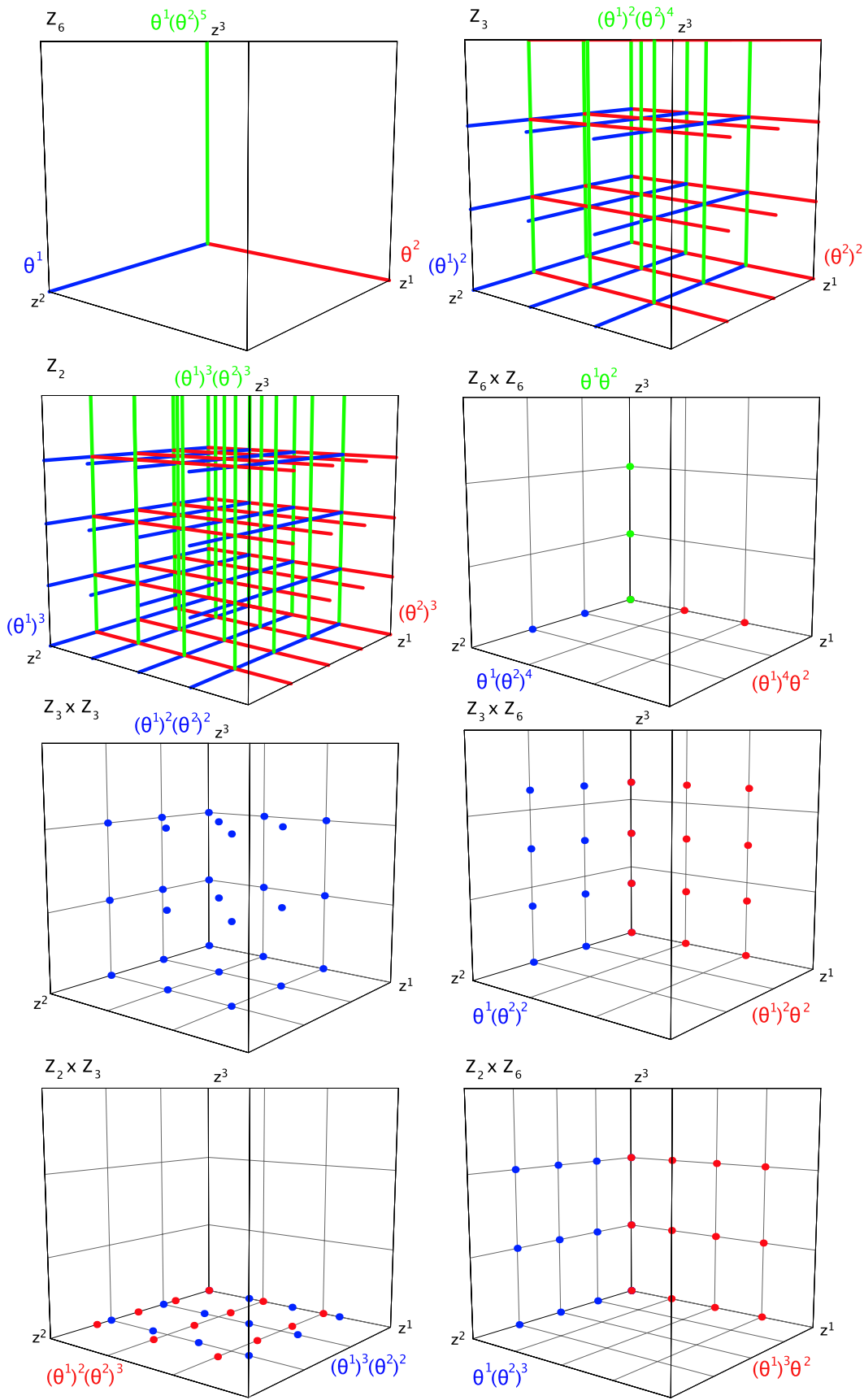}
\caption{Schematic picture of the fixed set configuration of $\IZ_6\times\IZ_6$}\label{ffixsixsix}
\end{center}
\end{figure}
Figure \ref{ffixsixsix} shows the schematic picture of the fixed set configuration. Again, it is the covering space that is shown, the representants of the equivalence classes are highlighted.

\begin{figure}[h!]
\begin{center}
\includegraphics[width=100mm]{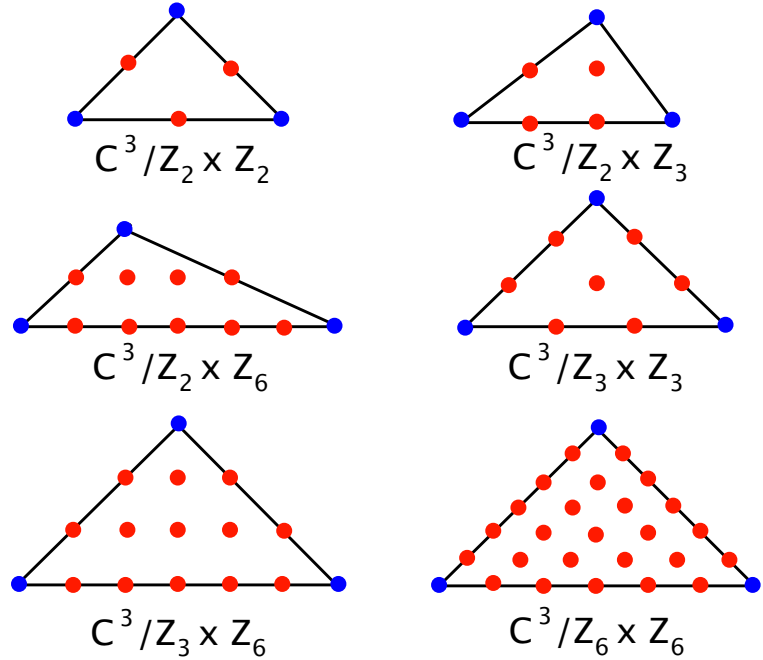}
\caption{Toric diagrams of patches for $T^6/\IZ_6\times\IZ_6$}\label{ingredients}
\end{center}
\end{figure}

Now we are ready to glue the patches together. Figure \ref{ingredients} schematically shows all the patches that will be needed in this example.
It is easiest to first look at the fixed lines. There are three $\IZ_6$ fixed lines, each contributing five exceptional divisors. Then there are twelve equivalence classes of $\IZ_3$ fixed lines, three of which coincide with the $\IZ_6$ fixed lines. The latter need not be counted, since they are already contained in the divisor count of the $\IZ_6$ fixed lines. The $\IZ_3$ fixed lines each contribute two exceptional divisors. Furthermore, there are twelve equivalence classes of $\IZ_2$ fixed lines, three of which again coincide with the $\IZ_6$ fixed lines. They give rise to one exceptional divisor each. From the fixed lines originate in total $3\cdot5+(12-3)\cdot2+(12-3)\cdot1=42$ exceptional divisors.

Now we study the fixed points. We associate the patches to the fixed points according to the intersection of fixed lines on which they sit. The exceptional divisors on the boundaries of their toric diagrams are identified with the divisors of the respective fixed lines. There is but one fixed point on the intersection of three $\IZ_6$ fixed lines. It is replaced by the resolution of the $\IC^2/\IZ_6\times\IZ_6$ patch, which contributes ten compact internal exceptional divisors. There are three equivalence classes of fixed points on the intersections of one $\IZ_6$ fixed line and two $\IZ_3$ fixed lines. They are replaced by the resolutions of the $\IC^2/\IZ_3\times\IZ_6$ patch, which contribute four compact exceptional divisors each. 
Then, there are five equivalence classes of fixed points on the intersections of three $\IZ_3$ fixed lines. They are replaced by the resolutions of the $\IC^2/\IZ_3\times\IZ_3$ patch, which contribute one compact exceptional divisors each. Furthermore, there are three equivalence classes of fixed points on the intersections of one $\IZ_6$ fixed line and two $\IZ_2$ fixed lines. They are replaced by the resolutions of the $\IC^2/\IZ_2\times\IZ_6$ patch, which contribute two compact exceptional divisors each. The rest of the fixed points sit on the intersections of one $\IZ_2$ and one $\IZ_3$ fixed line. There are six equivalence classes of them. They are replaced by the resolutions of the $\IC^2/\IZ_2\times\IZ_3$ patch, which is the same as the $\IC^2/\IZ_{6-II}$ patch, which contribute one compact exceptional divisors each. On the intersections of three $\IZ_2$ fixed lines sit resolved $\IC^2/\IZ_2\times\IZ_2$ patches, but since this patch has no internal points, it doesn't contribute any exceptional divisors which were not already counted by the fixed lines. The fixed points therefore yield $1\cdot10+3\cdot4+5\cdot1+3\cdot2+6\cdot1=42$ exceptional divisors. From fixed lines and fixed points together we arrive at 81 exceptional divisors.


\section{The inherited divisors}

So far, we have mainly spoken about the exceptional divisors which arise from the blow--ups of the singularities. In the local patches, the other natural set of divisors are the $D$--divisors, which descend from the local coordinates $\tilde z^i$ of the $\IC^3$--patch. On the compact space, i.e. the resolution of $T^6/\Gamma$, the $D$s are not the natural quantities anymore. The natural quantities are the divisors $R_i$ which descend from the covering space $T^6$ and are dual to the untwisted $(1,1)$--forms of the orbifold.
As discussed in Section \ref{sec:shape}, the three forms $dz^i\wedge d\ov z^i,\ i=1,2,3$ are invariant under all twists. For each pair $n_i=n_j$ in the twist (\ref{cplxtwist}), the forms $dz^i\wedge d\ov z^j$ and $dz^j\wedge d\ov z^i$ are invariant as well.
 
The inherited divisors $R_i$ together with the exceptional divisors $E_{k,\alpha,\beta,\gamma}$ form a basis for the divisor classes of the resolved orbifold. 

The $D$--divisors, which in the local patches are defined by $\tilde z^i=0$ are in the compact manifold defined by
\begin{equation}\label{defD}
D_{i\alpha} = \{ z^i = \zf{i}{\alpha} \},
\end{equation}
where $\alpha$ runs over the fixed loci in the $i$th direction. Therefore, they correspond to planes localized at the fixed points in the compact geometry.

The three "diagonal" $R_i$ dual to $dz^i\wedge d\ov z^i,\ i=1,2,3$ correspond to fixed planes parallel to the $D$s which can sit everywhere {\it except} at the loci of the fixed points. They are defined as $\{ z^i = c \not = \zf{i}{\alpha} \}$ and are "sliding" divisors in the sense that they can move away from the fixed point. $c$ corresponds to their position modulus.  
We need, however, to pay attention whether we use the local coordinates $\tilde z^i$ near the fixed point on the orbifold or the local coordinates $z^i$ on the cover. Locally, the map is $\tilde z^i = \left(z^i\right)^{n_i}$, where $n_i$ is the order of the group element that fixes the plane $D_i$. 
On the orbifold, the $R_i\ ,i=1,2,3$ are defined as
\begin{equation}\label{defR}
R_i=\{ \tilde z^i = c^{n_i} \},\quad c\neq \zf{i}{\alpha}.
\end{equation}
On the cover, they lift to a union of $n_i$ divisors $R_i = \bigcup_{k=1}^{n_i} \{ z^i = \varepsilon^k c\}$ with $\varepsilon^{n_i} =1$.

\begin{figure}[h!]
\begin{center}
\includegraphics[width=80mm]{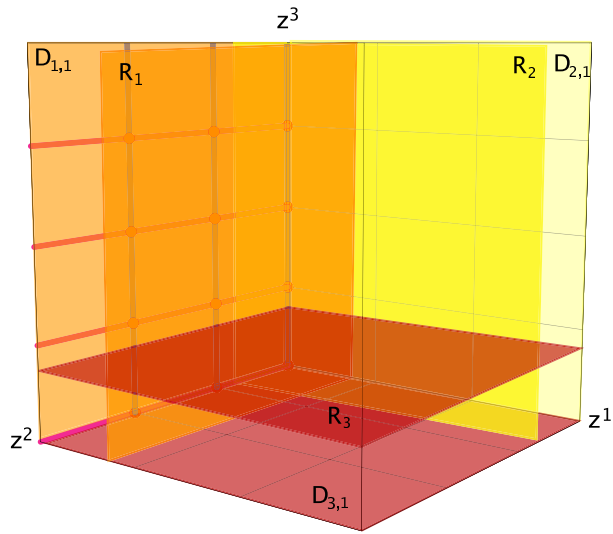}
\caption{Schematic picture of $D$- and $R$-divisors}
\label{fig:divisors}
\end{center}
\end{figure}
Figure \ref{fig:divisors} shows the schematic representation of three of the $D$ divisors and the three diagonal inherited divisors $R_i$. The figure shows the fixed set of $\IZ_{6-II}$ on $SU(2)\times SU(6)$, but this is not essential.


To relate the $R_i$ to the $D_i$, consider the local toric patch before blowing up. The fixed point lies at $c=\zf{i}{\alpha}$ and in the limit as $c$ approaches this point we find $R_i \sim n_i\,D_i$. This expresses the fact that the polynomial defining $R_i$ on the cover has a zero of order $n_i$ on $D_i$ at the fixed point. In the local toric patch $R_i \sim 0$, hence $n_i\,D_i \sim 0$. After blowing up, $R_i$ and $n_i\,D_i$ differ by the exceptional divisors $E_k$ which appear in the process of resolution. The difference is expressed precisely by the linear relation in the $i$th direction~(\ref{eq:linrels}) of the resolved toric variety $X_{\widetilde\Sigma}$ and takes the form 
\begin{equation}
  \label{eq:Reqlocal}
  R_i \sim n_i\,D_i + \sum_k E_k.
\end{equation}
This relation is independent from the chosen triangulation. Since such a relation holds for every fixed point $\zf{i}{\alpha}$, we add the label $\alpha$ which denotes the different fixed sets in the $i$--direction. Furthermore, we have to sum over all fixed sets which lie in the respective fixed plane $D_{i,\alpha}$:
\begin{equation}
  \label{eq:Reqglobal}
  R_i \sim n_i\,D_{i,\alpha} + \sum_{k, \beta} E_{k\alpha\beta} \qquad {\rm \;for\; all\; } \alpha {\rm \; and \; all\;} i,
\end{equation}
where $n_i$ is the order of the group element that fixes the plane $D_{i,\alpha}$. The precise form of the sum over the exceptional divisors depends on the singularities involved.

In general, an orbifold of the form $T^6/G$ has local singularities of the form $\IC^m/H$, where $H$ is some subgroup of index $p = [G:H]$ in $G$. If $H$ is a strict subgroup of $G$, the above discussion applies in exactly the same way and yields relations (\ref{eq:Reqlocal}) for divisors $R'_i$ with vanishing orders $n_i'$. In the end, however, it must be taken into account that $H$ is a subgroup, which means that the relations for the $R_i'$ with the action of $H$ must be embedded into those involving the $R_i$ with the action of $G$. The $R_i'$ are related to the $R_i$ by
\begin{equation}
  \label{eq:ReqGH}
  R_i = \frac{|G|}{|H|} R_i' = p\, R_i'.
\end{equation}

When a set is fixed only under a strict subgroup $H\subset G$, its elements are mapped into each other by the generator of the normal subgroup $G/H$. Therefore, the equivalence classes of invariant divisors must be considered. They are represented by $S = \sum_{\alpha} \widetilde{S}_{\alpha}$, where $\widetilde{S}_{\alpha}$ stands for any divisor $\Dt_{i\alpha}$ or $\Et_{k\alpha\beta}$ on the cover and the sum runs over the $p$ elements of the coset $G/H$. In this case, we can add up the corresponding relations:
$$
  \sum_\alpha R_i' \sim n_i'\sum_{\alpha} \Dt_{i\alpha} + \sum_{k,\beta} \sum_{\alpha} \Et_{k\alpha\beta}
$$
The left hand side is equal to $p\, R_i' = R_i$, therefore
\begin{equation}
  \label{eq:ReqH}
  R_i \sim n_i' D_i + \sum_{k,\beta} E_{k\beta}
\end{equation}
which is the same as the relation for $R_i'$. 

Something special happens if $n_i = n_j = n$ for $i \not= j$. In this situation, there are additional divisors on the cover, $R_{ij} = \bigcup_{k=1}^n \{z^i + \varepsilon^k z^j = \varepsilon^{k+k_0} c^{ij} \}$ for some integer $k_0$ and some constant $c_{ij}$, which descend to divisors on the orbifold. We have $\varepsilon^n = 1$ for even $n$, and $\varepsilon^{2n} = 1$ for odd $n$. Since the natural basis for $H^2(T^6)$ are the forms $h_{i\jbar}$ (see the previous subsection), we have to combine the various components of the $R_{ij}$ in a particular way in order to obtain divisors $R_{i\jbar}$ which are Poincar\'e dual to these forms. If we define the variables
\begin{align}
  \label{eq:zij}
  z^{ij}_{\pm} &= z^i \pm z^j, & z'_{\pm}{}^{ij} &= z^i \pm \varepsilon z^j,\\
  z^{ij}_{k} &= z^i + \varepsilon^k z^j, 
\end{align}
then 
\begin{align}
  \label{eq:Rij}
  R_{i\jbar} &= \{z^{ij}_+ + \bar{z}^{ij}_- = c^{ij} \} \cup \{z^{ij}_+ - \bar{z}^{ij}_- = c^{ij} \} \cup \{ z'_+{}^{ij} + \bar{z}'_-{}^{ij} = c^{ij} \} \cup \{ z'_+{}^{ij} - \bar{z}'_-{}^{ij} = c^{ij} \}.
\end{align}
These divisors again satisfy linear relations of the form~\eqref{eq:Reqglobal}:
\begin{align}
  \label{eq:Rijeqglobal}
  R_{i\jbar} \sim n D_{i\jbar\alpha} + \sum_{k, \beta, \gamma} E_{k\alpha\beta\gamma}.
\end{align}

\subsection{Example A: $T^6/\IZ_{6-I}$ on $G_2^2\times SU(3)$}\label{sec:relexA}

This example combines several complications: More than three inherited exceptional divisors, several kinds of local patches for the fixed points, and fixed sets which are in orbits with length greater than one.

The $D$--planes are $\Dt_{1,\alpha} = \{ z^1=\zf{1}{\alpha}\}$, $\alpha=1,\dots,6$, $\Dt_{2,\beta} = \{z^2=\zf{2}{\beta}\}$, $\beta=1,...,6$, and $\Dt_{3,\gamma} = \{z^3 = \zf{3}{\gamma}\}$, $\gamma=1,2,3$ on the cover. From these, we define the invariant combinations
\begin{align*}
  D_{1,1} &= \Dt_{1,1}, & D_{1,2} &= \Dt_{1,2} + \Dt_{1,4} + \Dt_{1,6}, &
  D_{1,3} &= \Dt_{1,3} + \Dt_{1,5}, \\
  D_{2,1} &= \Dt_{2,1} & D_{2,2} &= \Dt_{2,2} + \Dt_{2,4} + \Dt_{2,6} & D_{2,3} &= \Dt_{2,3} + \Dt_{2,5}    \\
  D_{3,\gamma} &= \Dt_{3,\gamma}.
\end{align*}

Now, we will construct the global linear relations~(\ref{eq:Reqglobal}). The $D_{1,1}$--plane contains three equivalence classes of $\IZ_{6-I}$--patches, three equivalence classes of $\IZ_{3}$--patches, and two equivalence classes of $\IZ_2$--fixed lines. From the local relations of $\IC^2/\IZ_n$ (\ref{eq:lineqsctwo}), we find the local relation to $R_1$ as in~(\ref{eq:ReqH}) (here, we already changed the labels of the divisors to match  the labels of the $\IZ_{6-I}$--patch):
\begin{equation}
  \label{eq:Z6IrelD12}
  R_1 = 2\,D_{{1,1}}+\, E_{{3,1}}.
\end{equation}
With this, the global relation is obtained from the local relation of $\IC^3/\IZ_{6-I}$ (\ref{eq:lineqsixI}), and the local relation of $\IC^3/\IZ_{3}$ (\ref{lineqthree}) :
\begin{equation}
  \label{eq:exArelD11} 
R_1=6\,D_{{1,1}}+\sum_{\gamma=1}^3E_{{1,\gamma}} +2\, \sum_{\mu=1}^2 \sum_{\gamma=1}^3 E_{{2,\mu,\gamma}}+3\, \sum_{\nu=1,2} E_{{3,\nu}}.
\end{equation}
The divisor $D_{1,2}$ only contains two equivalence classes of $\IZ_2$ fixed lines:
\begin{equation}
R_1=2\,D_{{1,2}}+\sum_{\nu=3}^6 E_{{3,\nu}}.
\end{equation}
Next, we look at the divisor $D_{1,3}$, which only contains $\IZ_3$ fixed points. The local linear equivalences~(\ref{lineqthree}) together with~(\ref{eq:ReqH}) lead to
\begin{equation}
  \label{eq:Z6Irel13}
  R_1 = 3\,D_{{1,3}}+\sum_{\mu=3}^5 \sum_{\gamma=1}^3 E_{{2,\mu,\gamma}}.
\end{equation}
The linear relations for $D_{2,\beta}$ are the same as those for $D_{1,\alpha}$: 
\begin{eqnarray}
  \label{eq:exarelD2}
  R_2&=&6\,D_{{2,1}}+\sum_{\gamma=1}^3 E_{{1,\gamma}}+2\, \sum_{\mu=1,3} \sum_{\gamma=1}^3 E_{{2,\mu,\gamma}}+3\, \sum_{\nu=1,3} E_{{3,\nu}},\nonumber\\
  R_2&=&2\,D_{{2,2}}+\, \sum_{\nu=2,4,5,6} E_{{3,\nu}}, \nonumber \\
  R_2&=&3\,D_{{2,3}}+\sum_{\mu=2,4,5} \sum_{\gamma=1}^3 E_{{2,\mu,\gamma}}.
\end{eqnarray}
Finally, the relations for $D_{3,\gamma}$ are again obtained from~(\ref{eq:lineqsixI}):
\begin{equation}
  \label{eq:exArelD3}
  R_3=3\,D_{{3,\gamma}}+2\, E_{1,\gamma} + \sum_{\mu=1}^5 E_{{2,\mu,\gamma}} \qquad \gamma=1,\dots,3.
\end{equation}


\section{The intersection ring}\label{sec:intersections}

There is a purely combinatorial way to determine the intersection ring of the resolved torus orbifold. This method is completely analogous to the one given in Section~\ref{sec:mori} for the local patches. Recall that first, the intersection numbers between three distinct divisors were determined, and then the linear relations were used to compute all the remaining intersection numbers. In the global situation we proceed in the same way. 

With the local and global linear relations worked out in the last section at our disposal, we can determine the intersection ring as follows: First we compute the intersection numbers including the $R_i$ between distinct divisors as well as the Stanley--Reisner ideal from a local compactification of the blown--up singularity. 
Then, we make use of the schematic picture of the fixed set configuration, see Section \ref{sec:schematic}, from which we can read off which of the divisors coming from different fixed sets never intersect. With the necessary input of all intersection numbers with three different divisors, all other intersection numbers can be determined by using the global linear equivalences (\ref{eq:Reqglobal}). 

To relate the inherited divisors to the divisors of a local patch, the patch must be compactified.
For this discussion, we focus on one specific patch, i.e. we fix $\alpha$ and drop it from the notation for the time being. 
For the compactification of the blow--up of $\IC^3$, we choose $\left( \IP^1 \right)^3$. Now we can again invoke the methods of toric geometry. 
We start with a lattice $N \cong \IZ^3$ with basis $f_i = m_i e_i$, where $e_i$ is the standard basis. The $m_i$ are positive integers that have to be chosen such that $m_1m_2m_3 = n_1n_2n_3 /|G|$ and the $n_i$ are the same as in~(\ref{eq:Reqlocal}). We construct an auxiliary polyhedron $\Delta^{(3)}$ by taking the cone $C_{\Delta^{(2)}}$ from Section~\ref{sec:mori} and rotating and rescaling it such that the vertices corresponding to the divisors $D_i$ lie at $v_{i+3} = n_i f_i$, $i=1,2,3$. Then we add the vertices $v_i = -f_i$ corresponding to the divisors $R_i$, $i=1,2,3$. The points $v_{k+6}$ corresponding to the exceptional divisors $E_k$ are now located on the face $\langle v_4, v_5, v_6 \rangle$. It is easy to check that the linear relations of the polyhedron $\Delta^{(3)}$ are precisely~(\ref{eq:Reqlocal}). We require the triangulation to be a star triangulation, i.e. all simplices contain the origin, and that the triangulation of the simplex $\langle 0, v_4, v_5, v_6 \rangle$ be induced from the triangulation of the cone $C_{\Delta^{(2)}}$. Computing the intersection numbers for three distinct divisors by determining the volume of the corresponding simplex with respect to the standard basis $e_i$ yields the local intersection numbers of the global orbifold. The local Stanley--Reisner ideal, i.e. the set of those divisors which do not intersect because they belong to different cones can be immediately read off from the auxiliary polyhedron.

Note, that this procedure equally applies to resolutions of fixed points and fixed lines. In the latter case, we start with the two--dimensional cone $C_{\Delta^{(1)}} \subset N'_{\IR} \cong \IR^2$ obtained from the resolution of the fixed line at the intersection of say $D_1$ and $D_2$. We extend the underlying lattice to $N = \IZ \oplus N' \cong \IZ^3$. Then we add the generator $v_3 = (1,0,0)$ corresponding to the divisor $D_3$ intersecting the fixed line in a point. (The indices of the $D_i$ have to be permuted according to the global coordinates of the singularity.) In this way, we obtain the cone $C_{\Delta^{(2)}} = \{0\} \times C_{\Delta^{(1)}} \cup v_3$ which is the input for the construction of $\Delta^{(3)}$ above. 

For the local patches corresponding to singularities of the form $\IC^m/H$ with $H$ a strict subgroup of $G$, the auxiliary polyhedron $\Delta^{(3)}_H$ is obtained by modifying the polyhedron $\Delta^{(3)}_G$ for $\IC^3/G$. For this, we observe that the exceptional divisors coming from the resolution of $\IC^m/H$ always form a subset of those coming from the resolution of $\IC^3/G$. Hence, we simply drop those points in $\Delta^{(3)}_G$ which do not correspond to an exceptional divisor coming from the resolution of $\IC^m/H$. 

If the equivalence class corresponding to the divisors $D_{i\alpha}$ or $E_{k\alpha\beta\gamma}$ has more than one element, we have two possibilities: Either we work with the representatives on the cover and plug in the invariant combination at the end of the calculation, or we work with the invariant divisors and modify the polyhedra accordingly. The second possibility reduces the calculational cost considerably, so we concentrate on this one. The second possibility reduces the calculations by a large amount, so we concentrate on this one. The modification of the polyhedron is determined by the linear relations~(\ref{eq:Reqlocal}) with $R_i=R_i'$ and $n_i=n_i'$ for the corresponding local singularity $\IC^m/H$. This amounts to dividing the $i$th component of $v_k$, $k\geq 4$, by $p$ such that the modified polyhedron also satisfies~(\ref{eq:Reqlocal}). If it happens that two or more conjugacy classes of the fixed point set, i.e. two or more exceptional divisor classes $E_{k,\alpha}$ lie at the same locus, we have to multiply the corresponding generator $v_{k+3}$ by the number of components. For fixed lines, we have to work with as many copies of the corresponding polyhedron as there are components.

We construct the auxiliary polyhedron $\Delta^{(3)}$ for every equivalence class of the fixed point set, and add the labels $\alpha, \beta, \gamma$ denoting the fixed point set to the divisors $D_i$ and $E_k$. The lattice $N$ is the same for all the polyhedra. In this way, we get all the intersection numbers $S_{abc}$ between between distinct divisors from the over-complete set $\{ S_a \} = \{R_i, D_{i,\alpha}, E_{k,\alpha,\beta,\gamma} \}$. The presence of the $R_i$ in all the auxiliary polyhedra ensures the correct relative normalizations of the intersection numbers in the different patches. The choice $m_i$ of the lattice basis fixes the overall normalization. In addition, we have the local Stanley--Reisner ideal. There is a global analogue of the Stanley--Reisner ideal. It is the set of all pairs of divisors with indices $i,\alpha$ and $i,\alpha'$ with $\alpha \not= \alpha'$. The divisors in such a pair never intersect since they lie at disjoint fixed point sets $\alpha$ and $\alpha'$, respectively.

Using the linear relations~(\ref{eq:Reqglobal}) which take the general form $\sum_{a} n_s S_a = 0$, we can construct a system of equations for the remaining intersection numbers involving two equal divisors $S_{aab}$ and three equal divisors $S_{aaa}$ by multiplying the linear relations by all possible products $S_bS_c$. This yields a highly overdetermined system of equations 
\begin{equation}\label{system}
\sum_a n_a S_{abc} = 0,
\end{equation} 
whose solution determines all the remaining intersection numbers. The information contained in the local and global Stanley--Reisner ideals simplifies this system greatly, since most of these equations are trivially satisfied after setting the corresponding intersections to zero.

The intersection ring can also be determined without solving the system of equations (\ref{system}). All that is needed are the intersection numbers obtained from the compactified local patches and the configuration of the fixed sets. If such a patch has no exceptional divisors on the boundary of the uncompactified toric diagram, the intersection numbers of these exceptional divisors remain unchanged in the global setting. 
If the intersection number involves exceptional divisors on the boundary of the toric diagram, the local intersection number must be multiplied with the number of patches which sit on the fixed line to which the exceptional divisor belongs.


As often the case, there is a more direct but equivalent way to obtain the intersection numbers which does not involve the polyhedra: The intersections between distinct divisors $D_{i\alpha}$ and $E_{k\alpha\beta\gamma}$ are those computed in the local patch, see Section~\ref{sec:mori}. The intersections between $R_j$ and $D_{i\alpha}$ are easily obtained from their defining polynomials on the cover. The intersection number between $R_1$, $R_2$, and $R_3$ is simply the number of solutions to $\{\left(\widetilde{z}^1\right)^{n_1} = c_1^{n_1}, \left(\widetilde{z}^2\right)^{n_2} = c_2^{n_2}, \left(\widetilde{z}^3\right)^{n_3} = c_3^{n_3}\}$ which is $n_1n_2n_3$. Taking into account that we calculated this on the cover, we need to divide by $|G|$ in order to get the result on the orbifold. Similarly, the divisors $D_{i\alpha}$ are defined by linear equations in the $\widetilde{z}^i$, hence we set the corresponding $n_i$ to 1. Therefore,
\begin{align}
  \label{eq:R1R2R3}
  R_1R_2R_3 &= \frac{1}{|G|} n_1n_2n_3 & R_iR_jD_{k\alpha} &= \frac{1}{|G|} n_in_j & R_iD_{j\alpha}D_{k\beta} = \frac{n_i}{|G|}
\end{align}
for $i,\, j,\, k$ pairwise distinct, and all $\alpha$ and $\beta$. Furthermore, $R_i$ and $D_{i\alpha}$ never intersect by definition. The only remaining intersection numbers involving both $R_j$ and $D_{i\alpha}$ are of the form $R_jD_{i\alpha}E_{k\alpha\beta\gamma}$. They vanish if $D_{i\alpha}$ and $E_{k\alpha\beta\gamma}$ do not intersect in the local toric patch, otherwise they are 1. Finally, there are the intersections between $R_i$ and the exceptional divisors. If the exceptional divisor lies in the interior of the toric diagram or on the boundary adjacent to $D_{i\alpha}$, it cannot intersect $R_i$. Also, $R_iR_jE_{k\alpha\beta\gamma} = 0$. The above can also be seen directly from a schematic picture such as Figure \ref{fig:divisors}, combined with the toric diagrams of the local patches. 

Using this procedure it is also straightforward to compute the intersection numbers involving the divisors $R_{i\jbar}$ and $D_{i\jbar}$. From the defining polynomials in~\eqref{eq:Rij} we find that the only non--vanishing intersection numbers are
\begin{align}
  \label{eq:R1R12R21}
  R_{i\jbar}R_{j\ibar}R_k &= -\frac{1}{|G|} n_i^2n_k, & D_{i\jbar\alpha}R_{j\ibar}R_k &= -\frac{1}{|G|} n_i n_k, & R_{i\jbar}R_{j\ibar}D_{k\alpha} &= -\frac{1}{|G|} n_i^2,\notag\\
  D_{i\jbar\alpha}D_{j\ibar\beta}R_k & = -\frac{1}{|G|} n_k, &  D_{i\jbar\alpha}R_{j\ibar}D_{k\beta} & = -\frac{1}{|G|} n_i, & D_{i\jbar\alpha}D_{j\ibar\beta}D_{k\gamma} &= -\frac{1}{|G|}, \notag\\
  R_{i\jbar}R_{j\kbar}R_{k\ibar} &= \frac{1}{|G|} n_i^3, &  R_{i\jbar}R_{j\kbar}D_{k\ibar\alpha} &= \frac{1}{|G|} n_i^2, &  R_{i\jbar}D_{j\kbar\alpha}D_{k\ibar\beta} &= \frac{1}{|G|} n_i, \notag\\
  D_{i\jbar\alpha}D_{j\kbar\beta}D_{k\ibar\gamma} &= \frac{1}{|G|},
\end{align}
for $i,\, j,\, k$ pairwise distinct, and all $\alpha$, $\beta$, and $\gamma$. The negative signs come from carefully taking into account the orientation reversal due to complex conjugation.

\subsection{Example A: $T^6/\IZ_{6-I}$ on $G_2^2\times SU(3)$}\label{exAint}

After the preparations of Section \ref{sec:relexA}, we are ready to compute the intersection ring for this example. First, we need to determine the basis for the lattice $N$ in which the auxiliary polyhedra will live. From~(\ref{eq:exArelD11}), (\ref{eq:exarelD2}), and~(\ref{eq:exArelD3}) we see that $n_1=n_2=6$, and $n_3=3$. Hence we can choose $m_1=m_2=3$, and $m_3=2$ and the lattice basis is $f_1=(3,0,0)$, $f_2=(0,3,0)$, $f_3=(0,0,2)$. We start with the polyhedron $\Delta_1^{(3)}$ for the $\IZ_{6-I}$ fixed points. Its lattice points are
\begin{align}
  \label{eq:Z6Ipoly}
  v_1 &= (-3,0,0), & v_2 &= (0,-3,0), & v_3 &= (0,0,-2), & v_4 &= (18,0,0), & v_5 &= (0,18,0),\notag\\
  v_6 &= (0,0,6), & v_7 &= (3,3,4), & v_8 &= (6,6,2), & v_9 &= (9,9,0),
\end{align}
corresponding to the divisors $R_1,R_2,R_3,D_1,D_2,D_3,E_1,E_2,E_3$ in that order. 
\begin{figure}[h!]
\begin{center}
\includegraphics[width=155mm]{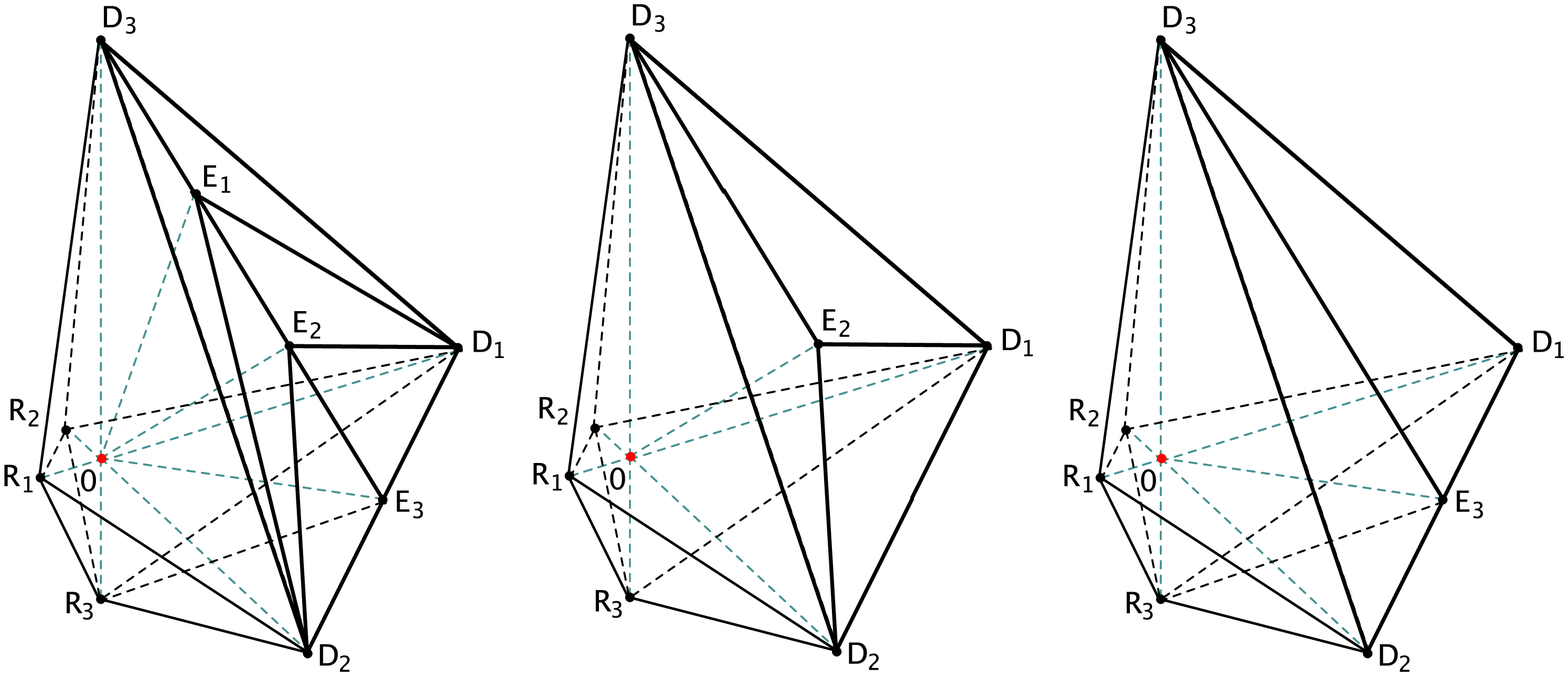}
\caption{The polyhedra of the local compactifications for the resolutions of $\IC^3/\IZ_{6-I}$, $\IC^3/\IZ_{3}$ and $\IC^2/\IZ_2\times\IC$.}
\label{fig:Z6I-cpt}
\end{center}
\end{figure}
By applying the methods described at the end of the last section, we obtain the following intersection numbers between three distinct divisors:
\begin{align}
  \label{eq:intZ6I}
  R_1R_2R_3 &= 18, & R_1R_2D_3 &= 6, & R_1R_3D_2 &=3, & R_1D_2D_3 &= 1, \notag\\
  R_2R_3D_1 &=  3, & R_2D_1D_3 &= 1, & R_3D_1E_3 &=1, & R_3D_2E_3 &= 1, \notag\\
  D_1E_1D_3 &=  1, & D_1E_1E_2 &= 1, & D_1E_2E_3 &=1, & D_2D_3E_1 &= 1, \notag\\
  D_2E_1E_2 &=  1, & D_2E_2E_3 &= 1,  
\end{align}
and the local Stanley--Reisner ideal
\begin{gather}
  \left\{R_iD_i = 0, R_iE_1 = 0, R_iE_2 = 0, R_1E_3 = 0, R_2E_3 = 0, \right.\notag\\
  \left. D_1D_2 = 0, D_3E_2 = 0, D_3E_3 = 0, E_1E_3 = 0,\; i=1,2,3 \right\}.  
  \label{eq:SRIZ6I}
\end{gather}
Now, we add the labels $\alpha,\beta,\gamma$ of the fixed points to the divisors: $D_i \to D_{i\alpha}$, $E_1 \to E_{1\gamma}$, $E_2 \to E_{2\alpha\beta\gamma}$, $E_3 \to E_{3\alpha}$, and set $\alpha=1, \beta=1, \gamma=1,2,3$. 

As explained in the last section, the polyhedra for the other patches are obtained from $\Delta_1^{(3)}$ by dropping or rescaling some of the points. For the $\IZ_3$ patches, we drop $v_7$ and $v_9$, for those at $\mu=2$ we set $v_5 = (0,9,0)$, $v_8 = (6,3,2)$. For those at $\mu=3$ we set $v_4 = (9,0,0)$, $v_8 = (3,6,2)$, and finally for those at $\mu=4,5$ we set $v_4 = (9,0,0), v_5 = (0,9,0), v_8 = (6,6,4)$. For the $\IZ_2$ fixed lines, we drop $v_7$ and $v_8$. For the fixed line at $\nu=3$ we set $v_5=(0,6,0), v_9=(3,9,0)$, while for those at $\nu=4,5,6$ we set $v_4=(6,0,0), v_5=(0,6,0), v_9=(9,9,0)$. The three polyhedra are depicted in Figure \ref {fig:Z6I-cpt}. Computing the analogues of~(\ref{eq:intZ6I}) and~(\ref{eq:SRIZ6I}) yields all the local information we need. The global information comes from the linear relations~(\ref{eq:Z6IrelD11}) to~(\ref{eq:Z6Irels3}) and the examination of Figure~\ref{fig:ffixedi} to determine those pairs of divisors which never intersect. Solving the resulting overdetermined system of linear equations then yields the intersection ring of $X$ in the basis $\{R_i, E_{k\alpha\beta\gamma}\}$:
\begin{align}
  \label{eq:ringZ6Ia}
  R_1R_2R_3 &= 18, & R_3E_{3,1}^2 &= -2, & R_3E_{3,\nu}^2 &= -6, & E_{1,\gamma}^3 &= 8, \notag\\
  E_{1,\gamma}^2 E_{2,1,\gamma} &= 2, & E_{1,\gamma}E_{2,1,\gamma}^2 &= -4, & E_{2,1\gamma}^3 &= 8, & E_{2,\mu,\gamma}^3 &= 9, \notag\\
  E_{2,1,\gamma}E_{3,1}^2 &= -2, & E_{3,1}^3 &= 8,
\end{align}
for $\mu=2,\dots,5$, $\nu=2,\dots,6$, $\gamma=1,2,3$.


\section{Divisor topologies, Part II}\label{sec:divtopII}

In Section \ref{sec:ExcepTop}, the topology of the compact factors of the exceptional divisors was determined in the setting of the local non--compact patches. Here, we discuss the divisor topologies in the compact geometry of the resolved toroidal orbifolds, i.e. in particular the topologies of the formerly non--compact $\IC$--factor of the semi--compact exceptional divisors and the topologies of the $D$--divisors about which we could not say anything in the local toric setting.

We begin by discussing the exceptional divisors. Their topology depends on the structure of the fixed point set they originate from. The following three situations can occur:
\renewcommand{\labelenumi}{E\theenumi)}
\begin{enumerate}
  \item Fixed points
  \label{item:E1}
  \item Fixed lines without fixed points
  \label{item:E2}
  \item Fixed lines with fixed points on top of them
  \label{item:E3}
\end{enumerate}
In addition, we must distinguish between equivalence classes of the fixed set consisting of a single element or more. We first discuss the case of a single element. The topology of the divisors in case~E\ref{item:E1}) has already been discussed in great detail in Section~\ref{sec:ExcepTop}. The local topology the divisors in the cases~E\ref{item:E2}) and~E\ref{item:E3}) has also been discussed in that section, and found to be (a blow--up of) $\IC\times\IP^1$. The $\IC$ factor is the local description of the $T^2/\IZ_k$ curve on which there were the $\IC^2/\IZ_m$ singularities whose resolution yielded the $\IP^1$ factor. 

For the determination of the topology of the resolved curves, it is necessary to know the topology of $T^2/\IZ_k$. This can be determined from the action of $\IZ_k$ on the respective fundamental domains. For $k=2$, there are four fixed points at $0, 1/2, \tau/2$, and $(1+\tau)/2$ for arbitrary $\tau$. The fundamental domain for the quotient can be taken to be the rhombus $[0,\tau,\tau+1/2,1/2]$ and the periodicity folds it along the line $[\tau/2,(1+\tau)/2]$. Hence, the topology of $T^2/\IZ_2$ without its singularities is that of a $\IP^1$ minus 4 points. For $k=3,4,6$ the value of $\tau$ is fixed to be $i, \exp(\frac{2\pi i}{3}), \exp(\frac{2\pi i}{6})$, respectively, and the fundamental domains are shown in Figure~\ref{fig:fundomains} (note that here the fundamental regions are in part chosen differently than those in the appendices). 

\begin{figure}[h!]
\begin{center}
\includegraphics[width=150mm]{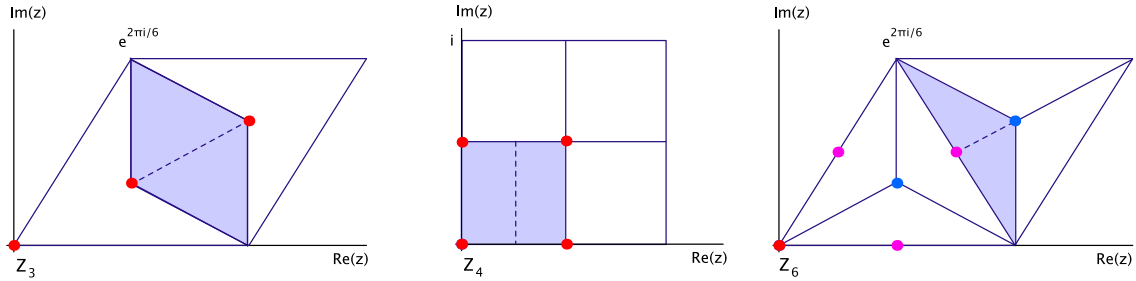}
\caption{The fundamental domains of $T^2/\IZ_k$, $k=3,4,6$. The dashed line indicates the folding.}
\label{fig:fundomains}
\end{center}
\end{figure}
From this figure, we see that the topology of $T^2/\IZ_k$ for $k=3,4,6$ is that of a $\IP^1$ minus 3, 2, 3 points, respectively. In the case~E\ref{item:E2}), there are no further fixed points, so the blow--up procedure merely glues points into this $\IP^1$. The topology of such an exceptional divisor is therefore the one of $\IF_0 = \IP^1 \times \IP^1$. In the case~E\ref{item:E3}), the topology further depends on the fixed points lying on these fixed lines. This depends on the choice of the root lattice for $T^6/G$, and can therefore only be discussed case by case. This will be done for the examples in the next subsection and the appendices. The general procedure consists of looking at the corresponding toric diagram. There will always be an exceptional curve whose line ends in the point corresponding to the exceptional divisor. 
This exceptional curve meets the $\IP^1$ (minus some points) we have just discussed in the missing points and therefore, the blow--up adds in the missing points. Any further lines ending in that point of the toric diagram correspond to additional blow--ups, i.e. additional $\IP^1$s that are glued in at the missing points. Therefore, for each fixed point lying on the fixed line and each additional line in the toric diagram there will be a blow--up of $\IF_0=\IP^1 \times \IP^1$.

If there are $p$ elements in the equivalence class of the fixed line, the topology is quite different for the case~E\ref{item:E2}). This is because the $p$ different $T^2/\IZ_k$'s are mapped into each other by the corresponding generator in such a way that the different singular points are permuted. When the invariant combinations are constructed by summing over all representatives, the singularities disappear and we are left with a $T^2$. Hence, in the case~E\ref{item:E2}) without fixed points, the topology of $E= \sum_{\alpha=1}^k \Et_\alpha$ is $\IP^1 \times T^2$. 

Similarly, the topology of the divisors $D_{i\alpha}$ depends on the structure of the fixed point sets lying in the divisor. We again treat first the case of $D$ lying in an equivalence class containing only one element. Recall that these divisors are defined by $D_{i\alpha} = \{z^i = \zf{i}{\alpha}\}$. The orbifold group $G$ acts on these divisors by $(z_j,z_k) \to (\varepsilon^{n_j}z_j,\varepsilon^{n_k} z_k)$ for $(z_j,z_k)\in D_{i\alpha}$ and $j\not=i\not=k$. Since $n_j+n_k = n - n_i < n$, the resolved space will not be a Calabi--Yau manifold anymore, but a rational surface. This happens because for resolutions of this type of action, the canonical class cannot be preserved. (In more mathematical terms, the resolution is not crepant.) In order to determine the topology, we will use a simplicial cell decomposition, remove the singular sets, glue in the smoothening spaces, i.e. perform the blow--ups, and use the additivity of the Euler number. This has to be done case by case. If, in particular, the fixed point set contains points, there will be a blow--up for each fixed point and for each line in the toric diagram of the fixed point which ends in the point corresponding to $D_i$. Another possibility is to apply the techniques of toric geometry given  in Section~\ref{sec:resolve} to singularities of the form $\IC^2/\IZ_n$ for which $n_1+n_2 \not= n$. 

When $D$ lies in an equivalence class with $p>1$ elements, the basic topology again changes to $\IP^1\times T^2$.

Note that when embedding the divisor $D$ into a (Calabi--Yau) manifold $X$ in general, not all the divisor classes of $D$ are realized as classes in $X$. In the case of resolved torus orbifolds, this happens because the underlying lattice of $D$ is not necessarily a sublattice of the underlying lattice of $X$. This means that the fixed point set of $D$ as a $T^4$--orbifold can be larger than the restriction of the fixed point set of the $T^6$--orbifold to $D$. In order to determine the topology of $D$, we have to work with the larger fixed point set of $D$ as a $T^4$--orbifold. It turns out that there is always a lattice defining a $T^6$--orbifold for which all divisor classes of $D$ are also realized in $X$. In fact, we observe that the topology of all those divisors which are present in several different lattices is independent of the lattice. 

The divisors $R_i$ contain by definition no component of the fixed point set. However, they can intersect fixed lines in points. If there are no fixed lines piercing them, the action of the orbifold group is free and their topology is that of a $T^4$. Otherwise, the intersection points have to be resolved in the same way as for the divisors $D_{i\alpha}$. In this case, the topology is always that of a K3 surface.

We can also use the intersection ring to study the topology of these divisors. If we describe the divisor $S$ (which can be of any type, i.e. $R$, $E$ or $D$ above) of the Calabi--Yau manifold $X$ by an embedding $i:S\longrightarrow X$, we have the associated short exact sequence for the tangent bundles $T_S$ and $T_X$ and the normal bundle $N_{S/X}$ of $S$ in $X$
\begin{equation}
  \label{eq:embedding}
  \shortexactseq{T_S}{}{T_X|_S}{}{N_{S/X}}.
\end{equation}
By the adjunction formula~\cite{Griffiths:1994ab} $N_{S/X} \cong \Oc(S)|_S$ we can relate the topology of $S$ to that of $X$ as follows. Expanding $\ch(T_X) = \ch(T_S)\ch(N_{S/X})$ and using the restriction formula $\int_S \omega = \int_X \omega \wedge S$ 
we obtain
\begin{align}
  \label{eq:c1(S)}
  \ch_1(S) &= -S \\
  \label{eq:c2(S)}
  \ch_1(N_{S/X})^2 &= S^2 \quad = \ch_2(X) - \ch_2(S) \\
  \label{eq:c2.S}
  \ch_2(X)\cdot S + S^3 &= \chi(S),
\end{align}
which gives the relation between the Chern classes of $S$ and the topological numbers of $X$. Furthermore, we have the holomorphic Euler characteristic of $S$
\begin{equation}
  \label{eq:holoEuler}
  \chi(\Oc_S) = 1 - h^{(1,0)}(S) + h^{(2,0)}(S).
\end{equation}
Noether's formula~\cite{Griffiths:1994ab} relates $\chi(\Oc_S)$ to the Chern classes of $S$:
\begin{equation}
  \label{eq:Noether}
  \chi(\Oc_S) = \frac{1}{12} \int_S \left(\ch_1(S)^2 + \ch_2(S)\right),
\end{equation}
from which we get
\begin{equation}
  \label{eq:S3}
  12\, \chi(\Oc_S) = S^3 + \chi(S).
\end{equation}
This equation can be used in two ways. Since we already determined the topologies of the divisors $R_i$, $D_{i\alpha}$ and $E_{k\alpha\beta\gamma}$, i.e. the values of $\chi(\Oc_S)$ and $\chi(S)$, we can cross--check them with the calculation for the self--intersection numbers in the intersection ring. We can also explicitly check the number of blow--ups of $S$. On the one hand, it is known~\cite{Griffiths:1994ab} that the holomorphic Euler characteristic is a birational invariant, i.e. it does not change under blow--ups. On the other hand, blowing up a surface adds a 2--cycle to it, hence increases the Euler number $\chi(S)$ by 1. Therefore, the self--intersection number $S^3$ is decreased by 1. Furthermore, $S^3$ restricted to $S$ becomes $S^2 = \ch_1(S)^2 = K_S^2$, where $K_S$ is the canonical divisor of $S$. Like $\chi(\Oc_S)$ and $\chi(S)$, $K_S^2$ is a characteristic quantity of a surface $S$. We have collected these three quantities for the basic topologies that we have found above in the following table:
\begin{equation}
  \label{eq:KS}
  \begin{array}{|c|c|c|c|c|}
    \hline
    S & \chi(S) & \chi(\Oc_S) & K_S^2 & h^{(1,0)}(S) \\
    \hline
    \IP^2 & 3 & 1 & 9 & 0\\
    \IF_n & 4 & 1 & 8 & 0\\
    \IP^1 \times T^2 & 0 & 0 & 0 & 1\\
    T^4 & 0 & 0 & 0 & 2\\
    \rm{K3} & 24 & 2 & 0 & 0 \\
    \hline
  \end{array}
\end{equation}
The invariants of the blow--ups of these surfaces are then obtained from the above observations. 

The second use of~(\ref{eq:S3}) is to determine $\ch_2\cdot S$ in~(\ref{eq:c2.S}) from the topology of $S$.

\subsection{Example A: $T^6/\IZ_{6-I}$ on $G_2^2\times SU(3)$}\label{sec:divsixiglobal}

Here, we discuss the topologies of the divisors of the resolution of $T^6/\IZ_{6-I}$ on $G_2^2\times SU(3)$. The topology of the compact exceptional divisors has been determined in Section~\ref{sec:exsixiaaa}: $E_{1,\gamma} = \IF_4$ and $E_{2,1,\gamma} = \IF_2$. By the remark at the end of appendix~\ref{app:rzthree}, $E_{2,\mu,\gamma}$, $\mu=2,\dots,5$, have the topology of a $\IP^2$. The divisor $E_{3,1}$ is of type~E\ref{item:E3}) and has a single representative, hence the basic topology is that of a $\IF_0$. There are 3 $\IZ_{6-I}$ fixed points on it, but there is only a single line ending in $E_3$ in the toric diagram of Figure~\ref{fig:fsixi}, which corresponds to the exceptional $\IP^1$, therefore there are no further blow--ups. The divisors $E_{3,\nu}$, $\nu=2,\dots,6$ are all of type~E\ref{item:E2}) with 3 representatives, hence their topology is that of $\IP^1\times T^2$.

The topology $D_{2,1}$ is determined as follows: The fixed point set of the action $\frac{1}{6}(1,4)$ agrees with the restriction of the fixed point set of $T^6/\IZ_{6-I}$ to $D_{2,1}$. The Euler number of $D_{2,1}$ minus the fixed point set is $(0-4\cdot 0-6\cdot 1)/6=-1$. The procedure of blowing up the singularities glues in 3 $\IP^1\times T^2$s at the $\IZ_2$ fixed lines which does not change the Euler number. The last fixed line is replaced by a $\IP^1 \times T^2$ minus 3 points, upon which there is still a free $\IZ_3$ action. Its Euler number is therefore $(0-3)/3 = -1$. The 6 $\IZ_3$ fixed points fall into 3 equivalence classes, furthermore we see from Figure~\ref{fig:frthree} that there is one line ending in $D_2$. Hence, each of these classes is replaced by a $\IP^1$, and the contribution to the Euler number is $3\cdot 2=6$. Finally, for the 3 $\IZ_{6-I}$ fixed points there are 2 lines ending in $D_2$ in the toric diagram in Figure~\ref{fig:fsixi}. At a single fixed point, the blow--up yields two $\IP^1$s touching in one point whose Euler number is $2\cdot 2 -1=3$. Adding everything up, the Euler number of $D_{2,1}$ is $-1 + 0 -1 + 6 + 3\cdot 3 = 13$ which can be viewed as the result of a blow--up of $\IF_0$ in 9 points. The same discussion as above also holds for $D_{1,1}$, however, there are no $\IZ_2$ fixed lines without fixed points. The topology of each representative of $D_{1,2}$ minus the fixed point set, viewed as a $T^4$ orbifold, is that of a $T^2 \times (T^2/\IZ_2 \setminus \{ 4\ \rm{pts} \})$. The representatives are permuted under the residual $\IZ_3$ action and the 12 points fall into 3 orbits of length 1 and 3 orbits of length 3. Hence, the topology of the class is still that of a $T^2 \times (T^2/\IZ_2 \setminus \{ 4 \ \rm{pts} \})$. After the blow--up it is therefore a $\IP^1 \times T^2$. The divisor $D_{2,2}$ has the same structure as $D_{1,2}$, therefore its topology is that of a $\IP^1\times T^2$. The topology of the divisors $D_{2,3}$ and $D_{1,3}$ is the same as the topology of $D_{i\alpha}$ in the $\IZ_3$ orbifold which is discussed in detail in Appendix~\ref{app:divtopZ3}. It can be viewed as a blow--up of $\IP^2$ in 12 points. Finally, there are the divisors $D_{3\gamma}$. The action $\frac{1}{6}(1,1)$ on $T^4$ has 24 fixed points, 1 of order 6, 15 of order 2, and 8 of order 3. The $\IZ_2$ fixed points fall into 5 orbits of length 3 under the $\IZ_3$ element, and the $\IZ_3$ fixed points fall into 4 orbits of length 2 under the $\IZ_2$ element. For each type of fixed point there is a single line ending in $D_3$ in the corresponding toric diagram, therefore the fixed points are all replaced by a $\IP^1$. The Euler number therefore is $(0-24)/6 + (1+5+4)\cdot 2 = 16$. Hence, $D_{3,\gamma}$ can be viewed as blow--up of $\IF_0$ in 12 points. 

The divisors $R_1$ and $R_2$ do not intersect any fixed lines lines, therefore they simply have the topology of $T^4$. The divisor $R_3$ has the topology of a K3. In Table~\ref{tab:TopZ6I}, we have summarized the topology of all the divisors. All the Euler numbers and types of surfaces we have determined above together with~(\ref{eq:ringZ6I}) agree with Noethers formula~(\ref{eq:S3}). 
\begin{table}[h!]
  \begin{center}
  $
  \begin{array}{c}
    \begin{array}{|c|c|c|c|c|}
      \hline
      E_{1\gamma} & E_{2,1\gamma} & E_{2\mu\gamma} & E_{3,1} & E_{3,2}         \\  
      \hline
      \IF_4       & \IF_2         & \IP^2          & \IF_0   & \IP^1\times T^2 \\
      \hline
    \end{array}
    \\
    \\
    \begin{array}{|c|c|c|c|c|c|c|c|c|}
      \hline
      D_{1,1}     & D_{1,2}         & D_{1,3}      & D_{2,1}     & D_{2,2}      & D_{3,\gamma} & R_1, R_2 & R_3 \\  
      \hline
      \Bl{9}\IF_n & \IP^1\times T^2 & \Bl{12}\IP^2 & \Bl{9}\IF_n & \Bl{12}\IP^2 & \Bl{12}\IF_n & T^4      & \rm{K3} \\
      \hline
    \end{array}
  \end{array}
  $
  \end{center}
  \caption{The topology of the divisors.}
  \label{tab:TopZ6I}
\end{table}
With the knowledge of the Euler numbers and the intersection ring we can determine the second Chern class $\ch_2$ on the basis $\{R_i,E_{k\alpha\beta\gamma}\}$ using~(\ref{eq:c2.S}):
\begin{align}
  \label{eq:c2Z6I}
  \ch_2\cdot E_{1,\gamma} &= -4, & \ch_2\cdot E_{2,1,\gamma} &= -4, & \ch_2 \cdot E_{2,\mu,\gamma} &= -6, & \ch_2\cdot E_{3,1} & = -4,\notag\\
  \ch_2\cdot E_{3,\nu} &= 0, & \ch_2 \cdot R_i &= 0, & \ch_2 \cdot R_3 = & 24. 
\end{align}
Since the second Chern class is a linear form on $H^2(X,\IZ)$ we can apply it to each of the linear relations in~(\ref{eq:Z6IrelD11}) to~(\ref{eq:Z6Irels3}) and again find complete agreement.

\subsection{Example B: $T^6/\IZ_{6-II}$ on $SU(2)\times SU(6)$}\label{sec:divsixiiglobal}

For the moment we only consider the triangulation a). The topology of the compact exceptional divisors was determined in Section~\ref{sec:divsixii}: $E_{1,\beta\gamma}$ was found to be $\Bl{2}\IF_1$. The remaining exceptional divisors are all of type~E\ref{item:E3}), hence the basic topology is $\IF_0$. Looking at Figure~\ref{fsixii}, we see that in the toric diagram of triangulation a) there is only one line ending in each of $E_2$, $E_3$, and $E_4$, which corresponds to the exceptional curve of $\IF_0$. Therefore there is no additional blow--up, and each of $E_{2\beta}$, $E_{3\gamma}$, and $E_{4\beta}$ has the topology of an $\IF_0$. 

The topology of $D_1$ is determined as follows: The action of $\frac{1}{6}(2,3)$ on $T^4$ factorizes and the topology of $D_1$ minus the fixed point set is that of $(T^2/\IZ_3 \setminus \{3 \ {\rm pts} \}) \times (T^2/\IZ_2 \setminus \{4 \ {\rm pts} \})$. Looking again at the toric diagram in Figure~\ref{fsixii}, we see that there is one line ending in $D_1$, hence the 12 singular points are replaced by a $\IP^1$. The topology of $D_1$ is therefore that of $\Bl{12}\IF_0$, and its Euler numbe is 16. (For the other triangulations there is no line ending in $D_1$, hence there is no blow--up and the topology is that of $\IF_0$.) For $D_{2\beta}$ the action of $\frac{1}{6}(1,3)$ on $T^4$ yields the $\IZ_3$ fixed line with 3 $\IZ_{6-II}$ fixed points on top of it as we can see in Figure~\ref{ffixsixiiaa}. In addition, there are 2 more $\IZ_3$ fixed lines which fall into an orbit of length 2 under the residual $\IZ_2$ action, as well as 12 $\IZ_2$ fixed points which fall into 4 orbits of length 3 under the residual $\IZ_3$ action. The latter two sets are not realized in the $T^6$--orbifold for this lattice. The Euler number of $D_{2\beta}$ minus the fixed point set is $(0-3\cdot 0-12)/6 = -2$. The blowing--up process glues in a $T^2 \times F$ at the class of the $\IZ_3$ fixed lines without fixed points, where $F$ are two $\IP^1$s intersecting in a point. There is no contribution to the Euler number from this space. The last fixed line is replaced by $T^2 \times F$ minus 4 points,  upon which there is still a free $\IZ_2$ action. Its Euler number is therefore $(0-4)/2 = -2$. For the 4 $\IZ_{6-II}$ fixed points on this fixed line, we see that in the corresponding toric diagram there is one line ending in $D_2$. (For triangulation d) there are two lines.) For a single fixed point, this contributes $\chi(\IP^1)=2$ to the Euler number. At the each of the four classes of $\IZ_2$ fixed points, we also glue in a $\IP^1$. Adding everything up, the Euler number of $D_{2\beta}$ is $-2 + 0 -2 + 4\cdot 2 + 4\cdot 2 = 12$, which can be viewed as the result of a blow--up of $\IF_0$ in 8 points. For $D_{3\gamma}$, the computation is similar. The action of $\frac{1}{6}(1,2)$ on $T^4$ yields the $\IZ_2$ fixed line with 4 $\IZ_{6-II}$ fixed points on top of it which we see in Figure~\ref{ffixsixiiaa}. In addition, there are 3 more $\IZ_2$ fixed lines which fall into an orbit of length 3 under the residual $\IZ_3$ action, as well as 6 $\IZ_3$ fixed points which fall into 3 orbits of length 2 under the residual $\IZ_2$ action. The latter two sets again are not realized in the $T^6$--orbifold for this lattice. The Euler number of $D_{2\beta}$ minus the fixed point set is $(0-4\cdot 0-6)/6 = -1$. The blowing--up process glues in a $T^2 \times \IP^1$ at the class of the $\IZ_2$ fixed line without fixed points. There is no contribution to the Euler number from this space. The last fixed line is replaced by $T^2 \times \IP^1$ minus 3 points, upon which there is still a free $\IZ_3$ action. Its Euler number is therefore $(0-3)/3 = -1$. For the 3 $\IZ_{6-II}$ fixed points on this fixed line, we see that in the corresponding toric diagram, there is one line ending in $D_2$. (For triangulation e) there are two lines.) For a single fixed point, this contributes $\chi(\IP^1)=2$ to the Euler number. At each of the three classes of $\IZ_3$ fixed points, we glue in two $\IP^1$s intersecting in one point whose Euler number is $2\cdot 2-1$. Adding everything up, the Euler number of $D_{3\gamma}$ is $-1 + 0 -1 + 3\cdot 2 + 3\cdot 3 = 13$, which can be viewed as the result of a blow--up of $\IF_0$ is 9 points.

The divisor $R_1$ does not intersect any fixed lines, therefore it simply has the topology of $T^4$. The divisors $R_2$ and $R_3$, on the other hand have the topology of a K3. In Table~\ref{tab:TopZ6II}, we have summarized the topology of all the divisors for all triangulations. All the Euler numbers and types of surfaces we have determined above together with~(\ref{eq:ringZ6I}) agree with Noethers formula~(\ref{eq:S3}). 
\begin{table}[h!]
  \begin{center}
  $
  \begin{array}{|c|c|c|c|c|c|c|c|c|c|}
    \hline
    {\rm Triang.} & E_{1,\beta\gamma} & E_{2,\beta}   & E_{3,\gamma}   & E_{4,\beta}   & D_1          & D_{2,\beta}  & D_{3,\gamma} & R_1 & R_2,R_3 \\  
    \hline
    {\rm a) }     & \Bl{2}\IF_1       & \IF_0         & \IF_0          & \IF_0         & \Bl{12}\IF_0 & \Bl{8}\IF_n  & \Bl{9}\IF_n  & T^4 & {\rm K3} \\
    {\rm b) }     & \IF_1             & \Bl{4}\IF_0   & \Bl{6}\IF_0    & \IF_0         & \IF_0        & \Bl{8}\IF_n  & \Bl{9}\IF_n  & T^4 & {\rm K3}\\ 
    {\rm c) }     & \Bl{1}\IF_1       & \IF_0         & \Bl{3}\IF_0    & \Bl{4}\IF_0   & \IF_0        & \Bl{8}\IF_n  & \Bl{9}\IF_n  & T^4 & {\rm K3}\\
    {\rm d) }     & \IF_1             & \IF_0         & \IF_0          & \Bl{8}\IF_0   & \IF_0        & \Bl{8}\IF_n  & \Bl{12}\IF_n & T^4 & {\rm K3}\\
    {\rm e) }     & \IP^2             & \IF_0         & \Bl{9}\IF_0    & \IF_0         & \IF_0        & \Bl{12}\IF_n & \Bl{9}\IF_n  & T^4 & {\rm K3}\\
    \hline
  \end{array}
  $
  \end{center}
  \caption{The topologies of the divisors for all triangulations.}
  \label{tab:TopZ6II}
\end{table}
With the knowledge of the Euler numbers and the intersection ring we can determine the second Chern class $\ch_2$ on the basis $\{R_i,E_{k\alpha\beta\gamma}\}$ by~(\ref{eq:c2.S}) (for triangulation a)):
\begin{align}
  \label{eq:c2Z6II}
  \ch_2\cdot E_{1\beta\gamma} &= 0, & \ch_2\cdot E_{2\beta} &= -4, & \ch_2 \cdot E_{3\gamma} &= -4, & \ch_2\cdot E_{4\beta} & = -4,\notag\\
  \ch_2 \cdot R_1 &= 0, & \ch_2 \cdot R_i = & 24. 
\end{align}
Since the second Chern class is a linear form on $H^2(X,\IZ)$, we can apply it to each of the linear relations in~(\ref{eq:Z6IrelD11}) to~(\ref{eq:Z6Irels3}) and again find complete agreement.


\section{The twisted complex structure moduli}
\label{sec:TwistedCplx}

There are two types of complex structure deformations: We can, loosely speaking, either deform the complex structure of the underlying torus, or the fixed point set. The former type of deformation is described by the untwisted complex structure moduli at the orbifold point. The latter is parametrized by the twisted complex structure moduli. 

To study the twisted complex structure moduli, we therefore have to look at the fixed sets. Isolated fixed points do not admit any complex structure deformations. Hence, only fixed lines need to be considered. However, if there are fixed points on them, their complex structure again cannot be deformed, it is constrained by the fixed points. So, we are left with fixed lines without fixed points on them. In the previous section, we have argued that the resolved singularities yield exceptional divisors which are ruled surfaces over a $\IP^1$ or a $T^2$. 

These ruled surfaces can also be viewed as an algebraic family of algebraic curves (here rational curves $\IP^1$) parametrized by the base curve $C$. For any smooth complex projective threefold $X$ with such a family of algebraic curves there is a map 
$$\varphi_*: H_1(C) \to H_3(X)$$
 which sends the 1--cycle $\gamma$ on $C$ to the 3--cycle $\varphi_*(\gamma)$ traced out by the fiber curve $E_t$ as $t$ traces out $\gamma$~\cite{Clemens:1972ab}. Since the fibers $E_t$ are algebraic cycles, the dual map on the cohomology respects the Hodge decomposition and yields a map 
 $$\varphi^*: H^{1,0}(C) \to H^{2,1}(X).$$ 
 For our varieties, $C$ is either a $\IP^1$ or a $T^2$. Since $h^{1,0}(\IP^1) = 0$ and $h^{1,0}(T^2) = 1$, only the ruled surfaces over $T^2$ give such a map. Hence we find that
\begin{equation}
  \label{eq:h21tw}
  h^{2,1}_{tw}(X) = \sum_{i} (n_i - 1) h^{1,0}(C_i)
\end{equation}
where the sum runs over the curves $C_i$ with topology $T^2$ parametrizing exceptional curves which come from the resolution of $\IC^2/\IZ_{n_i}$ singularities. To re-iterate in plain language, each equivalence class of order $n$ fixed lines without fixed points on them contributes as many twisted complex structure moduli as the fixed line has $\IP^1$--components in its resolution, namely $n-1$.

Note, that this situation has a well--known analogue for hypersurfaces in toric varieties. In this case, the complex structure deformations split into polynomial and nonpolynomial ones. The former correspond to deformations of the hypersurfaces while the latter correspond to deformations of the ambient toric variety. The nonpolynomial deformations also come from curves of $\IC^2/\IZ_n$ singularities. After resolution of the singularities, these become families of $\IP^1$s parametrized by the curve, in other words, ruled surfaces and a similar reasoning applies~\cite{Katz:1996ht}.


\chapter{From Calabi--Yau to Orientifold}\label{sec:orientifold}

The string theorist is often interested in type $II$ string theories with a setup with $D$-branes and ${\cal N}=1$ supersymmetry in four dimensions. Such a set-up necessarily leads to the orientifold theory.


\section[Yet another quotient]{Yet another quotient: The orientifold}

At the orbifold point, the orientifold projection is $\Omega\,I_6$, where $\Omega$ is the worldsheet orientation reversal and $I_6$ is an involution on the compactification manifold. In type IIB string theory with O3/O7--planes (instead of O5/O9), the holomorphic (3,0)--form $\Omega$ must transform as $\Omega \to -\Omega$. Therefore we choose
\begin{equation}
  \label{eq:I6}
  I_6:\ (z^1,z^2,z^3)\to (-z^1,-z^2,-z^3).
\end{equation}
Geometrically, this involution corresponds to taking a $\IZ_2$-quotient of the compactification manifold.

As long as we are at the orbifold point, all necessary information is encoded in (\ref{eq:I6}). To find the configuration of O3--planes, the fixed points under $I_6$ must be identified. On the covering space, $I_6$ always gives rise to 64 fixed points, i.e. 64 O3--planes. Some of them may be identified under the orbifold group $G$, such that there are less than 64 equivalence classes on the quotient. The O7--planes are found by identifying the fixed planes under the combined action of $I_6$ and the generators $\theta_{\IZ_2}$ of the $\IZ_2$ subgroups of $G$.
A point $x$ belongs to a fixed set, if it fulfills
\begin{equation}
  \label{eq:fixO}
  I_6\,\theta_{\IZ_2}\,x=x+a,\quad a\in \Lambda,
\end{equation}
where $\Lambda$ is the torus lattice.
Each $\IZ_2$ subgroup of $G$ gives rise to a stack of O7--planes. Therefore, there are none in the prime cases, one stack e.g. for $\IZ_{6-I}$ and three in the case of e.g. $\IZ_2\times \IZ_6$, which contains three $\IZ_2$ subgroups. The number of O7--planes per stack depends on the fixed points in the direction perpendicular to the O--plane and therefore on the particulars of the specific torus lattice.


\section{When the patches are not invariant: $h^{(1,1)}_-\neq 0$}\label{h11minus}

Whenever $G$ contains a subgroup $H$ of odd order, some of the fixed point sets of $H$ will not be invariant under the global orientifold involution $I_6$ and will fall into orbits of length two under $I_6$. This can also happen for groups of even order giving rise to fixed tori with non--trivial volume factors, see Section \ref{sec:nonstandardvols}. Some of these $I_6$--orbits may coincide with the $G$--orbits. In this case, no further effect arises. When $G$ contains in particular a $\IZ_2$ subgroup in each coordinate direction, all equivalence classes under $I_6$ and these subgroups coincide.
When certain fixed points or lines (which do not already form an orbit under $G$) are identified under the orientifold quotient, the second cohomology splits into an invariant and an anti--invariant part under $I_6$:
$$H^{1,1}(X)=H^{1,1}_{+}(X)\oplus H^{1,1}_{-}(X).$$
The geometry is effectively reduced by the quotient and the moduli associated to the exceptional divisors of the anti--invariant patches are consequently no longer geometric moduli. They take the form~\cite{Grimm:2005fa}
\begin{equation}
  \label{eq:newmoduli}
  G^a=C^a_2+S\,B^a_2.
\end{equation}
Tables \ref{table:hminusone} and \ref{table:hminustwo} give the values of $h^{(1,1}_{-}$ for all orbifolds.

\begin{table}[h!]
\begin{center}
\begin{center}
{\small
\begin{tabular}{|l|c|c|c|c|c|}
\hline
$\ \IZ_N$&Lattice $T^6$&$h_{(1,1)}$&$h_{(1,1)}^{-}$ \\ [2pt]
\hline
$\ \IZ_3 $    &\  $SU(3)^3 $          &36 &13\cr
$\  \IZ_4  $   &\ $SU(4)^2  $      &25 &6\cr
$\  \IZ_4 $    &\  $SU(2)\times SU(4)\times SO(5)$ &27&4\cr
$\  \IZ_4$      &\  $SU(2)^2\times SO(5)^2 $ &31  &0\cr
$\  \IZ_{6-I} $   & $(G_2\times SU(3)^{2})^{\flat}$  &25 &6\cr
$\  \IZ_{6-I}  $  &$ SU(3)\times G_2^2$ &29 &6\cr
$\  \IZ_{6-II} $  &$ SU(2)\times SU(6)$    &25  &6\cr
$\  \IZ_{6-II} $  &$SU(3)\times SO(8) $&29 &6\cr
$\  \IZ_{6-II} $  &$(SU(2)^2\times SU(3)\times SU(3))^{\sharp} $ &31 &8\cr
$\  \IZ_{6-II}  $ &$ SU(2)^2\times SU(3)\times G_2$  &35 &8\cr
$\  \IZ_7  $      &$ SU(7) $                             &24 &9\cr
$\  \IZ_{8-I}  $  &$ (SU(4)\times SU(4))^*$   &24 &5\cr
$\  \IZ_{8-I}  $  &$SO(5)\times SO(9)    $     &27 &0\cr
$\  \IZ_{8-II} $  &$ SU(2)\times SO(10)  $     &27 &4\cr
$\  \IZ_{8-II}  $   &$ SO(4)\times SO(9)$   &31 &0\cr
$\  \IZ_{12-I}$   &$ E_6 $  &25&6\cr
$\  \IZ_{12-I} $  &$ SU(3)\times F_4$  &29 &6\cr
$\  \IZ_{12-II} $  &$SO(4)\times F_4$  &31&0\cr
 \hline 
\end{tabular}}
\end{center}
\caption{Twists, lattices and $h_{(1,1)}^{-}$ for $\IZ_N$ orbifolds.}\label{table:hminusone}
\end{center}\end{table}

\begin{table}[h!]\begin{center}
\begin{center}
\begin{tabular}{|l|c|c|c|c|}
\hline
$\ \IZ_N $&Lattice $T^6$&
 $h_{(1,1)} $& $h_{(1,1)}^{-} $\\ [2pt]
 \hline
$\  \IZ_2 \times\IZ_2$     &$SU(2)^6$  &51 &0\cr
$\  \IZ_2 \times\IZ_4$    &$SU(2)^2\times SO(5)^2$ &61&0\cr
$\  \IZ_2 \times\IZ_6 $    &$ SU(2)^2\times SU(3)\times G_2$&51&0\cr
$\  \IZ_2 \times\IZ_{6'}$&$ SU(3)\times G_2^2$&36 &0\cr
$\  \IZ_3 \times\IZ_3  $   &$SU(3)^3$   &84 &37\cr
$\  \IZ_3 \times\IZ_6 $    &$SU(3)\times G_2^2 $&73 &22\cr
$\  \IZ_4 \times\IZ_4 $    &$SO(5)^3 $ &90&0\cr
$\  \IZ_6 \times\IZ_6  $   &$G_2^3$ &84&0
 \\ \hline 
\end{tabular}
\end{center}
\caption{Twists, lattices and $h_{(1,1)}^{-}$ for $\IZ_N\times \IZ_M$ orbifolds.}\label{table:hminustwo}
\end{center}\end{table}

\subsection{Example B: $T^6/\IZ_{6-II}$ on $SU(2)\times SU(6)$}

To determine the value of $h^{(1,1)}_{-}$ for this example, we must examine the configuration of fixed sets given in Figure \ref{ffixsixiiaa} and the resolution of the local patch, see Figure \ref{fig:sixiifive}, and determine the conjugacy classes of the fixed sets under the global involution $I_6:\,z^i\to -z^i$.
The fixed sets located at $z^2=0$ are invariant under $I_6$, those located at $z^2=1/3$ are mapped to $z^2=2/3$ and vice versa. Clearly, this is an example with $h^{(1,1)}_{-}\neq 0$. The divisors $E_{1,\beta\gamma},\, E_{2,\beta}$ and $E_{4,\beta}$ for $\beta=2,3$ are concerned here. Out of these twelve divisors, six invariant combinations can be formed: $E_{1,inv,\gamma}=\frac{1}{2}(E_{1,2\gamma}+E_{1,3\gamma}),\ E_{2,inv}=\frac{1} {2}(E_{2,2}+E_{2,3})$ and $E_{4,inv}=\frac{1}{2}(E_{4,2}+E_{4,3})$. With a minus sign instead of a plus sign, the combinations are anti-invariant, therefore $h^{(1,1)}_{-}=6$.

\section[The local orientifold involution]{The local orientifold involution on the resolved patches}

Now we want to discuss the orientifold action for the smooth Calabi--Yau manifolds $X$ resulting from the resolved torus orbifolds. For such a manifold $X$, we will denote its orientifold quotient $X/I_6$ by $B$ and the orientifold projection by $\pi:X \to B$. Away from the location of the resolved singularities, the orientifold involution retains the form~(\ref{eq:I6}). As explained above, the orbifold fixed points fall into two classes:
\renewcommand{\labelenumi}{O\theenumi)}
\begin{enumerate}
  \item The fixed point is invariant under $I_6$, i.e. its exceptional divisors are in $h^{1,1}_{+}$.
  \label{item:O1}
  \item The fixed point lies in an orbit of length two under $I_6$, i.e. is mapped to another fixed point. The invariant combinations of the corresponding exceptional divisors contribute to $h^{1,1}_{+}$, while the remaining linear combinations contribute to $h^{1,1}_{-}$. 
  \label{item:O2}
\end{enumerate}
\renewcommand{\labelenumi}{\arabic{\theenumi}}
The fixed points of class~O\ref{item:O1}) locally feel the involution: Let $\zf{}{\alpha}$ denote some fixed point. Since $\zf{}{\alpha}$ is invariant under~(\ref{eq:I6}),
\begin{equation}
  \label{eq:onei}
  (\zf{1}{\alpha}+\Delta z^1, \zf{2}{\alpha}+\Delta z^2, \zf{3}{\alpha}+\Delta z^3) \to (\zf{1}{\alpha}-\Delta z^1, \zf{2}{\alpha}-\Delta z^2, \zf{3}{\alpha}-\Delta z^3).
\end{equation}
In local coordinates centered around $\zf{}{\alpha}$, $I_6$ therefore acts as 
\begin{equation}
  \label{eq:locali}
  (z^1, z^2, z^3) \to (-z^1,-z^2, -z^3).
\end{equation}
In case~O\ref{item:O2}), the point $\zf{}{\alpha}$ is not fixed, but gets mapped to a different fixed point $\zf{}{\beta}$. So locally,
\begin{equation}
  \label{eq:oneii}
  (\zf{1}{\alpha}+\Delta z^1, \zf{2}{\alpha}+\Delta z^2, \zf{3}{\alpha}+\Delta z^3) \to (\zf{1}{\beta}-\Delta z^1, \zf{2}{\beta}-\Delta z^2, \zf{3}{\beta}-\Delta z^3).
\end{equation}
In the quotient, $\zf{}{\alpha}$ and $\zf{}{\beta}$ are identified, i.e. correspond the the same point.
In local coordinates centered around this point, $I_6$ therefore acts again as in (\ref{eq:locali}).

For the fixed lines, we apply the same prescription. The involution on fixed lines with fixed points on them is constrained by the involution on the fixed points.

What happens in the local patches after the singularities were resolved? 
A local involution $\cal I$ has to be defined in terms of the local coordinates, such that it agrees with the restriction of the global involution $I_6$ on $X$. Therefore, we require that $\cal I$ maps $z^i$ to $-z^i$. In addition to the three coordinates $z^i$ inherited from $\IC^3$, there are now also the new coordinates $y^k$ corresponding to the exceptional divisors $E_k$. For the choice of the action of $\cal I$ on the $y^k$ of an individual patch, there is some freedom. 

For simplicity we restrict the orientifold actions to be multiplications by $-1$ only. We do not take into account transpositions of coordinates or shifts by half a lattice vector. The latter have been considered in the context of toric Calabi--Yau hypersurfaces in~\cite{Berglund:1998va}. The allowed transpositions can be determined from the toric diagram of the local patch by requiring that the adjacencies of the diagram be preserved.

The only requirements ${\cal I}$ must fulfill are compatibility with the ${\IC}^*$--action of the toric variety, i.e. 
\begin{equation}
  \label{eq:Caction}
  (-z^1,-z^2,-z^3,(-1)^{\sigma_1}y^1,\dots,(-1)^{\sigma_n}y^n) = (\prod_{a=1}^r \lambda_1^{l_1^{(a)}} z^1,\dots,\prod_{a=1}^r \lambda_n^{l_n^{(a)}} y^n)
\end{equation}
where $l_i^{(a)}$ encode the linear relations~\eqref{eq:linrels} of the toric patch, and that subsets of the set of solutions to~\eqref{eq:Caction} must not be mapped to the excluded set of the toric variety and vice versa.\footnote{We thank Bogdan Florea for helpful correspondence on this point.} 

The fixed point set under the combined action of ${\cal I}$ and the scaling action of the toric variety gives the configuration of O3-- and O7--planes in the local patches. Care must be taken that only these solutions which do not lie in the excluded set are taken into account. We also exclude solutions which do not lead to solutions of the right dimension, i.e. do not lead to O3/O7--planes.

On an individual patch, we can in principle choose any of the possible involutions on the local coordinates. In the global model however, the resulting solutions of the individual patches must be compatible with each other. While O7--planes on the exceptional divisors in the interior of the toric diagram are not seen by the other patches, O7--plane solutions which lie on the $D$--planes or on the exceptional divisors on a fixed line must be reproduced by all patches which lie in the same plane, respectively on the same fixed line. This is of course also true for different types of patches which lie in the same plane. 

It is in principle possible for examples with many interior points of the toric diagram to choose different orientifold involutions on the different patches which lead to solutions that are consistent with each other. We choose the same involution on all patches, which for simple examples such as $\IZ_4$ or the $\IZ_6$ orbifolds is the only consistent possibility.

The solutions for the fixed sets under the combined action of ${\cal I}$ and the scaling action give also conditions to the $\lambda_i$ appearing in the scaling actions, they are set to $\pm 1$.
The O--plane solutions of the full patch descend to solutions on the restriction to the fixed lines on which the patch lies. For the restriction, we set the $\lambda_i$ which not corresponding to the Mori generators of the fixed line to $\pm1$ in accordance with the values of the $\lambda_i$ of the solution for the whole patch which lies on this fixed line.\footnote{We thank Domenico Orlando for an illuminating discussion on this point.}

A further global consistency requirement comes from the observation that the orientifold action commutes with the singularity resolution. A choice of the orientifold action on the resolved torus orbifold must therefore reproduce the orientifold action on the orbifold and yield the same fixed point set in the blow--down limit.

It turns out that it is not always possible to find an involution which reproduces the same O--plane configuration as at the orbifold point. Nevertheless we believe that the orientifold configuration in the resolved phase makes sense. It seems that in these examples the blow--up and the orientifold do not commute and that no smooth limit from the orientifold of the resolved Calabi--Yau manifold to the orientifold of the singular orbifold exists. It may also be that there is some further freedom in defining orientifold actions on orbifold CFTs which would commute with the blow--up. It would be very interesting to understand this point in more detail.

Given a consistent global orientifold action it might still happen that the model does not exist. This is the case if the tadpoles cannot be cancelled. We will explain how to compute the tadpoles from the topological data in Section~\ref{sec:Tadpoles}. We would like to mention that tadpole cancellation conditions can generically be different in different regions of the moduli space of the Calabi--Yau orientifold. This has first been observed on the two sides of the conifold singularity of Calabi-Yau hypersurfaces in toric varieties in~\cite{Brunner:2004zd}. In~\cite{DenefMM}, a similiar observation was made for the torus orbifold $T^6/\IZ_2\times\IZ_2$ where the tadpole cancellation condition at large radius differed from the one at the orbifold point.

\subsection{Example B: $T^6/\IZ_{6-II}$ on $SU(2)\times SU(6)$}\label{sec:Osixiiloc}

At the orbifold point, there are 64 O3--planes which fall into 16 conjugacy classes under the orbifold group. The combination $I_6\,\theta^3$ gives rise to one O7--plane at $z^2=0$ in the $(z^1,z^3)$--plane.

We now discuss the orientifold for the resolved case.
On the homogeneous coordinates $y^k$, several different local actions are possible. We give the eight possible actions which only involve sending coordinates to their negatives: 
\begin{align}
  \label{eq:invlocal}
  (1)\quad\Ic(z,y) &= (-z^1,-z^2,-z^3,y^1,y^2,y^3,y^4) \notag\\
 (2)\quad \Ic(z,y) &= (-z^1,-z^2,-z^3,y^1,y^2,-y^3,-y^4) \notag\\
  (3)\quad\Ic(z,y) &= (-z^1,-z^2,-z^3,y^1,-y^2,y^3,-y^4) \notag\\
  (4)\quad\Ic(z,y) &= (-z^1,-z^2,-z^3,y^1,-y^2,-y^3,y^4) \notag\\
  (5)\quad\Ic(z,y) &= (-z^1,-z^2,-z^3,-y^1,y^2,y^3,-y^4) \notag\\
  (6)\quad\Ic(z,y) &= (-z^1,-z^2,-z^3,-y^1,y^2,-y^3,y^4) \notag\\
 (7)\quad \Ic(z,y) &= (-z^1,-z^2,-z^3,-y^1,-y^2,y^3,y^4) \notag\\
 (8)\quad \Ic(z,y) &= (-z^1,-z^2,-z^3,-y^1,-y^2,-y^3,-y^4) 
\end{align}
In the orbifold limit, (\ref{eq:invlocal}) reduces to $I_6$. The combination of~(\ref{eq:invlocal}) and the scaling action of the resolved patch~(\ref{rescalessixii}) has the following fixed point sets:
\begin{eqnarray}
  \label{eq:Z6IIOplanes}
 (1),\,(8) &&\{z^2=0\} \cup \{y^4=0\} \cup \{z^1=y^1=y^3=0\} \cup \{z^3=y^1=y^3=0\}, \notag\\
 (2),\,(7) &&\{y^3=0\} \cup \{z^1=y^1=y^4=0\} \cup \{z^2=z^3=y^3=0\}\notag\\
 && \cup\ \{z^2=y^1=y^2=0\} \cup \{y^1=y^2=y^4=0\}, \notag\\
 (3),\,(6) &&\{z^1=0\} \cup \{z^3=0\} \cup \{y^2=0\}, \notag\\
  (4),\,(5) &&\{y^1=0\}.
  \end{eqnarray}
Note that the eight possible involutions only lead to four distinct fixed sets (but to different values for the $\lambda_i$).

We focus for the moment on the third possibility. 
With the scaling action
\begin{equation}\label{rescalessixiia}{(z^1,\,z^2,\,z^3,\,y^1,\,y^2,\,y^3,\,y^4) \to (\frac{\lambda_1\lambda_3}{\lambda_4}\,z^1,\,\lambda_2\,z^2,\,\lambda_3\,z^3,\, {1\over\lambda_4}\,y^1,\,\frac{\lambda_1}{\lambda_2^2}\,y^2,\,{\lambda_4\over \lambda_3^2}\,y^3, {\lambda_2\lambda_4\over \lambda_1^2}\,y^4)
}\end{equation}
we get the solutions
\begin{eqnarray}
{\rm (i)}. \quad z^1&=&0,\ \ \lambda_1=\lambda_2=\lambda_3=-1,\ \lambda_4=1,\cr
{\rm (ii)}.\quad z^3&=&0,\ \ \lambda_1=\lambda_2=-1,\ \lambda_3=\lambda_4=1,\cr
{\rm (iii)}.\quad y^2&=&0,\ \ \lambda_1=\lambda_4=1,\ \lambda_2=\lambda_3=-1.
\end{eqnarray}
This corresponds to an O7--plane wrapped on $D_{1}$, one on each of the four $D_{3,\gamma}$ and one wrapped on each of the two invariant $E_{2,\beta}$. No O3--plane solutions occur.

$\lambda_1$ and $\lambda_2$ correspond to the two Mori generators of the $\IZ_3$--fixed line. We restrict to it by setting $\lambda_3=-1,\ \lambda_4=1$ in accordance with solution (i) and (ii) which are seen by this fixed line. The scaling action thus becomes
\begin{equation}\label{rescalessixiizthree}{(z^1,\,z^2,\,z^3,\,y^1,\,y^2,\,y^3,\,y^4) \to (-\lambda_1\,z^1,\,\lambda_2\,z^2,-z^3,y^1,\,\frac{\lambda_1}{\lambda_2^2}\,y^2, y^3, {\lambda_2\over \lambda_1^2}\,y^4).
}\end{equation}
$y^1$ and $y^3$ do not appear in the fixed line, and the restriction makes sense only directly at the fixed point, i.e. for $z^3=0$. With this scaling action and the involution (3), we again reproduce the solutions (i) and (ii).

$\lambda_3$ corresponds to the Mori generator of the $\IZ_2$ fixed line. We restrict to it by setting $
\lambda_1=\lambda_2=-1,\ \lambda_4=1$. 
The scaling action becomes
\begin{equation}\label{rescalessixiiaztwo}
(z^1,\,z^2,\,z^3,\,y^1,\,y^2,\,y^3,\,y^4) \to (-\lambda_3\,z^1,-\,z^2,\,\lambda_3\,z^3,\,y^1,\,-y^2,\,{1\over \lambda_3^2}\,y^3, -\,y^4),
\end{equation}
which together with the involution (3) again reproduces the solutions (i) and (iii).
Global consistency is ensured since we only have one kind of patch on which we choose the same involution for all patches.


\section{The intersection ring}
\label{sec:Oring}

The intersection ring of the orientifold can be determined in two equivalent ways. The basis for both ways is the relation between the divisors on the Calabi--Yau manifold $X$ and the divisors on the orientifold $B$. The first observation is that the integral on $B$ is half the integral on $X$: 
\begin{equation}
  \label{eq:Ointegral}
  \int_B \widehat{S}_a \wedge \widehat{S}_b \wedge \widehat{S}_c = \frac{1}{2} \int_X S_a \wedge S_b \wedge S_c ,
\end{equation}
where the hat denotes the corresponding divisor on $B$. The second observation is that for a divisor $S_a$ on $X$ which is not fixed under $I_6$ we have $S_a = \pi^*\widehat{S}_a$. If, however, $S_a$ is fixed by $I_6$, we have to take $S_a=\frac{1}{2}\pi^*\widehat{S}_a$ because the volume of $S_a$ in $X$ is the same as the volume of $\widehat{S}_a$ on $B$. Applying these rules to the intersection ring obtained in Section~\ref{sec:intersections} immediately yields the intersection ring of $B$:  Triple intersection numbers between divisors which are not fixed under the orientifold involution become halved. If one of the divisors is fixed, the intersection numbers on the orientifold are the same as on the Calabi--Yau. If two (three) of the divisors are fixed, the intersection numbers on the orientifold must be multiplied by a factor of two (four).

The second way consists of applying these rules to the intersection ring of the local patches of the resolved singularities obtained in Section~\ref{sec:mori}, more precisely on the intersection ring of the auxiliary polyhedra $\Delta^{(3)}$ in Section~\ref{sec:intersections} and the global linear equivalences~(\ref{eq:Reqglobal}). This means that for each divisor which is fixed under $\Ic$, the corresponding coefficient in~(\ref{eq:Reqglobal}) is divided by 2. In the polyhedra, the distance to the origin of all those divisors which are fixed under the orientifold involution is halved. Then we solve the resulting system of equations for $\widehat{S}_a\widehat{S}_b\widehat{S}_c$ which we set up at the end of Section~\ref{sec:intersections}. Both methods give the same result.

\subsection{Example B: $T^6/\IZ_{6-II}$ on $SU(2)\times SU(6)$}\label{sec:sixiiOintersections}

If we choose the second method of determining the intersection ring, we have to modify the polyhedra and the linear equivalences.
\begin{figure}[h!]
\begin{center}
\includegraphics[width=140mm]{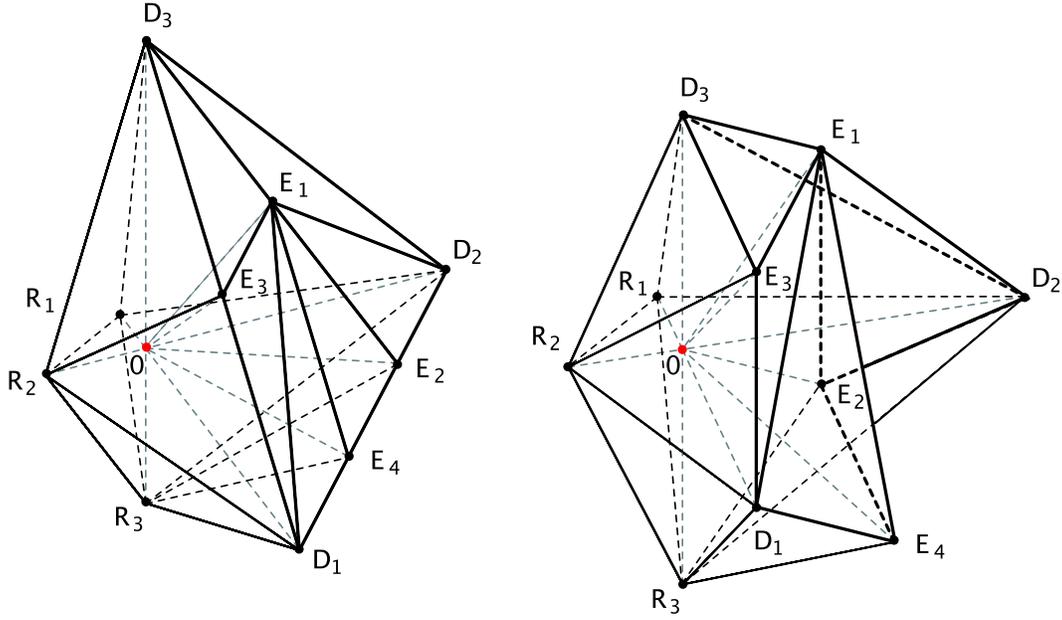}
\caption{Polyhedra of the compactified $\IC^3/\IZ_{6-II}$--patch for the Calabi--Yau and after the orientifold projection}
\label{fig:sixiiOpoly}
\end{center}
\end{figure}
Figure \ref{fig:sixiiOpoly} shows on the left hand side the polyhedron of the compactified $\IC^3/\IZ_{6-II}$--patch for the Calabi--Yau case. The right hand side shows the polyhedron after the local involution (\ref{eq:invlocal}, (3)). The distance to the origin of the vertices corresponding to $D_1,\ D_3$ and $E_2$ has been halved. The resulting polyhedron is not convex anymore. The global linear relations for the Calabi--Yau manifold are:
\begin{eqnarray}\label{eq:globalrelsixiiaO}
R_1&\sim&6\,D_{{1}}+3\,\sum_{\gamma=1}^4 E_{{3,\gamma}}+\sum_{\beta,\gamma}E_{{1,\beta\gamma}}+\sum_{\beta=1}^3[\,2\,E_{{2,\beta}}+4\,E_{{4,\beta}}],\cr
R_2&\sim&3\,D_{{2,\beta}}+\sum_{\gamma=1}^4E_{{1,\beta\gamma}}+2\,E_{{2,\beta}}+E_{{4,\beta}},\cr
R_3&\sim&2\,D_{{3,\gamma}}+\sum_{\beta=1}^3E_{{1,\beta\gamma}}+E_{{3,\gamma}}.
\end{eqnarray}
After the orientifold involution, they become
\begin{eqnarray}\label{eq:globalrelsixiiaOO}
R_1&\sim&3\,D_{{1}}+3\,\sum_{\gamma=1}^4 E_{{3,\gamma}}+\sum_{\beta,\gamma}E_{{1,\beta\gamma}}+\sum_{\beta=1}^2[\,E_{{2,\beta}}+4\,E_{{4,\beta}}],\cr
R_2&\sim&3\,D_{{2,\beta}}+\sum_{\gamma=1}^4E_{{1,\beta\gamma}}+E_{{2,\beta}}+E_{{4,\beta}},\cr
R_3&\sim&D_{{3,\gamma}}+\sum_{\beta=1}^2E_{{1,\beta\gamma}}+E_{{3,\gamma}}.
\end{eqnarray}

As explained in the last section, the intersection numbers can also be determined directly from the intersection numbers on the Calabi--Yau manifold without using the polyhedra. 
The intersection numbers of the Calabi--Yau are
\begin{align}\label{iz6ii}
R_1R_2R_3&=6,& R_3E_{2,\beta}E_{4,\beta}&=1,& E_{1,\beta\gamma}E_{2,\beta}E_{4,\beta}&=1,\cr 
R_2E_{3,\gamma}^2&=-2,& R_3E_{2,\beta}^2&=-2,& R_3E_{4,\beta}^2&=-2,\cr
E_{1,\beta\gamma}^3&=6,& E_{2,\beta}^3&=8,& E_{3,\gamma}^3&=8,\cr
E_{4,\beta}^3&=8, &E_{1,\beta\gamma}E_{2,\beta}^2&=-2,& E_{1,\beta\gamma}E_{3,\gamma}^2&=-2, \cr 
E_{1,\beta\gamma}E_{4,\beta}^2&=-2, &E_{2,\beta}^2E_{4,\beta}&=-2.& &\ & 
\end{align}
Intersection numbers which contain no factor of $E_{2,\beta}$ are halved for the orientifold. If the intersection number contains one factor of $E_{2,\beta}$, it remains the same. If two (three) factors $E_{2,\beta}$ are present, the number on the Calabi--Yau is multiplied by a factor of two (four). This leads to the following modified triple intersection numbers:
\begin{align}\label{iz6iiO}
R_1R_2R_3&=3,& R_3E_{2,\beta}E_{4,\beta}&=1,& E_{1,\beta\gamma}E_{2,\beta}E_{4,\beta}&=1,\cr 
R_2E_{3,\gamma}^2&=-1,& R_3E_{2,\beta}^2&=-4,& R_3E_{4,\beta}^2&=-1,\cr
E_{1,\beta\gamma}^3&=3,& E_{2,\beta}^3&=32,& E_{3,\gamma}^3&=4,\cr
E_{4,\beta}^3&=4, &E_{1,\beta\gamma}E_{2,\beta}^2&=-4,& E_{1,\beta\gamma}E_{3,\gamma}^2&=-1, \cr 
E_{1,\beta\gamma}E_{4,\beta}^2&=-1, &E_{2,\beta}^2E_{4,\beta}&=-4.& &\ &
\end{align}

\section{Global O--plane configuration and tadpole cancellation}
\label{sec:Tadpoles}

Those of the 64 $O3$--planes on the cover which are located away from the locations of the resolved patches resulting from the global involution remain the same in the orbifold of the resolved manifold. They are untouched by the process of resolving the singularities and the resulting modified local orientifold actions. 
The $O3$--plane solutions which coincide with orbifold fixed sets are replaced by the solutions of the corresponding resolved patch. The total number of $O3$--planes on the resolved orbifold quotient is obtained by counting the equivalence classes of $O3$--planes under the orbifold group and replacing those classes which coincide with resolved patches by the $O3$--plane solutions on these patches.
The $O3$--plane solutions are also reflected in the intersection ring. Take for example the solution $\{z^1=y^1=y^3=0\}$ given in (1) of (\ref{eq:Z6IIOplanes}). The corresponding intersection number is $D_1E_1E_3=\tfrac{1}{2}$, indicating the $\IZ_2$--singularity at the intersection point. Thus fractional intersection numbers are indicative of the presence of $O3$--planes. $O3$--planes which are located away from the fixed points and do not lie in the $D$--planes are reflected in the intersection numbers with the inherited divisors $R_i$, see for example $T^6/\IZ_3$ discussed in Appendix \ref{eq:Z3O}. If on the other hand the $O3$--planes lie in an $O7$--plane, their intersection numbers do not become fractional, since the effect of the orientifold involution is already captured by the $O7$--plane.

Since each $O7$--plane induces $-8$ units of $D7$--brane charge, a stack of 8 coincident D7--branes must be placed on top of each divisor fixed under the combination of the involution and the scaling action. Each such stack therefore carries an SO(8) gauge group. 

For the $D3$--brane charge, the case is a bit more involved. The contribution from the $O3$--planes is
\begin{equation}
  \label{othreetadpole}
  Q_3(O3)=-{1\over 4}\times n_{O3},
\end{equation}
where $n_{O3}$ denotes the number of $O3$--planes. The $D7$--branes also contribute to the $D3$--tadpole:
\begin{equation}
  \label{dseventadpole}
  Q_3(D7)=-\sum_a\,{n_{D7,a}\,\chi(S_a)\over 24},
\end{equation}
where $n_{D7,a}$ denotes the number of $D7$--branes in the stack located on the divisor $S_a$. As we have seen, the $S_a$ can be local $D$--divisors as well as exceptional divisors.
The last contribution to the $D3$--brane tadpole comes from the $O7$--planes:
\begin{equation}
  \label{oseventadpole}
  Q_3(O7)=-\sum_a\,{\chi(S_a)\over 6}.
\end{equation}
So the total $D3$--brane charge that must be cancelled is
\begin{equation}
  \label{totaltaddpole}
  Q_{3,tot}=-{n_{O3}\over4}-\sum_a\,{(n_{D7,a}+4)\,\chi(S_a)\over 24}.
\end{equation}
These are the values for the orientifold quotient, in the double cover this value must be multiplied by two.

\subsection{Example B: $T^6/\IZ_{6-II}$ on $SU(2)\times SU(6)$}\label{sec:sixiitadpoles}

In total, there are seven $O7$--planes. The 64 $O3$--planes which are the fixed points of $I_6$ on the covering space fall into 16 equivalence classes: The four $O3$--planes on the line $z^1=z^2=0$ are invariant, the remaining 12 on the plane $z^2=0$ fall into four equivalence classes with three elements each. The remaining 48 $O3$--planes lie in eight orbits of length six. The four $O3$--planes on the line $z^1=z^2=0$ coincide with the patches on the line $z^1=z^2=0$ and disappear, since there are no $O3$--solutions on the local patches. The remaining $O3$--planes all lie in the $O7$--planes wrapped on the $D_3,\gamma$ and are therefore not visible in the intersection numbers. In total, we are left with 12 $O3$--planes in the resolved orientifold quotient.

Now we treat the problem of tadpole cancellation.
On top of the $O7$--planes we place eight $D7$--branes to cancel the  $D7$--tadpole locally. This gives rise to a stack of $D7$--branes  with gauge group $SO(8)$ on each of the divisors $D_1,\ D_{3,\gamma}, \ E_{2,\beta}$.
The contribution to the $D3$--tadpole from each $O7/SO(8)$--stack is  $-\frac{12\,\chi(S)}{24}$. In the triangulation a), $E_2$ has the  topology of $\IF_0$, therefore $\chi(E_2)=4$. $D_1$ has the topology  of $\IF_0$ blown up in twelve points, therefore $\chi(D_1)=4+12=16$.  $D_3$ has the topology of a Hirzebruch surface blown up in nine  points, so $\chi(D_3)=4+9=13$. The total $D3$--tadpole from all $O7/SO (8)$--stacks is therefore
\begin{equation}\label{tad7sixii}
Q_{3}(O7/D7)=-\frac{1}{2}\left(\,\sum_{\beta=1,2}\chi(E_{2,\beta})+\chi (D_1)+\sum_{\gamma=1}^4\chi(D_3)\right)=-38.
\end{equation}
The $D3$--tadpole coming from the (non--exotic) $O3$--planes  themselves is
\begin{equation}\label{tad7sixiia}
Q_{3}(O3^-)=-\tfrac{1}{4}\,n_{O3}= -3.
\end{equation}
The total $D3$--tadpole therefore is $Q_{3_{\rm tot}}=-(38+3)=41$.


\part[Applications in String Phenomenology]{Applications in String\\[3pt] Phenomenology}

\chapter{Preliminaries}\label{chap:prelim}

In this section, the preliminary knowledge on type $IIB$ flux compactifications which is needed in later sections is presented.


\section{The type $IIB$ low energy effective action}

We briefly recall the massless spectrum and the 10D low energy effective action of type $IIB$ string theory.
The massless Neveu-Schwarz--Neveu-Schwarz sector consists of the scalar dilaton $\phi$, the metric or graviton $G_{\mu\nu}$ and the anti-symmetric tensor $B_{\mu\nu}$. The Ramond-Ramond sector consists of the  even forms $C_0,\ C_2, C_4$ etc. The field strengths of the form field $C_n$ is denoted by $F_{n+1}$, the field strength of $B_{\mu\nu}$ is $H_3$.  
The 10--dimensional  low energy effective action describing the bosonic massless degrees of freedom of the type $IIB$ superstring is \cite{joep}
\begin{eqnarray}
S_{IIB}&=&\frac{1}{2\,\kappa_{10}^2}\int d^{10}x\, (-G_E)^{1/2}\left(R_E-\frac{\partial_\mu \ov S\,\partial^\mu S}{2\,({\rm Im}\, S)^2}-\frac{{\cal M}_{ij}}{2}F_3^i\cdot F^j_3-\frac{1}{4}|\tilde F_5|^2\right)\cr
&&-\frac{\epsilon_{ij}}{8\,\kappa_{10}^2}\int C_4\wedge F^i_3\wedge F^j_3,
\end{eqnarray}
with $\kappa_{10}=2\pi\,g_{s}\,e^{-\phi_{10}}$ the ten-dimensional gravitational coupling ($g_{s}$ the string coupling), $G_{E,\mu\nu}=e^{-\phi/2}\,G_{\mu\nu}$ the metric in the Einstein frame,  $R_E$ the Ricci-tensor in the Einstein frame, $S=C_0+i\,e^{-\phi_{10}}$ the complexified dilaton, 
\begin{equation}
{\cal M}_{ij}=\frac{1}{{\rm Im}\,S}\left(\begin{array}{cc}|S|^2&-{\rm Re}\,S\\-{\rm Re}\,S&1\end{array}\right),\ \ F_3=\left(\begin{array}{c}H_3\\ F_3\end{array}\right).
\end{equation}
$\tilde F_5=F_5-\tfrac{1}{2}\,C_2\wedge H_3+\tfrac{1}{2}\,B_2\wedge F_3$ is the self-dual 5-form field strength, i.e. $*\tilde F_5=\tilde F_5$.
In this form, the effective action is manifestly $SL(2,\IR)$ invariant:
\begin{equation}
S'=\frac{a\,S+b}{c\,S+d},\quad {F_3'}=\left(\begin{array}{cc}d&c\\ b&a\end{array}\right)\left(\begin{array}{c}H_3\\ F_3\end{array}\right), \quad \tilde F_5'=\tilde F_5,\quad G_{E,\mu\nu}'=G_{E,\mu\nu}, 
\end{equation}
with $a,\,b,\,c,\,d \in \IR$ and $ad-bc=1$.

Now, the surplus 6 dimensions are being compactified on a Calabi--Yau manifold $X$. For the metric we use the following block-diagonal ansatz:
\begin{equation}\label{ansatzmetric}
ds^2=g^{(4)}_{\mu\nu}(x)\,dx^\mu dx^\nu+g_{mn}^{(6)}(y)\,dy^mdy^n,
\end{equation}
where $g^{(4)}_{\mu\nu}(x)$ is the four--dimensional Minkowski metric and $g_{mn}^{(6)}(y)$ is the metric of the internal Calabi--Yau space.
As explained in Section \ref{sec:shape}, a Calabi--Yau manifold has a moduli space which consists of $h^{(1,1)}(X)$ K\"ahler moduli $\Tc^i$ and $h^{(2,1)}(X)$ complex
structure moduli $\Uc^i$.
In addition, there is the complex dilaton field $S$.
The parameter space of $S$ is locally spanned by the coset
\begin{equation}\label{Scoset}{
\Mc_S=\fc{SU(1,1)}{U(1)}\ .}
\end{equation}
Furthermore, we have $e^{-\phi_{10}}=e^{-\phi_4}\ {\rm Vol}(X)^{-1/2}$, with ${\rm Vol}(X)$
the volume of the compactification manifold $X$ and $\phi_4$ the dilaton in four dimensions.

Without $D$--brane moduli, locally the closed string moduli space $\Mc$ is a direct product of
the complex dilaton field $S$, the K\"ahler $\Mc_K$ and complex structure moduli 

\begin{equation}\label{DIRECT}{
\Mc=\Mc_S\otimes \Mc_K\otimes \Mc_{CS} .}
\end{equation}
All factors are special K\"ahler manifolds on which a K\"ahler potential can be defined.

Before introducing the orientifold projection and $D$-branes, the theory has ${\cal N}=2$ supersymmetry in four dimensions, afterwards only ${\cal N}=1$.
An ${\cal N}=1$ supersymmetric theory is completely described by three quantities: The holomorphic superpotential, the K\"ahler potential, and the gauge kinetic function.

\section{The K\"ahler potential}

The total K\"ahler potential is a sum of the K\"ahler potential of the different factors of the moduli space (\ref{DIRECT}).
The K\"ahler potential for the dilaton is
\begin{equation}\label{Kdil}
K_S=-\ln (\tilde{S}-\ov {\tilde S}), 
\end{equation}
where ${\rm Im}\,\tilde S=e^{-\phi_{10}}\cdot {\rm Vol}(X)$.
For the complex structure moduli it is
\begin{equation}\label{Kcplx}
K_{CS}=-\ln \left(\int_X \Omega\wedge \ov\Omega\,\right),
\end{equation}
with $\Omega$ the Calabi--Yau $(3,0)$--form.
The K\"ahler potential for the K\"ahler moduli is
\begin{equation}\label{Kkm}
K_K=-\ln \,{\rm Vol}(X)=-\ln\,\left(\tfrac{1}{6}\int_X J\wedge J\wedge J\right),
\end{equation}
with $J$ the K\"ahler form. When the K\"ahler form is expressed through a basis of 2--forms dual to a basis of the divisor classes, the volume can be expressed in terms of the triple intersection numbers of the divisor basis.
The total K\"ahler potential is 
\begin{equation}\label{Khut}
\hat K=K_S+K_{CS}+K_K.
\end{equation}


\subsection{Example D: $T^6/\IZ_2\times\IZ_2$}\label{exKaehlerz2z2}

The simplest example is this of $T^6/\IZ_2\times\IZ_2$. The K\"ahler potential of the dilaton always has the form (\ref{Kdil}). For the K\"ahler potential of the complex structure moduli, we plug the complex coordinates given in (\ref{cpxtwotwo}) into $\Omega=dz^1\wedge dz^2\wedge dz^3$. (\ref{Kcplx}) then takes the simple form
\begin{equation}\label{Kz2z2cplx}
K_{CS}=-\ln\,(\Uc^1-\ov\Uc^1)-\ln\,(\Uc^2-\ov\Uc^2)-\ln\,(\Uc^3-\ov\Uc^3).
\end{equation}
At the orbifold point, the K\"ahler potential again takes a very simple form with only the three untwisted K\"ahler moduli contributing:
\begin{equation}\label{Kz2z2k}
K_{K}=-\ln\,(\Tc^1-\ov\Tc^1)-\ln\,(\Tc^2-\ov\Tc^2)-\ln\,(\Tc^3-\ov\Tc^3).
\end{equation}
After the transition to the smooth Calabi--Yau, we need to know all triple intersection numbers. With the methods described in Section \ref{sec:intersections}, we arrive at the triple intersection numbers given in Appendix \ref{app:intersz2z2}.

The K\"ahlerform can be parameterized as
\begin{equation}
J=\sum_{i=1}^3 r_{i}R_{i}-\sum_{,\beta,\gamma=1}^4t_{1\beta\gamma}E_{1\beta\gamma}-\sum_{\alpha,,\gamma=1}^4t_{2\alpha\gamma}E_{2\alpha\gamma}-\sum_{\alpha,\beta,=1}^4t_{3\alpha\beta}E_{3\alpha\beta}.
\end{equation}
The total volume becomes
\begin{eqnarray}
V&=&2\,r_{1}r_{2}r_{3}-r_1\sum_{\beta,\gamma}t_{1,\beta\gamma}^2-r_2\sum_{\alpha,\gamma}t_{2,\alpha\gamma}^2-r_3\sum_{\alpha,\beta}t_{3,\alpha\beta}^2-\sum_{\alpha,\beta,\gamma}t_{1,\beta\gamma}t_{2,\alpha\gamma}t_{3,\alpha\beta}\cr
&&+\frac{1}{2}\,\sum_{\alpha,\beta,\gamma}( t_{1,\beta\gamma}^2t_{2,\alpha\gamma}+ t_{1,\beta\gamma}t_{2,\alpha\gamma}^2+ t_{1,\beta\gamma}^2t_{3,\alpha\beta}+ t_{1,\beta\gamma}t_{3,\alpha\beta}^2+ t_{2,\alpha\gamma}^2t_{3,\alpha\beta}+t_{2,\alpha\gamma}t_{3,\alpha\beta}^2)\cr
&&-\frac{2}{3}\,\sum_{\alpha,\beta,\gamma}(t_{1,\beta\gamma}^3+t_{2,\alpha\gamma}^3+t_{3,\alpha\beta}^3), 
\end{eqnarray}
and the K\"ahler potential of the K\"ahler moduli is $-\ln V$.


\section{The orientifold action}

In part I, the orientifold was discussed in purely geometric terms.
To obtain an ${\cal N}=1$ (closed) string spectrum, one introduces an 
orientifold projection $\Om I_n$, with $\Om$ describing a reversal of the 
orientation of the closed string world--sheet and $I_n$ a reflection of $n$ internal
coordinates. For $\Om I_n$ to represent a symmetry of the original theory, $n$
has to be an even integer in \tb.
Generically, this projection produces orientifold fixed planes [$O(9-n)$--planes],
placed at the orbifold fixed points of $T^6/I_n$. They have negative tension, which 
has to be balanced by introducing positive tension objects.
Candidates for the latter may be collections of $D(9-n)$--branes and/or non--vanishing
three--form fluxes $H_3$ and $C_3$.
The orbifold group $\Gamma$ mixes with the orientifold group $\Om I_n$.
As a result, if the group $\Gamma$ contains $\IZ_2$--elements $\th$, 
which leave one complex plane fixed, we obtain additional $O(9-|n-4|)$-- or 
$O(3+|n-2|)$--planes from the element $\Om I_n \th$. 

In the following, only the two cases of $n=6$ ($O3$--planes) and $n=2$
($O7$--planes) will be relevant
to us. In type $IIB$ string theory, tadpoles cannot be completely canceled for all models given in Tables \ref{table:one} and \ref{table:two} without introducing torsion or vector structure or resolving the singularities.
The $\IZ_4$-- and $\IZ_8$--orbifolds, as well as $\IZ_{12-II}$, $\IZ_2\times\IZ_4$ and $\IZ_4\times\IZ_4$ must, at least at the orbifold point be dropped from the list \cite{AFIV,zwart}. 
This is to be contrasted with \ta intersecting $D6$--brane
constructions, where it has been shown that essentially all orbifold
groups $\Gamma$ allow for tadpole cancellation due to the appearance of only
untwisted and $\IZ_2$--twisted sector tadpoles 
\cite{BCS}.

For the dilaton, $S$ is used instead of $\tilde S$ in (\ref{Kdil}) with ${\rm Im}\,S=e^{-\phi_{10}}$.
The K\"ahler potential for the K\"ahler moduli (\ref{Kkm}) receives a factor of two due to the modified dilaton, i.e. it becomes
\begin{equation}\label{KkmO}
K_K=-2\,\ln \,{\rm Vol}(X)=-2\,\ln\,\left(\tfrac{1}{6}\int_X J\wedge J\wedge J\right).
\end{equation}


\section{Turning on background flux}

"Turning on fluxes" means giving non--vanishing vevs to some of the (untwisted) 
flux components $H_{ijk}$ and $F_{ijk}$, with $F_3=dC_2$, $H_3=dB_2$. 
The two $3$--forms $F_3,H_3$ are organized in the $SL(2,\IZ)_S$ covariant field:
\begin{equation}\label{fluxcomb}{
G_3=F_3-S\ H_3\ .}
\end{equation}

The most general $3$--form flux $G_3$ on $T^6$ has 20 components, which appear 
in the expansion
\begin{equation}\label{GC}{{1\over{(2\pi)^2\alpha'}}\ G_3=\sum_{i=0}^3 (A^i\omega_{A_i}+B^i\omega_{B_i})+
\sum_{j=1}^6(C^j\omega_{C_j}+D^j\omega_{D_j})}
\end{equation}
with respect to the complex $3$--form cohomology  
$H^3=H^{(3,0)}\oplus H^{(2,1)}\oplus H^{(1,2)}\oplus H^{(0,3)}$:
\begin{align}\label{cplxz}
\om_{A_0}&=d z^1\wedge dz^2\wedge d z^3\, ,\ \ &\om_{B_0}&=d\ov z^1\wedge d \ov z^2\wedge d \ov z^3\cr
 \om_{A_1}&=d\ov z^1\wedge
dz^2\wedge dz^3\, ,\ \ \ &\om_{B_1}&= dz^1\wedge d\ov z^2\wedge d\ov z^3\cr
\om_{A_2}&=dz^1\wedge
d\ov z^2\wedge dz^3\,,\ \ \ &\om_{B_2}&=d\ov z^1\wedge dz^2\wedge d\ov z^3\cr
\om_{A_3}&=dz^1\wedge dz^2\wedge d\ov z^3\, ,\ \ \ &\om_{B_3}&= d\ov z^1\wedge d\ov z^2\wedge dz^3\cr
\om_{C_1}&=dz^1\wedge d\ov z^1\wedge dz^2\, ,\ \ &\om_{D_1}&=dz^1\wedge d \ov z^1\wedge d \ov z^2\cr
 \om_{C_2}&=dz^1\wedge
d\ov z^1\wedge dz^3\, ,\ \ \ &\om_{D_2}&= dz^1\wedge d\ov z^1\wedge d\ov z^3\cr
\om_{C_3}&=dz^1\wedge
d z^2\wedge d\ov z^2\, ,\ \ \ &\om_{D_3}&=d\ov z^1\wedge dz^2\wedge d\ov z^2\cr
\om_{C_4}&=dz^2\wedge d\ov z^2\wedge d z^3\, ,\ \ \ &\om_{D_4}&= 
d z^2\wedge d\ov z^2\wedge d\ov z^3\cr 
\om_{C_5}&=dz^1\wedge
dz^3\wedge d\ov z^3\, ,\ \ \ &\om_{D_5}&=d\ov z^1\wedge dz^3\wedge d\ov z^3\cr
\om_{C_6}&=dz^2\wedge dz^3\wedge d\ov z^3\, ,\ \ \ &\om_{D_6}&= 
d\ov z^2\wedge d z^3\wedge d\ov z^3\, .
\end{align}
The $\om_{A_i}$,  $\om_{B_i}$ correspond to flux components 
with all one--forms coming from different complex coordinate directions, 
while the $\om_{C_i}$, $\om_{D_i}$ are flux components 
with two one--forms coming from the same complex coordinate directions. 
The latter were just written down for completeness, as they are projected out in 
all orbifolds.

In order to impose the flux quantization conditions on $G_3$, \ie
\begin{equation}\label{fluxqu}{
\fc{1}{(2\pi)^2\ap}\int_{C_3} F_3 \in n_0\ \IZ,\ \ \ 
\fc{1}{(2\pi)^2\ap}\int_{C_3} H_3 \in n_0\ \IZ ,}
\end{equation}
one has to transform the forms (\ref{cplxz})
into a real basis with the following 20 elements:
\begin{align}\label{realbase}
\alpha_0&=dx^1 \wedge dx^3  \wedge dx^5, &\beta^0=dx^2 \wedge dx^4 \wedge dx^4\ ,\cr
\alpha_1&=dx^2 \wedge dx^3  \wedge dx^5, &\beta^1=-dx^1 \wedge dx^4\wedge dx^6\ ,\cr
\alpha_2&=dx^1 \wedge dx^4  \wedge dx^5, &\beta^2=-dx^2 \wedge dx^3 \wedge dx^6\ ,\cr
\alpha_3&=dx^1 \wedge dx^2  \wedge dx^6, &\beta^3=-dx^2 \wedge dx^4 \wedge dx^5\ ,\cr
\gamma_1&=dx^1 \wedge dx^2  \wedge dx^3, &\delta^1=-dx^4 \wedge dx^5 \wedge dx^6\ ,\cr
\gamma_2&=dx^1 \wedge dx^2  \wedge dx^5, &\delta^2=-dx^3 \wedge dx^4 \wedge dx^6\ ,\cr
\gamma_3&=dx^1 \wedge dx^3  \wedge dx^6, &\delta^3=-dx^2 \wedge dx^5\wedge dx^6\ ,\cr
\gamma_4&=dx^2 \wedge dx^4  \wedge dx^5, &\delta^4=-dx^1 \wedge dx^2 \wedge dx^6\ ,\cr
\gamma_5&=dx^1 \wedge dx^5  \wedge dx^6, &\delta^5=-dx^2 \wedge dx^3 \wedge dx^4\ ,\cr
\gamma_6&=dx^2 \wedge dx^5  \wedge dx^6, &\delta^6=-dx^1 \wedge dx^2\wedge dx^4\ ,
\end{align}
The basis (\ref{realbase}) has the property $\int_{X_6} 
\alpha_i \wedge \beta^j=\delta^j_i,\ \int_{X_6} \gamma_i \wedge \delta^j=\delta^j_i$.
In real notation, the flux has the form:
\begin{equation}\label{GR}{{1\over{(2\pi)^2\alpha'}}{G_3}=\sum_{i=0}^{3} \lf[(a^i-S
c^i)\alpha_i+(b_i-S d_i)\beta^i\ri]+\sum_{j=1}^{6} \lf[(e^j-S
g^j)\gamma_j+(f_j-S h_j)\delta^j\ri]\ .}
\end{equation}
In this basis, the  $SL(2,\IZ)_S$--covariance of $G_3$ is manifest, while in the complex basis (\ref{cplxz}) the cohomology structure of $G_3$ is manifest.
The coefficients $a^i,\ b_i,\ c^i,\ f_i$ refer to the Ramond part of $G_3$, whereas the
coefficients $c^i,\ d_i,\ g^i,\ h_i$ refer to the Neveu-Schwarz part.

To pass from the complex basis (\ref{cplxz}) to the real basis (\ref{realbase}), one introduces complex
structures, i.e. the complex coordinates (cf. Section \ref{sec:moduli},  (\ref{ansatz})).
In many toroidal orbifolds, the complex structure is fixed completely by the twist. The remaining complex structure moduli are eventually fixed through the flux quantization condition.

Let us briefly comment on the integers $n_0$, introduced in the flux quantization conditions
(\ref{fluxqu}).
It has been pointed out in ref. \cite{FP},
 that there are subtleties for toroidal orientifolds  
due to additional $3$--cycles, which are not present in the covering space $T^6$.
If some integers are odd, additional discrete flux has to be
turned on in order to meet the quantization rule for those $3$--cycles.
We may bypass these problems in the $\IZ_N$ ($\IZ_N\times \IZ_M$)--orientifolds, 
if we choose the quantization numbers to be multiples of $n_0=2N$ ($n_0=2NM$)  
and do not allow for discrete flux at the orientifold planes 
\cite{BLT,CU,FontCY}.
Note, that for $h_{2,1}^{\rm twist.}\neq 0$, 
in addition to the untwisted flux components $H_{ijk}$ and $F_{ijk}$ there may
be also $NSNS$-- and $RR$--flux components from the twisted sector. We do not
consider them here. It is assumed, that their quantization rules freeze the twisted complex structure
moduli at the orbifold singularities. 

It should be mentioned that turning on background fluxes leads to a back--reaction of the geometry. Instead of (\ref{ansatzmetric}), we use the ansatz
\begin{equation}\label{ansatzwarpedmetric}
ds^2=e^{2A(y)}\,g^{(4)}_{\mu\nu}(x)\,dx^\mu dx^\nu+e^{-2A(y)}\,g_{mn}^{(6)}(y)\,dy^mdy^n,
\end{equation}
where $e^{2A(y)}$ is the so--called warp factor. For the class of fluxes we consider, the six--dimensional metric is related to the original Calabi--Yau metric via a conformal factor which is the inverse of the warp factor. In the large radius limit, the warp factor can be neglected.

\section{Invariant 3-forms}

On the torus $T^6$, there are 20+20 independent internal
components for $H_{ijk}$ and $F_{ijk}$. 
However, only a part of them is invariant under the orbifold group $\Gamma$.
More precisely, of the $20$ complex (untwisted) components of the flux $G_3$, only 
$2h^{untw.}_{(2,1)}(X_6)+2$ survive the orbifold twist. 
The orientifold action $\Om(-1)^{F_L}I_6$ does not give rise to any 
further restrictions. The allowed flux components are most conveniently found in the complex basis, in which
the orbifold group $\Gamma$ acts diagonally. 
In Tables \ref{table:fluxone} and \ref{table:fluxtwo}, the fluxes invariant under the different orbifold groups are listed.

\begin{table}[h!]\begin{center}
\begin{tabular}{|c||ccccccccc|}
\hline
$G_3$&$\IZ_3$
&$\IZ_4$&$\IZ_{6-I}$&$\IZ_{6-II}$&$\IZ_7$&$\IZ_{8-I}$&$\IZ_{8-II}$&$\IZ_{12-I}$&$\IZ_{12-II}$\cr
\hline
\noalign{\hrule}\noalign{\hrule}
$dz^1\wedge dz^2\wedge dz^3$& +&+&+&+&+&+&+&+&+\cr
$d\ov z^1\wedge dz^2\wedge dz^3$&$-$&$-$&$-$&$-$&$-$&$-$&$-$&$-$&$-$\cr
$dz^1\wedge d\ov z^2\wedge dz^3$& $-$&$-$&$-$&$-$&$-$&$-$&$-$&$-$&$-$\cr
$dz^1\wedge dz^2\wedge d\ov z^3$& $-$&+&$-$&+&$-$&$-$&+&$-$&+\cr
$d z^1\wedge d\ov z^2\wedge d\ov z^3$& $-$&$-$&$-$&$-$&$-$&$-$&$-$&$-$&$-$\cr
$d\ov z^1\wedge d z^2\wedge d\ov z^3$& $-$&$-$&$-$&$-$&$-$&$-$&$-$&$-$&$-$\cr
$d\ov z^1\wedge d\ov z^2\wedge d z^3$& $-$&+&$-$&+&$-$&$-$&+&$-$&+\cr
$d\ov z^1\wedge d\ov z^2\wedge d\ov z^3$& +&+&+&+&+&+&+&+&+\cr
\hline 
\end{tabular}
\caption{Invariant $3$--forms for point groups $\IZ_N$}\label{table:fluxone}
\end{center}\end{table}


\begin{table}[h!]\begin{center}
\resizebox{\linewidth}{!}{
\begin{tabular}{|c||cccccccc|}
\hline
$G_3$&$\IZ_2\times\IZ_2$&$\IZ_3\times \IZ_3$&$\IZ_2\times \IZ_4
$&$\IZ_4\times \IZ_4$&$\IZ_2\times \IZ_{6-I}$&$\IZ_2 \times \IZ_{6-II}$&$\IZ_3\times \IZ_6
$&$\IZ_6\times \IZ_6$\cr
\noalign{\hrule}\noalign{\hrule}
$dz^1\wedge dz^2\wedge dz^3$&+&+&+&+&+&+&+&+\cr
$d\ov z^1\wedge dz^2\wedge dz^3$& +&$-$&+&$-$&+&$-$&$-$&$-$\cr
$dz^1\wedge d\ov z^2\wedge dz^3$& +&$-$&$-$&$-$&$-$&$-$&$-$&$-$\cr
$dz^1\wedge dz^2\wedge d\ov z^3$& +&$-$&$-$&$-$&$-$&$-$&$-$&$-$\cr
$d z^1\wedge d\ov z^2\wedge d\ov z^3$& +&$-$&+&$-$&+&$-$&$-$&$-$\cr
$d\ov z^1\wedge d z^2\wedge d\ov z^3$& +&$-$&$-$&$-$&$-$&$-$&$-$&$-$\cr
$d\ov z^1\wedge d\ov z^2\wedge d z^3$& +&$-$&$-$&$-$&$-$&$-$&$-$&$-$\cr
$d\ov z^1\wedge d\ov z^2\wedge d\ov z^3$& +&+&+&+&+&+&+&+\cr
\hline 
\end{tabular}}
\caption{Invariant $3$--forms for point groups $\IZ_M\times \IZ_N$}\label{table:fluxtwo}
\end{center}\end{table}

The remaining twelve 3-forms of the form $dz^a\wedge d\ov z^a\wedge dz^b$ and 
$dz^a\wedge d\ov z^a\wedge d \ov z^b$, respectively are always projected out
and therefore do not appear in the tables. Note, 
that we have for completeness also listed the orbifold groups
$\Gamma\in \{\IZ_4,\,\IZ_{8-I},\,\IZ_{8-II},\,\IZ_{12-II}\}$, 
whose tadpoles may at the orbifold point only be cancelled in the more general orbifold setups with discrete
torsion or vector structure.

That the $(0,3)$ and $(3,0)$-forms always
survive is quite clear, as the $(3,0)$-form corresponds to the Calabi--Yau
3-form $\Omega$, which is always present, and the $(0,3)$-form to its
conjugate.

\subsection{Example D: $T^6/\IZ_2\times\IZ_2$}

We want to express the 3-forms allowed on $T^6/\IZ_2\times\IZ_2$ in the complex basis through the real basis (\ref{realbase}) and the complex structure moduli (\ref{cpxmodtwotwo}):
\begin{eqnarray}\label{cplxbasetwotwo}
\omega_{A0}&=& \alpha_0+ \sum_{i=1}^3 \alpha_i\,\ov{\Uc^i} -
\beta_1\,\ov{\Uc^2}\,\ov{\Uc^3} - \beta_2\,\ov{\Uc^1}\,\ov{\Uc^3} -
\beta_3\,\ov{\Uc^1}\ov{\Uc^2} + \beta_0\,\ov{\Uc^1}\ov{\Uc^2}\ov{\Uc^3} \cr 
\omega_{A1}& =& \alpha_0 +\alpha_1\,\ov{\Uc^1}+\alpha_2\,\Uc^{2}+\alpha_3\Uc^3 - \beta_1\,\Uc^2\Uc^3 -
\beta_2\,\ov{\Uc^1}\Uc^3 - \beta_3\,\ov{\Uc^1}\Uc^2 + \beta_0\,\ov{\Uc^1}\Uc^2\Uc^3 \cr
\omega_{A2}& =& \alpha_0 +\alpha_1\,\Uc^1+\alpha_2\,\ov{\Uc^2}+\alpha_3\,\Uc^3  - \beta_1\,\ov{\Uc^2}\Uc^3 -
\beta_2\,\Uc^{1}\Uc^3 - \beta_3\,\Uc^{1}\ov{\Uc^2} + \beta_0\,\Uc^{1}\ov{\Uc^2}\Uc^3 \cr
\omega_{A3}& =& \alpha_0 +\alpha_1\,\Uc^1+\alpha_2\Uc^{2}+\alpha_3\,\ov{\Uc^3}  - \beta_1\,\Uc^{2}\ov{\Uc^3} -
\beta_2\,\Uc^{1}\ov{\Uc^3} - \beta_3\,\Uc^{1}\Uc^{2} + \beta_0\,\Uc^{1}\Uc^{2}\ov{\Uc^3} \cr
\omega_{B0}&=&\alpha_0+ \sum_{i=1}^3 \alpha_i\,\Uc^i -
\beta_1\,\Uc^2\Uc^3 - \beta_2\,\Uc^1\Uc^3 - \beta_3\,\Uc^1\Uc^2 + \beta_0\,\Uc^1\Uc^2\Uc^3\cr
\omega_{B1}& =& \alpha_0 +\alpha_1\,\Uc^{1}+\alpha_2\,\ov{\Uc^2}+\alpha_3\,\ov{\Uc^3} -
\beta_1\,\ov{\Uc^2}\ov{\Uc^3} -
\beta_2\,\Uc^{1}\ov{\Uc^3} - \beta_3\,\Uc^{1}\ov{\Uc^2} + \beta_0\,\Uc^{1}\ov{\Uc^2}\ov{\Uc^3} \cr
\omega_{B2}& =& \alpha_0 +\alpha_1\,\ov{\Uc^1}+\alpha_2\,\Uc^{2}+\alpha_3\,\ov{\Uc^3}  - 
\beta_1\,\Uc^{2}\ov{\Uc^3} -
\beta_2\,\ov{\Uc^1}\ov{\Uc^3} - \beta_3\,\ov{\Uc^1}U^{2} + \beta_0\,\ov{\Uc^1}\Uc^2\ov{\Uc^3} \cr
\omega_{B3}& = &\alpha_0 +\alpha_1\,\ov{\Uc^1}+\alpha_2\,\ov{\Uc^2}+\alpha_3\,\Uc^3  - \beta_1\,\ov{\Uc^2}\Uc^{3} -
\beta_2\,\ov{\Uc^1}\Uc^{3} - \beta_3\,\ov{\Uc^1}\ov{\Uc^2} +
\beta_0\,\ov{\Uc^1}\ov{\Uc^2}\Uc^{3}.\cr
\end{eqnarray}
 $\omega_{B0}$
obviously corresponds to the $(3,0)$-part of the flux and the Calabi--Yau 3-form
$\Omega$ can be normalized to equal  $\omega_{B0}$.  $\omega_{A1}$,
$\omega_{A2}$ and  $\omega_{A3}$ are the $(2,1)$-components of the flux,
$\omega_{B1}$, $\omega_{B2}$ and $\omega_{B3}$ the $(1,2)$-components of the
flux, and $\omega_{A0}$ corresponds to the $(0, 3)$-part, i.e.~$\ov{\Omega}$.

Note that this basis is not normalized to one. It fulfills
\begin{eqnarray}
\int\om_{Ai}\wedge\om_{Bi}&=&\prod_{i=1}^3(\Uc^i-\ov \Uc^i),\quad i=0,\ldots,3\cr
\int\om_{Aj}\wedge\om_{Bk}&=&0,\quad j\neq k.
\end{eqnarray}
Expressed in this basis,  the $G_3$-flux takes the form
\begin{equation}
{{1\over{(2\pi)^2\alpha'}}G_3=\sum_{i=0}^3 (A^i\omega_{Ai}+B^i\omega_{Bi}).}
\end{equation}
By comparing the coefficients of $G_3$ expressed in the real basis and in the
complex basis and solving for the $\{A^i,\ B^i\}$, we can express the $\{A^i,\ B^i\}$ as a
function of $\{a^i, c^i, b_i, d_i\}$ and the moduli fields $\Sc, \Uc^i$.
By setting the respective coefficients to zero, we obtain equations for the
respective flux parts. This gives us constraints on the $\{a^i,
c^i, b_i, d_i\}$. What has to be taken into account as well is the fact that
the $\{a^i, c^i, b_i, d_i\}$ must be integer numbers. This requirement can
only be fulfilled for specific choices of the $\Uc^i$ and of $\Sc$, i.e. it
fixes the moduli.

\subsection{Example E: $T^6/\IZ_7$}

For illustration, we also present an example which differs from the commonly used $T^6/\IZ_2\times \IZ_2$ in several ways. First of all, there are no complex structure moduli.  Secondly, the complex coordinates (\ref{cpxzseven}) contain more than just two of the real coordinates, which leads to a non-trivial expansion of the flux in real coordinates.
In this example only the $(3,0)$-- and the $(0,3)$--flux component survive the $Z_7$--twist, so
$${1\over (2\pi)^2\alpha'}\,G_3=A_0\,\om_{A_0}+B_0\,\om_{B_0}.$$
With the complex coordinates given in (\ref{cpxzseven}) we find for the $(3,0)$--form:
\begin{eqnarray}
\om_{A_0}&=&-i\sqrt7\, \alpha_0+\half(7+i\sqrt7)\,\alpha_1+i\sqrt7\,\alpha_2-i\sqrt7\, \alpha_3-i\sqrt7 \,\beta_0+i\sqrt7\, \beta_1\cr
&&-i\sqrt7\, \beta_2+\half(7-i\sqrt7 )\,\beta_3-i\sqrt7\, \gamma_1+i\sqrt7\, \gamma_2+\half(-7+i\sqrt7 )\,\gamma_3\cr
&&-i\sqrt7\,\gamma_4+(7+i\sqrt7 )\,\gamma_5-(7-i\sqrt7 )\,\gamma_6-i\sqrt7\, \delta_1-(7+i\sqrt7 )\,\delta_2\cr
&&-i\sqrt7 \,\delta_3+(1-i\sqrt7)\,\delta_4+i\sqrt7 \,\delta_5-(7+i\sqrt7)\,\delta_6.
\end{eqnarray}
$\om_{B_0}$ is simply the complex conjugate of the above.
It is possible to express the two complex coefficients of the 3--form flux through four of the real coefficients which we may choose freely. The other real coefficients are constrained by the form of the flux.
For the complex coefficients we find
\begin{eqnarray}
A_0&=&\tfrac{1}{14}\,[\,(1+i\sqrt7 )\,a^0+2\,a^1+S\,((1+i\sqrt7 )\,c^0+2\,c^1)\,],\cr
B_0&=&\tfrac{1}{14}\,[\,(1-i\sqrt7 )\,a^0+2\,a^1-S\,((1-i\sqrt7 )\,c^0+2\,c^1)\,].
\end{eqnarray}
Expressed in real coordinates, the flux takes the form
\begin{eqnarray}
{1\over (2\pi)^2\alpha'}\,G_3&=&(a^0-S\,c^0)\,\alpha_0+(a^1-S\,c^1)\,\alpha_1+(-a^0+S\,c^0)\,\alpha_2+(a^0-S\,c^0)\,\alpha_3\cr
&&+(a^0-S\,c^0)\,\beta^0+(-a^0+S\,c^0)\,\beta^1+(a^0-S\,c^0)\,\beta^2\cr
&&+(a^0+a^1-S\,(c^0+c^1)))\,\beta^3+(a^0-S\,c^0)\,\gamma_1+(-a^0-S\,c^0)\,\gamma_2
\cr
&&+(-a^0-a^1+S\,(c^0+c^1))\,\gamma_3+(a^0-S\,c^0)\,\gamma_4\cr
&&+(a^1-S\,c^1)\,\gamma_5+(-a^0-a^1+S\,(c^0+c^1))\,\gamma_6-(a^1-S\,c^1)\,\delta^6\cr
&&-(-a^0-a^1+S\,(c^1+c^1))\,\delta^4-(a^0-S\,c^0)\,\delta^5-(a^1-S\,c^1)\,\delta^2\cr
&&-(-a^0+S\,c^0)\,\delta^3-(a^0-S\,c^0)\,\delta^1.
\end{eqnarray}


\section{The effective potential from fluxes}

After giving a vev to the field $G_3$, the Chern--Simons 
term\footnote{Throughout
this section, we work in the string--frame, i.e. with the Einstein term
$\fc{1}{(2\pi)^7\ \alpha'^4}\ \int d^{10}x\ \sqrt{-g_{10}}\
e^{-2\phi_{10}}\  R$.}
 \begin{equation}\label{CSbulk}{
\Sc_{CS}=\h\ \fc{1}{(2\pi)^7\ \ap^4}\ \int \fc{C_4\wedge
G_3\wedge\ov G_3}{S-\ov S}\ .}
\end{equation}
of the 
ten--dimensional effective \tb action gives rise to a tadpole for the 
$RR$ four--form $C_4$ (in units of $T_3$)\footnote{Here $T_p=(2\pi)^{-p}\ap^{-\h-\frac{p}{2}}$ is the $Dp$--brane tension \cite{joep} and
$\phi_4$ the dilaton
field in $D=4$.}:
\begin{equation}\label{Nflux}{
N_{flux}=\fc{1}{(2\pi)^{4}\ \ap^{2}}\ \int_{X_6} H_3\wedge F_3\ .}
\end{equation}
Hence in the presence of $3$--form fluxes, the tadpole cancellation condition is
\begin{equation}\label{modcfour}{
N_{flux}+N_{D3}=Q_{3,tot},}
\end{equation}
where $Q_{3,tot}$ is the total $O3$--plane charge which can receive contributions from the $O3$--plane tension, possible 2--form fluxes on the brane world--volume and curvature terms.

In order to satisfy the supergravity equations of motion, the flux combination (\ref{fluxcomb})
has to obey the imaginary self--duality (ISD) condition \cite{GKP}
\begin{equation}\label{ISD}
\star_6 G=+i\ G.
\end{equation}
When $G$ fulfills the above equation with a minus instead of a plus on the right hand side, it is called {\it imaginary anti-self dual} (IASD).
This condition ensures the existence of a solution for the metric and
$4$--form.
The $CP$ even analog of (\ref{Nflux}) originates from the piece 
$\fc{-1}{2\cdot 3!}\fc{1}{(2\pi)^7\ap^4}
\int d^{10}x\ \sqrt{-g_{10}}\ |G_3|^2$ of the $D=10$ \tb action and leads to
the potential term in $D=4$:
\begin{equation}\label{CPeven}{
\fc{1}{2\ (2\pi)^7\ \ap^4}\ \int_{X_6} d^6y\ G_3\wedge\star_6\ov G_3\ .}
\end{equation}
According to \cite{GKP,KST}, 
the latter may be split into a purely topological term $V_{top}$, 
independent of the moduli fields, and a second term $V_{flux}$, relevant for the
$F$--term contribution to the scalar potential.
After the decomposition $G_3=G^{ISD}+G^{IASD}$ one obtains 
\cite{GKP,KST, Wolf}:
\begin{eqnarray}\label{fluxdec}
V_{flux}&=&\fc{1}{(2\pi)^7\ \ap^4}\ \int_{X_6}\ G^{IASD}\wedge \star_6\ov G^{IASD}
\ ,\cr
V_{top}&=&-e^{-\phi_{10}}\ T_3\ N_{flux}\,.
\end{eqnarray}
Hence, the total contributions to the scalar potential are:
\begin{eqnarray}\label{total}
V&=&V_D+V_F,\cr
V_D&=&V_{D3/D7}+V_{O3/O7}+V_{top} ,\cr
V_F&=&V_{flux} .
\end{eqnarray}
The piece $V_D$ represents $D$--term contributions to the scalar potential
due to Fayet--Iliopolous terms, see 
\cite{CIM} 
for further details.
Only the last term corresponds to an $F$--term.
In the case that the tadpole conditions are met, Ramond tadpole contributions 
to $V$ are absent. If in addition only supersymmetric $2$--form
fluxes on the $D7$--brane world--volume are considered (see Section \ref{braneswithflux}), the first three terms add up to zero: 
$V_{D3/D7}+V_{O3/O7}+V_{top}=0$, \ie $V_D=0$. 
Let us remark that this condition may generically also fix some of the
K\"ahler moduli $\Tc^j$.
In the following, we shall assume that $V_D=0$ and study only the
$F$--term contribution $V_F=V_{flux}$ to the scalar potential $V$.
The potential $V_F$ displayed in  (\ref{total}) originates from the closed string sector only.

\section{String--theoretical K\"ahler moduli $\Tc^i$ vs.  field--theoretical fields $T^i$}

The imaginary part of the K\"ahler modulus $\Tc^i$ follows from the integral 
$\im(\Tc^i)=\int_{C^i} C_4$ of the Ramond $4$--form over a certain $4$--cycle $C^i$ of $X$. 
A $D7$--brane has the world--volume
Chern--Simons coupling $\int\ C_4\wedge F\wedge F$. 
Hence, a $D7$--brane wrapped around this $4$--cycle
$C_i$ gives rise to the $CP$--odd gauge term $\int\ \im(\Tc^i)\ F\wedge F$ in $D=4$.
On the other hand, the real parts of the moduli $\Tc^i$, 
which derive from the underlying string background 
do not yet  properly fit into complex scalars $T^i$ of ${\cal N}=1$ chiral 
multiplets in field theory.
According to the previous discussion, the real part $\re(T^i)$ of those scalar fields 
has to describe 
the gauge coupling of a $D7$--brane, which is wrapped around the  four--cycle $C_i$.
This coupling is measured by the volume of this $4$--cycle $C_i$.
More precisely, from the Born--Infeld action $e^{-\phi_{10}}\ \int\ d^8\xi\ 
\det(g+2\pi\ap F)^{1/2}$ 
we derive the $CP$--even gauge--coupling $\re(T^i):=e^{-\phi_{10}}\ \int_{C_i} d^4\xi\ 
\det(g)^{1/2}$.
In order for the $D7$--brane to respect $1/2$ of the supersymmetry
of the bulk theory, which is ${\cal N}=2$ in $D=4$, the internal $4$--cycle $C_i$ the $D7$ brane
is wrapped on has to fulfill the calibration condition \cite{Ruben}:
\begin{equation}\label{calibration}{
\int_{C_i} d^4\xi\  \det(g)^{1/2}=\h\ \int_{C_i}\ J\wedge J\ .}
\end{equation}
Note, that the r.h.s. just describes the volume of the $4$--cycle $C_i$.
Hence, the real part of the correct holomorphic modulus $T^i$ is:
\begin{equation}\label{calibrationi}{
\re(T^i):=e^{-\phi_{10}}\ \int_{C_i} d^4\xi\  \det(g)^{1/2}=\h\ e^{-\phi_{10}}\ 
\int_{C_i}\ J\wedge J\ .}
\end{equation}
More precisely, with $\omega_i$ the Poincar\'e dual
$2$--form of the $4$--cycle $C_i$, we have:
$\int\limits_{C_i}\ J\wedge J=\int\limits_{X_6} \omega_i \wedge J\wedge J$. 
With $J=\sum\limits_{j=1}^{h_{(1,1)}}\re(\Tc^j)\ \omega_j$ we may write
\begin{equation}\label{NICE}{
\re(T^i):=e^{-\phi_{10}}\ \fc{\p}{\p \Tc^i}\ \int_{X_6}\ J\wedge J\wedge J=
e^{-\phi_{10}}\ \fc{\p}{\p \Tc^i}\ e^{-K}.}
\end{equation}
Furthermore, with $\int_{X_6} J\wedge J\wedge J=d_{ijk}\ \re(\Tc^i)\ \re(\Tc^j)\ \re(\Tc^k)$, and 
$d_{ijk}$ the intersection form, we may also write
\begin{equation}\label{ALSONICE}{
\re(T^i):=e^{-\phi_{10}}\ d_{ijk}\ \re(\Tc^j)\ \re(\Tc^k)\ ,}
\end{equation}
which gives the volume of the $4$--cycle in string units.

\subsection{Example D: $T^6/\IZ_2\times\IZ_2$}

For the complex structure moduli, we do not need a
redefinition in \tb , so $\Uc^i=U^i$ as given in (\ref{cpxtwotwo}). 
The imaginary part of the K\"ahler moduli is given
by the coupling of the gauge fields on a $D7$-brane $g^{-2}_{D7,j}$, which is wrapped on the
tori $T^{2,k}$ and $T^{2,l}$.  So 
\begin{equation}\label{fieldT}{
T^j=a^j+i\,{e^{-\phi_4}\over {2\pi\alpha'^{1/2}}}
\sqrt{{\im\,\Tc^k\im\,\Tc^l\over{\im\,\Tc^j}}},}
\end{equation}
with $\Tc$ given in (\ref{kmodtwotwo}).
The imaginary part of the dilaton $S$ is given by the gauge coupling on the
$D3$--brane $g^{-2}_{D3}$:
\begin{equation}\label{fieldS}{
S=C_0+i\,{e^{-\phi_4}\over {2\pi}}{\alpha'^{3/2}
\over{\sqrt{\im\,\Tc^1\im\,\Tc^2\im\,\Tc^3}}}.}
\end{equation}

\section{Superpotential, $F$--terms and scalar potential}

It is was shown in \cite{TV} that turning on background fluxes leads to the following superpotential in the low energy effective theory:\footnote{The factor of $\lambda$ serves to obtain the 
correct mass dimension of 3 for the superpotential, i.e. $\lambda^{-1}=16\,\pi^5\ap^3$, such that $\kappa_{10}^{-2}=\fc{\lambda}{(2\pi)^2\ap}$} 
\begin{equation}\label{TVW}{
\hat W={\lambda\over{(2\pi)^2\alpha'}}\int_{X} G_3\wedge \Omega\ .}
\end{equation}

We are interested in the low energy effective potential with fluxes turned on. First, we
will look at those quantities that can be derived in the closed string sector.
The $F$-terms can be calculated from the K\"ahler potential and the
superpotential (\ref{TVW}), 
they only feel the
bulk:
\begin{equation}\label{Fterm}{\ov{F}^{\ov{I}} = e^{\kappa_4^2\hat{K}/2}\ \hat{K}^{\ov{I}J} \ 
(\partial_J\hat{W}+\kappa_4^2\ \hat{W}\ \partial_J\hat{K})\ ,}
\end{equation}
where the $I, \ J$ are taken to run over the dilaton $S$, the complex structure
moduli $U^i$  and the K\"ahler moduli $T^i$. With this, we can now calculate
the scalar potential:
\begin{equation}\label{Vhut}{\hat{V}=\hat{K}_{I\ov{J}}\ F^I\ov{F}^{\ov{J}}-3\ e^{\kappa_4^2\hat{K}}\ 
\kappa_4^2\ |\hat{W}|^2\ .}
\end{equation}
The explicit expressions for the $F$-terms are the following:
\begin{eqnarray}\label{ftermsT}
\ov{F}^{\ov{S}}&=&(S-\ov{S})\prod_M (M-\ov M)^{-1/2}
\int\ov{G}_3\wedge \Omega\cr
\ov{F}^{\ov{T}^i}&=&(T^i-\ov{T}^i)\prod_M (M-\ov M)^{-1/2}\kappa_4^2\
\hat{W}\cr
\ov{F}^{\ov{U}^i}&=&(U^i-\ov{U}^i)\prod_M (M-\ov M)^{-1/2}\ \kappa_4^2{\lambda\over{(2\pi)^2\alpha'}} \int G_3\wedge
\omega_{A_i}.
\end{eqnarray}
By looking at the $F$-terms, we see immediately that we only have a non-zero
$F^S$ if $G_3$ has a $(3,0)$-component. For $F^{T^i}$ to be non-zero, $G_3$
has to have a non-zero $(0,3)$-component. For the $F^{U^i}$ to be non-zero,
$G_3$ must have a $(1,2)$-component.

\subsection{Example D: $T^6/\IZ_2\times \IZ_2$}

Expressed with the coefficients of the real basis, we find $N_{\rm flux}$,
given in (\ref{Nflux}), to be
$$N_{\rm flux}=\sum_{i=0}^3 c^ib_i-\sum_{i=0}^3a^id_i.$$
We want to find the corresponding expression in complex language. We find 
\begin{eqnarray}\label{cplxNflux}
N_{\rm flux}&=&{1\over {(2\pi)^4(\alpha')^2}}{1\over {(S-\ov S)}}\int\ov
G_3\wedge G_3\cr
&=&-{{\prod\limits_{i=1}^3(U^i-\ov U^i)}\over{(S-\ov
S)}}\sum_{i=0}^3(|A^i|^2-|B^i|^2),
\end{eqnarray}
which is quite a nice expression. And it immediately teaches us something about
the behaviour of the different fluxes: The fluxes obeying the ISD-condition,
i.e. those having all $B_i=0$, have $N_{\rm flux}>0$, whereas the IASD-fluxes,
i.e. those with all $A_i=0$ have $N_{\rm flux}<0$.

With (\ref{cplxbasetwotwo}), it is now easy to write down the
superpotential explicitly for $T^6/\IZ_2\times \IZ_2$:
\begin{eqnarray}\label{WhT}
{1\over \lambda}\hat{W}&=&(a^0-Sc^0)U^1U^2U^3-\{(a^1-Sc^1)U^2U^3+
(a^2-Sc^2)U^1U^3+(a^3-Sc^3)U^1U^2\}\cr
&&-\sum_{i=1}^3(b_i-Sd_i)U^i-(b_0-Sd_0).
\end{eqnarray}
The explicit expressions for the $F$-terms (\ref{ftermsT}) are the following:
\begin{eqnarray}
\ov{F}^{\ov{S}}&=&\lambda\kappa_4^2(S-\ov{S})\prod_M(M-\ov M)^{-1/2}\times\{(a^0-\ov{S}c^0)U^1U^2U^3-[(a^1-\ov{S}c^1)U^2U^3\cr
&&+(a^2-\ov{S}c^2)U^1U^3+(a^3-\ov{S}c^3)U^1U^2]-\sum_{i=1}^3(b_i-\ov{S}d_i)U^i-(b_0-\ov{S}d_0)\} ,
\\
\ov{F}^{\ov{T}^i}&=&\lambda\kappa_4^2(T^i-\ov{T}^i)\prod_M(M-\ov M)^{-1/2}\times\{(a^0-Sc^0)U^1U^2U^3-[(a^1-Sc^1)U^2U^3\cr
&&+(a^2-Sc^2)U^1U^3+(a^3-Sc^3)U^1U^2]-\sum_{i=1}^3(b_i-Sd_i)U^i-(b_0-Sd_0)\},\\
\ov{F}^{\ov{U}^i}
&=&\kappa_4^2\,(U^i-\ov{U}^i)\prod_M(M-\ov M)^{-1/2}\times\,{\lambda\over{(2\pi)^2\alpha'}} \int G_3\wedge
\omega_{A_i}\ ,{\rm e.g.:}\cr
\ov{F}^{\ov{U}^1}&=&\lambda\kappa_4^2\,(U^1-\ov{U}^1)\prod_M(M-\ov M)^{-1/2}\times\{((a^0-Sc^0)\ov{U}^1U^2U^3-[(a^1-Sc^1)U^2U^3\cr
&&+(a^2-Sc^2)\ov{U}^1U^3+(a^3-Sc^3)
\ov{U}^1U^2]-[(b_1-Sd_1)\ov{U}^1+(b_2-Sd_2)U^2+\cr
&&(b_3-Sd_3)U^3]-(b_0-Sd_0)\},
\end{eqnarray}
where $M$ runs over all moduli $S,\,U^i,\,T^i$.
By looking at the $F$-terms, we see immediately that we only have a non-zero
$F^S$ if $G_3$ has a $(3,0)$-component. For $F^{T^i}$ to be non-zero, $G_3$
has to have a non-zero $(0,3)$-component. For the $F^{U^i}$ to be non-zero,
$G_3$ must have a $(1,2)$-component.

Now we are able to compute the expression for the scalar potential (\ref{Vhut}). The part
coming from the $T$-moduli cancels with $-3\,e^{\kappa_4^2\hat{K}}\kappa_4^2\,|\hat{W}|^2$, which is a generic property in no--scale models. So 
we are left with
\begin{eqnarray}\label{VhutT}
\hat{V}&=&\partial_S\partial_{\ov{S}}\hat{K}F^S\ov{F}^{\ov{S}}+
\sum_{i=1}^3\partial_{U_i}\partial_{\ov{U_i}}\hat{K}F^{U_i}\ov{F}^{\ov{U_i}}\cr
&=&{\lambda^2\kappa_4^2\over (2\pi)^4\alpha'^2}\prod_M\frac{1}{|M-\ov{M}|}\left(|\!\int\ov{G}_3\wedge
\Omega\,|^2+\sum_{i=1}^3\,|\!\int G_3\wedge \omega_{A_i}\,|^2\right).
\end{eqnarray}
We can see immediately that $\hat{V}$ is zero unless $G_3$ has a $(3,0)$- or a
$(1,2)$-part, i.e. is IASD. We also see immediately, that for IASD-fluxes,
$\hat V$ is strictly positive.
When we express this through the complex coefficients, this formula looks even
nicer:
\begin{equation}\label{VhutTc}
{\hat{V}=\lambda^2\kappa_4^2\ {\prod\limits_{i=1}^3|U^i-\ov{U}^i|\over|S-\ov{S}
|\prod\limits_{i=1}^3|T^i-\ov{T}^i|}\ \sum_{j=0}^3|B^j|^2.}
\end{equation}
Let us examine
eq. (\ref{CPeven}) : Expressed with our complex coefficients, we find
$${\int G_3\wedge \star_6\ov G_3\ \propto\  2\sum_{i=0}^3|B^i|^2+(\sum_{i=0}^3|A^i|^2-
\sum_{i=0}^3|B^i|^2),}$$
where the second term is obviously proportional to $N_{\rm flux}$, whereas
the first part corresponds to $V_{\rm flux}$, which is the contribution to the
scalar potential coming from the $F$-terms, which is exactly, what we have
calculated above.

\section{Supersymmetry conditions for the background flux}

For the flux to preserve ${\cal N}=1$ supersymmetry, the condition that all $F$--terms vanish must be imposed. This results in requiring 
\begin{eqnarray}\label{susycond}
\hat W&=&{\lambda\over{(2\pi)^2\alpha'}}\ \int G_3\wedge \Om=0,\cr
D_S\hat W&=&\partial_{S}\hat W+\kappa_4^2\ \hat W\ \partial_{S}\hat K=0,\cr
D_{U^i}\hat W&=&\partial_{U^i}\hat W+\kappa_4^2\ \hat W\ \partial_{U^i}\hat K=0,
\end{eqnarray}
where $D_M$ is the K\"ahler covariant derivative.
This corresponds to setting all $F$-terms to zero.
The above conditions can be rewritten as
\begin{equation}
\int G_3\wedge \Omega =0,\quad \int G_3\wedge \ov{\Omega} =0,\quad  \int G_3\wedge \omega_{A1}=\int G_3\wedge \omega_{A2}=\int
G_3\wedge\omega_{A3}=0.
\end{equation}
The first of the above equations if fulfilled for
$G_3$ not having a $(0,3)$-part. The second is fulfilled for
 $G_3$ not having a $(3,0)$-part and the remaining three equations are fulfilled for 
 $G_3$ not having a $(1,2)$-part. Therefore a supersymmetric flux can only have $(2,1)$--components and is automatically $ISD$.

\subsection{Example D: $T^6/\IZ_2\times\IZ_2$}

For this concrete example, the necessary equations are obtained by
setting the coefficients of the $(0,3)$-, $(1,2)$-, and $(3,0)$-flux to zero,
i.e. $A_0=B_0=B_1=B_2=B_3=0$. This corresponds to
\begin{eqnarray}
0&=& U^1U^2U^3(a^0-S c_0)-\sum_{i\neq j\neq k}(a^i-S c^i)U^jU^k-(b_0-S d_0)-
\sum_{i=1}^3(b_i-S d_i) U^i\cr
0&=& \ov{ U}^1U^2U^3(a^0-S c_0)-\{(a^1-Sc^1)U^2U^3+(a^2-S c^2)
\ov{U}^1U^3+(a^3-S c^3) \ov{U}^1U^2\}-\cr
&&-(b_0-Sd_0)-\{(b_1-S d_1) \ov{ U}^1+(b_2-S d_2) U^2+(b_3-S d_3) U^3\} \cr
0& =&\ov{U}^1\ov{U}^2\ov{U}^3(a^0-S c_0)-\sum_{i\neq j\neq k}(a^i-S c^i)\ov{U}^j\ov{U}^k-(b_0-S
d_0)-\sum_{i=1}^3(b_i-S d_i)\ov{ U}^i \cr
0&=& U^1 \ov{U}^2U^3(a^0-S c_0)-\{(a^1-S c^1)
\ov{U}^2U^3+(a^2-S c^2)U^1U^3+(a^3-S c^3)U^1 \ov{U}^2\}-\cr
&&-(b_0-S
d_0)-\{(b_1-S d_1) U^1+(b_2-S d_2) \ov{ U}^2+(b_3-S d_3) U^3\} \cr
0&=& U^1U^2 \ov{U}^3(a^0-S c_0)-\{(a^1-S c^1)U^2
\ov{U}^3+(a^2-S c^2)U^1 \ov{U}^3+(a^3-S c^3)U^1U^2\}-\cr
&&-(b_0-S
d_0)-\{(b_1-S d_1) U^1+(b_2-S d_2) U^2+(b_3-S d_3)  \ov{U}^3\},
\end{eqnarray}
with $i\neq j\neq k$.

Now one can solve for the $\{a^i, c^i,b_i, d_i\}$ and impose the constraint
that they be integer numbers to fulfill the quantization condition (\ref{fluxqu}). 
These constraints cannot be satisfied in
full generality, i.e. for arbitrary moduli and flux coefficients. By fixing
some of the moduli and/or flux coefficients, it is possible to obtain special
solutions. Here, we choose to fix the moduli to $U^1=U^2=U^3=S=i$.

One possible solution for the $(2,1)$-flux is
\begin{eqnarray}\label{eqzweieins}
{1\over{(2\pi)^2\alpha'}}\ G_{21}&=&[-d_0+i(d_1+d_2+d_3)]\ \alpha_0+[-d_1-i(-b_2-b_3+d_0)]\ 
\alpha_1\cr
&&+(-d_2-ib_2)\ \alpha_2+(-d_3-ib_3)\ \alpha_3+(-d_1-d_2-d_3-id_0)\ \beta^0\cr
&&+(-b_2-b_3+d_0-id_1)\ \beta^1+(b_2-id_2)\ \beta^2+(b_3-id_3)\ \beta^3,
\end{eqnarray}
where $b_2, b_3, d_0, d_1, d_2, d_3$ can be any integer number. As mentioned before, possible complications with flux quantization can be avoided, if
the flux is taken to have coefficients which are multiples of 8. This can be
achieved by simply taking  $b_2, b_3, d_0, d_1, d_2, d_3$ to be multiples of
8.

Expressed in the complex basis, the solution takes the form
\begin{eqnarray}\label{eqzweieinscplx}
{1\over{(2\pi)^2\alpha'}}\ G_{21}&=&\half\ [-b_2-b_3+i(d_2+d_3)]\ \om_{A1}+
\half\ [b_2-d_0+i(d_1+d_3)]\ \om_{A2}+\cr
&&+\half\ [b_3-d_0+i(d_1+d_2)]\ \om_{A3} .
\end{eqnarray}
For $N_{\rm flux}$ we find
\begin{eqnarray}\label{Nfluxsusy}
N_{\rm flux}&=&4\ (|A^1|^2+|A^2|^2+|A^3|^2)\cr
&=&2\ (\sum_{i=0}^3d_i^2+d_1d_2+d_1d_3+d_2d_3+b_2^2+b_3^2+b_2b_3-b_2d_0-b_3d_0).
\end{eqnarray}
If we require $N_{\rm flux}$ to have a certain value, this places quite
stringent constraints on our choice for the coefficients. The smallest
possible $N_{\rm flux}$ for our solution, the coefficients being multiples of
8, is $N_{\rm flux}=128$. To achieve this, we have several possibilities. We can
for example set either of the $d_i$ or $b_i$ to $\pm 8$, and all the other coefficients to
zero. For $d_0=8$ for example, all other coefficients being zero, this would amount to
$${1\over{(2\pi)^2\alpha'}}G_3=8\ (-\alpha_0-i\alpha_1-i\beta^0+\beta^1).$$
Another possible solution would be $b_2=8,\quad b_3=-8$, or the other way
round. This would result in
$${1\over{(2\pi)^2\alpha'}}G_3=8\ (-i\alpha_2+i\alpha_3+\beta^2-\beta^3).$$


\chapter{Braneworld Scenarios: Model building and soft SUSY breaking}

\begin{figure}[h!]
\begin{center}
\includegraphics[width=120mm]{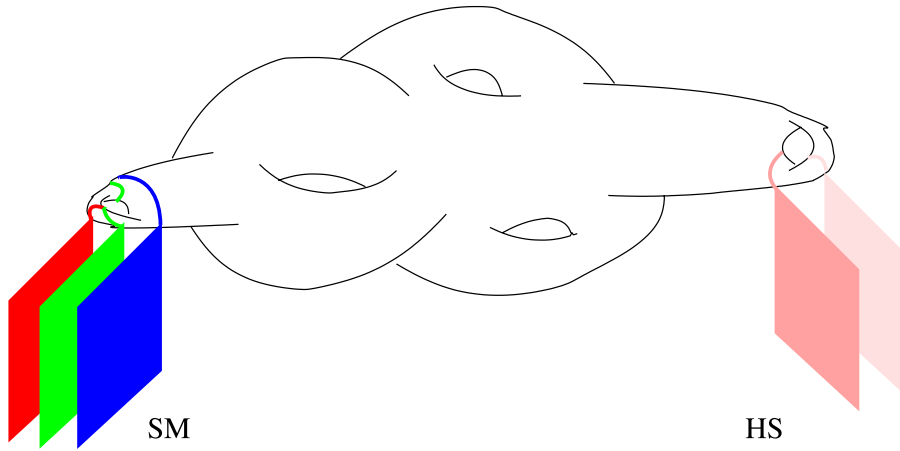}
\caption{Braneworld model with standard model and hidden sector $D$--branes wrapping cycles of a Calabi--Yau manifold}
\label{fig:cy_branes}
\end{center}
\end{figure}

One very promising way to reproduce the MSSM in string theory is to use
intersecting $D6$--branes in \ta orientifolds (for a  review see \cite{LustKS}).
The gauge degrees of freedom are due to open strings living
on each of the various stacks of $D6$--branes, whereas the chiral matter fields
are localized on the lower-dimensional intersection loci of the
$D6$--branes. More specifically, the $D6$--branes, all completely filling 
four-dimensional Minkowski space-time, are wrapped around supersymmetric
3-cycles in the internal space $X$, which generically intersect just
on points in $X$. Figure \ref{fig:cy_branes} schematically shows the set--up of space--time filling $D$--branes which constitute the standard model and hidden sectors wrapping some cycles on a Calabi--Yau manifold.

Note that the internal intersection numbers are
normally larger than one, a fact, which offers a nice explantion for
the family replication of the MSSM.
In order to preserve N=1 space-time supersymmetry 
in the open string sectors on the
intersecting $D$--branes, the intersection
angles must add up to $0\ \mod\,2\pi$. For a consistent string vacuum, all
Ramond and Neveu--Schwarz tadpoles must be cancelled. The tadpole conditions translate to conditions on the number of D--brane stacks and their arrangement.
Starting from the original work on non-supersymmetric
models \cite{BDL,BachasIK,BlumenhagenWH,AngelantonjHI,AldazabalDG}, 
several semirealistic MSSM-like models with intersecting
$D6$--branes
were constructed during the last years
\cite{uranga,DijkstraYM,BlumenhagenCG}.
However, for practical reasons, when turning on the SUSY-breaking fluxes, it is more convenient to use instead of the \ta
orientifolds with intersecting $D6$--branes the mirror ($T$--dual) \tb
orientifold description. After an appropriate mirror
transformation, the (supersymmetric) $D6$--branes are transformed into a system of
$D3$--branes plus supersymmetric $D7$--branes, where the non-trivial intersection
angles
in \ta become open string 2--form gauge fluxes (magnetic $f$-field
background fields) living on internal 4--cycles on the different
$D7$--brane world volumes. 
Note that $f$--fluxes are required at least on some
of the various stacks of $D7$--branes in order
to obtain realistic models with more than one {\sl chiral} generation
of quarks and leptons. Hence for getting chiral fermions, some of the $D7$--branes
must have mixed Dirichlet/Neumann boundary conditions in certain
internal directions. This means that they are a kind of
hybrid between $D3$-- and $D7$--branes.
This fact will become important for the structure of
the soft terms for the matter fields on the $D3/D7$--brane world volumes.

For this whole chapter, we will be in the $IIB$ orientifold theory compactified on $T^6/\IZ_2\times \IZ_2$ with three stacks of $D7$--branes, where each stack is wrapped on two of the three $T^2$--factors.

\section{Branes carrying internal gauge flux}\label{braneswithflux}

To obtain a chiral spectrum, we
must introduce (magnetic) two--form fluxes $F^j dx^j\wedge dy^j$ on the
internal part of the $D7$--brane world volume. 
Together with the internal $NS$ $B$--field\footnote{Note, that $b^j$ has to be quantized due to
the orientifold projection $\Om$ to the values $b^j=0$ or $b^j=\h$ \cite{Carlo}.} $b^j$
we combine the complete $2$--form flux into 
$\Fc=\sum\limits_{j=1}^3\Fc^j:=\sum\limits_{j=1}^3(b^j+2\pi \ap F^j)\ dx^j\wedge dy^j$. 
The latter gives rise to the total internal antisymmetric background
\begin{equation}\label{antisbg}{
\left(\begin{array}{cc}0&f^j\cr-f^j&0\end{array}\right),\ \ \ 
f^j=\fc{1}{(2\pi)^2}\ \int_{T^{2,j}} \Fc^j ,}
\end{equation}
with respect to the $j$--th internal plane.
The $2$--form fluxes $\Fc^j$ have to obey the quantization rule
\begin{equation}\label{fluxquant}{
m^j\fc{1}{(2\pi)^2\ap}\ \int_{T^{2,j}} \Fc^j=n^j ,\ \ \ n\in \IZ ,} 
\end{equation}
i.e. $f^j=\ap \fc{n^j}{m^j}$.
We obtain non--vanishing instanton numbers
\begin{equation}\label{instanton}{
m^j\ m^k\ \int\limits_{T^{2,j}\times T^{2,k}} \Fc\wedge \Fc =(2\pi)^4\ \ap^2\ n^j\ n^k}
\end{equation}
on the world--volume of a $D7$--brane, which is wrapped around the $4$--cycle 
$T^{2,j}\times T^{2,k}$ with the wrapping numbers $m^j$. Hence, through the
$CS$--coupling $T_7\ C_4\wedge \Fc\wedge \Fc$, a $D7$--brane may also induce contributions to the
$4$--form potential.
Note, that a $D3$--brane may be described by a $D7$--brane with $f^j\ra\infty$.
To cancel the tadpoles arising from the Ramond--Ramond forms $C_4$ and $C_8$, we introduce
$N_{D3}$ (space--time filling) $D3$--branes and $K$ stacks of $D7$--branes with internal fluxes. 
More concretely, $K^i$ stacks of $N_a^i$ $D7$--branes with internal $2$--form fluxes 
$\Fc^j,\Fc^k$ and wrapping numbers $m_a^j,m_a^k$ with respect to the $4$--cycle $T^{2,j}\times T^{2,k}$.
The cancellation condition for the tadpole arising from the $RR$ $4$--form $C_4$ is
\begin{equation}\label{cfour}{
N_{D3}+\fc{2}{(2\pi)^4\ap^2}\ \sum_{(i,j,k)\atop=\overline{(1,2,3)}}\sum_{a=1}^{K^i}\ N^i_a\ 
m_a^j\ m_a^k\ \int\limits_{T^{2,j}\times T^{2,k}} \Fc \wedge \Fc =32 ,}
\end{equation}
i.e. according to  (\ref{instanton})
\begin{equation}\label{ccfour}{
N_{D3}+2\  \sum_{(i,j,k)\atop=\overline{(1,2,3)}}\sum_{a=1}^{K^i}\ N^i_a\ n_a^j\ n_a^k=32.}
\end{equation}
Furthermore, the cancellation conditions for the $8$--form tadpoles yield:
\begin{eqnarray}\label{ceight}
2\ \sum_{a=1}^{K^3}\ N^3_a\ m_a^1\ m_a^2&=&-32,\cr
2\ \sum_{a=1}^{K^2}\ N^2_a\ m_a^1\ m_a^3&=&-32 ,\cr
2\ \sum_{a=1}^{K^1}\ N^1_a\ m_a^2\ m_a^3&=&-32.
\end{eqnarray}
The extra factor of $2$ in front of the sums over the $D7$--brane stacks accounts
for additional mirror branes. For each $D7$--brane with wrapping numbers $(m^i,m^j)$, we also
have to take into account its mirror $(-m^i,-m^j)$ in order to cancel induced
$RR$ $6$--form charges. The right hand sides of  (\ref{cfour}) and (\ref{ceight}) account for the 
contributions of the $O3$-- and $O7$--planes, respectively.
An $O3$--plane contributes $-\fc{1}{4}$ of a $D3$--brane charge $T_3$. Here, $T_p=(2\pi)^{-p}\ap^{-\h-\fc{p}{2}}$ is the $Dp$--brane tension \cite{joep} and
$\phi_4=\phi_{10}-\h\ln\lf[\im(\Tc^1)\im(\Tc^2) \im(\Tc^3)/\alpha'^3\ri]$ the dilaton
field in $D=4$.
In the covering space
the $64$ $O3$--planes are doubled, thus contributing $2\times 64\times (-\fc{1}{4})=-32$ on the
l.h.s. of (\ref{ccfour}). On the other hand, in $D7$--brane charge $T_7$ units, an $O7$--plane contributes
$4\,T_7$. Hence, in the covering space, four $O7$--planes contribute $2\times 4\times 4=32$
on the l.h.s. of (\ref{ceight}).

A $D3$--brane placed in the uncompactified $D=4$ space--time produces the
contribution
\begin{equation}\label{PDthree}
V_{D3}=T_3\ e^{-\phi_4}\ \fc{\ap^{3/2}}{\sqrt{\Tc_2^1 \Tc_2^2 \Tc_2^3}}
\end{equation}
to the total scalar potential $V$.
Furthermore, a $D7$--brane, wrapped around the $4$--cycle $T^{2,j}\times
T^{2,k}$
with wrapping numbers $m^j,m^k$ and internal gauge fluxes $f^k,f^l$ gives rise
to the potential
\begin{equation}\label{PDseven}{
V_{D7_j}=-T_7\ (2\pi)^4\ \ap^{3/2}\ e^{-\phi_4}\ m^k\ m^l\ \lf|1+i\fc{f^k}{\Tc_2^k}\ri|\
\lf|1+i\fc{f^l}{\Tc_2^l}\ri|\ \sqrt\fc{\Tc_2^k\ \Tc_2^l}{\Tc_2^j} .}
\end{equation}
In order that the $D7$--branes preserve some supersymmetry, their internal $2$--form fluxes 
$f^i,f^j$  must obey the supersymmetry condition \cite{BDL}:
\begin{equation}\label{susy}{f^i\over {\im\Tc^i}}=-{f^j\over{\im\Tc^j}} .
\end{equation}
In that case, the potential (\ref{PDseven}) simplifies:
\begin{equation}\label{PDseveni}{
V_{D7_j}=-T_7\ (2\pi)^4\ \ap^{3/2}\ e^{-\phi_4}
\ m^k\ m^l\ \lf(1-\fc{f^k\ f^l}{\Tc_2^k\ \Tc_2^l}\ri)
\ \sqrt\fc{\Tc_2^k\ \Tc_2^l}{\Tc_2^j} .}
\end{equation}
Hence, the presence of $N_{D3}$ space--time filling $D3$--branes and various stacks of 
$D7$--branes produces a positive potential\footnote{The extra factor of two in front of the 
$D7$--brane sum accounts for the mirror branes.}
\begin{equation}\label{potentialD}{
V_{D3/D7}=N_{D3}\ V_{D3}+2\ \sum_{j=1}^3 \sum_{a=1}^{K^j} N^j_a\ V_{D7_j} .}
\end{equation}
Furthermore, a negative potential is generated by the presence of the
$64$\ $O3$-- and $12$ $O7_j$--orientifold planes:
\begin{equation}\label{potentialDD}{
V_{O3/O7}=2  e^{-\phi_4}\, \ap^{3/2} \lf\{-\fc{64\,T'_3}{\sqrt{\Tc_2^1 \Tc_2^2 \Tc_2^3}}-
4\,T'_7\ (2\pi)^4\!\lf(\!\sqrt\fc{\Tc_2^2\ \Tc_2^3}{\Tc_2^1}+\sqrt\fc{\Tc_2^1\ \Tc_2^3}{\Tc_2^2}+
\sqrt\fc{\Tc_2^1\ \Tc_2^2}{\Tc_2^3}\ri)\!\ri\}.}
\end{equation}
Here, the orientifold tension for $Op$--planes is given by $T_p'=2^{p-5}T_p$ \cite{joep}.
The extra factor of $2$ is due to the covering space.
In the case of supersymmetric $D7$--branes, i.e. (\ref{susy}) holding for each brane, we have
\begin{equation}\label{susyi}{
V_{D3/D7}+V_{O3/O7}=0,}
\end{equation}
provided the tadpole conditions (\ref{cfour}) and (\ref{ceight}) are fulfilled.

The simplest solution to the equations (\ref{cfour}) and (\ref{ceight}) is represented by the following 
example: We take $32$ space--time filling $D3$--branes and place $8$ $D7$--branes on top of each 
of the $12$ $O7$--planes. This leads to a non--chiral spectrum and the $96$ $D7$--branes 
give rise to the gauge group $SO(8)^{12}$ \cite{GM}.

The requirement that a stack of branes $a$ with internal $2$--form fluxes $f^j_a$ is supersymmetric 
has the form
$${\sum_{j=1}^3\arctan\biggl(\fc{f^j_a}{\im({\cal T}^j)}\biggr)=0 .}$$
Furthermore, the condition, that a stack of branes $a$ with $2$--form fluxes $f^j_a$ and another stack
$b$ with $2$--form fluxes $f^j_b$ are
mutually supersymmetric is 
$$\sum_{j=1}^3\th_{ab}^j=0\ \mod\ 2 ,$$
with the relative ``flux'' $\th_{ab}^j$:
\begin{equation}\label{relativflux}
\th_{ab}^j=\fc{1}{\pi}\ \lf[\ \arctan\lf(\fc{f_b^j}{\im(\Tc^j)}\ri)-
\arctan\lf(\fc{f_a^j}{\im(\Tc^j)}\ri)\ \ri].
\end{equation}
These conditions will fix some of the K\"ahler moduli ${\cal T}^j$. Note that in the $T$--dual type $IIA$--picture,  the $\th_{ab}^j$ are exactly the angles between the two stacks of intersecting $D6$--branes.

\section{Open string low-energy effective action and soft terms}
    
There are several types of moduli fields in a type $IIB$ orientifold
compactification 
with $D$--branes. 
The closed string moduli fields arise from dimensional reduction of the 
bosonic part $(\phi,g_{MN},b_{MN},C_0,C_2,C_4)$ of the N=2 supergravity multiplet in $D=10$
after imposing the orientifold and orbifold action.
The spectrum has to be invariant under both
the orientifold action $\Om(-1)^{F_L}I_6$ and the orbifold group $\Gamma$.
Before applying the orbifold twist $\Gamma$, the untwisted sector consists of the states
invariant under $\Om(-1)^{F_L}I_6$: $\phi,g_{ij},b_{\mu i},C_0, C_{\mu i}, C_{ijkl},C_{\mu\nu ij},
C_{\mu\nu\rho\sigma}$. 
In addition, there are twisted
moduli consisting of the twisted $RR$--tensors.

Let us now come to the open string moduli fields. The massless untwisted moduli fields
originate from the $D=10$ gauge field $A_M$ reduced on the various $D$--branes.
The orientifold projection $\Om$ determines the allowed  Chan--Paton gauge degrees of 
freedom at the open string endpoints.
For a stack of space--time filling $D3$--branes, we obtain $6$ real scalars 
$\phi^i\ ,\ i=4,\ldots, 9$ in the adjoint of the 
gauge group of the respective stack. These scalars describe the transversal 
movement of the $D3$--branes, i.e. essentially the location of the $D3$--branes on the 
six--dimensional compactification manifold.
They may be combined into the three complex fields 
$C_i^3=\phi^{2i+2}+\Uc^i \phi^{2i+3}\ ,\ i=1,2,3$.
Furthermore, for a stack of $D7_3$--branes, which is wrapped around the $4$--cycle 
$T^{2,1}\times T^{2,2}$, we obtain the four Wilson lines $A_i\ ,\ i=4,5,6,7$ and 
two transversal coordinates $\phi^8,\phi^9$ in the adjoint representation. 
The latter describe the position of the $D7$--brane on the $2$--torus $T^{2,3}$.
Again, these six real fields may be combined into three complex fields $C^{7_3}_i\ ,\ i=1,2,3$.
After taking into account the other two $4$--cycles, 
on which other stacks of $D7$--branes may be wrapped, in total, 
we obtain the complex fields $C^{7_j}_i\ ,\ i,j=1,2,3$.
All the open  string fields described so far give rise to complex scalars of 
the untwisted sector with at least ${\cal N}=2$ supersymmetry.
A stack of  $D3$--branes gives rise to an ${\cal N}=4$ super Yang--Mills theory on its world--volume,
provided the stack does not sit at an orbifold singularity.
Hence, together with their world--volume gauge fields, the scalars $C_i^3$ are organized 
in ${\cal N}=4$ vector multiplets. The supersymmetry on the world--volume of a 
$D7$--brane, which is wrapped around a generic supersymmetric $4$--cycle of a
CY--space is ${\cal N}=2$. The scalar $C^{7_i}_i$ describes a (complex) scalar of a vectormultiplet.

Then there are ${\cal N}=2$ and chiral ${\cal N}=1$ fields in the so--called twisted sector which describes fields with endpoints on different brane stacks.
The twisted matter fields $C^{37_a}$ 
originate from open strings stretched between the $D3$-- and $D7$--branes from the $a$--th
stack. 
Generically, these fields respect ${\cal N}=2$ supersymmetry. However, there are twisted ${\cal N}=1$
matter fields $C^{7_a7_b}$ arising from open strings 
stretched between two different stacks $a$ and $b$ of $D7$--branes.

Let us now move on to the low--energy effective action describing the dynamics
of the various  moduli fields encountered above.
The complex scalars $S,T^j,U^j$ give rise to the closed string or bulk--moduli space. The untwisted open string moduli describe either the displacement 
transverse to the $D$--brane world--volume or the breaking of the gauge group by Wilson lines.
It is justified to expand the (full) K\"ahler potential $K$ and superpotential $W$
around this minimum $C_i=0$. 

The low--energy effective action for the massless open string sector of the
$D3/D7$--branes was computed by calculating string scattering amplitudes among open
string matter fields on the $D$-branes and bulk moduli fields  \cite{LMRS,LustFI}.
The charged matter fields $C$ enter the
K\"ahler potential at quadratic order as (for large K\"ahler moduli, which corresponds to the 
supergravity approximation under consideration): 
 \begin{eqnarray}\label{KK}
K(M,\ov M, C, \ov C)&=&\hat{K}(M, \ov M)+\sum_{a}\sum_{j=1}^3\sum_{i=1}^3 
G_{C^{7_a,j}_i\ov C^{7_a,j}_i}(M,\ov M)\ C^{7_a,j}_i\ov C^{7_a,j}_i\cr
&&+\sum_{a\neq b} G_{C^{7_a7_b}\ov C^{7_a7_b}}(M,\ov M)\ C^{7_a7_b}\ \ov C^{7_a7_b}+\Oc(C^4) .
\end{eqnarray}
Here, $\hat{K}(M, \ov M)$ is the closed string moduli K\"ahler potential (\ref{Khut}), discussed before.
The open string moduli fields $C$ summarize both untwisted $D7$--brane moduli $C^{7,j}_i$ 
and twisted matter fields $C^{7_a7_b}$. The fields $C^{7,j}_i$ 
account for the transverse $D7$--brane positions $C^{7,j}_j$ on the $j$--th subplane
and for the Wilson line moduli $C^{7,j}_i\ ,\ i\neq j$  on the $D7$--brane world volume. 
On the other hand, the fields $C^{7_a7_b}$ represent twisted matter fields originating 
from strings stretched between two stacks of $D7$--branes $a$ and $b$.
We have only displayed the $D7$--brane sector, as the $D3$--brane sector follows from
the latter by taking the limits $f^j\ra\infty$.
Furthermore, the holomorphic superpotential $W$ takes the form:
\begin{eqnarray}\label{super}
W(M, C)&=&\hat{W}(M)+\sum_{a=1}^3 C^{7_a}_1 C_2^{7_a} C_3^{7_a}+\sum_{a,b,j} d_{abj}\ 
C_j^{7_a} C^{7_{a}7_b} C^{7_a 7_b}\cr
&&+C^{7_17_2}C^{7_37_1}C^{7_27_3}+\sum_{I,J,K} Y_{IJK}(U^i)\ C_IC_JC_K+ \Oc(C^4) .
\end{eqnarray}
Again, $\hat{W}(M)$ is the closed string superpotential (\ref{TVW}), discussed before.
Finally, the coupling of the (closed string) moduli to the gauge fields is described by the
gauge kinetic functions. For the gauge fields living on the $D7$--branes,
wrapped around the $4$--cycle $T^{2,k}\times T^{2,l}$, these functions are given by 
\cite{LMRS,LustFI}
\begin{eqnarray}\label{gauge}
f_{D7_j}(S,T^j)&=&|m^k m^l|\lf(T^j-\ap^{-2} f^kf^lS \ri) , \ \ \
(j,k,l)=\overline{(1,2,3)},\cr
f_{D_3}(S)&=&S.
\end{eqnarray}
$m^k, m^l$ being the wrapping numbers.

The matter field metric for stings living on $D3$--branes is particularly simple:
\begin{equation}\label{Drei}{
G_{C^3_i\ov C^3_i}=\fc{-\kappa_4^{-2}}{(U^i-\ov U^i)\ (T^i-\ov T^i)\ } ,\ \ \ i=1,2,3 .}
\end{equation}
The metric for the untwisted matter fields living on the same stack of
$D7$--branes is the following:
\begin{eqnarray}\label{metricsfield}
G_{C^{7,j}_i\ov C^{7,j}_i}&=&{-\kappa_4^{-2}\over {(U^i-\ov U^i)\ (T^k-\ov{T}^k)}}\ 
{{|1+i\tilde{f}^k|}\over{|1+i\tilde{f}^i|}}\ ,\cr
G_{C^{7,j}_j\ov C^{7,j}_j}&=&{-\kappa_4^{-2}\over{(U^j-\ov U^j)\ (S-\ov{S})}}\ 
|1-\tilde{f}^i\tilde{f}^k|\ ,\quad i\neq k\neq j .
\end{eqnarray}
The matter field K\"ahler metric describing a 1/4 BPS sector 
is given by the following expression
\cite{LMRS,LustFI}:
\begin{equation}\label{siebensieben}{
G_{C^{7_a7_b}\ov C^{7_a7_b}}=\kappa_4^{-2}\, (S-\ov S)^{-\fc{1}{4}+\fc{3\beta}{2}+\gamma}
\prod_{j=1}^3 (T^j-\ov T^j)^{-\fc{1}{4}-\fc{\beta}{2}-\gamma(1-\theta_{ab}^j)}\ 
(U^j-\ov U^j)^{-\th_{ab}^j}\, 
\sqrt\fc{\Gamma(\th_{ab}^j)}{\Gamma(1-\th_{ab}^j)}.}
\end{equation}
To fix the constants $\beta,\gamma$,  
one has to calculate a 
four--point disk amplitude involving two twisted matter fields and two K\"ahler moduli $T^i$.
On the other hand, for twisted open string states from the $1/2\ BPS$--sector, the metric takes
a different form:
\begin{equation}\label{halfBPS}{
G_{C^{7_27_3}\ov C^{7_27_3}}={-\kappa_4^{-2}\over(S-\ov S)^{1/2}
(T^1-\ov T^1)^{1/2}}{1\over(U^2-\ov U^2)^{1/2}(U^3-\ov U^3)^{1/2}} .}
\end{equation}

\section{Structure of the soft supersymmetry breaking terms}

The effective low energy supergravity potential in the standard
limit with $M_{\rm Pl}\to \infty$
with $m_{3/2}$ fixed is for $N=1$ supersymmetry 
\cite{SOFTref,BIM}:
\begin{equation}\label{Veff}{
V^{{\rm eff}} = {1\over 2} D^2
+  G^{C_I\ov C_I}\, |\partial_I W^{({\rm eff})}|^2+\ m^2_{I\ov{I},{\rm
soft}}\, C_{I}\ov C_I+{1\over 3}A_{IJK}C_IC_JC_K + {\rm h.c.}\ ,
}
\end{equation}
with
\begin{eqnarray}\label{susyt}
D &=& - g_I\kappa_4^2\ G_{C_I\ov C_I}C_I\ov C_I ,\cr
W^{({\rm eff})} &=&{1\over 3}\ e^{\kappa_4^2\ \hat{K}/2}  Y_{IJK}\,C_IC_JC_K.
\end{eqnarray}
The $C_I$ are taken to run over the $C_i^3,\  C_i^{7a,j},\ C^{37,a},\ C^{7a7b}$,
where $i=1,2,3$, $a$ and $b$ run over the stacks of branes, and $j$ runs
over the  torus not being wrapped.
Furthermore, $g_I$ is the gauge coupling, which is related to the gauge kinetic function
$f_I(M)$ by $g_I^{-2}={\rm Im}(f_I(M))$. The respective gauge kinetic
functions are given in (\ref{gauge}). 

The diagonal structure of our metrics already results in some simplifications,
for example the purely diagonal structure of the scalar mass matrix. The fact
that we have $H_{ij}=0$ results in even more drastic simplifications: In our case, no $B$-term
$B_{IJ}C_IC_J$ 
appears in the effective scalar potential, and also no $\mu$-term $\half\,
\mu_{IJ}C_IC_J$ is generated
in $W^{({\rm eff})}$.

The gravitino mass is given by
\begin{equation}\label{gravi}
m_{3/2}=e^{\kappa_4^2\hat{K}/2}\kappa_4^2\
|\hat{W}|.
\end{equation}
The soft supersymmetry breaking terms are
\begin{eqnarray}\label{susybr}
m^2_{I \ov{I},{\rm soft}}&=& \kappa_4^2\ [\ (|m_{3/2}|^2 +\kappa_4^2\ \hat{V})\,G_{C_I\ov C_I} -
F^\rho \ov F^{\ov \sigma } R_{\rho {\ov \sigma} I \ov{I}}\ ]  ,\cr
A_{IJK} &=& F^\rho D_\rho\ (e^{\kappa_4^2\hat{K}/2} Y_{IJK}),
\end{eqnarray}
where the Greek indices are running over $S,\ T^i,\ U^i$ and
\begin{eqnarray}\label{Ri}
R_{\rho \ov \sigma I \ov{I}} &=& {{\partial^4 K}\over{\partial
C_I\partial\ov C_I\partial M_\rho\partial\ov M_\sigma}}-{{\partial^3 K}\over{\partial
C_I\partial
M_\rho\partial\ov C_K}}G^{\ov C^KC^K}{{\partial^3K}\over{\partial\ov C_I\partial
\ov M_\sigma\partial C_K}}\ ,\cr
D_\rho(e^{\kappa_4^2\ \hat{K}/2} Y_{IJK}) &=&
\partial_\rho(e^{\kappa_4^2\hat{K}/2}  Y_{IJK}) + {1\over 2}
\kappa_4^2\hat{K}_\rho\ (e^{\kappa_4^2\hat{K}/2}
Y_{IJK})\cr
&&-e^{\kappa_4^2\hat{K}/2}G^{\ov C_IC_I}\partial_\rho G_{C_I (\ov C_I}Y_{JK)I}.
\end{eqnarray} 
The gaugino mass is
\begin{equation}\label{gaugino}{m_{gI}=F^\rho\ \partial_\rho\log({\rm Im}f_I),}
\end{equation}
$f_I(M)$ being the gauge kinetic function.

The soft SUSY-breaking terms for a semi-realisitic set-up of intersecting branes/branes with fluxes have been worked out in \cite{Lust:2004dn}. For earlier results for soft terms on $D3$--branes, see \cite{softIL}

We first look at $W^{({\rm eff})}$. We find
\begin{equation}{W^{({\rm eff})}={1\over
3}[(S-\ov{S})\prod_{i=1}^3(T^i-\ov{T}^i)\prod_{i=1}^3(U^i-\ov{U}^i)]^{-1/2}
\ Y_{IJK}\ C_IC_JC_K.}
\end{equation}
From eq. (\ref{super})  we know that $Y_{IJK}=\epsilon_{IJK}$ in the case of the
untwisted matter fields $C_i^3,C_i^{7,a}$ and the combination $\sum_a
C^{7_{a,1}7_{a,2}}C^{7_{a,2}7_{a,3}}C^{7_{a,3}7_{a,1}}$.

Before the expressions for the scalar masses $m_{I\ov{I},{\rm soft}}$ can be calculated, we must
first find the explicit expressions for the curvature tensor. For this, the reader is referred to Appendix A of \cite{Lust:2004dn}. 
The scalar masses are
\begin{eqnarray}
(m^{33}_{i\ov i})^2&=&\kappa_4^2\ \lf[(|m_{3/2}|^2+\kappa_4^2\hat{V})\ G_{C^3_i\ov
C^3_i}-|F^{U^i}|^2R^{3}_{U^i\ov U^ii\ov{i}}-|F^{T^i}|^2R^{3}_{T^i\ov T^ii\ov{i}}\ri] ,\cr
(m^{7,j}_{i\ov i})^2&=&\kappa_4^2\ \lf[(|m_{3/2}|^2+\kappa_4^2\hat{V})\ G_{C^{7,j}_i\ov
C^{7,j}_i}-\sum_{M,N}F^M\ov F^{\ov N}R^{7,j}_{M\ov N i\ov i}\ri] ,\cr
(m^{37_a})^2&=&\kappa_4^2\ \lf[(|m_{3/2}|^2+\kappa_4^2\hat{V})\ G_{C^{37_a}\ov
C^{37_a}}-\sum_{M,N}F^M\ov F^{\ov N}R^{37_a}_{M\ov N}\ri] ,\cr
(m^{7a7b})^2&=&\kappa_4^2\ \lf[(|m_{3/2}|^2+\kappa_4^2\hat{V})\ G_{C^{7a7b}\ov
C^{7a7b}}-\sum_{M,N}F^M\ov F^{\ov N}R^{7a7b}_{M\ov N}\ri],
\end{eqnarray}
where $M,\ N$ run over $S,\ T^i,\ U^i$.
Let elucidate these scalar mass terms by giving the very simple example of a stack of $D7$--branes which does not carry any 2--form flux and only the $(3,0)$-- and the $(0,3)$--flux component turned on. The metric (\ref{metricsfield}) simplifies and this results in the following scalar masses:
\begin{eqnarray}\label{simpleex}
(m^{7}_{1\ov 1})^2&=&{\lambda^2\ \kappa_4^6\over (2\pi)^4\alpha'^2}\ 
\fc{G_{C_1\ov C_1}^{7,2}}{|\prod_M (M-\ov M)|}\ 
\lf|\ \int\ov G_3\wedge \Omega\ \ri|^2 ,\nonumber\\[2pt]
(m^{7}_{2\ov 2})^2&=&{\lambda^2\ \kappa_4^6\over (2\pi)^4\alpha'^2}\ 
\fc{G_{C_2\ov C_2}^{7,2}}{|\prod_M (M-\ov M)|}\ 
\lf|\ \int G_3\wedge \Omega\ \ri|^2 ,\nonumber\\[2pt]
(m^{7}_{3\ov 3})^2&=&{\lambda^2\ \kappa_4^6\over (2\pi)^4\alpha'^2}\ 
\fc{G_{C_3\ov C_3}^{7,2}}{|\prod_M (M-\ov M)|}\ 
\lf|\ \int\ov G_3\wedge \Omega\ \ri|^2.
\end{eqnarray}
Note that to $(m^{7}_{1\ov 1})^2$ and $(m^{7}_{3\ov 3})^2$, the IASD $(3,0)$--component contributes, while  to $(m^{7}_{2\ov 2})^2$ the ISD $(0,3)$--component contributes. For non--trivial 2--form flux on the $D7$--brane world--volume, both ISD and IASD flux components contribute to the scalar mass terms. The scalar mass term for strings on $D3$--branes has the same form as $(m^{7}_{1\ov 1})^2$ in (\ref{simpleex}) and is therefore also non--zero only for IASD flux. 
A contribution from the supersymmetric $(2,1)$--component does not arise in this set--up. When the superpotential is modified such that it also depends on the open string position moduli, a supersymmetric mass terms appears \cite{Lust:2005bd}.

The trilinear coupling is
\begin{eqnarray}
A_{IJK}&=&i\prod_M(M-\ov M)^{-1}{\lambda\kappa_4^2\over
(2\pi)^2\alpha'}\left\{Y_{IJK}\int G_3\wedge \ov\Om+3\ Y_{IJK}\int \ov G_3\wedge \ov
\Om\right.\cr
&&+\left.\sum_i\int \ov G_3\wedge \ov\om_{A_i}\ [Y_{IJK}-(U^i-\ov U^i)\ \p_{U^i}Y_{IJK}]\right\}\cr
&&-i\prod_M(M-\ov M)^{-1/2}\ F^\rho\ G^{\ov C_IC_I}\ \partial_\rho
G_{C_I(\ov C_I}Y_{JK)I} .
\end{eqnarray}
The term $-\sum_i\int \ov G_3\wedge \ov\om_{A_i}(U^i-\ov U^i)\p_{U^i}Y_{IJK}$
appears because general $Y_{IJK}$ may depend on the complex structure moduli.

Note the case where the $I,\  J,\  K$ refer to the 3-brane matter fields $C^3_i$: Then
the last term cancels the terms $3\int \ov G_3\wedge \ov
\Om$ and $\sum_i\int\ov G_3\wedge \ov\om_{A_i}$, and we are left with
$$A_{IJK}=i\, \epsilon_{IJK}\prod_M(M-\ov M)^{-1}{\lambda\kappa_4^2\over
(2\pi)^2\alpha'}\int G_3\wedge \ov\Om,$$
i.e. we only get a trilinear coupling from the $(3,0)$-flux, which agrees with
the results of \cite{CIU,GGJL}. This is not true
for the other matter fields, as their metrics have a more complicated
dependence on the moduli.

The gauge couplings have been given in eqs (\ref{gauge}). Through them, we obtain 
the gaugino masses:
\begin{eqnarray}
m_{g, D7j}&=&F^S\ {{-\alpha'^{-2}f^kf^l}\over{(T^j-\ov{T}^j)-\alpha'^{-2}f^kf^l(S-\ov{S})}}+
F^{T^j}\ {{1}\over{(T^j-\ov{T}^j)-\alpha'^{-2}f^kf^l(S-\ov{S})}},\cr
m_{g, D3}&=&F^S{1\over(S-\ov S)}=-i\prod_M(M-\ov
M)^{-1/2}{\lambda\kappa_4^2\over (2\pi)^2\alpha'}\int G_3\wedge \ov\Om,
\end{eqnarray}
with $k\neq l\neq j,\ $ $j$ being the torus not wrapped by the 7-brane.

To end this section, we give an estimate for the gravitino mass (\ref{gravi}).
The  gravitino mass may be rewritten as
\begin{eqnarray}\label{gravmass}
m_{3/2}&=&\fc{1}{\sqrt 2\ (2\pi)^6}\ \ \fc{M_{\rm string}^8}{M_{\rm Planck}^2}\ 
\fc{1}{\im(S)^{1/2}\ \prod\limits_{j=1}^3\im(T^j)^{1/2}\im(U^j)^{1/2}}\ 
\lf|\int_{X_6} G_3\wedge \Omega\ \ri|\cr
&=&\fc{g_{\rm string}^2}{\sqrt 2\ (2\pi)^4}\ \ \fc{M_{\rm string}^2}{M_{\rm Planck}^2}\   
\fc{\prod\limits_{j=1}^3\im(U^j)^{-1/2}}{{\rm Vol}(X_6)}\ \ 
\lf|\int_{X_6} G_3\wedge \Omega\ \ri| ,
\end{eqnarray}
with the \tb string coupling constant $g_{\rm string}=e^{\phi_{10}}=(2\pi\ \im S)^{-1}$.
The latter is assumed to be small in order to justify a perturbative orientifold construction.
The factor ${\rm Vol}(X_6)=\im(\Tc^1)\im(\Tc^2)\im(\Tc^3)$ is the volume\footnote{The following
relations have been used: 
$\phi_4=\phi_{10}-\h\ln[\im(\Tc^1)\im(\Tc^2) \im(\Tc^3)/\ap^3]$,\ $\kappa_{10}^{-2}=
\fc{2}{(2\pi)^7}\ \ap^{-4}$, and $e^{\kappa_4^2\hat K}=
\fc{(2\pi)^4\ e^{4\phi_4}}{2^7\ \prod\limits_{j=1}^3 \im U^j}$. 
Moreover, we have: $\fc{\im(S)^3}{
\prod\limits_{j=1}^3 \im(T^j)}=\fc{\ap^6}{{\rm Vol}(X_6)^2}$. Consult {\it Ref.} \cite{LMRS} for more 
details.} 
of the six--dimensional compactification manifold $X$, measured in string units $\ap^3$.
The relation between the string scale $\ap=M_{\rm string}^{-2}$ and the four--dimensional 
Planck mass $M_{\rm Planck}$ is given by: 
\begin{equation}\label{Planck}
M_{\rm Planck}=2^{3/2}\ \ g_{\rm string}^{-1}\ M_{\rm string}^4\ \sqrt{{\rm Vol}(X_6)}.
\end{equation}
Qualitatively, the integral $|\int_{X_6} G_3\wedge \Omega\,|$ is of
order $\fc{M_{\rm Planck}}{M_{\rm string}^6}$. Since the moduli fields $S,T^j,U^j$ are 
dimensionsless, we deduce from the first line of (\ref{gravmass}): 
$m_{3/2}\sim\fc{M_{\rm string}^2}{M_{\rm Planck}}$.

In the following, 
let us assume an isotropic compactification of radius $R$, i.e.
${\rm Vol}(X_6)=R^6$ and $U^j=i$.
The flux quantization condition 
$\fc{1}{(2\pi)^2\ap}\ \int_{C_3}G^{(0,3)}_3=\xi_1\in \IZ$ for a $(0,3)$--form flux component
of $G_3$ essentially yields the estimate:
\begin{equation}\label{fluxq}
G_3^{(0,3)} \sim (2\pi)^2\ \fc{\xi_1\ \ap}{R^3}.
\end{equation}
With this information and
$$\lf|\ \int_{X_6}G_3\wedge\Omega\ \ri|=(2\pi)^8\ \xi_1\ \ap\ R^3=2^{-3/2}\ (2\pi)^8\ g_{\rm string}\ 
\fc{M_{\rm Planck}}{M_{\rm string}^6}\ \ \xi_1 ,$$
we obtain for the gravitino mass $m_{3/2}$:
\begin{equation}\label{obtain}{
m_{3/2}=\pi^2\ \ \fc{1}{\im(S)^{1/2}\im(T)^{3/2}}\ \ \fc{M_{\rm string}^2}{M_{\rm Planck}}
\ \ \xi_1.}
\end{equation}
Since the physical moduli fields $T$ are dimensionless, we have:
\begin{equation}\label{WEhave}{
m_{3/2}\ \sim\ g_{\rm string}^{1/2}\ \ \fc{M_{\rm string}^2}{M_{\rm Planck}}\ \ \xi_1 .}
\end{equation}


\chapter{Moduli stabilization in the framework of KKLT}

In the low energy effective theory, the moduli of the compactification manifold used in string theory correspond to massless fields or flat directions in the effective potential. Since these fields would give rise to a fifth force of roughly gravitational strength, they are clearly in conflict with experiment. In a realistic string theory model, they should not be present, therefore one is interested in mechanisms which generate a potential for these massless scalars, which "lifts" them, i.e. gives them a vev and a mass. As explained in Chapter \ref{chap:prelim},  turning on background fluxes presents such a mechanism for the dilaton and the complex structure moduli. Since the flux superpotential (\ref{TVW}) does not depend on the K\"ahler moduli, a different mechanism must be used to fix them.


\section{Introducing the KKLT-scenario}

\begin{figure}[h!]
\begin{center}
\includegraphics[width=110mm]{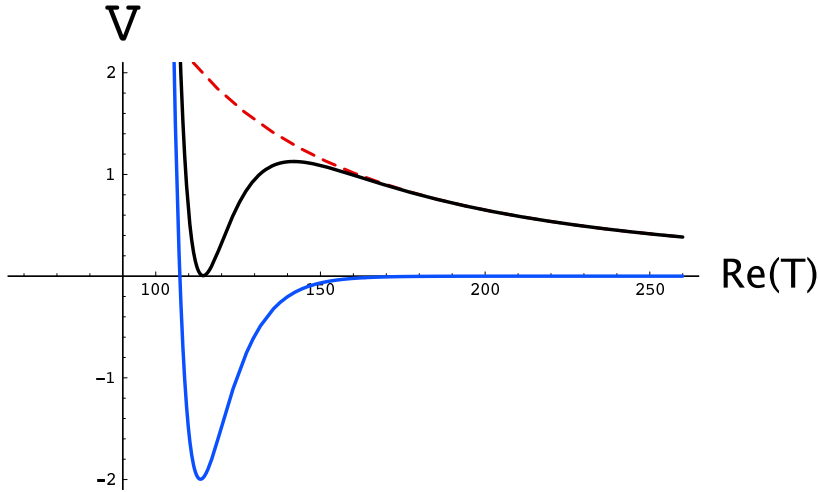}
\caption{Scalar potentials for a KKLT--model with one K\"ahler modulus}
\label{fig:kklt}
\end{center}
\end{figure}

Kachru, Kallosh, Linde and Trivedi \cite{KachruAW} have proposed a way to obtain a string theory vacuum with small positive cosmological constant, which is at least meta-stable.
The proposed mechanism consists of two steps. In the first step, the geometric moduli and the dilaton of a type IIB string compactification are fixed. The complex structure moduli and the dilaton are fixed by turning on a supersymmetric or at least imaginary self--dual background flux. The K\"ahler moduli are fixed via non-perturbative effects. Once all moduli are fixed, one has a stable, supersymmetric anti-deSitter vacuum. In the second step, supersymmetry is broken by adding $\ov{D3}$-branes. This results in a positive contribution to the scalar potential and gives rise to a meta-stable deSitter vacuum, which can be tuned to have the small positive cosmological constant that is called for by experiment. Figure \ref{fig:kklt} shows the scalar potential for the original toy model with only one overall K\"ahler modulus. The blue curve is the stable AdS vacuum
after step one. The red dashed line is the contribution to the potential from the $\ov{D3}$--branes, the black curve shows the resulting meta--stable dS vacuum.

Attempts of providing explicit models in which at least step one of the KKLT proposal is successfully realized include \cite{Racetrack, DenefMM}. A working example is based on the resolution of the $T^6/\IZ_2\times \IZ_2$ orbifold \cite{DenefMM}.


\section{The origin of the non-perturbative superpotential}\label{sec:origin}

There are basically two possible origins for a non--perturbative superpotential which depends on the K\"ahler moduli: Euclidean $D3$--brane instantons wrapping four--cycles on $X$ or gaugino condensation on the world--volume of $D7$-branes which again wrap four-cycles on $X$.
In both cases, the non--perturbative superpotential has the form
\begin{equation}\label{supo}{
W_{\rm np}\sim g_i~e^{-h_iV_i}\ ,}
\end{equation}
where $V_i$ denotes the volume of the wrapped divisor $S_i$.
The constant $h_i$ in the exponent depends on which mechanism, $D3$--brane instantons
or gaugino condensation, is responsible
for the generation of $W_{\rm np}$.
We will first discuss the case of $D3$--brane instantons as the origin of $W_{\rm np}$.
In an ${\cal N}=1$ supersymmetric Yang-Mills gauge theory,
the contribution from a single $D3$--instanton to the superpotential
is given by 
$$W_{\rm np}\sim \Lambda^{3b}.$$
Here, $b$ is the $\beta$-function coefficient of the corresponding gauge group, and $\Lambda$ is the dynamical scale
of the gauge theory:
$$\Lambda^3= e^{-{8\pi^2\over bg^2}}.$$
After relating  the gauge coupling to the volume of $S_i$, 
\begin{equation}\label{gvol}
{4\pi\over g^2}=V_i\ ,
\end{equation}
one obtains
\begin{equation}\label{supoinsa}{
W_{\rm np}\sim \, g_i~e^{-2\pi V_i}.}
\end{equation}
So $h_i=2\pi$.
Instead of wrapping $D3$--branes on $S_i$, we may also consider
a stack of $N$ space-time-filling $D7$--branes
wrapped on $S_i$. 
In general orientifold compactifications, the existence of
the $D7$--branes is in fact forced by the tadpole
cancellation conditions.
Consider the open string spectrum on
the $D7$--branes. 
It is in general given by an effective ${\cal N}=1$ supersymmetric
$U(N)$ gauge theory with some additional matter fields.

First consider pure ${\cal N}=1$ Yang--Mills theory with gauge group
$G$ without any matter fields.
Gaugino condensation generates a nonperturbative superpotential
$$W_{\rm np}\sim \Lambda^{3}=e^{-{8\pi^2\over bg^2}},$$
with eq. (\ref{gvol}) we then get
\begin{equation}\label{supoinsab}
W_{\rm np}\sim \, g_i~e^{-{2\pi V_i\over b}},
\end{equation}
and hence $h_i=2\pi/b$ for pure  SQCD.

Now consider  ${\cal N}=1$ SQCD with gauge group $G=SU(N_C)$ and with
$N_F$ 
matter fields $Q$, $\tilde Q$ in the
fundamental plus anti--fundamental representations
$N_F(\underline N_C\oplus \underline{\bar N_C})$.
For $N_F<N_C$, there is a dynamically generated
superpotential (for a review see e.g. \cite{IntriligatorAU})
\begin{equation}\label{suposeiberg}{
W_{\rm np}=(N_c-N_f)\ \biggl({\Lambda^{3N_C-N_F}\over\det (Q\tilde Q)}\biggr)^{1/(N_c-N_f)}.}
\end{equation}
Here, $b=3\,N_C-N_F$ is the ${\cal N}=1$ $\beta$--function coefficient of
SQCD. 
The vacuum expectation values of the meson superfields $M\sim Q\tilde Q$
break the gauge group $SU(N_C)$ to the non-Abelian subgroup
$SU(N_C-N_F)$.
If $N_F=N_C-1$, the superpotential is generated by gauge instantons.
On the other hand,
 the superpotential arises due to the gaugino 
condensation in the unbroken $SU(N_C-N_F)$ gauge group.
Therefore the gaugino condensate is determined by the scale of the
unbroken gauge group,
$\langle \lambda\lambda\rangle\sim \Lambda^3_{N_C-N_F}$, where the
scale $\Lambda_{N_C-N_F}$ of the low-energy $SU(N_C-N_F)$ gauge theory
can be associated to the scale $\Lambda$ of the high-energy
gauge theory as $\Lambda^{3(N_C-N_F)}_{N_C-N_F}=\Lambda^{3N_C-N_F}/\det M$.
This precisely yields the effective superpotential eq. (\ref{suposeiberg}).
Finally, for $N_F\geq N_C$ there is no dynamically generated superpotential
of this type.


\section{Witten's criterion and the index on $D3$--branes}\label{sec:index}

Which of the divisors present in our compactification manifold give rise to a non-perturbative superpotential? The prefactor $g_i$ to the superpotential (\ref{supo}) generically depends on the complex structure moduli and comes from a fermionic one-loop determinant. Unfortunately it is so far an unsolved question how to compute it for the general case\footnote{In \cite{hep-th/0504058}, $g_i$ was determined for a special case with the help of a chain of string dualities.}.

The best one can do is decide whether $g_i=0$ or not.
In  the framework of $M/F$ theory, Witten has shown \cite{WittenBN} that Euclidean $M5$-brane instantons wrapping a divisor $\tilde S$ in a Calabi--Yau four--fold $X_4$ give rise to a non--perturbative superpotential if the holomorphic Euler characteristic $\chi ({\cal O}_{\tilde S})$ of the divisor fulfills
\begin{equation}\label{chid}{
\chi({\cal O}_{\tilde S})=h^{0,0}(\tilde S)-h^{0,1}(\tilde S)+h^{0,2}(\tilde S)-h^{0,3}(\tilde S)=1.}
\end{equation}
One arrives at this criterion by studying the fermionic zero modes of the Dirac operator on the world--volume of the Euclidean $M5$--brane. The criterion (\ref{chid}) is fulfilled if exactly two fermionic zero modes are present.

Later on it was realized that turning on background flux can have the effect of lifting zero-modes. It can therefore happen that even if there are too many zero-modes in the original geometry to fulfill (\ref{chid}), a non-perturbative superpotential may be generated  \cite{hep-th/0504058, hep-th/0501081, hep-th/0503072, hep-th/0503125, hep-th/0503138, hep-th/0504041}.

Witten's criterion can be used for models in type $IIB$ string theory if their lift to $F$-theory is known. 
Conditions for the generation of the superpotential directly for type $IIB$--orientifolds 
without the detour of analyzing 
the $M/F$--theory case first have been worked out in \cite{hep-th/0507069, hep-th/0507091}, where an index $\chi_{D3}$ for the Dirac operator on the $D3$--brane was proposed.
In terms of this index, the condition for the generation of a non--perturbative superpotential for the wrapped divisor $S$ takes the form
\begin{equation}\label{indexD3}
\chi_{D3}(S)={1\over 2}\left(N_+-N_-\right)=1,
\end{equation}
where $N_\pm$ is the number of fermionic zero modes with $U(1)$ charge $\pm\tfrac{1}{2}$ in the direction normal to $S$. The presence of background fluxes can give rise to zero modes of mixed chirality, in which case the index changes and is not of purely geometric nature anymore. The fermionic zero modes on the world--volume of the $D3$--brane can be related to the Hodge numbers $h^{(0,0)},\,h^{(1,0)},\,h^{(2,0)}$ of $S$ by mapping the spinors to $(0,p)$--forms.
The spinors living on the word--volume of the $D3$-brane can locally be expressed as
\begin{eqnarray}\label{spinors}
\epsilon_+&=&\phi\, |\Omega> +\phi_{\bar a} \gamma^{\bar a} |\Omega>+\phi_{\overline{ab}}\gamma^{\overline{ab}}|\Omega> , \cr
\epsilon_-&=&\phi_{\bar z}\gamma^{\bar z} |\Omega> +\phi_{\overline {az}} \gamma^{\overline {az}} |\Omega>+\phi_{\overline{zab}}\gamma^{\overline{zab}}|\Omega> .
\end{eqnarray}
Here, $\epsilon_+$ ($\epsilon_-$) denotes the spinor with positive (negative) chirality 
with respect to the structure group $SO(2)$ of the normal bundle of the divisor $S$ inside the compact space. $|\Omega>$ denotes the fermionic Clifford--vacuum, while the $\gamma$s are products of $\gamma$--matrices. The $\gamma$--matrices with indices $\ov a, \ov b, \ov z$ etc. act as creation operators on the Clifford vacuum.  $a,b$ label the directions in the $D3$--brane world--volume, $z$ denotes the direction in the Calabi--Yau manifold normal to the wrapped divisor. Note that $\epsilon_+$ and $\epsilon_-$ also carry an $SO(1,3)$ spinor index, so the number of zero modes is doubled. The zero--modes of negative chirality have one leg in the direction normal to the divisor $S$. They can be related to the zero--modes on the world--volume using Serre--duality:
$$g^{z\bar z}\Omega_{\overline{abz}}\phi_{z}=\tilde\phi_{\overline{ab}} ,\quad
g^{z\bar z}g^{a\bar a}\Omega_{\overline{abz}}\phi_{az}=\tilde\phi_{\overline a} , \quad
g^{z\bar z}g^{a\bar a}g^{b\bar b}\Omega_{\overline{abz}}\phi_{abz}=\tilde\phi, $$
where $\Omega$ is again the $(3,0)$--form of the Calabi--Yau.
We see thus that we have $h^{(1,0)}$ zero modes $\phi_{\ov a}$ of positive chirality corresponding to $(0,1)$--forms and another $h^{(1,0)}$ zero modes of negative chirality coming from $\phi_{\ov az}$, also corresponding to $(0,1)$--forms via Serre duality. Analogously, we have $h^{(0,0)}=1$ scalar zero modes of positive chirality ($\phi$) and  negative chirality ($\phi_{\ov{abz}}$), and $h^{(2,0)}$ $(0,2)$--form zero modes of positive ($\phi_{\ov{ab}}$) and negative chirality ($\phi_{\ov z}$).

On the $D3$--brane, the gauge fixing of the $\kappa$--symmetry must be chosen such that it is compatible with the orientifold projection. Some of the zero modes are pure gauge and are annihilated by the $\kappa$--symmetry projector. The orientifold action projects out part of the zero modes. 
We can distinguish three cases regarding the position of the $O7$--planes in relation to the divisor $S$ wrapped by the Euclidean $D3$--brane:
\begin{itemize}
\item[(a).]\label{item:ontop} The $O7$--plane wraps the divisor $S$.
\item[(b).]\label{item:intersects} The $O7$--plane intersects $S$ along one complex dimension.
\item[(c).]\label{item:parallel}The $O7$--plane is parallel to $S$.
\end{itemize}
The analysis of the action of the projectors of $\kappa$--symmetry and orientifold on the zero modes \cite{pheno} is summarized in Table \ref{table:zero}.
\begin{table}[h!]
  \begin{center}
  \begin{tabular}{|c|cc|cc|cc|}
    \hline
   Case &\multicolumn{2}{c|}{(a)} & \multicolumn{2}{c|}{(b)}& \multicolumn{2}{c|}{(c)}\cr
    \hline
    Chirality & + & $-$ &+ & $-$ & + & $-$\cr
    \hline
    $h^{(0,0)}$& $\phi$ &-- & $\phi$ &-- & $\phi$ &$\phi_{\ov{abz}}$\cr
   $h^{(1,0)}$&-- &$\phi_{\ov{az}}$&$[\phi_{\ov a}]$&$\phi_{\ov{az}}$&$[\phi_{\ov a}]$&$\phi_{\ov{az}}$\cr
   $h^{(2,0)}$&$[\phi_{\ov{ab}}]$&-- &-- &$\phi_{\ov z}$ &$[\phi_{\ov{ab}}]$&$\phi_{\ov z}$ \\[2pt]
   \hline
   $\chi_{D3}(S)$&\multicolumn{2}{c|}{$1-h^{(1,0)}_{(-)}+[h^{(2,0)}_{(+)}]$}&\multicolumn{2}{c|}{$1+[h^{(1,0)}_{(+)}]$}&\multicolumn{2}{c|}{$[h^{(1,0)}_{(+)}]-h^{(1,0)}_{(-)}$}\cr
   \ &\ & \ &\multicolumn{2}{c|}{$\ \ -h^{(1,0)}_{(-)}-h^{(2,0)}_{(-)}$} &\multicolumn{2}{c|}{$\ +[h^{(2,0)}_{(+)}]-h^{(2,0)}_{(-)}$}\\[3pt]
   \hline
  \end{tabular}
  \caption{Surviving zero modes after $\kappa$--fixing and orientifold projection}
  \label{table:zero}
  \end{center}
\end{table}
The zero modes given in square brackets are the ones that are in general lifted by background flux. Furthermore, $h^{(i,0)}_{(+)}$ ($h^{(i,0)}_{(-)}$) denotes the positive (negative) chirality zero modes corresponding to this Hodge number. Note that in case (a), where the divisor $S$ feels the full orientifold projection, $\chi_{D3}(S)\equiv \chi({\cal O}_S)$, i.e. the index reduces to the holomorphic Euler characteristic. This happens because of each type of zero modes, only one chirality survives the orientifold projection, which matches the situation on the fourfold. In case (c) on the other hand, $S$ does not feel the effect of the orientifold projection at all, therefore (unless zero modes are lifted by flux) the positive and negative chirality zero modes compensate each other, such that the index equals zero. This is analogous to the case of a Calabi--Yau manifold without orientifold for which no non--perturbative superpotential is generated. 

To decide whether there are enough contributions to the non-perturbative superpotential to stabilize all moduli, it is of prime importance to know the Hodge numbers $h^{(1,0)}$ and $h^{(2,0)}$ for a complete set of divisors in our compactification manifold. The geometric methods described in part I of this thesis yield all necessary information for the case of resolved orbifolds of type $T^6/\IZ_n$ and $T^6/\IZ_n\times \IZ_m$.


\section{Vacuum structure and stability}\label{sec:structure}

In this section,  the vacuum structure of \tb toroidal orientifold compactifications is discussed. 
The discussion is based on the effective ${\cal N}=1$
superpotential
\begin{equation}\label{fullsupo}
 W=W_{\rm flux}(S,U^j)+W_{\rm np}(T^i, U^i) ,
 \end{equation}
with $W_{\rm flux}$ as given in (\ref{TVW}) and
\begin{eqnarray}
W_{\rm np}(T^i, U^i)&=&\sum_{i=1}^{h^{(1,1)}_{+}}  {g^i(U^i)} \, e^{-h^iT^i},
\end{eqnarray}
where the sum only runs over the $h^{(1,1)}_{+}$ geometric moduli.
The first term depends on the dilaton field $S$ and the complex structure moduli $U^j$.
The second term is
of non-perturbative nature as discussed in the last section and depends on the K\"ahler moduli
$T^i$ and the complex structure moduli. 

The vacua of the effective ${\cal N}=1$ supergravity theory are determined
by the associated scalar potential \cite{WessCP}
\begin{equation}\label{scalarpotnp}{
V=e^{\kappa^2_4 K}\Biggl(|D_{S}W|^2+\sum_{i=1}^{h^{(1,1)}_{+}}|D_{T^i}W|^2+
\sum_{j=1}^{h^{(2,1)}_{-}}|D_{U^j}W|^2-3\ |W|^2\Biggr).}
\end{equation}

The extrema of (\ref{scalarpotnp}) are determined by imposing the supersymmetry conditions (\ref{susycond}) on the full superpotential (\ref{fullsupo}). Since the twisted complex structure moduli are not well understood, we will in the following concentrate on cases with $h^{(2,1)}_{tw}=0$.

There is an issue concerning the stability of the obtained 
extrema after imposing the supersymmetry conditions.
The stability of AdS vacua in gravity coupled to scalar fields has been investigated 
in \cite{BF}. Stability is guaranteed, if all scalar masses fulfill the Breitenlohner--Freedman
(BF) bound \cite{BF}, i.e. if their mass eigenvalues do not fall below a certain minimal bound.
The latter is a negative number related to the scalar potential at the minimum.
It can be shown completely model independently that 
all scalars have masses above this bound in any ${\cal N}=1$ supersymmetric AdS vacuum in 
supergravity theories, even if it is a saddle point or a maximum (c.f. \cite{Duff} and Appendix C of  \cite{Lukas}).
For the AdS case, there is no need to worry. But since KKLT propose to lift the vacuum to deSitter space, stability becomes an issue. 
Here, one has to require the absence of any tachyonic
scalar fields, i.e. the (mass)$^2$ of all scalars must be positive.
This means that all eigenvalues of the scalar field mass matrix
${\partial ^2V\over\partial \phi_\alpha\partial \ov\phi_\beta}$ 
($\phi_\alpha,\phi_\beta=S,U^j,T^i$) must be
positive. In this way, some severe constraints can be derived on the possible orbifolds leading to stable vacua, a question which was raised in \cite{ChoiSX}.

We present here a condition, which, if fulfilled, excludes the existence of a (meta--) stable minimum in the sense explained above, i.e. such that we are still at a minimum {\it after} the uplift to de Sitter space.

Case by case studies on toroidal orbifold models \cite{Lust:2005dy} have suggested that the existence of complex structure moduli is a necessary condition for a model to allow stable vacua. 
To verify this conjecture, we start with a setting with only a dilaton and an arbitrary form of the K\"ahler potential for the K\"ahler moduli (i.e. $h^{(2,1)}=0$). In this set--up, the superpotential (\ref{fullsupo}) takes the general form
\begin{equation}\label{SUPP}{
W=B+A\ S+\lambda\ \sum_{i=1}^{h^{(1,1)}_+(X)} g_i\ e^{a_i\ T^i} .}
\end{equation}
Generically, $A,B \in \IC$, $g_i\in \IC$, and $a_i \in \IR_{-}$.
In addition, $\lambda\in \IR$ is a real parameter accounting for a possible so--called
K\"ahler gauge, as discussed in \cite{KachruA}. 
A stable vacuum is excluded when one or more of the eigenvalues of the scalar mass matrix are negative. After a rather tedious series of reformulations  (see \cite{pheno} for the full derivation), it is possible to phrase this condition purely in terms of the K\"ahler moduli of those divisors of the original Calabi--Yau manifold which contribute to the superpotential and the intersection form
\begin{equation}\label{Intersections}
\Kc_{ij}=\int_{X} \omega_i\wedge\omega_j\wedge J=\Kc_{ijk}\ \Tc^k ,\ \ \  
\Kc_{ijk}=\int_{X} \omega_i\wedge\omega_j\wedge\omega_k,
\end{equation}
with the K\"ahler form parametrized as 
$$J=\sum\limits_{j=1}^{h^{(1,1)}_+} \Tc^j\ \omega_j.$$
The condition for the {\it exclusion} of a stable vacuum then reads
\begin{equation}\label{Boxii}
 \exists\ \ {i\in\{1,\ldots,h^{(1,1)}_+\}}\  : \ \Tc^i>0\  \ \ \land\ \ \ \Kc_{ii} \geq 0.
\end{equation}
Expressed in words, if at least one of the geometric K\"ahler moduli is larger than zero and the diagonal element of the intersection form (\ref{Intersections}) corresponding to this modulus is larger than or equal to zero, no stable vacuum exists.
For toroidal orientifolds and their resolutions, the first part of the condition (\ref{Boxii}) is always satisfied by the untwisted K\"ahler moduli and also the second part can be verified. Like this, all toroidal orbifolds and their resolutions which have $h^{(2,1)}=0$ are excluded as candidate models for the KKLT proposal.


\section{Resolved toroidal orbifolds as candidate models for KKLT}

In \cite{DenefMM}, a toroidal orbifold model, namely type $IIB$ string theory compactified on the orientifold of the resolved $T^6/\IZ_2\times\IZ_2$, was checked for its suitability as a compactification manifold for the KKLT proposal. Since the $F$-theory lift of this example is known, Witten's criterion could be checked directly and the results of \cite{DenefMM} strongly indicate that in this model, all geometric moduli can be fixed.

The methods to obtain a smooth Calabi--Yau manifold from a toroidal orbifold and to subsequently pass to the corresponding orientifold as described in the first part of this thesis enable us to explicitly check other toroidal orbifolds for their suitability as candidate models for the KKLT proposal.

The requirement that the scalar mass matrix be positive, discussed in Section \ref{sec:structure}, places severe constraints on the list of possible models. Those orbifolds without complex structure moduli do not give rise to stable vacua after the uplift to dS space. Thus $\IZ_3,\,\IZ_7,\,\IZ_{8-I}$ on $SU(4)^2$, $\IZ_2\times\IZ_{6'}$, $\IZ_3\times\IZ_{3}$, $\IZ_4\times\IZ_{4}$ and $\IZ_6\times\IZ_{6}$ are excluded from the list of possible models given in Tables \ref{table:one} and \ref{table:two}.

Since the stabilization of twisted complex structure moduli via 3--form flux is not well understood yet, the models with $h_{(2,1)}^{twist}\neq 0$ cannot be checked explicitly. Yet considerations regarding the topology of their divisors suggest that they might not be suitable candidate models anyway.

The only models which are not already excluded and are directly amenable to our methods are $T^6/\IZ_4  $  on $SU(4)^2  $, $T^6/ \IZ_{6-II} $  on $ SU(2)\times SU(6)$, the above mentioned $T^6/\IZ_2 \times\IZ_2$, and $T^6/\IZ_2 \times\IZ_4$.

The question one would like to answer is: Do enough of the divisors of the above models contribute to the non-perturbative superpotential that all K\"ahler moduli can be fixed?

To answer this question, the topologies of the divisors must be studied. 
In Section \ref{sec:divtopII} we have seen that there are four basic topologies for the divisors of the resolved toroidal orbifolds: The inherited divisors $R_i$ have the topology of either {\bf (i)} $K3$ or {\bf (ii)} $T^4$. 
The exceptional divisors $E_i$  can be birationally equivalent to either 
{\bf (iii)} a rational surface (i.e. $\IP^2$ or $\IF_n$) or 
{\bf (iv)} $\IP^1\times T^2$. 
The same is true for the $D$--divisors, which are linear combinations of the $R$s and $E$s. The rational surfaces have $h^{(1,0)}=h^{(2,0)}=0$ and therefore $\chi({\cal O}_{S})=1$. Since $h^{(1,0)}$ and $h^{(2,0)}$ are birational invariants, the number of blow--ups which depends on the triangulation of the resolution is irrelevant here. $\IP^1\times T^2$ has $h^{(1,0)}=1,\ h^{(2,0)}=0$,  $T^4$ has  $h^{(1,0)}=2,\ h^{(2,0)}=1$, which both results in $\chi({\cal O}_{S})=0$. $K3$ has $h^{(1,0)}=0,\ h^{(2,0)}=1$ and therefore $\chi({\cal O}_{S})=2$.

Since except for $T^6/\IZ_2 \times\IZ_2$, the $F$-theory lifts of these models are not known, it must be determined directly in type $IIB$ which divisors contribute to the non--perturbative superpotential. Here, we make use of the index for the Dirac operator on the world--volume of the Euclidean $D3$--brane (\ref{indexD3}). In \cite{hep-th/0507069}, it was shown that only the fermionic zero modes associated to $h^{(1,0)}$ and $h^{(2,0)}$ can be lifted by background flux. In the cases with $h^{(1,0)}=h^{(2,0)}=0$, the effect of the fluxes can therefore be disregarded. The values of the index for the four divisor topologies arising from resolutions of toroidal orbifolds are given in Table \ref{table:indexfour}.
\begin{table}[h!]
  \begin{center}
  \begin{tabular}{|c|c|c|c|}
    \hline
   Topology &{(a)} &{(b)}&{(c)}\cr
    \hline
    $K3$ & $2/[1]$ &$0$ &$0/[-1]$ \cr
    $T^4$& $0/[-1]$ &$0/[-2]$ & $0/[-3]$\cr
   $\IP^1\times T^2$&$0$ &$1/[0]$&$0/[-1]$\cr
   $\IP^2,\,\IF_n$&$1$&$1$&$0$\cr
   \hline
    \end{tabular}
  \caption{Index $\chi_{D3}$ for the four basic topologies}
  \label{table:indexfour}
  \end{center}
\end{table}
The numbers in square brackets are the values of the index in the case that the corresponding zero modes have been lifted by flux, cf. Table \ref{table:zero}. We see thus that for the case (c), we never get a contribution, so we better seek an orientifold action which leads to many $O7$--plane solutions. $K3$ can contribute in case (a) if the $h^{(2,0)}_{(+)}$ zero modes are lifted by flux. In our set--up, case (a) cannot arise, since only the inherited divisors $R_i$ can have the topology of $K3$, and these divisors are never wrapped by $O7$--planes. A divisor with the topology of $T^4$ can likewise never contribute. $\IP^2\times T^2$ can contribute in case (b) if {\it no} zero modes are lifted by flux. The rational surfaces contribute in the cases (a) and (b) irrespective of the background flux. To summarize: All those models are likely to allow the stabilization of all geometric moduli for which 
\begin{itemize}
\item[(i).] the fixed points and fixed lines are all in equivalence classes with only one member, giving rise to $E$ and $D$ divisors which are birationally equivalent to rational surfaces and 
\item[(ii).] an orientifold action exists which gives rise to enough $O7$--plane solutions that each divisor intersects an $O7$--plane in at least one complex dimension.
\end{itemize}
When these conditions are met, it is likely that all geometric moduli will be stabilized when the full scalar potential (\ref{scalarpotnp}) is minimized.

Requirements (i) and (ii) are both met by $T^6/\IZ_4$  on $SU(4)^2$, $T^6/ \IZ_{6-II}$ on $SU(2)\times SU(6)$, $T^6/\IZ_2 \times\IZ_2$ and $T^6/\IZ_2 \times\IZ_4$, therefore we expect that all geometric moduli can be stabilized in these cases.
In the case of $T^6/ \IZ_{6-II} $, $h^{(1,1)}_{-}=6$. These six moduli are not geometric anymore after the orientifold--projection and thus cannot be stabilized by Euclidean $D3$--brane instantons. This example will be discussed in more detail in the next subsection.

Models with fixed lines without fixed points on them which lie in orbits of length greater than one do not satisfy criterion (i) since the divisors corresponding to these fixed lines have the basic topology of $\IP^1\times T^2$. These are exactly the models with $h^{(2,1)}_{twist}\neq0$. Unless an elaborate configuration of $O$--planes can be chosen such that all these divisors fall into category (b), these examples in general allow only for a partial stabilization of the geometric moduli via Euclidean $D3$--brane instantons.
It should be stressed that examples like these are still not completely hopeless since additional effects might lead to the complete stabilization of all moduli.
On the other hand, this survey again confirms the old suspicion that manifolds with the right geometrical properties to allow the stabilization of all K\"ahler moduli by Euclidean $D3$--brane instantons or gaugino condensates are not very generic.

\subsection{Example B: $T^6/ \IZ_{6-II}$ on $SU(2)\times SU(6)$}\label{Z6IIcandidate}

In this section, we will explicitly check the suitability of $T^6/ \IZ_{6-II}$ on $SU(2)\times SU(6)$ as a candidate model for the KKLT proposal.
In Table \ref{tab:TopZ6II}, the topologies of the exceptional and $D$--divisors were given. We see that all of them are rational surfaces and will therefore contribute to the non--perturbative superpotential for the cases (a) and (b). Condition (i) is therefore met.

 The triple intersection numbers are given in (\ref{iz6ii}).
 The K\"ahler form can be parametrized as
 \begin{equation}
 J=\sum_{i=1}^3r_i R_i-\sum_{\beta,\,\gamma} t_{1,\beta\gamma}E_{1,\beta\gamma}-\sum_{\beta}\,(t_{2,\beta}E_{2,\beta}+t_{4,\beta}E_{4,\beta})-\sum_\gamma t_{3,\gamma}E_{3,\gamma}.
 \end{equation}
With (\ref{iz6ii}), we arrive at the total volume
 \begin{eqnarray}\label{volsixii}
 V&=&6\,r_1r_2r_3+r_3\sum_{\beta=1}^3 t_{2,\beta}t_{4,\beta}-\sum_{\beta,\gamma}t_{1,\beta\gamma}t_{2,\beta}t_{4,\beta}-r_2\sum_{\gamma=1}^4 t_{3,\gamma}^2-r_3\sum_{\beta=1}^3(\,t_{2,\beta}^2+t_{4,\beta}^2)\cr
 &&-\sum_{\beta,\gamma}t_{1,\beta\gamma}^3+\sum_{\beta=1}^3t_{2,\beta}^2t_{4,\beta}-{4\over 3}\left[\sum_{\beta=1}^3(\,t_{2,\beta}^3+t_{4,\beta}^3)+\sum_{\gamma=1}^4 t_{3,\gamma}^3\right]\notag\\[2pt]
 &&+\sum_{\beta,\gamma}(\,t_{1,\beta\gamma}t_{2,\beta}^2+t_{1,\beta\gamma}t_{3,\gamma}^2+t_{1,\beta\gamma}t_{4,\beta}^2).
 \end{eqnarray}
In the next step, we perform the orientifold projection, which was discussed for this specific model in the example sections of Chapter \ref{sec:orientifold}. 
As discussed there, not all fixed point sets are invariant under the global involution $I_6$, which leads to $h^{(1,1)}_{-}=6$.
Out of the eight possibilities for the orientifold involution on the resolved patches given in (\ref{eq:Z6IIOplanes}), we have chosen the one which leads to $O$--plane solutions satisfying criterion (ii), namely
\begin{equation}\label{invlocal}
(z^1,z^2,z^3,y^1,y^2,y^3,y^4) \to (-z^1,-z^2,-z^3, y^1, -y^2, y^3, -y^4).
\end{equation}
This orientifold action gives rise to an $O7$--plane wrapped on $D_{1}$, one on each of the four $D_{3,\gamma}$ and one wrapped on each of the two invariant $E_{2,\beta}$. On the local patches, no $O3$--plane solutions occur. Examination of Figure \ref{fsixii} shows that all $D$-- and $E$--divisors intersect one of the divisors carrying an $O7$--plane in at least one complex dimension.
In total, there are seven $O7$--planes and 12 $O3$--planes, as discussed in Section \ref{sec:sixiitadpoles}.

On top of the $O7$--planes we place eight $D7$--branes to cancel the $D7$--tadpole locally. This gives rise to a stack of $D7$--branes with gauge group $SO(8)$ on each of the divisors $D_1,\ D_{3,\gamma},\ E_{2,\beta}$.
In Section \ref{sec:sixiitadpoles}, the total $D3$--tadpole was determined to be $Q_{3_{\rm tot}}=-41$.
We choose to saturate the $D3$--tadpole completely with flux since we want to avoid mobile $D3$--branes, i.e. we need $N_{flux}=41$.

The modified intersection numbers according to Section \ref{sec:sixiiOintersections} lead to the modified overall volume
 \begin{eqnarray}\label{volsixiiO}
 V&=&3\,r_1r_2r_3+r_3\sum_{\beta=1}^2 t_{2,\beta}t_{4,\beta}-\sum_{\beta,\gamma}t_{1,\beta\gamma}t_{2,\beta}t_{4,\beta}-\tfrac{1}{2}\,r_2\sum_{\gamma=1}^4 t_{3,\gamma}^2-r_3\sum_{\beta=1}^2(2\,t_{2,\beta}^2+\tfrac{1}{2}\,t_{4,\beta}^2)\cr
 &&-\tfrac{1}{2}\,\sum_{\beta,\gamma}t_{1,\beta\gamma}^3+2\sum_{\beta=1}^2t_{2,\beta}^2t_{4,\beta}-{4\over 3}\left[\sum_{\beta=1}^2(4\,t_{2,\beta}^3+\tfrac{1}{2}\,t_{4,\beta}^3)+\tfrac{1}{2}\,\sum_{\gamma=1}^4 t_{3,\gamma}^3\right]\notag\\[2pt]
 &&+\sum_{\beta,\gamma}(\,2\,t_{1,\beta\gamma}t_{2,\beta}^2+\tfrac{1}{2}\,t_{1,\beta\gamma}t_{3,\gamma}^2+\tfrac{1}{2}\,t_{1,\beta\gamma}t_{4,\beta}^2).
 \end{eqnarray}
The total K\"ahler potential becomes thus
\begin{equation}\label{Ksixii}
\hat K=-\ln (S-\ov S)-\ln(U^3-\ov U^3)-2\,\ln\,V.
\end{equation}
We get contributions for Euclidean $D3$--brane instantons for all exceptional and $D$--divisors. For those divisors on which the $O7$--planes are wrapped, something special happens: Since a stack of $D7$--branes is wrapped on them, they contribute to the non--perturbative superpotential through gaugino condensates, see Section \ref{sec:origin}.
Since these divisors do not intersect each other, no bifundamental matter is present, while adjoint matter is excluded because $h^{(1,0)}=h^{(2,0)}=0$ for these divisors.
The total superpotential becomes thus 
\begin{equation}\label{supersixii}
W_{\rm tot}=W_{flux}+\sum_{S_i} g_{S_i}\,e^{-2\pi\,V(S_i)}+\sum_{S'_i} g_{S'_i}\,e^{-2\pi\,V(S'_i)/6},
\end{equation}
where $S_i\in \{E_{1,\beta\gamma},\, E_{3,\gamma},\, E_{4,\beta},\, D_{2,\beta}\}$ and $S'_i\in \{D_{1},\, D_{3,\gamma},\, E_{2,\beta}\}$. We thus get contributions for all exceptional divisors. Since the $D$s are linear combinations of the $E$s and the $R$s, enough contributions are available to stabilize $h^{(1,1)}_{+}$ independent moduli.
 
With this, we have collected all necessary ingredients to numerically minimize the scalar potential, once a suitable flux is turned on which satisfies the tadpole cancellation condition (\ref{modcfour}).


\chapter{Conclusions and Summary}

Toroidal orbifolds have found wide use in string compactifications because they provide simple, yet non-trivial toy models which reproduce a number of phenomenologically desirable properties, such as ${\cal N}=1$ supersymmetry and chiral fermions (in heterotic string theory), and family repetition.

In the first part of this thesis, the geometry of toroidal orbifolds was presented in detail. At the orbifold point, i.e. in the singular case, a complete parametrization of the untwisted moduli in terms of the radii and angles of the torus lattice was given for all commonly used orbifolds (see Tables \ref{table:one} and \ref{table:two}). 
In Chapter 3, the systematic resolution of the quotient singularities was discussed. The treatment of the singular orbifold has been well--known for some time already, although the results are very scattered in the literature and a systematic and complete treatment as given in Appendix B of this thesis was lacking. The systematic treatment of the resolution of toroidal orbifolds however was undertaken only recently \cite{geometry}. 

First, the basics of toric geometry were introduced and used to resolve orbifold singularities via blow--ups in a local, non--compact geometry. Appendix A gives the resolutions, linear relations, Mori generators, local intersection properties and divisor topologies for all models involved. Then, the procedure of gluing the resolved patches together to arrive at the smooth Calabi--Yau manifold was explained. Here, the global knowledge derived from the covering space $T^6$ was heavily used. The method to calculate the complete intersection ring of the smooth Calabi--Yau was given, as well as the topologies of the divisors in the compact manifold. In Chapter 4, the transition to the orientifold quotient was discussed. 

It should be stressed that the class of smooth manifolds obtained from maximally resolved toroidal orbifolds is one of the few classes of Calabi--Yau manifolds which are well--understood and allow a number explicit calculations. The systematic transition to their orientifold quotients is also a big asset. The importance of resolved toroidal orbifolds for concrete applications in string theory should therefore not be underestimated.
The methods presented in the first part of this thesis provide a powerful toolbox which can be exploited for a wide variety of calculations in string theory. Apart from their practical use, these constructions, simple though they are, or maybe exactly because of their simplicity are very appealing and have a charm of their own.

In the second part of this thesis, the geometrical treatment of part I was motivated by two main examples which make use of this geometrical knowledge. Both examples belong to the subject of string compactifications with background flux, which on the one hand present a mechanism to break supersymmetry and one the other hand stabilize at least part of the moduli. The first main example treated a string compactification on the singular $T^6/\IZ_2\times\IZ_2$ orbifold. The effective potential from the background flux and the supersymmetry breaking soft terms, i.e. scalar mass terms, trilinear couplings etc. were calculated. Although the set--up of intersecting branes/D--branes with fluxes compactified on $T^6/\IZ_2\times\IZ_2$ is but a toy model, calculations of this type can yield estimates which may in principle even serve to falsify specific string models in future experiments. 

The second example attempted moduli stabilization along the lines of the KKLT proposal. Here, extensive use was made of the methods to resolve a singular orbifold and make the transition to the orientifold quotient. By a general consideration concerning the stability of the vacuum solutions after the uplift, several orbifold models, namely the ones without complex structure moduli were excluded as candidate models for the KKLT proposal. Most of the remaining ones seem to allow only a partial stabilization of the geometric moduli without recourse to additional mechanisms. One of the conclusions of this second example is therefore that at least in the restricted class of orientifolds originating from resolved toroidal orbifolds, KKLT candidate models are not as generic as one would have hoped.

For the future, there is a variety of paths that could be pursued starting from the present state of knowledge in the subject of resolved toroidal orbifolds. One possibility would be to attempt the construction of the corresponding mirror manifolds. 

A very interesting task would be to determine the variation of the Hodge structure and the period integrals of the resolved toroidal orbifolds. Once this is known, the genus zero world--sheet instantons could be calculated for topological string theory on this type of manifolds. 

In a construction analogous to the one used in Section 3 of \cite{DenefMM}, the Calabi--Yau fourfolds corresponding to the resolved threefolds could be constructed, yielding the $F$--theory lifts for the type $IIB$ models.

On the more phenomenological side, a task for the future would be to investigate the stabilization of the twisted complex structure moduli through fluxes. This is at present not well understood because the contribution of the twisted $(2,1)$--forms to the K\"ahler potential of the complex structure moduli is not known.

Moduli stabilization could also be attempted for those models which in a pure KKLT setting only allow for a partial stabilization of all moduli, e.g. by considering $D$--term effects.

In the existing literature, toroidal orbifolds have already lent themselves to many more applications than we could present here. Now, that the methods to produce smooth Calabi--Yau manifolds and their orientifold quotients from their singular orbifold limits are fully understood, we expect them to find many more applications in string theory models.


\part{Appendix}

\appendix


\chapter{Resolutions of local orientifold singularities}\label{appA}

In this appendix,  the resolutions of the ${\IC}^3/\IZ_N$, ${\IC}^3/\IZ_N\times\IZ_M$ and ${\IC}^2/\IZ_N$--orbifolds not yet treated in the main text are discussed.

\section{Resolution of ${\IC}^3/\IZ_{3}$}\label{app:rzthree}
\label{sec:Z3}

$\IZ_3$ acts as follows on ${\IC}^3$:
\begin{equation*}{\theta:\ (z^1,\, z^2,\, z^3) \to (\varepsilon\, z^1, \varepsilon\, z^2, \varepsilon\, z^3),\quad \varepsilon=e^{2 \pi i/3}.
}\end{equation*}
To find the components of the $v_i$, we have to solve $(v_1)_i+(v_2)_i+(v_3)_i=0\ \mod\,3$. This leads to the following three generators of the fan (or some other linear combination thereof):
\begin{equation*}{v_1=(-1,-1,1),\ v_2=(1,0,1),\ v_3=(0,1,1).
}\end{equation*}
To resolve the singularity, we find that only $\theta$ fulfills (\ref{eq:criterion}). This leads to one  new generator:
\begin{equation*}{
w={1\over 3}\,v_1+{1\over 3}\,v_2+{1\over 3}\,v_3=(0,0,1).}
\end{equation*}
In this case, the triangulation is unique. 
\begin{figure}[h!]
\begin{center}
\includegraphics[width=120mm]{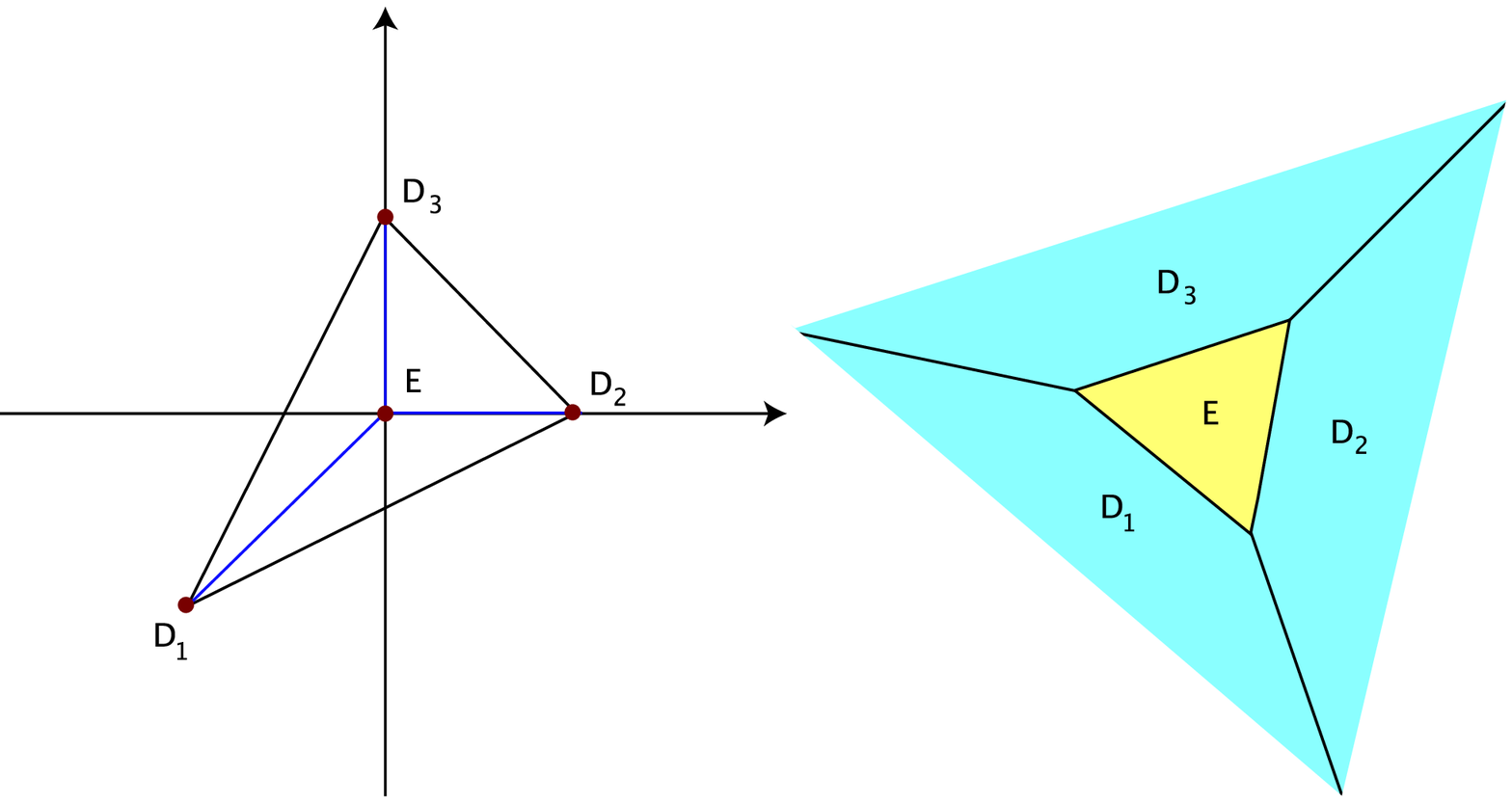}
\caption{Toric diagram of the resolution of ${\IC}^3/\IZ_{3}$ and dual graph}
\label{fig:frthree}
\end{center}
\end{figure}
Figure \ref{fig:frthree} shows the corresponding toric diagram and its dual graph.
We have now three three-dimensional cones: $(D_1,\,E,\,D_2)$, $(D_1,\,E,\,D_3)$ and $(D_2,\,E,\,D_3).$
Let us identify the blown--up geometry. The $\tilde U_i$ are
\begin{equation}
{\tilde U_1={z^2\over z^1},\quad \tilde U_2={z^3\over z^1},\quad \tilde U_3=z^1z^2z^3y.
}
\end{equation}
The rescaling that leaves the $\tilde U_i$ invariant is 
\begin{equation}\label{rescalesthree}{(z^1,z^2,z^3,y) \to (\lambda\,z^1,\,\lambda\,z^2,\,\lambda\,z^3,\,\lambda^{-3}\,y).
}
\end{equation}
Thus the blown--up geometry corresponds to
\begin{equation*}
  X_{\tilde\Sigma}=(\IC^4\setminus F_{\tilde\Sigma})/\IC^*.
\end{equation*}
The excluded set is $F_{\tilde\Sigma}=\{\,(z^1,z^2,z^3)=0\}$, the action of ${\IC}^{*}$ is given by (\ref{rescalesthree}).
It turns out that $X_{\tilde\Sigma}$ corresponds to the line bundle ${\cal O}(-3)$ over ${\IP}^2$. The exceptional divisor $E$ is identified with the zero section of this bundle.
(\ref{rescalesthree}) corresponds to the linear relation between our divisors
\begin{equation*}\label{linrelthree}{
D_1+D_2+D_3-3\,E=0.
}\end{equation*}
With this, we are ready to write down $(P\,|\,Q)$:
\begin{equation}{(P\,|\,Q)=\left( \begin{array}{cccccc}
D_1&\!\!-1&\!\!-1&1&|&1\cr
D_2&1&0&1&|&1\cr
D_3&0&1&1&|&1\cr
E&0&0&1&|&\!\!-3\end{array}\right).
}
\end{equation}
This immediately yields the following linear equivalences:
\begin{equation}\label{lineqthree}{
0 \sim 3\,D_i + E,\qquad i=1,\dots,3.
}
\end{equation}
The curve $C$ corresponding to the single column of $Q$ generates the Mori cone. We find that $C=\,D_1\cdot E=D_2\cdot E=D_3\cdot E$. Furthermore, $E^3=9$.

We will now discuss the topology of $E$. The star of $E$ is the whole toric diagram. Its Mori generator is exactly that of ${\IP}^2$, so it has the topology of $E$.

\section{Resolution of ${\IC}^3/\IZ_{4}$}\label{app:rzfour}

$\IZ_{4}$ acts as follows on ${\IC}^3$:
\begin{equation*}
{\theta:\ (z^1,\, z^2,\, z^3) \to (\varepsilon\, z^1, \varepsilon\, z^2, \varepsilon^2\, z^3),\quad \varepsilon=e^{2 \pi i/4}.
}
\end{equation*}
To find the components of the $v_i$, we have to solve $(v_1)_i+(v_2)_i+2\,(v_3)_i=0\ \mod\,4$. This leads to the following three generators of the fan:
\begin{equation*}{v_1=(2,0,1),\ v_2=(0,2,1),\ v_3=(-1,1,1).
}
\end{equation*}
To resolve the singularity, we find that $\theta$ and $\theta^2$ fulfill (\ref{eq:criterion}). This leads to two new generators:
\begin{eqnarray*}
w_1&=&{1\over 4}\,v_1+{1\over 4}\,v_2+{1\over 2}\,v_3=(0,1,1),\\ [2pt]
w_2&=&{1\over 2}\,v_1+{1\over 2}\,v_2=(1,1,1).
\end{eqnarray*}
In this case, there is again but one triangulation. 
\begin{figure}[h!]
\begin{center}
\includegraphics[width=120mm]{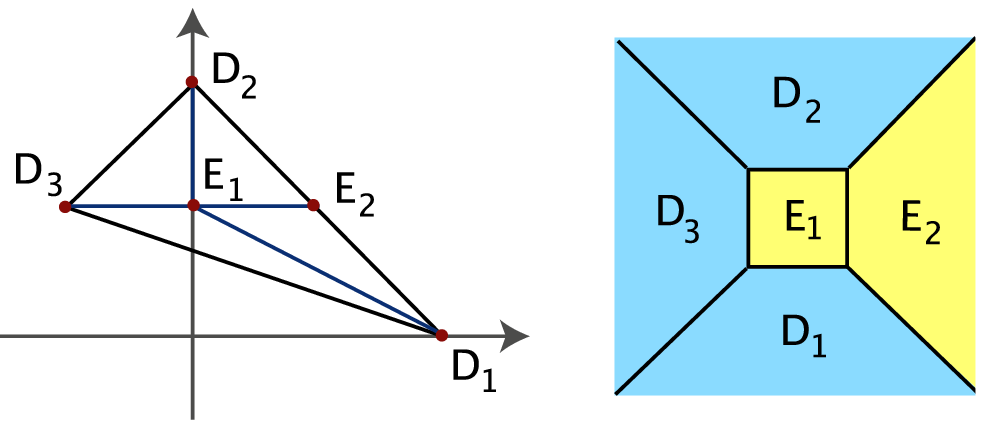}
\caption{Toric diagram of the resolution of ${\IC}^3/\IZ_{4}$ and dual graph}\label{ffour}
\end{center}
\end{figure}
Figure \ref{ffour} shows the toric diagram and its dual graph.

The $\tilde U_i$ of the resolved geometries are
\begin{equation}{
\tilde U_1={(z^1)^2(z^3)^{-1}y^2},\quad \tilde U_2={(z^2)^2z^3y^1y^2},\quad \tilde U_3=z^1z^2z^2y^1y^2.
}\end{equation}
The rescalings that leave the $\tilde U_i$ invariant are
\begin{equation}\label{rescalesfour}{(z^1,\,z^2,\,z^3,\,y^1,\,y^2) \to (\lambda_1\,z^1,\,\lambda_1\,z^2,\,\lambda_1^2\lambda_2\,z^3,\, {1\over\lambda_1^4\lambda_2^2}\,y^1,\,\lambda_2\,y^2).
}
\end{equation}
According to (\ref{eq:blowup}), the new blown-up geometry is
\begin{equation*}{
X_{\tilde\Sigma}=\,({\IC}^{5}\setminus F_{\tilde\Sigma})/({\IC}^*)^2,
}
\end{equation*}
where the action of $({\IC}^*)^2$ is given by (\ref{rescalesfour}).

We have the following four three-dimensional cones: $(D_1,\,D_3,\,E_1),\,(D_1,\,E_1,\,E_2)$, $(D_2,\,E_1,\,E_2),\,(D_2,\,D_3,\,E_1)$.
We identify the two generators of the Mori cone and write them for the columns of $Q$:
\begin{equation}{(P\,|\,Q)=\left(\begin{array}{cccccc}
2&0&1&|&0&1\cr
0&2&1&|&0&1\cr
\!\!\!-1&1&1&|&1&0\cr
0&1&1&|&\!\!\!-2&0 \cr
1&1&1&|&1&\!\!\!-2\end{array}\right).
}\end{equation}
{From} $Q$, we can determine the linear equivalences:
\begin{eqnarray}\label{lineqfour}
0&\sim& 4\,D_{{1}}+E_{{1}}+2\,E_{{2}},\cr
0&\sim& 4\,D_{{2}}+E_{{1}}+2\,E_{{2}},\cr
0&\sim& 2\,D_{{3}}+E_{{1}}.
\end{eqnarray}
There are four compact curves in our geometry, which are related to the $C_i$ as follows: $C_1=D_1\cdot E_1=D_2\cdot E_1,\  C_2=E_1\cdot E_2,\  E_1\cdot D_3=2\,C_1+C_2$. Furthermore, $E_1^3=8$.

From the Mori generators of the star of $E_1$, we find $E_1$ to be an ${\IF}_2$. $E_2$ corresponds to $\IP^1\times \IC$.

\section{Resolution of ${\IC}^3/\IZ_{6-I}$}

See Sections \ref{sec:exzsixi}, \ref{sec:rzsixi}, \ref{sec:exsixiaa} and  \ref{sec:exsixiaaa}.

\section{Resolution of ${\IC}^3/\IZ_{6-II}$}\label{app:rzsixii}

This example allows several resolutions to illustrate the differences in the intersection numbers. 
$\IZ_{6-II}$ acts as follows on ${\IC}^3$:
\begin{equation*}
{\theta:\ (z^1,\, z^2,\, z^3) \to (\varepsilon\, z^1, \varepsilon^2\, z^2, \varepsilon^3\, z^3),\quad \varepsilon=e^{2 \pi i/6}.
}\end{equation*}
To find the components of the $v_i$, we have to solve $(v_1)_i+2\,(v_2)_i+3\,(v_3)_i=0\ \mod\,6$. This leads to the following three generators of the fan (or some other linear combination thereof):
\begin{equation*}{v_1=(-2,-1,1),\ v_2=(1,-1,1),\ v_3=(0,1,1).
}\end{equation*}

To resolve the singularity, we find that $\theta,\,\theta^2,\,\theta^3$ and $\theta^4$ fulfill (\ref{eq:criterion}). This leads to four  new generators:
\begin{eqnarray*}
w_1&=&{1\over 6}\,v_1+{2\over 6}\,v_2+{3\over 6}\,v_3=(0,0,1),\\ [2pt]
w_2&=&{2\over 6}\,v_1+{4\over 6}\,v_2=(0,-1,1),\\ [2pt]
w_3&=&{3\over 6}\,v_1+{3\over 6}\,v_3=(-1,0,1),\\ [2pt]
w_4&=&{4\over 6}\,v_1+{2\over 6}\,v_2=(-1,-1,1)
.\end{eqnarray*}
In this case, there are five  triangulations. 
\begin{figure}[h!]
\begin{center}
\includegraphics[width=140mm]{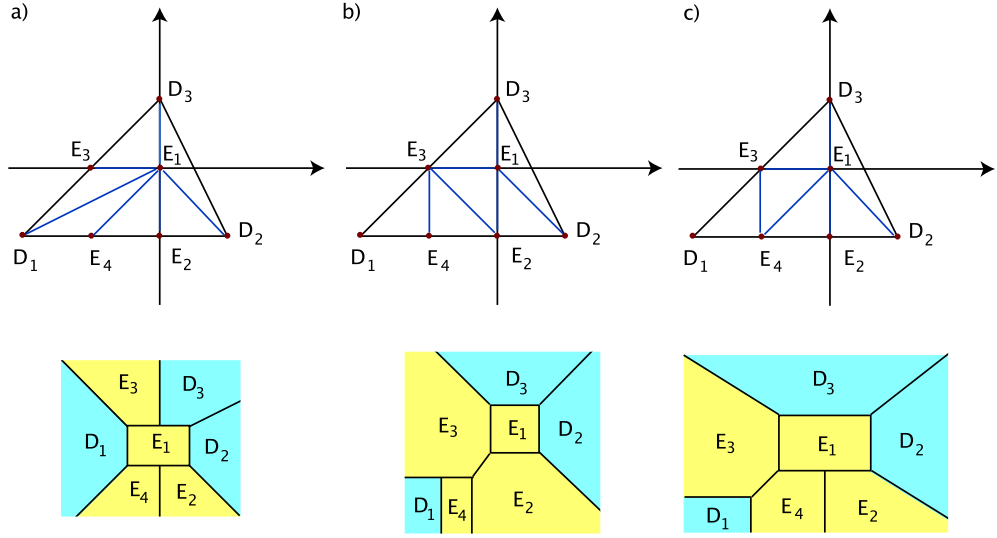}
\caption{Toric diagrams of the resolutions of ${\IC}^3/\IZ_{6-II}$ and dual graphs}\label{fsixii}
\end{center}
\end{figure}
Figure \ref{fsixii} shows three of the corresponding toric diagrams and their dual graphs.
All five diagrams are given in Figure \ref{fig:sixiifive}.

The $\tilde U_i$ of the resolved geometries are
\begin{eqnarray}
\tilde U_1&=&{z^2\over (z^1)^2y^3y^4},\quad \tilde U_2={z^3\over z^1z^2y^2y^4},\quad \tilde U_3=z^1z^2z^2y^1y^2y^3y^4.
\end{eqnarray}
The rescalings that leave the $\tilde U_i$ invariant are
\begin{equation}\label{rescalessixii}{(z^1,\,z^2,\,z^3,\,y^1,\,y^2,\,y^3,\,y^4) \to (\lambda_1\,z^1,\,\lambda_2\,z^2,\,\lambda_1\lambda_2\lambda_3\lambda_4\,z^3,\, {1\over\lambda_2^3\lambda_3^2\lambda_4}\,y^1,\,\lambda_3\,y^2,\,{\lambda_2\over \lambda_1^2\lambda_4}\,y^3, \lambda_4\,y^4).
}\end{equation}
According to (\ref{eq:blowup}), the new blown-up geometry is
\begin{equation*}{
X_{\tilde\Sigma}=\,({\IC}^{7}\setminus F_{\tilde\Sigma})/({\IC}^*)^4,
}\end{equation*}
where the action of $({\IC}^*)^4$ is given by (\ref{rescalessixii}). The five different resolutions of ${\IC}^3/\IZ_{6-II}$ only differ from each other by the excluded set. We must identify it for each case separately. So is for example $(z^1, y^1)=0$ in the excluded set for the cases b) and c), but not for a). We will not write down the three excluded sets explicitly. In what follows, we will only treat the cases depicted in Figure \ref{fsixii}.

\vskip0.5cm
\noindent{\it Case a)}
\vskip0.5cm

In this case, we have the following six three-dimensional cones: $(D_1,\,E_4,\,E_1)$, $(D_1,\,E_1,\,E_3)$, $(D_2,\,E_2,\,E_1),\,(D_2,\,E_1,\,D_3)$, $(D_3,\,E_1,\,E_3),\,(E_1,\,E_2,\,E_4)$.
We identify the four generators of the Mori cone and write them for the columns of $Q$:
\begin{equation}{(P\,|\,Q)=\left(\begin{array}{cccccccc}
\!\!-2&\!\!-1&1&|&1&\!\!\!-1&1&0\cr
1&\!\!-1&1&|&0&0&0&1\cr
0&1&1&|&1&0&0&0\cr
0&0&1&|&0&\!\!\!-1&0&0 \cr
0&\!\!-1&1&|&0&0&1&\!\!\!-2\cr
\!\!-1&0&1&|&\!\!\!-2&1&0&0\cr
\!\!-1&-1&1&|&0&1&\!\!\!-2&1\end{array}\right).
}\end{equation}
{From} $Q$, we can determine the linear equivalences:
\begin{eqnarray}\label{lineqsixii}
0&\sim& 6\,D_{{1}}+E_{{1}}+2\,E_{{2}}+3\,E_{{3}}+4\,E_{{4}},\cr
0&\sim& 3\,D_{{2}}+E_{{1}}+2\,E_{{2}}+E_{{4}},\cr
0&\sim& 2\,D_{{3}}+E_{{1}}+E_{{3}}.
\end{eqnarray}
There are six compact curves in our geometry, which are related to the $C_i$ as follows: $C_1=E_1\cdot E_3,\  C_2=E_1\cdot D_1,\  C_3=E_1\cdot E_4,\ C_4=E_1\cdot E_2,\ E_1\cdot D_3=C_1+3\,C_2+2\,C_3+C_4,\ E_1\cdot D_2=C_1+2\,C_2+C_3$. Furthermore, $E_1^3=6$.

\vskip0.5cm
\noindent{\it Case b)}
\vskip0.5cm

In this case, we have the following six three-dimensional cones: $(D_2,\,E_1,\,D_3)$, $(D_3,\,E_1,\,E_3)$, $(D_1,\,E_3,\,E_4),\,(E_4,\,E_1,\,E_3)$, $(E_1,\,E_2,\,E_4),\,(E_1,\,E_2,\,D_2)$. We identify the four generators of the Mori cone and write them for the columns of $Q$:
\begin{equation}{(P\,|\,Q)=\left(\begin{array}{cccccccc}
\!\!-2&\!\!-1&1&|&1&0&0&0\cr
1&\!\!-1&1&|&0&0&0&1\cr
0&1&1&|&0&0&1&0\cr
0&0&1&|&1&\!\!\!-1&\!\!\!-1&0 \cr
0&\!\!-1&1&|&0&1&0&\!\!\!-2\cr
\!\!-1&0&1&|&\!\!\!-1&1&\!\!\!-1&0\cr
\!\!-1&-1&1&|&\!\!\!-1&\!\!\!-1&1&1\end{array}\right).
}\end{equation}
The linear equivalences are the same as in case a).
There are again six compact curves in our geometry, which are related to the $C_i$ as follows: $C_1=E_3\cdot E_4,\  C_2=E_1\cdot E_4,\  C_3=E_1\cdot E_3,\ C_4=E_1\cdot E_2,\ E_1\cdot D_2=C_2+C_3, \ E_1\cdot D_3=2\,C_2+C_3+C_4$. Here, $E_1^3=7$.

\vskip0.5cm
\noindent{\it Case c)}
\vskip0.5cm

In this case, we have the following six three-dimensional cones: $(D_2,\,E_1,\,D_3)$, $(D_3,\,E_1,\,E_3)$, $(D_1,\,E_3,\,E_4),\,(E_4,\,E_2,\,E_3)$, $(E_1,\,E_2,\,E_3),\,(E_1,\,E_2,\,D_2)$.
We identify the four generators of the Mori cone and write them for the columns of $Q$:
\begin{equation}{(P\,|\,Q)=\left(\begin{array}{cccccccc}
\!\!-2&\!\!-1&1&|&0&0&0&1\cr
1&\!\!-1&1&|&0&0&1&0\cr
0&1&1&|&1&0&0&0\cr
0&0&1&|&\!\!\!-2&1&\!\!\!-1&0 \cr
0&\!\!-1&1&|&1&\!\!\!-1&\!\!\!-1&1\cr
\!\!-1&0&1&|&0&\!\!\!-1&1&0\cr
\!\!-1&-1&1&|&0&1&0&\!\!\!-2\end{array}\right).
}\end{equation}
The linear equivalences between the divisors remain the same as in case a).
There are again six compact curves in our geometry, which are related to the $C_i$ as follows: $C_1=E_1\cdot D_2=E_1\cdot E_3,\  C_2=E_2\cdot E_3,\  C_3=E_1\cdot E_2,\ C_4=E_3\cdot E_4,\ D_3\cdot E_1=C_1+C_3$. Here, $E_1^3=8$.

The topologies of the exceptional divisors are discussed in Section \ref{sec:divsixii}.


\section{Resolution of ${\IC}^3/\IZ_{7}$}\label{app:rzseven}

$\IZ_{7}$ acts as follows on ${\IC}^3$:
\begin{equation*}
{\theta:\ (z^1,\, z^2,\, z^3) \to (\varepsilon\, z^1, \varepsilon^2\, z^2, \varepsilon^4\, z^3),\quad \varepsilon=e^{2 \pi i/7}.
}\end{equation*}
To find the components of the $v_i$, we have to solve $(v_1)_i+2\,(v_2)_i+4\,(v_3)_i=0\ \mod\,7$. This leads to the following three generators of the fan (or some other linear combination thereof):
\begin{equation*}{v_1=(-2,0,1),\ v_2=(1,-2,1),\ v_3=(0,1,1).
}\end{equation*}
To resolve the singularity, we find that $\theta,\,\theta^2$ and $\theta^4$ fulfill (\ref{eq:criterion}). This leads to three new generators:
\begin{eqnarray*}
w_1&=&{1\over 7}\,v_1+{2\over 7}\,v_2+{4\over 7}\,v_3=(0,0,1),\\ [2pt]
w_2&=&{2\over 7}\,v_1+{4\over 7}\,v_2+{1\over 7}\,v_3=(0,-1,1),\\ [2pt]
w_3&=&{4\over 7}\,v_1+{1\over 7}\,v_2+{2\over 7}\,v_3=(-1,0,1).
\end{eqnarray*}
In this case, the triangulation is unique.
\begin{figure}[h!]
\begin{center}
\includegraphics[width=120mm]{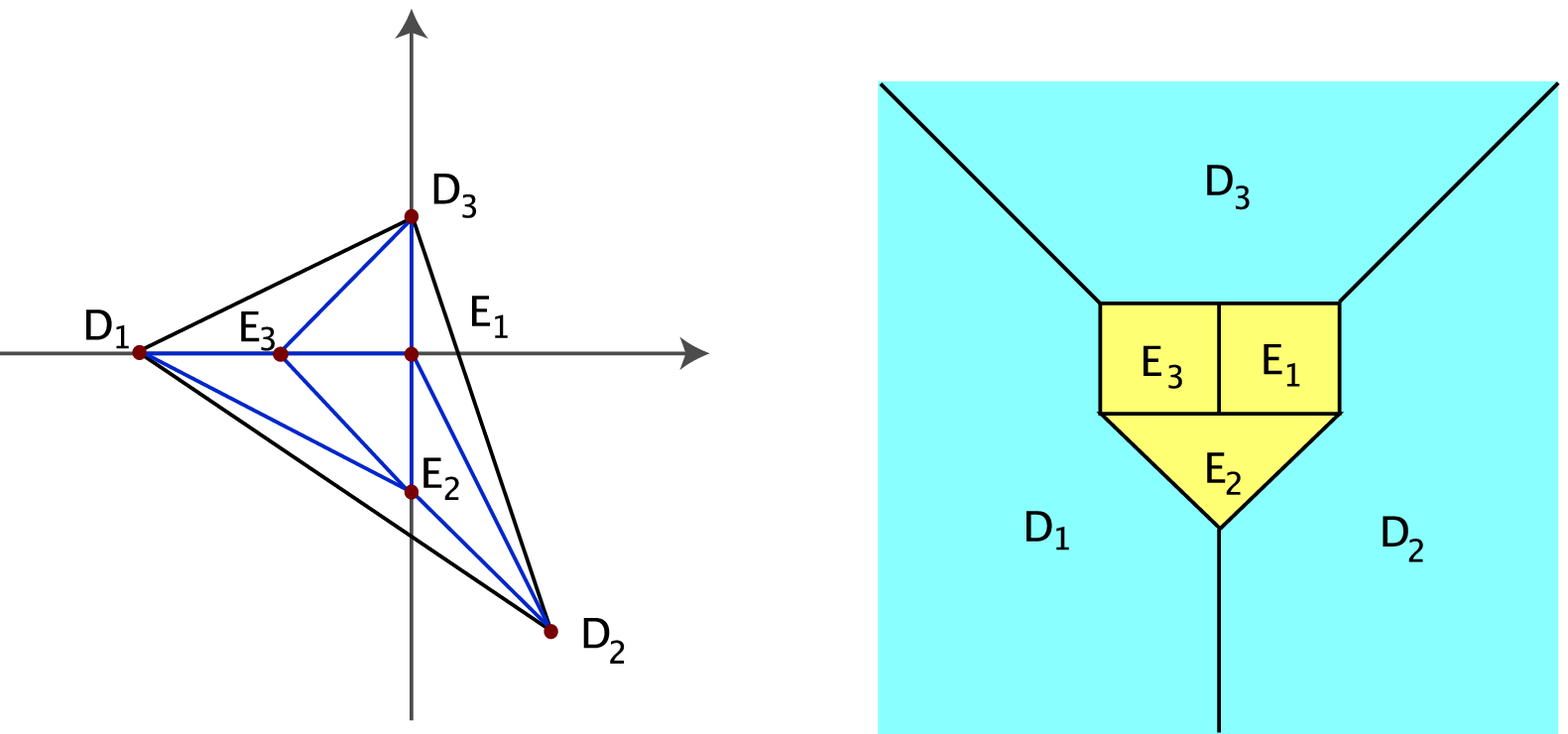}
\caption{Toric diagram of the resolution of ${\IC}^3/\IZ_{7}$ and dual graph}\label{frseven}
\end{center}
\end{figure}
Figure \ref{frseven} shows the corresponding toric diagram and its dual graph.
We have now seven three-dimensional cones: $(D_1,\,E_3,\,E_2)$, $(D_1,\,E_3,\,D_3),\ (D_1,\,E_2,\,D_2)$, $(D_2,\,E_2,\,E_1)$, $(D_2,\,E_1,\,D_3)$, $(E_1,\,E_2,\,E_3)$, and $(E_1,\,E_3,\,D_3)$.
Let us identify the blown--up geometry. The $\tilde U_i$ are
\begin{equation}
{\tilde U_1={z^2\over (z^1)^2y^3},\quad \tilde U_2={z^3\over (z^2)^2y^2},\quad \tilde U_3=z^1z^2z^3y^1y^2y^3.
}\end{equation}
The rescaling that leaves the $\tilde U_i$ invariant is 
\begin{equation}\label{rescalesseven}{(z^1,z^2,z^3,y^1,y^2,y^3) \to (\lambda_1\,z^1,\,\lambda_2\,z^2,\,\lambda_2^2\lambda_3\,z^3,\,{\lambda_1\over\lambda_2^{4}\lambda_3^2}\,y^1, \lambda_3\,y^2,\,{\lambda_2\over \lambda_1^2}\,y^3).
}\end{equation}
Thus the blown-up geometry corresponds to
\begin{equation*}{
X_{\tilde\Sigma}=({\IC}^6\setminus F_{\tilde\Sigma})/({\IC}^*)^3.
}\end{equation*}
We refrain from giving the excluded set of simultaneous zeros of coordinates not belonging to the same cone explicitly. The action of $({\IC}^{*})^3$ is given by (\ref{rescalesseven}).
There are three generators of the Mori cone:
\begin{equation*}{
C_1=(1,0,0,1,0,-2),\quad C_2=(0,1,0,0,-2,1),\quad C_3=(0,0,1,-2,1,0),
}\end{equation*}
With this, we are ready to write down $(P\,|\,Q)$:
\begin{equation}{(P\,|\,Q)=\left(\begin{array}{cccccccc}
D_1&\!\!\!-2&0&1&|&1&0&0\cr
D_2&1&\!\!\!-2&1&|&0&1&0\cr
D_3&1&1&1&|&0&0&1\cr
E_1&0&0&1&|&1&0&\!\!\!-2\cr
E_2&0&\!\!\!-1&1&|&0&\!\!\!-2&1\cr
E_3&\!\!\!-1&0&1&|&\!\!\!-2&1&0
\end{array}\right).
}\end{equation}
This immediately yields the following linear equivalences:
\begin{eqnarray}\label{lineqseven}
0&\sim&7\,D_1+E_1+2\,E_2+4\,E_3,\cr
0&\sim&7\,D_2+2\,E_1+4\,E_2+E_3,\cr 
0&\sim&7\,D_3+4\,E_1+E_2+2\,E_3.
\end{eqnarray}

We have nine internal lines in our toric diagram corresponding to compact curves. We find that $C_1=\,E_2\cdot E_3=D_3\cdot E_3,\quad C_2=E_1\cdot E_2=D_1\cdot E_2,\quad C_3=E_1\cdot D_2=E_1\cdot E_3,\quad D_1\cdot E_3=2\,C_2+C_3,\quad D_2\cdot E_2=C_1+2\,C_2$ and $D_3\cdot E_1=C_2+2\,C_3$. Furthermore, $E_1^3=E_2^3=E_3^3=8$.

By studying the stars of the three compact exceptional divisors $E_1,\,E_2,\,E_3$, we find that all of them correspond to a Hirzebruch surface ${\IF}_2$.


\section{Resolution of ${\IC}^3/\IZ_{8-I}$}\label{app:rzeighti}

$\IZ_{8-I}$ acts as follows on ${\IC}^3$:
\begin{equation*}
{\theta:\ (z^1,\, z^2,\, z^3) \to (\varepsilon\, z^1, \varepsilon^2\, z^2, \varepsilon^5\, z^3),\quad \varepsilon=e^{2 \pi i/8}.
}\end{equation*}
To find the components of the $v_i$, we have to solve $(v_1)_i+2\,(v_2)_i+5\,(v_3)_i=0\ \mod\,8$. This leads to the following three generators of the fan:
\begin{equation*}{v_1=(3,1,1),\ v_2=(0,2,1),\ v_3=(1,-1,1).
}\end{equation*}
To resolve the singularity, we find that $\theta,\,\theta^2,\,\theta^4$ and $\theta^5$ fulfill (\ref{eq:criterion}). This leads to four new generators:
\begin{equation*}{
w_1=(1,0,1),\,w_2=(1,1,1),\,w_3=(2,0,1),\,w_4=(2,1,1).}\end{equation*}
In this case, there are 4 triangulations. 
\begin{figure}[h!]
\begin{center}
\includegraphics[width=120mm]{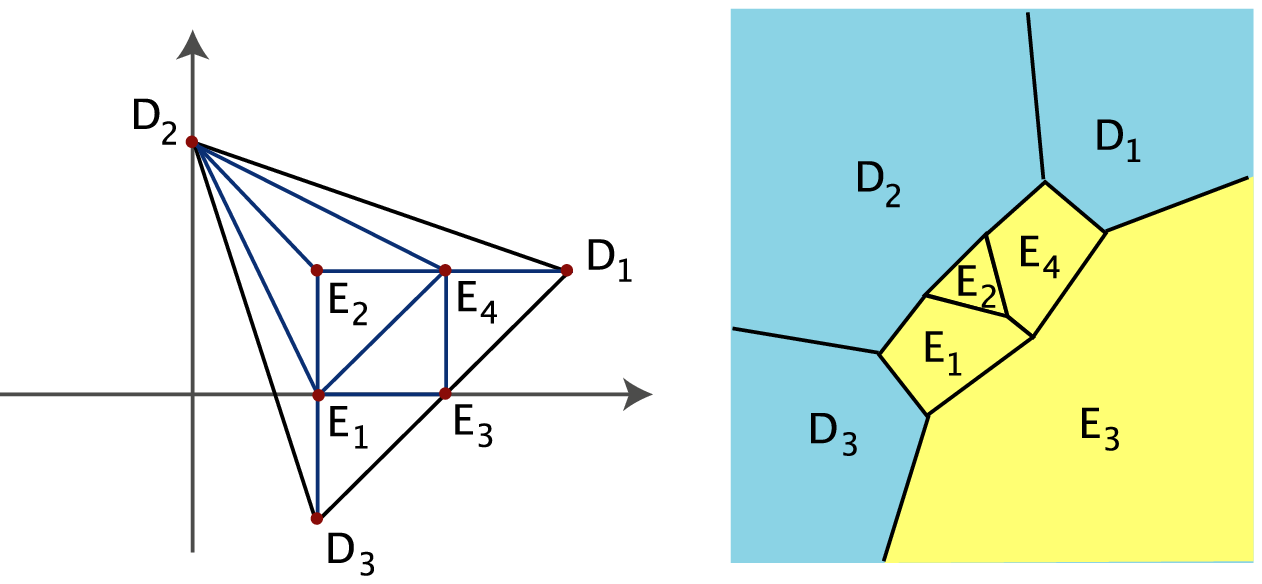}
\caption{Toric diagram of the resolution of ${\IC}^3/\IZ_{8-I}$ and dual graph}\label{freighti}
\end{center}
\end{figure}
Figure \ref{freighti} shows the toric diagram of one of them and its dual graph.
We have now eight three-dimensional cones: $(D_1,\,D_2,\,E_4)$, $(D_1,\,E_3,\,E_4),\ (D_2,\,E_2,\,E_4)$, $(D_2,\,E_1,\,E_2),\ (D_2,\,D_3,\,E_1),\ (D_3,\,E_1,\,E_3)$, $(E_1,\,E_3,\,E_4)$, $(E_1,\,E_2,\,E_4)$.
Let us identify the blown--up geometry. The $\tilde U_i$ are
\begin{eqnarray}
\tilde U_1&=&{(z^1)^3z^3y^1y^2(y^3)^2(y^4)^2},\cr
\tilde U_2&=&{z^1(z^2)^2(z^3)^{-1}y^2y^4},\cr
\tilde U_3&=&z^1z^2z^3y^1y^2y^3y^4.
\end{eqnarray}
The rescaling that leaves the $\tilde U_i$ invariant is 
\begin{eqnarray}\label{rescaleseighti}&&(z^1,z^2,z^3,y^1,y^2,y^3,y^4) \to \cr
&&(\lambda_1\,z^1,\,\lambda_1^3\lambda_3\lambda_4\,z^2,\,\lambda_1^5\lambda_2\lambda_3^2\lambda_4^3\,z^3,\,{1\over\lambda_1^{8}\lambda_2^2\lambda_3^4\lambda_4^5}\,y^1, \lambda_2\,y^2,\,{\lambda_3}\,y^3,\lambda_4y^4).
\end{eqnarray}
Thus the blown-up geometry corresponds to
\begin{equation*}{
X_{\tilde\Sigma}=({\IC}^{7}\setminus F_{\tilde\Sigma})/({\IC}^*)^4.
}\end{equation*}
We refrain from giving the excluded set of simultaneous zeros of coordinates not belonging to the same cone explicitly. The action of $({\IC}^{*})^4$ is given by (\ref{rescaleseighti}).
Using the method discussed in Chapter 3, we find four  generators of the Mori cone, which form the columns of $Q$:
\begin{equation}{(P\,|\,Q)=\left(\begin{array}{ccccccccc}
D_1&3&2&1&        |&0      &1        &0      &0 \cr
D_2&0&2&1&|&1      &0        &0     &0 \cr
D_3&1&\!\!\!-1&1&|&0       &0         &0     &1 \cr
E_1&1&0&1&        |&1       &1        &\!\!\!-1      &\!\!\!-1  \cr
E_2&1&1&1&        |&\!\!\!-3       &0        &1      &0 \cr
E_3&2&0&1&        |&0       &\!\!\!-1        &1       &\!\!\!-1 \cr
E_4&2&1&1&        |&1       &\!\!\!-1        &\!\!\!-1       &1\end{array}\right).
}\end{equation}
This immediately yields the following linear equivalences:
\begin{eqnarray}\label{lineqeighti}
0&\sim&8\,D_{{1}}+E_{{1}}+2\,E_{{2}}+4\,E_{{3}}+5\,E_{{4}},\cr
0&\sim&4\,D_{{2}}+E_{{1}}+2\,E_{{2}}+E_{{4}},\cr
0&\sim&8\,D_{{3}}+5\,E_{{1}}+2\,E_{{2}}+4\,E_{{3}}+E_{{4}}.
\end{eqnarray}

We have ten internal lines in our toric diagram corresponding to compact curves. We find that $C_1=\,D_2\cdot E_2=E_2\cdot E_4=E_1\cdot E_2,\quad C_2=E_3\cdot E_4,\quad C_3=E_1\cdot E_4,\quad C_4=E_1\cdot E_3,\quad D_2\cdot E_1=C_3+C_4,\quad D_3\cdot E_1=C_1+3\,C_3+2\,C_4$. Here, $E_1^3=7,\ E_2^3=9,\ E_4^3=7$.

We will now briefly discuss the topology of the exceptional divisors. $E_2$ corresponds to a $\IP^2$. Do identufy the other, we flop the curve $(E_1
\cdot E_4)$ to $(E_2\cdot E_3)$. Now $E_1$ and $E_4$ correspond to $\IF_2$, while $E_3$ is $\IP^1\times \IC$ blown up in two points.


\section{Resolution of ${\IC}^3/\IZ_{8-II}$}\label{app:rzeightii}

$\IZ_{8-II}$ acts as follows on ${\IC}^3$:
\begin{equation*}
{\theta:\ (z^1,\, z^2,\, z^3) \to (\varepsilon\, z^1, \varepsilon^3\, z^2, \varepsilon^4\, z^3),\quad \varepsilon=e^{2 \pi i/8}.
}\end{equation*}
To find the components of the $v_i$, we have to solve $(v_1)_i+3\,(v_2)_i+4\,(v_3)_i=0\ \mod\,8$. This leads to the following three generators of the fan:
\begin{equation*}{v_1=(1,-2,1),\ v_2=(1,2,1),\ v_3=(-1,1,1).
}\end{equation*}
To resolve the singularity, we find that $\theta,\,\theta^2,\,\theta^3,\,\theta^4$ and $\theta^6$ fulfill (\ref{eq:criterion}). This leads to five new generators:
\begin{equation*}{
w_1=(0,1,1),\,w_2=(1,1,1),\,w_3=(0,0,1),\,w_4=(1,0,1),\,w_5=(1,-1,1).}
\end{equation*}
In this case, there are 6  triangulations. 
\begin{figure}[h!]
\begin{center}
\includegraphics[width=120mm]{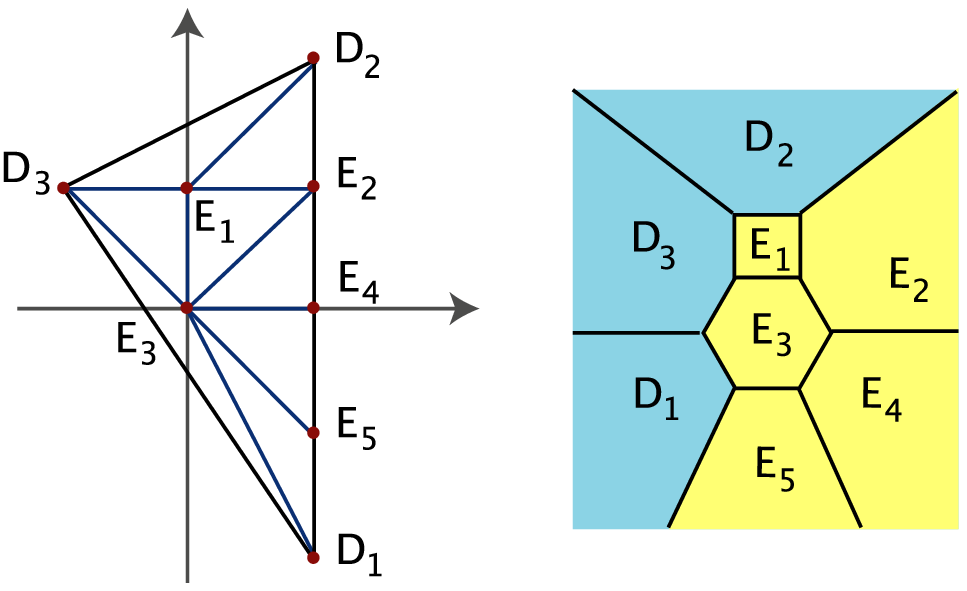}
\caption{Toric diagram of the resolution of ${\IC}^3/\IZ_{8-II}$ and dual graph}\label{freightii}
\end{center}
\end{figure}
Figure \ref{freightii} shows the toric diagram of one of them and its dual graph.
We have now eight three-dimensional cones: $(D_1,\,D_3,\,E_3)$, $(D_1,\,E_3,\,E_5),\ (D_2,\,D_3,\,E_1)$, $(D_2,\,E_1,\,E_2),\ (D_3,\,E_1,\,E_3),\ (E_1,\,E_2,\,E_3)$, $(E_2,\,E_3,\,E_4)$, $(E_3,\,E_4,\,E_5)$.
Let us identify the blown--up geometry. The $\tilde U_i$ are
\begin{eqnarray}
\tilde U_1&=&{z^1z^2(z^3)^{-1}zy^2y^4y^5},\cr
\tilde U_2&=&{(z^1)^{-1}(z^2)^2z^3y^1y^2(y^5)^{-1}},\cr
\tilde U_3&=&z^1z^2z^3y^1y^2y^3y^4y^5.
\end{eqnarray}
The rescaling that leaves the $\tilde U_i$ invariant is 
\begin{eqnarray}\label{rescaleseightii}&&(z^1,z^2,z^3,y^1,y^2,y^3,y^4) \to \cr
&&(\lambda_1\,z^1,\,\lambda_2\,z^2,\,\lambda_3\,z^3,\,{\lambda_1^2\lambda_5\over\lambda_2^{2}\lambda_3\lambda_4}\,y^1, \lambda_4\,y^2,\,{\lambda_2^2\lambda_4\over \lambda_1^2\lambda_3\lambda_5}\,y^3,{\lambda_3\over\lambda_1\lambda_2\lambda_4\lambda_5}\,y^4,\lambda_5\,y^5).
\end{eqnarray}
Thus the blown-up geometry corresponds to
\begin{equation*}{
X_{\tilde\Sigma}=({\IC}^{8}\setminus F_{\tilde\Sigma})/({\IC}^*)^5.
}\end{equation*}
We refrain from giving the excluded set of simultaneous zeros of coordinates not belonging to the same cone explicitly. The action of $({\IC}^{*})^5$ is given by (\ref{rescaleseightii}).
Using the method discussed in Chapter 3, we find five  generators of the Mori cone, which form the columns of $Q$:
\begin{equation}{(P\,|\,Q)=\left(\begin{array}{cccccccccc}
D_1&1&\!\!\!-2&1&        |&1     &0        &0      &0 &0 \cr
D_2&1&2&1&|&0      &0        &0     &0  &1\cr
D_3&\!\!\!-1&1&1&|&0       &0         &0     &1  &0\cr
E_1&0&1&1&        |&0       &0        &1      &\!\!\!-2 &\!\!\!-1  \cr
E_2&1&1&1&        |&0      &0        &\!\!\!-1      &1 &\!\!\!-1\cr
E_3&0&0&1&        |&0       &0        &\!\!\!-1       &0 &1\cr
E_4&1&0&1&        |&1       &\!\!\!-2        &1       &0 &0\cr
E_5&1&\!\!\!-1&1&|&\!\!\!-2&1&0&0&0\end{array}\right).
}\end{equation}
This immediately yields the following linear equivalences:
\begin{eqnarray}\label{lineqeightii}
0&\sim&8\,D_{{1}}+E_{{1}}+2\,E_{{2}}+3\,E_{{3}}+4\,E_{{4}}+6\,E_5,\cr
0&\sim&8\,D_{{2}}+3\,E_{{1}}+6\,E_{{2}}+E_{{3}}+4\,E_4+2\,E_5,\cr
0&\sim&2\,D_{{3}}+E_{{1}}+E_{{3}}.
\end{eqnarray}

We have ten internal lines in our toric diagram corresponding to compact curves. We find that $C_1=\,E_3\cdot E_5,\quad C_2=E_3\cdot E_4,\quad C_3=E_2\cdot E_3,\quad C_4=E_1\cdot E_3=D_2\cdot E_1,\quad C_5=E_1\cdot E_2,\quad D_3\cdot E_1=C_4+C_5,\quad D_1\cdot E_3=C_2+2\,C_3+C_4,\quad D_3\cdot E_3=C_1+2\,C_2+3\,C_3+C_4$. Here, $E_1^3=8,\ E_3^3=6$.

We will now briefly discuss the topology of the exceptional divisors. $E_1$ corresponds to an $\IF_1$. For $E_3$ we must work a little harder and perform two flops to find that it is also an $\IF_1$. $E_2$ (after one flop), $E_4$ and $E_5$ are all $\IP^1\times \IC$.


\section{Resolution of ${\IC}^3/\IZ_{12-I}$}\label{app:rztwelvei}

$\IZ_{12-I}$ acts as follows on ${\IC}^3$:
\begin{equation}\label{twisttwelvei}{\theta:\ (z^1,\, z^2,\, z^3) \to (\varepsilon\, z^1, \varepsilon^4\, z^2, \varepsilon^7\, z^3),\quad \varepsilon=e^{2 \pi i/12}.
}\end{equation}
To find the components of the $v_i$, we have to solve $(v_1)_i+4\,(v_2)_i+7\,(v_3)_i=0\ \mod\,12$. This leads to the following three generators of the fan (or some other linear combination thereof):
\begin{equation}{v_1=(4,3,1),\ v_2=(-1,1,1),\ v_3=(0,-1,1).
}\end{equation}
To resolve the singularity, we find that $\theta,\,\theta^2,\,\theta^3,\,\theta^4,\,\theta^6,\,\theta^7,\,$ and $\theta^9$ fulfill (\ref{eq:criterion}). This leads to seven new generators:
\begin{eqnarray}
w_1&=&(0,0,1),\,w_2=(0,1,1),\,w_3=(1,0,1),\,w_4=(1,1,1),\, w_5=(2,1,1),\cr
w_6&=&(2,2,1),\,w_7=(3,2,1)
.\end{eqnarray}
In this case, there are 35 triangulations. 
\begin{figure}[h!]
\begin{center}
\includegraphics[width=140mm]{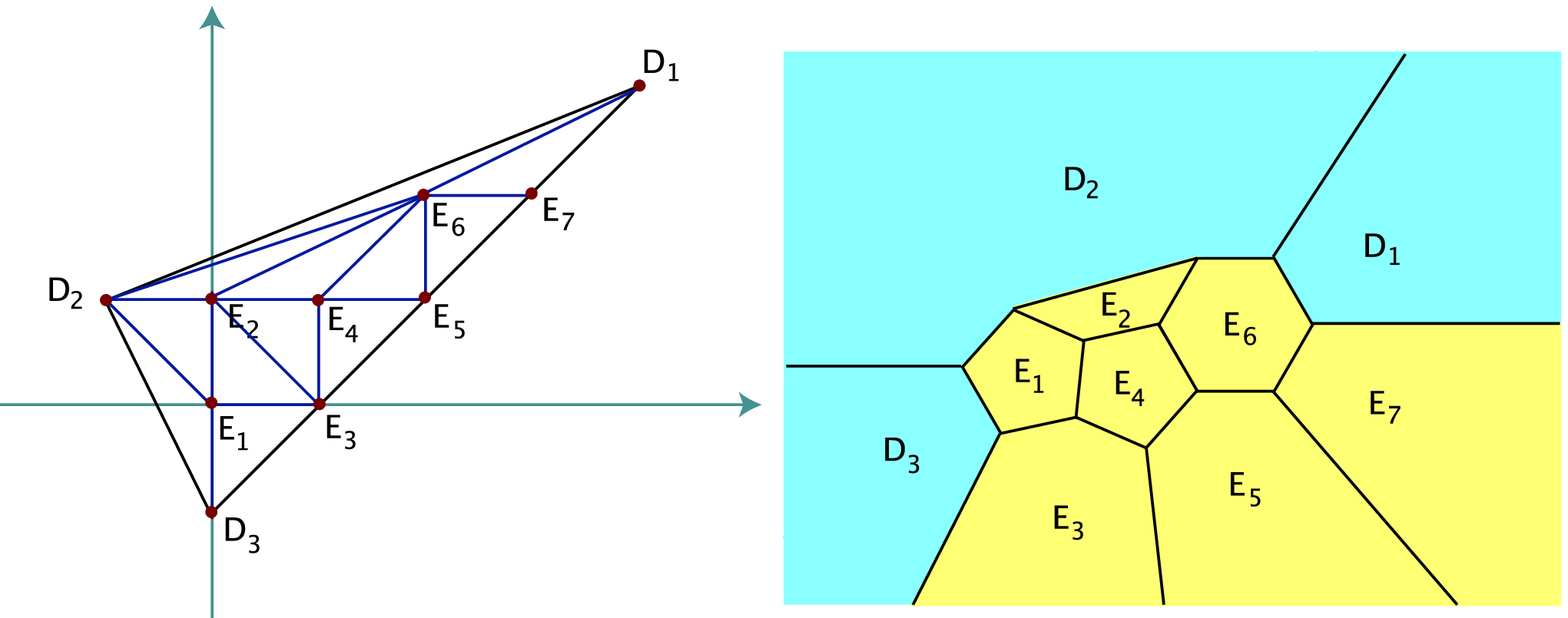}
\caption{Toric diagram of the resolution of ${\IC}^3/\IZ_{12-I}$ and dual graph}\label{frtwelve}
\end{center}
\end{figure}
Figure \ref{frtwelve} shows the toric diagram of one of them and its dual graph.
We have now twelve three-dimensional cones: $(D_1,\,D_2,\,E_6)$, $(D_1,\,E_6,\,E_7),\ (E_4,\,E_6,\,E_7)$, $(E_4,\,E_5,\,E_7),\ (E_3,\,E_4,\,E_5)$, $(E_2,\,E_4,\,E_6)$, $(D_2,\,E_2,\,E_6)$, $(E_2,\,E_3,\,E_4),\ (E_1,\,E_2,\,E_3)$,  $(D_3,\,E_1,\,E_2)$, $(D_2,\,D_3,\,E_1)$ and $(D_2,\,E_1,\,E_2)$.
Let us identify the blown--up geometry. The $\tilde U_i$ are
\begin{eqnarray}\label{tildeUtwelvei}
\tilde U_1&=&{(z^1)^4(z^2)^{-1}y^3y^4(y^5)^2(y^6)^2(y^7)^2},\cr
\tilde U_2&=&{(z^1)^3z^2(z^3)^{-1}y^2y^4y^5(y^6)^2(y^7)^2},\cr
\tilde U_3&=&z^1z^2z^3y^1y^2y^3y^4y^5y^6y^7.
\end{eqnarray}
The rescaling that leaves the $\tilde U_i$ invariant is 
\begin{eqnarray}\label{rescalestwelve}(z^1,z^2,z^3,y^1,\ldots,y^7) &\to& 
(\lambda_1\,z^1,\,\lambda_2\,z^2,\,\lambda_1^3\lambda_2\lambda_3\lambda_4\lambda_5\lambda_6^2\lambda_7^2\,z^3,\,{1\over\lambda_2^{3}\lambda_3^2\lambda_4\lambda_6}\,y^1,\cr
&& \lambda_3\,y^2,\,{\lambda_2\over \lambda_1^4\lambda_4\lambda_5^2\lambda_6^2\lambda_7^3}\,y^3,\lambda_4\,y^4,\lambda_5\,y^5,\lambda_6\,y^6,\lambda_7\,y^7).
\end{eqnarray}
Thus the blown-up geometry corresponds to
\begin{equation}\label{blowuptwelve}{
X_{\tilde\Sigma}=({\IC}^{10}\setminus F_{\tilde\Sigma})/({\IC}^*)^7.
}\end{equation}
We refrain from giving the excluded set of simultaneous zeros of coordinates not belonging to the same cone explicitly. The action of $({\IC}^{*})^7$ is given by (\ref{rescalestwelve}).
Using the method discussed in Chapter 3, we find seven generators of the Mori cone, which form the columns of $Q$:
\begin{equation}{(P\,|\,Q)=\left(\begin{array}{cccccccccccc}
D_1&4&3&1&        |&1      &0        &0      &0                 &0      &0        &0\cr
D_2&\!\!\!-1&1&1&|&0       &0        &0     &0                 &0      &0        &1\cr
D_3&0&\!\!\!-1&1&|&0       &0         &0     &0                &0      &1        &0\cr
E_1&0&0&1&        |&0       &0        &0      &0               &1       &\!\!\!-2&\!\!\!-1\cr
E_2&0&1&1&        |&0       &0        &0      &1               &\!\!\!-1&1        &\!\!\!-1\cr
E_3&1&0&1&        |&0       &0        &1       &0             &\!\!\!-1&0        &1\cr
E_4&1&1&1&        |&0       &1        &\!\!\!-1       &\!\!\!-2 &1       &0        &0\cr
E_5&2&1&1&        |&1       & \!\!\!-1&\!\!\!-1&1            &0        &0        &0\cr
E_6&2&2&1&        |&0       &\!\!\!-  1&1      &0              &0        &0        &0\cr
E_7&3&2&1&        |&\!\!\!-2&1         &0      &0             &0        &0        &0\end{array}\right).
}\end{equation}
This immediately yields the following linear equivalences:
\begin{eqnarray}\label{lineqtwelve}
0&\sim&12\,D_{{1}}+E_{{1}}+2\,E_{{2}}+3\,E_{{3}}+4\,E_{{4}}+6\,E_{{5}}+7\,E_6+9\,E_7,\cr
0&\sim&3\,D_{{2}}+E_{{1}}+2\,E_{{2}}+E_{{4}}+E_{{6}},\cr
0&\sim&12\,D_{{3}}+7\,E_{{1}}+2\,E_{{2}}+9\,E_{{3}}+4\,E_{{4}}+6\,E_{{5}}+E_6+3\,E_7.
\end{eqnarray}

We have fifteen internal lines in our toric diagram corresponding to compact curves. We find that $C_1=\,E_6\cdot E_7,\quad C_2=E_5\cdot E_6,\quad C_3=E_4\cdot E_5,\quad C_4=E_4\cdot E_6=E_3\cdot E_4,\quad C_5=E_2\cdot E_3,\quad C_6=D_2\cdot E_1=E_1\cdot E_3,\quad C_7=E_1\cdot E_2,\quad E_2\cdot E_6=C_5+C_7,\quad E_2\cdot E_4=C_3+C_4,\quad D_2\cdot E_2=C_3+C_4+3\,C_5+2\,C_7,\quad D_3\cdot E_1=C_6+C_7$. Here, $E_1^3=8,\ E_2^3=7,\ E_4^3=8,\ E_6^3=6$.

We will now briefly discuss the topology of the exceptional divisors.  We find that $E_1$ is an ${\IF}_1$, $E_2$ corresponds to a ${\IP}^2$ blown up in two points, $E_4$ is another ${\IF}_1$. For $E_6$, we have to do a little more work, since the topology cannot be read off directly from the fan. We find it to be birationally equivalent to a $\IP^2$; to find this, we must perform a sequence of three flops.

Looking at the non-compact exceptional divisors, we find $E_3$ to be equivalent to a $\IP^1$ after two flops, after one flop transition, $E_5$ is a $\IP^1\times\IC$, and $E_7$ can be seen to be a $\IP^1\times\IC$ directly.


\section{Resolution of ${\IC}^3/\IZ_{12-II}$}\label{app:rztwelveii}

$\IZ_{12-II}$ acts as follows on ${\IC}^3$:
\begin{equation*}
{\theta:\ (z^1,\, z^2,\, z^3) \to (\varepsilon\, z^1, \varepsilon^5\, z^2, \varepsilon^6\, z^3),\quad \varepsilon=e^{2 \pi i/12}.
}\end{equation*}
To find the components of the $v_i$, we have to solve $(v_1)_i+5\,(v_2)_i+6\,(v_3)_i=0\ \mod\,12$. This leads to the following three generators of the fan (or some other linear combination thereof):
\begin{equation*}{v_1=(1,5,1),\ v_2=(1,-1,1),\ v_3=(-1,0,1).
}\end{equation*}
To resolve the singularity, we find that $\theta,\,\theta^2,\,\theta^3,\,\theta^4,\,\theta^5,\,\theta^6,\,\theta^8,\,$ and $\theta^{10}$ fulfill (\ref{eq:criterion}). This leads to eight new generators:
\begin{eqnarray*}
w_1&=&(0,0,1),\,w_2=(1,0,1),\,w_3=(0,1,1),\,w_4=(1,1,1),\, w_5=(0,2,1),\cr
w_6&=&(1,2,1),\,w_7=(1,3,1),\,w_8=(1,4,1)
.\end{eqnarray*}
In this case, there are 39 triangulations. 
\begin{figure}[h!]
\begin{center}
\includegraphics[width=100mm]{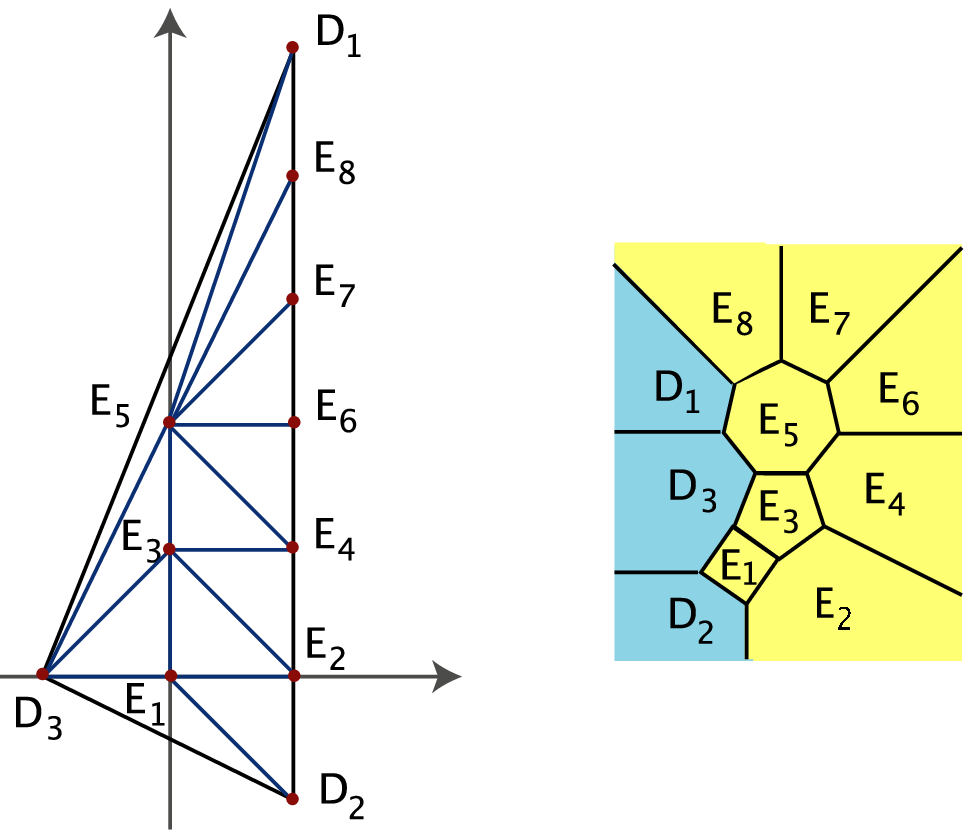}
\caption{Toric diagram of the resolution of ${\IC}^3/\IZ_{12-II}$ and dual graph}\label{frtwelveii}
\end{center}
\end{figure}
Figure \ref{frtwelveii} shows the toric diagram of one of them and its dual graph.
We have now twelve three-dimensional cones: $(D_1,\,D_3,\,E_5)$, $(D_1,\,E_5,\,E_8),\ (E_5,\,E_7,\,E_8)$, $(E_5,\,E_6,\,E_7),\ (E_4,\,E_5,\,E_6),\ (E_3,\,E_4,\,E_5)$, $(D_3,\,E_3,\,E_5)$, $(E_1,\,E_2,\,E_3)$, $(D_3,\,E_1,\,E_3)$,  $(D_2,\,E_1,\,E_2),\ (D_2,\,D_3,\,E_1)$ and $(E_2,\,E_3,\,E_4)$.
Let us identify the blown--up geometry. The $\tilde U_i$ are
\begin{eqnarray}\label{tildeUtwelveii}
\tilde U_1&=&{z^1z^2(z^3)^{-1}y^2y^3y^4y^6y^7y^8},\cr
\tilde U_2&=&{(z^1)^5(z^2)^{-1}y^4(y^5)^2(y^6)^2(y^7)^3(y^8)^4},\cr
\tilde U_3&=&z^1z^2z^3y^1y^2y^3y^4y^5y^6y^7y^8.
\end{eqnarray}
The rescaling that leaves the $\tilde U_i$ invariant is 
\begin{eqnarray}\label{rescalestwelveii}&&(z^1,z^2,z^3,y^1,y^2,y^3,y^4,y^5,y^6,y^7,y^8) \to \cr
&&(\lambda_1\,z^1,\,\lambda_1^5\lambda_4\lambda_5^2\lambda_6^2\lambda_7^3\lambda_8^4\,z^2,\,\lambda_1^6\lambda_2\lambda_3\lambda_4^2\lambda_5^5\lambda_6^3\lambda_7^4\lambda_8^5\,z^3,\,{1\over\lambda_1^{12}\lambda_2^2\lambda_3^2\lambda_4^4\lambda_5^5\lambda_6^6\lambda_7^8\lambda_8^{13}}\,y^1, \lambda_2\,y^2,\cr
&&\,\lambda_3\,y^3,\lambda_4y^4,\lambda_5y^5,\lambda_6y^6,\lambda_7y^7,\lambda_8\,y^8).
\end{eqnarray}
Thus the blown-up geometry corresponds to
\begin{equation*}
{
X_{\tilde\Sigma}=({\IC}^{11}\setminus F_{\tilde\Sigma})/({\IC}^*)^8.
}\end{equation*}
We refrain from giving the excluded set of simultaneous zeros of coordinates not belonging to the same cone explicitly. The action of $({\IC}^{*})^8$ is given by (\ref{rescalestwelveii}).
Using the method discussed in Chapter 3, we find eleven generators of the Mori cone, which form the columns of $Q$:
\begin{equation}{(P\,|\,Q)=\left(\begin{array}{cccccccccccccccc}
D_1&1&5&1&        |&0      &1        &1      &0                 &0      &0        &0&0&0&0&0\cr
D_2&1&\!\!\!-1&1&|&0       &0        &0     &0                 &0      &0        &0&0&0&0&1\cr
D_3&\!\!\!-1&0&1&|&1       &1         &0     &0                &0      &0        &1&0&0&1&0\cr
E_1&0&0&1&        |&0       &0        &0      &0               &  0     &0          &0&1&1&\!\!\!-2&\!\!\!-1\cr
E_2&1&0&1&        |&0&0&0&0&0&0&0&0&\!\!\!-1&1&\!\!\!-1\cr
E_3&0&1&1&        |&0       &1        &0      &0               &0&1        &\!\!\!-3&\!\!\!-2&\!\!\!-1&0&1\cr
E_4&1&1&1&        |&0       &0        &0       &0             &0&\!\!\!-1        &1&0&1&0&0\cr
E_5&0&2&1&        |&\!\!\!-2       &\!\!\!-3        &0       &0&0       &\!\!\!-1        &1&1&0&0&0\cr
E_6&1&2&1&        |&0       &0&0      &1              &\!\!\!-2        &1        &0&0&0&0&0\cr
E_7&1&3&1&        |&0&0         &1      &\!\!\!-2             &1        &0        &0&0&0&0&0\cr
E_8&1&4&1&      |&1&0&\!\!\!-2&1&0&0&0&0&0&0&0
\end{array}\right).
}\end{equation}
This immediately yields the following linear equivalences:
\begin{eqnarray}\label{lineqtwelveii}
0&\sim&12\,D_{{1}}+E_{{1}}+2\,E_{{2}}+3\,E_{{3}}+4\,E_{{4}}+5\,E_{{5}}+6\,E_6+8\,E_7+10\,E_8,\cr
0&\sim&12\,D_{{2}}+5\,E_{{1}}+10\,E_{{2}}+3\,E_3+8\,E_{{4}}+E_{{5}}+6\,E_6+4\,E_7+2\,E_8,\cr
0&\sim&2\,D_{{3}}+E_{{1}}+E_{{3}}+E_{{5}}.
\end{eqnarray}

We have fourteen internal lines in our toric diagram corresponding to compact curves. We find that $C_1=\,D_1\cdot E_5,\quad C_2=D_3\cdot E_5,\quad C_3=E_5\cdot E_8,\quad C_4=E_5\cdot E_7,\quad C_5=E_5\cdot E_6,\quad C_8=D_3\cdot E_3,\quad C_7=E_1\cdot E_2,\quad E_2\cdot E_6=C_5+C_7,\quad E_2\cdot E_4=C_3+C_4,\quad D_2\cdot E_2=C_3+C_4+3\,C_5+2\,C_7,\quad D_3\cdot E_1=C_6+C_7$. Here, $E_1^3=8,\ E_3^3=7,\ E_5^3=5$.

We will now discuss the topologies of the exceptional divisors. $E_1,\ E_3$ and $E_5$ are each $\IF_1$s. To see this, we must perform one flop for $E_3$ and three for $E_5$. $E_2,\ E_4$ and $E_6$ are each $\IP^1\times \IC$ with one blow--up, whereas $E_7$ and $E_8$ are $\IP^1\times \IC$.


\section{Resolution of ${\IC}^3/(\IZ_{2}\times\IZ_2)$}\label{app:rztwotwo}

$(\IZ_{2}\times\IZ_2)$ acts as follows on ${\IC}^3$:
\begin{eqnarray}\label{twisttwotwo}
\theta^1:\ (z^1,\, z^2,\, z^3)& \to& (\varepsilon\, z^1, \, z^2, \varepsilon\, z^3),\cr
\theta^2:\ (z^1,\, z^2,\, z^3)& \to& (\, z^1, \, \varepsilon\,z^2, \varepsilon\, z^3),\cr
\theta^1\theta^2:\ (z^1,\, z^2,\, z^3)& \to& (\varepsilon\, z^1, \, \varepsilon\,z^2,  z^3),
\end{eqnarray}
with $\varepsilon=e^{2 \pi i/2}$.
To find the components of the $v_i$, we have to solve 
\begin{eqnarray}
(v_1)_i+(v_3)_i&=&0\ \mod\,2,\cr
(v_2)_i+(v_3)_i&=&0\ \mod\,2,\cr
(v_1)_i+(v_3)_i&=&0\ \mod\,2.
\end{eqnarray}
This leads to the following three generators of the fan:
\begin{equation*}{v_1=(0,2,1),\ v_2=(0,0,1),\ v_3=(2,0,1).
}\end{equation*}
To resolve the singularity, we find that $\theta^1,\,\theta^2$  and $\theta^1\theta^2$ fulfill (\ref{eq:criterion}). This leads to three new generators:
\begin{eqnarray*}
w_1&=&(1,0,1),\ w_2=(1,1,1),\ w_3=(0,1,1).\end{eqnarray*}
In this case, there are four distinct triangulations. 
\begin{figure}[h!]
\begin{center}
\includegraphics[width=120mm]{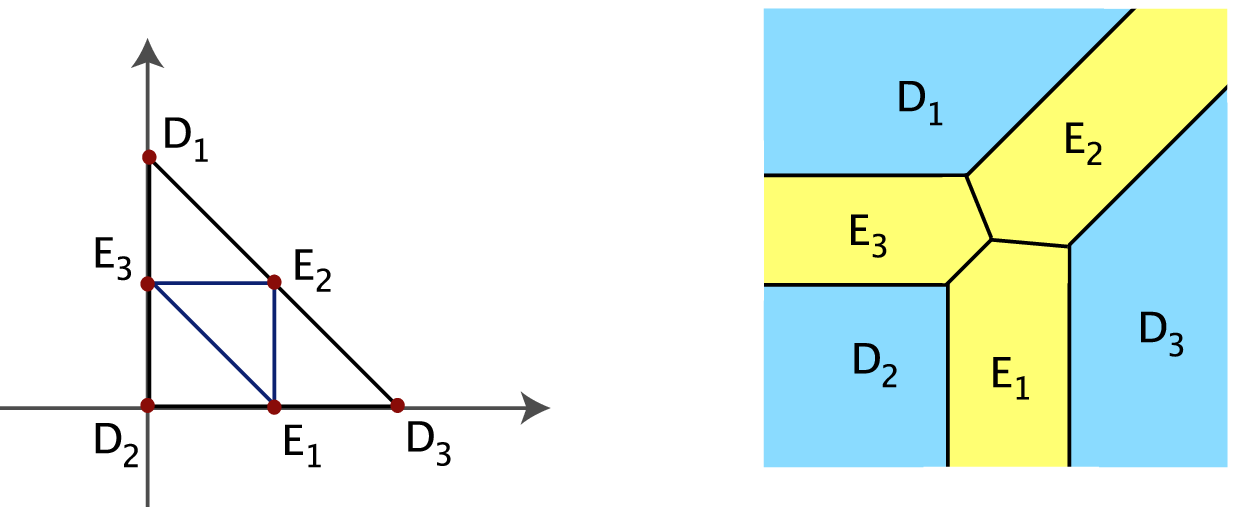}
\caption{Toric diagram of resolution of ${\IC}^3/\IZ_{2}\times\IZ_2$ and dual graph}\label{frtwotwo}
\end{center}
\end{figure}
Figure \ref{frtwotwo} shows two of them.
Let us identify the blown--up geometry. The $\tilde U_i$ are
\begin{eqnarray}\label{tildeUtwotwo}
\tilde U_1&=&{(z^3)^2y^1y^2},\cr
\tilde U_2&=&{(z^1)^2y^2y^3},\cr
\tilde U_3&=&z^1z^2z^3y^1y^2y^3.
\end{eqnarray}
The rescaling that leaves the $\tilde U_i$ invariant is 
\begin{eqnarray}\label{rescalestwotwo}&&(z^1,z^2,z^3,y^1,y^2,y^3) \to \cr
&&(\lambda_1\,z^1,\,\lambda_1\lambda_2\lambda_3\,z^2,\,\lambda_2\,z^3,\,{1\over \lambda_2^2\lambda_3}\,y^1, {\lambda_3}\,y^2,\,{1\over\lambda_1^2\lambda_3}\,y^3).
\end{eqnarray}
Thus the blown-up geometry corresponds to
\begin{equation*}{
X_{\tilde\Sigma}=({\IC}^{6}\setminus F_{\tilde\Sigma})/({\IC}^*)^3.
}\end{equation*}
The excluded sets differ for the different resolutions. We refrain from giving them explicitly. The action of $({\IC}^{*})^3$ is given by (\ref{rescalestwotwo}).

The $2\cdot 2=4$ three-dimensional cones are in this case $(D_1,E_2,E_3),\ (D_2,E_1,E_3)$,  $(D_3,E_1,E_2)$, and $(E_1,E_2,E_3)$.
We find three generators of the Mori cone and write them as columns of $Q$:
\begin{equation}{(P|\,Q)=\left(\begin{array}{ccccccccccc}
D_1&0&2&1&|&1&0&0\cr
D_2&0&0&1&|&0&1&0\cr
D_3&2&0&1&|&0&0&1\cr
E_1&1&0&1&|&1&\!\!\!-1&\!\!\!-1\cr
E_2&1&1&1&|&\!\!\!-1&1&\!\!\!-1\cr
E_3&0&1&1&|&\!\!\!-1&\!\!\!-1&1
\end{array}\right)}\end{equation}
This leads to the following linear equivalences between the divisors:
\begin{eqnarray}\label{lineqtwotwo}
0&\sim&2\,D_{{1}}+E_{{2}}+E_{{3}},\cr
0&\sim&2\,D_{{2}}+E_{{1}}+E_{3},\cr
0&\sim&2\,D_{{3}}+E_{{1}}+E_{{2}}.
\end{eqnarray}

From the intersection numbers, we find the following relations between the Mori generators and the nine compact curves of our geometry: $C_1=E_2\cdot E_3,\ C_2=E_1\cdot E_3,\ C_3=E_1\cdot E_2$.

We will now discuss the topologies of the exceptional divisors. They are all semi-compact and correspond to  $\IP^1\times \IC$.


\section{Resolution of ${\IC}^3/(\IZ_{2}\times\IZ_4)$}\label{app:rztwofour}

$(\IZ_{2}\times\IZ_4)$ acts as follows on ${\IC}^3$:
\begin{eqnarray}\label{twisttwofour}
\theta^1:\ (z^1,\, z^2,\, z^3)& \to& (\varepsilon^2\, z^1, \, z^2, \varepsilon^2\, z^3),\cr
\theta^2:\ (z^1,\, z^2,\, z^3)& \to& (\, z^1, \, \varepsilon\,z^2, \varepsilon^3\, z^3),\cr
\theta^1\theta^2:\ (z^1,\, z^2,\, z^3)& \to& (\varepsilon^2\, z^1, \, \varepsilon\,z^2, \varepsilon\, z^3),
\end{eqnarray}
with $\varepsilon=e^{2 \pi i/4}$.
To find the components of the $v_i$, we have to solve 
\begin{eqnarray}
2\,(v_1)_i+2\,(v_3)_i&=&0\ \mod\,4,\cr
(v_2)_i+3\,(v_3)_i&=&0\ \mod\,4,\cr
2\,(v_1)_i+(v_2)_i+(v_3)_i&=&0\ \mod\,4.
\end{eqnarray}
This leads to the following three generators of the fan:
\begin{equation*}{v_1=(1,-1,1),\ v_2=(-3,1,1),\ v_3=(1,1,1).
}\end{equation*}
To resolve the singularity, we now have many more possibilities for new vertices. We find that $\theta^1,\,\theta^2, \,(\theta^2)^2,\,(\theta^2)^3,\,\theta^1\theta^2$  and $\theta^1(\theta^2)^2$ fulfill (\ref{eq:criterion}). This leads to six new generators:
\begin{eqnarray*}
w_1&=&(1,0,1),\ w_2=(0,1,1),\ w_3=(-1,1,1),\ w_4=(-2,-1,1),\ w_5=(0,0,1),\cr
w_6&=&(-1,0,1)
.\end{eqnarray*}
In this case, there are 24 distinct triangulations. 
\begin{figure}[h!]
\begin{center}
\includegraphics[width=120mm]{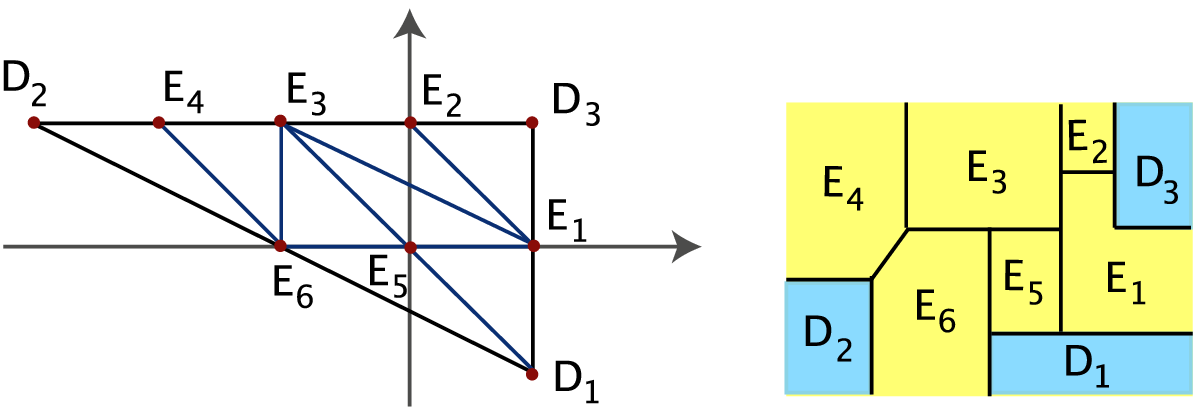}
\caption{Toric diagram of resolution of ${\IC}^3/\IZ_{2}\times\IZ_4$ and dual graph}\label{frtwofour}
\end{center}
\end{figure}
Figure \ref{frtwofour} shows one of them. It was chosen to be compatible with the triangulation of $\IZ_2\times\IZ_2$. 
Let us identify the blown--up geometry. The $\tilde U_i$ are
\begin{eqnarray}\label{tildeUtwofour}
\tilde U_1&=&{z^1z^3y^1\over(z^2)^3y^3(y^4)^2y^6},\cr
\tilde U_2&=&{z^2z^3y^2y^3y^4\over z^1},\cr
\tilde U_3&=&z^1z^2z^3y^1y^2y^3y^4y^5y^6y^7.
\end{eqnarray}
The rescaling that leaves the $\tilde U_i$ invariant is 
\begin{eqnarray}\label{rescalestwofour}(z^1,z^2,z^3,y^1,..,y^6) &\to& 
(\lambda_1\,z^1,\,\lambda_2\,z^2,\,\lambda_3\,z^3,\,{\lambda_2^3\lambda_4\lambda_5^2\lambda_6\over\lambda_1\lambda_3}\,y^1, {\lambda_1\over \lambda_2\lambda_3\lambda_4\lambda_5}\,y^2,\cr
&&{\lambda_4}\,y^3,\lambda_5\,y^4,{\lambda_3\over \lambda_1\lambda_2^3\lambda_4\lambda_5^2\lambda_6^2}\,y^5, \lambda_6\, y^6).
\end{eqnarray}
Thus the blown-up geometry corresponds to
\begin{equation*}{
X_{\tilde\Sigma}=({\IC}^{9}\setminus F_{\tilde\Sigma})/({\IC}^*)^6.
}\end{equation*}
The excluded sets differ for the different resolutions. We refrain from giving them explicitly. The action of $({\IC}^{*})^6$ is given by (\ref{rescalestwofour}).

The $2\cdot 4=8$ three-dimensional cones are in this case $(D_2,E_4,E_6),\ (E_3,E_4,E_6)$,  $(E_3,E_5,E_6)$, $(D_3,E_1,E_2),\ (D_1,E_1,E_5),\ (D_1,E_5,E_6),\ (E_2,E_3,E_1),\ (D_3,E_1,E_2)$.
We find six generators of the Mori cone and write them as columns of $Q$:
\begin{equation}{(P|\,Q)=\left(\begin{array}{ccccccccccc}
D_1&1&\!\!\!\-1&1&|&0&0&1&0&0&0\cr
D_2&\!\!\!-3&1&1&|&0&0&0&0&0&1\cr
D_3&1&1&1&|&1&0&0&0&0&0\cr
E_1&1&0&1&|&0&\!\!\!-1&0&1&0&0\cr
E_2&0&1&1&|&\!\!\!-2&1&0&0&0&0\cr
E_3&\!\!\!-1&1&1&|&1&\!\!\!-1&1&0&\!\!\!-1&1\cr
E_4&\!\!\!-2&1&1&|&0&0&0&0&1&\!\!\!-2\cr
E_5&0&0&1&|&0&1&\!\!\!-2&\!\!\!-2&1&0\cr
E_6&\!\!\!-1&0&1&|&0&0&0&1&\!\!\!-1&0
\end{array}\right)}\end{equation}
This leads to the following linear equivalences between the divisors:
\begin{eqnarray}\label{lineqtwofour}
0&\sim&2\,D_{{1}}+E_{{1}}+E_{{5}}+E_{{6}},\cr
0&\sim&4\,D_{{2}}+E_{{2}}+2\,E_{{3}}+3\,E_{{4}}+E_{5}+2\,E_{{6}},\cr
0&\sim&4\,D_{{3}}+2\,E_{{1}}+3\,E_{{2}}+2\,E_{{3}}+E_{{4}}+E_{{5}}.
\end{eqnarray}

From the intersection numbers, we find the following relations between the Mori generators and the nine compact curves of our geometry: $C_1=E_1\cdot E_2,\ C_2=E_1\cdot E_3,\ C_3=E_1\cdot E_5=E_5\cdot E_6,\ C_4=E_3\cdot E_5=D_1\cdot E_5,\ C_5=E_3\cdot E_6,\ C_6=E_4\cdot E_6$. Furthermore, $E_5^3=8$.

We will now discuss the topologies of the exceptional divisors. $E_5$ is an $\IF_1$. $E_2$ and $E_4$ are $\IP^1\times \IC$, $E_1,\ E_3$ and $E_6$ are $\IP^1\times \IC$ with two blow--ups.


\section{Resolution of ${\IC}^3/(\IZ_2\times\IZ_6)$}\label{app:rztwosix}

$\IZ_{2}\times\IZ_6$ acts as follows on ${\IC}^3$:
\begin{eqnarray}\label{twisttwosix}
\theta^1:\ (z^1,\, z^2,\, z^3)& \to& (\varepsilon^3\, z^1, \, z^2, \varepsilon^3\, z^3),\cr
\theta^2:\ (z^1,\, z^2,\, z^3)& \to &(\, z^1, \, \varepsilon\,z^2, \varepsilon^5\, z^3),\cr
\theta^1\theta^2:\ (z^1,\, z^2,\, z^3)& \to& (\varepsilon^3\, z^1, \, \varepsilon\,z^2, \varepsilon^2\, z^3),
\end{eqnarray}
with $\varepsilon=e^{2 \pi i/6}$.
To find the components of the $v_i$, we have to solve 
\begin{eqnarray}
3\,(v_1)_i+3\,(v_3)_i&=&0\ \mod\,6,\cr
(v_2)_i+5\,(v_3)_i&=&0\ \mod\,6,\cr
3\,(v_1)_i+(v_2)_i+2\,(v_3)_i&=&0\ \mod\,6.
\end{eqnarray}
This leads to the following three generators of the fan:
\begin{equation*}{v_1=(1,0,1),\ v_2=(-1,-4,1),\ v_3=(-1,2,1).
}\end{equation*}
To resolve the singularity, we find that $\theta^1,\,\theta^2, \,(\theta^2)^2,\,(\theta^2)^3,\,(\theta^2)^4,\,(\theta^2)^5,\,\theta^1\theta^2, \,\theta^1(\theta^2)^2$  and $\,(\theta^1)^2\theta^3$ fulfill (\ref{eq:criterion}). This leads to nine new generators:
\begin{eqnarray*}
w_1&=&(0,1,1),\ w_2=(-1,1,1),\ w_3=(-1,0,1),\ w_4=(-1,-1,1),\ w_5=(-1,-2,1),\cr
w_6&=&(-1,-3,1),\ w_7=(0,0,1),\,w_8=(0,-1,1),\,w_9=(0,-2,1)
.\end{eqnarray*}
In this case, there are 156 distinct triangulations. 
\begin{figure}[h!]
\begin{center}
\includegraphics[width=120mm]{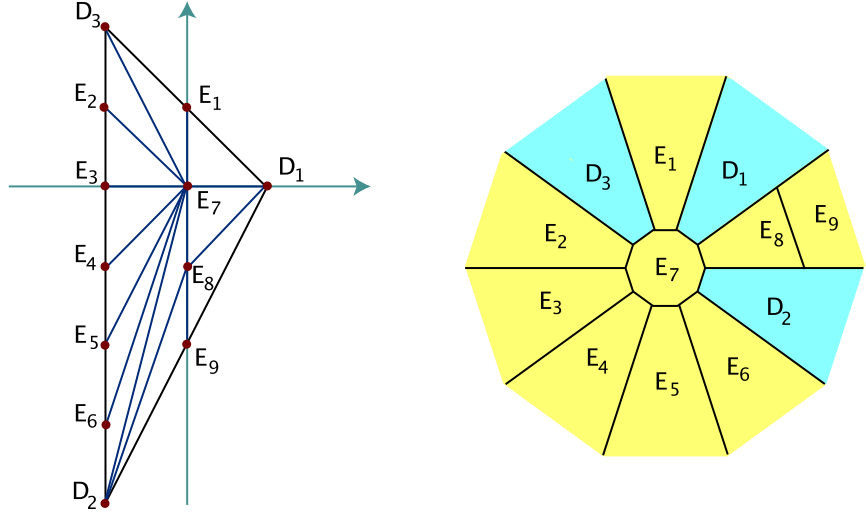}
\caption{Toric diagram of one of the resolutions of ${\IC}^3/\IZ_{2}\times\IZ_6$ and dual graph}\label{ftwosix}
\end{center}
\end{figure}
Figure \ref{ftwosix} shows one of them. Let us identify the blown--up geometry. The $\tilde U_i$ are
\begin{eqnarray}\label{tildeUtwosix}
\tilde U_1&=&{z^1\over z^2z^3y^2y^3y^4y^5y^6},\quad \tilde U_2={(z^3)^2y^1y^2\over(z^2)^4y^4(y^5)^2(y^6)^3y^8(y^9)^2},\cr
\tilde U_3&=&z^1z^2z^3y^1...y^9.
\end{eqnarray}
The rescalings that leave the $\tilde U_i$ invariant are
\begin{eqnarray}\label{rescalestwosix}
(z^1,z^2,z^3,y^1,...,y^9)& \to\ &(\lambda_1\lambda_2...\lambda_7\,z^1,\,\lambda_1\,z^2,\,\lambda_2\,z^3,\, {\lambda_1^4\lambda_5\lambda_6^2\lambda_7^3\lambda_8\lambda_9^2\over\lambda_2^2\lambda_3}\,y^1,\lambda_3\,y^2,\,\lambda_4\,y^3,\cr
&&\lambda_5\,y^4,\lambda_6\,y^5,\lambda_7\,y^6,\,{1\over\lambda_1^6\lambda_3\lambda_4^2\lambda_5^3\lambda_6^4\lambda_7^5\lambda_8^2\lambda_9^3}\,y^7,\lambda_8\,y^8,\,\lambda_9\,y^9).
\end{eqnarray}
According to (\ref{eq:blowup}), the new blown-up geometry is
\begin{equation*}{
X_{\tilde\Sigma}=\,({\IC}^{12}\setminus F_{\tilde\Sigma})/({\IC}^*)^9,
}\end{equation*}
where the action of $({\IC}^*)^9$ is given by (\ref{rescalestwosix}). We refrain from giving the excluded set explicitly here. As can readily be seen in the dual graph, we have 13 compact curves in $X_{\tilde\Sigma}$. Eight of them are exceptional.

We have now $2\cdot6=12$ three-dimensional cones: $(D_3,\,E_1,\,E_7),\ (D_3,\,E_2,\,E_7)$, $(E_2,\,E_3,\,E_7)$,  $(D_1,\,E_1,\,E_7),\ (E_3,\,E_4,\,E_7)$, $(D_1,\,E_7,\,E_8),\ (E_4,\,E_5,\,E_7)$, $(E_5,\,E_6,\,E_7)$, $(D_2,\,E_6,\,E_7)$, $(D_2,\,E_7,\,E_8),\ (D_1,\,E_8,\,E_9)$, and $(D_2,\,E_8,\,E_9)$.
With the method illustrated in the last paragraph, we find ten generators of the Mori cone, which form the columns of $Q$:
\begin{equation}{(P\,|\,Q)=\left(\begin{array}{ccccccccccccccc}
D_1&1&0&1&|&0&1&0&0&0&0&0&0&0&1\cr
D_2&\!\!\!-1&\!\!\!-4&1&|&0&0&0&0&0&0&0&1&\!\!\!-1&1\cr
D_3&\!\!\!-1&2&1&|&\!\!\!-1&1&1&0&0&0&0&0&0&0\cr
E_1&0&1&1&|&1&\!\!\!-2&0&0&0&0&0&0&0&0\cr
E_2&\!\!\!-1&1&1&|&1&0&\!\!\!-2&1&0&0&0&0&0&0\cr
E_3&\!\!\!-1&0&1&|&0&0&1&\!\!\!-2&1&0&0&0&0&0\cr
E_4&\!\!\!-1&\!\!\!-1&1&|&0&0&0&1&\!\!\!-2&0&1&0&0&0\cr
E_5&\!\!\!-1&\!\!\!-2&1&|&0&0&0&0&1&0&\!\!\!-2&1&0&0\cr
E_6&\!\!\!-1&\!\!\!-3&1&|&0&0&0&0&0&0&1&\!\!\!-2&1&0\cr
E_7&0&0&1&|&\!\!\!-1&0&0&0&0&1&0&0&\!\!\!-1&0\cr
E_8&0&\!\!\!-1&1&|&0&0&0&0&0&\!\!\!-2&0&0&1&0\cr
E_9&0&\!\!\!-2&1&|&0&0&0&0&0&1&0&0&0&\!\!\!-2\end{array}\right).
}\end{equation}
{From} the rows of $Q$, we can read off the linear equivalences:
\begin{eqnarray}\label{lineqtwosixI}
0 &\sim& -6\,D_{{1}}-E_{{5}}-E_{{2}}-2\,E_{{3}}-5\,E_{{9}}-2\,E_{{6}}-3\,E_{{7}}-4\,E_{{8}}-3\,E_{{4}},\cr 
0 
&\sim& -2\,D_{{2}}-E_{{4}}-E_{{1}}-E_{{2}}-E_{{3}},\cr 
0 &\sim& -6\,D_{{3}}-5\,E_{{5}}-3\,E_{{1}}-2\,E_{{2}}-E_{{3}}-E_{{9}}-4\,E_{{6}}-3\,E_{{7}}-2\,E_{{8}}.
\end{eqnarray}
The matrix elements of $Q$ contain the intersection numbers of the $C_i$ with the $D_i,\,E_i$. From the linear equivalences between the divisors, we find the following relations between the curves $C_i$ and the compact curves of our geometry: 
$C_1=D_3\cdot E_7,\ C_2=E_1\cdot E_7,\ C_3=E_2\cdot E_7,\ C_4=E_3\cdot E_7,\ C_5=E_4\cdot E_7,\   C_6=D_1\cdot E_8=D_2\cdot E_8,\ C_7=E_5\cdot E_7,\ C_8=E_6\cdot E_7,\ C_9=D_2\cdot E_7, \ D_1\cdot E_7=C_1+C_3+C_4+C_5+C_7+C_8+C_9,\ E_7\cdot E_8=2\,C_1+C_2+C_3-C_5-2\,C_7-3\,C_8-4\,C_9,\ E_8\cdot E_9=2\,C_1+C_2+C_3-C_5-2\,C_6-2\,C_7-3\,C_8-4\,C_9$. For the triple self--intersections of the compact exceptional divisors, we find $E_7^3=2,\ E_8^3=8$.
Looking at the stars of the expectional divisors, we see that $E_i$, $i=1,\dots,6$ and $E_9$ have the topology $\IP^1 \times \IC$. $E_7$ is an $\IF_4$ blown-up in six points and $E_8$ is an $\IF_2$.

\section{Resolution of ${\IC}^3/(\IZ_{2}\times\IZ_{6'})$}\label{app:rztwosixp}

$\IZ_{2}\times\IZ_{6'}$ acts as follows on ${\IC}^3$:
\begin{eqnarray}\label{twisttwosixi}
\theta^1:\ (z^1,\, z^2,\, z^3)& \to& (\varepsilon^3\, z^1, \, z^2, \varepsilon^3\, z^3),\cr
\theta^2:\ (z^1,\, z^2,\, z^3)& \to& (\varepsilon\, z^1, \, \varepsilon\,z^2, \varepsilon^4\, z^3),\cr
\theta^1\theta^2:\ (z^1,\, z^2,\, z^3)& \to& (\varepsilon^4\, z^1, \, \varepsilon\,z^2, \varepsilon\, z^3),
\end{eqnarray}
with $\varepsilon=e^{2 \pi i/6}$.
To find the components of the $v_i$, we have to solve 
\begin{eqnarray}
3\,(v_1)_i+3\,(v_3)_i&=&0\ \mod\,6,\cr
(v_1)_i+(v_2)_i+4\,(v_3)_i&=&0\ \mod\,6,\cr
4\,(v_1)_i+(v_2)_i+(v_3)_i&=&0\ \mod\,6.
\end{eqnarray}
This leads to the following three generators of the fan:
\begin{equation*}{v_1=(0,-2,1),\ v_2=(-2,0,1),\ v_3=(2,2,1).
}\end{equation*}
Now we resolve the singularity. We find that $\theta^1,\,\theta^2, \,(\theta^2)^2, \,(\theta^2)^3,\,\theta^1\theta^2, \,\theta^1(\theta^2)^3$  and $\,\theta^1(\theta^2)^4$ fulfill (\ref{eq:criterion}). This leads to seven new generators:
\begin{eqnarray*}
w_1&=&(1,0,1),\ w_2=(1,1,1),\ w_3=(0,0,1),\ w_4=(-1,-1,1),\ w_5=(0,-1,1),\cr
w_6&=&(0,1,1),\ w_7=(-1,0,1)
.\end{eqnarray*}
In this case, there are 80 distinct triangulations.
\begin{figure}[h!]
\begin{center}
\includegraphics[width=120mm]{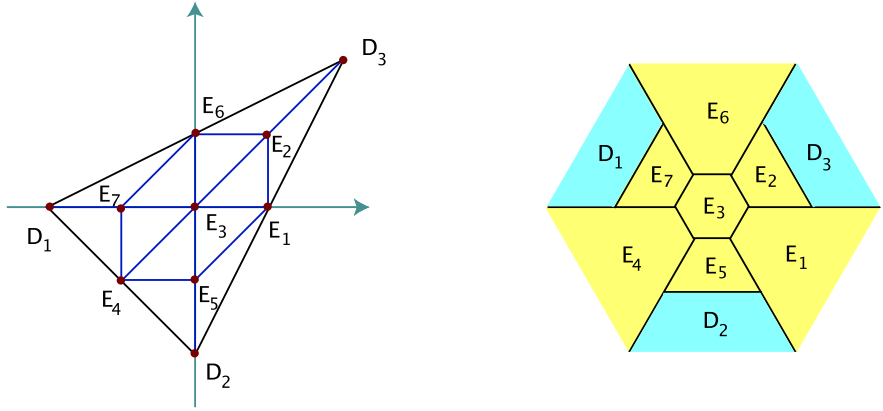}
\caption{Toric diagram of one of the resolutions of ${\IC}^3/\IZ_{2}\times\IZ_{6'}$ and dual graph}\label{frtwosixp}
\end{center}
\end{figure}
Figure \ref{frtwosixp} shows one of them.
Let us identify the blown--up geometry. The $\tilde U_i$ are
\begin{eqnarray}\label{tildeUtwosixp}
\tilde U_1&=&{(z^3)^2y^1y^2\over(z^2)^{2}y^4y^7},\quad \tilde U_2={(z^3)^2y^2y^6\over (z^1)^2y^4y^5},\quad \tilde U_3=z^1z^2z^3y^1y^2y^3y^4y^5y^6y^7.
\end{eqnarray}
The rescaling that leaves the $\tilde U_i$ invariant is 
\begin{eqnarray}\label{rescalestwosixp}
(z^1,z^2,z^3,y^1,...,y^7) &\to& 
(\lambda_1\,z^1,\,\lambda_2\,z^2,\,\lambda_3\,z^3,\,{\lambda_2^2\lambda_5\lambda_7\over\lambda_3^2\lambda_4}\,y^1, \lambda_4\,y^2,\,{\lambda_3^3\lambda_4\over \lambda_2^3\lambda_2^3\lambda_5^3\lambda_6^2\lambda_7^2}\,y^3,\cr
&&\lambda_5\,y^4,\lambda_6\,y^5,{\lambda_1^2\lambda_5\lambda_6\over\lambda_3^2\lambda_4}\,y^6,\lambda_7\,y^7).
\end{eqnarray}
Thus the blown-up geometry corresponds to
\begin{equation*}{
X_{\tilde\Sigma}=({\IC}^{10}\setminus F_{\tilde\Sigma})/({\IC}^*)^7.
}\end{equation*}
The excluded sets differ for the different resolutions. We refrain from giving them explicitly. The action of $({\IC}^{*})^7$ is given by (\ref{rescalestwosixp}).

The $2\cdot 6=12$ three-dimensional cones are in this case $(D_3,E_2,E_6),\ (D_3,E_1,E_2)$, $(E_2,E_3,E_6)$, $(E_1,E_2,E_3),\ (D_1,E_6,E_7),\ (E_3,E_6,E_7)$, $ (E_1,E_3,E_5)$, $(D_2,E_1,E_5)$, $(D_1,E_4,E_7)$, $(E_3,E_4,E_7)$, $(E_3,E_4,E_5),\ (D_2,E_4,E_5)$.
We find nine generators of the Mori cone and write them as columns of $Q$:
\begin{equation}{(P|\,Q)=\left(\begin{array}{cccccccccccccc}
D_1&0&\!\!\!-2&1&|&0&0&0&0&1&0&0&0&0\cr
D_2&\!\!\!-2&0&1&|&0&0&0&0&0&1&0&1&\!\!\!-1\cr
D_3&2&2&1&|&0&0&1&0&\!\!\!-2&\!\!\!-1&0&0&1\cr
E_1&1&0&1&|&0&0&0&0&0&0&1&0&0\cr
E_2&1&1&1&|&0&0&0&1&0&0&\!\!\!-2&\!\!\!-1&1\cr
E_3&0&0&1&|&1&\!\!\!-1&\!\!\!-1&\!\!\!-1&1&\!\!\!-1&1&\!\!\!-1&\!\!\!-1\cr
E_4&\!\!\!-1&\!\!\!-1&1&|&0&1&\!\!\!-1&0&0&1&0&0&0\cr
E_5&0&\!\!\!-1&1&|&0&1&0&\!\!\!-1&0&0&0&1&0\cr
E_6&0&1&1&|&\!\!\!-2&\!\!\!-1&1&1&0&0&0&0&0\cr
E_7&\!\!\!-1&0&1&|&1&0&0&0&0&0&0&0&0
\end{array}\right)}\end{equation}
This leads to the following linear equivalences between the divisors:
\begin{eqnarray}\label{lineqwosixII}
0&\sim&6\,D_{{1}}+4\,E_{{2}}+E_{{3}}+2\,E_{{4}}+E_{{7}}+3\,E_{{1}}+3\,E_{{5}},\cr 
0&\sim&6\,D_{{2}}+3\,E_{{1}}+E_{{2}}+4\,E_{{3}}+2\,E_{{4}}+3\,E_{{6}}+E_{{7}},\cr 
0&\sim&6\,D_{{3}}+E_{{2}}+E_{{3}}+2\,E_{{4}}+3\,E_{{6}}+4\,E_{{7}}+3\,E_{{5}}.
\end{eqnarray}

From the intersection numbers, we find the following relations between the Mori generators and the fifteen compact curves of our geometry: $C_1=E_4\cdot E_6=E_5\cdot E_6,\ C_2=E_3\cdot E_6,\ C_3=E_3\cdot E_4,\ C_4=E_3\cdot E_5,\ C_5=D_3\cdot E_4=D_2\cdot D_3,\ C_6=D_3\cdot E_3,\ C_7=D_2\cdot E_2=E_2\cdot E_5, \ E_6\cdot E_7=C_1+C_2,\ D_1\cdot D_3=C_5+C_6, \ E_2\cdot E_3=C_3-C_4+C_6,\ E_1\cdot E_2=C_3-C_4+C_6+C_7,\ D_2\cdot E_3=C_2+C_4-C_6$.

Let us now discuss the topologies of the compact exceptional divisors. From the Mori generators of the star of $E_3$, we cannot directly read off the topology, but after the two flop transitions $(E_2,E_3)\to (E_1,E_6),\ (E_3,E_4)\to (E_5,E_7)$, we find again a very simple star whose two Mori generators are those of an ${\IF}_0$. So $E_3$ is birationally equivalent to a Hirzebruch surface ${\IF}_0$. The cases of $E_2, E_5$ and $E_7$ are simpler. From their stars we can see directly that they correspond to ${\IF}_1$s.

The non-compact exceptional divisors $E_1,\, E_4$ and $E_6$ all turn out to be $\IP^1\times\IC$ blown up in two points.


\section{Resolution of ${\IC}^3/(\IZ_{3}\times\IZ_3)$}\label{app:rzthreethree}

$(\IZ_{3}\times\IZ_3)$ acts as follows on ${\IC}^3$:
\begin{eqnarray}\label{twistthreethree}
\theta^1:\ (z^1,\, z^2,\, z^3)& \to& (\varepsilon\, z^1, \, z^2, \varepsilon^2\, z^3),\cr
\theta^2:\ (z^1,\, z^2,\, z^3)& \to& (\, z^1, \, \varepsilon\,z^2, \varepsilon^2\, z^3),\cr
\theta^1\theta^2:\ (z^1,\, z^2,\, z^3)& \to& (\varepsilon\, z^1, \, \varepsilon\,z^2, \varepsilon\, z^3),
\end{eqnarray}
with $\varepsilon=e^{2 \pi i/3}$.
To find the components of the $v_i$, we have to solve 
\begin{eqnarray}
(v_1)_i+2\,(v_3)_i&=&0\ \mod\,3,\cr
(v_2)_i+2\,(v_3)_i&=&0\ \mod\,3,\cr
(v_1)_i+(v_2)_i+(v_3)_i&=&0\ \mod\,3.
\end{eqnarray}
This leads to the following three generators of the fan:
\begin{equation*}{v_1=(-2,2,1),\ v_2=(-2,-1,1),\ v_3=(1,-1,1).
}\end{equation*}
To resolve the singularity, we now have many more possibilities for new vertices. We find that $\theta^1,\,(\theta^1)^2,\,\theta^2, \,(\theta^2)^2,\,\theta^1\theta^2, \,\theta^1(\theta^2)^2$  and $\,(\theta^1)^2\theta^2$ fulfill (\ref{eq:criterion}). This leads to seven new generators:
\begin{eqnarray*}
w_1&=&(0,0,1),\ w_2=(-1,1,1),\ w_3=(0,-1,1),\ w_4=(-1,-1,1),\ w_5=(-1,0,1),\cr
w_6&=&(-2,0,1),\ w_7=(-2,1,1)
.\end{eqnarray*}
In this case, there are 79 distinct triangulations. 
\begin{figure}[h!]
\begin{center}
\includegraphics[width=120mm]{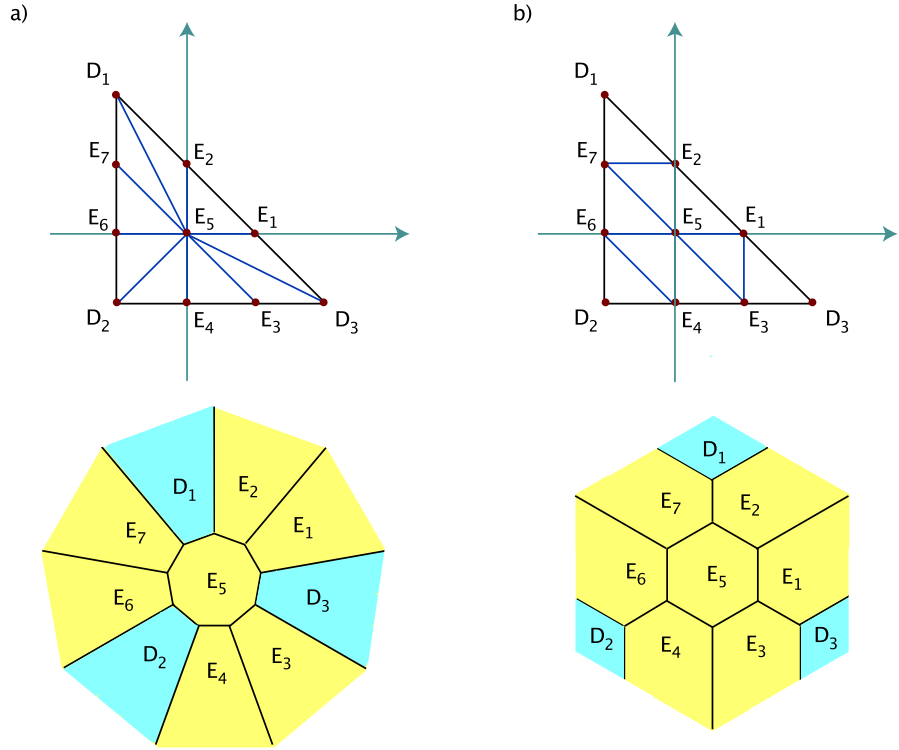}
\caption{Toric diagram of two of the resolutions of ${\IC}^3/\IZ_{3}\times\IZ_3$ and dual graphs}\label{frthreethree}
\end{center}
\end{figure}
Figure \ref{frthreethree} shows two of them.
Let us identify the blown--up geometry. The $\tilde U_i$ are
\begin{eqnarray}\label{tildeUthreethree}
\tilde U_1&=&{z^3\over(z^1)^2(z^2)^{2}y^2y^4y^5(y^6)^2(y^7)^2},\cr
\tilde U_2&=&{(z^1)^2y^2y^7\over z^2z^3y^3y^4},\cr
\tilde U_3&=&z^1z^2z^3y^1y^2y^3y^4y^5y^6y^7.
\end{eqnarray}
The rescaling that leaves the $\tilde U_i$ invariant is 
\begin{eqnarray}\label{rescalesthreethree}
(z^1,z^2,z^3,y^1,...,y^7) &\to&
(\lambda_1\,z^1,\,\lambda_2\,z^2,\,\lambda_1^2\lambda_2^2\lambda_3\lambda_4\lambda_5\lambda_6^2\lambda_7^2\,z^3,\,{1\over\lambda_1^{3}\lambda_3^2\lambda_5\lambda_6\lambda_7^2}\,y^1,\cr
&& \lambda_3\,y^2,\,{1\over \lambda_2^3\lambda_4^2\lambda_5\lambda_6^2\lambda_7}\,y^3,\lambda_4y^4,\lambda_5y^5,\lambda_6y^6,\lambda_7y^7).
\end{eqnarray}
Thus the blown-up geometry corresponds to
\begin{equation}\label{eq:blowupthree}
X_{\tilde\Sigma}=({\IC}^{10}\setminus F_{\tilde\Sigma})/({\IC}^*)^7.
\end{equation}
The excluded sets differ for the different resolutions. We refrain from giving them explicitly. The action of $({\IC}^{*})^7$ is given by (\ref{rescalesthreethree}).

Let us now give the intersection properties for the two resolutions shown in the figure.

\vskip0.5cm
\noindent{\it Case a)}
\vskip0.5cm

The $3\cdot 3=9$ three-dimensional cones are in this case $(D_3,E_1,E_5),\ (E_1,E_2,E_5)$, $(D_1,E_2,E_5)$, $(D_1,E_5,E_7),\ (E_5,E_6,E_7),\ (D_2,E_5,E_6),\ (D_2,E_4,E_5),\ (E_3,E_4,E_5)$, $(D_3,E_3,E_5)$.
We find nine generators of the Mori cone and write them as columns of $Q$:
\begin{equation}{(P|\,Q)=\left(\begin{array}{cccccccccccccc}
D_1&\!\!\!-2&2&1&|&0&0&1&\!\!\!-1&1&0&0&0&0\cr
D_2&\!\!\!-2&\!\!\!-1&1&|&0&0&0&0&0&1&\!\!\!-1&1&0\cr
D_3&1&\!\!\!-1&1&|&1&\!\!\!-1&0&0&0&0&0&0&1\cr
E_1&0&0&1&|&\!\!\!-2&1&1&0&0&0&0&0&0\cr
E_2&\!\!\!-1&1&1&|&1&0&\!\!\!-2&1&0&0&0&0&0\cr
E_3&0&\!\!\!-1&1&|&0&1&0&0&0&0&0&1&\!\!\!-2\cr
E_4&\!\!\!-1&\!\!\!-1&1&|&0&0&0&0&0&0&1&\!\!\!-2&1\cr
E_5&\!\!\!-1&0&1&|&0&\!\!\!-1&0&\!\!\!-1&0&0&\!\!\!-1&0&0\cr
E_6&\!\!\!-2&0&1&|&0&0&0&0&1&\!\!\!-2&1&0&0\cr
E_7&\!\!\!-2&1&1&|&0&0&0&1&\!\!\!-2&1&0&0&0
\end{array}\right)}\end{equation}
This leads to the following linear equivalences between the divisors:
\begin{eqnarray}\label{lineqthreethree}
0&\sim3&\,D_{{1}}+E_{{1}}+2\,E_{{2}}+E_{{5}}+E_{{6}}+2\,E_{{7}},\cr
0&\sim3&\,D_{{2}}+E_{{3}}+2\,E_{{4}}+E_{{5}}+2\,E_{6}+E_{{7}},\cr
0&\sim3&\,D_{{3}}+2\,E_{{1}}+E_{{2}}+2\,E_{{3}}+E_{{4}}+E_{{5}}.
\end{eqnarray}

From the intersection numbers, we find the following relations between the Mori generators and the nine compact curves of our geometry: $C_1=E_1\cdot E_5,\ C_2=D_3\cdot E_5,\ C_3=E_2\cdot E_5,\ C_4=D_1\cdot E_5,\ C_5=E_5\cdot E_7,\ C_6=E_5\cdot E_6,\ C_7=D_2\cdot E_5, \ E_4\cdot E_5=C_1+C_2+C_4-C_6-2\,C_7,\ E_3\cdot E_5=-C_1-2\,C_2+C_4+C_5+C_6+C_7$. Furthermore, $E_5^3=3$.

\vskip0.5cm
\noindent{\it Case b)}
\vskip0.5cm

Here, we have the following 9 tree-dimensional cones: $(D_1,E_2,E_7),\ (E_5,E_6,E_7)$, $(E_2,E_5,E_7)$,  $(E_1,E_2,E_5),\,(D_2,E_4,E_6),\ (E_4,E_5,E_6),\ (E_3,E_4,E_5),\ (E_1,E_3,E_5),\ (D_3,E_1,E_3)$.
$(P|\,Q)$ takes the following form:
\begin{equation}{(P|\,Q)=\left(\begin{array}{cccccccccccccc}
D_1&\!\!\!-2&2&1&|&1&0&0&0&0&0&0&0&0\cr
D_2&\!\!\!-2&\!\!\!-1&1&|&0&0&0&0&0&1&0&0&0\cr
D_3&1&\!\!\!-1&1&|&0&0&0&0&0&0&0&0&1\cr
E_1&0&0&1&|&0&0&0&1&\!\!\!-1&0&0&1&\!\!\!-1\cr
E_2&\!\!\!-1&1&1&|&\!\!\!-1&1&0&\!\!\!-1&1&0&0&0&0\cr
E_3&0&\!\!\!-1&1&|&0&0&0&0&1&0&1&\!\!\!-1&\!\!\!-1\cr
E_4&\!\!\!-1&\!\!\!-1&1&|&0&0&1&0&0&\!\!\!-1&\!\!\!-1&1&0\cr
E_5&\!\!\!-1&0&1&|&1&\!\!\!-1&\!\!\!-1&\!\!\!-1&\!\!\!-1&1&\!\!\!-1&\!\!\!-1&1\cr
E_6&\!\!\!-2&0&1&|&0&1&\!\!\!-1&0&0&\!\!\!-1&1&0&0\cr
E_7&\!\!\!-2&1&1&|&\!\!\!-1&\!\!\!-1&1&1&0&0&0&0&0\end{array}\right)}\end{equation}
The linear equivalences remain of course the same. The relations between the nine compact curves and the $C_i$ are $C_1=E_2\cdot E_7,\ C_2=E_5\cdot E_7,\ C_3=E_5\cdot E_6,\ C_4=E_2\cdot E_5,\ C_5=E_1\cdot E_5,\ C_6=E_4\cdot E_6,\ C_9=E_1\cdot E_3,\ E_4\cdot E_5=-C_3+C_4+C_5,\ E_3\cdot E_5=C_2+C_3-C_5.$ Here, $E_5^3=6$.

We will now discuss the topologies of the exceptional divisors.  The only compact exceptional divisor is $E_5$. In the triangulation a), the star of $E_5$ corresponds to the whole toric diagram. Unfortunately, one cannot read off the topology directly from the Mori generators. The triangulations a) and b) are connected by three flop transitions: $(E_1,E_3)\to(D_3,E_5),\ (E_2,E_7)\to (D_1,E_5)$ and $(E_4,E_6)\to (D_2,E_5)$. For the triangulation b), $D_1, D_2$ and $D_3$ are not part of the star. Unfortunately, the Mori generators of this star aren't helpful either. If we perform two flop-transitions, we end up in a very simple case. We flop the curve $(E_5, E_7)$ to $(E_2, E_6)$, furthermore, we flop $(E_3, E_5)$ to $(E_1, E_4)$. Thus, we arrive at a star which contains only $E_1, E_2, E_4, E_5, E_6$; it corresponds to an ${\IF}_0$. So both triangulations are birationally equivalent to ${\IF}_0$. Therefore, $h^{(1,0)}=h^{(2,0)}=0$, since the $h^{(p,0)}$ are birational invariants.

In triangulation a), all non-compact exceptional divisors have the topology of $\IP^1\times \IC$. 
In triangulation b), they are all $\IP^1\times \IC$ blown up in one point.

\section{Resolution of ${\IC}^3/(\IZ_{3}\times\IZ_6)$}\label{app:rzthreesix}

$(\IZ_{3}\times\IZ_6)$ acts as follows on ${\IC}^3$:
\begin{eqnarray}\label{twistthreesix}
\theta^1:\ (z^1,\, z^2,\, z^3)& \to& (\varepsilon^2\, z^1, \, z^2, \varepsilon^4\, z^3),\cr
\theta^2:\ (z^1,\, z^2,\, z^3)& \to& (\, z^1, \, \varepsilon\,z^2, \varepsilon^5\, z^3),\cr
\theta^1\theta^2:\ (z^1,\, z^3,\, z^2)& \to& (\varepsilon^2\, z^1, \, \varepsilon\,z^2, \varepsilon^3\, z^3),
\end{eqnarray}
with $\varepsilon=e^{2 \pi i/6}$.
To find the components of the $v_i$, we have to solve 
\begin{eqnarray}
2\,(v_1)_i+4\,(v_3)_i&=&0\ \mod\,6,\cr
(v_2)_i+5\,(v_3)_i&=&0\ \mod\,6,\cr
2\,(v_1)_i+(v_2)_i+3\,(v_3)_i&=&0\ \mod\,6.
\end{eqnarray}
This leads to the following three generators of the fan:
\begin{equation*}{v_1=(1,5,1),\ v_2=(-5,-1,1),\ v_3=(1,-1,1).
}\end{equation*}
To resolve the singularity, we now have many more possibilities for new vertices. We find that $\theta^1,\,(\theta^1)^2,\,\theta^2, \,(\theta^2)^2,\,(\theta^2)^3,\,(\theta^2)^4,\,(\theta^2)^5,\,\theta^1\theta^2, \,\theta^1(\theta^2)^2,\,\theta^1(\theta^2)^3,\,\theta^1(\theta^2)^4,\,(\theta^1)^2\theta^2$  and $(\theta^1)^1(\theta^2)^2$ fulfill (\ref{eq:criterion}). This leads to thirteen new generators:
\begin{eqnarray*}
w_1&=&(1,1,1),\ w_2=(1,3,1),\ w_3=(0,-1,1),\ w_4=(-1,-1,1),\ w_5=(-2,-1,1),\cr
w_6&=&(-3,-1,1),\ w_7=(-4,-1,1),\ w_8=(0,1,1),\ w_9=(-1,1,1),\ w_{10}=(-2,1,1),\cr
w_{11}&=&(-3,1,1),\ w_{12}=(0,3,1),\ w_{13}=(-1,3,1)
.\end{eqnarray*}
In this case, there are 14303 possible triangulations\footnote{This number was obtained using the package TOPCOM.}.
\begin{figure}[h!]
\begin{center}
\includegraphics[width=140mm]{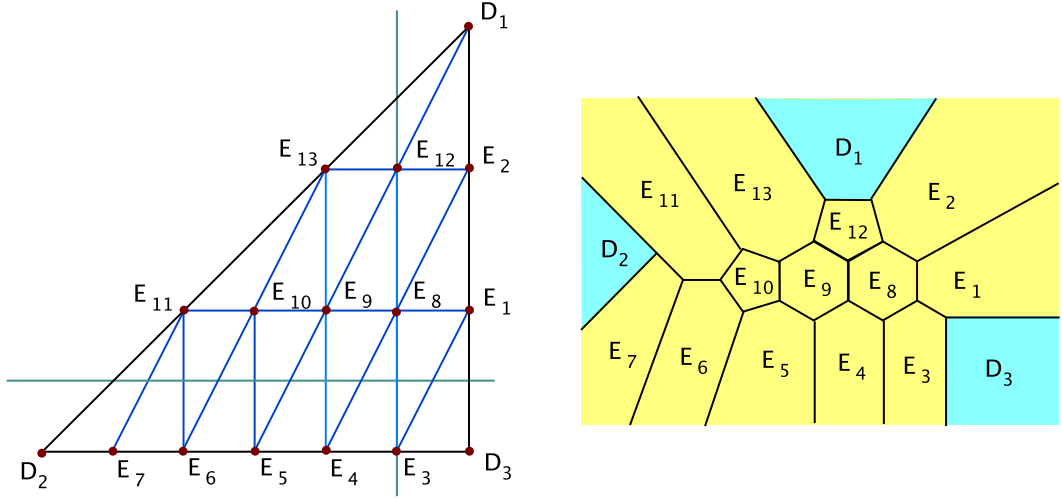}
\caption{Toric diagram of one of the resolutions of ${\IC}^3/(\IZ_{3}\times\IZ_6)$ and dual graph}\label{frthreesix}
\end{center}
\end{figure}
Figure \ref{frthreesix} shows one of them.

Let us identify the blown--up geometry. The $\tilde U_i$ are
\begin{eqnarray}\label{tildeUthreesix}
\tilde U_1&=&{z^1z^3y^1y^2\over(z^2)^{5}y^4(y^5)^2(y^6)^3(y^7)^4y^9(y^{10})^2(y^{11})^3},\cr\tilde U_2&=&{(z^1)^5y^1(y^2)^3y^8y^9y^{10}y^{11}(y^{12})^3(y^{13})^3\over z^2z^3y^3y^4y^5y^6y^7},\cr
\tilde U_3&=&z^1z^2z^3y^1\ldots y^{13}.
\end{eqnarray}
The rescaling that leaves the $\tilde U_i$ invariant is 
\begin{eqnarray}\label{rescalesthreesix}
&&(z^1,z^2,z^3,y^1,\ldots,y^{13}) \to \cr
&&(\lambda_1\,z^1,\,\lambda_2\,z^2,\,{\lambda_2^5\lambda_5\lambda_6^2\lambda_7^3\lambda_8^4\lambda_9\lambda_{10}^2\lambda_{11}^3\lambda_{13}\over\lambda_1\lambda_3\lambda_4}\,z^3,\,{\lambda_3}\,y^1, \lambda_4\,y^2,-{\lambda_1^3\lambda_3\lambda_4^2\lambda_{12}\over \lambda_2^6\lambda_5^2\lambda_6^3\lambda_7^4\lambda_8^5\lambda_9\lambda_{10}^2\lambda_{11}^3}\,y^3,\cr
&&\lambda_5\,y^4,\lambda_6\,y^5,{\lambda_7}\,y^6,\lambda_8\,y^7,\,{-1\over\lambda_1^3\lambda_3\lambda_4^2\lambda_9\lambda_{10}\lambda_{11}\lambda_{12}^2\lambda_{13}^2}\,y^8, \lambda_9\,y^9,...,\lambda_{13}\,y^{13}).
\end{eqnarray}
Thus the blown-up geometry corresponds to
\begin{equation*}{
X_{\tilde\Sigma}=({\IC}^{16}\setminus F_{\tilde\Sigma})/({\IC}^*)^{13}.
}\end{equation*}
The excluded sets differ for the different resolutions. We refrain from giving them explicitly. The action of $({\IC}^{*})^{13}$ is given by (\ref{rescalesthreesix}).

There are $3\cdot 6=18$ three-dimensional cones.
We find 17 generators of the Mori cone and write them as columns of $Q$:
\begin{equation}{Q=\left({\small\begin{array}{ccccccccccccccccc}
 1& 0& 0& 0& 0& 0& 0& 0& 0& 0& 0& 0& 0& 0& 0& 0& 0\cr
 0& 0& 0& 0& 0& 0& 0& 0& 0& 0& 0& 0& 0& 0& 0& 0& 1\cr
0&0&0&0&0&0&0&0&1&0&0&0&0&0&0&0&0\cr
0&1&-1&0&0&0&0&0&-1&1&0&0&0&0&0&0&0\cr
-1&-1&1&1&0&0&0&0&-1&1&0&0&0&0&0&0&0\cr
0&0&1&0&0&0&0&0&-1&-1&1&0&0&0&0&0&0\cr
0&0&0&0&0&1&0&0&0&1&-1&-1&1&0&0&0&0\cr
0&0&0&0&0&0&0&1&0&0&0&1&-1&-1&1&0&0\cr
0&0&0&0&0&0&0&0&0&0&0&0&0&1&-1&-1&1\cr
0&0&0&0&0&0&0&0&0&0&0&0&0&0&0&1&-2\cr
1&-1&-1&-1&1&-1&0&0&1&-1&-1&1&0&0&0&0&0\cr
0&0&0&1&-1&-1&-1&-1&0&0&1&-1&-1&1&0&0&0\cr
0&0&0&0&0&0&1&-1&0&0&0&0&1&-1&-1&1&0\cr
0&0&0&0&0&0&0&0&0&0&0&0&0&0&1&-1&0\cr
-1&1&0&-1&-1&1&1&0&0&0&0&0&0&0&0&0&0\cr
0&0&0&0&1&0&-1&1&0&0&0&0&0&0&0&0&0
\end{array}}\right).}\end{equation}
This leads to the following linear relations:
\begin{eqnarray}\label{lineqthreesix}
0&\sim&3\,D_{{1}}+E_1+2\,E_{{2}}+E_8+E_{{9}}+E_{{10}}+E_{{11}}+2\,E_{{12}}+E_{{13}},\cr 
0 &\sim&6\,D_{{2}}+E_{{3}}+2\,E_{{4}}+3\,E_{{5}}+4\,E_{{6}}+5\,E_{{7}}+E_{{8}}+2\,E_{{9}}+3\,E_{{10}}\cr
&&+4\,E_{{11}}+E_{{12}}+2\,E_{{13}},\cr 
0&\sim&6\,D_{{3}}+4\,E_{{1}}+2\,E_{{2}}+5\,E_{{3}}+4\,E_{{4}}+3\,E_{{5}}+2\,E_{{6}
}+E_{{7}}+3\,E_{{8}}\cr
&&+2\,E_{{9}}+E_{{10}}+E_{{12}}.
\end{eqnarray}
From the intersection numbers, we find the following relations between the Mori gen- 
erators and the twenty-one compact curves of our geometry:
\begin{eqnarray}\label{curverel}
C_1&=&E_2\cdot E_{12},\ C_2=E_2\cdot E_8,\ C_3=E_1\cdot E_8,\ C_4=E_8\cdot E_{12},\ C_5=E_9\cdot E_{12}\cr
C_6&=&E_8\cdot E_9,\ C_7=E_9\cdot E_{13},\ C_8=E_9\cdot E_{10},\ C_9=E_1\cdot E_3,\ C_{14}=E_5\cdot E_{10},\cr
C_{15}&=&E_6\cdot E_{10},\ C_{16}=E_6\cdot E_{11},\ C_{17}=E_7\cdot E_{11},\ D_1\cdot E_{12}=C_4+C_5,\cr
&& E_{12}\cdot E_{13}=C_1+C_4,\ E_{10}\cdot E_{13}=C_{14}+C_{15},\ E_{10}\cdot E_{11}=C_8+C_{14},\cr
&& E_3\cdot E_8=-C_3+C_4+C_6,\ E_4\cdot E_8=C_2+C_3-C_6,\ 
E_4\cdot E_9=-C_6+C_7+C_8,\cr
&& E_5\cdot E_9=C_5+C_6-C_8.
\end{eqnarray}
Furthermore, $E_8^3=E_9^3=6,\ E_{10}^3=E_{12}^3=7$.
Let us now discuss the topologies of the exceptional divisors.
We find the compact exceptional divisors, after performing one, respectively two flops,  to be birationally equivalent to $\IF_1$.
The non-compact exceptional divisors $E_1,...,E_6$ are $\IP^1\times \IC$s after one flop, $E_7$ is a $\IP^1$ and $E_{11}, \, E_{12}$ are $\IP^1\times \IC$ blown up in two points.

\section{Resolution of ${\IC}^3/(\IZ_{4}\times\IZ_4)$}\label{app:rzfourfour}

$(\IZ_{4}\times\IZ_4)$ acts as follows on ${\IC}^3$:
\begin{eqnarray}\label{twistfourfour}
\theta^1:\ (z^1,\, z^2,\, z^3)& \to& (\varepsilon\, z^1, \, z^2, \varepsilon^3\, z^3),\cr
\theta^2:\ (z^1,\, z^2,\, z^3)& \to& (\, z^1, \, \varepsilon\,z^2, \varepsilon^3\, z^3),\cr
\theta^1\theta^2:\ (z^1,\, z^3,\, z^2)& \to& (\varepsilon\, z^1, \, \varepsilon\,z^2, \varepsilon^2\, z^3),
\end{eqnarray}
with $\varepsilon=e^{2 \pi i/4}$.
To find the components of the $v_i$, we have to solve 
\begin{eqnarray}
(v_1)_i+3\,(v_3)_i&=&0\ \mod\,4,\cr
(v_2)_i+3\,(v_3)_i&=&0\ \mod\,4,\cr
(v_1)_i+(v_2)_i+2\,(v_3)_i&=&0\ \mod\,4.
\end{eqnarray}
This leads to the following three generators of the fan:
\begin{equation*}{v_1=(1,-3,1),\ v_2=(-3,1,1),\ v_3=(1,1,1).
}\end{equation*}
To resolve the singularity, we now have many more possibilities for new vertices. We find that $\theta^1,\,(\theta^1)^2,\,(\theta^1)^3,\,\theta^2, \,(\theta^2)^2,\,(\theta^2)^3,\,\theta^1\theta^2, \,\theta^1(\theta^2)^2,\,\theta^1(\theta^2)^3,\,(\theta^1)^2\theta^2,\,(\theta^1)^3\theta^2$  and $(\theta^1)^2(\theta^2)^2$ fulfill (\ref{eq:criterion}). This leads to twelve new generators:
\begin{eqnarray*}
w_1&=&(1,0,1),\ w_2=(1,-1,1),\ w_3=(1,-2,1),\ w_4=(0,1,1),\ w_5=(-1,1,1),\cr
w_6&=&(-2,1,1),\ w_7=(0,0,1),\ w_8=(-1,0,1),\ w_9=(-2,0,1),\ w_{10}=(0,-1,1),\cr
w_{11}&=&(0,-2,1),\ w_{12}=(-1,-1,1)
.\end{eqnarray*}
In this case, there are 7424 possible triangulations\footnote{This number was obtained using the package TOPCOM.}.
\begin{figure}[h!]
\begin{center}
\includegraphics[width=120mm]{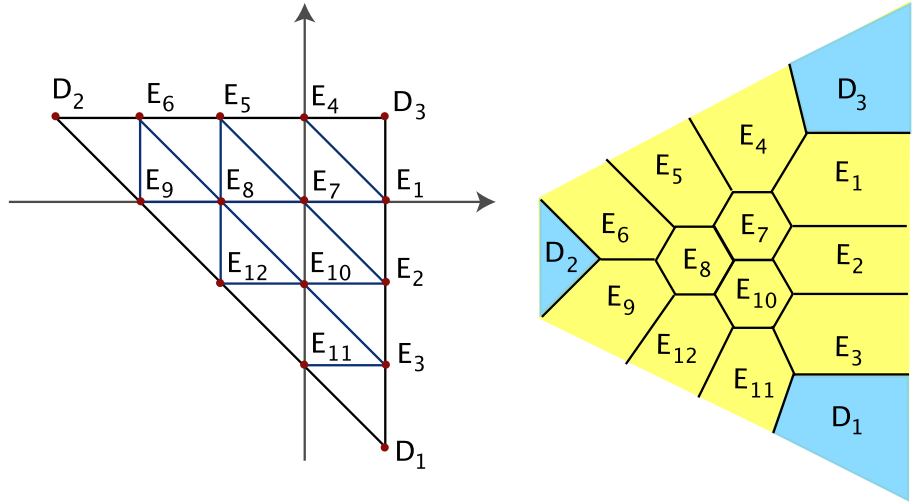}
\caption{Toric diagram of one of the resolutions of ${\IC}^3/(\IZ_{4}\times\IZ_4)$ and dual graph}\label{frfourfour}
\end{center}
\end{figure}
Figure \ref{frfourfour} shows one of them.

Let us identify the blown--up geometry. The $\tilde U_i$ are
\begin{eqnarray}\label{tildeUfourfour}
\tilde U_1&=&{z^1z^3y^1y^2y^3\over(z^2)^{3}y^5(y^6)^2y^8(y^9)^2y^{12}},\quad \tilde U_2={z^2z^3y^4y^5y^6\over (z^1)^3y^2(y^3)^2y^{10}(y^{11})^2y^{12}},\cr
\tilde U_3&=&z^1z^2z^3y^1\ldots y^{12}.
\end{eqnarray}
The rescaling that leaves the $\tilde U_i$ invariant is 
\begin{eqnarray}\label{rescalesfourfour}
&&(z^1,z^2,z^3,y^1,\ldots,y^{13}) \to \cr
&&(\lambda_1\,z^1,\,\lambda_2\,z^2,\,{\lambda_3}\,z^3,\,{\lambda_2^3\lambda_6\lambda_7^2\lambda_8\lambda_9^2\lambda_{12}\over\lambda_1\lambda_3\lambda_4\lambda_5}\,y^1, \lambda_4\,y^2,\lambda_5\,y^3,{\lambda_1^3\lambda_4\lambda_5\lambda_{10}\lambda_{11}^2\lambda_{12}\over\lambda_2\lambda_3\lambda_6\lambda_7}\,y^4, \lambda_6\,y^5,\cr
&&{\lambda_7}\,y^6,{\lambda_3\over \lambda_1^3\lambda_2^3\lambda_4\lambda_5^2\lambda_6\lambda_7^2\lambda_8^2\lambda_9^3\lambda_{10}^2\lambda_{11}^3\lambda_{12}^3}\,y^7,\lambda_8\,y^8, \lambda_9\,y^9,\lambda_{10}\,y^{10},\lambda_{11}\,y^{11},\lambda_{12}\,y^{12}).
\end{eqnarray}
Thus the blown-up geometry corresponds to
\begin{equation*}{
X_{\tilde\Sigma}=({\IC}^{15}\setminus F_{\tilde\Sigma})/({\IC}^*)^{12}.
}\end{equation*}
The excluded sets differ for the different resolutions. We refrain from giving them explicitly. The action of $({\IC}^{*})^{12}$ is given by (\ref{rescalesfourfour}).

There are $4\cdot 4=16$ three-dimensional cones.
We find eighteen generators of the Mori cone and write them as columns of $Q$:
\begin{equation}{Q=\left(\begin{array}{cccccccccccccccccc}
 0& 0& 0& 0& 0& 0& 0& 0& 0& 0& 0& 0& 0& 0& 0& 0& 0&1\cr
 0& 0& 0& 0& 0& 0& 0& 1& 0& 0& 0& 0& 0& 0& 0& 0& 0&0\cr
1&0&0&0&0&0&0&0&0&0&0&0&0&0&0&0&0&0\cr
\!\!\!-1&1&\!\!\!-1&0&0&0&0&0&0&1&0&0&0&0&0&0&0&0\cr
0 &0&1  &0&0&0&0&0&0&\!\!\!-1&1&\!\!\!-1&0&0&0&1&0&0\cr
0&0&0   &0&0&0&0&0&0& 0&0& 1 &0&0&0&\!\!\!-1&1&\!\!\!-1\cr
\!\!\!-1&\!\!\!-1&1&1&0&0&0&0&0&0&0&0&0&0&0&0&0&0\cr
0 & 1&0&\!\!\!-1&\!\!\!-1&1&1&0&0&0&0&0&0&0&0&0&0&0\cr
0 & 0&0& 0&1&0&\!\!\!-1&\!\!\!-1&1&0&0&0&0&0&0&0&0&0\cr
1&\!\!\!-1&\!\!\!-1&\!\!\!-1&1&\!\!\!-1&0&0&0&\!\!\!-1&\!\!\!-1&1&1&0&0&0&0&0\cr
0& 0& 0 & 1&\!\!\!-1&\!\!\!-1&\!\!\!-1&1&\!\!\!-1&0&1&0&\!\!\!-1&\!\!\!-1&1&0&0&0\cr
0&0&0&0&0&0&1&\!\!\!-1&\!\!\!-1&0&0&0&0&1&0&0&0&0\cr
0&0&0&0&0&1&0&0&0&1&\!\!\!-1&\!\!\!-1&\!\!\!-1&1&\!\!\!-1&\!\!\!-1&\!\!\!-1&1\cr
0&0&0&0&0&0&0&0&0&0&0&0&0&0&1&1&\!\!\!-1&\!\!\!-1\cr
0&0&0&0&0&0&0&0&1&0&0&0&1&\!\!\!-1&\!\!\!-1&0&1&0
\end{array}\right)}\end{equation}
This leads to the following linear relations:
\begin{eqnarray}\label{lineqfourfour}
0&\sim&4\,D_{{1}}+E_1+2\,E_{{2}}+3\,E_3+E_{{7}}+E_{{8}}+E_{{9}}+2\,E_{{10}}+3\,E_{{11}}+2\,E_{12},\cr 
0 &\sim&4\,D_{{2}}+E_{{4}}+2\,E_{{5}}+3\,E_{{6}}+E_{{7}}+2\,E_{{8}}+3\,E_{{9}}+E_{{10}}+E_{{11}}+2\,E_{{12}}\cr 
0&\sim&4\,D_{{3}}+3\,E_{{1}}+2\,E_{{2}}+E_{{3}}+3\,E_{{4}}+2\,E_{{5}}+E_{{6}
}+2\,E_{{7}}+E_{{8}}+E_{{10}}.
\end{eqnarray}
From the intersection numbers, we find the following relations between the Mori gen- 
erators and the eighteen compact curves of our geometry:
\begin{eqnarray}\label{curverelfourfour}
&&C_1=E_1\cdot E_{4},\ C_2=E_4\cdot E_7,\ C_3=E_1\cdot E_7,\ C_4=E_5\cdot E_{7},\ C_5=E_5\cdot E_{8},\cr
&&C_6=E_7\cdot E_8,\ C_7=E_6\cdot E_{8},\ C_8=E_6\cdot E_{9},\ C_9=E_8\cdot E_9,\ C_{12}=E_2\cdot E_{9},\cr 
&&C_{15}=E_{10}\cdot E_{12},\ C_{18}=E_3\cdot E_{11},\ 
E_2\cdot E_{7}=-C_3+C_4+C_6,\cr
&&\ E_{3}\cdot E_{10}=-C_6+C_{7}+C_9-C_{12}+C_{15},\cr
&& E_{8}\cdot E_{10}=-C_6+C_7+C_9,\ E_{8}\cdot E_{12}=C_{5}+C_{6}-C_9,\cr
&&E_7\cdot E_{10}=C_2+C_3-C_6,\ E_{10}\cdot E_{11}=C_2+C_3-C_6+C_{12}-C_{15}.
\end{eqnarray}
Furthermore, $E_7^3=E_8^3=E_{10}^3=6$.
Let us now discuss the topologies of the exceptional divisors.
We find the compact exceptional divisors, i.e. $E_7,\,E_8,\,E_{10}$, after performing two flops,  to be birationally equivalent to $\IF_1$.
The non-compact exceptional divisors $E_1,...,E_6$ are $\IP^1\times \IC$s after with one bow-up, $E_9,\,E_{11},\,E_{12}$ are $\IP^1\times \IC$ after one flop.

\section{Resolution of ${\IC}^3/(\IZ_{6}\times\IZ_6)$}\label{app:rzsixsix}

$(\IZ_{6}\times\IZ_6)$ acts as follows on ${\IC}^3$:
\begin{eqnarray}\label{twistsixsix}
\theta^1:\ (z^1,\, z^2,\, z^3)& \to& (\varepsilon^1\, z^1, \, z^2, \varepsilon^5\, z^3),\cr
\theta^2:\ (z^1,\, z^2,\, z^3)& \to& (\, z^1, \, \varepsilon\,z^2, \varepsilon^5\, z^3),\cr
\theta^1\theta^2:\ (z^1,\, z^3,\, z^2)& \to& (\varepsilon\, z^1, \, \varepsilon\,z^2, \varepsilon^4\, z^3),
\end{eqnarray}
with $\varepsilon=e^{2 \pi i/6}$.
To find the components of the $v_i$, we have to solve 
\begin{eqnarray}
(v_1)_i+5\,(v_3)_i&=&0\ \mod\,6,\cr
(v_2)_i+5\,(v_3)_i&=&0\ \mod\,6,\cr
(v_1)_i+(v_2)_i+4\,(v_3)_i&=&0\ \mod\,6.
\end{eqnarray}
This leads to the following three generators of the fan:
\begin{equation*}{v_1=(5,-6,1),\ v_2=(5,6,1),\ v_3=(-1,0,1).
}\end{equation*}
To resolve the singularity, we have many possibilities for new vertices. We find that $\theta^1,...,\,(\theta^1)^5,\,\theta^2, ... ,\,(\theta^2)^5,\,\theta^1\theta^2, ...,(\theta^1)^5\theta^2,\,\theta^1(\theta^2)^2,...\,\theta^1(\theta^2)^5$, $(\theta^1)^2(\theta^2)^2,...,\,(\theta^1)^2(\theta^2)^4$,\  $(\theta^1)^3(\theta^2)^2$, $ (\theta^1)^3(\theta^2)^3$  and $(\theta^1)^4(\theta^2)^2$ fulfill (\ref{eq:criterion}). This leads to 25 new generators:
\begin{eqnarray*}
w_1&=&(0,-1,1),\ w_2=(1,-2,1),...,\ w_5=(4,-5,1),\ w_6=(0,1,1),\ w_7=(1,2,1),...,\cr
w_{10}&=&(4,5,1),\ w_{11}=(1,0,1),\ w_{12}=(2,-1,1),...,\, w_{15}=(5,-4,1),\ w_{16}=(2,1,1),\cr
w_{17}&=&(3,1,1),...,\, w_{19}=(5,4,1),\ w_{20}=(3,0,1),...,\,w_{22}=(5,2,1),\ w_{23}=(4,-1,1),\cr
w_{24}&=&(5,0,1),\ w_{25}=(5,-2,1)
.\end{eqnarray*}

\begin{figure}[h!]
\begin{center}
\includegraphics[width=140mm]{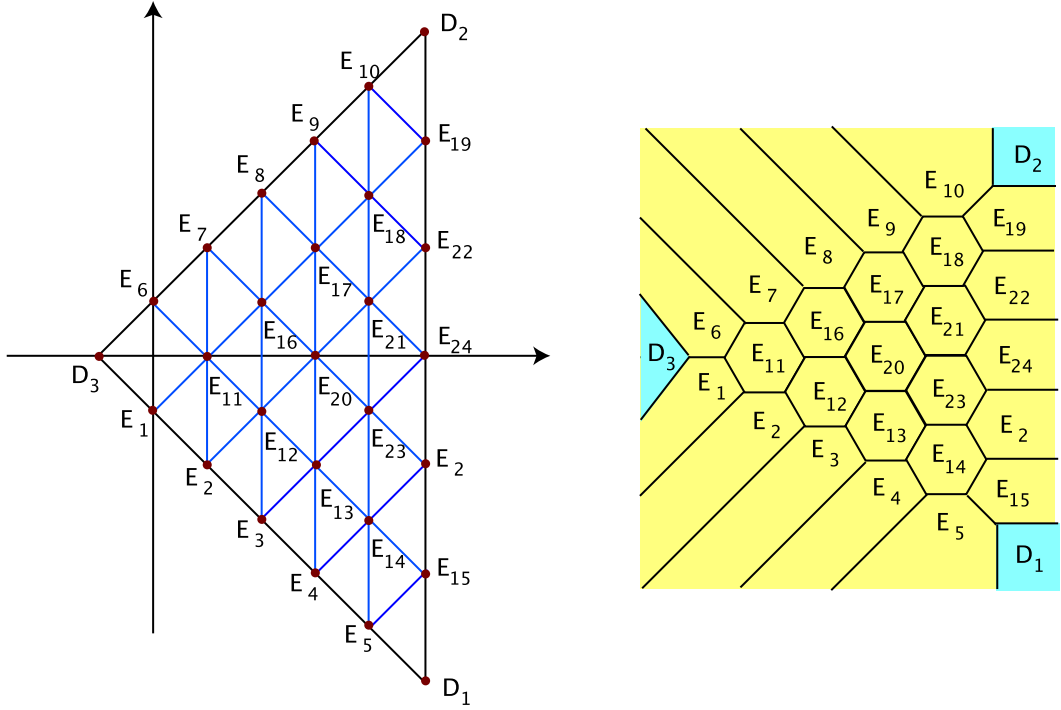}
\caption{Toric diagram of one of the resolutions of ${\IC}^3/\IZ_{6}\times\IZ_{6}$ and dual graph}\label{frsixsix}
\end{center}
\end{figure}
In this case, there are several thousands of distinct triangulations. A very simple one is shown in Figure \ref{frsixsix}.

Let us identify the blown--up geometry. The $\tilde U_i$ are
\begin{eqnarray}\label{tildeUsixsix}
\tilde U_1&=&(z^1)^5(z^2)^5(z^3)^{-1}y^2(y^3)^2(y^4)^3(y^5)^4y^7(y^8)^2(y^9)^3(y^{10})^4y^{11}(y^{12})^2(y^{13})^3(y^{14})^4(y^{15})^5\cr
&&(y^{16})^2(y^{17})^3(y^{18})^4(y^{19})^5(y^{20})^3(y^{21})^4(y^{22})^5(y^{23})^4(y^{24})^5(y^{25})^5,\cr
\tilde U_2&=&{(z^2)^6y^6(y^7)^2(y^8)^3(y^9)^4(y^{10})^5y^{16}(y^{17})^2(y^{18})^3(y^{19})^4y^{21}(y^{22})^2\over (z^1)^6y^1(y^2)^2(y^3)^3(y^4)^4(y^5)^5y^{12}(y^{13})^2(y^{14})^3(y^{15})^4y^{23}(y^{25})^2},\cr
\tilde U_3&=&z^1z^2z^3y^1\ldots y^{25}.
\end{eqnarray}
The rescaling that leaves the $\tilde U_i$ invariant is 
\begin{eqnarray}\label{rescalessixsix}
&&(z^1,z^2,z^3,y^1,\ldots,y^{13}) \to \cr
&&(\lambda_1\,z^1,\,\lambda_2\,z^2,\,{\lambda_1^5\lambda_2^5\lambda_3\lambda_4^2\lambda_5^3\lambda_6^4\lambda_7\lambda_8^2\lambda_9^3\lambda_{10}^4\lambda_{11}\lambda_{12}^2\lambda_{13}^3\lambda_{14}^4\lambda_{15}^5\lambda_{16}^2\lambda_{17}^3\lambda_{18}^4\lambda_{19}^5\lambda_{20}^3\lambda_{21}^4\lambda_{22}^5\lambda_{23}^4\lambda_{24}^5\lambda_{25}^5}\,z^3,\cr
&&-(\lambda_1^6\lambda_3^2\lambda_4^3\lambda_5^4\lambda_6^5\lambda_{11}\lambda_{12}^2\lambda_{13}^3\lambda_{14}^4\lambda_{15}^5\lambda_{16}\lambda_{17}\lambda_{18}\lambda_{19}\lambda_{20}^2\lambda_{21}^2\lambda_{22}^2\lambda_{23}
^3\lambda_{24}^3\lambda_{25}^4)^{-1}\,y^1, \lambda_3\,y^2,\lambda_4\,y^3, \lambda_5\,y^4,\cr
&&\lambda_6\,y^5,-(\lambda_2^6\lambda_7^2\lambda_8^3\lambda_9^4\lambda_{10}^5\lambda_{11}\lambda_{12}\lambda_{13}\lambda_{14}\lambda_{15}\lambda_{16}^2\lambda_{17}^3\lambda_{18}^4\lambda_{19}^5\lambda_{20}^2\lambda_{21}^3\lambda_{22}^4\lambda_{23}^2\lambda_{24}^3\lambda_{25}^2)^{-1}\,y^6,\lambda_7\,y^7,\,\lambda_8\,y^8,\cr
&& \lambda_9\,y^9,\lambda_{10}\,y^{10},\lambda_{11}\,y^{11},\lambda_{12}\,y^{12},\lambda_{13}\,y^{13}).
\end{eqnarray}
Thus the blown-up geometry corresponds to
\begin{equation*}{
X_{\tilde\Sigma}=({\IC}^{28}\setminus F_{\tilde\Sigma})/({\IC}^*)^{25}.
}\end{equation*}
The excluded sets differ for the different resolutions. We refrain from giving them explicitly. The action of $({\IC}^{*})^{25}$ is given by (\ref{rescalessixsix}).

There are $6\cdot 6=36$ three-dimensional cones.
We find 45 generators of the Mori cone, in fact, all 45 relations of type (\ref{eq:Moripairs}) turn out to be Mori generators, none can be eliminated. For the sake of brevity, we refrain from writing them down.

We find the following linear equivalences between the divisors:
\begin{eqnarray}\label{linrelsixsix}
0&\sim&\,6\,D_1+E_1+ 2\,E_2+ 3\,E_3+ 4\,E_4+ 5\,E_5,+E_{11}+2\,E_{12}+ 3\,E_{13} +4\,E_{14}+ 5\,E_{15}\cr
&&+E_{16}+E_{17}+E_{18}+E_{19}+ 2\,E_{20}+ 2\,E_{21}+ 2\,E_{22}+3\,E_{23}+ 3\,E_{24}+ 4\,E_{25},\cr
0&\sim&\,6\,D_2+E_6+2\,E_7,+3\,E_8+ 4\,E_9+ 5\,E_{10}+ E_{11}+E_{12}+E_{13}+E_{14}+E_{15}\cr
&&+ 2\,E_{16} +3\,E_{17}+ 4\,E_{18}+ 5\,E_{19}+ 2\,E_{20}+ 3\,E_{21}+4\,E_{22}+
2\,E_{23}+ 3\,E_{24}+ 2\,E_{25},\cr
0&\sim&\,6\,D_3+ 5\,E_1+ 4\,E_2+ 3\,E_3+ 2\,E_4+E_5+ 5\,E_6+ 4\,E_7+ 3\,E_8+ 2\,E_9+E_{10}\cr
&&+ 4\,E_{11}+ 3\,E_{12}+ 2\,E_{13} +E_{14}+ 3\,E_{16}+ 2\,E_{17}+E_{18}+ 2\,E_{20}+E_{21}+E_{23}.\end{eqnarray}
For all compact exceptional divisors we find $E^3=6$.

We will now discuss the divisor topologies. The compact exceptional divisors all have the same stars, therefore the same topologies. After two flop transitions, they are $\IF_0$s. The non-compact exceptional divisors are all $\IP^1\times \IC$ after one flop.

\section{Resolution of ${\IC}^2/\IZ_n$-type orbifolds}\label{app:rzctwo}

\begin{figure}[h!]
\begin{center}
\includegraphics[width=85mm]{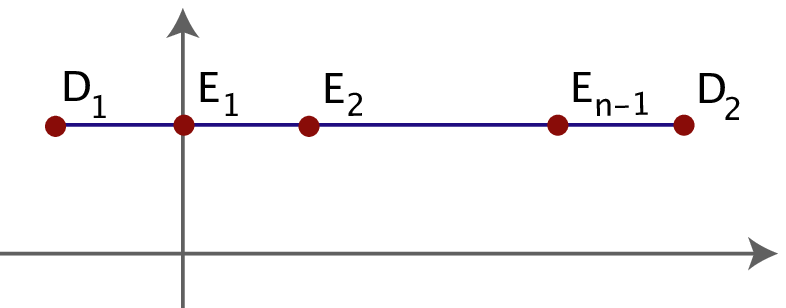}
\caption{Toric diagram of the resolution of  $\IC^2/\IZ_n$}\label{fig:ctwo}
\end{center}
\end{figure}

Here, we treat the patches of the form ${\IC}^2/\IZ_n$, associated to fixed lines. We will be rather brief. As mentioned already in the main text, these singularities are rational double points of type 
$A_{n-1}$.
The action of $\IZ_n$ on ${\IC}^2$ is:
\begin{equation}\label{twistctwo}{\theta:\ (z^1,\, z^2) \to (\varepsilon\, z^1, \varepsilon^{n-1}\, z^2),\quad \varepsilon=e^{2 \pi i/n}.
}\end{equation}
To find the components of the $v_i$, we have to solve $(v_1)_i+(n-1)\,(v_2)_i=0\ \mod\,n$. This leads to the following two generators of the fan:
\begin{equation}{v_1=(n-1,1),\quad v_2=(-1,1).
}\end{equation}
All $\theta^i$, $i=1,...,n-1$ fulfill (\ref{eq:criterion}), so we get $n-1$ new generators:
\begin{equation}{
w_i=\frac{i}{n}v_1 + \frac{n-i}{n}v_2 = (i-1,1), \qquad i=1,\dots,n-1}\end{equation}
The $\tilde U_i$ are
 \begin{equation}\label{tildeUctwo}{\tilde U_1={(z^1)^{n-1}y^2(y^3)^2...\,(y^{n-1})^{n-2}\over z^2},\quad \tilde U_2=z^1z^2y^1y^2...\,y^{n-1}.
 }\end{equation}
 The toric diagram of the resolution (Hirzebruch-Jung sphere tree) of $\IC^2/\IZ_n$ can be found in Figure \ref{fig:ctwo}. 

 The rescaling that leaves the $\tilde U_i$ invariant is 
 \begin{eqnarray}\label{rescalesctwo}
&& (z^1,z^2,y^1,y^2, ..., y^{n-1}) \to \cr
 &&(\lambda_1\,z^1,\,\lambda_1^{n-1}\lambda_2\lambda_3^2...\lambda_{n-1}^{n-2}\,z^2,\,(\lambda_1^{n}\lambda_2^{2}\lambda_3^3...\lambda_{n-1}^{n-1})^{-1}\,y^1,\lambda_2\,y^2, \lambda_3\,y^3,...,\lambda_{n-1}\,y^{n-1}).
 \end{eqnarray}
The blown--up geometry is
\begin{equation}\label{blowupcn}{
X_{\tilde\Sigma}=({\IC}^{n+1}\setminus F_{\tilde\Sigma})/\left({\IC}^*\right)^{n-1},
}\end{equation}
where the $\left({\IC}^{*}\right)^{n-1}$ action is determined in the following $(P|Q)$--matrix
\begin{equation}{(P\,|\,Q)=\left(\begin{array}{ccccccccccc}
D_1    &n-1&1&|& 1& 0&0&...&0& 0& 0\cr
E_1    &  0&1&|&-2& 1&0&...&0& 0& 0\cr
E_2    &  1&1&|& 1&-2&1&...&0& 0& 0\cr
...    &   &1&|&  &  & &...& &  &  \cr
E_{n-2}&  1&1&|& 0& 0&0&...&1&-2& 1\cr
E_{n-1}&  1&1&|& 0& 0&0&...&0& 1&-2\cr
D_2    & -1&1&|& 0& 0&0&...&0& 0& 1\cr
\end{array}\right)}\end{equation}
We observe that the Q matrix is nothing but the Cartan matrix for $A_{n-1}$. With it, we obtain the following linear equivalences:
\begin{equation}\label{eq:lineqsctwo}{E_1 \sim 2D_1, \qquad 2E_i \sim E_{i-1} + E_{i+1}, i=2,...,n-2, \qquad E_{n-1} \sim 2D_2}\end{equation}


\chapter{Calabi--Yau manifolds from resolved orbifolds}

In this appendix, the details of all models listed in Tables \ref{table:one} and \ref{table:two} are collected.
The metric, antisymmetric tensor, complex structure, parametrization of the K\"ahler moduli as well as the fixed set configuration and a description of how the resolved patches must be glued is given for all models. For a few selected examples, also the intersection ring, divisor topologies and orientifold of the resolution are worked out.


\section{The $\IZ_3$--orbifold}

\subsection{Metric, complex structure and moduli}

The $\IZ_3$-orbifold lives on the root lattice of $SU(3)^3$ and acts on it as follows:
\begin{eqnarray}
Q\ e_1&=&e_2,\quad Q\ e_2=-e_1-e_2,\quad Q\ e_3=e_4,\cr 
Q\ e_4&=&-e_3-e_4,\quad Q\ e_5=e_6 ,\quad Q\ e_6=-e_5-e_6\ .\end{eqnarray}
$Q^tg\,Q=g$ and $Q^tb\,Q=b$ lead to:
{\arraycolsep2pt
\begin{equation}{
g\!=\!\left(\!\!\begin{array}{cccccc}
R_1^2&-\frac{1}{2}R_1^2&R_1R_2\cos\theta_{13}&x&R_1R_3\cos\theta_{26}&y\cr
-\frac{1}{2}R_1^2&R_1^2&R_1R_2\cos\theta_{23}&R_1R_2\cos\theta_{13}&R_1R_2\cos\theta_{25}&R_1R_2\cos\theta_{26}\cr
R_1R_2\cos\theta_{13}&R_1R_2\cos\theta_{23}&R_2^2&-\frac{1}{2}R_2^2&R_2R_3\cos\theta_{46}&z\cr
x&R_1R_2\cos\theta_{13}&-\frac{1}{2}R_2^2&R_2^2&R_2R_3\cos\theta_{45}&R_2R_3\cos\theta_{46}\cr
R_1R_3\cos\theta_{26}&R_1R_3\cos\theta_{25}&R_2R_3\cos\theta_{46}&R_2R_3\cos\theta_{45}&R_3^3&-\frac{1}{2}R_3^2\cr
y&R_1R_3\cos\theta_{26}&z&R_1R_3\cos\theta_{46}&-\frac{1}{2}R_3^2&R_3^2\end{array}\!\!\right),
}\end{equation}}
with $x=-R_1R_2(\cos\theta_{13}+\cos\theta_{23})$, $y=-R_1R_3\,(\cos\theta_{25}+\cos\theta_{26})$, $z=-R_2R_3(\cos\theta_{45}+\cos\theta_{46})$ and the nine real parameters $R_1^2,\ R_2^2,\ R_3^2,\ \theta_{13},\ \theta_{23}$, $\theta_{25},\ \theta_{26},\ \theta_{45},\ \theta_{46}$.  
For $b$ we find
\begin{equation}{
b=\left(\begin{array}{cccccc}
0&b_1&b_5&-b_4-b_5&b_7&-b_6-b_7\cr
-b_1&0&b_4&b_5&b_6&b_7\cr
-b_5&-b_4&0&b_2&b_9&-b_8-b_9\cr
b_4+b_5&-b_5&-b_2&0&b_8&b_9\cr
-b_7&-b_6&-b_9&-b_8&0&b_3\cr
b_6+b_7&-b_7&b_8+b_9&-b_9&-b_3&0\end{array}\right)}\end{equation}

Following \ref{ansatz} we introduce complex structures and the 
complex coordinates $z^1,z^2,z^3$
\begin{eqnarray}
z^1&=&{3^{-1/4}}\ (x^1+e^{2\pi i/3}\,x^2),\\
z^2&=&{3^{-1/4}}\ (x^3+e^{2\pi i/3}\,x^4),\\
z^3&=&{3^{-1/4}}\ (x^5+e^{2\pi i/3}\,x^6).
\end{eqnarray}
The nine twist-invariant 2--forms in the real cohomology are
\begin{eqnarray}
\om_1 &=& dx^1\wedge dx^2,\quad \om_2=dx^3\wedge dx^4,\quad \om_3=dx^5\wedge dx^6,\cr
\om_4 &=&-dx^1\wedge dx^4+dx^2\wedge dx^3,\ \ \om_5=-dx^1\wedge dx^4+dx^2\wedge dx^4\cr
\om_6&=&-dx^1\wedge dx^6+dx^2\wedge dx^5\cr
\om_7 &=&-dx^3\wedge dx^6+dx^4\wedge dx^5\cr
\om_8&=&dx^1\wedge dx^5-dx^1\wedge dx^6+dx^2\wedge dx^6\cr
\om_9 &=&dx^3\wedge dx^5-dx^3\wedge dx^6+dx^4\wedge dx^6,
\end{eqnarray}
so we can write $B=\sum_{i=1}^9b_i\,\om_i$. From the pairing ${\cal T}^i\,\om_i=B+i\,J$ we find
\begin{eqnarray}
\Tc^i&=&b_1+i\,\tfrac{\sqrt3}{2}\,R_1^2,\quad \Tc^2=b_2+i\,\tfrac{\sqrt3}{2}\,R_2^2,\quad \Tc^3=b_3+i\,\tfrac{\sqrt3}{2}\,R_3^2,\nonumber\\[1pt]
\Tc^4&=&b_4-i\,\sqrt3\,R_1R_2\,\cos\theta_{12},\quad 
\Tc^5=b_5+i\,\sqrt3\,R_1R_2\,\cos\theta_{23},\nonumber\\[1pt]
\Tc^6&=&b_6-i\,\sqrt3\,R_1R_3\cos\theta_{26}, \Tc^7=b_7-i\,\sqrt3\,R_2R_3\,\cos_{46},\nonumber\\[1pt]
\Tc^8&=&b_8+i\,\tfrac{1}{\sqrt3}\,R_1R_3\,(5\,\cos\theta_{25}+\cos\theta_{26})\nonumber\\[1pt]
\Tc^9&=&b_9+i\,\tfrac{1}{\sqrt3}\,R_2R_3\,(5\,\cos\theta_{45}+\cos\theta_{46}).
\end{eqnarray}

\subsection{Fixed sets}

This is a prime orbifold and therefore a very easy case. There are 27 isolated quotient singularities. 
\begin{table}[h!]
  \begin{center}
  \begin{tabular}{|c|c|c|c|}
    \hline
    \ Group el.&\ Order &\ Fixed Set& Conj. Classes\cr
    \hline
    \ $\theta$ & 3      &\ 27 fixed  points & 27\cr
    \hline
  \end{tabular}
  \caption{Fixed point set for $\IZ_3$.}
  \label{tab:fsthree}
  \end{center}
\end{table}
\begin{figure}[h!]
  \begin{center}
  \includegraphics[width=140mm]{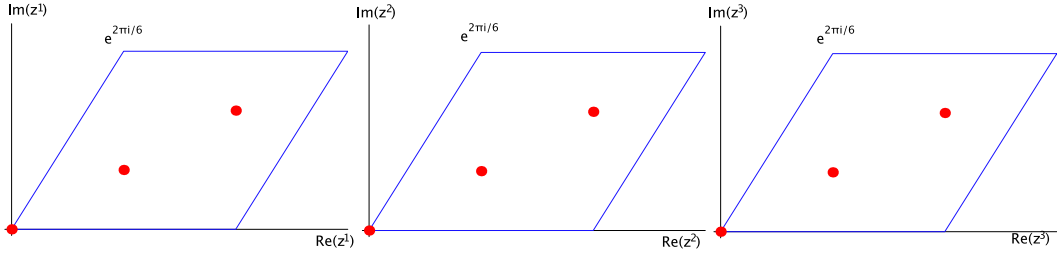}
  \caption{Fundamental regions for the $\IZ_3$--orbifold}
  \label{fig:ffuthree}
  \end{center}
\end{figure}
The $T^6$ factorizes into $(T^2)^3$. Figure~\ref{fig:ffuthree} shows the fundamental regions of the three tori corresponding to $z^1,\,z^2,\,z^3$ and their fixed points. The fundamental regions are all the same in this case and the three fixed points of the  $\IZ_3$--twist are $z^1_{{\rm fixed},1}=z^2_{{\rm fixed},1}=z^3_{{\rm fixed},1}=0,\ z^1_{{\rm fixed},2}=z^2_{{\rm fixed},2}=z^3_{{\rm fixed},2}=1/\sqrt3\,e^{\pi i/6}$, and $z^1_{{\rm fixed},3}=z^2_{{\rm fixed},3}=z^3_{{\rm fixed},3}=1+i/\sqrt3$.
\begin{figure}[h!]
  \begin{center}
  \includegraphics[width=85mm]{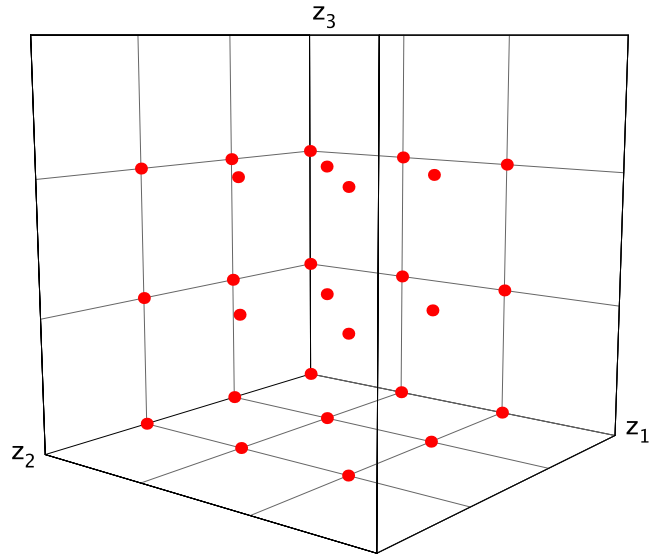}
  \caption{Schematic picture of the fixed set configuration of the $\IZ_3$--orbifold}
  \label{fig:ffixthree}
  \end{center}
\end{figure}
Figure~\ref{fig:ffixthree} shows the configuration of the fixed points in a schematic way, where each complex coordinate is shown as a coordinate axis and the opposite faces of the resulting cube of length 1 are identified.

\subsection{The gluing procedure}

On each of the 27 isolated $\IZ_3$ fixed points we put a local patch, which each contributes one compact exceptional divisor. We denote these divisors by $E_{\alpha\beta\gamma}$, where $\alpha$ denotes the 3 fixed points on the $z^1$--axis, $\beta$ those on the $z^2$--axis and $\gamma$ those of the $z^3$--axis. 
There are 9 fixed planes for this example, i.e. $z^i=\zf{i}{\alpha}, \ z^j,z^k$ free. To each of these we associate a divisor which we denote them by $D_{i\alpha}, \ i,\alpha=1,2,3$. $D_{1\alpha}$ are the divisors associated to $z^1=\zf{1}{\alpha}$, where $\alpha$ labels the fixed point. Furthermore, there are 9 divisors $R_{mn}$ which are inherited from $T^6$. 

Since there are no fixed lines in this example, $h^{(2,1)}_{tw}=0$.

\subsection{The intersection ring}

From the relations~(\ref{lineqthree}), the global linear relations~(\ref{eq:Reqglobal}) become:
\begin{equation}
  \label{eq:Reqthree}
  R_{i} \sim 3\,D_{i, \alpha}+\sum_{\beta,\gamma=1}^3 E_{{\alpha\beta\gamma}}.
\end{equation}
The calculation of the intersection ring is very simple, since all exceptional divisors are compact and neither intersect each other nor the $R_{mn}$. From~(\ref{eq:R1R2R3}) we see that $R_1R_2R_3 = 9$ and from Appendix~\ref{app:rzthree} we find that $E_{\alpha\beta\gamma}^3 = 9$.


\subsection{Divisor topologies}\label{app:divtopZ3}

As noted in Appendix~\ref{app:rzthree}, the $E_{\alpha\beta\gamma}$ have the topology of a ${\IP}^2$. All divisors $D_{i\alpha}$ have exactly the same properties, therefore it is enough to look at one of them and determine its Euler number. After removing the 9 fixed points from the $T^4$ the $\IZ_3$ action is free, hence the quotient has Euler number $(0-9)/3 = -3$. Resolving the singularities corresponds to gluing in 9 $P^1$. Therefore, $\chi(D) = -3 + 2\cdot9 = 15$. From~(\ref{eq:Reqthree}) and the intersection numbers given above, we find $D_{i\alpha}^3 = -3$, and using~(\ref{eq:S3}) its holomorphic Euler characteristic is $\chi(O_{D_{i\alpha}}) = 1$. Applying~(\ref{eq:c2.S}) to both $D_{i\alpha}$ and $E_{\alpha\beta\gamma}$ and then plugging into~(\ref{eq:Reqthree}) we can also determine the second Chern class to be
\begin{eqnarray}
  \label{eq:Z3c2}
  {\rm c}_2\cdot R_i &= &0, \ \quad \quad {\rm c}_2\cdot E_{\alpha\beta\gamma} = -6.
\end{eqnarray}

\subsection{The orientifold}
\label{eq:Z3O}

The O--plane configuration at the orbifold point is very simple, we have 64 O3--planes, located at the $I_6$ fixed points in each direction. They fall into 22 conjugacy classes, apart from $z_{\rm fixed}=(0,0,0)$, which is invariant, all other points fall into orbits of length 3. Only at $z_{\rm fixed}=(0,0,0)$, a fixed point coincides with one of the O3--planes.

All the exceptional divisors $E_{\alpha\beta\gamma}$ except $E_{111}$ fall into orbits of length two under $I_6$. Therefore, this example has $h^{1,1}_{-}=13$.

We now consider the orientifold of the resolved orbifold. There are two possible choices for the local involution leading to O3-- and O7--planes:
\begin{eqnarray}
(1)\quad {\cal I}(z,y)&=&(-z^1,-z^2, -z^3, y),\cr
(2)\quad {\cal I}(z,y)&=&(-z^1,-z^2, -z^3, -y).
\end{eqnarray}
We choose $(1)$ and have to solve (see~(\ref{rescalesthree}))
\begin{equation}
  \label{eq:othree}
  (-z^1,-z^2,-z^3,y)=(\lambda\, z^1, \lambda\,z^2, \lambda\,z^3, \lambda^{-3}\,y),
\end{equation}
leading to the solution $y=0,\ \lambda=-1$. This corresponds to an O7--plane located on the exceptional divisor $E$. Since only the divisor $E$ located at the fixed points itself appears in the solution, no further global consistency conditions need to be considered. Since only at $(0,0,0)$ a fixed point of the orbifold group coincides with a fixed point of $I_6$, we see that the O3--plane present at this point in the orbifold phase has disappeared.

The O3--planes away from the patches that we found in the orbifold phase are also present in the resolved case since they lie away from the resolved patches. We can see them by looking at the intersection ring of the orientifold and interpreting certain intersection numbers as number of O3--planes as discussed in Section~\ref{sec:Oring}. There are three cases. The fixed points of the orientifold action that lie at $z=(0,0,\frac{1}{2}),\, (0,0,\frac{\tau}{2},\, (0,0,\frac{1}{2}(1+\tau))$ form an equivalence class and correspond to the intersection of $D_{1,1} = \{ z^1=0 \},\, D_{2,1} = \{ z^2 =0 \}$, and $R_3 = \{z^3=c, c\not = 0,\,\frac{1}{3},\,\frac{2}{3} \}$. This intersection number is $D_{1,1}D_{2,1}R_3 = \frac{1}{2}$, hence we have a single O3--plane sitting there. By permuting the coordinates we get a total of 3 O3--planes. Then, there are the fixed points that lie at $z=(0,\frac{1}{2},\frac{1}{2}),\, (0,\frac{1}{2},\frac{\tau}{2},\dots, (0,\frac{1}{2}(1+\tau),\frac{\tau}{2}, (0,0,\frac{1}{2}(1+\tau))$. The corresponding intersection number is $D_{1,1}R_2R_3 = \frac{3}{2}$, hence we have 3 O3--planes. Taking into account the permutations of the coordinates yields a total of 9 O3-planes. Finally, the remaining orientifold fixed points correspond to the intersection $R_1R_2R_3 =\frac{9}{2}$, thus there are 9 more O3--planes. This makes a grand total of 21 O3--planes which agrees with the number of conjugacy classes.


\section{The $Z_{4}$ orbifold on $SU(4)^2$}
\label{sec:Z4orbSU4SU4} 

\subsection{Metric, complex structure and moduli}

On the root lattice of $SU(4)^2$, the twist $Q$ has the following action:
\begin{eqnarray}
Q\ e_1&=&e_2,\quad Q\ e_2=e_3,\quad Q\ e_3=-e_1-e_2-e_3,\cr 
Q\ e_4&=&e_5,\quad Q\ e_5=e_6 ,\quad Q\ e_6=-e_4-e_5-e_6\ .\end{eqnarray}
The twist $Q$ allows for seven independent real deformations of the metric $g$
and five real deformations  of the
anti--symmetric tensor $b$. These results follow from solving the equations
$Q^tg\,Q=g$ and $Q^tb\,Q=b$:
{\arraycolsep2pt
\begin{equation}{
g\!=\!\left(\!\!\begin{array}{cccccc}
R_1^2&R_1^2\cos\theta_{23}&x&R_1R_2\cos\theta_{36}&y&R_1R_2\cos\theta_{34}\cr
R_1^2\cos\theta_{23}&R_1^2&R_1^2\cos\theta_{23}&R_1R_2\cos\theta_{35}&R_1R_2\cos\theta_{36}&y\cr
x&R_1^2\cos\theta_{23}&R_1^2&R_1R_2\cos\theta_{34}&R_1R_2\cos\theta_{35}&R_1R_2\cos\theta_{36}\cr
R_1R_2\cos\theta_{36}&R_1R_2\cos\theta_{35}&R_1R_2\cos\theta_{34}&R_2^2&R_2^2\cos\theta_{56}&z\cr
y&R_1R_2\cos\theta_{36}&R_1R_2\cos\theta_{35}&R_2^2\cos\theta_{56}&R_2^2&R_2^2\cos\theta_{56}\cr
R_1R_2\cos\theta_{34}&y&R_1R_2\cos\theta_{36}&z&R_2^2\cos\theta_{56}&R_2^2\end{array}\!\!\right),
}\end{equation}}
with $x=-R_1^2(1+2\,\cos\theta_{23})$, $y=-R_1R_2\,(\cos\theta_{34}+\cos\theta_{35}+\cos\theta_{36})$, $z=-R_2^2(1+2\,\cos\theta_{56})$ and the seven real parameters $R_1^2,\ R_2^2,\ \theta_{23}$, $\theta_{34},\ \theta_{35},\ \theta_{36},\,\theta_{56}$.  
For $b$ we find
\begin{equation}{
b=\left(\begin{array}{cccccc}
0&b_1&0&b_5&-b_3-b_4-b_5&b_3\cr
-b_1&0&b_1&b_4&b_5&-b_3-b_4-b_5\cr
0&-b_1&0&b_3&b_4&b_5\cr
-b_5&-b_4&-b_3&0&b_2&0\cr
b_3+b_4+b_5&-b_5&-b_4&-b_2&0&b_2\cr
-b_3&b_3+b_4+b_5&-b_5&0&-b_2&0\end{array}\right)}\end{equation}
with the five real parameters $b_1,\ b_2,\ b_3,\ b_4,\ b_5$. We see that we get 5 untwisted K\"ahler moduli and one untwisted complex structure modulus in this orbifold.

With (\ref{ansatz}) and (\ref{solveeq}) we arrive at the following complex structure:
\begin{eqnarray}
z^1&=&\tfrac{1}{\sqrt2}\,(x^1+i\,x^2-x^3),\cr
z^2&=&\tfrac{1}{\sqrt2}\,(x^4+i\,x^5-x^6),\cr
z^3&=&\tfrac{1}{2\sqrt{u_2}}\,[x^1-x^2+x^3+{\cal U}\,(x^4-x^5+x^6)],
\end{eqnarray}
with 
\begin{equation}
\Uc=-\frac{R_2}{2\,R_1}\sec\theta_{23}(\cos\theta_{34}+\cos\theta_{36}+i\,\sqrt{-(\cos\theta_{34}+\cos\theta_{36})^2+4\,\cos\theta_{23}\cos\theta_{56}}).
\end{equation}
The five untwisted real 2--forms that are invariant under this orbifold twist are
\begin{eqnarray}
\om_1&=&dx^1\wedge dx^2+dx^2\wedge dx^3,\cr 
\om_2&=&dx^1\wedge dx^4-dx^1\wedge dx^5+dx^2\wedge dx^5-dx^2\wedge dx^6+dx^3\wedge dx^6,\cr
\om_3&=&-dx^1\wedge dx^5+dx^1\wedge dx^6-dx^2\wedge dx^6+dx^3\wedge dx^4,\cr
\om_4&=&-dx^1\wedge dx^5+dx^2\wedge dx^4-dx^2\wedge dx^6+dx^3\wedge dx^5,\cr
\om_5&=&dx^4\wedge dx^5+dx^5\wedge dx^6.
\end{eqnarray}
Via $B+i\,J={\cal T}^i\,\om_i$ the K\"ahler moduli are:
\begin{eqnarray}
{\cal T}^1&=&b_1+i\,2\,R_1^2\,(1+\cos\theta_{23}),\cr
{\cal T}^2&=&b_2+i\,\tfrac{1}{2}R_1R_2(5\cos\theta_{34}+6\cos\theta_{35}+\cos\theta_{36}+5\sqrt{-(\cos\theta_{34}+\cos\theta_{36})^2+4\cos\theta_{23}\cos\theta_{56}},\cr
{\cal T}^3&=&b_3+i\,2R_1R_2(-\cos\theta_{35}-\cos\theta_{36}+\sqrt{-(\cos\theta_{34}+\cos\theta_{36})^2+4\cos\theta_{23}\cos\theta_{56}},\nonumber\\[1pt]
{\cal T}^4&=&b_4+i\,2\,R_1R_2\,(\cos\theta_{34}-\cos\theta_{36}),\nonumber\\[1pt]
{\cal T}^5&=&b_5+i\,2\,R_1^2\,(1+\cos\theta_{56}).
\end{eqnarray}

\subsection{Fixed sets}

For the $\IZ_4$--twist, we need to look only at the $\theta$-, and $\theta^2$-twisted sectors.

Table \ref{fsfoura} summarizes the important data of the fixed sets. The invariant subtorus under $\theta^2$ is $(x^3,0,x^3,x^6,0,x^6)$, corresponding to $z^3=\,$ invariant.

\begin{table}[h!]\begin{center}
\begin{tabular}{|c|c|c|c|}
\hline
Group el.& Order & Fixed Set& Conj. Classes \cr
\hline
\noalign{\hrule}\noalign{\hrule}
$\theta$& 4     &\ 16 fixed points & 16\cr
$\theta^2$& 2     &\ 4 fixed lines & 4\cr
\hline
\end{tabular}
\caption{Fixed point set for $\IZ_{4}$--orbifold on  $SU(4)^2$.}\label{fsfoura}
\end{center}\end{table}

\begin{figure}[h!]
\begin{center}
\includegraphics[width=85mm]{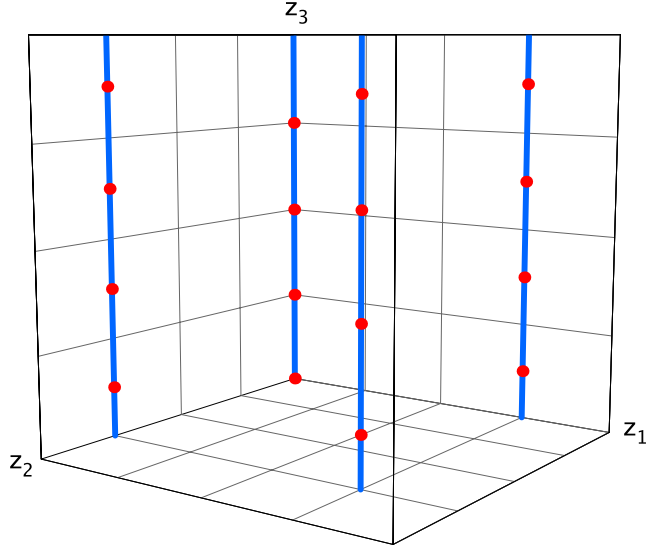}
\caption{Schematic picture of the fixed set configuration of $\IZ_{4}$ on $SU(4)^2$}\label{ffixfoura}
\end{center}
\end{figure}
Figure \ref{ffixfoura} shows the configuration of the fixed sets in a schematic way, where each complex coordinate is shown as a coordinate axis and the opposite faces of the resulting cube of length 1 are identified. 

This orbifold has special properties due to the $SU(4)^2$ lattice which leads to a non--standard volume factor for the fixed torus in $z^3$--direction (see Table \ref{table:vol}) and changed periodicities of the real lattice. Instead of the usual real lattice shift of one unit, the shift is $1/2$ for the coordinates entering $z^3$. The fixed points in $z^3$--direction do not all lie in the same four $D_3$ planes as usual but the points at $(z^1, z^2)\neq (0,0)$ are shifted up by $1/4$.

\subsection{The gluing procedure}

In this model, there are 16 local $\IZ_{4}$--patches which, four of each sit on one of the four $\IZ_2$--fixed lines. Each of the $\IZ_{4}$--patches contributes one exceptional divisor, and so does each of the fixed lines, therefore we get $16\cdot1+4\cdot1=20$ exceptional divisors in total.

Since there are no fixed lines without fixed points on them in this example, there are no twisted complex structure moduli.

\subsection{The intersection ring}

\begin{figure}[h!]
\begin{center}
\includegraphics[width=50mm]{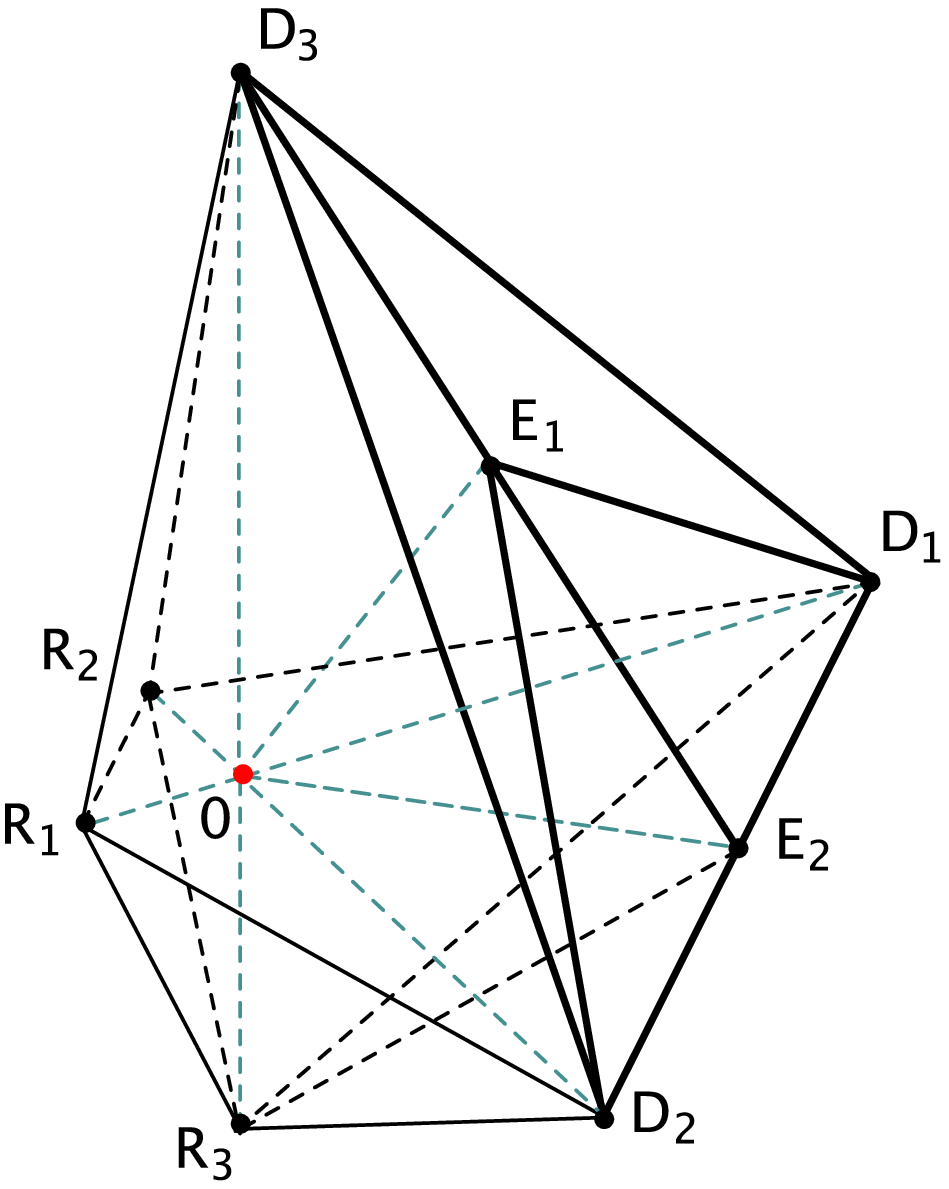}
\caption{The polyhedron $\Delta_1^{(3)}$ describing the local compactification of the resolution of $\IC^3/\IZ_{6-I}$.}
\label{fig:Z4-cpt}
\end{center}
\end{figure}

From the local linear relations~(\ref{lineqfour}), we find the following global linear relations:
\begin{eqnarray}
  \label{eq:Reqfour}
  R_{1} &\sim& 4\,D_{1,1}+\sum_{\gamma=1}^4 \left(E_{1,1,1,2\gamma-1}+E_{1,1,2,2\gamma}\right) + 2\sum_{\beta=1,2}E_{2,1,\beta},\cr
  R_{1} &\sim& 4\,D_{1,2}+\sum_{\beta=1,2}  \sum_{\gamma=1}^4 E_{1,2,\beta,2\gamma}  + 2\sum_{\beta=1,2}  E_{2,2,\beta},\cr
  R_{2} &\sim& 4\,D_{2,1}+\sum_{\gamma=1}^4 \left(E_{1,1,1,2\gamma-1}+E_{1,2,1,2\gamma}\right) + 2\sum_{\alpha=1,2}E_{2,\alpha,1},\cr
  R_{2} &\sim& 4\,D_{2,2}+\sum_{\alpha=1,2} \sum_{\gamma=1}^4 E_{1,\alpha,2,2\gamma} + 2\sum_{\alpha=1,2} E_{2,\alpha,2},\cr
  R_{3} &\sim& 2\,D_{3, 2\gamma-1}+E_{1,1,1,2\gamma-1}, \cr
  R_{3} &\sim& 2\,D_{3, 2\gamma}+E_{1,1,2\gamma} + \sum_{\beta=1,2} E_{1,2,\beta,2\gamma}, 
\end{eqnarray}
where $\gamma=1,...,4$.
To compute the intersection ring, we need to determine the basis for the lattice $N$ in which the auxiliary polyhedron will live. From~(\ref{eq:Reqfour}) we see that $n_1=n_2=4$, and $n_3=2$. Hence we can choose $m_1=m_2=m_3=2$, and the lattice basis is $f_1=(2,0,0)$, $f_2=(0,2,0)$, $f_3=(0,0,2)$. The lattice points of the polyhedron $\Delta^{(3)}$ for the local compactification of the $\IZ_4$ fixed points are
\begin{align}
  \label{eq:Z4poly}
  v_1 &= (-2,0,0), & v_2 &= (0,-2,0), & v_3 &= (0,0,-2), & v_4 &= (8,0,0), & v_5 &= (0,8,0),\notag\\
  v_6 &= (0,0,4), & v_7 &= (2,2,2), & v_8 &= (4,4,0),
\end{align}
corresponding to the divisors $R_1,R_2,R_3,D_1,D_2,D_3,E_1,E_2$ in that order. The polyhedron is shown in Figure~\ref{fig:Z4-cpt}. 
From the intersection ring of the 16 polyhedra and the linear relations~(\ref{eq:Reqfour}) we obtain the following nonvanishing intersection numbers of $X$ in the basis $\{R_i,E_{k\alpha\beta\gamma}\}$:
\begin{align}
  R_1R_2R_3 &=8, & R_{3}E_{2,\alpha\beta}^2 &=-2, & E_{1,\alpha\beta\gamma}E_{2,\alpha\beta}^2&=-2, & E_{1,\alpha,\beta,\gamma}^3&=8, & E_{2,\alpha,\beta}^3&=8.
\end{align}

\subsection{Divisor topologies}

$E_{1\alpha\beta\gamma}$ are $\IF_2$. $E_{2,\alpha\beta}$ are $\IF_0$, $D_{1,\alpha}$ and $D_{2,\beta}$ are $\Bl{8}\IF_n$. For $D_{3,\gamma}$ we start with the restriction of the $\IZ_4$ action to $z^3=\zf{3}{\gamma}$, which is $\frac{1}{4}(1,1)$ on $T^4$. The Euler number of $D_{3,\gamma}$ minus the 16 fixed points is $(0-16)/4=-4$. The fixed points fall into 10 classes. For odd $\gamma$ 3 of these classes are not resolved and the corresponding singularities are replaced by points. At the remaining fixed points we glue in a $\IP^1$ as usual. Therefore the Euler number of $D_{3,\gamma}$, $\gamma$ odd, is $-4+3\cdot 1 + 7\cdot 2 =13$. For even $\gamma$ only one of the ten classes is not blown up and the Euler number is $-4+1+9\cdot 2=15$. We conclude that the topology of $D_{3,\gamma}$ is $\Bl{9}\IF_0$ and $\Bl{11}\IF_0$ for $\gamma$ odd and even, respectively.
$R_1$ and $R_2$ are $T^4$ and $R_3$ is a K3. The second Chern class is
\begin{align}
  \ch_2\cdot E_{1,\alpha\beta\gamma} &= -4, & \ch_2 \cdot E_{2,\alpha\beta} &= -4, & \ch_2 \cdot R_i &= 0, & \ch_2 \cdot R_3 &= 24.
\end{align}

\subsection{The orientifold}
\label{sec:Z4orbSU4SU4_O}

The O--plane configuration at the orbifold point is very simple, we have 64 O3--planes, located at the $I_6$ fixed points in each direction. They fall into 22 conjugacy classes, apart from $(0,0,0)$, which is invariant. Under the combination $I_6\,\theta^2$, one O7--plane is fixed located at $z^3=0$ in the $(z^1,z^2)$--plane.

The $h^{1,1}_{-}=6$ of this example is due to the 12 $\IC^3/\IZ_4$ patches which are shifted in the $z^3$--direction by $1/4$, which are mapped to each other by $I_6$ pairwise on each fixed line.

Now we discuss the orientifold of the resolved case. There are two possibilities for $\cal I$ on the $\IC^3/\IZ_4$ patch which lead to a solution with O3-- and O7--planes: 
\begin{eqnarray}(1)\quad {\cal I}(z,y)&=&(-z^1,-z^2,-z^3,y^1,y^2),\label{ZfourOi}\cr
(2)\quad {\cal I}(z,y)&=&(-z^1,-z^2,-z^3,-y^1,-y^2).
\end{eqnarray}
For the choice (1), we solve
\begin{equation}\label{ofouri}{(-z^1,-z^2,-z^3,y^1,y^2)=(\lambda_1\, z^1, \lambda_1\,z^2, \lambda_2\,z^3, \frac{1}{\lambda_2^2}\,y^1,\frac{\lambda_2}{\lambda_1^2}\,y^2),}
\end{equation}
leading to the following two solutions: 
$$y^2=0,\ \lambda_1=\lambda_2=-1,\ \ \ z^3=0,\ \lambda_1=-1,\lambda_2=1.$$ 
This corresponds to an O7--plane located on $E_2$ and one on $D_3$. 
The choice (2) leads to
$$y^1=0,\ \lambda_1=\lambda_2=-1.$$
We will concentrate on the choice (1) for $\cal I$. The restriction of the scaling action to the $\IC^2/\IZ_2$ fixed line which is compatible with the solution $y^2=0$ consists of $D_1,\, D_2$ and $E_2$ is given by setting $\lambda_2=-1$:
\begin{equation}
(-z^1, -z^2, -z^3,y^1,y^2)=(\lambda_1\, z^1, \lambda_1\,z^2,-z^3, y^1,-\frac{1}{\lambda_1^2}\,y^2).
\end{equation}
This obviously leads to the solution
$$y^2=0,\ \lambda_1=-1.$$
In the resolved case, we have thus 4 O7--planes on the $E_{2\alpha\beta}$, and 8 O7--planes on the 8 $D_{3\gamma}$. Four of the $\IC^3/\IZ_4$ patches coincide with location of O3--planes before the blow--up. Since no O3--plane solutions appear in the resolved patches, these O3--planes are not present in the resolved case and we are left with 18 O3--planes located away from the resolved patches.

The modified intersection intersection numbers are
\begin{eqnarray}
&&R_{1}R_{2}R_{3}=4,\ R_{3}E_{2,\alpha\beta}^2=-4,\cr
&&E_{1,\alpha\beta\gamma}E_{2,\alpha\beta}^2=-4,\ E_{1,\alpha,\beta,\gamma}^3=4,\ E_{2,\alpha,\beta}^3=32.
\end{eqnarray}


\section{The $Z_{4}$ orbifold on $SU(2)\times SO(5)\times SU(4)$}
\label{sec:Z4onSU4xSU2xSO5} 
\subsection{Metric, complex structure and moduli}

The twist $Q$ has the following action on the root lattice of $SU(2)\times SO(5)\times SU(4)$:
\begin{eqnarray}
Q\ e_1&=&e_2,\quad Q\ e_2=e_3,\quad Q\ e_3=-e_1-e_2-e_3,\cr 
Q\ e_4&=&e_4+2\,e_5,\quad Q\ e_5=-e_4-e_5 ,\quad Q\ e_6=-e_6\ .\end{eqnarray}
The form of metric and anti-symmetric tensor follow from solving the equations
$Q^tg\,Q=g$ and $Q^tb\,Q=b$:
{\arraycolsep1pt
\begin{equation}{
g\!=\!\!\left(\!\!\begin{array}{cccccc}
R_1^2&R_1^2\cos\theta_{23}&x&-R_1R_2\cos\theta_{34}&-R_1R_2\cos\theta_{35}&R_1R_3\cos\theta_{36}\cr
R_1^2\cos\theta_{23}&R_1^2&R_1^2\cos\theta_{23}&y&-y&-R_1R_3\cos\theta_{36}\cr
x&R_1^2\cos\theta_{23}&R_1^2&R_1R_2\cos\theta_{34}&R_1R_2\cos\theta_{35}&R_1R_2\cos\theta_{36}\cr
-R_1R_2\cos\theta_{34}&y&R_1R_2\cos\theta_{34}&2\,R_2^2&-R_2^2&0\cr
-R_1R_2\cos\theta_{35}&-y&R_1R_2\cos\theta_{35}&-R_2^2&R_2^2&0\cr
R_1R_3\cos\theta_{36}&-R_1R_3\cos\theta_{36}&R_1R_3\cos\theta_{36}&0&0&R_3^2\end{array}\!\!\right),
}\end{equation}}
with $x=-R_1^2(1+2\,\cos\theta_{23})$, $y=R_1R_2\,(\cos\theta_{34}+2\,\cos\theta_{35})$. The seven real parameters $R_1^2,\ R_2^2,\ R_3^2, \theta_{23}$, $\theta_{34},\ \theta_{35},\ \theta_{36}$.  
For $b$ we find
\begin{equation}{
b=\left(\begin{array}{cccccc}
0&b_1&0&-b_2&-b_3&b_4\cr
-b_1&0&b_1&b_2+2\,b_3&-b_2-b_3&-b_4\cr
0&-b_1&0&b_2&b_3&b_4\cr
b_2&-b_2-2\,b_3&-b_2&0&b_5&0\cr
b_3&b_2+b_3&-b_3&-b_5&0&0\cr
-b_4&b_4&-b_4&0&0&0\end{array}\right)}\end{equation}
with the five real parameters $b_1,\ b_2,\ b_3,\ b_4,\ b_5$. We see that we get 5 untwisted K\"ahler moduli and one untwisted complex structure modulus in this orbifold.
With (\ref{ansatz}) and (\ref{solveeq}) we arrive at the following complex structure:
\begin{eqnarray}
z^1&=&\frac{1}{\sqrt2}\,(x^1+i\,x^2-x^3),\cr
z^2&=&x^4+\left(\frac{1}{2}-\frac{i}{2}\right)\,x^5,\cr
z^3&=&\frac{1}{2\sqrt{2\,u_2}}\,(x^1-x^2+x^3+2\,{\cal U}\,x^6).
\end{eqnarray}
with 
\begin{equation}
\Uc=-\frac{R_3}{2\,R_1}\sec\theta_{23}(\cos\theta_{36}+i\,\sqrt{-\cos\theta_{23}-\cos\theta_{36}^2}).
\end{equation}
The five untwisted real 2--forms that are invariant under this orbifold twist are
\begin{eqnarray}
\om_1&=&dx^1\wedge dx^2+dx^2\wedge dx^3,\cr 
\om_2&=&-dx^1\wedge dx^4+dx^2\wedge dx^4-dx^2\wedge dx^5+dx^3\wedge dx^4,\cr
\om_3&=&-dx^1\wedge dx^5+2\,dx^2\wedge dx^4-dx^2\wedge dx^5+dx^3\wedge dx^5,\cr
\om_4&=&dx^1\wedge dx^6-dx^2\wedge dx^6+dx^3\wedge dx^6,\cr
\om_5&=&dx^4\wedge dx^5.
\end{eqnarray}
Via $B+i\,J={\cal T}^i\,\om_i$ the K\"ahler moduli are:
\begin{eqnarray}
{\cal T}^1&=&b_1+i\,2\,R_1^2\,(1+\cos\theta_{23}),\nonumber\\[1pt]
{\cal T}^2&=&b_2-i\,R_1R_2\,(\cos\theta_{34}+5\cos\theta_{35}),\nonumber\\[1pt]
{\cal T}^3&=&b_3+i\,R_1R_2\,(4\,\cos\theta_{34}+\cos\theta_{35}),\cr
{\cal T}^4&=&b_4+i\,3\,R_1R_3\,\sqrt{-\cos\theta_{23}-\cos\theta_{36}^2},\cr
{\cal T}^5&=&b_5+i\,R_2^2.
\end{eqnarray}

\subsection{Fixed sets}

This being another $\IZ_4$--twist, we again need to look only at the $\theta$-, and $\theta^2$-twisted sectors.

Table \ref{fsfouraa} summarizes the important data of the fixed sets. The invariant subtorus under $\theta^2$ is $(x^3,0,x^3,0,0,x^6)$, corresponding to $z^3=\,$ invariant.

\begin{table}[h!]\begin{center}
\begin{tabular}{|c|c|c|c|}
\hline
Group el.& Order &Fixed Set& Conj. Classes \cr
\hline
\noalign{\hrule}\noalign{\hrule}
$\theta$& 4     &16  fixed points & 16\cr
$\theta^2$& 2   &8 fixed lines & 6\cr
\hline
\end{tabular}
\caption{Fixed point set for $\IZ_{4}$--orbifold on  $SU(2)\times SO(5)\times SU(4)$.}\label{fsfouraa}
\end{center}\end{table}

\begin{figure}[h!]
\begin{center}
\includegraphics[width=85mm]{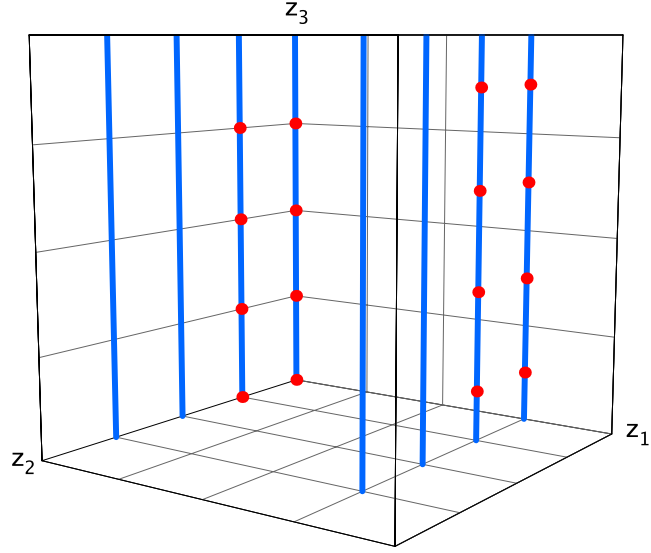}
\caption{Schematic picture of the fixed set configuration of $\IZ_{4}$ on $SU(2)\times SO(5)\times SU(4)$}\label{ffixfouraa}
\end{center}
\end{figure}
Figure \ref{ffixfouraa} shows the configuration of the fixed sets in a schematic way, where each complex coordinate is shown as a coordinate axis and the opposite faces of the resulting cube of length 1 are identified. 
Since in this case only one $SU(4)$ factor is present in the lattice, the periodicity change occurs only in one real direction and only the fixed points with $z^1\neq0$ are shifted up by $1/4$.

\subsection{The gluing procedure}

In this model, there are 16 local $\IZ_{4}$--patches which, four of each sit on a $\IZ_2$--fixed line. In total, there are six equivalence classes of $\IZ_2$--fixed lines. Each of the $\IZ_{4}$--patches contributes one exceptional divisor, and so does each of the fixed lines, therefore we get $16\cdot1+6\cdot1=22$ exceptional divisors in total. 

There are two $\IZ_2$ fixed lines without fixed points on them, so $h^{(2,1)}_{tw}=2$.

\subsection{The intersection ring}

From the local linear relations~(\ref{lineqfour}), we find the following global relations:
\begin{eqnarray}
  \label{eq:Reqfour2}
  R_{1} &\sim& 4\,D_{1,1}+\sum_{\beta=1,2} \sum_{\gamma=1}^4 E_{1,1,\beta,2\gamma-1}  + 2\sum_{\beta=1}^3 E_{2,1,\beta},\cr
  R_{1} &\sim& 4\,D_{1,2}+\sum_{\beta=1,2} \sum_{\gamma=1}^4 E_{1,2,\beta,2\gamma}    + 2\sum_{\beta=1}^3 E_{2,2,\beta},\cr
  R_{2} &\sim& 4\,D_{2,\beta}+\sum_{\gamma=1}^4 \left(E_{1,1,\beta,2\gamma-1}+E_{1,2,\beta,2\gamma}\right) + 2\sum_{\alpha=1,2}E_{2,\alpha,\beta},\cr
  R_{2} &\sim& 2\,D_{2, 3} + \sum_{\alpha=1}^2 E_{2,\alpha,3},\cr
  R_{3} &\sim& 2\,D_{3, 2\gamma-1}+\sum_{\beta=1,2} E_{1,1,\beta,2\gamma-1}, \cr
  R_{3} &\sim& 2\,D_{3, 2\gamma}  +\sum_{\beta=1,2} E_{1,2,\beta,2\gamma}, 
\end{eqnarray}
where $\alpha,\beta=1,2,\ \gamma=1,...,4$.
The polyhedron the $\IZ_2$ fixed lines is obtained from~(\ref{eq:Z4poly}) by dropping $v_7$. We obtain the following nonvanishing intersection numbers of $X$ in the basis $\{R_i,E_{k\alpha\beta\gamma}\}$:
\begin{align}
  R_1R_2R_3 &=8, & R_{3}E_{2,\alpha\beta}^2 &=-2, & E_{1,\alpha\beta\gamma}E_{2,\alpha\beta}^2&=-2, \notag\\
  E_{1,\alpha,\beta,\gamma}^3&=8, & E_{2,\alpha,\beta}^3&=8, & E_{2,\alpha,3}^2R_3 &= -4,
\end{align}
for $\alpha,\beta=1,2$ and all $\gamma$.

\subsection{Divisor topologies}

The topology of those divisors which were already present in the model in Appendix~\ref{sec:Z4orbSU4SU4} does not change except for $D_{3\gamma}$. The difference is that for both even and odd $\gamma$ two of the ten classes of fixed points of $T^4\/IZ_4$ are not resolved and the corresponding singularities are replaced by points. The Euler number therefore is $-4 + 2\cdot 1 + 8\cdot 2 = 14$ and the topology of $D_{3,\gamma}$ is $\Bl{10}\IF_0$. According to Section~\ref{sec:Topology} the new divisors $E_{2,\alpha,3}$ are of type E\ref{item:E2}), hence their topology is $\IP^1\times T^2$. By a similar argument as for $D_{1,2}$ in Section~\ref{sec:divsixiglobal} we find that the topology of $D_{2,3}$ is also that of a $\IP^1 \times T^2$. Finally, the second Chern class is
\begin{align}
  \ch_2\cdot E_{1,\alpha\beta\gamma} &= -4, & \ch_2 \cdot E_{2,\alpha\beta} &= -4, & \ch_2\cdot E_{2,\alpha,3} &= 0, & \ch_2 \cdot R_i &= 0, \notag\\ \ch_2 \cdot R_3 &= 24.
\end{align}

\subsection{The orientifold}
\label{sec:Z4orbSU4SU2SO5_O}

In this example, we have at the orbifold point 64 O3--planes which fall into 28 conjugacy classes under the orbifold group. Under $I_6\,\theta^2$, we have two O7--planes in the $(z^1,z^2)$--plane located at $z^3=0$ and $z^3=\tfrac{1}{2}U$.
$h^{1,1}_{-}=4$ is due to the 8 $\IC^3/\IZ_4$ patches which are shifted in the $z^3$--direction by $1/4$ and are mapped pairwise onto each other under $I_6$. There are two classes of fixed lines with fixed points which are invariant under the orientifold action, hence $h^{2,1}_+=2$.

The local involution on the resolved patches is the same as in Appendix~\ref{sec:Z4orbSU4SU4_O}. Since 8 of the O3--planes coincide with the local patches, they are not present in the resolved case. This means that we are left with 20 O3--planes, which are located away from the fixed points. Choosing the involution (1) of~(\ref{ZfourOi}), we have 4 O7--planes on the $E_{2\alpha\beta}$ and 8 on the $D_{3\gamma}$--planes.

The modified intersection numbers are
\begin{align}
  R_1R_2R_3 &=4, & R_{3}E_{2,\alpha\beta}^2 &=-4, & E_{1,\alpha\beta\gamma}E_{2,\alpha\beta}^2&=-4, \notag\\
  E_{1,\alpha,\beta,\gamma}^3&=4, & E_{2,\alpha,\beta}^3&=32, & E_{2,\alpha,3}^2R_3 &= -16,
\end{align}
for $\alpha,\beta=1,2$ and all $\gamma$.


\section{The $Z_{4}$ orbifold on $SU(2)^2\times SO(5)^2$}

\subsection{Metric, complex structure and moduli}

On the root lattice of $SU(4)^2$, the twist $Q$ has the following action:
\begin{eqnarray}
Q\ e_1&=&e_1+2\,e_2,\quad Q\ e_2=-e_1-e_2,\quad Q\ e_3=e_3+2\,e_4,\cr 
Q\ e_4&=&-e_3-e_4,\quad Q\ e_5=-e_5 ,\quad Q\ e_6=-e_6\ .\end{eqnarray}
The form of metric and anti-symmetric tensor follow from solving the equations
$Q^tg\,Q=g$ and $Q^tb\,Q=b$:
{\arraycolsep3pt
\begin{equation}{
g\!=\!\left(\!\!\begin{array}{cccccc}
2R_1^2&-R_1^2&2R_1R_2\cos\theta_{24}&x&0&0\cr
-R_1^2&R_1^2&R_1R_2\cos\theta_{23}&R_1R_2\cos\theta_{24}&0&0\cr
2R_1R_2\cos\theta_{24}&R_1R_2\cos\theta_{23}&2\,R_2^2&-R_2^2&0&0\cr
x&R_1R_2\cos\theta_{24}&-R_2^2&R_2^2&0&0\cr
0&0&0&0&R_3^2&R_3R_4\cos\theta_{56}\cr
0&0&0&0&R_3R_4\cos\theta_{56}&R_4^2\end{array}\!\!\right),
}\end{equation}}
with $x=-R_1R_2\,(\cos\theta_{23}+2\,\cos\theta_{24})$. The seven real parameters $R_1^2,\ R_2^2,\ R_3^2, \ R_4^2,\ \theta_{23}$, $\theta_{24},\ \theta_{56}$.  
For $b$ we find
\begin{equation}{
b=\left(\begin{array}{cccccc}
0&b_1&2\,b_4&-b_4-2\,b_5&0&0\cr
-b_1&0&b_4&b_5&0&0\cr
-2\,b_4&-b_4&0&b_2&0&0\cr
-b_4-2\,b_5&-b_5&-b_2&0&0&0\cr
0&0&0&0&0&b_3\cr
0&0&0&0&-b_3&0\end{array}\right)}\end{equation}
with the five real parameters $b_1,\ b_2,\ b_3,\ b_4,\ b_5$. We see that we get 5 untwisted K\"ahler moduli and one untwisted complex structure modulus in this orbifold.

With (\ref{ansatz}) and (\ref{solveeq}) we arrive at the following complex coordinates:
\begin{eqnarray}
z^1&=&x^1+\left(\frac{1}{2}-\frac{i}{2}\right)\,x^2,\cr
z^2&=&x^3+\left(\frac{1}{2}-\frac{i}{2}\right)\,x^4,\cr
z^3&=&\frac{1}{\sqrt{2\,{\rm Im}\,{\cal U}}}\,(x^5+{\cal U}\,x^6),
\end{eqnarray}
with  $\Uc=\frac{R_4}{R_3}\,e^{i\theta_{56}}$.
The five untwisted real 2--forms that are invariant under this orbifold twist are
\begin{eqnarray}
\om_1&=&dx^1\wedge dx^2,\quad \om_2=dx^3\wedge dx^4,\quad \om_3=dx^5\wedge dx^6,\cr
\om_4&=&-dx^1\wedge dx^4+dx^2\wedge dx^3,\cr
\om_5&=&2\,dx^1\wedge dx^5-2\,dx^1\wedge dx^4+dx^2\wedge dx^4.
\end{eqnarray}
Via $B+i\,J={\cal T}^i\,\om_i$ the K\"ahler moduli are:
\begin{eqnarray}
{\cal T}^1&=&b_1+i\,R_1^2,\quad {\cal T}^2=b_2+i\,R_2^2,\quad {\cal T}^3=b_3+i\,R_3R_4\,\sin\theta_{56},\cr
{\cal T}^4&=&b_4-i\,2\,R_1R_2\,\cos\theta_{24},\quad
{\cal T}^5=b_5+i\,R_1R_2\,(7\,\cos\theta_{23}+5\,\cos\theta_{24}).
\end{eqnarray}

\subsection{Fixed sets}

This being another $\IZ_4$--twist, we again need to look only at the $\theta$-, and $\theta^2$-twisted sectors.

Table \ref{fsfouraaa} summarizes the important data of the fixed sets. The invariant subtorus under $\theta^2$ is $(0,0,0,0,x^5,x^6)$, corresponding to $z^3=\,$ invariant.

\begin{table}[h!]\begin{center}
\begin{tabular}{|c|c|c|c|}
\hline
Group el.& Order &Fixed Set& Conj. Classes \cr
\hline
\noalign{\hrule}\noalign{\hrule}
$\theta$&4    &16 fixed points & 16\cr
$\theta^2$& 2    &16 fixed lines & 10\cr
\hline
\end{tabular}
\caption{Fixed point set for $\IZ_{4}$--orbifold on  $SU(4)^2\times SO(5)^2$.}\label{fsfouraaa}
\end{center}\end{table}

\begin{figure}[h!]
\begin{center}
\includegraphics[width=85mm]{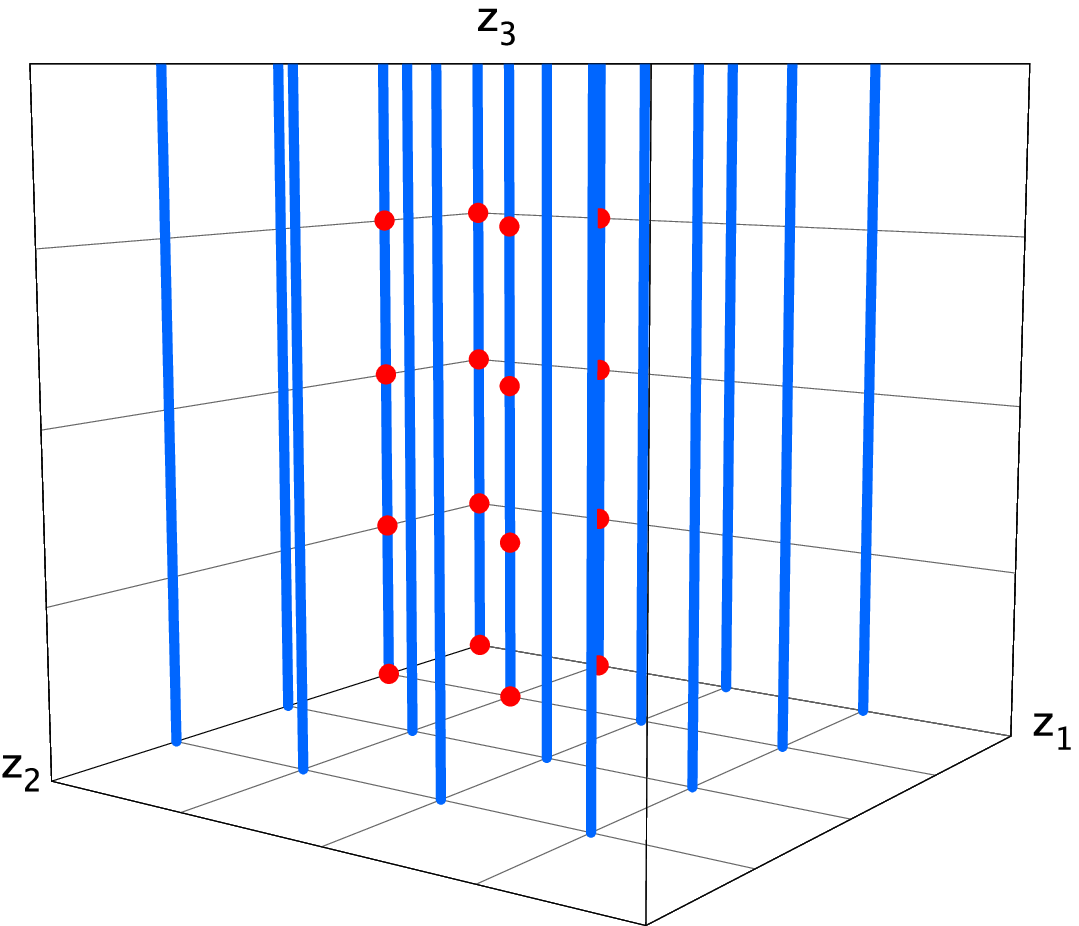}
\caption{Schematic picture of the fixed set configuration of $\IZ_{4}$ on $SU(4)^2\times SO(5)^2$}\label{ffixfouraaa}
\end{center}
\end{figure}
Figure \ref{ffixfouraaa} shows the configuration of the fixed sets in a schematic way, where each complex coordinate is shown as a coordinate axis and the opposite faces of the resulting cube of length 1 are identified.

\subsection{The gluing procedure}

In this model, there are 16 local $\IZ_{4}$--patches which, four of each sit on a $\IZ_2$--fixed line. In total, there are ten equivalence classes of $\IZ_2$--fixed lines. Each of the $\IZ_{4}$--patches contributes one exceptional divisor, and so does each of the fixed lines, therefore we get $16\cdot1+10\cdot1=26$ exceptional divisors in total.

In this lattice, six $\IZ_2$ fixed lines without fixed points on them appear, therefore $h^{(2,1)}_{tw}=6$.

\subsection{The intersection ring}

As we have seen, there are 16 local $\IC^3/\IZ_{4}$ patches which sit in groups of four on a $\IC^2/\IZ_2$ fixed line. They yield the same exceptional divisors as in Section~\ref{sec:Z4onSU4xSU2xSO5}, however, according the different labeling of the fixed points here, we denote them by $E_{1,\alpha\beta\gamma}$ and $E_{2,\alpha,\beta}$, $\alpha,\beta=1,2$, $\gamma = 1,\dots,4$. The remaining 12 $\IC^2/\IZ_2$ fixed lines fall into six equivalence classes. 
The invariant divisors are 
\begin{align}
  E_{2,1} &= \Et_{2,1,1}, & E_{2,2} &= \Et_{2,1,2}, & E_{2,3} &= \Et_{2,1,3} + \Et_{2,1,4}, \notag\\ 
  E_{2,4} &= \Et_{2,2,1}, & E_{2,5} &= \Et_{2,2,2}, & E_{2,6} &= \Et_{2,2,3} + \Et_{2,2,4}, \notag\\
  E_{2,7} &= \Et_{2,3,1} + \Et_{2,4,1}, & E_{2,8} &= \Et_{2,3,2} + \Et_{2,4,2}, & E_{2,9} &= \Et_{2,3,3} + \Et_{2,4,4}, \notag\\
  E_{2,10} &= \Et_{2,3,4} + \Et_{2,4,3}.
\end{align}
where $\Et_{2,\alpha,\beta}$ are the representatives on the cover. 

From the local linear relations~(\ref{lineqfour}), we find the following global relations:
\begin{align}
  \label{eq:Reqfour3}
  R_{1} &= 4\,D_{1, \alpha}+\sum_{\beta=1}^2\sum_{\gamma=1}^4 E_{1,{\alpha\beta\gamma}}+2\sum_{\mu=1}^3 E_{2,3\alpha-3+\mu}, && \alpha=1,2, \notag\\
  R_{1} &= 2\,D_{1, 3} + \sum_{\mu=7}^{10} E_{2,\mu},\notag\\
  R_{2} &= 4\,D_{2, \beta}+\sum_{\alpha=1}^2\sum_{\gamma=1}^4 E_{1,{\alpha\beta\gamma}}+2\sum_{\mu=0,3,6}E_{2,\mu+\beta},&& \beta=1,2, \notag\\
  R_{2} &= 2\,D_{2, 3} + \sum_{\mu=3,6,9,10} E_{2,\mu},\notag\\
  R_{3} &= 2\,D_{3, \gamma}+\sum_{\alpha=1}^2 \sum_{\beta=1}^2 E_{1,{\alpha\beta\gamma}}, && \gamma=1,\dots,4, 
\end{align}
The polyhedron the $\IZ_2$ fixed lines are is obtained from~(\ref{eq:Z4poly}) by dropping $v_7$. We obtain the following nonvanishing intersection numbers:
\begin{align}
  R_1R_2R_3 &=8, & R_{3}E_{2,\mu}^2 &=-2, & E_{1,\alpha\beta\gamma}E_{2,\mu}^2&=-2, & \notag\\
  E_{1,\alpha,\beta,\gamma}^3&=8, & E_{2,\mu}^3&=8, & & \notag\\
  \intertext{for $\mu=1,2,4,5$ and} 
  E_{2,\mu}^2R_3 &= -4,
\end{align}
for $\mu=3,6,\dots,10$ and $\alpha,\beta=1,2$. 

\subsection{Divisor topologies}

The topology of those divisors which were already present in the model in Appendix~\ref{sec:Z4onSU4xSU2xSO5} does not change except for $D_{3\gamma}$. The difference is that all of the ten classes of fixed points of $T^4\/IZ_4$ are resolved. The Euler number therefore is $-4 + 10\cdot 2 = 16$ and the topology of $D_{3,\gamma}$ is $\Bl{12}\IF_0$. All the new divisors $E_{2,\mu}$ and $D_{3,\beta}$ have the topology of a $\IP^1 \times T^2$. Finally, the second Chern class is
\begin{align}
  \ch_2\cdot E_{1,\alpha\beta\gamma} &= -4, & \ch_2 \cdot R_i &= 0, & \ch_2 \cdot R_3 &= 24, & \ch_2 \cdot E_{2,\mu} &= -4, \notag\\
  \intertext{for $\mu=1,2,4,5$ and} 
  \ch_2\cdot E_{2,\mu} &= 0,
\end{align}
for $\mu=3,6,\dots,10$.

\subsection{The orientifold}
\label{sec:Z4orbSU2SO52_O}

In this example, we have at the orbifold point 64 O3--planes which fall into 40 conjugacy classes under the orbifold group. Under $I_6\,\theta^2$, we have four O7--planes in the $(z^1,z^2)$--plane located at $z^3=0,\,\tfrac{1}{2},\,\tfrac{1}{2}\,{\cal U}^3$ and $z^3=\tfrac{1}{2}\,(1+{\cal U}^3)$.
$h^{1,1}_{-}=0$ since all $\IC^3/\IZ_4$ patches are fixed under $I_6$. There are six classes of fixed lines with fixed points which are invariant under the orientifold action, hence $h^{2,1}_+=6$.

The local involution on the resolved patches is the same as for the $SU(4)^2$--lattice in Appendix~\ref{sec:Z4orbSU4SU4_O}. Since all of the local patches coincide with the locations of the O3--planes, these 16 O3--planes are not present in the resolved case. This means that we are left with 24 O3--planes, which are located away from the fixed points. Choosing the involution (1) of (\ref{ZfourOi}), we have 4 O7--planes on the $E_{2\alpha\beta}$ and 4 on the $D_{3\gamma}$--planes.

The modified intersection numbers are
\begin{align}
  R_1R_2R_3 &=4, & R_{3}E_{2,\mu}^2 &=-4, & E_{1,\alpha\beta\gamma}E_{2,\mu}^2&=-4, & \notag\\
  E_{1,\alpha,\beta,\gamma}^3&=4, & E_{2,\mu}^3&=32, & & \notag\\
  \intertext{for $\mu=1,2,4,5$ and} 
  E_{2,\mu}^2R_3 &= -16,
\end{align}
for $\mu=3,6,\dots,10$ and $\alpha,\beta=1,2$.


\section{The $Z_{6-I}$ orbifold on $G_2\times SU(3)^2$}


\subsection{Metric, complex structure and moduli}

On the root lattice of $G_2\times SU(3)^2$, we act with the generalized Coxeter twist $Q=S_1S_2S_3S_4P_{36}P_{45}$, as explained in Section 3. It has the following action:
\begin{eqnarray}
Q\ e_1&=&2\,e_1+3\,e_2 ,\ \ \ Q\ e_2=-e_1 -e_2\ ,\cr
Q\ e_3&=&e_6,\ \ \ Q\ e_4=e_5\ ,\cr
Q\ e_5&=&-e_3-e_4,\ \ \ Q\ e_6=e_4.\end{eqnarray}
As before, the twist $Q$ allows for five independent real deformations of the metric $g$
and five real deformations  of the
anti--symmetric tensor $b$:
\begin{equation}\label{metricsixii}{
g=\left(\begin{array}{cccccc}
R_1^2&-{1\over 2}R_1^2&x&-x&-x&x\cr
-{1\over 2}R_1^2&{1\over 3}R_1^2&-y &x&y&z\cr
x&-y&R_5^2&-{1\over 2}R_5^2&-2\,R_5^2\cos\theta_{46}&R_5^2\cos\theta_{46}\cr
-x&x&-{1\over 2}R_5^2& R_5^2&R_5^2\cos\theta_{46}&R_5^2\cos\theta_{46}\cr
-x&y&-2\,R_5^2\cos\theta_{46}&R_5^2\cos\theta_{46}&R_5^2&-\half R_5^2\cr
x&z&R_5^2\cos\theta_{46}&R_5^2\cos\theta_{46}&-\half R_5^2&R_5^2
\end{array}\right),}\end{equation}
with $x={1\over\sqrt3}(R_1R_5(\cos\theta_{25}+\cos\theta_{26}),\, y={1\over\sqrt3}R_1R_5\cos\theta_{25}, \,z={1\over\sqrt3}R_1R_5\cos\theta_{26}$ and the five real parameters $R_1^2,\ R_5^2,\ \theta_{25},\ \theta_{26}$ and $\theta_{46}$. 
For $b$ we find
\begin{equation}\label{bfieldsixiI}{
b=\left(\begin{array}{cccccc}
0&b_1&2\,b_2+b_3&-b_2-2\,b_3&-2\,b_2-b_3&b_2-b_3\cr
-b_1&0&-b_2&b_2+b_3&b_2&b_3\cr
-2\,b_2-b_3&b_2&0&-b_5&0&b_4\cr
b_2+2\,b_3&-b_2-b_3&b_5&0&b_4&-b_4\cr
2\,b_2+b_3&-b_2&0&-b_4&0&b_5\cr
-b_2+b_3&-b_2&-b_4&b_4&-b_5&0\end{array}\right)
}\end{equation}
with the five real parameters $b_1,\ b_2,\ b_3,\ b_4,\ b_5$. 
$a,\,b,\,c,\,d$ and $e$ are constants left unfixed by the twist. We choose them such that we get
\begin{eqnarray}\label{cplxzsixiIsim}
z^1&=&3^{-1/4}\,(x^1+\tfrac{1}{3}\,e^{5\pi i/6}\,x^2),\cr
z^2&=&\tfrac{1}{\sqrt2}\,(x^3+e^{2\pi i/3}\,x^4-x^5+e^{2\pi i/6}\,x^6),\cr
z^3&=&\tfrac{1}{\sqrt2}\,(x^3-e^{2\pi i/6}\,x^4+x^5+e^{2\pi i/3}\,x^6).
\end{eqnarray}
The five real untwisted 2--forms that are invariant under this orbifold twist are
\begin{eqnarray}\label{invformsixiI}
\om_1&=&dx^1\wedge dx^2,\cr 
\om_2&=&2\,dx^1\wedge dx^3-dx^1\wedge dx^4-2\,dx^1\wedge dx^5+dx^1\wedge dx^6-dx^2\wedge dx^3\cr
&&+dx^2\wedge dx^4+dx^2\wedge dx^5,\cr
\om_3&=&dx^1\wedge dx^3-2\,dx^1\wedge dx^4-dx^1\wedge dx^5-dx^1\wedge dx^6+dx^2\wedge dx^4+dx^2\wedge dx^6,\cr
\om_4&=&dx^3\wedge dx^6+ dx^4\wedge dx^5-dx^4\wedge dx^6,\cr
\om_5&=&dx^5\wedge dx^6.
\end{eqnarray}
The $B$--field (\ref{bfieldsixiI}) has the simple form
$B=b_1\ \om_1+b_2\ \om_2+b_3\ \om_3+b_4\ \om_4+b_5\ \om_5$.
Examination of the K\"ahler form yields
\begin{eqnarray}
{\cal T}^1&=&b_1+i\,\tfrac{1}{2\sqrt3}\,R_1^2,\quad {\cal T}^2=b_2+i\,\tfrac{1}{3}\,R_1R_5\,(\cos\theta_{25}+20\,\cos\theta_{26}),\cr
{\cal T}^3&=&b_3-i\,R_1R_5\,(4\,\cos\theta_{25}-\cos\theta_{26}),\cr
{\cal T}^4&=&b_4+i\,\tfrac{3\sqrt3}{2}\,R_5^2,\quad {\cal T}^5=b_5-i\,\sqrt3\,R_5^2.
\end{eqnarray}

\subsection{Fixed sets}

In order to find the fixed point sets, we need to look at the $\theta$--, $\theta^2$-- and $\theta^3$--twists. $\theta^4$ and $\theta^5$ yield no new information, since they are simply the anti--twists of $\theta^2$ and, $\theta$. The action of the twist $\theta$ on the lattice $G_2\times SU(3)^2$ was given in (A.12).
\begin{figure}[h!]
  \begin{center}
  \includegraphics[width=140mm]{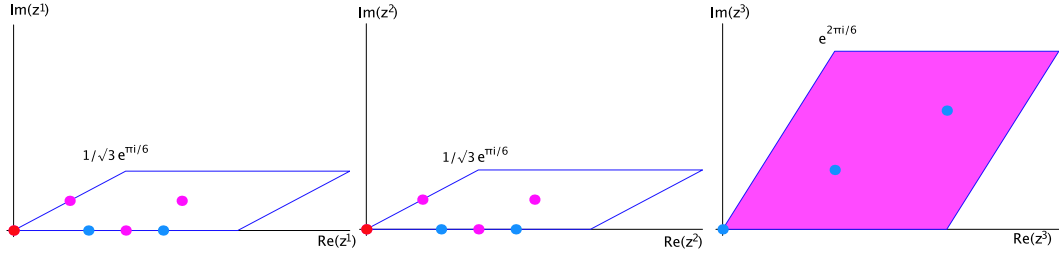}
  \caption{Fundamental regions for the $\IZ_{6-I}$--orbifold}
  \label{fig:fzsixif}
  \end{center}
\end{figure}
Figure~\ref{fig:fzsixif} shows the fundamental regions of the three tori corresponding to $z^1,\,z^2,\,z^3$ and their fixed points in the different sectors. The $\IZ_{6-I}$--twist has only one fixed point in each torus, namely $\zf{1}{1}=\zf{2}{1}=\zf{3}{1} = 0$. The $\IZ_3$--twist has three fixed points, namely $\zf{1}{\alpha} = \zf{2}{\beta} = 0,1/3, 2/3$ for $\alpha,\beta=1,3,5$ and $\zf{3}{\gamma}=0, 1/\sqrt3 \,e^{\pi i/6}, 1+i/\sqrt3$ for $\gamma=1,2,3$. The $\IZ_2$--twist, which arises in the $\theta^3$-twisted sector, has four fixed points, corresponding to $\zf{1}{\alpha} = 0,\half,\half \tau,\half(1+\tau)$, $\alpha=1,2,4,6$ for the respective modular parameter $\tau$. 

The equivalence classes of fixed point set are described as follows: We first look at the $z^1$ and $z^2$--directions. The two $\IZ_3$--fixed points at $1/3$ and $2/3$ are mapped to each other by $\theta$ and form orbits of length two. We choose to represent this orbit by $\zf{i}{2}$, $i=1,2$. The three $\IZ_3$ fixed points in the $z^3$--direction each form a separate conjugacy class. Therefore, we obtain the 15 conjugacy classes of $\IZ_3$--fixed points, 5 in each plane $z^3=\zf{3}{\gamma}$, $\gamma=1,2,3$:
\begin{align}
  \label{eq:zthreeconj}
  \mu=1:\; &(0,0, \zf{3}{\gamma}) & & &\notag\\  
  \mu=2:\; &(0,\tfrac{1}{3},\zf{3}{\gamma}),\ (0,\tfrac{2}{3},\zf{3}{\gamma}) & \mu=3:\; & (\tfrac{1}{3},0,\zf{3}{\gamma}),\ (\tfrac{2}{3},0,\zf{3}{\gamma})\notag\\ 
  \mu=4:\; &(\tfrac{1}{3},\tfrac{1}{3},\zf{3}{\gamma}),\ (\tfrac{2}{3},\tfrac{2}{3}, \zf{3}{\gamma}) & \mu=5:\; & (\tfrac{1}{3}, \tfrac{2}{3},\zf{3}{\gamma}),\ (\tfrac{2}{3},\tfrac{1}{3},\zf{3}{\gamma}). 
\end{align}
The $\IZ_2$--fixed points in the $z^1$--direction form the two orbits under $\theta^2$, namely $0$, and $\half \to \half(1+\tau) \to \half \tau$. The corresponding two conjugacy classes will be represented by $\zf{i}{3}$, $i=1,2$.
\begin{table}[h!]
  \begin{center}
  \begin{tabular}{|c|c|c|c|}
    \hline
    Group el.& Order & Fixed Set& Conj. Classes \cr
    \hline 
    \noalign{\hrule}
    $\ \theta  $ & 6 &  3\ {\rm fixed\ points} &\  3\cr
    $\ \theta^2$ & 3 & 27\ {\rm fixed\ points} &\ 15\cr
    $\ \theta^3$ & 2 &  4\ {\rm fixed\  lines} &\  2\cr
    \hline
  \end{tabular}
  \caption{Fixed point sets for $\IZ_{6-I}$ on $G_2\times SU(3)^2$.}
  \label{tab:fssixigsuii}
\end{center} 
\end{table}
Table~\ref{tab:fssixigsuii} summarizes the relevant data of the fixed sets. The invariant subtorus under $\theta^3$ is $(0,0,x^5-x^6,-x^6,x^5,x^6)$, corresponding to the complex $z^3$--coordinate being invariant.
\begin{figure}[h!]
\begin{center}
\includegraphics[width=85mm]{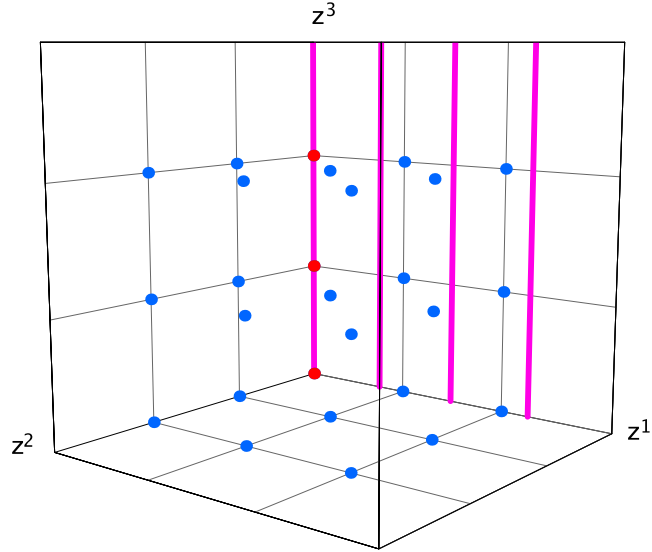}
\caption{Schematic picture of the fixed point set of the $\IZ_{6-I}$--orbifold on $G_2\times SU(3)^2$}
\label{fig:ffixsixi}
\end{center}
\end{figure}
Figure~\ref{fig:ffixsixi} shows the configuration of the fixed point set in a schematic way, where each complex coordinate is shown as a coordinate axis and the opposite faces of the resulting cube of length 1 are identified.

\subsection{The gluing procedure}

From Table~\ref{tab:fssixi} and Figure~\ref{fig:ffixedi}, we see that there are three fixed points with a $\IZ_{6-I}$ singularity. Resolving this singularity amounts to replacing the $\IC^3/\IZ_{6-I}$ patch by $X_{\widetilde{\Sigma}}$ in~(\ref{eq:blowupsixi}). By Figure~\ref{fig:fsixi}, each resolution contributes two exceptional divisors $E_{1,\gamma}$, and $E_{2,1,\gamma}$, $\gamma=1,2,3$, from the interior of the diagram and one exceptional divisor from the boundary of the diagram, respectively The latter will be considered in the next paragraph.

Returning to Table~\ref{tab:fssixi} and Figure~\ref{fig:ffixedi}, we have furthermore 15 conjugacy classes of $\IZ_3$ fixed points. Blowing them up replaces each of them locally by $X_{\widetilde{\Sigma}}$ in~(\ref{eq:blowupthree}) and contributes one exceptional divisor as can be seen from Figure~\ref{fig:frthree}. Since three of these fixed points sit at the location of the $\IZ_{6-I}$ fixed points which we have already taken into account ($E_{2,1,\gamma}$), we only count 12 of them, and denote the resulting divisors by $E_{2,\mu,\gamma},\ \mu = 2,\dots,5, \, \gamma=1,2,3$. The invariant divisors are built according to the conjugacy classes in~(\ref{eq:zthreeconj})
\begin{align}
  \label{eq:E2conj}
  E_{2,2,\gamma} &= \Et_{2,1,2,\gamma} +  \Et_{2,1,3,\gamma}, & E_{2,3,\gamma} &=  \Et_{2,3,1,\gamma} +  \Et_{2,5,1,\gamma},\notag\\
  E_{2,4,\gamma} &= \Et_{2,3,2,\gamma} +  \Et_{2,5,3,\gamma}, & E_{2,5,\gamma} &=  \Et_{2,3,3,\gamma} +  \Et_{2,5,2,\gamma}.
\end{align}
where $\Et_{2,\alpha,\beta,\gamma}$ are the representatives on the cover.

Then, we finally have 2 conjugacy classes of fixed lines of the form $\IC^2/\IZ_2$. We see that after the resolution, each class contributes one exceptional divisor $E_{3,\alpha}, \alpha=1,2$. On the fixed line at $\zf{1}{1}=\zf{2}{1}=0$ sit the three $\IZ_{6-I}$ fixed points. The divisor coming from the blow--up of this fixed line, $E_{3,1}$, is identified with the three exceptional divisors corresponding to the points on the boundary of the toric diagram of the resolution of $\IC^3/\IZ_{6-I}$ that we mentioned above. The other exceptional divisor is the invariant combination $E_{3,2} = \sum_{\alpha=2,4,6} \Et_{3,\alpha}$, where $\Et_{3,\alpha}$ are the representatives on the cover. Consequently, $E_{3,1}$ could have a different topology than $E_{3,2}$.

This results in $3\cdot 2+12\cdot 1+2\cdot 1=20$ exceptional divisors.  There is one $\IZ_2$ fixed line without fixed points on it, therefore, by~(\ref{eq:h21tw}), $h^{(2,1)}_{tw}=1$.

\subsection{The intersection ring}

Furthermore, we have fixed planes $\Dt_{1,\alpha} = \{ z^1=\zf{1}{\alpha}\}$, $\alpha=1,\dots,6$, $\Dt_{2,\beta} = \{z^2=\zf{2}{\beta}\}$, $\beta=1,2,3$, and $\Dt_{3,\gamma} = \{z^3 = \zf{3}{\gamma}\}$, $\gamma=1,2,3$ on the cover. From these we define the invariant combinations
\begin{align*}
  D_{1,1} &= \Dt_{1,1}, & D_{1,2} &= \Dt_{1,2} + \Dt_{1,4} + \Dt_{1,6}, &
  D_{1,3} &= \Dt_{1,3} + \Dt_{1,5}, \\
  D_{2,1} &= \Dt_{2,1}, & D_{2,2} &= \Dt_{2,2} + \Dt_{2,3}, & D_{3,\gamma} &= \Dt_{3,\gamma}.
\end{align*}
Next, we need the global linear relations~(\ref{eq:Reqglobal}) in order to determine the intersection ring. The relation for $D_{1,1}$ is obtained from (\ref{eq:lineqsixI}) :
\begin{equation}
  \label{eq:Z6IrelD11} 
  R_1=6\,D_{{1,1}}+\sum_{\gamma=1}^3E_{{1,\gamma}} +2\, \sum_{\mu=1}^2 \sum_{\gamma=1}^3 E_{{2,\mu,\gamma}}+3\, E_{{3,1}}.
\end{equation}
The divisor $D_{1,2}$ only contains a single equivalence class of $\IZ_2$ fixed line. 
From the local relations~(\ref{eq:lineqsctwo}), we find the local relation to $R_1$ as in~(\ref{eq:ReqH}) (here, we already changed the labels of the divisors to match  the labels of the $\IZ_{6-II}$--patch):
\begin{equation}
  \label{eq:Z6IrelD122}
  R_1 = 2D_{{1,2}}+\, E_{{3,2}}.
\end{equation}
Next, we look at the divisor $D_{1,3}$, which only contains $\IZ_3$ fixed points. The local linear equivalences~(\ref{lineqthree}) and~(\ref{eq:ReqH}) lead to
\begin{equation}
  \label{eq:Z6Irel133}
  R_1 = 3\,D_{{1,3}}+\sum_{\mu=3}^5 \sum_{\gamma=1}^3 E_{{2,\mu,\gamma}}
\end{equation}
The linear relations for $D_{2,\beta}$ are the same as those for $D_{1,\alpha}$ except that the one coming from the $\IZ_2$ fixed line is absent: 
\begin{eqnarray}
  \label{eq:Z6Irels2}
  R_2&=&6\,D_{{2,1}}+\sum_{\gamma=1}^3 E_{{1,\gamma}}+2\, \sum_{\mu=1,3} \sum_{\gamma=1}^3 E_{{2,\mu,\gamma}}+3\, \sum_{\alpha=1}^2 E_{{3,\alpha}},\nonumber\\
  R_2&=&3\,D_{{2,2}}+\sum_{\mu=2,4,5} \sum_{\gamma=1}^3 E_{{2,\mu,\gamma}}.
\end{eqnarray}
Finally, the relations for $D_{3,\gamma}$ are again obtained from~(\ref{eq:lineqsixI}):
\begin{equation}
  \label{eq:Z6Irels3}
  R_3=3\,D_{{3,\gamma}}+2\, E_{1,\gamma} + \sum_{\mu=1}^5 E_{{2,\mu,\gamma}} \qquad \gamma=1,\dots,3.
\end{equation}
Now, we are ready to compute the intersection ring. The polyhedra are those given in Section \ref{exAint}.

Solving the overdetermined system of linear equations then yields the intersection ring of $X$ in the basis $\{R_i, E_{k\alpha\beta\gamma}\}$:
\begin{align}
  \label{eq:ringZ6I}
  R_1R_2R_3 &= 18, & R_3E_{3,1}^2 &= -2, & R_3E_{3,2}^2 &= -6, & E_{1,\gamma}^3 &= 8, \notag\\
  E_{1,\gamma}^2 E_{2,1,\gamma} &= 2, & E_{1,\gamma}E_{2,1,\gamma}^2 &= -4, & E_{2,1\gamma}^3 &= 8, & E_{2,\mu,\gamma}^3 &= 9, \notag\\
  E_{2,1,\gamma}E_{3,1}^2 &= -2, & E_{3,1}^3 &= 8,
\end{align}
for $\mu=2,\dots,5$, $\gamma=1,2,3$. Here we have given only the nonvanishing intersection numbers and those involving the $D_{i\alpha}$ can be obtained using the linear relations~(\ref{eq:Z6IrelD11}) to~(\ref{eq:Z6Irels3}).

\subsection{Divisor topologies}

The divisor topologies are essentially the same as those of  $\IZ_{6-I}$ on $G_2^2\times SU(3)$, see Section \ref{sec:divsixiglobal}.

\subsection{The orientifold}

We examine first the orbifold phase. We have 64 O3--planes from the action of $I_6$. From the $\IZ_2$--twist $I_6\,\theta^3$, we get one O7--plane at $z^3=0$. The 64 O3--planes fall into 22 conjugacy classes under the orbifold group. $(0,0,0)$ is alone in its equivalence class, while all other points are in orbits of length 3. 

Comparison with the fixed set configuration (Figure~\ref{fig:ffixedi}) shows that the O3--planes sit on top of the $\theta^3$--fixed lines. The fixed points located in the $z^3=0$ plane lie on an $O7$ plane, the others do not.

For this example we have $h^{1,1}_{-}=6$. The fixed points except for the one at $(0,0,0)$ fall into equivalence classes of length two under $I_6$. They partly coincide with the equivalence classes under the orbifold group. Two of the divisors in $H^{1,1}_{-}$ arise from the two $\IC^3/\IZ_{6-I}$--patches at $z^3\neq0$ which are being interchanged, the remaining four come from the $\IC^3/\IZ_3$ patches in these two planes. The fixed line without fixed point is invariant under the orientifold action, hence $h^{2,1}_+=1$.

We will now discuss the orientifold of the resolved orbifold. For the local involution $\cal I$ in the $\IC^3/\IZ_{6-I}$--patches, there are four possibilities which lead to solutions of the right dimensionality (i.e. O3-- and O7--planes):
\begin{eqnarray}
(1)\quad {\cal I}(z,y)&=&(-z^1,-z^2, -z^3, y^1,y^2,y^3),\cr
(2)\quad {\cal I}(z,y)&=&(-z^1,-z^2, -z^3, y^1,-y^2,-y^3),\cr
(3)\quad {\cal I}(z,y)&=&(-z^1,-z^2, -z^3, -y^1,y^2,-y^3),\cr
(4)\quad {\cal I}(z,y)&=&(-z^1,-z^2, -z^3, -y^1,-y^2,y^3).
\end{eqnarray}
They lead to the following two distinct solutions:
\begin{eqnarray}
(1),\ (3)\quad \{y^2=0\} \cup \{z^3=0\},\cr
(2),\ (4)\quad \{y^1=0\} \cup \{y^3=0\}.
\end{eqnarray}
Choosing the simplest possibility for $\cal I$, namely $(1)$, we have to solve
\begin{equation}
  \label{OZsixi}
  {(-z, y)=(\lambda_1\,z^1,\,\lambda_1\,z^2,\,\lambda_2\,z^3,\, {\lambda_3\over\lambda_2^2}\,y^1,\,{\lambda_2\over \lambda_3^2}\,y^2,\,{\lambda_3\over\lambda_1^2}\,y^3).}
\end{equation}
We find two solutions which are in the allowed set of the toric variety:
\begin{equation}{a)\quad y^2=0\ {\rm with}\ \lambda_1=\lambda_2=-1,\, \lambda_3=1.}\end{equation}
This corresponds to the whole exceptional divisor $E_2$ being fixed, therefore this gives an O7--plane. This solution does not lead to any global consistency conditions since the other patches do not see it.
\begin{equation}{b)\quad z^3=0\ {\rm with}\ \lambda_1=-1,\, \lambda_2= \lambda_3=1.}\end{equation}
This corresponds an O7--plane on the divisor $D_3$. This solution must be compatible with the solutions of the other patches which lie in the same plane.

We will now check the consistency of this solution with the restriction of the $\IC^3/\IZ_{6-I}$ patch to the $\IC^2/\IZ_2$ fixed line. The latter is described by the coordinates $z^1,\,z^2$ and $y^3$. Choosing e.g. $\lambda_2=\lambda_3=1$ in the scaling action produces the restriction. Since neither of the two coordinates appearing in the above solutions are contained in the fixed line, no restrictions on the $\lambda_i$ arise.
The equation we must solve for the fixed line is
\begin{equation}\label{OZsixitwo}{(-z^1, -z^2, y^3)=(\lambda_1\,z^1,\,\lambda_1\,z^2,\,\frac{1}{\lambda_1^2}\,y^3).}\end{equation}
It is trivially fulfilled by
$$\lambda_1=-1,$$
and therefore also does not lead to any more restrictions on the solutions for the $\IC^3/\IZ_{6-I}$ patch. The only thing left to check is the compatibility of the O--plane solutions of the other patches with the O7--planes on the $D_{3, \gamma}$. 
For this we must examine the $\IC^3/\IZ_3$--patches. The details of their resolution can be found in Appendix \ref{app:rzthree}. 
There are two possible choices for the local involution:
\begin{eqnarray}
(1)\quad {\cal I}(z,y)&=&(-z^1,-z^2, -z^3, y),\cr
(2)\quad {\cal I}(z,y)&=&(-z^1,-z^2, -z^3, -y).
\end{eqnarray}
$(1)$ leads to the equation (see~(\ref{rescalesthree}))
\begin{equation}\label{OZthreea}{(-z^1,-z^2,-z^3, y)=(\lambda\,z^1,\,\lambda\,z^2,\,\lambda\,z^3,\, {1\over\lambda^3}\,y).}\end{equation}
with the solution 
$$y=0,\quad \lambda=-1.$$
$(2)$ leads to the equation
\begin{equation}\label{OZthree}{(-z^1,-z^2,-z^3, -y)=(\lambda\,z^1,\,\lambda\,z^2,\,\lambda\,z^3,\, {1\over\lambda^3}\,y).}\end{equation}
For this choice of involution, (\ref{OZthree}) is trivially fulfilled by
$$\lambda=-1,$$
without any restriction on the coordinates. 
Since both solutions do not lead to any further restriction of $z^3$, they are also consistent with $z^3=0$, i.e. an O7--plane on $D_3$.

For the fixed lines without fixed points on them, we simply solve
\begin{equation}\label{OZtwo}{(-z^1, -z^2,-z^3, y)=(\lambda\,z^1,\,\lambda\,z^2,\,z^3,\, {1\over\lambda^2}\,y).}\end{equation}
This gives one allowed solution, $z^3=0,\ \lambda=-1$, corresponding to an O7--plane at the locations of the fixed points in $z^3$ direction.

In total, there are three O7--planes on the $D_{3,\gamma}$--planes and three O7--planes on the $E_{2, \gamma}$ divisors. In comparison with the O--plane configuration at the orbifold point, we see that there is no continuous limit since the two O7--planes on $D_{3,2}$ and $D_{3,3}$ do not appear at the orbifold point.
Before the blow--up, we had 22 equivalence classes of O3--planes. The one at $(0,0,0)$ coincides with a fixed point and is not present after the blow--up, since no O3--brane solutions appear in the resolved patch. The others lie away from the local patches. Therefore we are only left with 21 O3--planes.

The modified intersection numbers are (cf. (\ref{eq:ringZ6I}))
\begin{align}
  \label{eq:ringZ6IO}
  R_1R_2R_3 &= 9, & R_3E_{3,1}^2 &= -1, & R_3E_{3,2}^2 &= -3, & E_{1,\gamma}^3 &= 4, \notag\\
  E_{1,\gamma}^2 E_{2,1,\gamma} &= 2, & E_{1,\gamma}E_{2,1,\gamma}^2 &= -8, & E_{2,1\gamma}^3 &=32, & E_{2,\mu,\gamma}^3 &= 9/2, \notag\\
  E_{2,1,\gamma}E_{3,1}^2 &= -2, & E_{3,1}^3 &= 3.
\end{align}


\section{The $Z_{6-I}$ orbifold on $G_2^2\times SU(3)$}


\subsection{Metric, complex structure and moduli}

See Sections \ref{sec:exa}, \ref{sec:exai}, and \ref{sec:exaii}. 

\subsection{Fixed sets}

See Section \ref{sec:schematic}.

\subsection{The gluing procedure}

See Section \ref{sec:exsixiglue}.

\subsection{The intersection ring}

See Sections \ref{sec:relexA}   and \ref{exAint}.

\subsection{Divisor topologies}

See Section \ref{sec:divsixiglobal}

\subsection{The orientifold}

This case differs from $\IZ_{6-I}$ on $G_2\times SU(3)^2$ only in the number of $\IZ_2$ fixed lines. All that was said in the last subsection applies in this case as well. There are two differences: One arises for the O7--planes at the orbifold point, where we have four which fall into two equivalence classes. The other for the fixed lines without fixed points. There are now five of them which are invariant under the orientifold action, hence $h^{2,1}_+=5$.


\section{The $\IZ_{6-II}$--orbifold on  $SU(2)\times SU(6)$}\label{sec:Z6IIonSU2xSU6}


\subsection{Metric, complex structure and moduli}\label{app:z6iisu2su6m}

On the root lattice of
$SU(6)\times SU(2)$, 
the twist $Q$ acts on the six roots $e_i$ in the following way:
\begin{eqnarray}\label{q6iisu2su6}
Q\ e_i&=&e_{i+1},\ \ \ i=1,\ldots 4 ,\cr
Q\ e_5&=&-e_1-e_2-e_3-e_4-e_5 ,\cr
Q\ e_6&=&-e_6 .\end{eqnarray}
The twist $Q$ allows for five independent real deformations of the metric $g$
and three real deformations  of the
anti--symmetric tensor $b$. As before, these results follow from solving the equations
$Q^tg\,Q=g$ and $Q^tb\,Q=b$ which leads to:
{\arraycolsep2pt
\begin{equation}
g=\!\left(\begin{array}{cccccc}
R_1^2&R_1^2\cos\theta_{45}& R_1^2\cos\theta_{35}&x&R_1^2\cos\theta_{35}&R_1R_6\cos\theta_{56}\cr
R_1^2\cos\theta_{45}&R_1^2&R_1^2\cos\theta_{45}&R_1^2\cos\theta_{35}&x&-R_1R_6\cos\theta_{56}\cr
R_1^2\cos\theta_{35}&R_1^2\cos\theta_{45}&R_1^2&R_1^2\cos\theta_{45}&R_1^2\cos\theta_{35}&R_1R_6\cos\theta_{56}\cr
x&R_1^2\cos\theta_{35}&R_1^2\cos\theta_{45}&R_1^2&R_1^2\cos\theta_{45}&-R_1R_6\cos\theta_{56}\cr
R_1^2\cos\theta_{35}&x&R_1^2\cos\theta_{35}&R_1^2\cos\theta_{45}&R_1^2&R_1R_6\cos\theta_{56}\cr
R_1R_6\cos\theta_{56}&-R_1R_6\cos\theta_{56}&R_1R_6\cos\theta_{56}&-R_1R_6\cos\theta_{56}&R_1R_6\cos\theta_{56}&R_6^2
\end{array}\right),\end{equation}}
$x=-R_1^2\,(1+2\,\cos\theta_{35}+2\,\cos\theta_{45}),$ 
with the arbitrary real parameters $R_1^2,R_6^2,\theta_{35},\theta_{45},\theta_{56}$  and 
\begin{equation}
b=\left(\begin{array}{cccccc}
0&b_1&b_2&0&-b_2&b_3\cr
-b_1&0&b_1&b_2&0&-b_3\cr
-b_2&-b_1&0&b_1&b_2&b_3\cr
0&-b_2&-b_1&0&b_1&-b_3\cr
b_2&0&-b_2&-b_1&0&b_3\cr
-b_3&b_3&-b_3&b_3&-b_3&0\end{array}\right)\end{equation}
with the arbitrary real parameters $b_1,b_2,b_3$.
The complex coordinates are
\begin{eqnarray}\label{worki}
z^1&=&\tfrac{1}{\sqrt3}\,(x^1+e^{2\pi i/6} x^2+e^{2\pi i/3}\, x^3-x^4+e^{2\pi i/3}\,x^5),\cr
z^2&=&\tfrac{1}{2\sqrt2}\,(x^1+e^{2\pi i/3} x^2+e^{-2\pi i/3}\,x^3+x^4+e^{2\pi i/3}\,x^5) ,\cr
z^3&=&\tfrac{1}{\sqrt{\im\,\Uc^3}}\ \lf[
\tfrac{1}{3}\,(x^1-x^2+x^3-x^4+x^5)+\Uc^3\, x^6\ \ri],\end{eqnarray}
with the complex structure modulus $\Uc^3$:
\begin{equation}\label{complexstri}{
\Uc^3=\fc{R_6}{R_1}\,\fc{\cos\theta_{56}+i\,\fc{1}{\sqrt3}
\sqrt{1+2\,\cos\theta_{35}-3\, \cos\theta_{56}^2}}{1+2\,\cos\theta_{35}}.}\end{equation}
The three invariant 2-forms of the real cohomology are in this case
\begin{eqnarray}\label{invrealb}
\om_1&=&dx^1\wedge dx^2+dx^2\wedge dx^3+dx^3\wedge dx^4+dx^4\wedge dx^5,\cr
\om_2&=&dx^1\wedge dx^3-dx^1\wedge dx^5+dx^2\wedge dx^4+dx^3\wedge dx^5,\cr
\om_3&=&dx^1\wedge dx^6-dx^2\wedge dx^6+dx^3\wedge dx^6-dx^4\wedge dx^6+dx^5\wedge dx^6.
\end{eqnarray}
The three K\"ahler moduli $\Tc^i$ are
\begin{eqnarray}\label{z6iisu2su6k}
\Tc^1&=&b_1-i\tfrac{4}{\sqrt3}\,R_1^2\,(-1+\cos\theta_{35})\,,\nonumber\\[2pt]
\Tc^2&=&b_2+i\,\tfrac{4}{\sqrt3}\,R_1^2\,(1+2\,\cos\theta_{35}+3\,\cos\theta_{45}) ,\nonumber\\[2pt]
\Tc^3&=&b_3+i\,\tfrac{5}{\sqrt6}\,R_1R_6\,\sqrt{-1+4\,\cos\theta_{45}-3\,\cos2\,\theta_{56}}.
\end{eqnarray}

\subsection{Fixed sets}

This being another $\IZ_6$--twist, we need to look again only at the $\theta$-, $\theta^2$- and $\theta^3$-twisted sectors.

Table \ref{fssixiia} summarizes the important data of the fixed sets. The invariant subtorus under $\theta^2$ is $(x^5,0,x^5,0,x^5,x^6)$, corresponding to $z^3=\,$ invariant,  the invariant subtorus under $\theta^3$ is $(x^4,x^5,0,x^4,x^5,0)$, corresponding to $z^2=\,$ invariant.

\begin{table}[h!]\begin{center}
\begin{tabular}{|c|c|c|c|}
\hline
\ Group el.& Order & Fixed Set& Conj. Classes \cr
\hline
\noalign{\hrule}\noalign{\hrule}
$\theta$&6&12\ {\rm fixed\ points} &\ 12\cr
$\theta^2$&3      &3\ {\rm fixed\ lines} &\ 3\cr
$\theta^3$&2      &4\ {\rm fixed\ lines} &\ 4\cr
\hline
\end{tabular}
\caption{Fixed point set for $\IZ_{6-II}$--orbifold on  $SU(2)\times SU(6)$.}\label{fssixiia}
\end{center}\end{table}

\begin{figure}[h!]
\begin{center}
\includegraphics[width=85mm]{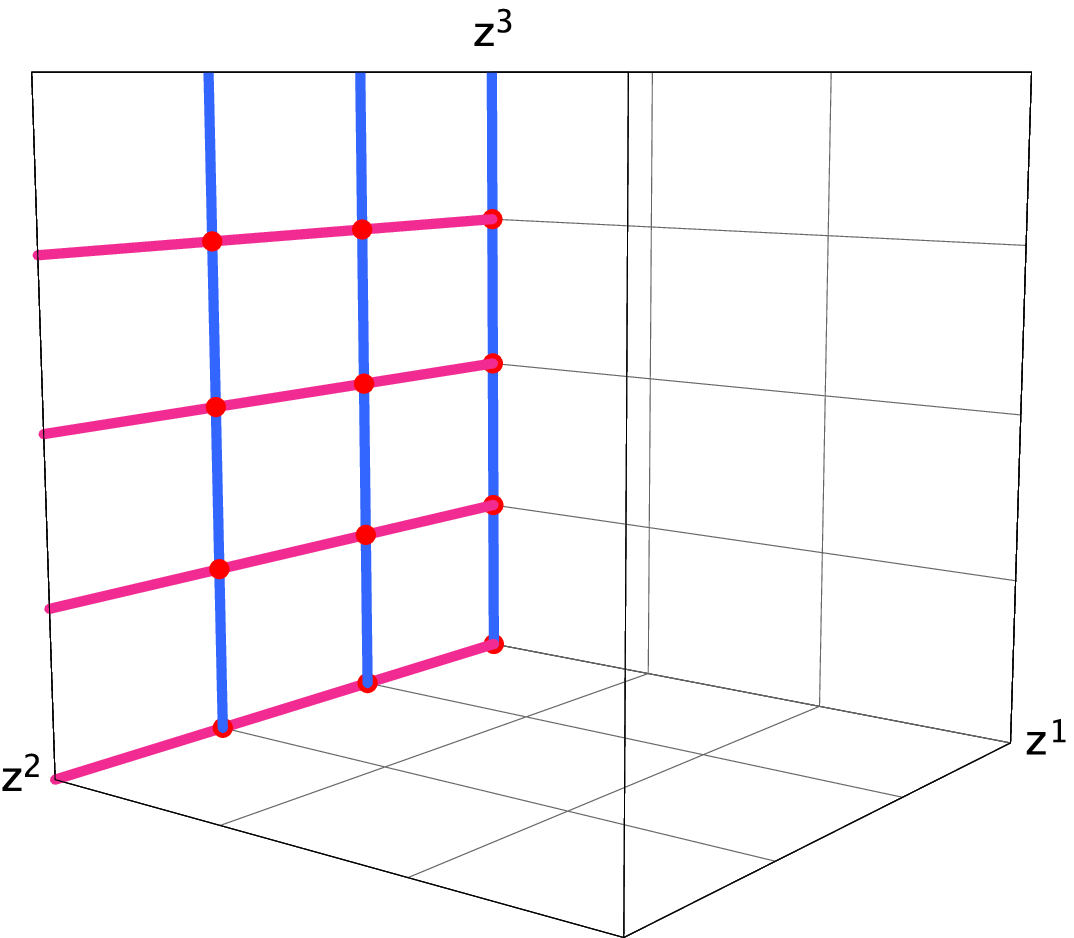}
\caption{Schematic picture of the fixed set configuration of $\IZ_{6-II}$ on $SU(2)\times SU(6)$}\label{ffixsixiiaa}
\end{center}
\end{figure}
Figure \ref{ffixsixiiaa} shows the configuration of the fixed sets in a schematic way, where each complex coordinate is shown as a coordinate axis and the opposite faces of the resulting cube of length 1 are identified. 


\subsection{The gluing procedure}

In this model, there are 12 local $\IZ_{6-II}$--patches which each sit at the intersection of two fixed lines, 3 fixed lines originating from $\IZ_3$ and 4 fixed lines under the $\IZ_2$--element. 
For the labeling of the exceptional divisors, we need Figure \ref{fsixii}.  For each of the 12 patches, we get one compact divisor $E_1$, we label them $E_{1,\alpha\beta\gamma}$, with $\alpha=1,\,\beta=1,2,3,\ \gamma=1,...,4$. The divisor $E_3$ is identified with the exceptional divisor on the ${\IC}^2/\IZ_2$--patch, therefore there are four of them: $E_{3,\alpha\gamma}$. Of the divisors $E_2,\ E_4$, we get three each: $E_{2,\alpha\beta},\ E_{4,\alpha\beta}$. They are identified with the two exceptional divisors of the ${\IC}^2/\IZ_3$--patch. Since in this lattice $\alpha=1$, we suppress the label $\alpha$ in the following.
In total, there are $12\cdot 1+3\cdot 2+4\cdot 1=22$ exceptional divisors.
There are 8 fixed planes with their associated divisors: $D_1,\,D_{2,\beta},\ \beta=1,2,3,\, D_{3,\gamma},\ \gamma=1,...,4$. The inherited divisors are $R_i= \{(z^i)^6 = c^6\},\ i=1,2,3$. 

On this lattice, there are no fixed lines without fixed points on them, so $h^{(2,1)}_{tw}=0$.

\subsection{The intersection ring}

The intersection ring is discussed together with the intersection ring of the orientifold in Section \ref{sec:sixiiOintersections}.

\subsection{Divisor topologies}

See Sections \ref{sec:divsixii} and \ref{sec:divsixiiglobal}.

\subsection{The orientifold}

See examples in Chapter  \ref{sec:orientifold}.


\section{The $\IZ_{6-II}$ orbifold on $SU(3)\times SO(8)$}\label{sec:Z6IIonSU3xSO8}


\subsection{Metric, complex structure and moduli}

On the root lattice of
$SU(3)\times SO(8)$, 
the twist $Q$ acts on the six roots $e_i$ in the following way:
\begin{eqnarray}
Q\ e_1&=&e_2,\quad Q\ e_2=e_1+e_2+e_3+e_4 ,\cr
Q\ e_3&=&-e_1-e_2-e_3,\quad Q\ e_4=-e_1-e_2-e_4,\cr
Q\ e_5&=&e_6,\quad Q\ e_6=-e_5-e_6\ .\end{eqnarray}
The twist $Q$ allows for five independent real deformations of the metric $g$
and three real deformations  of the
anti--symmetric tensor $b$:
\begin{equation}{
g=\left(\begin{array}{cccccc}
x&y&R_1^2\cos\theta_{13}&R_3^2+R_1^2(-1+\cos\theta_{13})&0&0\cr
y&x&z&z&0&0\cr
R_1^2\cos\theta_{13}&z&R_1^2&R_1R_3\cos\theta_{34}&0&0\cr
R_3^2+R_1^2(-1+\cos\theta_{13})&z&R_1R_3\cos\theta_{34}&R_3^2&0&0\cr
0&0&0&0&R_5^2&-\half R_5^2\cr
0&0&0&0&-\half R_5^2&R_5^2\end{array}\right),}\end{equation}
with $x=R_1^2\cos\theta_{13}+R_3(R_3+R_1\cos\theta_{34}),\ y=-\half (R_1^2-2R_3^2-R_1(3R_1\,\cos\theta_{13}+R_3\cos\theta_{34})),\ z=-\half R_1(R_1+R_1\cos\theta_{13}+R_3\cos\theta_{34})$ 
and the arbitrary real parameters $R_1^2,\, R_3^2\,R_5^2,\,\theta_{13 },\,\theta_{34}$  and 
\begin{equation}{
b=\left(\begin{array}{cccccc}
0&b_1+b_2&-b_2&b_2&0&0\cr
-b_1-b_2&0&b_1+2\,b_2&b_1&0&0\cr
b_2&-b_1-2\,b_2&0&b_2&0&0\cr
-b_2&-b_1&-b_2&0&0&0\cr
0&0&0&0&0&b_3\cr
0&0&0&0&-b_3&0\end{array}\right).}\end{equation}
This leads to the complex structure
\begin{eqnarray}\label{cplxabcd}
z^1&=&x^1+e^{2\pi i/6}\,x^2-x^3-x^4,\quad z^2=x^5+e^{2\pi i/3}\,x^6,\cr
z^3&=&{1\over\sqrt{{\rm Im}\,{\cal U}^3}}\,(x^1-x^2+x^3+{\cal U}^3\,(x^1-x^2+x^4)),
\end{eqnarray}
with 
\begin{eqnarray}\label{U}
{\cal U}^3&=&-2+{1\over R_1\,(2+\cos\theta_{13}-R_3\cos\theta_{34})}\left[\,3 \,R_1\,(1+\cos\theta_{13})-i\,\sqrt{3}\,(R_1)^{-1/2}\times\right.\notag\\[1mm]
&&\times(\,2\,R_1R_3^2-R_1^3+R_1\cos\theta_{13}\,(R_1^2+R_3^2-2R_2R_3\cos\theta_{34})\notag\\[2mm]
&&\left.-R_3\cos\theta_{34}(R_3^2-R_1^2+R_1R_3\cos\theta_{34}))^{1/2}\right].
\end{eqnarray}
The invariant 2--forms of the real cohomology are
\begin{eqnarray}\label{invreal}
\om_1&=&dx^1\wedge dx^2+dx^2\wedge dx^3+dx^2\wedge dx^4,\cr
\om_2&=&dx^1\wedge dx^2-dx^1\wedge dx^3+dx^1\wedge dx^4+2\,dx^2\wedge dx^3+dx^3\wedge dx^4,\cr
\om_3&=&dx^5\wedge dx^6.
\end{eqnarray}
With the invariant 2--forms above, we find the following K\"ahler moduli:
\begin{eqnarray}
{\cal T}^1&=&\tfrac{\sqrt3}{2}\,R_1\,(R_1\,(1-\cos\theta_{13})+R_3\,\cos\theta_{34})),\cr
{\cal T}^2&=&\tfrac{1}{2\sqrt3 }\,(-3\,R_1^2\,(-1+\cos\theta_{13})+3\,R_1R_3\cos\theta_{34}+10\sqrt{R_1}\times\cr
&&\times(2\,R_1R_3^2-R_1^3+R_1\cos\theta_{13}\,(R_1^2+R_3^2-2R_2R_3\cos\theta_{34})\cr
&&-R_3\cos\theta_{34}(R_3^2-R_1^2+R_1R_3\cos\theta_{34}))^{1/2},\cr
{\cal T}^3&=&\tfrac{\sqrt3}{2}\,R_5^2.
\end{eqnarray}

\subsection{Fixed sets}

\begin{figure}[h!]
\begin{center}
\includegraphics[width=85mm]{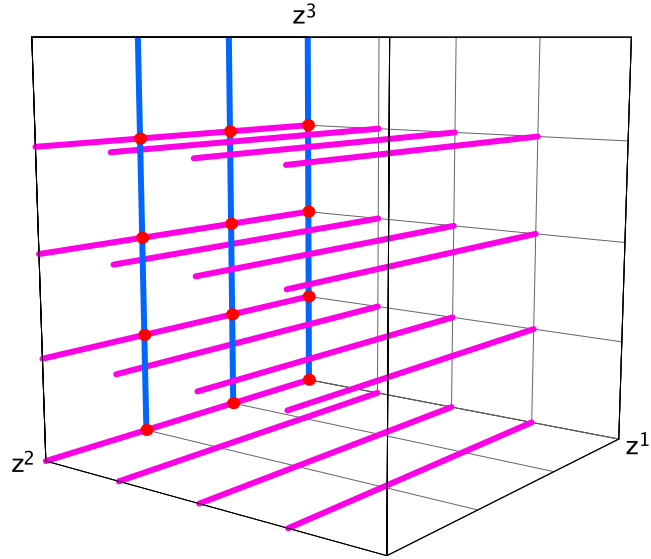}
\caption{Schematic picture of the fixed set configuration of $\IZ_{6-II}$ on $SU(3)\times SO(8)$}\label{ffixsixiiaaa}
\end{center}
\end{figure}

Here, the analysis of the fixed point set is very similar to the previous example and we will only point out the differences.  As for the fixed point set, the only change occurs in the $z^1$- direction. Apart from $\zf{1}{1}=0$, we now have $\zf{1}{2}=\half({1\over\sqrt3}\,e^{\pi i/6}),\ \zf{1}{4}=\half,\ \zf{1}{6}=\half(1+{1\over\sqrt3}\,e^{\pi i/6})$, at which we have further $\IZ_2$ fixed lines in the $z^2$ direction. In addition, the order three element maps these points as $\half({1\over\sqrt3}\,e^{\pi i/6}) \to 1/2 \to \half(1+{1\over\sqrt3}\,e^{\pi i/6}) \to \half({1\over\sqrt3}\,e^{\pi i/6})$. The resulting conjugacy classes are
\begin{align}
  \label{eq:classessixiiii}
  \alpha=1:\; & \notag\\
  \gamma=1:\; & (0, z^2, 0)\quad \gamma=2:\;\ (0,z^2, \tfrac{1}{2})\quad \gamma=3:\;\ (0,z^2,\tfrac{1}{2} U^3)\quad \gamma=4:\; \ (0,z^2,\tfrac{1}{2}(1+U^3))\notag\\
  \alpha=2:\; & \notag\\
  \gamma=1:\; & (\tfrac{1}{2}, z^2, 0),\ (\tfrac{1}{2\sqrt3}\,e^{\pi i/6}, z^2,0),\ (\tfrac{1}{2}\,(1+\tfrac{1}{\sqrt3}\,e^{\pi i/6}), z^2, 0)\notag\\
  \gamma=2:\; & (\tfrac{1}{2}, z^2,  \tfrac{1}{2}),\ (\tfrac{1}{2\sqrt3}\,e^{\pi i/6}, z^2, \tfrac{1}{2}),\ (\tfrac{1}{2}\,(1+\tfrac{1}{\sqrt3}\,e^{\pi i/6}), z^2,  \tfrac{1}{2})\notag\\
  \gamma=3:\; & (\tfrac{1}{2}, z^2,\tfrac{1}{2} U^3),\ (\tfrac{1}{2\sqrt3}\,e^{\pi i/6}, z^2,\tfrac{1}{2} U^3),\ (\tfrac{1}{2}\,(1+\tfrac{1}{\sqrt3}\,e^{\pi i/6}), z^2,\tfrac{1}{2} U^3)\notag\\
  \gamma=4:\; & (\tfrac{1}{2}, z^2, \tfrac{1}{2}(1+U^3)),\ (\tfrac{1}{2\sqrt3}\,e^{\pi i/6}, z^2,\tfrac{1}{2}(1+U^3)),\ (\tfrac{1}{2}\,(1+\tfrac{1}{\sqrt3}\,e^{\pi i/6}), z^2, \tfrac{1}{2}(1+U^3)).
\end{align}
Table~\ref{fssixiaaa} summarizes the relevant data of the fixed point set. The invariant subtori under $\theta^2$ and $\theta^3$ are $(x^3+x^4,0,x^3,x^4,0,0)$ and $(0,0,0,0,x^5,x^6)$, respectively.
\begin{table}[h!]\begin{center}
\begin{tabular}{|c|c|c|c|}
\hline
Group el.& Order &Fixed Set& Conj. Classes \cr
\hline
\noalign{\hrule}\noalign{\hrule}
$ \theta$& 6   &12  fixed points & 12\cr
$\theta^2$& 3   &3  fixed lines & 3\cr
$\theta^3$& 2 &16  fixed lines & 8\cr
\hline
\end{tabular}
\caption{Fixed point set for $\IZ_{6-II}$ orbifold on $SU(3)\times SO(8)$}\label{fssixiaaa}
\end{center}\end{table}
Figure \ref{ffixsixiiaaa} shows the configuration of the fixed point set in a schematic way.

\subsection{The gluing procedure}

On this lattice, there are four $\IZ_2$ fixed lines without fixed points on them, therefore $h^{(2,1)}_{tw}=4$.
Here we have again 12 $\IC^3/\IZ_{6-II}$ patches which each sit at the intersection of two fixed lines, and in addition 3 fixed lines originating from the order 3 element as well as 8 classes of fixed lines from the order two element. The fixed points yield the same exceptional divisors $E_{1,\beta\gamma}, E_{2,\beta}, E_{3,\alpha\gamma}, E_{4,\beta}$, $\alpha=1$, $\beta=1,2,3$, $\gamma=1,\dots,4$, as in Appendix~\ref{sec:Z6IIonSU2xSU6}. Moreover, there are four exceptional divisors $E_{3,2\gamma}$ coming from the additional $\IC^2/\IZ_2$ fixed lines. These are the invariant combinations $E_{3,2\gamma}=\sum_{\alpha=2,4,6} \Et_{3,\alpha\gamma}$, where $\Et_{3,\alpha\gamma}$ are the representatives on the cover. This gives a total of $12\cdot 1+3\cdot 2+8\cdot 1=26$ exceptional divisors. On this lattice there are four classes of $\IC^2/\IZ_2$ fixed lines without fixed points on them, therefore by~\eqref{eq:h21tw} we have $h^{2,1}_{\rm twist.}=4$. We have 9 fixed planes with their associated divisors: $D_{1\alpha},\ \alpha=1,2,\,D_{2\beta},\ \beta=1,2,3,\, D_{3\gamma},\ \gamma=1,...,4$. Here, $D_{1,2}$ is the invariant combination $D_{1,2}=\sum_{\alpha=2,4,6} \Dt_{1\alpha}$ of the representatives $\Dt_{1\alpha}$ on the cover.

\subsection{The intersection ring}

The global linear relations are the same as those in~\eqref{eq:globalrelsixiiaO} except for a new relation involving $R_1$ and $D_{1,2}$, as well as an additional term involving $E_{3,2,\gamma}$ in the relations for $R_3$:
\begin{align}
  \label{eq:globalrelsixiib}
  R_1&=6\,D_{{1,1}}+3\,\sum_{\gamma=1}^4 E_{{3,1,\gamma}}+\sum_{\beta=1}^3\sum_{\gamma=1}^4 E_{{1,\beta\gamma}}+2\sum_{\beta=1}^3 \left( 2\,E_{{2,\beta}}+4\,E_{{4,\beta}}\right), \cr
  R_1&=2\,D_{{1,2}}+\sum_{\gamma=1}^4 E_{{3,2,\gamma}},\cr
  R_2&=3\,D_{{2,\beta}}+\sum_{\gamma=1}^4E_{{1,\beta\gamma}}+2\,E_{{2,\beta}}+E_{{4,\beta}}, \qquad \beta=1,2,3, \cr
  R_3&=2\,D_{{3,\gamma}}+\sum_{\beta=1}^3E_{{1,\beta\gamma}}+\sum_{\alpha=1}^2 E_{{3,\alpha\gamma}}, \quad\qquad \gamma=1,\dots,4.
\end{align}
We obtain the following nonvanishing intersection numbers of $X$ in the basis $\{R_i,E_{k\alpha\beta\gamma}\}$:
\begin{align}
  R_1R_2R_3&=6, & R_2E_{3,1,\gamma}^2 &= -2, & R_2E_{3,2,\gamma}^2 &= -6, & R_3E_{2,\beta}^2 &= -2, \notag\\
  R_3E_{4,\beta} &= -2, & R_3E_{2,\beta}E_{4,\beta} &= 1, & E_{1,\beta\gamma}^3&=6, & E_{2,\beta}^3& =8, \notag\\
  E_{3,1,\gamma}^3&=8, & E_{4,\beta}^3&=8, & E_{1,\beta\gamma}E_{2\beta}^2 &= -2, & E_{1,\beta\gamma}E_{3,1,\gamma}^2 &= -2, \notag\\
  E_{1,\beta\gamma}E_{4,\beta}^2 &= -2, & E_{1,\beta\gamma}E_{2,\beta}E_{4,\beta} &= 1, &  E_{2,\beta}^2E_{4,\beta} &= -2,
\end{align}

\subsection{Divisor topologies}

For the topology of the divisors there are only a few changes with respect to the lattice $SU(2)\times SU(6)$. First of all, the topology of the divisors $E_{1,\beta\gamma}$, $E_{2,\beta}$, $E_{4,\beta}$, $D_{2,\beta}$, and $D_{3,\gamma}$ are the same as in Table~\ref{tab:TopZ6II}. The divisors $E_{3,1,\gamma}$ and $D_{1,1}$ have the same topology as $E_{3,\gamma}$ and $D_1$, respectively, in that table. The topology of the new divisors $E_{3,2\gamma}$ and $D_{1,2}$ are as follows: The divisors are of type~E\ref{item:E2}) with 3 representatives, hence their topology is that of $\IP^1 \times T^2$. The topology of each representative of $D_{1,2}$ minus the fixed point set, viewed as a $T^4$ orbifold, is that of a $T^2 \times (T^2/\IZ_2 \setminus \{ 4\ \rm{pts} \})$. They are permuted under the residual $\IZ_3$ action and the 12 points fall into 3 orbits of length 1 and 3 orbits of length 3. Hence, the topology of the class is still that of a $T^2 \times (T^2/\IZ_2 \setminus \{ 4 \ \rm{pts} \})$. After the blow--up it is therefore a $\IP^1 \times T^2$. For both, $E_{3,2\gamma}$ and $D_{1,2}$, the topology is obviously independent of the choice of resolution of $\IC^3/\IZ_{6-II}$. For completeness, we display the second Chern classes in the basis $\{R_i, E_{k\alpha\beta\gamma}\}$ (for triangulation a)):
\begin{align}
  \label{eq:c2Z6IIb}
  \ch_2\cdot E_{1\beta\gamma} &= 0, & \ch_2\cdot E_{2\beta} &= -4, & \ch_2 \cdot E_{3,1\gamma} &= -4, & \ch_2\cdot E_{3,2\gamma} &= 0, & \notag\\
  \ch_2\cdot E_{4\beta} & = -4, &\ch_2 \cdot R_1 &= 0, & \ch_2 \cdot R_i &=  24. 
\end{align}

\subsection{The orientifold}
\label{sec:Z6IISU3xSO8_O}

At the orbifold point, there are 64 O3--planes which fall into 24 conjugacy classes under the orbifold group. Under $I_6\,\theta^3$ we find four O7--planes in the $(z^1,z^3)$--plane which fall into two conjugacy classes: The one at $z^2=0$ and those at $z^2=\frac{1}{2},\, \frac{\tau}{2},\, \frac{1}{2}(1+\tau)$. $h^{(1,1)}_{-}=6$ is obtained in the same way as for the $SU(2)\times SU(6)$--lattice. Similarly, everything we found for the resolved orbifold in that appendix applies here as well. There are four classes of fixed lines with fixed points which are invariant under the orientifold action, hence $h^{2,1}=4$.


\section{The  $\IZ_{6-II}$ orbifold on $SU(2)^2\times SU(3)^2$}\label{Z6IIonSU2xSU2xSU3xSU3}

\subsection{Metric, complex structure and moduli}

This time, we associate $e_1$ and $e_2$ with $\IZ_2$. The generalized Coxeter element $Q=S_1S_2S_3P_{36}P_{45}$ contains transpositions of the roots of the $SU(3)$--factors. Using (\ref{Weyl}) and the Cartan matrix of $SU(3)$, we find the following for the total twist:
\begin{eqnarray}\label{twistactionsixrii}
Q\, e_1&= &-e_1,\quad Q\,e_2=-e_2,\cr
Q\,e_3&=&-e_5,\quad Q\,e_4=-e_5+e_6,\cr
Q\,e_5&=&e_4,\quad Q\,e_6=e_3.
\end{eqnarray}
{From} $Q^tg\,Q=g$ we find the following $g$:
\begin{equation}\label{gsixii}{ g=\left(\begin{array}{cccccc}
R_1^2&R_1R_2\cos\theta_{12}&0&0&0&0\cr
R_1R_2\cos\theta_{12}& R_3^2&0&0&0&0\cr
0&0&R_2^2&-\half R_3^2&-2\,R_3^2\cos\theta_{46}&R_3^2\cos\theta_{46}\cr
0&0&-\half R_3^2&R_3^2&R_3^2\cos\theta_{46}&R_3^2\cos\theta_{46}\cr
0&0&-2\,R_3^2\cos\theta_{46}&R_3^2\cos\theta_{46}&R_3^2&-\half R_3^2\cr
0&0&R_3^2\cos\theta_{46}&R_3^2\cos\theta_{46}&-\half R_3^2&R_3^2\end{array}\right),}\end{equation}
$R_1,\, R_2,\,R_3,\,\theta_{12}$ and $\theta_{46}$ being its five real deformation parameters.
For the antisymmetric tensor $b$, we get
\begin{equation}\label{bfieldsixii}{
b=\left(\begin{array}{cccccc}
0&b_1&0&0&0&\cr
-b_1&0&0&0&0&0\cr
0&0&0&-b_3&0&b_2\cr
0&0&b_3&0&b_2&-b_2\cr
0&0&0&-b_2&0&b_3\cr
0&0&-b_2&b_2&-b_3&0\end{array}\right),}\end{equation}
with the three real parameters  $b_1,\,b_2,\,b_3$. 
This leads to the complex coordinates
\begin{eqnarray}\label{cplxabc}
z^1&=&\tfrac{1}{\sqrt3}\,(x^3+e^{2\pi i/3}\,x^4-x^5+e^{2\pi i/6}\,x^6),\nonumber\\
z^2&=&\tfrac{1}{2}\,(x^3-e^{2\pi i/6}\,x^4+x^5+e^{2\pi i/3}\,x^6),\nonumber\\
z^3&=&\tfrac{1}{\sqrt{2\,{\rm Im}\,{\cal U}^3}}(\,x^1+{\cal U}^3\,x^2\,),
\end{eqnarray}
with ${{\cal U}^3}=R_2/R_1\,e^{i\theta_{12}}$.
The invariant real 2--forms are
\begin{eqnarray}
\om_1&=&dx^1\wedge dx^2,\quad \om_2=dx^3\wedge dx^6+dx^4\wedge dx^5-dx^4\wedge dx^6,\cr
\om_3&=&-dx^3\wedge dx^4+dx^5\wedge dx^6.
\end{eqnarray}
Via $B+i\,J={\cal T}^i\,\om_i$, we find
\begin{equation}
{{\cal T}^1=b_1+R_1R_2\,\sin\theta_{12},\quad {\cal T}^2=b_2+i\,\tfrac{3\sqrt3}{2}\,R_3^2,\quad {\cal T}^3=b_3-i\,2\sqrt3\,R_3^2\,\cos\theta_{46}.}\end{equation}

\subsection{Fixed sets}

\begin{figure}[h!]
\begin{center}
\includegraphics[width=85mm]{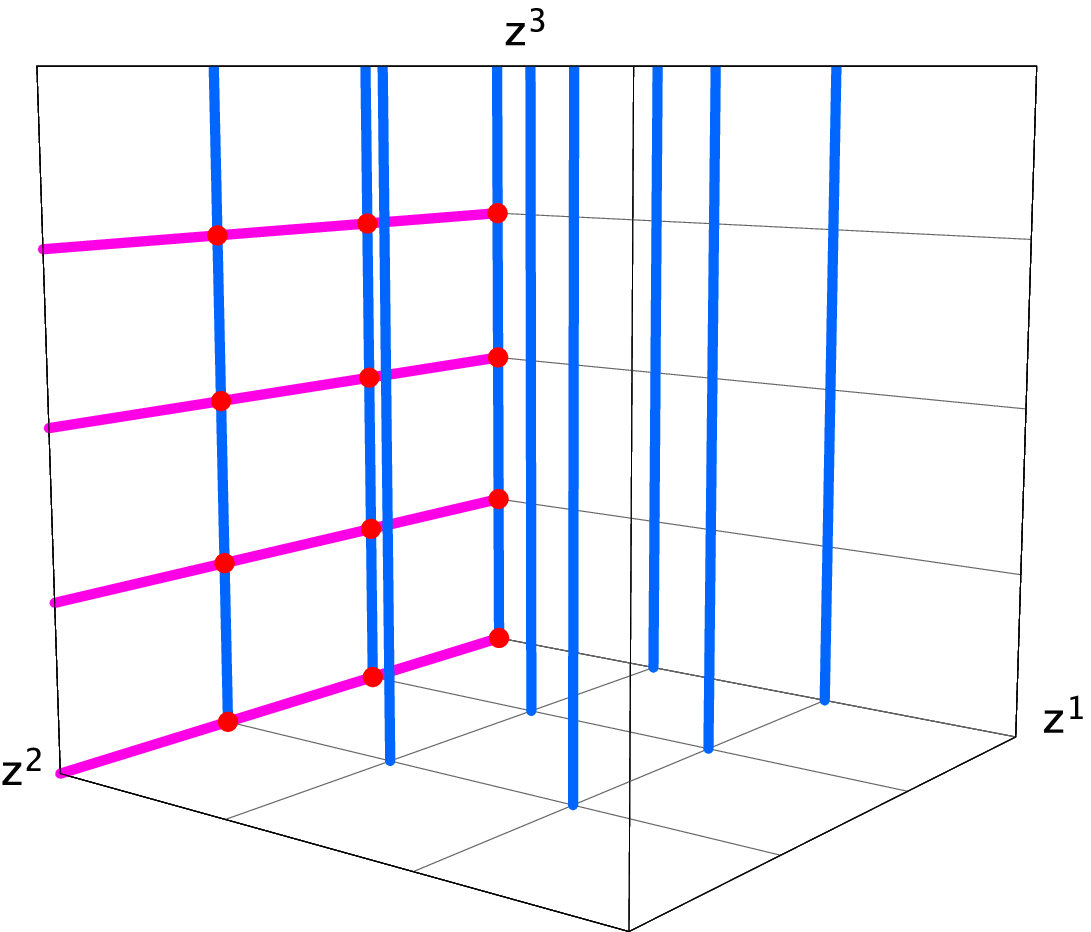}
\caption{Schematic picture of the fixed set configuration of $\IZ_{6-II}$ on $SU(2)^2\times SU(3)^2$}
\label{ffixsixiig}
\end{center}
\end{figure}

Here, the analysis of the fixed point set is very similar to the example in Appendix~\ref{sec:Z6IIonSU3xSO8} and we will only point out the differences.  The only change is that instead of $\IC^2/\IZ_2$ fixed lines in the $z^2$ direction, we now have $\IC^2/\IZ_3$ fixed lines in the $z^3$ direction which lie, apart from $\zf{1}{1}=0$, at $\zf{1}{3}=1/3$ and $\zf{1}{5}=2/3$. In addition, the order two element maps the latter two points into each other. The resulting conjugacy classes are
\begin{align}
  \label{eq:classsixii}
  \mu=1:\; &(0,0,z^3) & \mu=2:\; & (0,\tfrac{1}{\sqrt3}\,e^{\pi i/6}, z^3)\notag\\
  \mu=3:\; &(0,1+\tfrac{i}{\sqrt3}, z^3) & \mu=4:\; &(\tfrac{1}{3}, 0, z^3),\ (\tfrac{2}{3}, 0, z^3)\notag\\
  \mu=5:\; &(\frac{1}{3}, \tfrac{1}{\sqrt3}\,e^{\pi i/6}, z^3),\ (\tfrac{2}{3}, \tfrac{1}{\sqrt3}\,e^{\pi i/6}, z^3) & \mu=6:\; &(\tfrac{1}{3}, 1+\tfrac{i}{\sqrt3}, z^3),\ (\tfrac{2}{3}, 1+\tfrac{i}{\sqrt3}, z^3).
\end{align}
Table~\ref{fssixiiaaa} summarizes the important data of the fixed sets. The invariant subtori under $\theta^2$ and $\theta^3$ are $(x^1,x^2,0,0,0,0)$ and $(0,0,x^5-x^6,-x^6,x^5,x^6)$, respectively.
\begin{table}[h!]\begin{center}
\begin{tabular}{|c|c|c|c|}
\hline
Group el.& Order & Fixed Set& Conj. Classes \cr
\noalign{\hrule}\noalign{\hrule}
$\theta$& 6    &12 fixed points & 12\cr
$ \theta^2$&3      &9  fixed lines &6\cr
$ \theta^3$&2      &4 fixed lines &4\cr
\hline
\end{tabular}
\caption{Fixed point set for $\IZ_{6-II}$ on $SU(2)^2\times SU(3)^2$.}
\label{fssixiig}
\end{center}\end{table}
Figure~\ref{ffixsixiig} shows the configuration of the fixed sets in a schematic way. 

\subsection{The gluing procedure}

The fixed point set yields the same exceptional divisors $E_{1,\beta\gamma}, E_{2,\alpha\beta}, E_{3,\gamma}, E_{4,\alpha\beta}$, $\alpha=1$, $\beta=1,2,3$, $\gamma=1,\dots,4$, as in Appendix~\ref{sec:Z6IIonSU2xSU6}. Moreover, there are three pairs of exceptional divisors $E_{2,3\beta}$, $E_{4,3\beta}$ coming from the additional $\IC^2/\IZ_3$ fixed lines. These are the invariant combinations $E_{i,3\beta}=\sum_{\alpha=3,5} \Et_{i,\alpha\gamma\beta}$, $i=2,4$, where $\Et_{i,\alpha\beta}$ are the representatives on the cover. This gives a total of $12\cdot 1+6\cdot 2+4\cdot 1=28$ exceptional divisors. On this lattice there are three classes of $\IC^2/\IZ_3$ fixed lines without fixed points on them, therefore by~\eqref{eq:h21tw} we have $h^{2,1}_{\rm twist.}=6$. We have 9 fixed planes with their associated divisors: $D_{1\alpha},\ \alpha=1,3,\,D_{2\beta},\ \beta=1,2,3,\, D_{3\gamma},\ \gamma=1,...,4$. Here, $D_{1,3}$ is the invariant combination $D_{1,3}=\sum_{\alpha=3,5} \Dt_{1\alpha}$ of the representatives $\Dt_{1\alpha}$ on the cover.

\subsection{The intersection ring}

The global linear relations are the same as those in~\eqref{eq:globalrelsixiiaO} except for a new relation involving $R_1$ and $D_{1,3}$, as well as additional terms involving $E_{i,3\beta}$, $i=2,4$ in the relations for $R_2$:
\begin{align}
  \label{eq:globalrelsixiic}
  R_1&=6\,D_{{1,1}}+3\,\sum_{\gamma=1}^4 E_{{3,1,\gamma}}+\sum_{\beta=1}^3\sum_{\gamma=1}^4 E_{{1,\beta\gamma}}+\sum_{\beta=1}^3 \left( 2\,E_{{2,1,\beta}}+4\,E_{{4,1,\beta}}\right), \cr
  R_1&=3\,D_{{1,3}}+\sum_{\beta=1}^3 \left( E_{{2,3,\beta}} + 2\,E_{4,3,\beta} \right), \cr
  R_2&=3\,D_{{2,\beta}}+\sum_{\gamma=1}^4E_{{1,\beta\gamma}}+\sum_{\alpha=1,3} \left(2\,E_{{2,\alpha\beta}}+ E_{{4,\alpha\beta}}\right), \qquad \beta = 1,2,3 \cr
  R_3&=2\,D_{{3,\gamma}}+\sum_{\beta=1}^3E_{{1,\beta\gamma}}+ E_{{3,1,\gamma}}, \qquad \gamma=1,\dots,4.
\end{align}
For the nonvanishing intersection numbers of $X$ in the basis $\{R_i,E_{k\alpha\beta\gamma}\}$ we find:
\begin{align}
  R_1R_2R_3&=6, & R_2E_{3,1,\gamma}^2 &= -2, & R_3E_{2,1,\beta}^2 &= -2, & R_3E_{2,3,\beta}^2 &= -4, \notag\\
  R_3E_{4,1,\beta} &= -2, & R_3E_{4,3,\beta} &= -4, & R_3E_{2,1,\beta}E_{4,1,\beta} &= 1, & R_3E_{2,3,\beta}E_{4,3,\beta} &= 2, \notag\\  
  E_{1,\beta\gamma}^3&=6, & E_{2,1,\beta}^3& =8, & E_{3,1,\gamma}^3&=8, & E_{4,1,\beta}^3&=8, \notag\\
  E_{1,\beta\gamma}E_{2,1,\beta}^2 &= -1, & E_{1,\beta\gamma}E_{3,1,\gamma}^2 &= -1, & E_{1,\beta\gamma}E_{4,1,\beta}^2 &= -1, & E_{1,\beta\gamma}E_{2,1,\beta}E_{4,1,\beta} &= 1, \notag\\
  E_{2,1,\beta}^2E_{4,1,\beta} &= -2,
\end{align}

\subsection{Divisor topologies}

The topology of the divisors $E_{1,\beta\gamma}$, $E_{3,1,\gamma}$, $D_{2,\beta}$, and $D_{3,\gamma}$ are the same as in Table~\ref{tab:TopZ6II}. The divisors $E_{2,1,\beta}$, $E_{4,1,\beta}$, and $D_{1,1}$ have the same topology as $E_{2,\beta}$, $E_{4,\beta}$, and $D_1$, respectively, in that table. The topology of the new divisors $E_{2,3,\beta}$, $E_{4,3,\beta}$, and $D_{1,3}$ are $\IP^1 \times T^2$ and $\Bl{8}\IF_n$, respectively, independent of the choice of resolution of $\IC^3/\IZ_{6-II}$. For completeness, we display the second Chern classes in the basis $\{R_i, E_{k\alpha\beta\gamma}\}$ (for triangulation a)):
\begin{align}
  \label{eq:c2Z6IIc}
  \ch_2\cdot E_{1\beta\gamma} &= 0, & \ch_2\cdot E_{2,1\beta} &= -4, & \ch_2\cdot E_{2,3\beta} &= 0, & \ch_2 \cdot E_{3,\gamma} &= -4, & \notag\\
  \ch_2\cdot E_{4,1\beta} & = -4, &\ch_2\cdot E_{4,3\beta} & = 0, &\ch_2 \cdot R_1 &= 0, & \ch_2 \cdot R_i &=  24. 
\end{align}

\subsection{The orientifold}
\label{sec:Z6IISU2xSU3_O}

At the orbifold point, there are 64 O3--planes which fall into 16 conjugacy classes under the orbifold group, as for the lattice $SU(2)\times SU(6)$. Under $I_6\,\theta^3$ we find four O7--planes in the $(z^1,z^3)$--plane which fall into two conjugacy classes: The one at $z^2=0$ and those at $z^2=\frac{1}{2},\, \frac{\tau}{2},\, \frac{1}{2}(1+\tau)$. $h^{(1,1)}_{-}=8$ is obtained in the same way as for the lattice $SU(2)xSU(6)$ with two additional divisors coming from the new divisors $E_{2,3,\beta}$, $E_{4,3,\beta}$. Since these divisors come from fixed lines without fixed point, and latter also contribute to the twisted complex structure moduli according to the discussion in Section~\ref{sec:TwistedCplx}, some of these moduli are projected out by the induced orientifold action on $H^{2,1}_{+}$. In fact, since the three fixed lines at $(\zf{1}{2},\zf{2}{\beta})$, $\beta=1,2,3$ are identified with those at $(\zf{1}{3},\zf{2}{\beta})$, we find that $h^{2,1}_{+}=3$. Similarly, everything we found for the resolved orbifold in that appendix applies here as well.

\section{The $\IZ_{6-II}$ orbifold on  $SU(2)^2\times SU(3)\times G_2$}

\subsection{Metric, complex structure and moduli}

The twist $Q$ acts on the six roots $e_i$ in the following way:
\begin{eqnarray}\label{twistactionsixr}
Q\, e_1&=& 2\,e_1+3\,e_2,\quad Q\,e_2=-e_1-e_2,\cr
Q\,e_3&=&e_4,\quad Q\,e_4=-e_3-e_4,\cr
Q\,e_i&=&-e_i,\quad i=5,6.
\end{eqnarray}
Now we solve for the metric and antisymmetric tensor, using $Q^tg\,Q=g$ and $Q^tb\,Q=b$:
\begin{equation}{
g=\ \left(\begin{array}{cccccc}
R_1^2&-1/2R_1^2&0&0&0&0\cr
-1/2R_1^2&1/3 R_1^2&0&0&0&0\cr
0&0&R_3^2&-1/2R_3^3&0&0\cr
0&0&-1/2R_3^2&R_3^2&0&0\cr
0&0&0&0&R_5^2&R_5R_6\cos\theta_{56}\cr
0&0&0&0&R_5R_6\cos\theta_{56}&R_6^2\end{array}\right).}\end{equation}
As can be seen, $R_1,\,R_3,\, R_5,\, R_6$ and $\theta_{56}$ are (real) free parameters. The $\IZ_2$--twists leave their part of the metric completely undetermined. 
For the antisymmetric tensor $b$, we get
\begin{equation}\label{bfield}{
b=\left(\begin{array}{cccccc}
0&b_1&0&0&0&\cr
-b_1&0&0&0&0&0\cr
0&0&0&b_2&0&0\cr
0&0&-b_2&0&0&0\cr
0&0&0&0&0&b_3\cr
0&0&0&0&-b_3&0\end{array}\right),}\end{equation}
with the three real parameters  $b_1,b_2,b_3$. Three of the free parameters of $g$ can be combined with the three free parameters of $b$ into three complex K\"ahler moduli, the remaining two free parameters of $g$ form one complex structure modulus.

Using the ansatz (\ref{ansatz}) we find the following complex coordinates:
\begin{eqnarray}\label{dzsix}
z^1&=&x^1+\tfrac{1}{\sqrt3}e^{5\pi i/6}x^2,\nonumber\\[2pt]
z^2&=&x^3+e^{2\pi i/3}x^4,\nonumber\\[2pt]
z^3&=&\tfrac{1}{ \sqrt{{2\,\rm Im}\, {\cal U}^3}}(\,x^5+{\cal U}^3x^6),\quad{\rm with}\quad {\cal U}^3={R_6\over R_5}e^{i\theta_{56}}
\end{eqnarray}
the complex structure modulus.
The invariant 2-forms are in this simple case $\om_1=dx^1\wedge dx^2, \ \om_2=dx^3\wedge dx^4$ and $\om_3=dx^5\wedge dx^6$. 
Via $B+i\,J={\cal T}^i\,\om_i$, we find the three K\"ahler moduli to take the following form:
\begin{eqnarray}
\Tc^1&=&b_1+i\,\tfrac{1}{2\sqrt3}\,R_1^2,\nonumber\\[1pt]
\Tc^2&=&b_2+i\,\tfrac{\sqrt3}{2}\,R_3^2,\nonumber\\[1pt]
\Tc^3&=&b_3+i\,R_5R_6\sin\theta_{56}.\end{eqnarray}

\subsection{Fixed sets}

\begin{figure}[h!]
\begin{center}
\includegraphics[width=140mm]{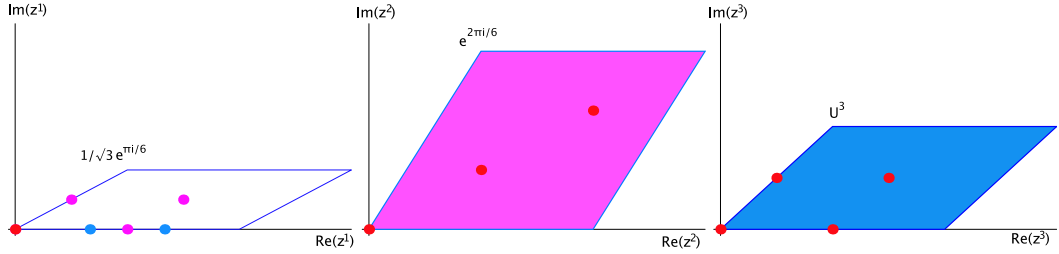}
\caption{Fundamental regions for the $\IZ_{6-II}$ orbifold on $SU(2)^2\times SU(3)\times G_2$}\label{ffusixi}
\end{center}
\end{figure}

\begin{figure}[h!]
\begin{center}
\includegraphics[width=85mm]{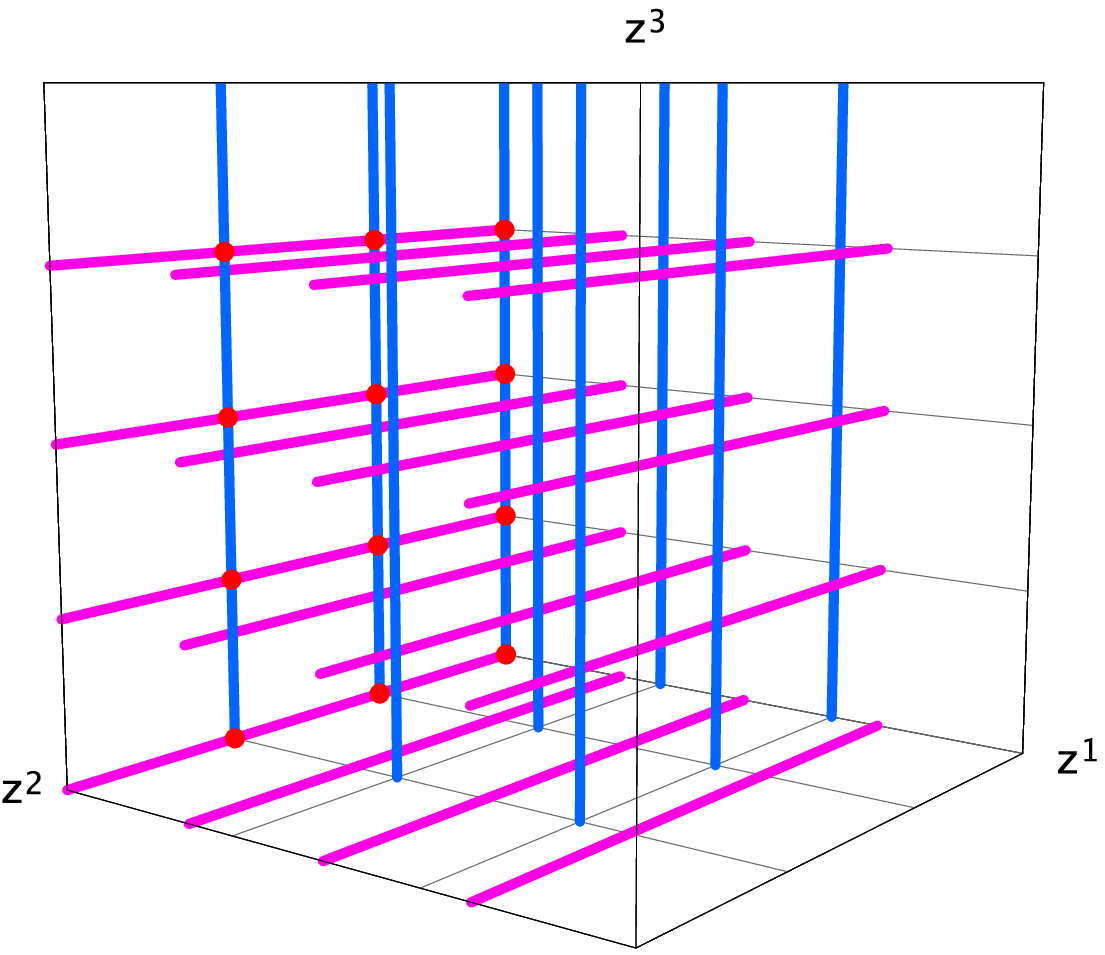}
\caption{Schematic picture of the fixed set configuration of $\IZ_{6-II}$ on $SU(2)^2\times SU(3)\times G_2$}
\label{ffixsixiia}
\end{center}
\end{figure}

Here, the analysis of the fixed point set is a combination of those in the Appendices~\ref{sec:Z6IIonSU3xSO8} and~\ref{Z6IIonSU2xSU2xSU3xSU3}. The main difference to the lattices in the Appendices~\ref{sec:Z6IIonSU2xSU6} to~\ref{Z6IIonSU2xSU2xSU3xSU3} is that the torus now factorizes into $(T^2)^3$. Figure~\ref{ffusixi} shows the fundamental regions of the three tori corresponding to $z^1,\,z^2,\,z^3$ and their fixed points. 
For $z^1$, we have 0 as the fixed point of the $\IZ_6$--twist, $0,1/3,2/3$ as the fixed points of the $\IZ_3$--twist arising in the second twisted sector and the four fixed points of the $\IZ_2$--twist arising in the third twisted sector. For $z^2$ we get the usual three fixed points of the $\IZ_3$--twist, namely $0, 1/\sqrt3 \,e^{\pi i/6}$ and $1+i/\sqrt3$, and for $z^3$ we find the four fixed points $0,\half, \half U^3, \half(1+U^3)$. Therefore, apart from $\zf{1}{1}=0$, we now have both $\zf{1}{2}=\half({1\over\sqrt3}\,e^{\pi i/6}),\ \zf{1}{4}=\half,\ \zf{1}{6}=\half(1+{1\over\sqrt3}\,e^{\pi i/6})$, at which we have further $\IZ_2$ fixed lines in the $z^2$ direction, and $\zf{1}{3}=1/3$ and $\zf{1}{5}=2/3$, at which we have further $\IZ_3$ fixed lines in the $z^3$ direction. The conjugacy classes of these fixed lines were given in~\eqref{eq:classessixiiii} and~\eqref{eq:classsixii}. Table~\ref{fssixiiaaa} summarizes the relevant data of the fixed point set. The invariant subtori under $\theta^2$ and $\theta^3$ are $(0,0,0,0,x^5,x^6)$ and $(0,0,x^3,x^4,0,0)$, respectively. 
\begin{table}[h!]\begin{center}
\begin{tabular}{|c|c|c|c|}
\hline
Group el.& Order & Fixed Set& Conj. Classes \cr
\noalign{\hrule}\noalign{\hrule}
$\theta$& 6    &12 fixed points &\ 12\cr
$ \theta^2$&3      &9  fixed lines &\ 6\cr
$ \theta^3$&2      &16 fixed lines &\ 8\cr
\hline
\end{tabular}
\caption{Fixed point set for $\IZ_{6-II}$ on $SU(2)^2\times SU(3)\times G_2$.}\label{fssixiiaaa}
\end{center}\end{table}
Figure \ref{ffixsixiia} shows the configuration of the fixed sets in a schematic way.

\subsection{The gluing procedure}

The fixed point set yields the same exceptional divisors $E_{1,\beta\gamma}, E_{2,\alpha\beta}, E_{3,\alpha\gamma}, E_{4,\alpha\beta}$, $\alpha=1,2,3$, $\beta=1,2,3$, $\gamma=1,\dots,4$, as in the Appendices~\ref{sec:Z6IIonSU3xSO8} and~\ref{Z6IIonSU2xSU2xSU3xSU3}. Here, $\alpha=1,2$ for $E_{3,\alpha\gamma}$ and $\alpha=1,3$ for $E_{2,\alpha\beta}$ and $E_{4,\alpha\beta}$. This gives a grand total of $12\cdot 1+6\cdot 2+8\cdot 1=32$ exceptional divisors. On this lattice, there are four classes of $\IC^2/\IZ_2$ fixed lines and three classes of $\IC^2/\IZ_3$ fixed lines without fixed points on them, therefore by~\eqref{eq:h21tw}, we have $h^{2,1}_{\rm twist.}=10$.

\subsection{The intersection Ring}

The global linear relations are obtained by combining~\eqref{eq:globalrelsixiib} and~\eqref{eq:globalrelsixiic}:
\begin{align}
  \label{eq:globalrelsixiid}
  R_1&=6\,D_{{1,1}}+3\,\sum_{\gamma=1}^4 E_{{3,1,\gamma}}+\sum_{\beta=1}^3\sum_{\gamma=1}^4 E_{{1,\beta\gamma}}+\sum_{\beta=1}^3 \left( 2\,E_{{2,1,\beta}}+4\,E_{{4,1,\beta}}\right), \cr
  R_1&=2\,D_{{1,2}}+\sum_{\gamma=1}^4 E_{{3,2,\gamma}}, \cr
  R_1&=3\,D_{{1,3}}+\sum_{\beta=1}^3 \left( E_{{2,3,\beta}} + 2\,E_{4,3,\beta} \right),\cr
  R_2&=3\,D_{{2,\beta}}+\sum_{\gamma=1}^4E_{{1,\beta\gamma}}+\sum_{\alpha=1,3} \left(2\,E_{{2,\alpha\beta}}+ E_{{4,\alpha\beta}}\right), \qquad \beta = 1,2,3 \cr
  R_3&=2\,D_{{3,\gamma}}+\sum_{\beta=1}^3E_{{1,\beta\gamma}}+ \sum_{\alpha=1}^2 E_{{3,\alpha,\gamma}}, \qquad \gamma=1,\dots,4.
\end{align}
We obtain the following nonvanishing intersection numbers of $X$ in the basis $\{R_i,E_{k\alpha\beta\gamma}\}$:
\begin{align}
  R_1R_2R_3&=6, & R_2E_{3,1,\gamma}^2 &= -2, & R_2E_{3,2,\gamma}^2 &= -6,& R_3E_{2,1,\beta}^2 &= -2, \notag\\
  R_3E_{2,3,\beta}^2 &= -4, & R_3E_{4,1,\beta} &= -2, & R_3E_{4,3,\beta} &= -4, & R_3E_{2,1,\beta}E_{4,1,\beta} &= 1, \notag\\ 
  R_3E_{2,3,\beta}E_{4,3,\beta} &= 2, & E_{1,\beta\gamma}^3&=6, & E_{2,1,\beta}^3& =8, & E_{3,1,\gamma}^3&=8, \notag\\
  E_{4,1,\beta}^3&=8, & E_{1,\beta\gamma}E_{2,1,\beta}^2 &= -1, & E_{1,\beta\gamma}E_{3,1,\gamma}^2 &= -1, & E_{1,\beta\gamma}E_{4,1,\beta}^2 &= -1, \notag\\
  E_{1,\beta\gamma}E_{2,1,\beta}E_{4,1,\beta} &= 1, & E_{2,1,\beta}^2E_{4,1,\beta} &= -2,
\end{align}

\subsection{Divisor topologies}

The topology of all the divisors has already been determined in one of the Appendices~\ref{sec:Z6IIonSU2xSU6} to~\ref{Z6IIonSU2xSU2xSU3xSU3}. The second Chern class in the basis $\{R_i, E_{k\alpha\beta\gamma}\}$ (for triangulation a)) reads:
\begin{align}
  \label{eq:c2Z6IId}
  \ch_2\cdot E_{1\beta\gamma} &= 0, & \ch_2\cdot E_{2,1\beta} &= -4, & \ch_2\cdot E_{2,3\beta} &= 0, & \ch_2 \cdot E_{3,1\gamma} &= -4, & \notag\\
  \ch_2 \cdot E_{3,2\gamma} &= 0, & \ch_2\cdot E_{4,1\beta} & = -4, &\ch_2\cdot E_{4,3\beta} & = 0, &\ch_2 \cdot R_1 &= 0, & \notag\\
  \ch_2 \cdot R_i &=  24. 
\end{align}

\subsection{The orientifold}

At the orbifold point, there are 64 O3--planes which fall into 24 conjugacy classes under the orbifold group as for the lattice $SU(3)\times SO(8)$ in Appendix~\ref{sec:Z6IISU3xSO8_O}. Under $I_6\,\theta^3$ we find four O7--planes in the $(z^1,z^3)$--plane which fall into two conjugacy classes: The one at $z^2=0$ and those at $z^2=\frac{1}{2},\, \frac{\tau}{2},\, \frac{1}{2}(1+\tau)$. $h^{(1,1)}_{-}=8$ is obtained in the same way as in Appendix~\ref{Z6IIonSU2xSU2xSU3xSU3} with two additional divisors coming from the new divisors $E_{2,3,\beta}$, $E_{4,3,\beta}$. Since these divisors come from fixed lines without fixed point, and latter also contribute to the twisted complex structure moduli according to the discussion in Section~\ref{sec:TwistedCplx}, some of these moduli are projected out by the induced orientifold action on $H^{2,1}_{+}$. In fact, the three fixed lines at $(\zf{1}{2},\zf{2}{\beta})$, $\beta=1,2,3$ are identified with those at $(\zf{1}{3},\zf{2}{\beta})$. Moreover, there are four classes of $\IC^2/\IZ_2$ fixed lines without fixed points, therefore we find that $h^{2,1}_{+}=7$. Similarly, everything we found for the resolved orbifold on the $SU(2)\times SU(6)$ lattice applies here as well.

\section{The $\IZ_7$ orbifold}

\subsection{Metric, complex structure and moduli}

Here, the (only possible) torus lattice for the $\IZ_7$--orbifold is the root lattice of $SU(7)$, with
the twist $Q$ acting on the six roots $e_i$ in the following way:
\begin{eqnarray}
Q\ e_i&=&e_{i+1} ,\ \ \ i=1,\ldots 5\ ,\cr
Q\ e_6&=&-e_1-e_2-e_3-e_4-e_5-e_6.\end{eqnarray}
The twist $Q$ allows for three independent real deformations of the metric $g$
and three  real deformations  of the
anti--symmetric tensor $b$:
\begin{equation}\label{gseven}{
g=R^2\, \left(\begin{array}{cccccc}
1&\cos\theta_{12}&\cos\theta_{13}&x&x&\cos\theta_{13}\cr
\cos\theta_{12}&1&\cos\theta_{12}&\cos\theta_{13}&x&x\cr
\cos\theta_{13}&\cos\theta_{12}&1&\cos\theta_{12}&\cos\theta_{13}&x\cr
x&\cos\theta_{13}&\cos\theta_{12}&1&\cos\theta_{12}&\cos\theta_{13}\cr
x&x&\cos\theta_{13}&\cos\theta_{12}&1&\cos\theta_{12}\cr
\cos\theta_{13}&x&x&\alpha^2_{13}&\cos\theta_{12}&1\end{array}\right),}\end{equation}
with $x=-\h-\cos\theta_{12}-\cos\theta_{13}$ and the three real parameters $R^2,\,\cos\theta_{12},\,\cos\theta_{13}$ and  
\begin{equation}{
b=\left(\begin{array}{cccccc}
0&b_1&b_2&b_3&-b_3&-b_2\cr
-b_1&0&b_1&b_2&b_3&-b_3\cr
-b_2&-b_1&0&b_1&b_2&b_3\cr
-b_3&-b_2&-b_1&0&b_1&b_2\cr
b_3&-b_3&-b_2&-b_1&0&b_1\cr
b_2&b_3&-b_3&-b_2&-b_1&0
\end{array}\right),}\end{equation}
with the three real parameters  $b_1,\,b_2,\,b_3$.
\begin{eqnarray}\label{cpxzseven}
z^1&=&x^1+(-1)^{2/7}\, x^2+(-1)^{4/7}\, x^3+(-1)^{6/7}\, x^4-(-1)^{1/7}\, x^5-
(-1)^{3/7}\, x^6 ,\cr
z^2&=&x^1+(-1)^{4/7}\, x^2-(-1)^{1/7}\, x^3-(-1)^{5/7}\, x^4+(-1)^{2/7}\, x^5+
(-1)^{6/7}\, x^6 ,\cr
z^3&=&x^1-\!(-1)^{1/7}\, x^2+(-1)^{2/7}\, x^3-\!(-1)^{3/7}\, x^4+(-1)^{4/7}\, x^5-\!
(-1)^{5/7}\, x^6.\end{eqnarray}

The three invariant 2--forms of the real cohomology are
\begin{eqnarray}\label{realseven}
\om_1&=&dx^1\wedge dx^2+dx^2\wedge dx^3+dx^3\wedge dx^4+dx^4\wedge dx^5+dx^5\wedge dx^6,\cr
\om_2&=&dx^1\wedge dx^3-dx^1\wedge dx^6+dx^2\wedge dx^4+dx^3\wedge dx^5+dx^4\wedge dx^6,\cr
\om_3&=&dx^1\wedge dx^4-dx^1\wedge dx^5+dx^2\wedge dx^5-dx^2\wedge dx^6+dx^3\wedge dx^6.\end{eqnarray}
Because of the calculational cost, we resort to another method of obtaining the K\"ahler moduli.
As a first step, the metric (\ref{gseven}) may be expressed through the sechsbein $e$, i.e.
$g=e^te$.
This may be obtained from \cite{structure}, where the lattice vectors $e_i$ are 
expressed as a linear combination of a set of six real orthonormal 
basis vectors $\tilde e_i$
\begin{equation}\label{sechsbein}{
e_i=\sum_{j=1,3,5} R_j\ \lf\{\cos[(i-1)\kappa_j\alpha+\phi_j]\ \tilde e_j+
\sin[(i-1)\kappa_j\alpha+\phi_j]\ \tilde e_{j+1}\ri\},}
\end{equation}
with
\begin{eqnarray}
R_1^2&=&R^2\ 
[\cos\theta_{12}\ (\alpha_5^2-\alpha_1^2)+\cos\theta_{13}\ (\alpha_5^2-\alpha_3^2)+\tfrac{1}{2}\ \alpha_5^2] ,\nonumber\\[1pt]
R_3^2&=&R^2\ 
[\cos\theta_{12}\ (\alpha_1^2-\alpha_3^2)+\cos\theta_{13}\ (\alpha_1^2-\alpha_5^2)+\tfrac{1}{2}\ \alpha_1^2] ,\nonumber\\[1pt]
R_5^2&=&R^2\ 
[\cos\theta_{12}\ (\alpha_3^2-\alpha_5^2)+\cos\theta_{13}\ (\alpha_3^2-\alpha_1^2)+\tfrac{1}{2}\ \alpha_3^2] ,\nonumber\\[1pt]
\alpha_i^2&=&\tfrac{4}{7}\ [1-\cos(b_i\alpha)] ,\ \ \ i=1,3,5 ,
\end{eqnarray}
and $\alpha=\fc{2\pi}{7}$, $\kappa_1=1,\ \kappa_3=2$ and $\kappa_5=4$.
The angles $\phi_i$ are arbitrary reflecting the freedom of how
to embed our six--dimensional lattice into the orthonormal system 
$\{\tilde e_i\}_{i=1,\ldots,6}$.
We have set the three free angles $\phi_i$ to zero after having realized that they act on each 
$z^i$ just as an overall phase. 
Examination of the K\"ahler form yields
\begin{eqnarray}\label{moduli}
\Tc^1&=&R_1^2+\fc{4}{7}\ i\ \lf[b_3\ \sin\lf(\fc{\pi}{7}\ri)+b_1\ \sin\lf(\fc{2\pi}{7}\ri)+b_2\ 
\sin\lf(\fc{3\pi}{7}\ri)\ri] ,\nonumber\\[2pt]
\Tc^2&=&R_3^2-\fc{4}{7}\ i\ \lf[b_2\ \sin\lf(\fc{\pi}{7}\ri)+b_3\ \sin\lf(\fc{2\pi}{7}\ri)-b_1\ 
\sin\lf(\fc{3\pi}{7}\ri)\ri] ,\nonumber\\[2pt]
\Tc^3&=&R_5^2-\fc{4}{7}\ i\ \lf[b_1\ \sin\lf(\fc{\pi}{7}\ri)-b_2\ \sin\lf(\fc{2\pi}{7}\ri)+b_3\ 
\sin\lf(\fc{3\pi}{7}\ri)\ri].\end{eqnarray}

\subsection{Fixed sets}

This is another prime orbifold, where we only need to look at the first twisted sector. We find seven isolated fixed points, see Table \ref{fsseven}.

\begin{table}[h!]\begin{center}
\begin{tabular}{|c|c|c|c|}
\hline
Group el.& Order &Fixed Set& Conj. Classes \cr
\hline
\noalign{\hrule}\noalign{\hrule}
$\theta$& 7     &7 fixed points& 7\cr
\hline
\end{tabular}
\caption{Fixed point set for $\IZ_7$.}\label{fsseven}
\end{center}\end{table}

\subsection{The gluing procedure}

Here, we have seven isolated fixed points. The corresponding $\IZ_7$--patches each have three compact exceptional divisors, so in total we have 21.
There are the usual three inherited divisors $R_i = \{(z^i)^7 = c^7\},\ i=1,2,3$.

In this example, there are no fixed lines. Therefore $h^{(2,1)}_{tw}=0$.


\section{The $\IZ_{8-I}$ orbifold on $SU(4)^2\times SU(4)^2$}

\subsection{Metric, complex structure and moduli}

On the root lattice of $SU(4)^2$, the twist $Q$ has the following action:
\begin{eqnarray}
Q\ e_1&=&e_6,\quad Q\ e_2=e_5,\quad Q\ e_3=e_4,\cr 
Q\ e_4&=&-e_1-e_2-e_3,\quad Q\ e_5=e_3 ,\quad Q\ e_6=e_2\ .\end{eqnarray}
The form of metric and anti-symmetric tensor follow from solving the equations
$Q^tg\,Q=g$ and $Q^tb\,Q=b$:
{\arraycolsep3pt
\begin{equation}{
g\!=\!\left(\!\!\begin{array}{cccccc}
R^2&R^2\cos\theta_{56}&x&R^2\cos\theta_{36}&R^2\cos\theta_{36}&-R^2\cos\theta_{36}\cr
R^2\cos\theta_{56}&R^2&R^2\cos\theta_{56}&R^2\cos\theta_{36}&-R^2\cos\theta_{36}&-R^2\cos\theta_{36}\cr
x&R^2\cos\theta_{56}&R^2&-R^2\cos\theta_{36}&-R^2\cos\theta_{36}&R^2\cos\theta_{36}\cr
R^2\cos\theta_{36}&R^2\cos\theta_{36}&-R^2\cos\theta_{36}&R^2&R^2\cos\theta_{56}&x\cr
R^2\cos\theta_{36}&-R^2\cos\theta_{36}&-R^2\cos\theta_{36}&R^2\cos\theta_{56}&R^2&R^2\cos\theta_{56}\cr
-R^2\cos\theta_{36}&-R^2\cos\theta_{36}&R^2\cos\theta_{36}&x&R^2\cos\theta_{56}&R^2\end{array}\!\!\right),
}\end{equation}}
with $x=-R^2\,(1+2\,\cos\theta_{56})$. The three real parameters $R^2,\ \theta_{36},\ \theta_{56}$.  
For $b$ we find
\begin{equation}{
b=\left(\begin{array}{cccccc}
0&-b_3&0&b_2&-b_2&b_1\cr
b_3&0&-b_3&-b_2&b_1&-b_1\cr
0&b_3&0&b_1&-b_1&b_2\cr
-b_2&b_2&-b_1&0&b_3&0\cr
b_2&-b_1&b_1&-b_3&0&b_3\cr
-b_1&b_1&-b_2&0&-b_3&0\end{array}\right)}\end{equation}
with the three real parameters $b_1,\ b_2,\ b_3$. We see that we get 3 untwisted K\"ahler moduli while the complex structure is completely fixed in this orbifold.
The complex coordinates are
\begin{eqnarray}
z^1&=&\tfrac{1}{\sqrt2}\,(x^1+i\,x^2-x^3-e^{2\pi i/8}\,x^4+e^{6\pi i/8}\,x^5+e^{2\pi i/8}\,x^6),\cr
z^2&=&\tfrac{1}{4\sqrt2}\,(x^1-x^2+x^3-i\,x^4+i\,x^5-i\,x^6),\cr
z^3&=&\tfrac{1}{\sqrt2}\,(x^1-i\,x^2-x^3-e^{6\pi i/8}\,x^4+e^{2\pi i/8}\,x^5+e^{6\pi i/8}\,x^6).
\end{eqnarray}
The three invariant 2--forms in the real cohomology are
\begin{eqnarray}
\om_1&=&dx^1\wedge dx^6+dx^2\wedge dx^5-dx^2\wedge dx^6-dx^3\wedge dx^5,\cr
\om_2&=&dx^1\wedge dx^4-dx^1\wedge dx^5-dx^2\wedge dx^4+dx^3\wedge dx^6,\cr
\om_3&=&-dx^1\wedge dx^2-dx^2\wedge dx^3+dx^4\wedge dx^5+dx^5\wedge dx^6.
\end{eqnarray}
Inspection of the K\"ahler form yields
\begin{eqnarray}
\Tc^1&=&b_1+i\,2\,R_1^2\,(\sqrt2+(-2+\sqrt2)\,\cos\theta_{56}),\nonumber\\[1pt]
\Tc^2&=&b_2-i\,2\,R_1^2(\sqrt2+(2+\sqrt2)\,\cos\theta_{56}),\nonumber\\[1pt]
\Tc^3&=&b_3+i\,4\sqrt2\,R_1^2\,\cos\theta_{36}  .
\end{eqnarray}

\subsection{Fixed sets}

Table \ref{fseighti} summarizes the important data of the fixed sets. The invariant subtorus under $\theta^4$ is $(x^3,0,x^3,x^6,0,x6)$, which corresponds to $z^2$ being invariant.

\begin{table}[h!]\begin{center}
\begin{tabular}{|c|c|c|c|}
\hline
Group el.& Order & Fixed Set& Conj. Classes \cr
\hline
\noalign{\hrule}\noalign{\hrule}
$\theta  $&8     & 4 fixed points&4\cr
$\theta^2  $&4  & 16 fixed points& 10\cr
$\theta^3  $&8   & 4 fixed points&4\cr
$\theta^4  $&2   & 4 fixed lines&3\cr
\hline
\end{tabular}
\caption{Fixed point set for $\IZ_{8-I}$ on $SU(4)^2$.}\label{fseighti}
\end{center}\end{table}

\begin{figure}[h!]
\begin{center}
\includegraphics[width=85mm]{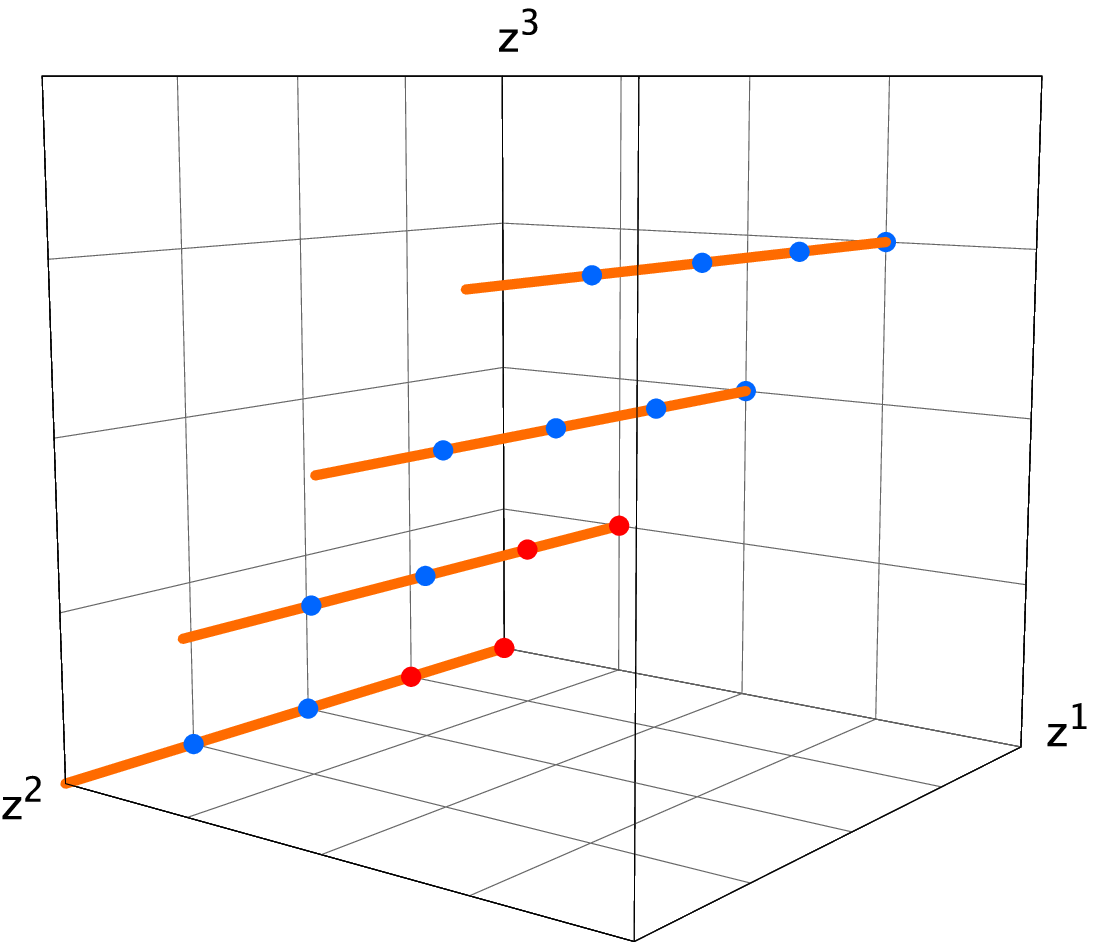}
\caption{Schematic picture of the fixed set configuration of $\IZ_{8-I}$ on $SU(4)^2$}\label{ffixeighti}
\end{center}
\end{figure}
Figure \ref{ffixeighti} shows the configuration of the fixed sets in a schematic way, where each complex coordinate is shown as a coordinate axis and the opposite faces of the resulting cube of length 1 are identified.

\subsection{The gluing procedure}

There are four $\IZ_{8-I}$--patches, which each contribute three internal exceptional divisors. The exceptional divisor on the boundary of the toric diagram Figure \ref{freighti} is identified with the divisor of the resolution of the fixed torus the patch is located on. Of the ten $\IZ_4$--patches, we only need to count six, because four were already counted by the  $\IZ_{8-I}$--patches. They each contribute one exceptional divisor, see Figure \ref{ffour}.  Furthermore, there are three $\IZ_2$--fixed lines which each contribute one exceptional divisor. In total, there are $4\cdot 3+6\cdot1+3\cdot1=21$ exceptional divisors.

On this lattice, there are no fixed lines without fixed points on them, therefore $h^{(2,1)}_{tw}=0$.


\section{The $\IZ_{8-I}$ orbifold on $SO(5)\times SO(9)$}

\subsection{Metric, complex structure and moduli}

On the root lattice of $SO(5)\times SO(9)$, the twist $Q$ has the following action:
\begin{eqnarray}
Q\ e_1&=&e_2,\quad Q\ e_2=e_3,\quad Q\ e_3=e_1+e_2+e_3+2\,e_4,\cr 
Q\ e_4&=&-e_1-e_2-e_3-e_4,\quad Q\ e_5=e_5+2\,e_6 ,\quad Q\ e_6=-e_5-e_6\ .\end{eqnarray}
The form of metric and anti-symmetric tensor follow from solving the equations
$Q^tg\,Q=g$ and $Q^tb\,Q=b$:
{
\begin{equation}{
g=\left(\begin{array}{cccccc}
-2R_1^2\cos\theta_{34}&x&0&y&0&0\cr
x&-2R_1^2\cos\theta_{34}&x&y&0&0\cr
0&x&-2R_1^2\cos\theta_{34}&R_1^2\cos\theta_{34}&0&0\cr
y&y&R_1^2\cos\theta_{34}&R_1^2&0&0\cr
0&0&0&0&2\,R_2^2&-R_2^2\cr
0&0&0&0&-R_2^2&R_2^2\end{array}\right),
}\end{equation}}
with $x=R_1^2\,(1+2\,\cos\theta_{34})$, $y=-R_1^2\,(1+\cos\theta_{34})$. The three real parameters $R_1^2,\ R_2^2,\ \theta_{36}$.  
For $b$ we find
\begin{equation}{
b=\left(\begin{array}{cccccc}
0&b_1+b_2&-2\,b_1&b_1&0&0\cr
-b_1-b_2&0&b_1+b_2&-b_1&0&0\cr
2\,b_1&-b_1-b_2&0&b_2&0&0\cr
-b_1&b_1&-b_2&0&0&0\cr
0&0&0&0&0&b_3\cr
0&0&0&0&-b_3&0\end{array}\right)}\end{equation}
with the three real parameters $b_1,\ b_2,\ b_3$. We see that we get 3 untwisted K\"ahler moduli while the complex structure is completely fixed in this orbifold.
The complex coordinates are
\begin{eqnarray}
z^1&=&\tfrac{1}{\sqrt2}\,(x^1+^{2\pi i/8}\,x^2+i\,x^3-(\tfrac{1}{\sqrt{2}}+\tfrac{1}{2}\,(1+i)\,x^4)),\cr
z^2&=&\tfrac{1}{\sqrt2}\,(x^5-\tfrac{1}{2}\,(1+i)\,x^6),\cr
z^3&=&\tfrac{1}{\sqrt2}\,(x^1+^{6\pi i/8}\,x^2-i\,x^3+(\tfrac{1}{\sqrt{2}}-\tfrac{1}{2}\,(1+i)\,x^4)).
\end{eqnarray}
The three invariant 2--forms in the real cohomology are
\begin{eqnarray}
\om_1&=&dx^1\wedge dx^2-2\,dx^1\wedge dx^3-dx1\wedge dx^4+dx^2\wedge dx^3-dx^1\wedge dx^4,\cr
\om_2&=&dx^1\wedge dx^2+dx^2\wedge dx^3+dx^3\wedge dx^4,\cr
\om_3&=&dx^5\wedge dx^6.
\end{eqnarray}
Inspection of the K\"ahler form yields
\begin{equation}
\Tc^1=b_1+i\,-2\sqrt2\,R_1^2\,(1+3\,\cos\theta_{34})  ,\quad \Tc^2=b_2+i\,\frac{1}{\sqrt2}\,R_1^2(1-4,\cos\theta_{34})  ,\quad \Tc^3=b_3+i\,R_2^2 .
\end{equation}

\subsection{Fixed sets}

Table \ref{fseightia} summarizes the important data of the fixed sets. The invariant subtorus under $\theta^4$ is $(0,0,0,0,x^5,x^6)$, which corresponds to $z^2$ being invariant.

\begin{table}[h!]\begin{center}
\begin{tabular}{|c|c|c|c|}
\hline
Group el.& Order & Fixed Set& Conj. Classes \cr
\hline
\noalign{\hrule}\noalign{\hrule}
$\theta  $&8     &\ 4 fixed points&4\cr
$\theta^2  $&4  &\ 16 fixed points& 10\cr
$\theta^3  $&8   &\ 4 fixed points&4\cr
$\theta^4  $&2   &\ 16 fixed lines&6\cr
\hline
\end{tabular}
\caption{Fixed point set for $\IZ_{8-I}$ on $SO(5)\times SO(9)$.}\label{fseightia}
\end{center}\end{table}

\begin{figure}[h!]
\begin{center}
\includegraphics[width=85mm]{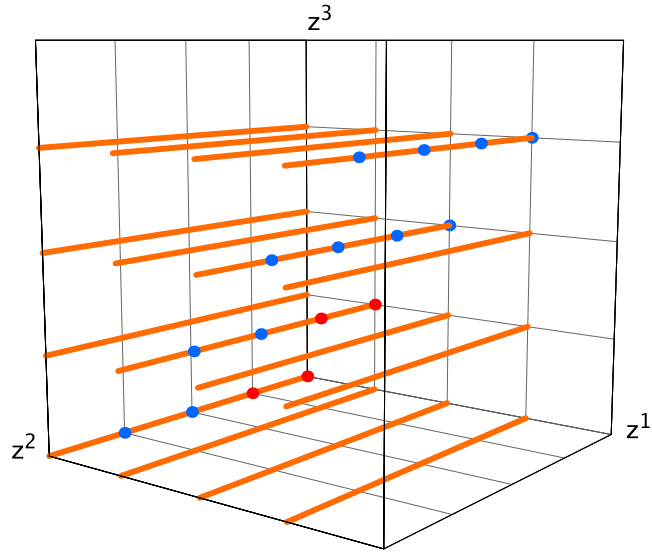}
\caption{Schematic picture of the fixed set configuration of $\IZ_{8-I}$ on $SO(5)\times SO(9)$}\label{ffixeightia}
\end{center}
\end{figure}
Figure \ref{ffixeightia} shows the configuration of the fixed sets in a schematic way, where each complex coordinate is shown as a coordinate axis and the opposite faces of the resulting cube of length 1 are identified.

\subsection{The gluing procedure}

There are four $\IZ_{8-I}$--patches, which each contribute three internal exceptional divisors. The exceptional divisor on the boundary of the toric diagram Figure \ref{freighti} is identified with the divisor of the resolution of the fixed torus the patch is located on. Of the ten $\IZ_4$--patches, we only need to count six, because four were already counted by the  $\IZ_{8-I}$--patches. They each contribute one exceptional divisor, see Figure \ref{ffour}.  Furthermore, there are six $\IZ_2$--fixed lines which each contribute one exceptional divisor. In total, there are $4\cdot 3+6\cdot1+6\cdot1=24$ exceptional divisors.

On this lattice, there are three $\IZ_2$ fixed lines without fixed points on them, therefore $h^{(2,1)}_{tw}=3$.


\section{The $\IZ_{8-II}$ orbifold on $SU(2)\times SO(10)$}

\subsection{Metric, complex structure and moduli}

On the root lattice of $SU(2)\times SO(10)$, the twist $Q$ has the following action:
\begin{eqnarray}
Q\ e_1&=&e_2,\quad Q\ e_2=e_3,\quad Q\ e_3=e_1+e_2+e_3+e_4+e_5,\cr 
Q\ e_4&=&-e_1-e_2-e_3-e_4,\quad Q\ e_5=-e_1-e_2-e_3-e_5 ,\quad Q\ e_6=-e_6\ .\end{eqnarray}
The form of metric and anti-symmetric tensor follow from solving the equations
$Q^tg\,Q=g$ and $Q^tb\,Q=b$:
{
\begin{equation}{
g=\left(\!\!\begin{array}{cccccc}
R_1^2&x&0&y&y&0\cr
x&R_1^2&x&y&y&0\cr
0&x&R_1^2&-\frac{1}{2}\,R_1^2&-\frac{1}{2}\,R_1^2&0\cr
y&y&-\frac{1}{2}\,R_1^2&R_2^2&R_2^2\cos\theta_{45}&-R_2R_3\cos\theta_{56}\cr
y&y&-\frac{1}{2}\,R_1^2&R_2^2\cos\theta_{45}&R_2^2&R_2R_3\cos\theta_{56}\cr
0&0&0&-R_2R_3\cos\theta_{56}&R_2R_3\cos\theta_{56}&R_3^2\end{array}\right),
}\end{equation}}
with $x=-R_1^2+\frac{1}{2}R_2^2\,(1+\cos\theta_{45})$, $y=\frac{1}{2}R_1^2+\frac{1}{2}R_2^2\,(1+\cos\theta_{45})$. The five real parameters $R_1^2,\ R_2^2,\ R_3^2\ \theta_{36},\ \theta_{56}$.  
For $b$ we find
\begin{equation}{
b=\left(\begin{array}{cccccc}
0&-b_1+b_2&2\,b_1&-b_1&-b_1&0\cr
b_1-b_2&0&-b_1+b_2&b_1&b_1&0\cr
-2\,b_1&b_1-b_2&0&b_2&b_2&0\cr
b_1&-b_1&-b_2&0&0&-b_3\cr
b_1&-b_1&-b_2&0&0&b_3\cr
0&0&0&b_3&-b_3&0\end{array}\right)}\end{equation}
with the three real parameters $b_1,\ b_2,\ b_3$. We see that we get three untwisted K\"ahler moduli and one untwisted complex structure modulus in this orbifold.
The complex coordinates are
\begin{eqnarray}
z^1&=&2^{-3/4}\,(x^1+e^{2\pi i/8}\,x^2+i\,x^3-\half\,(1+\sqrt2+i)\,(x^4+x^5),\nonumber\\[2pt]
z^2&=&2^{-3/4}\,(x^1+e^{6\pi i/8}\,x^2-i\,x^3+\half\,(-1+\sqrt2+i)\,(x^4+x^5),\cr
z^3&=&\frac{1}{2\sqrt{2\,{\rm Im}\,\Uc}}\,(-x^4+x^5+2\,\Uc\,x^6),
\end{eqnarray}
with $\Uc=\frac{R_3}{R_2}\,\frac{\cos\theta_{56}+i\,\frac{1}{\sqrt{2}}\sqrt{-\cos\theta_{45}-\cos\theta_{56}}}{1-\cos\theta_{45}}$.
The three invariant 2--forms in the real cohomology are
\begin{eqnarray}
\om_1&=&dx^1\wedge dx^2+2\,dx^1\wedge dx^3-dx^1\wedge dx^4-dx^1\wedge dx^5,\cr
\om_2&=&dx^1\wedge dx^2+dx^2\wedge dx^3+dx^3\wedge dx^4+dx^3\wedge dx^5,\cr
\om_3&=&-dx^4\wedge dx^6+dx^5\wedge dx^6.
\end{eqnarray}
Inspection of the K\"ahler form yields
\begin{eqnarray}
\Tc^1&=&b_1+i\,\tfrac{1}{\sqrt2}\,(-7\,R_1^2+3\,R_2^2\,(1+\cos\theta_{45}))  ,\cr
\Tc^2&=&b_2+i\,\tfrac{1}{\sqrt2}\,(2\,R_1^2+R_2^2\,(1+\cos\theta_{45}))  ,\cr
\Tc^3&=&b_3+i\,\sqrt2\,R_2R_3\,\sqrt{-\cos\theta_{45}-\cos2\,\theta_{56}}  .
\end{eqnarray}

\subsection{Fixed sets}

For the $\IZ_{8-II}$--twist, we need to look at the $\theta,...,\theta^4$--twisted sectors.
The fixed torus under $\theta^2$ and $\theta^4$ is $(0,0,0,0,x^5,x^6)$, which corresponds to $z^3$ being invariant.

Table \ref{fseightiia} summarizes the important data of the fixed sets.
\begin{table}[h!]\begin{center}
\begin{tabular}{|c|c|c|c|}
\hline
Group el.& Order &Fixed Set& Conj. Classes \cr
\hline
\noalign{\hrule}\noalign{\hrule}
$ \theta$&8& \ 8  fixed points& 8\cr
$ \theta^2$&4   &\ 2  fixed lines&2\cr
$ \theta^3 $&8&\ 8 fixed points&8\cr
$ \theta^4  $&2 &\ 8  fixed lines&4\cr
\hline
\end{tabular}
\caption{Fixed point set for $\IZ_{8-II}$ on $SU(2)\times SO(10)$.}\label{fseightiia}
\end{center}\end{table}

\subsection{The gluing procedure}

The eight $\IZ_{8-II}$--patches each contribute two internal exceptional divisors, see Figure \ref{freightii}. The three exceptional divisors on the boundary of the toric diagram are identified with the exceptional divisors of the resolution of the $\IZ_4$ fixed lines on top of which the patch is located. The two $\IZ_4$ fixed lines contribute each three exceptional divisors, the two $\IZ_2$ fixed lines contribute one each (the other two are already counted by the $\IZ_4$--patch. In total, there are $8\cdot3+2\cdot3+2\cdot1=24$ exceptional divisors.

On this lattice, there are two $\IZ_2$ fixed lines without fixed points on them, therefore $h^{(2,1)}_{tw}=2$.


\section{The $\IZ_{8-II}$ orbifold on $SO(4)\times SO(9)$}

\subsection{Metric, complex structure and moduli}

On the root lattice of $SO(4)\times SO(9)$, the twist $Q$ has the following action:
\begin{eqnarray}
Q\ e_1&=&e_2,\quad Q\ e_2=e_3,\quad Q\ e_3=e_1+e_2+e_3+2\,e_4,\cr 
Q\ e_4&=&-e_1-e_2-e_3-e_4,\quad Q\ e_5=-e_5 ,\quad Q\ e_6=-e_6\ .\end{eqnarray}
The form of metric and anti-symmetric tensor follow from solving the equations
$Q^tg\,Q=g$ and $Q^tb\,Q=b$:
{\arraycolsep3pt
\begin{equation}{
g=\left(\!\!\begin{array}{cccccc}
-2R_1^2\cos\theta_{34}&x&0&y&0&0\cr
x&-2R_1^2\cos\theta_{34}&x&y&0&0\cr
0&x&-2R_1^2\cos\theta_{34}&R_1^2\cos\theta_{34}&0&0\cr
y&y&R_1^2\cos\theta_{34}&R_1^2&0&0\cr
0&0&0&0&R_2^2&R_2R_3\cos\theta_{56}\cr
0&0&0&0&R_2R_3\cos\theta_{56}&R_3^2\end{array}\right),
}\end{equation}}
with $x=R_1^2\,(1+2\,\cos\theta_{34})$, $y=-R_1^2\,(1+\cos\theta_{34})$. The five real parameters $R_1^2,\ R_2^2,\ R_3^2\ \theta_{36},\ \theta_{56}$.  
For $b$ we find
\begin{equation}{
b=\left(\begin{array}{cccccc}
0&b_1+b_2&-2\,b_1&b_1&0&0\cr
-b_1-b_2&0&b_1+b_2&-b_1&0&0\cr
2\,b_1&-b_1-b_2&0&b_2&0&0\cr
-b_1&b_1&-b_2&0&0&0\cr
0&0&0&0&0&b_3\cr
0&0&0&0&-b_3&0\end{array}\right)}\end{equation}
with the three real parameters $b_1,\ b_2,\ b_3$. We see that we get three untwisted K\"ahler moduli and one untwisted complex structure modulus in this orbifold.
The complex coordinates are
\begin{eqnarray}
z^1&=&2^{-3/4}\,(x^1+e^{2\pi i/8}\,x^2+i\,x^3-\tfrac{1}{2}\,(1+\sqrt2+i)\,x^4),\nonumber\\[1pt]
z^2&=&2^{-3/4}\,(x^1+e^{6\pi i/8}\,x^2-i\,x^3+\tfrac{1}{2}\,(-1+\sqrt2+i)\,x^4),\nonumber\\[1pt]
z^3&=&\tfrac{1}{\sqrt{2\,{\rm Im}\,\Uc}}\,(x^5+\Uc\,x^6),
\end{eqnarray}
with $\Uc=\frac{R_3}{R_2}\,e^{i\theta_{56}}$.
The three invariant 2--forms in the real cohomology are
\begin{eqnarray}
\om_1&=&dx^1\wedge dx^2-2\,dx^1\wedge dx^3+dx^1\wedge dx^4+dx^2\wedge dx^3-dx^2\wedge dx^4,\cr
\om_2&=&dx^1\wedge dx^2+dx^2\wedge dx^3+dx^3\wedge dx^4,\cr
\om_3&=&dx^5\wedge dx^6.
\end{eqnarray}
Inspection of the K\"ahler form yields
\begin{eqnarray}
\Tc^1&=&b_1-i\,{\sqrt2}\,R_1^2\,(3+8\,\cos\theta_{34})  ,\cr
\Tc^2&=&b_2+i\,\tfrac{1}{\sqrt2}\,R_1^2\,(1-4\,\cos\theta_{34})  ,\cr
\Tc^3&=&b_3+i\,R_2R_3\,\sin\theta_{56}  .
\end{eqnarray}

\subsection{Fixed sets}

For the $\IZ_{8-II}$--twist, we need to look at the $\theta,...,\theta^4$--twisted sectors.
The fixed torus under $\theta^2$ and $\theta^4$ is $(0,0,0,0,x^5,x^6)$, which corresponds to $z^3$ being invariant.

Table \ref{fseightiiaa} summarizes the important data of the fixed sets.
\begin{table}[h!]\begin{center}
\begin{tabular}{|c|c|c|c|}
\hline
Group el.& Order & Fixed Set& Conj. Classes \cr
\hline
\noalign{\hrule}\noalign{\hrule}
$\theta$&8       &\ 8  fixed points& 8\cr
$\theta^2$&4  &\ 4  fixed lines&3\cr
$ \theta^3 $&8 &\ 8 fixed points&8\cr
$ \theta^4  $&2  &\ 16  fixed lines&6\cr
\hline
\end{tabular}
\caption{Fixed point set for $\IZ_{8-II}$ on $SO(4)\times SO(9)$.}\label{fseightiiaa}
\end{center}\end{table}

\subsection{The gluing procedure}

The eight $\IZ_{8-II}$--patches each contribute two internal exceptional divisors, see Figure \ref{freightii}. The three exceptional divisors on the boundary of the toric diagram are identified with the exceptional divisors of the resolution of the $\IZ_4$ fixed lines on top of which the patch is located. The three $\IZ_4$ fixed lines contribute each three exceptional divisors, the three $\IZ_2$ fixed lines contribute one each (the other three are already counted by the $\IZ_4$--patch. In total, there are $8\cdot3+3\cdot3+3\cdot1=28$ exceptional divisors.

On this lattice, there are 3 $\IZ_2$ fixed lines and one $\IZ_4$ fixed line without fixed points on them, therefore $h^{(2,1)}_{tw}=3\cdot1+1\cdot3=6$.

\section{The $\IZ_{12-I}$--orbifold on $E_6$}

\subsection{Metric, complex structure and moduli}


When we choose the $E_6$--lattice for the torus lattice, the Coxeter twist acts as
\begin{eqnarray}
Qe_1&=&e_2,\quad Qe_2=e_3,\quad Qe_3=e_1+e_2+e_3+e_4+e_6,\cr
Qe_4&=&e_5,\quad Qe_5=-e_1-e_2-e_3-e_4-e_5,\cr
Qe_6&=&-e_1-e_2-e_3-e_6.\end{eqnarray}
The twist allows for 3 real deformations of the metric, which has the form
\begin{equation}{
g=\left(\begin{array}{cccccc}
R_5^2&x&\half(R_6^2-R_5^2)&R_5^2-R_6^2&y&-z\cr
x&R_5^2&x&y&R_5^2-R_6^2&-z\cr
\half(R_6^2-R_5^2)&x&R_5^2&-\half R_5^2&y&-\half R_6^2\cr
R_5^2-R_6^2&y&-\half R_5^2&R_5^2&x&z\cr
y&R_5^2-R_6^2&y&x&R_5^2&z\cr
-R_5R_6\cos\theta_{56}&-z&-\half R_6^2&z&z&R_6^2
\end{array}\right),}\end{equation}
with $x=-\half R_5^2+R_5R_6 \cos\theta_{56},\ y=\half(R_6^2-R_5^2)-R_5R_6\cos\theta_{56},\,z=R_5R_6\cos\theta_{56}$
and $R_5,\,R_6$ and $\cos\theta_{56}$ arbitrary free parameters. Also for the $B$--field, the twist allows 3 real deformations:
\begin{equation}{
b=\left(\begin{array}{ccccccc}
0& b_1&b_2-b_1-b_3&0&b_1-b_2&b_3\cr
-b_1&0&b_1&b_2-b_1&0&-b_3\cr
-b_2+b_1+b_3&-b_1&0&b_1-b_3&b_2-b_1&b_2\cr
0&-b_2+b_1&-b_1+b_3&0&b_1&-b_3\cr
b_2-b_1&0&-b_2+b_1&-b_1&0&b_3\cr
-b_3&b_3&-b_2&b_3&-b_3&0\end{array}\right),}\end{equation}
with $b_1,\,b_2,\,b_3$ the arbitrary parameters, so we have as expected three untwisted K\"ahler moduli and no complex structure moduli. 
The invariant 2--form of the real cohomology are
\begin{eqnarray}\label{realtwelveii}
\om_1&=&dx^1\wedge dx^2-dx^1\wedge dx^3+dx^1\wedge dx^5+dx^2\wedge dx^3-dx^2\wedge dx^4+dx^3\wedge dx^4\cr
&&-dx^3\wedge dx^5+dx^4\wedge dx^5,\cr
\om_2&=&dx^1\wedge dx^3-dx^1\wedge dx^5+dx^2\wedge dx^4+dx^3\wedge dx^5+dx^3\wedge dx^6,\cr
\om_3&=&-dx^1\wedge dx^3+dx^1\wedge dx^6-dx^2\wedge dx^6-dx^3\wedge dx^4\cr
&&-dx^4\wedge dx^6+dx^5\wedge dx^6.
\end{eqnarray}
The complex structure is
\begin{eqnarray}\label{cplxtwelveii}
z^1&=&\tfrac{1}{\sqrt6}\,(x^1+e^{2\pi i/12}\,x^2+e^{2\pi i/6}\,x^3-x^4+e^{-10\pi i/12}\,x^5-(1-i\,)\,e^{2\pi i/6}\,x^6)\cr
z^2&=&\tfrac{1}{2}\,(x^1+e^{2\pi i/3}\,x^2+e^{-2\pi i/3}\,x^3+x^4+e^{2\pi i/3}\,x^5),\cr
z^3&=&\tfrac{1}{3^{1/4}\sqrt6}(x^1+e^{10\pi i/12}x^2-e^{2\pi i/3}x^3-x^4-e^{10\pi i/12}x^5+(1\!-\!i)e^{2\pi i/3}x^6).
\end{eqnarray}
Again we pair $B+i\,J={\cal T}^i\,\om_i$ in the real cohomology and get the following K\"ahler moduli:
\begin{eqnarray}\label{kaehlertwelveii}
{\cal T}^1&=&b_1+i\,2\sqrt3\,(2\,R_5^2-R_6^2),\nonumber\\[1pt]
{\cal T}^2&=&b_2-i\,(\sqrt3\,R_5^2+(1+\sqrt3)\,R_6^2+R_5R_6\,\cos\theta_{56}),\nonumber\\[1pt]
{\cal T}^3&=&b_3+i\,(\tfrac{5}{6}\,R_6^2-6\,R_5R_6\,\cos\theta_{56})
.\end{eqnarray}

\subsection{Fixed sets}

For the $\IZ_{12-I}$--twist, we need to look at the $\theta$-, $\theta^2, \ldots,\theta^6$--twisted sectors.
Here, the fixed points of several of the group elements end up in the same place, as for example the fixed points of $\theta,\,\theta^2,\,\theta^5$ and three of the 27 fixed points of $\theta^4$.

Table \ref{fstwelveia} summarizes the important data of the fixed sets. The invariant subtorus under $\theta^3$ and $\theta^6$ is $(x^4,x^5,0,x^4,x^5,0)$, which corresponds to $z^2$ being invariant.

\begin{table}[h!]\begin{center}
\begin{tabular}{|c|c|c|c|}
\hline
Group el.& Order & Fixed Set& Conj. Classes \cr
\hline
\noalign{\hrule}\noalign{\hrule}
$\theta$&12      &\ 3  fixed points& 3\cr
$ \theta^2$&6  &\ 3  fixed points&3\cr
$ \theta^3 $&4  &\ 1  fixed line&1\cr
$ \theta^4  $&3&\ 27  fixed points& 9\cr
$  \theta^5$&12  &\ 3 fixed points& 3\cr
$ \theta^6  $&2&\ 4 fixed lines &2\cr
\hline
\end{tabular}
\caption{Fixed point set for $\IZ_{12-I}$ on $E_6$.}\label{fstwelveia}
\end{center}\end{table}

\begin{figure}[h!]
\begin{center}
\includegraphics[width=85mm]{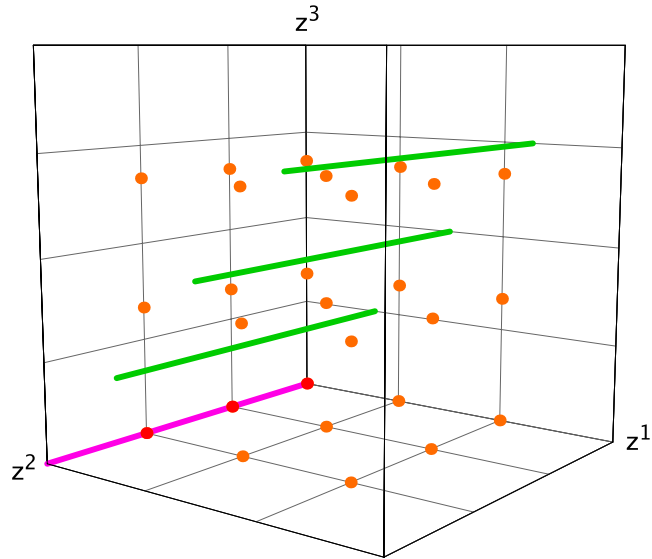}
\caption{Schematic picture of the fixed set configuration of $\IZ_{12-I}$ on $E_6$}\label{ffixtwelvei}
\end{center}
\end{figure}
Figure \ref{ffixtwelvei} shows the configuration of the fixed sets in a schematic way, where each complex coordinate is shown as a coordinate axis and the opposite faces of the resulting cube of length 1 are identified.

\subsection{The gluing procedure}

Here, we get three local $\IZ_{12-I}$--patches with each four compact exceptional divisors and three exceptional divisors on the boundary $(D_1,D_3)$. We need not count the fixed points under $\theta^2=\IZ_{6-I},\,\theta^4=\IZ_3$ and $\theta^5=\overline{\IZ_{12-I}}$ that have the same locations as they are already incorporated in the $\IZ_{12-I}$--patch. So only six of the nine conjugacy classes of $\IZ_3$ must be counted with each one compact exceptional divisor. The fixed line under the $\IZ_4$-element contributes three exceptional divisors, and of the two conjugacy classes of fixed lines under $\IZ_2$, we count one. This gives us $3\cdot4+1\cdot3+6\cdot1+1\cdot1=22$ exceptional divisors.

On this lattice, there is one $\IZ_2$ fixed line without fixed points on it, therefore $h^{(2,1)}_{tw}=1$.


\section{The $\IZ_{12-I}$ orbifold on $SU(3)\times F_4$}


\subsection{Metric, complex structure and moduli}

Here, the torus lattice is the root lattice of $SU(3)\times F_4$. The action of the twist upon the roots is
\begin{eqnarray}\label{twelve}
Qe_1&=&e_2,\quad Qe_2=-e_1-e_2,\quad Qe_3=e_4,\cr
Qe_4&=&e_3+3_4+2e_5,\quad Qe_5=e_6,\quad Qe_6=-e_3-e_4-e_5-e_6.\end{eqnarray}
The twist allows for 3 real deformations of the metric, which has the form
\begin{equation}{
g=\left(\begin{array}{cccccc}
R_1^2&-\half R_1^2&0&0&0&0\cr
\half R_1^2& R1^2&0&0&0&0\cr
0&0&R_3^2&R_3^2\,\cos\theta_{34}&x&x\cr
0&0&R_3^2\,\cos\theta_{34}&R_3^2& -\half R_3^2&x\cr
0&0&x&-\half R_3^2&\half R_3^2&\half R_3^2\,\cos\theta_{34}\cr
0&0&x&x&\half R_3^2\,\cos\theta_{34}&\half R_3^2
\end{array}\right),}\end{equation}
with $x=R_3^2\,\cos\theta_{34}$ and $R_1,\,R_3$ and $\cos\theta_{34}$ arbitrary free parameters. Also for the $B$--field, the twist allows 3 real deformations:
\begin{equation}{
b=\left(\begin{array}{cccccc}
0& b_1&0&0&0&0\cr
-b_1&0&0&0&0&0\cr
0&0&0&2\,b_3&b_2&-b_2\cr
0&0&-2\,b_3&0&2\,b_3&b_2\cr
0&0&-b_2&-2\,b_3&0&b_3\cr
0&0&b_2&-b_2&-b_3&0\end{array}\right),}\end{equation}
with $b_1,\,b_2,\,b_3$ the arbitrary parameters, so we have three untwisted K\"ahler moduli and no complex structure moduli. 
The three invariant 2--forms in the real cohomology are
\begin{eqnarray}\label{realtwelvei}
\om_1&=&dx^1\wedge dx^2,\cr
\om_2&=&dx^3\wedge dx^5-dx^3\wedge dx^6+dx^4\wedge dx^6,\cr
\om_3&=&2\,dx^3\wedge dx^4+2\,dx^4\wedge dx^5+dx^5\wedge dx^6
.\end{eqnarray}
The complex coordinates are:
\begin{eqnarray}\label{fcplx}
z^1&=&3^{-1/4}\,(x^3+e^{2\pi i/6}\,x^4+\tfrac{1}{\sqrt2}[e^{11\pi i/12}\,x^5+e^{\pi i/12}\,x^6]),\cr
z^2&=&3^{-1/4}\,(x^1+e^{2\pi i/3}\,x^2),\cr
z^3&=&3^{-1/4}\,(x^3+e^{10 \pi i/12}\,x^4+\tfrac{1}{\sqrt2}[\,e^{-5\pi i/12}\,x^5+e^{5\pi i/12}\,x^6]).\end{eqnarray}
For the K\"ahler moduli, we find
\begin{equation}
{\cal T}^1=b_1+i\,\tfrac{\sqrt3}{2}\,R_1^2,\quad {\cal T}^2=b_2+i\,\tfrac{3}{4}\,R_3^2\,(-1+2\,\cos\theta_{34}),\quad
{\cal T}^3=b_3+i\,\tfrac{9}{4}\,R_3^2.
\end{equation}

\subsection{Fixed sets}

Table \ref{fstwelveiaa} summarizes the important data of the fixed sets. The invariant subtorus under $\theta^3$ and $\theta^6$ is $(x^1,x^2,0,x^4,x^5,0)$, which corresponds to $z^2$ being invariant.

\begin{table}[h!]\begin{center}
\begin{tabular}{|c|c|c|c|}
\hline
Group el.& Order &Fixed Set& Conj. Classes \cr
\hline
\noalign{\hrule}\noalign{\hrule}
$\theta  $&12  &3 fixed points&3\cr
$\theta^2  $&6   &3 fixed points& 3\cr
$\theta^3  $&4  &4 fixed lines&2\cr
$\theta^4  $&3  &27 fixed points&9\cr
$ \theta^5$&12&3 fixed points& 3\cr
$ \theta^6   $&2 &16 fixed lines& 4\cr
\hline
\end{tabular}
\caption{Fixed point set for $\IZ_{12-I}$ on $SU(3)\times F_4$.}\label{fstwelveiaa}
\end{center}\end{table}

\begin{figure}[h!]
\begin{center}
\includegraphics[width=85mm]{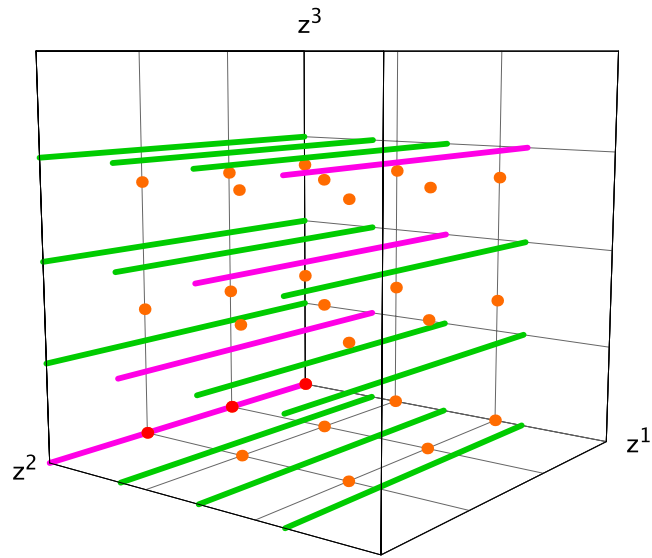}
\caption{Schematic picture of the fixed set configuration of $\IZ_{12-I}$ on $SU(3)\times F_4$}\label{ffixtwelveii}
\end{center}
\end{figure}
Figure \ref{ffixtwelveii} shows the configuration of the fixed sets in a schematic way, where each complex coordinate is shown as a coordinate axis and the opposite faces of the resulting cube of length 1 are identified.

\subsection{The gluing procedure}

There are three $\IZ_{12-I}$--patches with each four compact exceptional divisors and three exceptional divisors on the boundary $(D_1,D_3)$ and again only six of the nine conjugacy classes of $\IZ_3$ must be counted with each one compact exceptional divisor. The two fixed lines under the $\IZ_4$-element contribute each three exceptional divisors, and of the four conjugacy classes of fixed lines under $\IZ_2$, we count two.
This gives us $3\cdot4+2\cdot3+6\cdot1+2\cdot1=26$ exceptional divisors.

On this lattice, there are one $\IZ_4$ and two $\IZ_2$ fixed lines without fixed points on them, therefore $h^{(2,1)}_{tw}=1\cdot3+2\cdot1=5$.


\section{The $\IZ_{12-II}$ orbifold on $SO(4)\times F_4$}


\subsection{Metric, complex structure and moduli}

Here, the torus lattice is the root lattice of $SO(4)\times F_4$. The action of the twist upon the roots is
\begin{eqnarray}\label{twelveii}
Qe_1&=&e_2,\quad Qe_2=e_1+e_2+2\,e_3,\quad Qe_3=e_4,\cr
Qe_4&=&-e_1-e_2-e_3-e_4,\quad Qe_5=-e_5,\quad Qe_6=-e_6.\end{eqnarray}
The twist allows for 3 real deformations of the metric, which has the form
\begin{equation}{
g=\left(\begin{array}{cccccc}
2\,R_1^2&2\,R_1^2\cos\theta_{34}&x&x&0&0\cr
2\, R_1^2\cos\theta_{34}&2\, R1^2&-R_1^2&x&0&0\cr
x&-R_1^2&R_1^2&R_1^2\cos\theta_{34}&0&0\cr
x&x&R_1^2\cos\theta_{34}&R_1^2&0&0\cr
0&0&0&0&R_2^2& R_2R_3\,\cos\theta_{56}\cr
0&0&0&0&R_2R_3\,\cos\theta_{56}&R_3^2
\end{array}\right),}\end{equation}
with $x=-R_1^2\,(\frac{1}{2}+\cos\theta_{34})$ and $R_1,\,R_2,\ R_3$ and $\theta_{34},\ \theta{56}$ arbitrary free parameters. Also for the $B$--field, the twist allows 3 real deformations:
\begin{equation}{
b=\left(\begin{array}{cccccc}
0&2\, b_2&-b_1&b_1&0&0\cr
-2\,b_2&0&2\,b_2&-b_1&0&0\cr
b_1&-2\,b_2&0&b_2&0&0\cr
-b_1&b_1&-b_2&0&0&0\cr
0&0&0&0&0&b_3\cr
0&0&0&0&-b_3&0\end{array}\right),}\end{equation}
with $b_1,\,b_2,\,b_3$ the arbitrary parameters, so we have three untwisted K\"ahler moduli and one untwisted complex structure modulus. 
The complex coordinates are:
\begin{eqnarray}
z^1&=&3^{-1/4}\,(x^1+e^{2\pi i/12}\,x^2+\tfrac{1}{ \sqrt2}[e^{11\pi i/12}\,x^3-e^{\pi i/12}\,x^4]),\cr
z^2&=&3^{-1/4}\,(x^1+e^{10\pi i/12}\,x^2+\tfrac{1}{\sqrt2}[e^{-5\pi i/12}\,x^3+e^{5\pi i/12}\,x^4]),\cr
z^3&=&\tfrac{1}{\sqrt{2\,{\rm Im}\,\Uc}}\,(x^5+\Uc\,x^6).\end{eqnarray}
The three invariant 2--forms in the real cohomology are
\begin{eqnarray}
\om_1&=&-dx^1\wedge dx^3+dx^1\wedge dx^4-dx^2\wedge dx^4,\cr
\om_2&=&2\,dx^1\wedge dx^2+2\,dx^2\wedge dx^3+dx^3\wedge dx^4,\cr
\om_3&=&dx^5\wedge dx^6
.\end{eqnarray}
For the K\"ahler moduli, we find
\begin{equation}
{\cal T}^1=b_1+i\,\tfrac{3}{2}\,R_1^2\,(1-2\,\cos\theta_{34}),\quad
{\cal T}^2=b_2+i\,\tfrac{9}{2}\,R_1^2,\quad
{\cal T}^3=b_3+i\,R_2R_3\,\sin\theta_{56}.
\end{equation}

\subsection{Fixed sets}

Table \ref{fstwelveii} summarizes the important data of the fixed sets. The invariant subtorus under $\theta^2,\,\theta^4$ and $\theta^6$ is $(0,0,0,0,x^5,x^6)$, which corresponds to $z^3$ being invariant.

\begin{table}[h!]\begin{center}
\begin{tabular}{|c|c|c|c|}
\hline
Group el.& Order &Fixed Set& Conj. Classes \cr
\hline
\noalign{\hrule}\noalign{\hrule}
$\theta  $&12  &4 fixed points&4\cr
$\theta^2  $&6   &1 fixed line& 1\cr
$\theta^3  $&4  &16 fixed points&8\cr
$\theta^4  $&3  &9 fixed lines&3\cr
$ \theta^5$&12&4 fixed points& 4\cr
$ \theta^6   $&2 &16 fixed lines& 4\cr
\hline
\end{tabular}
\caption{Fixed point set for $\IZ_{12-II}$ on $SO(4)\times F_4$.}\label{fstwelveii}
\end{center}\end{table}

\subsection{The gluing procedure}

There are four  $\IZ_{12-II}$--patches which contribute three internal exceptional divisors each, see Figure \ref{frtwelveii}. They sit on a $\IZ_6$ fixed line, which contributes five exceptional divisors. Of the eight $\IZ_4$ fixed points, we count only four (because the locations of the other four coincide with the $\IZ_{12-II}$ fixed points), they contribute one exceptional divisor each, see Figure \ref{ffour}. There are three $\IZ_3$ fixed lines, one coincides with the $\IZ_6$ fixed line, so only the other two count and contribute two exceptional divisors. Lastly, there are four $\IZ_2$-fixed lines, where again only three count with one exceptional divisor each. In total, this adds up to $4\cdot3+1\cdot5+4\cdot1+2\cdot2+3\cdot1=28$.

As for the twisted complex structure moduli, the two $\IZ_3$ fixed lines contribute two each and the two $\IZ_2$ fixed lines on which no fixed points sit contribute one each, which gives a total of 6.


\section{The $\IZ_2\times\IZ_2$--orbifold}\label{app:z2z2}

\subsection{Metric, complex structure and moduli}

The torus factorizes into $(T^2)^3$ under the combined twists, where the $T^2$ are not constrained. The twists act on the lattice basis:
\begin{eqnarray}
Q_1\, e_1&=&- e_1,\quad Q_1\,e_2=-e_2,\quad Q_1\,e_3=e_3,\quad Q_1\,e_4=e_4,\cr
Q_1\,e_5&=&-e_5,\quad Q_1\,e_6=-e_6,\cr
Q_2\, e_1&=& e_1,\quad Q_2\,e_2=e_2,\quad Q_2\,e_3=-e_3,\quad Q_2\,e_4=-e_4,\cr
Q_2\,e_5&=&-e_5,\quad Q_2\,e_6=-e_6.
\end{eqnarray}
The combined twist $Q_3$ has the form
\begin{eqnarray}
Q_3\, e_1&=& -e_1,\quad Q_3\,e_2=-e_2,\cr
Q_3\,e_3&=&-e_3,\quad Q_3\,e_4=-e_4,\cr
Q_3\,e_5&=&e_5,\quad Q_3\,e_6=e_6.
\end{eqnarray}
We require the metric to be invariant under all three twists, i.e. we impose the three conditions $Q_i^Tg\,Q_i=g,\quad i=1,2,3$. This leads to the following solution:
{\arraycolsep1pt
\begin{equation}{g=\left(\begin{array}{cccccc}
R_1^2&R_1R_2\,\cos\theta_{12}&0&0&0&0\cr
R_1R_2\,\cos\theta_{12}&R_1^2&0&0&0&0\cr
0&0&R_3^2&R_3R_4\,\cos\theta_{34}&0&0\cr
0&0&R_3R_4\,\cos\theta_{34}&R_4^2&0&0\cr
0&0&0&0&R_5^2&R_5R_6\,\cos\theta_{56}\cr
0&0&0&0&R_5R_6\,\cos\theta_{56}&R_6^2\end{array}\right).}\end{equation}
}
The solution for $b$ matches the pattern:
\begin{equation}\label{znzmb}{b=\left(\begin{array}{cccccc}
0&b_1&0&0&0&0\cr
-b_1&0&0&0&0&0\cr
0&0&0&b_2&0&0\cr
0&0&-b_2&0&0&0\cr
0&0&0&0&0&-b_3\cr
0&0&0&0&-b_3&0\end{array}\right),}\end{equation}
we therefore know to have three K\"ahler moduli and three untwisted complex structure moduli. 
The $b$-field (\ref{znzmb}) has the form which is typical for all $\IZ_n\times\IZ_m$ orbifolds.

The complex structure turns out to be as follows:
\begin{equation}\label{cpxtwotwo}
z^1=x^1+{\cal U}^1\,x^2,\ z^2=x^3+{\cal U}^2\,x^4,\ z^3=x^5+{\cal U}^3\,x^6, 
\end{equation}
with 
\begin{equation}\label{cpxmodtwotwo}
{\cal U}^1=\frac{R_2}{R_1}\,e^{i\theta_{12}},\quad {\cal U}^2=\frac{R_4}{R_3}\,e^{i\theta_{34}},\quad
{\cal U}^3=\frac{R_6}{R_5}\,e^{i\theta_{56}}.
\end{equation}
 For the K\"ahler moduli, we find
\begin{equation}\label{kmodtwotwo}
{\cal T}^1=b_1+i\,{R_1R_2}\,\cos\theta_{12},\quad
{\cal T}^2=b_2+i\,{R_3R_4}\,\cos\theta_{34},\quad
{\cal T}^3=b_3+i\,{R_5R_6}\,\cos\theta_{56}.
\end{equation}

\subsection{Fixed sets}

We need to examine the $\theta^1,\ \theta^2$ and $\theta^1\theta^2$--twists. Table \ref{fstwotwo} gives the particulars of the fixed sets.

The fixed torus associated to the $\theta^1$--twist is $(0,0,x^3,x^4,0,0)$ corresponding to $z^2$ being invariant; the torus that remains fixed under $\theta^2$ and $(\theta^2)^2$ is $(x^1,x^2,0,0,0,0)$, corresponding to $z^1$ being invariant; the torus that is fixed by $\theta^1(\theta^2)^2$  is $(0,0,0,0,x^5,x^6)$, corresponding to $z^3$ being invariant.

The fundamental regions are as depicted in Figure \ref{fundamental}.

\begin{table}[h!]\begin{center}
\begin{tabular}{|c|c|c|c|}
\hline
Group el.& Order& Fixed Set& Conj. Classes \cr
\hline
\noalign{\hrule}\noalign{\hrule}
$\theta^1   $&2 &16  fixed lines& 16\cr
$\theta^2   $&2   &16  fixed lines&16\cr
$ \theta^1\theta^2   $&$2\times2 $     &16  fixed lines& 16\cr
\hline
\end{tabular}
\caption{Fixed point set for $\IZ_2\times \IZ_2$.} \label{fstwotwo}
\end{center}\end{table}

\begin{figure}[h!]
\begin{center}
\includegraphics[width=85mm]{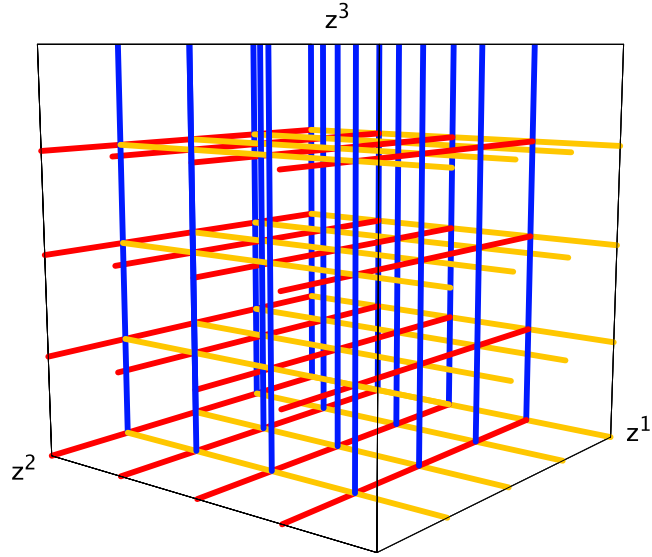}
\caption{Schematic picture of the fixed set configuration of $\IZ_2\times \IZ_2$}\label{ffixtwotwo}
\end{center}
\end{figure}
Figure \ref{ffixtwotwo} shows the configuration of the fixed sets in a schematic way, where each complex coordinate is shown as a coordinate axis and the opposite faces of the resulting cube of length 1 are identified.

\subsection{The gluing procedure}

Each of the $3\cdot 16=48$  $\IZ_2$--fixed lines contributes one exceptional divisor. At each of the 64 points where three fixed lines intersect sits a resolved $\IC^2/\IZ_2\times \IZ_2$--patch.

Since in this example, there are no fixed lines without fixed points on them, $h^{(2,1)}_{tw}=0$.

\subsection{The intersection ring}\label{app:intersz2z2}

From the local linear equivalences (\ref{lineqtwotwo}) we arrive at the following global relations:
\begin{eqnarray}\label{globalreltwotwo}
R_i&\sim&2\,D_{i,\alpha}+\sum_{\beta,\gamma=1}^4 E_{j,\alpha\beta\gamma}+\sum_{\beta,\gamma=1}^4 E_{k,\alpha\beta\gamma},\quad i\neq j\neq k.
\end{eqnarray}
From (\ref{eq:ReqGH}) we know that $R_{1}R_{2}R_{3}=2$. From the toric diagrams of the compactified patches, we can read off directly the intersection numbers with three distinct divisors:
$$E_{1,\beta\gamma}E_{2,\alpha\gamma}E_{3,\alpha\beta}=1\ \ {\rm with} \ \alpha,\beta,\gamma=1,...,4.$$
We find the following intersection numbers:
\begin{eqnarray}
&&R_1E_{1,\beta\gamma}^2=-2,\ R_2E_{2,\alpha\gamma}^2=-2,\ R_3E_{3,\alpha\beta}^2=-2,\cr
&&E_{1,\beta\gamma}^2E_{2,\alpha\gamma}=-1,\ E_{1,\beta\gamma}E_{2,\alpha\gamma}^2=-1,\ E_{1,\beta\gamma}^2E_{3,\alpha\beta}=-1,\cr
&& E_{1,\beta\gamma}E_{3,\alpha\beta}^2=-1,\ E_{2,\alpha\gamma}^2E_{3,\alpha\beta}=-1,\ E_{2,\alpha\gamma}E_{3,\alpha\beta}^2=-1,\cr
&&E_{1,\beta\gamma}^3=4,\ E_{2,\alpha\gamma}^3=4,\ E_{3,\alpha\beta}^3=4.
\end{eqnarray}


\section{The $\IZ_2\times\IZ_4$--orbifold}


\subsection{Metric, complex structure and moduli}

The torus factorizes into $(T^2)^3$ under the combined twists, where the first of the $T^2$ is not contrained. The twists act on the lattice basis:
\begin{eqnarray}
Q_1\, e_1&=&- e_1,\quad Q_1\,e_2=-e_2,\quad Q_1\,e_3=e_3,\quad Q_1\,e_4=e_4,\cr
Q_1\,e_5&=&-e_5,\quad Q_1\,e_6=-e_6,\cr
Q_2\, e_1&=& e_1,\quad Q_2\,e_2=e_2,\quad Q_2\,e_3=e_3+2\,e_4,\quad Q_2\,e_4=-e_3-e_4,\cr
Q_2\,e_5&=&e_5+2\,e_6,\quad Q_2\,e_6=-e_5-e_6.
\end{eqnarray}
The combined twist $Q_3$ has the form
\begin{eqnarray}
Q_3\, e_1&=& -e_1,\quad Q_3\,e_2=-e_2,\cr
Q_3\,e_3&=&e_3+2\,e_4,\quad Q_3\,e_4=-e_3-e_4,\cr
Q_3\,e_5&=&-e_5-2\,e_6,\quad Q_3\,e_6=e_5+e_6.
\end{eqnarray}
We require the metric to be invariant under all three twists, i.e. we impose the three conditions $Q_i^Tg\,Q_i=g,\quad i=1,2,3$. This leads to the following solution:
\begin{equation}{g=\left(\begin{array}{cccccc}
R_1^2&R_1R_2\,\cos\theta_{12}&0&0&0&0\cr
R_1R_2\,\cos\theta_{12}&R_1^2&0&0&0&0\cr
0&0&2\,R_3^2&-R_3^2&0&0\cr
0&0&-R_3^2&R_3^2&0&0\cr
0&0&0&0&2\,R_5^2&-R_5^2\cr
0&0&0&0&-R_5^2&R_5^2\end{array}\right).}\end{equation}
The solution for $b$ matches the pattern of (\ref{znzmb}), we therefore know to have three K\"ahler moduli and three untwisted complex structure moduli. 
For the complex structure we get
\begin{equation}
z^1=\tfrac{1}{\sqrt{2\,{\rm Im}\,{\cal U}^3}}\,(x^1+{\cal U}^3\, x^2),\quad
z^2=x^3-\tfrac{1}{2}(1-i)\,x^4,\quad
z^3=x^5-\tfrac{1}{2}(1-i)\,x^6,\end{equation}
with ${\cal U}^3=R_2/R_1\, e^{i\theta_{12}}$.
Examination of the K\"ahler form yields
\begin{equation}
\Tc^1=b_1+i\,R_1R_2\sin\theta_{12},\ \ 
\Tc^2=b_2+i\,R_3^2,\ \ 
\Tc^3=b_3+i\,R_5^2.\end{equation}

\subsection{Fixed sets}

We need to examine the $\theta^1,\ \theta^2,\ (\theta^2)^2\ \theta^1\theta^2$ and $\theta^1(\theta^2)^2$--twists. Table \ref{fstwofour} gives the particulars of the fixed sets.

The fixed torus associated to the $\theta^1$--twist is $(0,0,x^3,x^4,0,0)$ corresponding to $z^2$ being invariant; the torus that remains fixed under $\theta^2$ and $(\theta^2)^2$ is $(x^1,x^2,0,0,0,0)$, corresponding to $z^1$ being invariant; the torus that is fixed by $\theta^1(\theta^2)^2$  is $(0,0,0,0,x^5,x^6)$, corresponding to $z^3$ being invariant.

\begin{figure}[h!]
\begin{center}
\includegraphics[width=140mm]{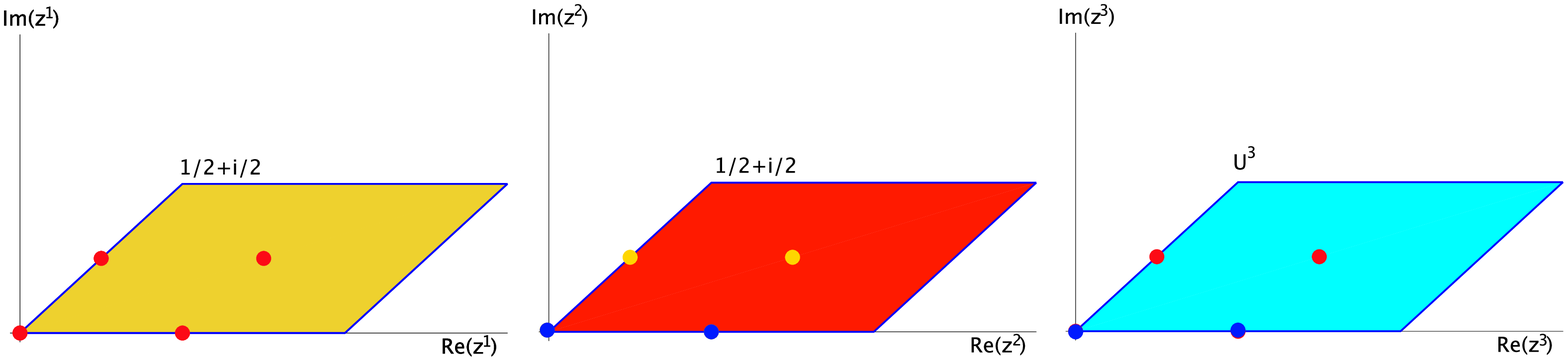}
\caption{Fundamental regions for the $\IZ_2\times\IZ_4$--orbifold}\label{ffutwofour}
\end{center}
\end{figure}
Figure \ref{ffutwofour} shows the fundamental regions of the three tori corresponding to $z^1,\,z^2,\,z^3$ and their fixed points. In each of them, we get  the usual four fixed points of the $\IZ_2$--twist.

\begin{table}[h!]\begin{center}
\begin{tabular}{|c|c|c|c|}
\hline
Group el.& Order &Fixed Set& Conj. Classes \cr
\hline
\noalign{\hrule}\noalign{\hrule}
$\theta^1   $&2     &16  fixed lines& 12\cr
$\theta^2   $&4  &4  fixed lines&4\cr
$ (\theta^2)^2   $&2      &16  fixed lines& 10\cr
$ \theta^1\theta^2   $&$2\times4 $     &16  fixed points& 16\cr
$ \theta^1(\theta^2)^2   $&2  &16  fixed lines& 12\cr
\hline
\end{tabular}
\caption{Fixed point set for $\IZ_2\times \IZ_4$.} \label{fstwofour}
\end{center}\end{table}

\begin{figure}[h!]
\begin{center}
\includegraphics[width=85mm]{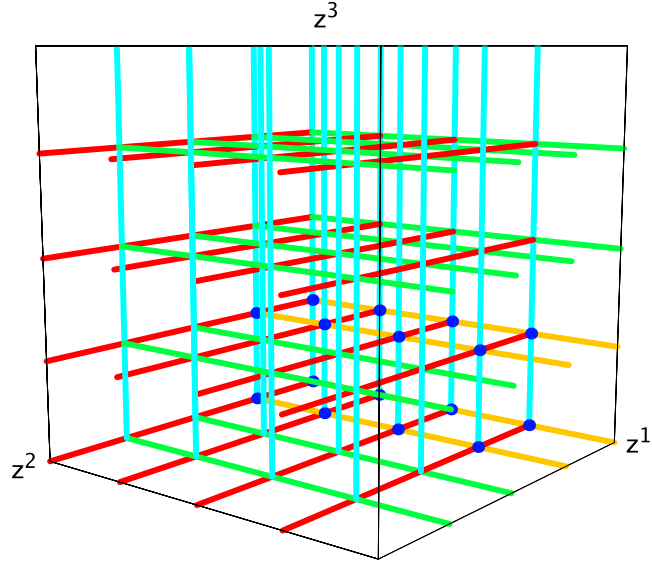}
\caption{Schematic picture of the fixed set configuration of $\IZ_2\times \IZ_4$}\label{ffixtwofour}
\end{center}
\end{figure}
Figure \ref{ffixtwofour} shows the configuration of the fixed sets in a schematic way, where each complex coordinate is shown as a coordinate axis and the opposite faces of the resulting cube of length 1 are identified.

\subsection{The gluing procedure}

From the 16 $\IZ_2\times\IZ_4$--patches, see Figure \ref{frtwofour}, we get each one compact exceptional divisor, $E_{5,\alpha\beta\gamma},\ \alpha=1,...,4,\ \beta,\gamma=1,2$. From the 4 $\IZ_4$--fixed lines, we get three each: $E_{2,\beta\gamma},\, E_{3,\beta\gamma},\, E_{4,\beta\gamma},\ \beta,\gamma=1,2$. From the $12+12+(10-4)=30$ $\IZ_2$--fixed lines each one: $E_{1,\alpha\gamma}, \, E_{6,\alpha\beta},\ i=\alpha,...,4,\, \beta,\gamma=1,2,3$ in the $z^2$--, respectively $z^3$--direction and in the $z^1$--direction $E_{7,\beta\gamma},\ (\beta,\gamma)=(1,3),(3,1),(2,3),(3,2),(3,3),(4,3)$. This adds up to $16\cdot1+4\cdot3+30\cdot1=58$ exceptional divisors. At the intersection points of three $\IZ_2$ fixed lines sit the resolved $\IC^3/\IZ_2\times\IZ_2$ patches.

Since in this example, there are no fixed lines without fixed points on them, $h^{(2,1)}_{tw}=0$.

\subsection{The intersection ring}

Since the $\IC^3/\IZ_2\times\IZ_2$ fixed points fall into conjugacy classes, we have to form the corresponding invariant divisors:
\begin{align}
  \label{eq:Z2xZ4invdivs}
  D_{2,1} &= \Dt_{2,1}, & D_{2,2} &= \Dt_{2,2}, & D_{2,3} &= \Dt_{2,3} + \Dt_{2,4}, \notag\\
  D_{3,1} &= \Dt_{3,1}, & D_{3,2} &= \Dt_{3,2}, & D_{3,3} &= \Dt_{3,3} + \Dt_{3,4}, \notag\\
  E_{1,\alpha,1} &= \Et_{1,\alpha,1}, & E_{1,\alpha,2} &= \Et_{1,\alpha,2}, & E_{1,\alpha,3} &= \Et_{1,\alpha,3} + \Et_{1,\alpha,4}, \notag\\
  E_{6,\alpha,1} &= \Et_{6,\alpha,1}, & E_{6,\alpha,2} &= \Et_{6,\alpha,2}, & E_{6,\alpha,3} &= \Et_{6,\alpha,3} + \Et_{6,\alpha,4} ,\notag\\  
  E_{3,1} &= \Et_{3,1,1}, & E_{3,2} &= \Et_{3,1,2}, & E_{3,3} &= \Et_{3,1,3} + \Et_{3,1,4}, \notag\\
  E_{3,4} &= \Et_{3,2,1}, & E_{3,5} &= \Et_{3,2,2}, & E_{3,6} &= \Et_{3,2,3} + \Et_{3,2,4}, \notag\\
  E_{3,7} &= \Et_{3,3,1} + \Et_{3,4,1}, & E_{3,8} &= \Et_{3,3,2} + \Et_{3,4,2}, & E_{3,9} &= \Et_{3,3,3} + \Et_{3,4,4}, \notag\\
  E_{3,10} &= \Et_{3,3,4} + \Et_{3,4,3},
\end{align}
where $\Dt_{2,\beta}$, $\Dt_{3,\gamma}$, $\Et_{1,\alpha,\gamma}$, $\Et_{6,\alpha,\beta}$, $\Et_{3,\beta\gamma}$ are the divisors on the cover. From the local linear relations~(\ref{lineqtwofour}) and~(\ref{lineqtwotwo}) we arrive at the following global relations:
\begin{eqnarray}
  \label{globalreltwofour}
  R_1&\sim&2\,D_{1,\alpha}+\sum_{\gamma=1}^3 E_{1,\alpha\gamma}+\sum_{\beta,\gamma=1,2} E_{5,\alpha\beta\gamma}+\sum_{\beta=1}^3 E_{6,\alpha\beta},\quad \alpha=1,..,4,\cr
  R_2&\sim&4\,D_{2,\beta}+\sum_{\gamma=1,2}[\,E_{2,\beta\gamma}+2\,E_{3,\beta\gamma}+3\,E_{4,\beta\gamma}]+\sum_{\alpha=1}^4\sum_{\gamma=1,2}E_{5,\alpha\beta\gamma}\cr
     &    &+2\sum_{\alpha=1}^4E_{6,\alpha\beta}+2\,E_{3,\mu},\ \ \beta=1,2,\cr
  R_2&\sim&2\,D_{2,3}+\sum_{\alpha=1}^4E_{6,\alpha3}+\sum_{\mu=7}^{10} E_{3,\mu},\cr
  R_3&\sim&4\,D_{3,\gamma}+2\sum_{\alpha=1}^4E_{1,\alpha\gamma}+\sum_{\beta=1,2}[\,3\,E_{2,\beta\gamma}+2\,E_{3,\beta\gamma}+E_{4,\beta\gamma}]+\sum_{\alpha=1}^4\sum_{\beta=1,2}E_{5,\alpha\beta\gamma}\cr
     &    &+2\,E_{3,\mu},\quad \gamma=1,2,\cr
  R_3&\sim&2\,D_{3,3}+\sum_{\alpha=1}^4E_{1,\alpha3}+\sum_{\mu=3,6,9,10} E_{3,\mu},
\end{eqnarray}
where we set in the second line $\mu=3,6$ for $\beta=1,2$, respectively, and in the fourth line $\mu=7,8$ for $\gamma=1,2$, respectively. Furthermore for compactness of the display in these two lines, we have written $E_{3,\mu}$, $\mu=1,\dots,4$ instead of $E_{3,\beta\gamma}$, $\beta,\gamma=1,2$ according to their origin on the cover as in~(\ref{eq:Z2xZ4invdivs}).
\begin{figure}[h!]
\begin{center}
\includegraphics[width=50mm]{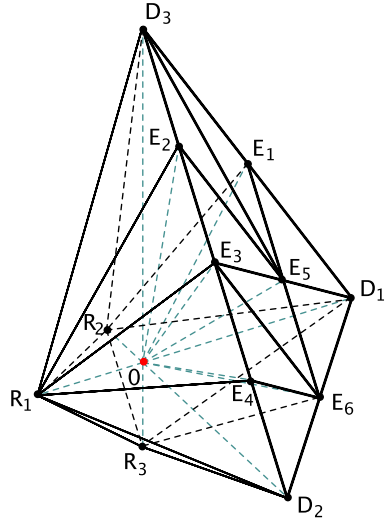}
\caption{The polyhedron $\Delta^{(3)}$ describing the local compactification of the resolution of $\IC^3/\IZ_2\times\IZ_4$.}
\label{fig:Z2Z4-cpt}
\end{center}
\end{figure}
We obtain the following nonvanishing intersection numbers of $X$ in the basis $\{R_i,E_{k\alpha\beta\gamma}\}$:
\begin{align}
  R_1R_2R_3 &=4, & R_1E_{2\beta\gamma}^2 &=-2, & R_1E_{2\beta\gamma}E_{3\mu} &=1, & R_1E_{3\mu'}^2 &=-2, \notag\\
  R_1E_{3\mu}E_{4\beta\gamma} &= 1, & R_1E_{4\beta\gamma}^2 &=-2, & R_2E_{1\alpha\gamma}^2 &=-2, & R_2E_{1\alpha3}^2 &= -4, \notag\\
  R_3E_{6\alpha\beta}^2 &=-2, & R_3E_{6\alpha3}^2 &= -4, & E_{1\alpha\gamma}^3 &=3, & E_{1\alpha3}^3 &=4, \notag\\
  E_{1\alpha\gamma'}^2E_{3\mu'} &= -1, & E_{1\alpha\gamma}^2E_{6\alpha3} &= -1, & E_{1\alpha3}^2E_{6\alpha\beta} &= -1, & E_{1\alpha3}^2E_{6\alpha3} &= -2, \notag\\ 
  E_{1\alpha\gamma}E_{2\beta\gamma}^2 &=-2, & E_{1\alpha\gamma}E_{2\beta\gamma}E_{3\mu} &= 1, & E_{1\alpha\gamma'}E_{3\mu'}^2 &= -1, & E_{1\alpha\gamma}E_{3\mu}E_{5\alpha\beta\gamma} &= 1, \notag\\
  E_{1\alpha3}E_{3\mu_1}E_{6\alpha\beta} &=1, & E_{1\alpha\gamma}E_{3\mu_2}E_{6\alpha3} &=1, & E_{1\alpha3}E_{3\mu_3}E_{6\alpha3} &=1, & E_{1\alpha\gamma}E_{5\alpha\beta\gamma}^2 &= -2, \notag\\
  E_{1\alpha3}E_{6\alpha\beta}^2 &=-1, & E_{1\alpha\gamma}E_{6\alpha3}^2 &=-1, & E_{1\alpha3}E_{6\alpha3}^2 &=-2, & E_{2\beta\gamma}^3 &= 8, \notag\\
  E_{2\beta\gamma}^2E_{3\mu} &= -4, & E_{2\beta\gamma}E_{3\mu}^2 &= 2, & E_{3\mu''}^3 &= 4, & E_{3\mu}^2E_{4\beta\gamma} &=2, \notag\\
  E_{3\mu'}^2E_{6\alpha\beta'} &= -1, & E_{3\mu}E_{4\beta\gamma}^2 &= -4, & E_{3\mu}E_{4\beta\gamma}E_{6\alpha\beta} &= 1, & E_{3\mu}E_{5\alpha\beta\gamma}^2 &=-2, \notag\\
  E_{3\mu}E_{5\alpha\beta\gamma}E_{6\alpha\beta} &=1, & E_{3\mu'}E_{6\alpha\beta'}^2 &= -1, & E_{4\beta\gamma}^3 &= 8, & E_{4\beta\gamma}^2E_{6\alpha\beta} &= -2, \notag\\
  E_{5\alpha\beta\gamma}^3 &= 8, & E_{5\alpha\beta\gamma}^2E_{6\alpha\beta} & =-2, & E_{6\alpha\beta}^3 &=3, & E_{6\alpha3}^3 &= 4,
\end{align}
where $\alpha=1,\dots,4$, $\beta,\gamma=1,2$, $\beta',\gamma'=1,\dots,3$, $\mu=1,2,4,5$, $\mu'=1,\dots,10$, $\mu''=3,6,\dots,10$, $\mu_1 = 3,6$, $\mu_2=7,8$, and $\mu_3=9,10$. The intersection numbers involving $\beta,\beta',\gamma,\gamma',\mu$, and $\mu'$ only have the given value for appropriate values of the labels, otherwise they vanish. As an example, $R_1E_{2\beta\gamma}E_{3\mu} =1$ has to be understood as $R_1E_{2,1,1}E_{3,1} = 1, R_1E_{2,1,2}E_{3,2} =1, R_1E_{2,2,1}E_{3,4} =1, R_1E_{2,2,2}E_{3,5} = 1$. Which values of $\mu$ fit to the values of $\beta,\gamma$ can be determined by looking at the first summand in the definition of $E_{3,\mu}$ in~(\ref{eq:Z2xZ4invdivs}). 

\subsection{Divisor topologies}

The topology of the exceptional divisors $E_{5,\alpha\beta\gamma}$ was determined in Appendix~\ref{sec:localZ2xZ4} to be an $\IF_1$. According to Section~\ref{sec:Topology} the remaining divisors are of type E\ref{item:E3}). Since the there is only one line ending at $E_{2\beta\gamma}$ and $E_{4,\beta\gamma}$ in the toric diagram in Figure~\ref{frtwofour}, both of them have the topology of an $\IF_0$. The divisors $E_{1,\alpha\gamma}$ and $E_{6,\alpha\beta}$ have two $\IC^3/\IZ_2\times\IZ_4$ and one $\IC^3/\IZ_2\times\IZ_2$ fixed points lying on them, with three and two lines ending at them in the toric diagram in the respective Figures~\ref{frtwofour} and~\ref{frtwotwo}. Hence, their topology is $\Bl{5}\IF_0$. The divisors $E_{3\mu}$, $\mu=1,2,4,5$ have four $\IC^3/\IZ_2\times\IZ_4$ fixed points lying on them, with three lines ending at them in the corresponding toric diagram. Therefore their topology is that of $\Bl{8}\IF_0$. Similarly, the divisors $E_{1,\alpha3}$, $E_{3\mu}$, $\mu=3,6,\dots,10$, and $E_{6,\alpha3}$ have four $\IC^3/\IZ_2\times\IZ_2$ patches lying on them, with two lines ending at them in the toric diagram. Therefore their topology is that of $\Bl{4}\IF_0$. The topology of the divisors $D_{i\alpha}$ is obtained from that of $T^2\setminus \{\textrm{4 pts} \} \times T^2\setminus \{\textrm{4 pts} \}$. Since in Figure~\ref{frtwofour} there is one line ending at $D_1$ and none at $D_2$ and $D_3$, the topology of $D_{1\alpha}$ is that of $\Bl{4}\IF_0$, while $D_{2\beta'}$ and $D_{3\gamma'}$ have the topology of $\IF_0$. Finally, the $R_i$ are K3 surfaces.

Finally, the second Chern class is
\begin{align}
  \ch_2\cdot E_{1,\alpha\gamma} &= 6, & \ch_2\cdot E_{1,\alpha3} &= 4, & \ch_2\cdot E_{2,\beta\gamma} &= -4, & \ch_2\cdot E_{3\mu} &= 12, \notag\\
  \ch_2\cdot E_{3\mu"} &= 4 & \ch_2\cdot E_{4,\beta\gamma} &= -4, & \ch_2\cdot E_{5,\alpha\beta\gamma} &= -4, & \ch_2\cdot E_{6\alpha\beta} &= 6, \notag\\
  \ch_2\cdot E_{6,\alpha3} &=4, & \ch_2 \cdot R_i &= 24,
\end{align}
for $\mu=1,2,4,5$ and $\mu"=3,6,\dots,10$.

\subsection{The orientifold}

At the orbifold point, we get the following configuration of O--planes:
The fixed points under $I_6$ are 64 O3--planes which fall into 40 conjugacy classes under the orbifold group. From the combination $I_6\,\theta^1$ arise four O7--planes in the $(z^1,z^3)$--plane. They fall into the conjugacy classes $z^2=0,\,z^2=1/2$ and $z^2=1/2\,\tau,\ 1/2\,(1+\tau)$. Under $I_6\,(\theta^2)^2$ are four O7--planes fixed in the $(z^2, z^3)$--plane, which are all in separate conjugacy classes. Under $I_6\,\theta^1(\theta^2)^3$, another four O7--planes arise, this time in the $(z^1,z^2)$--plane in the three conjugacy classes $z^3=0,\,z^3=1/2$ and $z^3=1/2\,\tau,\ 1/2\,(1+\tau)$.

For this example, $h^{(1,1)}_{-}=0$ since all the fixed points under the orbifold group lie on $\IZ_2$ fixed points and are therefore invariant under $I_6$.

After the blow--up, we have to deal with two different patches, the $\IC^3/\IZ_2\times\IZ_2$--patch and the $\IC^3/\IZ_2\times\IZ_4$--patch. For both cases, we choose the simplest possibility for $\cal I$, namely sending $z^i\to -z^i$ while leaving the $y^i$ unchanged.
The fixed sets under the combination of the scaling action of $\IC^3/\IZ_2\times\IZ_4$ (\ref{rescalestwofour}) and $\cal I$ are
\begin{itemize}
\item $z^1=0,\ \lambda_1=1,\,\lambda_2=\lambda_3=-1,\,\lambda_4=\lambda_5=\lambda_6=1,$
\item $z^2=0,\ \lambda_1=-1,\,\lambda_2=1,\,\lambda_3=-1,\ \lambda_4=\lambda_5=\lambda_6=1,$
\item $z^3=0,\ \lambda_1=\lambda_2=-1,\, \lambda_3=\lambda_4=\lambda_5=\lambda_6=1,$
\item $y^3=0,\ \lambda_1=\lambda_2=\lambda_3=\lambda_4=-1,\,\lambda_5=\lambda_6=1.$
\end{itemize}
The fixed sets under the combination of the scaling action of $\IC^3/\IZ_2\times\IZ_2$ (\ref{rescalestwofour}) and $\cal I$ are
\begin{itemize}
\item $z^1=0,\ \lambda_1=\lambda_2=-1,\,\lambda_3=1,$
\item $z^2=0,\ \lambda_1=1,\,\lambda_2=-1,\,\lambda_3=1,$
\item $z^3=0,\ \lambda_1=-1,\,\lambda_2=\lambda_3=1.$
\end{itemize}
These solutions correspond to O7--planes on the four $D_1$--planes, on the three equivalence classes of $D_2$ and $D_3$--planes and on the four exceptional divisors $E_{3,1},\,E_{3,2},\,E_{3,4}$ and $E_{3,5}$ which arise from the resolution of the four $\IZ_4$ fixed lines. This amounts to a total of 14 O7--planes. In the blown down limit, the O7--planes on the $E_3$ disappear and we recover the 10 O7--planes of the orbifold limit. There are no O3--plane solutions in the local patches, and since such a patch sits at every location of an O3--plane in the orbifolds limit, no O3--planes arise in the blown up case.

Intersection numbers involving divisors not fixed under the orientifold involution are halved:
\begin{align}
  R_1R_2R_3 &=2, & R_1E_{2\beta\gamma}^2 &=-1, & R_1E_{3\mu''}^2 &=-1, \notag\\
   R_1E_{4\beta\gamma}^2 &=-1, & R_2E_{1\alpha\gamma}^2 &=-1, & R_2E_{1\alpha3}^2 &= -2, \notag\\
  R_3E_{6\alpha\beta}^2 &=-1, & R_3E_{6\alpha3}^2 &= -2, & E_{1\alpha\gamma}^3 &=3/2, \notag \\
  E_{1\alpha3}^3 &=2, & E_{1\alpha\gamma'}^2E_{3\mu''} &= -1/2, & E_{1\alpha\gamma}^2E_{6\alpha3} &= -1/2,\notag \\
   E_{1\alpha3}^2E_{6\alpha\beta} &= -1/2, & E_{1\alpha3}^2E_{6\alpha3} &= -1, &
  E_{1\alpha\gamma}E_{2\beta\gamma}^2 &=-1, \notag \\
  E_{1\alpha\gamma'}E_{3\mu''}^2 &= -1/2, &
  E_{1\alpha3}E_{3\mu_1}E_{6\alpha\beta} &=1/2, & E_{1\alpha\gamma}E_{3\mu_2}E_{6\alpha3} &=1/2,\notag \\
   E_{1\alpha3}E_{3\mu_3}E_{6\alpha3} &=1/2, & E_{1\alpha\gamma}E_{5\alpha\beta\gamma}^2 &= -1, &
  E_{1\alpha3}E_{6\alpha\beta}^2 &=-1/2,\notag \\
   E_{1\alpha\gamma}E_{6\alpha3}^2 &=-1/2, & E_{1\alpha3}E_{6\alpha3}^2 &=-1, & E_{2\beta\gamma}^3 &= 4, \notag\\
  E_{3\mu''}^3 &= 2, &
  E_{3\mu''}^2E_{6\alpha\beta'} &= -1/2, &
   E_{3\mu''}E_{6\alpha\beta'}^2 &= -1/2, \notag \\
   E_{4\beta\gamma}^3 &= 4, & E_{4\beta\gamma}^2E_{6\alpha\beta} &= -1, &
  E_{5\alpha\beta\gamma}^3 &= 4, \notag \\
  E_{5\alpha\beta\gamma}^2E_{6\alpha\beta} & =-1, & E_{6\alpha\beta}^3 &=3/2, & E_{6\alpha3}^3 &= 2.
\end{align}
The intersection numbers involving the $E_{3,\mu}, \ \mu=1,2,4,5$, which are fixed are
\begin{align}
R_1E_{3\mu}E_{4\beta\gamma} &= 1, & R_1E_{2\beta\gamma}E_{3\mu} &=1, & E_{1\alpha\gamma'}^2E_{3\mu} &= -1,\notag \\
 E_{1\alpha\gamma}E_{2\beta\gamma}E_{3\mu} &= 1, & E_{1\alpha\gamma'}E_{3\mu}^2 &= -2,& E_{1\alpha\gamma}E_{3\mu}E_{5\alpha\beta\gamma} &= 1,\notag \\
  E_{2\beta\gamma}^2E_{3\mu} &= -4, & E_{2\beta\gamma}E_{3\mu}^2 &= 4, & E_{3\mu}^2E_{4\beta\gamma} &=4,\notag \\
  E_{3\mu}^2E_{6\alpha\beta'} &= -2,& E_{3\mu}E_{4\beta\gamma}^2 &= -4, & E_{3\mu}E_{4\beta\gamma}E_{6\alpha\beta} &= 1,\notag\\
  E_{3\mu}E_{5\alpha\beta\gamma}^2 &=-2, & E_{3\mu}E_{5\alpha\beta\gamma}E_{6\alpha\beta} &=1, & E_{3\mu}E_{6\alpha\beta'}^2 &= -1. 
\end{align}


\section{The $\IZ_2\times \IZ_6$--orbifold}

\subsection{Metric, complex structure and moduli}

The root lattice of $SU(2)^2\times SU(3)\times G_2$ is the one compatible to the point group.
The twists act on the lattice basis as follows:
\begin{eqnarray}
Q_1\, e_1&=&- e_1,\quad Q_1\,e_2=-e_2,\quad Q_1\,e_3=e_3,\quad Q_1\,e_4=e_4,\cr
Q_1\,e_5&=&-e_5,\quad Q_1\,e_6=-e_6,\cr
Q_2\, e_1&=& e_1,\quad Q_2\,e_2=e_2,\quad Q_2\,e_3=2\,e_3+3\,e_4,\quad Q_2\,e_4=-e_3-e_4,\cr
Q_2\,e_5&=&-e_6,\quad Q_2\,e_6=e_5+e_6.
\end{eqnarray}
The twists reproduce the correct eigenvalues and the conditions $Q_1^2=1,\ Q_2^6=1$. While the other twists are the usual Coxeter-twists, the $Q_2$-twist on $e_5,\,e_6$ is a generalized Coxeter--twist on $SU(3)$, namely $S_1P_{12}$. The combined twist $Q_3$ has the form
\begin{eqnarray}
Q_3\, e_1&=& -e_1,\quad Q_3\,e_2=-e_2,\cr
Q_3\,e_3&=&2\,e_3+3\,e_4,\quad Q_3\,e_4=-e_3-e_4,\cr
Q_3\,e_5&=&e_6,\quad Q_3\,e_6=-e_5-e_6,
\end{eqnarray}
and also reproduces the required eigenvalues. We require the metric to be invariant under all three twists, i.e. we impose the three conditions $Q_i^Tg\,Q_i=g,\quad i=1,2,3$. This leads to the following solution:
\begin{equation}{g=\left(\begin{array}{cccccc}
R_1^2&R_1R_2\,\cos\theta_{12}&0&0&0&0\cr
R_1R_2\,\cos\theta_{12}&R_1^2&0&0&0&0\cr
0&0&R_3^2&-\half R_3^2&0&0\cr
0&0&-\half R_3^2&{1\over3} R_3^2&0&0\cr
0&0&0&0&R_5^2&-\half R_5^2\cr
0&0&0&0&-\half R_5^2&R_5^2\end{array}\right).}\end{equation}
The solution for $b$ matches the pattern of (\ref{znzmb}), we therefore know to have three K\"ahler moduli whereas the complex structure is completely fixed. For the complex structure we get
\begin{equation}
z^1=\tfrac{1}{\sqrt{2\,{\rm Im}\,{\cal U}^3}}\,(x^1+{\cal U}^3\, x^2),\quad
z^2=x^3+\tfrac{1}{\sqrt3}e^{10\pi i/12}\,x^4,\quad
z^3=x^5+e^{2\pi i/3}\,x^6,\end{equation}
with ${\cal U}^3=R_2/R_1 e^{i\theta_{12}}$.
Examination of the K\"ahler form yields
\begin{equation}
\Tc^1=b_1+i\,R_1R_2\,\sin\theta_{12},\quad
\Tc^2=b_2+i\,\tfrac{1}{2\sqrt3}\,R_3^2,\quad
\Tc^3=b_3+i\,\tfrac{\sqrt3}{2}\,R_5^2.\end{equation}

\subsection{Fixed sets}

We need to examine the $\theta^1,\ \theta^2,\ ( \theta^2)^2,\  (\theta^2)^3,\ \theta^1\theta^2, \theta^1(\theta^2)^2$ and $\theta^1(\theta^2)^2$--twists. Table \ref{fstwosix} gives the particulars of the fixed sets.

The fixed torus associated to the $\theta^1$--twist is $(0,0,x^3,x^4,0,0)$ corresponding to $z^2$ being invariant; the torus that remains fixed under $\theta^2,\,(\theta^2)^2$ and $(\theta^2)^3$  is $(x^1,x^2,0,0,0,0)$, corresponding to $z^1$ being invariant; the torus that is fixed by $\theta^1(\theta^2)^3$  is $(0,0,0,0,x^5,x^6)$, corresponding to $z^3$ being invariant.

\begin{figure}[h!]
\begin{center}
\includegraphics[width=140mm]{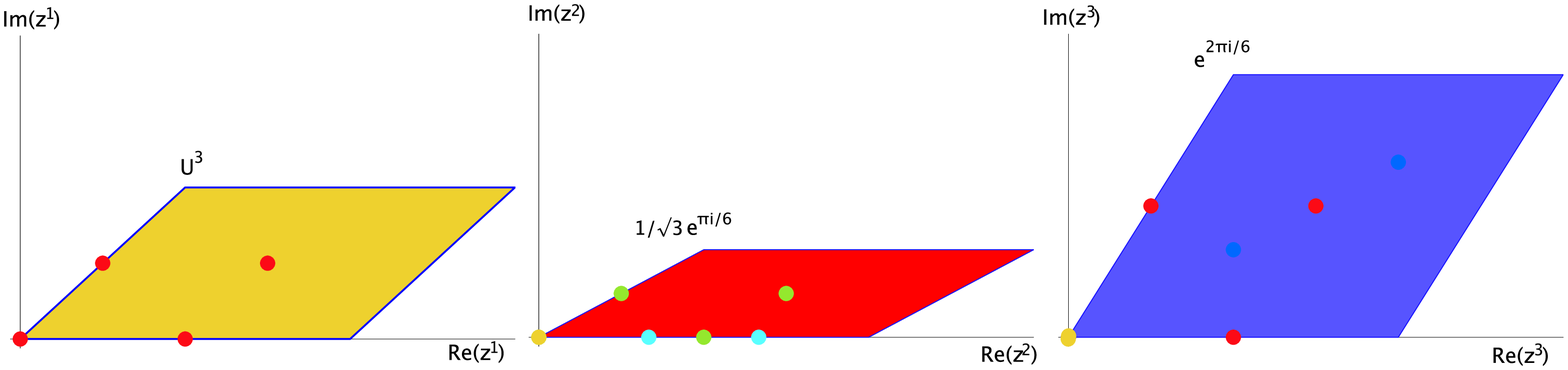}
\caption{Fundamental regions for the $\IZ_2\times\IZ_6$--orbifold}\label{ffixedii}
\end{center}
\end{figure}
Figure \ref{ffixedii} shows the fundamental regions of the three tori corresponding to $z^1,\,z^2,\,z^3$ and their fixed points. In many cases, fixed points under different group elements sit on the same spots, so it isn't possible to show them all in different colors.

\begin{table}[h!]\begin{center}
\begin{tabular}{|c|c|c|c|}
\hline
Group el.& Fixed Set&Order & Conj. Classes \cr
\hline
\noalign{\hrule}\noalign{\hrule}
$\theta^1  $&2   &16  fixed lines&8\cr
$\theta^2   $&6 &1  fixed line&1\cr
$(\theta^2)^2  $&3    &9  fixed lines& 4\cr
$(\theta^2)^3  $&2   &16  fixed lines& 6\cr
$ \theta^1\theta^2   $&$2\times6 $     &12  fixed points& 8\cr
$ \theta^1(\theta^2)^2   $&${2}\times3 $    &12  fixed points& 8\cr
$ \theta^1(\theta^2)^3  $&2    &16  fixed lines& 8\cr
\hline
\end{tabular}
\caption{Fixed point set for $\IZ_2\times\IZ_6$.}\label{fstwosix}
\end{center}\end{table}
\begin{figure}[h!]
\begin{center}
\includegraphics[width=85mm]{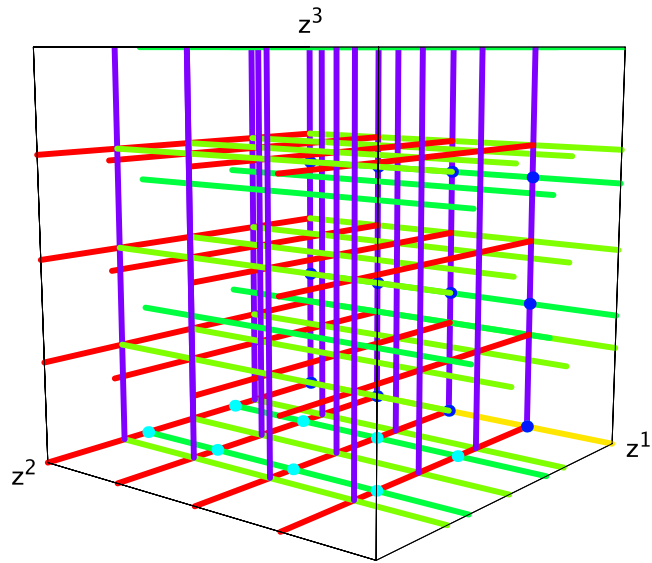}
\caption{Schematic picture of the fixed point configuration of the $\IZ_2\times\IZ_6$--orbifold}\label{fixedtwosix}
\end{center}
\end{figure}
Figure \ref{fixedtwosix} shows the configuration of the fixed sets in a schematic way, where each complex coordinate is shown as a coordinate axis and the opposite faces of the resulting cube of length 1 are identified. Note that the covering space and not the quotient is being shown, part of the fixed sets are identified by the group action.

The $\IZ_2 \times \IZ_6$ orbifold is one of the cases mentioned before, where we additionally have the fixed points at the intersections of the $\IZ_2$ fixed lines.

\subsection{The gluing procedure}

We will first look at the fixed lines.  According to Table \ref{fstwosix}, there are three sets of 8, 8 and 6 order two fixed lines, one oder six fixed line and 4 order three fixed lines. Special attention must be paid to the fixed line at $z^2=z^3=0$. In this locus, there are three coincident fixed lines (order six, order three and order two). Of these, we only count the order 6 fixed line (5 exceptional divisors). Therefore, we only count 3 of the four order three fixed lines (which contribute each 2 exceptional divisors), and one less of the corresponding $\IZ_2$--fixed lines. This gives in total $1\cdot5+3\cdot2+21\cdot1=32$ exceptional divisors from fixed lines.

Note that the 8 fixed points of $\theta^1\theta^2$ and the 8 of $\theta^1(\theta^2)^2$ share the four fixed points on the line $z^2=z^3=0$, so we are left with 12 individual fixed points (instead of the naive 16). The 4 fixed points one the line $z^2=z^3=0$ come from the $\IZ_2\times\IZ_6$-patch, which contributes 2 compact exceptional divisors. They sit at a triple intersection of one order six and 2 order two fixed lines, which is reflected by the exceptional divisors on the boundary of this patch.
The remaining 8 fixed points come from $\IZ_{6-II}$--patches, they sit at the intersections of one order two and one order three fixed lines and contribute 1 compact exceptional divisor. The four points on the line $z^2=z^3=0$ are fixed under two group elements, so we count them twice, while the remaining 8 points are fixed only under one group element, so we count them once. This yields the same result.
So we get $4\cdot2+8\cdot1=16$ divisors from fixed points, which makes together with the 32 exceptional divisors form the fixed lines a total of 48 exceptional divisors.

In this example, there is one $\IZ_3$ fixed line without fixed points on it, so $h^{(2,1)}_{tw}=2$.


\section{The $\IZ_2\times \IZ_{6'}$--orbifold}

\subsection{Metric, complex structure and moduli}

The root lattice of $SU(3)\times G_2^2$ is the one compatible to the point group.
The twists act on the lattice basis as follows:
\begin{eqnarray}
Q_1\, e_1&=&- e_1,\quad Q_1\,e_2=-e_2,\quad Q_1\,e_3=e_3,\quad Q_1\,e_4=e_4,\cr
Q_1\,e_5&=&-e_5,\quad Q_1\,e_6=-e_6,\cr
Q_2\, e_1&=& -e_2,\quad Q_2\,e_2=e_1+e_2,\quad Q_2\,e_3=2\,e_3+3\,e_4,\quad Q_2\,e_4=-e_3-e_4,\cr
Q_2\,e_5&=&-2\,e_5-3\,e_6,\quad Q_2\,e_6=e_5+e_6.
\end{eqnarray}
Here, the $Q_2$--twist on $e_5,\,e_6$ is minus the usual Coxeter--twist on $SU(3)$. The twists reproduce the correct eigenvalues and the conditions $Q_1^2=1,\ Q_2^6=1$. The combined twist $Q_3$ has the form
\begin{eqnarray}
Q_3\, e_1&=& e_2,\quad Q_3\,e_2=-e_1-e_2,\cr
Q_3\,e_3&=&2\,e_3+3\,e_4,\quad Q_3\,e_4=-e_3-e_4,\cr
Q_3\,e_5&=&2\,e_5+3\,e_6,\quad Q_3\,e_6=-e_5-e_6,
\end{eqnarray}
and also reproduces the required eigenvalues. We require the metric to be invariant under all three twists, i.e. we impose the three conditions $Q_i^Tg\,Q_i=g,\quad i=1,2,3$. This leads to the following solution:
\begin{equation}g=\left(\begin{array}{cccccc}
R_1^2&-\half R_1^2&0&0&0&0\cr
-\half R_1^2&R_1^2&0&0&0&0\cr
0&0&R_3^2&-\half R_3^2&0&0\cr
0&0&-\half R_3^2&{1\over3} R_3^2&0&0\cr
0&0&0&0&R_5^2&-\half R_5^2\cr
0&0&0&0&-\half R_5^2&{1\over3} R_5^2
\end{array}\right).\end{equation}
The solution for $b$ matches the pattern of (\ref{znzmb}),
we therefore know to have three K\"ahler moduli whereas the complex structure is completely fixed. For the complex structure we get
\begin{equation}
z^1=3^{1/4}\,(x^1+e^{2\pi i/3}\, x^2),\quad
z^2=x^3+\tfrac{1}{\sqrt3}e^{10\pi i/12}\,x^4,\quad
z^3=x^5+\tfrac{1}{\sqrt3}e^{10\pi i/12}\,x^6.\end{equation}
Examination of the K\"ahler form yields
\begin{equation}
\Tc^1=b_1+i\,\tfrac{\sqrt3}{2}\, R_1^2,\quad
\Tc^2=b_2+i\,\tfrac{1}{2\sqrt3}\,R_3^2,\quad
\Tc^3=b_3+i\,\tfrac{1}{2\sqrt3}\,R_5^2.\end{equation}

\subsection{Fixed sets}

We need to examine the $\theta^1,\ \theta^2,\ (\theta^2)^2,\ (\theta^2)^3,\ \theta^1\theta^2, \ \theta^1(\theta^2)^2$ and $\theta^1(\theta^2)^3$--twists. Table \ref{fstwosixp} gives the particulars of the fixed sets.

The fixed torus associated to the $\theta^1$--twist is $(0,0,x^3,x^4,0,0)$ corresponding to $z^2$ being invariant; the torus that remains fixed under $(\theta^2)^3$ is $(0,0,0,0,x^5,x^6)$, corresponding to $z^3$ being invariant; the torus that is fixed by $\theta^1(\theta^2)^3$  is $(x^1,x^2,0,0,0,0)$, corresponding to $z^1$ being invariant.

\begin{figure}[h!]
\begin{center}
\includegraphics[width=140mm]{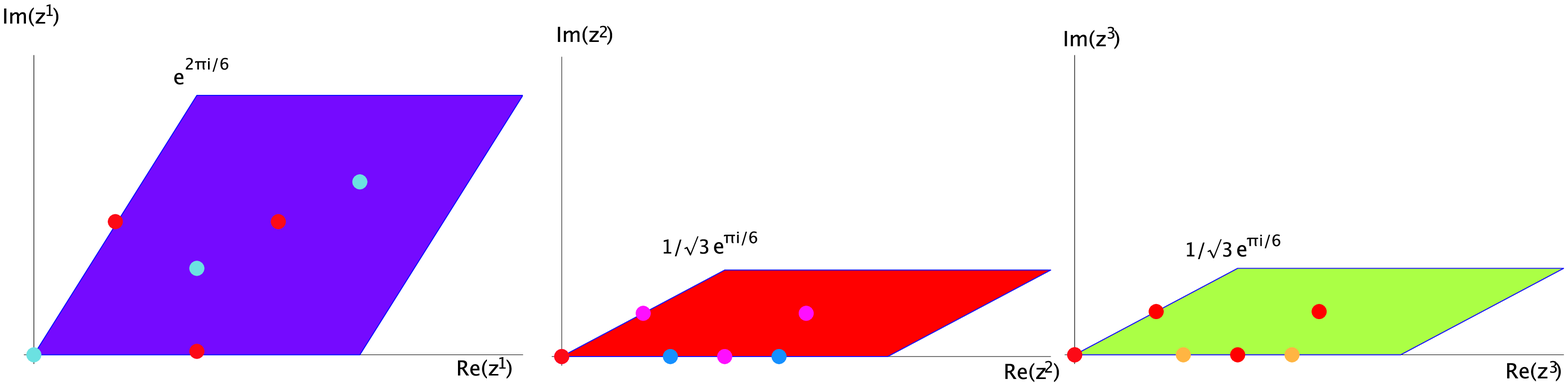}
\caption{Fundamental regions for the $\IZ_2\times\IZ_{6'}$--orbifold}\label{ffutwosixp}
\end{center}
\end{figure}
Figure \ref{ffutwosixp} shows the fundamental regions of the three tori corresponding to $z^1,\,z^2,\,z^3$ and their fixed points. 

\begin{table}[h!]\begin{center}
\begin{tabular}{|c|c|c|c|}
\hline
Group el.& Order &Fixed Set& Conj. Classes \cr
\hline
\noalign{\hrule}\noalign{\hrule}
$\theta^1  $&2       &16\ {\rm fixed\ lines} &\ 6\cr
$ \theta^2  $&6   &3\ {\rm fixed\ points} &\ 2\cr
$ (\theta^2)^2  $&3   &27 \ {\rm fixed\ points} &\ 9\cr
$ (\theta^2)^3  $&2  &16 \ {\rm fixed\ lines} &\ 6\cr
$ \theta^1\theta^2   $&${2}\times6 $    &3\ {\rm fixed\ points} &\ 2\cr
$ \theta^1(\theta^2)^2  $&${2}\times3$      &3\ {\rm fixed\ points} &\ 2\cr
$ \theta^1(\theta^2)^3  $&2   &16\ {\rm fixed\ lines} &\ 6\cr
\hline
\end{tabular}
\caption{Fixed point set for $\IZ_2\times \IZ_{6'}$.}\label{fstwosixp}
\end{center}\end{table}

\begin{figure}[h!]
\begin{center}
\includegraphics[width=85mm]{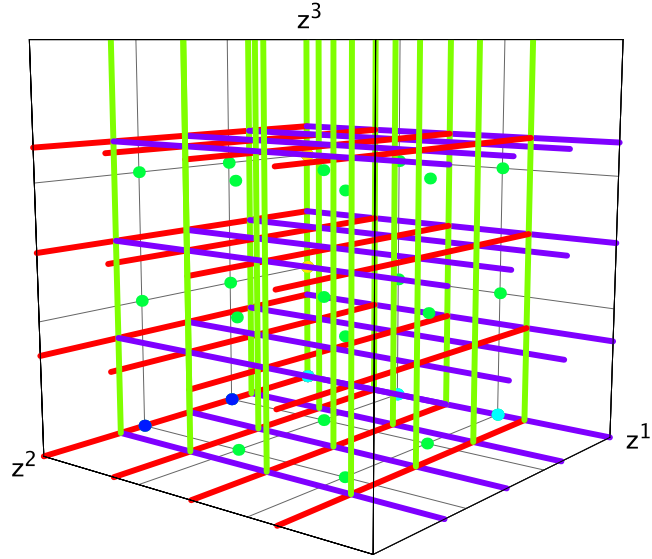}
\caption{Schematic picture of the fixed set configuration of $\IZ_2\times\IZ_{6'}$}\label{ffixtwosixp}
\end{center}
\end{figure}
Figure \ref{ffixtwosixp} shows the schematic fixed set configuration. Note that the covering space is shown, some of the fixed sets are identified under the orbifold group.

\subsection{The gluing procedure}

Here we have in each coordinate direction 6 order 2 fixed lines, which gives us a total of 18 exceptional divisors from fixed lines. The fixed point at $(0,0,0)$ is the only one that sits at a triple intersection of fixed lines, it is associated with the $\IZ_2\times\IZ_{6'}$--patch, which has 4 internal points. The other fixed point on $z^1=\,$fixed, $z^2=z^3=0$, on $z^2=\,$fixed, $z^1=z^3=0$ and $z^3=\,$fixed, $z^1=z^2=0$ come from $\IZ_{6-I}$--patches which each contribute 2 compact exceptional divisors. Of the remaining fixed points that do not lie on any fixed line, we  count 5; they each come from a $\IZ_3$--patch. In total, we get $1\cdot4+3\cdot2+5\cdot1=15$ exceptional divisors from fixed points, which makes together with the 18 from fixed line 33 in total.

Since in this example, there are no fixed lines without fixed points on them, $h^{(2,1)}_{tw}=0$.



\section{The $\IZ_3\times \IZ_3$--orbifold}

\subsection{Metric, complex structure and moduli}

The root lattice of $SU(3)\times SU(3)\times SU(3)$ is compatible with the point group.
The twists act on the lattice basis as follows:
\begin{eqnarray}
Q_1\, e_1&=& e_2,\quad Q_1\,e_2=-e_1-e_2,\quad Q_1\,e_3=e_3,\quad Q_1\,e_4=e_4,\cr
Q_1\,e_5&=&e_6,\quad Q_1\,e_6=-e_5-e_6,\cr
Q_2\, e_1&=& e_1,\quad Q_2\,e_2=e_2,\quad Q_2\,e_3=e_4,\quad Q_2\,e_4=-e_3-e_4,\cr
Q_2\,e_5&=&e_6,\quad Q_2\,e_6=-e_5-e_6.
\end{eqnarray}
The twists are the usual Coxeter--twists on $SU(3)$ and reproduce the correct eigenvalues and the condition $Q^3=1$. The combined twist $Q_3$ has the form
\begin{eqnarray}
Q_3\, e_1&=& e_2,\quad Q_3\,e_2=-e_1-e_2,\cr
Q_3\,e_3&=&e_4,\quad Q_3\,e_4=-e_3-e_4,\cr
Q_3\,e_5&=&-e_5-e_6,\quad Q_3\,e_6=e_5,
\end{eqnarray}
and also reproduces the required eigenvalues. The twist on $e_5,\,e_6$ is the anti-twist of the usual Coxeter--twist. We require the metric to be invariant under all three twists, i.e. we impose the three conditions $Q_i^Tg\,Q_i=g,\quad i=1,2,3$. This leads to the following solution:
\begin{equation}{g=\left(\begin{array}{cccccc}
R_1^2&-\half R_1^2&0&0&0&0\cr
-\half R_1^2&R_1^2&0&0&0&0\cr
0&0&R_3^2&-\half R_3^2&0&0\cr
0&0&-\half R_3^2&R_3^2&0&0\cr
0&0&0&0&R_5^2&-\half R_5^2\cr
0&0&0&0&-\half R_5^2&R_5^2\end{array}\right).}\end{equation}
This corresponds exactly to the metric of $SU(3)^3$ without any extra degrees of freedom.
The solution for $b$ matches the pattern of (\ref{znzmb}), we therefore know to have three K\"ahler moduli whereas the complex structure is completely fixed (recall that in the simple $Z_3$--twist, we had nine K\"ahler moduli). For the complex structure we get
\begin{equation}
z^1=3^{1/4}\,(x^1+e^{2\pi i/3}\,x^2),\quad
z^2=3^{1/4}\,(x^3+e^{2\pi i/3}\,x^4),\quad
z^3=3^{1/4}\,(x^5+e^{2\pi i/3}\,x^6).\end{equation}
Examination of the K\"ahler form yields
\begin{equation}
\Tc^1=b_1+i\,\tfrac{\sqrt3}{2}\,R_1^2,\quad
\Tc^2=b_2+i\,\tfrac{\sqrt3}{2}\,R_3^2,\quad
\Tc^3=b_3+i\,\tfrac{\sqrt3}{2}\,R_5^2.\end{equation}

\subsection{Fixed sets}

This is a combination of two prime orbifolds, therefore the conjugacy classes are in one-to-one correspondence with the fixed points. We need to examine the $\theta^1,\ \theta^2,\ \theta^1\theta^2$ and $\theta^1(\theta^2)^2$--twists ($\,(\theta^1)^2\theta^2$ is the anti-twist of $\theta^1(\theta^2)^2$). Table \ref{fsthreethree} gives the particulars of the fixed sets.

The fixed torus associated to the $\theta^1$--twist is $(0,0,x^3,x^4,0,0)$ corresponding to $z^2$ being invariant; the torus that remains fixed under $\theta^2$ is $(x^1,x^2,0,0,0,0)$, corresponding to $z^1$ being invariant; the torus that is fixed by $\theta^1(\theta^2)^2$  is $(0,0,0,0,x^5,x^6)$, corresponding to $z^3$ being invariant.

\begin{figure}[h!]
\begin{center}
\includegraphics[width=140mm]{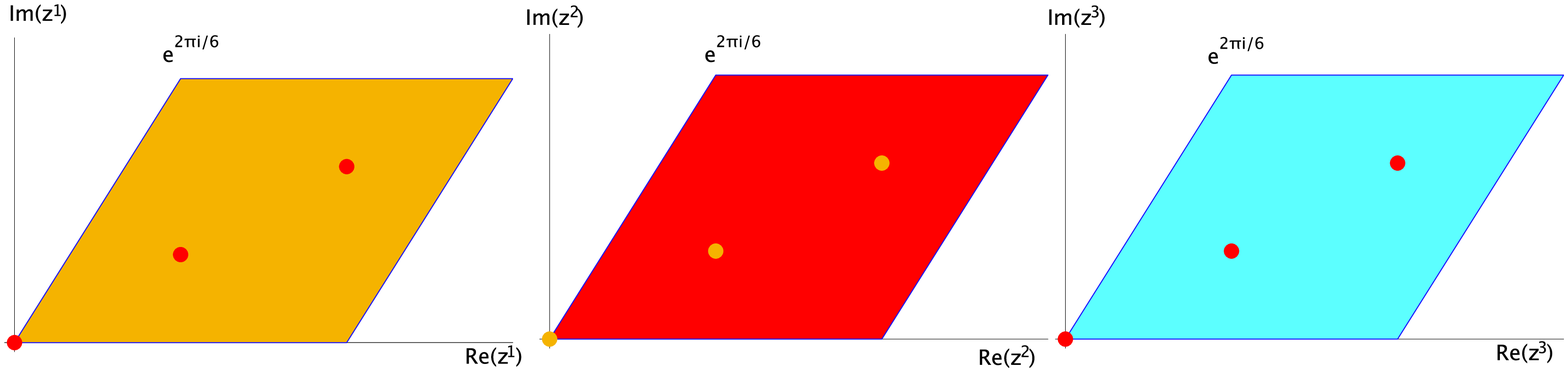}
\caption{Fundamental regions for the $\IZ_3\times\IZ_3$--orbifold}\label{ffuthreethree}
\end{center}
\end{figure}
Figure \ref{ffuthreethree} shows the fundamental regions of the three tori corresponding to $z^1,\,z^2,\,z^3$ and their fixed points. In each of them, we get  the usual three fixed points of the $\IZ_3$--twist, namely $z^1_{{\rm fixed},1}=z^2_{{\rm fixed},1}=z^3_{{\rm fixed},1}=0,\ z^1_{{\rm fixed},2}=z^2_{{\rm fixed},2}=z^3_{{\rm fixed},2}=1/\sqrt3 \,e^{\pi i/6}$ and $z^1_{{\rm fixed},3}=z^2_{{\rm fixed},3}=z^3_{{\rm fixed},3}=1+i/\sqrt3$.

\begin{table}[h!]\begin{center}
\begin{tabular}{|c|c|c|c|}
\hline
Group el.& Order & Fixed Set& Conj. Classes \cr
\hline
\noalign{\hrule}\noalign{\hrule}
$ \theta^1   $&3      &9  fixed lines& 9\cr
$\theta^2   $&3   &9  fixed lines& 9\cr
$ \theta^1\theta^2   $&${3}\times3  $    &27  fixed points&27\cr
$ \theta^1(\theta^2)^2   $&3      &9  fixed lines& 9\cr
\hline
\end{tabular}
\caption{Fixed point set for $\IZ_3\times\IZ_3$.}\label{fsthreethree}
\end{center}\end{table}

\begin{figure}[h!]
\begin{center}
\includegraphics[width=85mm]{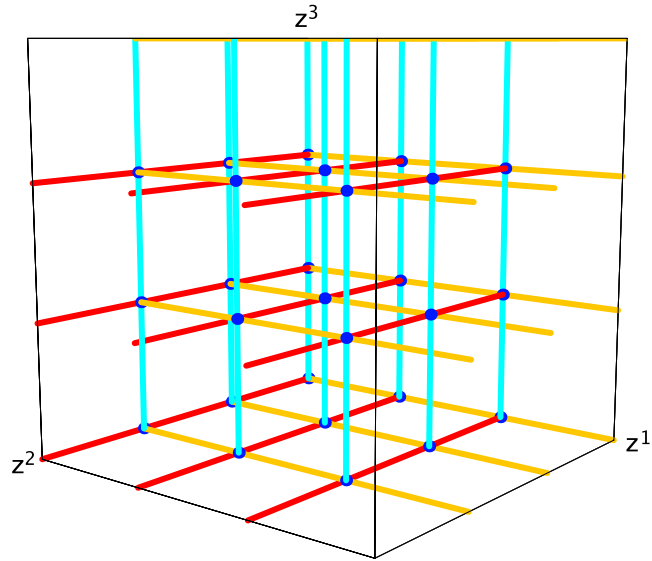}
\caption{Schematic picture of the fixed set configuration of $\IZ_3\times \IZ_3$}\label{fig:ffixthreethree}
\end{center}
\end{figure}
Figure \ref{fig:ffixthreethree} shows the configuration of the fixed sets in a schematic way, where each complex coordinate is shown as a coordinate axis and the opposite faces of the resulting cube of length 1 are identified.

\subsection{The gluing procedure}

Being a combination of prime orbifolds, this is again a relatively simple example.
The three different ${\IC}^2/\IZ_3$--patches each contribute 2 exceptional divisors. Of each kind of $\IZ_3$--fixed lines, we have 9. The $\IZ_3\times \IZ_3$--element has 27 isolated fixed points. The $\IZ_3\times \IZ_3$--patch has one compact exceptional divisor and two on each of the boundaries which get identified with those of the fixed lines, on whose intersections the fixed points sit. So we have $9\cdot2+9\cdot2+9\cdot2+27=81$ exceptional divisors. 

The compact exceptional divisors, we denote by $E_{5,\alpha\beta\gamma},\ \alpha,\beta,\gamma=1,2,3$. Of the non-compact exceptional divisors, we have nine each, $E_{1,2,\alpha \gamma},\ E_{3,4\beta\gamma},\, E_{6,7 \alpha\beta}$. 

In each of the coordinate planes we have 3 fixed planes with associated divisors $D_{i\alpha},\ i=1,2,3$. 

Since in this example, there are no fixed lines without fixed points on them, $h^{(2,1)}_{tw}=0$.

\subsection{The intersection ring}

\begin{figure}[h!]
\begin{center}
\includegraphics[width=140mm]{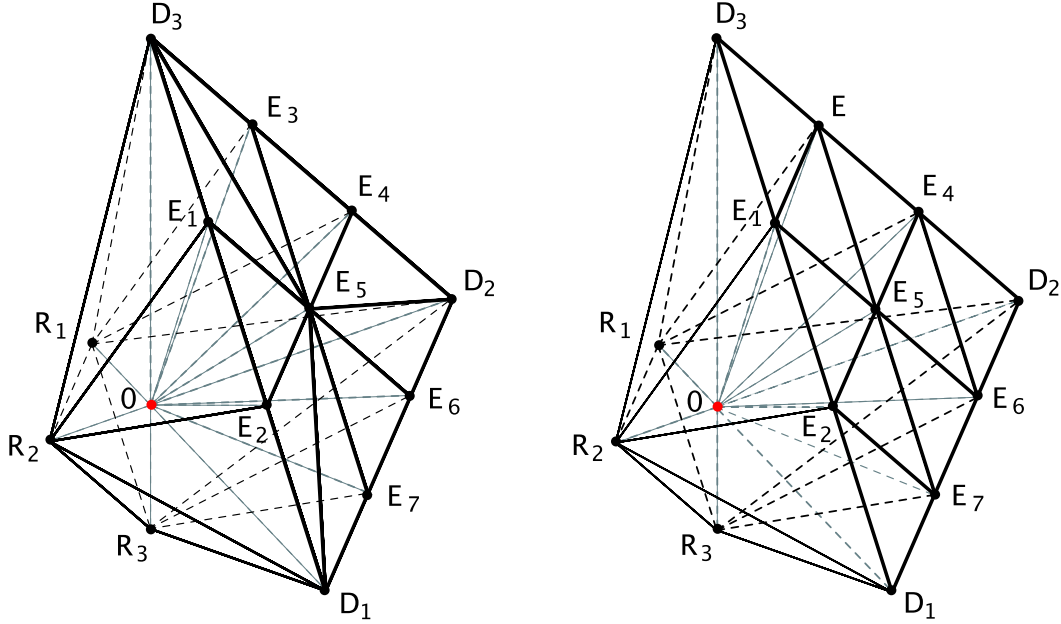}
\caption{The polyhedra $\Delta_1^{(3)}$ describing the local compactification of the resolution of $\IC^3/\IZ_{3}\times\IZ_3$.}
\label{fig:Z3Z3-cpt}
\end{center}
\end{figure}

We do the calculation for the triangulation a) in Figure \ref{frthreethree}.
From the local linear equivalences (\ref{lineqthreethree}) we arrive at the following global relations:
\begin{eqnarray}\label{globalrelthreethree}
R_1&=&3\,D_{{1,\alpha}}+\sum_{\beta=1}^3( E_{{6,\alpha\beta}}+2\,E_{{7,\alpha\beta}})+\sum_{\gamma=1}^3(E_{{1, \alpha\gamma}}+2\,E_{{2, \alpha\gamma}})+\sum_{\beta,\gamma=1}^3E_{{5,\alpha\beta\gamma}},\cr
R_2&=&3\,D_{{2,\beta}}+\sum_{\alpha=1}^3(2\,E_{{6,\alpha\beta}}+E_{{7,\alpha\beta}})+\sum_{\gamma=1}^3(2\,E_{{4,\beta\gamma}}+E_{{3,\beta\gamma}})+\sum_{\alpha,\gamma=1}^3 E_{{5,\alpha\beta\gamma}},\cr
R_3&=&3\,D_{{3,\gamma}}+\sum_{\alpha=1,}^3(2\,E_{{1,\alpha\gamma}}+E_{{2,\alpha\gamma}})+\sum_{\beta=1}^3(2\,E_{{3,\beta\gamma}}+E_{{4,\beta\gamma}})+\sum_{\alpha,\beta=1}^3 E_{{5,\alpha\beta\gamma}}.
\end{eqnarray}
The basis for the lattice $N$ from~(\ref{globalrelthreethree}) to be $f_1=(3,0,0)$, $f_2=(0,1,0)$, $f_3=(0,0,1)$. The lattice points of the polyhedron $\Delta^{(3)}$ for the local compactification of the $\IZ_3\times\IZ_3$ fixed points are
\begin{align}
  \label{eq:Z3Z3apoly}
  v_1 &= (-3,0,0), & v_2 &= (0,-1,0), & v_3 &= (0,0,-1), & v_4 &= (9,0,0), & v_5 &= (0,3,0), \notag\\
  v_6 &= (0,0,3), & v_7 &= (1,0,2), & v_8 &= (2,0,1), & v_9 &= (0,1,2), & v_{10} &= (0,2,1), \notag\\
  v_{11} &= (1,1,1), & v_{12} &= (1,2,0), & v_{13} &= (2,1,0), 
\end{align}
corresponding to the divisors $R_1,R_2,R_3,D_1,D_2,D_3,E_1,\dots,E_7$ in that order.  Figure~\ref{fig:Z3Z3-cpt} shows the polyhedron of the compactified $\IC^3/\IZ_{3}\times\IZ_3$--patch for both triangulations treated in Appendix \ref{app:rzthreethree}. 
From (\ref{eq:ReqGH}) we know that $R_{1}R_{2}R_{3}=3$. From the toric diagrams of the compactified patches, we can read off directly the intersection numbers with three distinct divisors:
\begin{eqnarray}
&&R_1E_{3,\beta\gamma}E_{4,\beta\gamma}=1,\ R_2E_{1,\alpha\gamma}E_{2,\alpha\gamma}=1,\ R_3E_{6,\alpha\beta}E_{7,\alpha\beta}=1,\cr 
&& E_{1,\alpha\gamma}E_{2,\alpha\gamma}E_{5,\alpha\beta\gamma}=1,\ E_{5,\alpha\beta\gamma}E_{6,\alpha\beta}E_{7,\alpha\beta}=1,\ E_{3,\beta\gamma}E_{4,\beta\gamma}E_{5,\alpha\beta\gamma}=1.
\end{eqnarray}
We find the following intersection numbers with the inherited divisors:
\begin{eqnarray}
&&R_1E_{3,\beta\gamma}^2=-2,\ R_1E_{4,\beta\gamma}^2=-2,\ R_2E_{1,\alpha\gamma}^2=-2,\ R_2E_{2,\alpha\gamma}^2=-2,\cr
&&R_3E_{6,\alpha\beta}^2=-2,\ R_3E_{7,\alpha\beta}^2=-2.
\end{eqnarray}
For the other triple intersection, where not all divisors are distinct, we find
\begin{eqnarray}\label{tithreethree}
&&E_{1,\alpha\gamma}^2E_{5,\alpha\beta\gamma}=-2,\ E_{2,\alpha\gamma}^2E_{5,\alpha\beta\gamma}=-2,\ E_{3,\beta\gamma}^2E_{5,\alpha\beta\gamma}=-2,\cr
&&E_{4,\beta\gamma}^2E_{5,\alpha\beta\gamma}=-2,\ E_{5,\alpha\beta\gamma}E_{6,\alpha\beta}^2=-2,\ E_{5,\alpha\beta\gamma}E_{7,\alpha\beta}^2=-2,\cr
&&E_{1,\alpha\gamma}^2E_{2,\alpha\gamma}=-1,\ E_{1,\alpha\gamma}E_{2,\alpha\gamma}^2=-1,\ E_{3,\beta\gamma}^2E_{4,\beta\gamma}=-1,\cr
&&E_{3,\beta\gamma}E_{4,\beta\gamma}^2=-1,\ E_{6,\alpha\beta}^2E_{7,\alpha\beta}=-1,\ E_{6,\alpha\beta}E_{7,\alpha\beta}^2=-1,\cr 
&&E_{1,\alpha\gamma}^3=8,\ E_{2,\alpha\gamma}^3=8,\ E_{3,\beta\gamma}^3=8,\ E_{4,\beta\gamma}^3=8,\cr 
&&E_{5,\alpha\beta\gamma}^3=3,\ E_{6,\alpha\beta}^3=8,\ E_{7,\alpha\beta}^3=8.
\end{eqnarray}
The K\"ahlerform can be parameterized as
\begin{eqnarray}\label{Jthreethree}
J&=&\sum_{i=1}^3 r_{i}R_{i}-\sum_{\alpha,\beta=1}^3 (t_{6,\beta\gamma}E_{6,\alpha\beta}+t_{7,\beta\gamma}E_{7,\alpha\beta})-\sum_{\alpha,\gamma=1}^3 (t_{1,\alpha,\gamma}E_{1,\alpha\gamma}+t_{2,\alpha,\gamma}E_{2,\alpha\gamma})\cr
&&-\sum_{\beta,\gamma=1}^3 (t_{3,\beta\gamma}E_{3,\beta\gamma}+t_{4,\beta\gamma}E_{4,\beta\gamma})-\sum_{\alpha\beta,\gamma=1}^3t_{5,\alpha\beta\gamma}E_{5,\alpha\beta\gamma}.
\end{eqnarray}
With (\ref{tithreethree}) and (\ref{Jthreethree}), the total volume becomes
\begin{eqnarray}
V&=&3\,r_{1}r_{2}r_{3}+r_1\sum_{\beta\gamma} (t_{3,\beta\gamma}t_{4,\beta\gamma}-t_{3,\beta\gamma}^2-t_{4,\beta\gamma}^2)\cr
&&+r_2\sum_{\alpha\gamma} (t_{1,\alpha\gamma}t_{2,\alpha\gamma}-t_{1,\alpha\gamma}^2-t_{2,\alpha\gamma}^2)+r_3\sum_{\alpha\beta} (t_{6,\alpha\beta}t_{7,\alpha\beta}-t_{6,\alpha\beta}^2-t_{7,\alpha\beta}^2)\cr
&&+\sum_{\beta\gamma} \lf[\frac{1}{2}(t_{3,{\beta\gamma} }^2t_{4,\beta\gamma}+t_{3,\beta\gamma}t_{4,\beta\gamma}^2)-\frac{4}{3}(t_{3,\beta\gamma}^3+t_{4,\beta\gamma}^3)\ri]\cr
&&+\sum_{\alpha\gamma} \lf[\frac{1}{2}(t_{1,{\alpha\gamma} }^2t_{2,\alpha\gamma}+t_{1,\alpha\gamma}t_{2,\alpha\gamma}^2)-\frac{4}{3}(t_{1,\alpha\gamma}^3+t_{2,\alpha\gamma}^3)\ri]\cr
&&+\sum_{\alpha\beta} \lf[\frac{1}{2}(t_{6,{\alpha\beta} }^2t_{7,\alpha\beta}+t_{6,\alpha\beta}t_{7,\alpha\beta}^2)-\frac{4}{3}(t_{6,\alpha\beta}^3+t_{7,\alpha\beta}^3)\ri]\cr
&&+\sum_{\alpha\beta\gamma}t_{5,\alpha\beta\gamma}(-t_{1,\alpha\gamma}t_{2,\alpha\gamma}-t_{3,\beta\gamma}t_{4,\beta\gamma}-t_{6,\alpha\beta}t_{7,\alpha\beta}\cr
&&+t_{1,{\alpha\gamma} }^2+t_{2,\alpha\gamma}^2+t_{3,{\beta\gamma} }^2+t_{4,\beta\gamma}^2+t_{6,{\alpha\beta} }^2+t_{7,{\alpha\beta} }^2)-\frac{1}{2}t_{5,\alpha\beta\gamma}^3.
\end{eqnarray}

\subsection{Divisor topologies}

As we have seen in Appendix \ref{app:rzthreethree}, $E_5$ is birationally equivalent to an $\IF_0$, it is an $\IF_0$ with five blow--ups, as can also be seen from the triple self-intersection number $E_5^3=3=8-5$. 
After gluing, the other exceptional divisors all have the topology of $\IP^1\times\IP^1$, the number of blown up points depends on the triangulation. For triangulation a), there are no blow--ups, as is reflected in their triple self-intersections being equal to eight, for b) we have $\Bl{3}(\IP^1\times\IP^1)$, one blow--up for each time the fixed line hits a fixed plane.
The $D$s without the exceptional divisors have the topology of $T^2/\IZ_3 \setminus 3\ {\rm pts}\times T^2/\IZ_3\setminus 3\ {\rm pts}$, which corresponds to $\IP^1 \setminus 3\ {\rm pts}\times \IP^1\setminus 3\ {\rm pts}$. In triangulation b), the points are simply put back in after the blow--up, in a), we get $\Bl{9}(\IP^1\times \IP^1)$, which is again nicely reflected in the triple self-intersection numbers: $D_i^3=-1=8-9,\ i=1,2,3$.  Finally, the $R_i$ are K3 surfaces.

Finally, the second Chern class is
\begin{align}
  \ch_2\cdot E_{1,\alpha\gamma} &= -4, & \ch_2\cdot E_{2,\alpha\gamma} &= -4, & \ch_2\cdot E_{3,\beta\gamma} &= -4, & \ch_2\cdot E_{4,\beta\gamma} &= -4, \notag\\
  \ch_2\cdot E_{5\alpha\beta\gamma} &= 6 & \ch_2\cdot E_{6,\alpha\beta} &= -4, & \ch_2\cdot E_{7,\alpha\beta} &= -4, & \ch_2\cdot R_i &= 24.
\end{align}

\subsection{The orientifold}

At the orbifold point, there are 64 O3--planes which fall into 18 conjugacy classes under the orbifold action. Since $\IZ_3\times\IZ_3$ has no $\IZ_2$ subgroups, no O7--planes appear.

For this example, $h^{1,1}_{-}=37$. 13 of these divisors come from the $E_{5\alpha\beta\gamma}$, which except for the one located at $(0,0,0)$ fall into conjugacy classes of length two under $I_6$. The remaining 24 come from the two exceptional divisors of the $\IC^2/\IZ_3$ fixed lines, which except for those three which pass the origin also fall into equivalence classes of length two.

Now we discuss the orientifold for the resolved case. For the local involution, we choose the simplest possibility, i.e. $z^i\to -z^i$ while on the $y^i$ nothing happens. Looking for the fixed points of the combination of the involution and the scaling action of the resolved patch (\ref{rescalesthreethree}), we find 
$$y^5=0,\quad \lambda_1= \lambda_2= \lambda_5=-1,\  \lambda_3= \lambda_4= \lambda_6= \lambda_7=1,$$
i.e. an O7--plane wrapped on $E_5$. In total, we arrive at 14 O7--planes wrapped on the remaining linear combinations of the $E_{5\alpha\beta\gamma}$s in $H^{1,1}_{+}$.
Only the resolved patch at $(0,0,0)$ coincides with an O3--plane location at the orbifold point. This O3--plane is not present in the resolved case. We are therefore left with 17 equivalence classes of O3--planes which are located away from the resolved patches. In this case, the O--plane configuration of the orbifold point is reproduced in the blown--down limit.

In the global relations (\ref{globalrelthreethree}), the coefficient of $E_5$ is changed to $1/2$ and the intersection numbers change as follows:

\begin{align}
  R_{1}R_{2}R_{3}&=\frac{3}{2}, & R_1E_{3,\beta\gamma}^2&=-1, & R_1E_{3,\beta\gamma}E_{4,\beta\gamma}&=\frac{1}{2}, \notag\\
  R_1E_{4,\beta\gamma}^2&=-1, & R_2E_{1,\alpha\gamma}^2&=-1, & R_2E_{1,\alpha\gamma}E_{2,\alpha\gamma}&=\frac{1}{2}, \notag\\
  R_2E_{2,\alpha\gamma}^2&=-1, & R_3E_{6,\alpha\beta}^2&=-1, & R_3E_{6,\alpha\beta}E_{7,\alpha\beta} &=\frac{1}{2}, &\notag\\
  R_3E_{7,\alpha\beta}^2&=-1, & E_{1,\alpha\gamma}^3&=4, & E_{1,\alpha\gamma}^2E_{2,\alpha\gamma}&=-\frac{1}{2},\notag\\
  E_{1,\alpha\gamma}E_{2,\alpha\gamma}^2&=-\frac{1}{2}, & E_{1,\alpha\gamma}E_{2,\alpha\gamma}E_{5,\alpha\beta\gamma}&=1, & E_{1,\alpha\gamma}^2E_{5,\alpha\beta\gamma}&=-2, &\notag\\
  E_{2,\alpha\gamma}^3&=4, & E_{2,\alpha\gamma}^2E_{5,\alpha\beta\gamma}&=-2, & E_{3,\beta\gamma}^3&=4, \notag\\
  E_{3,\beta\gamma}^2E_{4,\beta\gamma}&=-\frac{1}{2}, & E_{3,\beta\gamma}E_{4,\beta\gamma}^2&=-\frac{1}{2}, & E_{3,\beta\gamma}E_{4,\beta\gamma}E_{5,\alpha\beta\gamma}&=1, &\notag\\
  E_{3,\beta\gamma}^2E_{5,\alpha\beta\gamma}&=-2, & E_{4,\beta\gamma}^3&=4, & E_{4,\beta\gamma}^2E_{5,\alpha\beta\gamma}&=-2, \notag\\
  E_{5,\alpha\beta\gamma}^3&=12, & E_{5,\alpha\beta\gamma}E_{6,\alpha\beta}^2&=-2, & E_{5,\alpha\beta\gamma}E_{6,\alpha\beta}E_{7,\alpha\beta}&=1, \notag\\ 
  E_{5,\alpha\beta\gamma}E_{7,\alpha\beta}^2&=-2, & E_{6,\alpha\beta}^3&=4, & E_{6,\alpha\beta}^2E_{7,\alpha\beta}&=-\frac{1}{2}, \notag\\
  E_{6,\alpha\beta}E_{7,\alpha\beta}^2&=-\frac{1}{2} & E_{7,\alpha\beta}^3&=4.
\end{align}


\section{The $\IZ_3\times \IZ_{6}$--orbifold}

\subsection{Metric, complex structure and moduli}

The root lattice of  $SU(2)^2\times SU(3)\times G_2$ is compatible with the point group.
The twists act on the lattice basis as follows:
\begin{eqnarray}\label{twistactionththlii}
Q_1\, e_1&=& e_2,\quad Q_1\,e_2=-e_1-e_2,\quad Q_1\,e_3=e_3,\quad Q_1\,e_4=e_4,\cr
Q_1\,e_5&=&e_5+3\,e_6,\quad Q_1\,e_6=-e_5-2\,e_6,\cr
Q_2\, e_1&=& e_1,\quad Q_2\,e_2=e_2,\quad Q_2\,e_3=2\,e_3+3\,e_4,\quad Q_2\,e_4=-e_3-e_4,\cr
Q_2\,e_5&=&2\,e_5+3\,e_6,\quad Q_2\,e_6=-e_5-e_6.
\end{eqnarray}
The twist on $e_5,\,e_6$ in $Q_1$ is minus the anti-twist of the usual Coxeter--twist on $G_2$, while all other twists are the usual $\IZ_2,\,\IZ_3$ and $\IZ_6$--twists on their respective lattices.
The twists reproduce the correct eigenvalues and the conditions $Q_1^3=1,\ Q_2^6=1$. 
The combined twist $Q_3$ has the form
\begin{eqnarray}\label{twicoii}
Q_3\, e_1&=& e_2,\quad Q_3\,e_2=-e_1-e_2,\cr
Q_3\,e_3&=&2\,e_3+3\,e_4,\quad Q_3\,e_4=-e_3-e_4,\cr
Q_3\,e_5&=&-e_5,\quad Q_3\,e_6=-e_6,
\end{eqnarray}
which is as just mentioned a $\IZ_{6-II}$--twist, namely the one on the lattice $SU(2)^2\times SU(3)\times G_2$. As before, we require the metric to be 
invariant under all three twists, i.e. we impose the three conditions $Q_i^Tg\,Q_i=g,\quad i=1,2,3$. 
This leads to the following solution:
\begin{equation}\label{gii}{g=\left(\begin{array}{cccccc}
R_1^2&-\half R_1^2&0&0&0&0\cr
-\half R_1^2&R_1^2&0&0&0&0\cr
0&0&R_3^2&-\half R_3^2&0&0\cr
0&0&-\half R_3^2&{1\over 3} R_3^2&0&0\cr
0&0&0&0&R_5^2&-\half R_5^2\cr
0&0&0&0&-\half R_5^2&{1\over 3} R_5^2\end{array}\right).}
\end{equation}
This corresponds exactly to the metric of $SU(3)\times (G_2)^2$ without any extra degrees of freedom.
The corresponding solution for $b$ has the form of (\ref{znzmb}).
We have three K\"ahler moduli while the complex structure is completely fixed. We find the following complex coordinates:
\begin{equation}\label{dzsii}
z^1=3^{1/4}\,(x^1+e^{2\pi i/3}\,x^2),\quad
z^2=x^3+\tfrac{1}{\sqrt3}\,e^{5\pi i/6}\,x^4 ,\quad
z^3=x^5+\tfrac{1}{\sqrt3}\,e^{5\pi i/6}\,x^6 .
\end{equation}
The invariant 2-forms in the real cohomology are simply $dx^1\wedge dx^2, \ dx^3\wedge dx^4$ and $dx^5\wedge dx^6$. 
Via the pairing $J+i\,B={\cal T}^i\,\om_i$ in the real cohomology, we find the 
three K\"ahler moduli
\begin{equation}\label{ksii}
\Tc^1=b_1+i\,\tfrac{\sqrt3}{2}\,R_1^2 ,\quad
\Tc^2=b_2+i\,\tfrac{1}{2\sqrt3}\,R_3^2,\quad
\Tc^3=b_3+i\,\tfrac{1}{2\sqrt3}\,R_5^2.
\end{equation}

\subsection{Fixed sets}

Here, we need to examine the group elements $\theta^1,\,\theta^2,\,(\theta^2)^2,\,(\theta^2)^3,\,\theta^1(\theta^2)^2,\,\theta^1(\theta^2)^3,\,\theta^1(\theta^2)^4$ and $\theta^1(\theta^2)^5$.

The fixed torus associated to the $\theta^1$--twist is $(0,0,x^3,x^4,0,0)$ corresponding to $z^2$ being invariant; the torus that remains fixed under $,\theta^2,\,(\theta^2)^2,\,(\theta^2)^3$ is $(x^1,x^2,0,0,0,0)$, corresponding to $z^1$ being invariant; the torus that is fixed by $\theta^1(\theta^2)^4$  is $(0,0,0,0,x^5,x^6)$, corresponding to $z^3$ being invariant.

\begin{figure}[h!]
\begin{center}
\includegraphics[width=140mm]{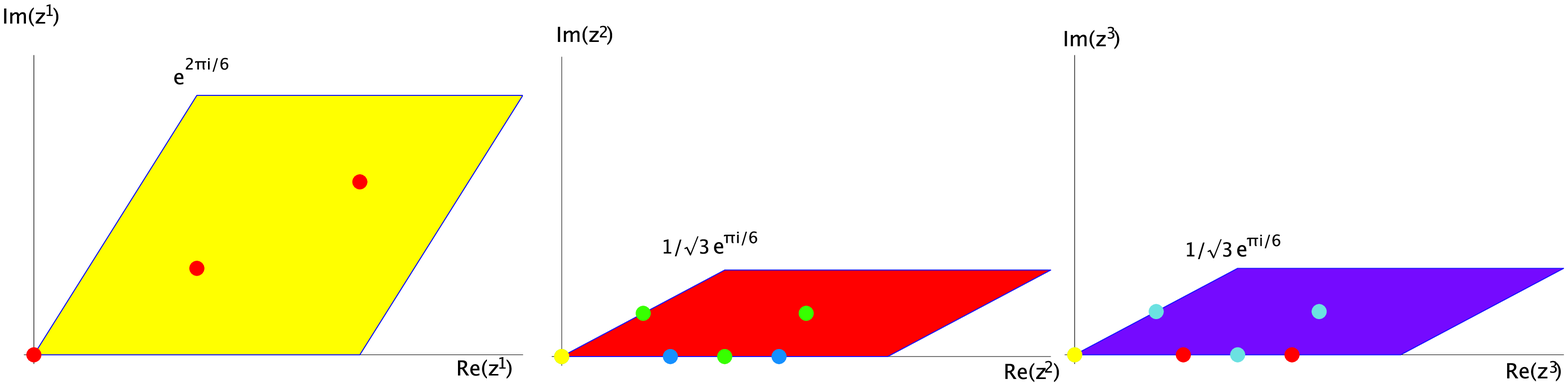}
\caption{Fundamental regions for the $\IZ_3\times\IZ_{6}$--orbifold}\label{ffuthreesix}
\end{center}
\end{figure}
Figure \ref{ffuthreesix} shows the fundamental regions of the three tori corresponding to $z^1,\,z^2,\,z^3$ and their fixed points.

\begin{table}[h!]\begin{center}
\begin{tabular}{|c|c|c|c|}
\hline
Group el.& Order & Fixed Set& Conj. Classes \cr
\hline
\noalign{\hrule}\noalign{\hrule}
$ \theta^1  $&3      &9\ {\rm fixed\ lines} &\ 6\cr
$ \theta^2   $&6   &1\ {\rm fixed\ line} &\ 1\cr
$ (\theta^2)^2  $&3     &9 \ {\rm fixed\ lines} &\ 5\cr
$ (\theta^2)^3  $&2 &16 \ {\rm fixed\ lines} &\ 4\cr
$ \theta^1\theta^2   $&$3\times6 $     &12\ {\rm fixed\ points} &\ 6\cr
$ \theta^1(\theta^2)^2  $&${3}\times3$    &27\ {\rm fixed\ points} &\ 15\cr
$ \theta^1(\theta^2)^3  $&${3}\times2$   &12\ {\rm fixed\ points} &\ 6\cr
$ \theta^1(\theta^2)^4   $&3   &9\ {\rm fixed\ lines} &\ 6\cr
$ \theta^1(\theta^2)^5   $&${3}\times6 $   &3\ {\rm fixed\ points} &\ 3\cr
\hline
\end{tabular}
\caption{Fixed point set for $\IZ_3\times \IZ_6$.}\label{fsthreesix}
\end{center}\end{table}

\begin{figure}[h!]
\begin{center}
\includegraphics[width=85mm]{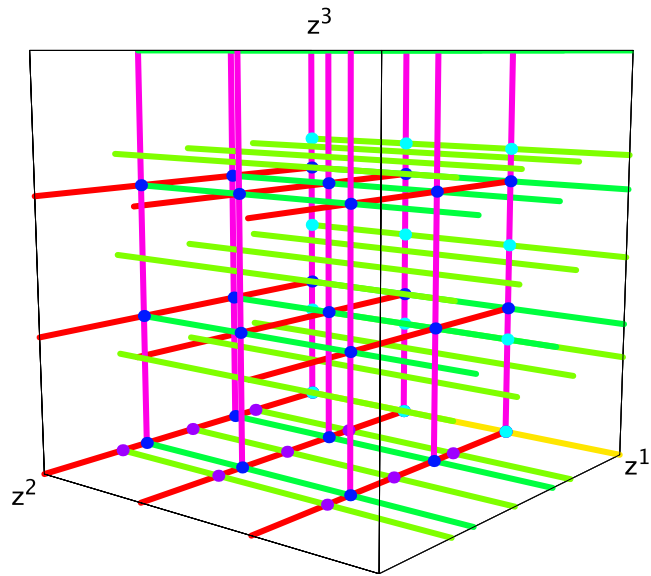}
\caption{Schematic picture of the fixed set configuration of the $\IZ_3\times\IZ_6$--orbifold}\label{ffixthreesix}
\end{center}
\end{figure}
Figure \ref{ffixthreesix} shows the schematic picture of the fixed set configuration. Note that the covering space is shown, some of the fixed sets are identified under the orbifold group.

\subsection{The gluing procedure}

From the fixed lines, we get the following number of exceptional divisors: $6\cdot2+1\cdot5+4\cdot2+3\cdot1+6\cdot2=40$. The fixed points at $z^1=\,$fixed, $z^2=z^3=0$ belong to the $\IZ_3\times\IZ_6$--patch and contribute 4 exceptional divisors each. The remaining 12 fixed points on triple intersections of fixed lines come from $\IZ_3\times\IZ_3$--patches; additionally, we have 6 fixed points on the intersection of two fixed lines which belong to $\IZ_{6-II}$--patches. So there are altogether $3\cdot4+12\cdot1+6\cdot1=30$ exceptional divisors from fixed points, which gives 70 in total.

In this example, there is one $\IZ_2$ fixed line without fixed points on it, so $h^{(2,1)}_{tw}=1$.


\section{The $\IZ_4\times \IZ_4$--orbifold}

\subsection{Metric, complex structure and moduli}

The root lattice of $SO(5)^3$ accommodates the combined twists.
The twists acting on the lattice basis are:
\begin{eqnarray}
Q_1\, e_1&=& e_1+2\,e_2,\quad Q_1\,e_2=-e_1-e_2,\quad Q_1\,e_3=e_3,\quad Q_1\,e_4=e_4,\cr
Q_1\,e_5&=&e_5+2\,e_6,\quad Q_1\,e_6=-e_5-e_6,\cr
Q_2\, e_1&=& e_1,\quad Q_2\,e_2=e_2,\quad Q_2\,e_3=e_3+2\,e_4,\quad Q_2\,e_4=-e_3-e_4,\cr
Q_2\,e_5&=&e_5+2\,e_6,\quad Q_2\,e_6=-e_5-e_6.
\end{eqnarray}
The twists are the usual Coxeter--twists on $SO(5)$ and reproduce the correct eigenvalues and the condition $Q^4=1$. The combined twist $Q_3$ has the form
\begin{eqnarray}
Q_3\, e_1&=& e_1+2\,e_2,\quad Q_3\,e_2=-e_1-e_2,\cr
Q_3\,e_3&=&e_3+2\,e_4,\quad Q_3\,e_4=-e_3-e_4,\cr
Q_3\,e_5&=&-e_5,\quad Q_3\,e_6=-e_6.
\end{eqnarray}
We require the metric to be invariant under all three twists, i.e. we impose the three conditions $Q_i^Tg\,Q_i=g,\quad i=1,2,3$. This leads to the following solution:
\begin{equation}{g=\left(\begin{array}{cccccc}
2\,R_1^2&-R_1^2&0&0&0&0\cr
- R_1^2&R_1^2&0&0&0&0\cr
0&0&2\,R_3^2&-R_3^2&0&0\cr
0&0&-R_3^2&R_3^2&0&0\cr
0&0&0&0&2\,R_5^2&-R_5^2\cr
0&0&0&0&-R_5^2&R_5^2\end{array}\right).}\end{equation}
This corresponds exactly to the metric of $SO(5)^3$ without any extra degrees of freedom.
The solution for $b$ matches the pattern of (\ref{znzmb}),
we therefore know to have three K\"ahler moduli whereas the complex structure is completely fixed.
For the complex structure we get
\begin{equation}
z^1=x^1-\tfrac{1}{2}(1-i)\,x^2,\quad z^2=x^3-\tfrac{1}{2}(1-i)\,x^4,\quad z^3=x^5-\tfrac{1}{2}(1-i)\,x^6.\end{equation}
Examination of the K\"ahler form yields
\begin{equation}
\Tc^1=b_1+i\,R_1^2,\quad \Tc^2=b_2+i\,R_3^2,\quad \Tc^3=b_3+i\,R_5^2.
\end{equation}

\subsection{Fixed sets}

We need to examine the $\theta^1,\ (\theta^1)^2,\ \theta^2,\ ( \theta^2)^2,\ \theta^1\theta^2,\ (\theta^1)^2\theta^2,\ \theta^1(\theta^2)^2,\ (\theta^1)^2(\theta^2)^2$ and $(\theta^1)^3(\theta^2)^2$--twists. Table 4 gives the particulars of the fixed sets.

The fixed torus associated to the $\theta^1$ and $(\theta^1)^2$--twists is $(0,0,x^3,x^4,0,0)$ corresponding to $z^2$ being invariant; the torus that remains fixed under $\theta^2$ and $(\theta^2)^2$   is $(x^1,x^2,0,0,0,0)$, corresponding to $z^1$ being invariant; the torus that is fixed by $(\theta^1)^2(\theta^2)^2$ and $(\theta^1)^3(\theta^2)^2$  is $(0,0,0,0,x^5,x^6)$, corresponding to $z^3$ being invariant.

\begin{figure}[h!]
\begin{center}
\includegraphics[width=140mm]{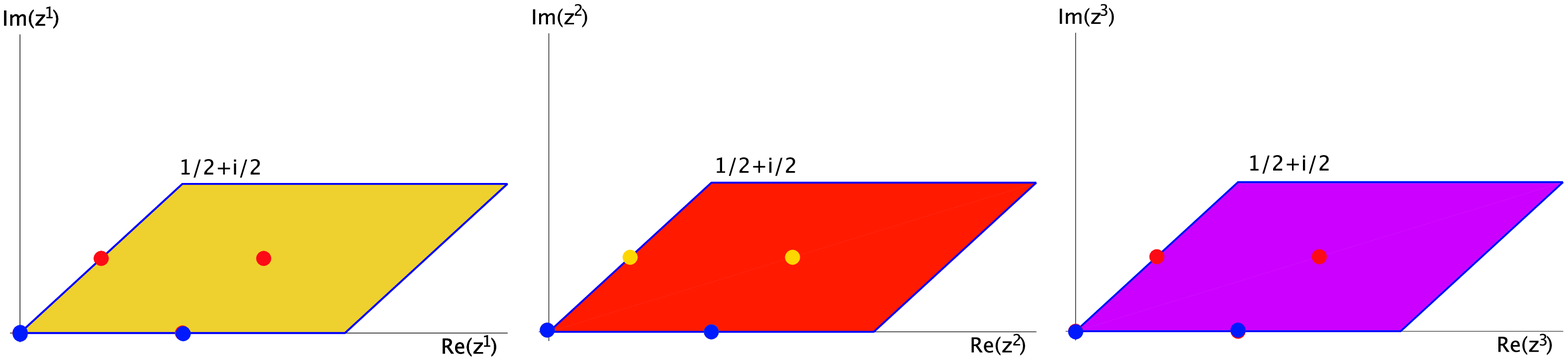}
\caption{Fundamental regions for the $\IZ_4\times\IZ_4$--orbifold}\label{fffourfour}
\end{center}
\end{figure}
Figure \ref{fffourfour} shows the fundamental regions of the three tori corresponding to $z^1,\,z^2,\,z^3$ and their fixed points. In many cases, fixed points under different group elements sit on the same spots, so it isn't possible to show them all in different colors.

\begin{table}[h!]\begin{center}
\begin{tabular}{|c|c|c|c|}
\hline
Group el.& Order & Fixed Set& Conj. Classes \cr
\hline
\noalign{\hrule}\noalign{\hrule}
$\theta^1  $&4   &4\ {\rm fixed\ lines} &\ 4\cr
$\theta^2   $&4 &4\ {\rm fixed\ line} &\ 4\cr
$ (\theta^1)^2  $&2     &16 \ {\rm fixed\ lines} &\ 9\cr
$ (\theta^2)^2  $&2   &16 \ {\rm fixed\ lines} &\ 9\cr
$ \theta^1\theta^2   $&${4}\times4 $    &16\ {\rm fixed\ points} &\ 12\cr
$ (\theta^1)^2\theta^2   $&$2\times4 $    &16\ {\rm fixed\ points} &\ 12\cr
$ \theta^1(\theta^2)^2   $&${2}\times4 $    &16\ {\rm fixed\ points} &\ 12\cr
$ (\theta^1)^2(\theta^2)^2  $&2     &16\ {\rm fixed\ lines} &\ 9\cr
$ (\theta^1)^3(\theta^2)^2  $&4     &4\ {\rm fixed\ lines} &\ 4\cr
\hline
\end{tabular}
\caption{Fixed point set for $\IZ_4\times \IZ_4$.}\label{fsfourfour}
\end{center}\end{table}
\begin{figure}[h!]
\begin{center}
\includegraphics[width=85mm]{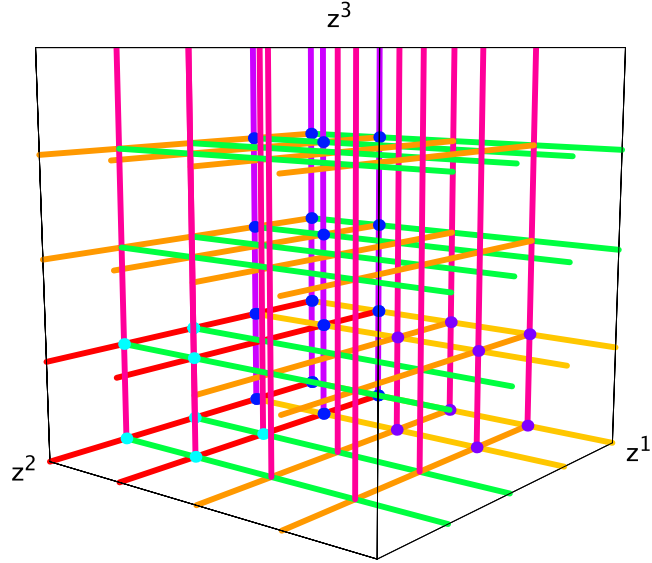}
\caption{Schematic picture of the fixed point configuration of the $\IZ_4\times\IZ_4$--orbifold}\label{fixedfourfour}
\end{center}
\end{figure}
Figure \ref{fixedfourfour} shows the configuration of the fixed sets in a schematic way, where each complex coordinate is shown as a coordinate axis and the opposite faces of the resulting cube of length 1 are identified. Note that the covering space and not the quotient is being shown, part of the fixed sets are identified by the group action.

\subsection{The gluing procedure}

At each of the eight fixed points that sit at the intersection of four $\IZ_4$ fixed lines are $\IZ_4\times \IZ_4$--patches. One such patch contributes three internal exceptional divisors, see Figure \ref{frfourfour}. The patches at fixed points which sit at the intersection of one $\IZ_4$ and two $\IZ_2$ fixed lines are of type $\IZ_2\times \IZ_4$. There are twelve of them and each contributes one internal exceptional divisor, see Figure \ref{frtwofour}. There are twelve $\IZ_4$ fixed lines, each contributing three exceptional divisors and 15 $\IZ_2$ fixed lines, each contributing one exceptional divisor. In total, this gives $8\cdot3+12\cdot1+12\cdot3+15\cdot1=87$ exceptional divisors.

Since in this example, there are no fixed lines without fixed points on them, $h^{(2,1)}_{tw}=0$.


\section{The $\IZ_6\times \IZ_{6}$--orbifold}

\subsection{Metric, complex structure and moduli}

The root lattice of $G_2\times G_2\times G_2$ is compatible with the point group.
The twists acts on the lattice basis as follows:
\begin{eqnarray}
Q_1\, e_1&=&2\,e_1+3\,e_2,\quad Q_1\,e_2=-e_1-e_2,\quad Q_1\,e_3=e_3,\quad Q_1\,e_4=e_4,\cr
Q_1\,e_5&=&2\,e_5+3\,e_6,\quad Q_1\,e_6=-e_5-e_6,\cr
Q_2\, e_1&=& e_1,\quad Q_2\,e_2=e_2,\quad Q_2\,e_3=2\,e_3+3\,e_4,\quad Q_2\,e_4=-e_3-e_4,\cr
Q_2\,e_5&=&2\,e_5+3\,e_6,\quad Q_2\,e_6=-e_5-e_6.
\end{eqnarray}
The twists reproduce the correct eigenvalues and the conditions $Q_1^6=1,\ Q_2^6=1$. The combined twist $Q_3$ has the form
\begin{eqnarray}
Q_3\, e_1&=&2\, e_1+3\,e_3,\quad Q_3\,e_2=-e_1-e_2,\cr
Q_3\,e_3&=&2\,e_3+3\,e_4,\quad Q_3\,e_4=-e_3-e_4,\cr
Q_3\,e_5&=&e_5+3\,e_6,\quad Q_3\,e_6=-e_5-2\,e_6,
\end{eqnarray}
 where the twist on $e_5,\,e_6$ is twice the Coxeter--twist on $G_2$. $Q_3$ also reproduces the required eigenvalues. We require the metric to be invariant under all three twists, i.e. we impose the three conditions $Q_i^Tg\,Q_i=g,\quad i=1,2,3$. This leads to the following solution:
\begin{equation}{g=\left(\begin{array}{cccccc}
R_1^2&-\half R_1^2&0&0&0&0\cr
-\half R_1^2&{1\over3}R_1^2&0&0&0&0\cr
0&0&R_3^2&-\half R_3^2&0&0\cr
0&0&-\half R_3^2&{1\over3} R_3^2&0&0\cr
0&0&0&0&R_5^2&-\half R_5^2\cr
0&0&0&0&-\half R_5^2&{1\over3} R_5^2\end{array}\right).}\end{equation}
The solution for $b$ matches the pattern of (\ref{znzmb}),
we therefore know to have three K\"ahler moduli whereas the complex structure is completely fixed. 
For the complex structure we get
\begin{eqnarray}
z^1&=&3^{1/4}\,(x^1+\tfrac{1}{\sqrt3}e^{10\pi i/12}\,x^2),\quad
z^2=3^{1/4}\,(x^3+\tfrac{1}{\sqrt3}e^{10\pi i/12}\,x^4),\cr
z^3&=&3^{1/4}\,(x^5+\tfrac{1}{\sqrt3}e^{10\pi i/12}\,x^6).
\end{eqnarray}
Examination of the K\"ahler form yields
\begin{equation}
\Tc^1=b_1+i\,\tfrac{1}{2\sqrt3}\, R_1^2,\quad
\Tc^2=b_2+i\,\tfrac{1}{2\sqrt3}\,R_3^2,\quad
\Tc^3=b_3+i\,\tfrac{1}{2\sqrt3}\,R_5^2.
\end{equation}

\subsection{Fixed sets}\label{app:sixsix}

Here, we must take 19 group elements into account, namely $\theta^1,\,(\theta^1)^2,\,(\theta^1)^3$, $\theta^2,\,(\theta^2)^2,\,(\theta^2)^3$, $\theta^1\theta^2$, $\theta^1(\theta^2)^2,\,\theta^1(\theta^2)^3$, $\theta^1(\theta^2)^4,\,\theta^1(\theta^2)^5$, $(\theta^1)^2\theta^2,$ $(\theta^1)^3\theta^2$, $(\theta^1)^4\theta^2,\,(\theta^1)^2(\theta^2)^2$, $(\theta^1)^2(\theta^2)^3$, $(\theta^1)^2(\theta^2)^4$, $(\theta^1)^3(\theta^2)^2$ and $(\theta^1)^3(\theta^2)^3$.

The fixed torus associated to the $\theta^1,\,(\theta^1)^2,\,(\theta^1)^3$--twists is $(0,0,x^3,x^4,0,0)$ corresponding to $z^2$ being invariant; the torus that remains fixed under $,\theta^2,\,(\theta^2)^2,\,(\theta^2)^3$ is $(x^1,x^2,0,0,0,0)$, corresponding to $z^1$ being invariant; the torus that is fixed by $\theta^1(\theta^2)^5,\, (\theta^1)^2(\theta^2)^4,\, (\theta^1)^3(\theta^2)^3$  is $(0,0,0,0,x^5,x^6)$, corresponding to $z^3$ being invariant.

\begin{figure}[h!]
\begin{center}
\includegraphics[width=140mm]{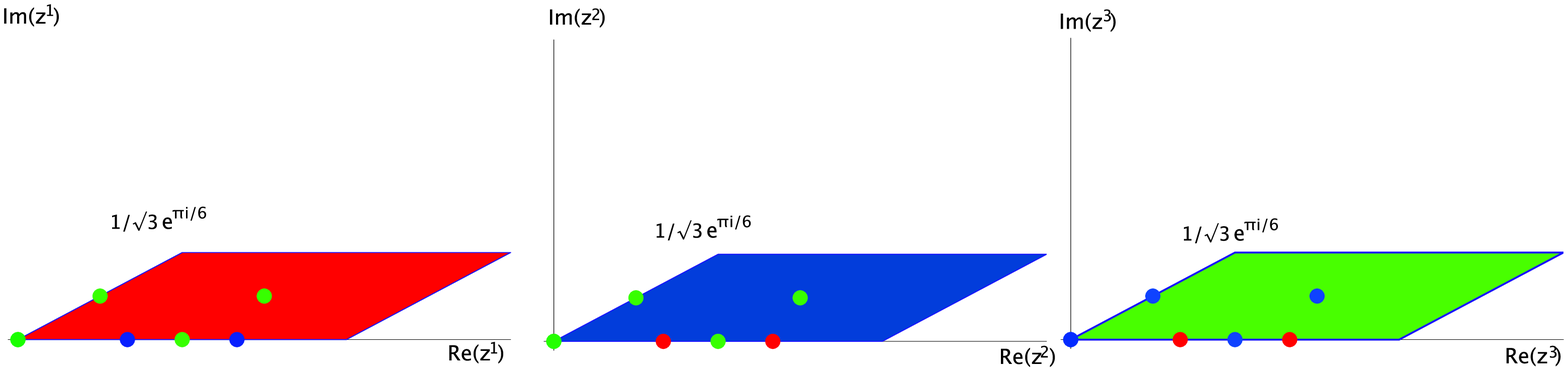}
\caption{Fundamental regions for the $\IZ_6\times\IZ_{6}$--orbifold}\label{ffusixsix}
\end{center}
\end{figure}
Figure \ref{ffusixsix} shows the fundamental regions of the three tori corresponding to $z^1,\,z^2,\,z^3$ and their fixed points. 

Table \ref{fssixsix} in Section \ref{sec:exsixsix} summarizes the data of the fixed sets.
Figure \ref{ffixsixsix} shows the schematic picture of the fixed set configuration. Again, it is the covering space that is shown.

\subsection{The gluing procedure}

See Section \ref{sec:exsixsix}.

Since in this example, there are no fixed lines without fixed points on them, $h^{(2,1)}_{tw}=0$.


\chapter{Cartan matrices of the relevant Lie groups}\label{AppC}

$A_n/SU(n+1)$
\begin{equation}\label{sun}{A=\left(\begin{array}{cccccccc}
2&-1&0&.&.&.&0&0\cr
-1&2&-1&.&.&.&0&0\cr
0&-1&2&-1&.&.&0&0\cr
.&.&.&.&.&.&.&.\cr
0&0&0&.&.&-1&2&-1\cr
0&0&0&.&.&0&-1&2\end{array}\right)}\end{equation}

\noindent$B_n/SO(2n+1)$
\begin{equation}\label{sonnone}{A=\left(\begin{array}{cccccccc}
2&-1&0&.&.&.&0&0\cr
-1&2&-1&.&.&.&0&0\cr
0&-1&2&-1&.&.&0&0\cr
.&.&.&.&.&.&.&.\cr
0&0&0&.&.&-1&2&-2\cr
0&0&0&.&.&0&-1&2\end{array}\right)}\end{equation}

\noindent$D_n/SO(2n)$
\begin{equation}\label{sonn}{A=\left(\begin{array}{cccccccc}
2&-1&0&.&.&.&0&0\cr
-1&2&-1&.&.&.&0&0\cr
0&-1&2&-1&.&.&0&0\cr
.&.&.&.&.&.&.&.\cr
0&0&0&.&.&2&-1&-1\cr
0&0&0&.&.&-1&2&0\cr
0&0&0&.&.&-1&0&2\end{array}\right)}\end{equation}

\noindent$E_6$
\begin{equation}\label{esix}{A=\left(\begin{array}{cccccc}
2&-1&0&0&0&0\cr
-1&2&-1&0&0&0\cr
0&-1&2&-1&0&-1\cr
0&0&-1&2&-1&0\cr
0&0&0&-1&2&0\cr
0&0&-1&0&0&2\end{array}\right)}\end{equation}

\noindent$F_4$
\begin{equation}\label{ff}{A=\left(\begin{array}{cccc}
2&-1&0&0\cr
-1&2&-2&0\cr
0&-1&2&-1\cr
0&0&-1&2\end{array}\right)}\end{equation}

\noindent$G_2$
\begin{equation}\label{gtwo}{A=\left(\begin{array}{cc}
2&-3\cr-1&2\end{array}\right)}\end{equation}


\end{document}